%% file: multiquark.tex
\documentclass[twoside,12pt]{article}
\usepackage{epsfig}
\usepackage{graphics}
\usepackage{bm}
\usepackage{cancel}
\usepackage{color}
\usepackage{graphicx}
\usepackage{dcolumn}
\usepackage{multirow}
\usepackage{amsmath}
\usepackage{slashed}
\usepackage{amsmath,amssymb}
\usepackage{array}                                                   
\usepackage{indentfirst}

\makeatletter
\newcommand{\thickhline}{\noalign {\ifnum 0=`}\fi \hrule height 1pt\futurelet \reserved@a \@xhline}
\newcolumntype{"}{@{\hskip\tabcolsep\vrule width 1pt\hskip\tabcolsep}}                             
\makeatother

\usepackage{graphicx} 

\usepackage[colorlinks,
            citecolor=blue,
            anchorcolor=red,
            menucolor=red,
            linkcolor=red,
            filecolor=red,
            runcolor=red,
            urlcolor=blue,
            frenchlinks=red]{hyperref}

\newcommand{\be}{\begin{equation}}
\newcommand{\ee}{\end{equation}}
\newcommand{\bea}{\begin{eqnarray}}
\newcommand{\eea}{\end{eqnarray}}

\begin{document}

\title{ \vspace{1cm} Pentaquark and Tetraquark states}

\author{Yan-Rui Liu,$^{1}$\footnotemark[1]~~~Hua-Xing Chen,$^2$\footnotemark[1]~~~Wei Chen,$^3$\footnotemark[1]~~~Xiang Liu,$^{4,5}$\footnotemark[2]~~~Shi-Lin Zhu$^{6,7,8}$\footnotemark[3]\\
\\
$^1$School of Physics, Shandong University, Jinan 250100, China \\
$^2$School of Physics, Beihang University, Beijing 100191, China \\
$^3$School of Physics, Sun Yat-Sen University, Guangzhou 510275, China \\
$^4$School of Physical Science and Technology, \\ Lanzhou University, Lanzhou 730000, China\\
$^5$Research Center for Hadron and CSR Physics, Lanzhou University and \\ Institute of Modern Physics of CAS, Lanzhou 730000, China\\
$^6$School of Physics and State Key Laboratory of Nuclear Physics and Technology, \\ Peking University, Beijing 100871, China \\
$^7$Collaborative Innovation Center of Quantum Matter, Beijing 100871, China \\
$^8$Center of High Energy Physics, Peking University, Beijing 100871, China}

\renewcommand{\thefootnote}{\fnsymbol{footnote}}
\footnotetext[1]{These authors equally contribute to this work.}
\footnotetext[2]{Corresponding author: xiangliu@lzu.edu.cn}
\footnotetext[3]{Corresponding author: zhusl@pku.edu.cn}

\maketitle

\begin{abstract}
The past seventeen years have witnessed tremendous progress on the
experimental and theoretical explorations of the multiquark states.
The hidden-charm and hidden-bottom multiquark systems were reviewed
extensively in Ref.~\cite{Chen:2016qju}. In this article, we shall
update the experimental and theoretical efforts on the hidden
heavy flavor multiquark systems in the past three years. Especially
the LHCb collaboration not only confirmed the existence of the
hidden-charm pentaquarks but also provided strong evidence of the
molecular picture. Besides the well-known $XYZ$ and $P_c$ states, we
shall discuss more interesting tetraquark and pentaquark systems
either with one, two, three or even four heavy quarks. Some very
intriguing states include the fully heavy exotic tetraquark states 
$QQ\bar Q\bar Q$ and doubly heavy tetraquark states $QQ\bar q \bar
q$, where $Q$ is a heavy quark. The $QQ\bar Q\bar Q$ states may be
produced at LHC while the $QQ\bar q \bar q$ system may be searched
for at BelleII and LHCb. Moreover, we shall pay special attention to
various theoretical schemes such as the chromomagnetic interaction
(CMI), constituent quark model, meson exchange model, heavy quark
and heavy diquark symmetry, QCD sum rules, Faddeev equation for the
three body systems, Skyrme model and the chiral quark-soliton model,
and the lattice QCD simulations. We shall emphasize the
model-independent predictions of various models which are
truly/closely related to Quantum Chromodynamics (QCD). For example,
the chromomagnetic interaction arises from the gluon exchange which
is fundamental and universal in QCD and responsible for the mixing
and mass splitting of the conventional mesons and baryons within the
same multiplet. The same CMI mechanism shall also be responsible for
the mixing of the different color configurations and mass splittings
within the multiplets in the multiquark sector. There have also
accumulated many lattice QCD simulations through multiple channel
scattering on the lattice in recent years, which provide deep
insights into the underlying structure/dynamics of the $XYZ$ states.
In terms of the recent $P_c$ states, the lattice simulations of the
charmed baryon and anti-charmed meson scattering are badly needed.
We shall also discuss some important states which may be searched
for at BESIII, BelleII and LHCb in the coming years.
\end{abstract}


\tableofcontents

\input{section1.tex}

\input{section2.tex}

\input{section3.tex}

\input{section4.tex}

\input{section5.tex}

\input{section6.tex}

\input{section7.tex}

\input{section8.tex}

\input{section9.tex}

\input{section10.tex}

\input{section11.tex}

\vspace{0.2cm}

\section*{Acknowledgments}

We would like to express our gratitude to all the collaborators and
colleagues who contributed to the investigations presented here, in
particular to Dian-Yong Chen, Rui Chen, Xiao-Lin Chen, Er-Liang Cui,
Wei-Zhen Deng, Jun He, Ning Li, Shi-Yuan Li, Xiao-Hai Liu, Yu-Nan Liu, Zhi-Gang Luo,
Li Ma, Takayuki Matsuki, Zong-Guo Si, T. G. Steele, Zhi-Feng Sun,
Guan-Juan Wang, Jing Wu, Lu Zhao, Zhi-Yong Zhou.
We thank Er-Liang Cui for helping prepare some relevant
documents. This project is supported by the National Natural Science
Foundation of China under Grants No. 11722540, No. 11775132,
No. 11575008, No. 11621131001 and National Key Basic Research
Program of China (2015CB856700), the China National Funds for
Distinguished Young Scientists under Grant No. 11825503, the
Fundamental Research Funds for the Central Universities, and the
National Program for Support of Top-notch Young Professionals.

\bibliographystyle{elsarticle-num}
\bibliography{ref}

\end{document}

%% file: section1.tex
%
\section{Introduction}
\label{sect:1}

In the past decades, hadron physicists show great interest in
hunting for evidences of the multiquark states. As a new form of
matter beyond conventional mesons ($\bar q q$) and baryons ($qqq$),
multiquark states containing more than three quarks are of special
importance in the hadron family. Especially, with the observations
of various charmonium-like $XYZ$ states, the study of multiquark
states has entered upon a new era.

In this article, we will give a concise review on the research
progress of the tetraquark ($\bar q \bar q q q$) and pentaquark
($\bar q q q q q$) states, which are typical multiquark matters.
Before doing that, let us first look back on the history of
particle/hadron physics. It is well known that the development of
quantum mechanics is closely related to the study of atomic
spectroscopy, which reveals the mysterious veil of the atom's
microstructure. Here, we find a similar situation that the study of
hadron spectroscopy is bringing us new insights into the internal
structure and dynamics of hadrons.

In the early 1960s, many strongly interacting particles were
observed in particle/nucleon experiments, which were named as
``hadron'' by L.~B.~Okun later~\cite{Okun:1962kca}. Based on these
observations, M.~Gell-Mann and G.~Zweig independently developed the
classification scheme for hadrons---the quark
model~\cite{GellMann:1964nj,Zweig:1964jf,Zweig:1981pd}.
Especially, the name ``quark'' was given by M.~Gell-Mann 
to denote the substructure of hadrons. The quark model achieves a
great success, and it is a milestone in the development of particle
physics. A well-known example is the prediction of the baryon with
three $strange$ quarks, that is the $\Omega$. It was discovered in
1964~\cite{Barnes:1964pd} after its existence, mass, and decay
products had been predicted in 1961 independently by
M.~Gell-Mann~\cite{GellMann:1961ky} and
Y.~Ne'eman~\cite{Neeman:1961jhl}.

Then we would like to mention the establishment of the Cornell
model. In 1974, as the first state in the charmonium family, the
$J/\psi$ was discovered by the E598
Collaboration~\cite{Aubert:1974js} and the SLAC-SP-017
Collaboration~\cite{Augustin:1974xw}, which confirmed the existence
of the $charm$ quark~\cite{Glashow:1970gm}. After that, a series of
charmonia were discovered, such as the $\psi(3686)$
\cite{Abrams:1974yy}, $\psi(3770)$ \cite{Rapidis:1977cv},
$\psi(4040)$ \cite{Goldhaber:1977qn}, $\psi(4160)$
\cite{Brandelik:1978ei}, and $\psi(4415)$ \cite{Siegrist:1976br},
etc. Based on these experimental observations, the Cornell model was
proposed~\cite{Eichten:1974af,Eichten:1978tg}, which uses the
Cornell potential $V(r)=-k/r+r/a^2$~\cite{Eichten:1979ms} to depict
the interaction between the $charm$ and $anti$-$charm$ quarks. This
potential consists of both the Coulomb-type and linear potentials,
which can well reproduce masses of the above observed charmonia at
that time. Inspired by the Cornell model, various potential models
were
developed~\cite{Barbieri:1975jd,Stanley:1980zm,Carlson:1983rw,Richardson:1978bt,Buchmuller:1980bm,Buchmuller:1980su,Martin:1980rm,Bhanot:1978mj,Quigg:1977dd,Fulcher:1991dm,Gupta:1993pd,Zeng:1994vj,Ebert:2002pp,Godfrey:1985xj},
among which the Godfrey-Isgur model~\cite{Godfrey:1985xj} is quite
popular. This model contains the semi-relativistic kinetic and
potential energy terms, which can be applied to quantitatively
describe not only meson spectra \cite{Godfrey:1985xj} but also
baryon spectra \cite{Capstick:1986bm}. These phenomenological
models, inspired by the charmonium family, greatly improve our
understanding of the internal structure and dynamics of conventional
mesons and baryons.

Nowadays several hundreds of hadrons have been observed in particle
experiments, most of which can be categorized into two families:
baryons made of three quarks and mesons made of one quark and one
antiquark~\cite{Tanabashi:2018oca}. They are formed by the $up$,
$down$, $strange$, $charm$, and $bottom$ quarks/antiquarks (the
$top$ quark has a very short lifetime and therefore does not form
hadrons), which are governed by the strong interaction. Although the
mechanism of the color confinement remains one of the most difficult
problems in particle physics, various phenomenological models were
proposed to quantitatively describe the hadron spectroscopy. Due to
the joint efforts from the particle/hadron theorists and
experimentalists, the meson and baryon families have been nearly
complete, and almost all the ground-state hadrons have been
discovered. Especially, the doubly charmed
baryon~$\Xi_{cc}^{++}(3621)$ was recently discovered by LHCb
Collaboration~\cite{Aaij:2017ueg}.

Other than hundreds of conventional mesons and baryons, there were
only tens of multiquark candidates observed in particle experiments,
although the concept of the multiquark states appeared together with
the birth of the quark model. For example, M.~Gell-Mann
indicated~\cite{GellMann:1964nj}: "{\it Baryons can now be
constructed from quarks by using the combinations $(qqq)$,
$(qqqq\bar{q})$, etc., while mesons are made out of $(q\bar{q})$,
$(qq\bar{q}\bar{q})$, etc}", and G.~Zweig~\cite{Zweig:1981pd} also
wrote: "{\it In general, we would expect that baryons are built not
only from the product of these aces, $AAA$, but also from
$\bar{A}AAAA$, $\bar{A}\bar{A}AAAAA$, etc., where $\bar{A}$ denotes
an anti-ace. Similarly, mesons could be formed from $\bar{A}A$,
$\bar{A}\bar{A}AA$, etc}". Due to the constraint of reliable
experimental and theoretical techniques, the study of the multiquark
states was far from satisfactory before 2003. At that time,
theorists mainly focused on a) the light scalar mesons $\sigma$,
$\kappa$, $a_0(980)$, and $f_0(980)$, b) the scalar mesons
$f_0(1370)$, $f_0(1500)$, and $f_0(1710)$, and c) the
$\Lambda(1405)$, and discussed their possible interpretations as
multiquark states and glueballs (see review
articles~\cite{Klempt:2009pi,Klempt:2007cp} for details).

\begin{figure}[htbp]
\centering
\includegraphics[width=500pt]{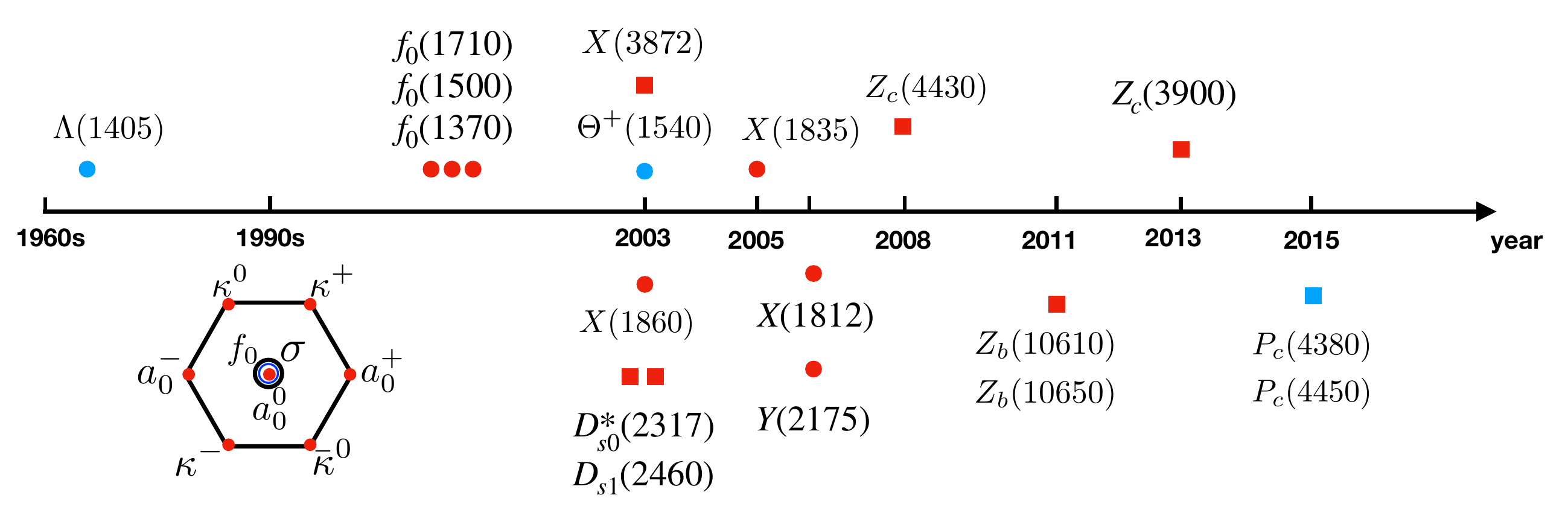}
\caption{Observations of some typical exotic hadronic
states.}\label{fig-sec1-timeline}
\end{figure}

2003 is an important year in the history of the multiquark states.
Since 2003, there has been continuous progress in this field. With
the accumulation of experimental data, a series of charmonium-like
$XYZ$ states were reported (see Fig. \ref{fig-sec1-timeline}), which
stimulated us to reveal their exotic inner structures. We shall try
our best to convey the progress and excitement to the readers in the
present review. Here are several typical examples:
\begin{itemize}

\item The BES and BESIII Collaborations reported a series of
light-flavor multiquark candidates, including the $X(1860)$ observed
in the $J/\psi\to \gamma p\bar{p}$ decay \cite{Bai:2003sw}, the
$X(1835)$ in $J/\psi\to\gamma\eta^\prime\pi^+\pi^-$
\cite{Ablikim:2005um,Ablikim:2010au}, the $X(1812)$ in $J/\psi\to
\gamma \omega\phi$ \cite{Ablikim:2006dw}, the $Y(2175)$ in
$J/\psi\to \eta\phi f_0(980)$ \cite{Ablikim:2007ab,Ablikim:2014pfc},
and so on. These observations inspired extensive discussions, and
various exotic interpretations such as multiquark states were
proposed. Since the present review mainly focuses on the
heavy-flavor multiquark states, interested readers may consult
Ref.~\cite{Klempt:2007cp} for more information.

\item In 1997, Diakonov {\it et al.} predicted the existence of
a light-flavor pentaquark with the chiral soliton model
\cite{Diakonov:1997mm}, which has a mass around 1530 MeV, spin
$1/2$, isospin $0$ and strangeness $+1$. In 2003, LEPS Collaboration
claimed that a narrow $\Theta^+(1540)$ particle consistent with the
above prediction was observed in the $\gamma n\to K^+K^-n$ reaction
\cite{Nakano:2003qx}. Very quickly, the $\Theta^+(1540)$, as a
pentaquark candidate, became a super star at that
time~\cite{Zhu:2004xa}. Unfortunately, it was not confirmed by the
subsequent series of high precise experiments~\cite{Liu:2014yva}.
The $\Theta^+(1540)$ is probably not a genuine
resonance~\cite{MartinezTorres:2010zzb}. The rise and fall of the
$\Theta^+(1540)$ unveiled our ignorance on the non-perturbative
behavior of quantum chromodynamics. The lessons and experience with
the $\Theta^+(1540)$ have been helpful in the search of the
pentaquark and tetraquark states.

\item In 2003, BaBar Collaboration observed a narrow heavy-light state $D_{s0}^*(2317)$
in the $D_{s}^+\pi^0$ invariant mass spectrum \cite{Aubert:2003fg}.
Since the mass of the $D_{s0}^*(2317)$ is about $100$ MeV lower than
the quark model prediction of the $P$-wave charmed-strange meson
with $J^P=0^+$ \cite{Godfrey:1985xj}, its tetraquark explanation was
proposed in Refs.
\cite{Barnes:2003dj,Cheng:2003kg,Szczepaniak:2003vy}. A similar
situation happened to the $D_{s1}(2460)$ observed by CLEO
Collaboration \cite{Besson:2003cp}. The strong channel coupling
between the S-wave DK($K^\ast$) scattering states and bare quark
model $c\bar s$ states plays a pivotal role in lowering the masses
of the $D_{s0}^*(2317)$ and $D_{s1}(2460)$
\cite{vanBeveren:2003kd,Dai:2006uz}.

\item Still in 2003, $X(3872)$, as the first particle in charmonium-like
$XYZ$ family, was discovered by Belle Collaboration
\cite{Choi:2003ue}. Since the mass of the $X(3872)$ is close to the
threshold of the $D\bar{D}^*$ pair, its assignment as the
$D\bar{D}^*$ molecular state is very popular. We note that its
nature is still under heated debates today. Similar to the
$D_{s0}^*(2317)$ and $D_{s1}(2460)$, there also exists a low mass
puzzle related to the $X(3872)$. Its mass is much lower than the
quark model prediction of the charmonium $\chi_{c1}(2P)$ state
\cite{Godfrey:1985xj}. Again, the coupled channel effect may mediate
this difference \cite{Meng:2007cx}.

\end{itemize}

\begin{figure}[htbp]
\centering
\includegraphics[width=500pt]{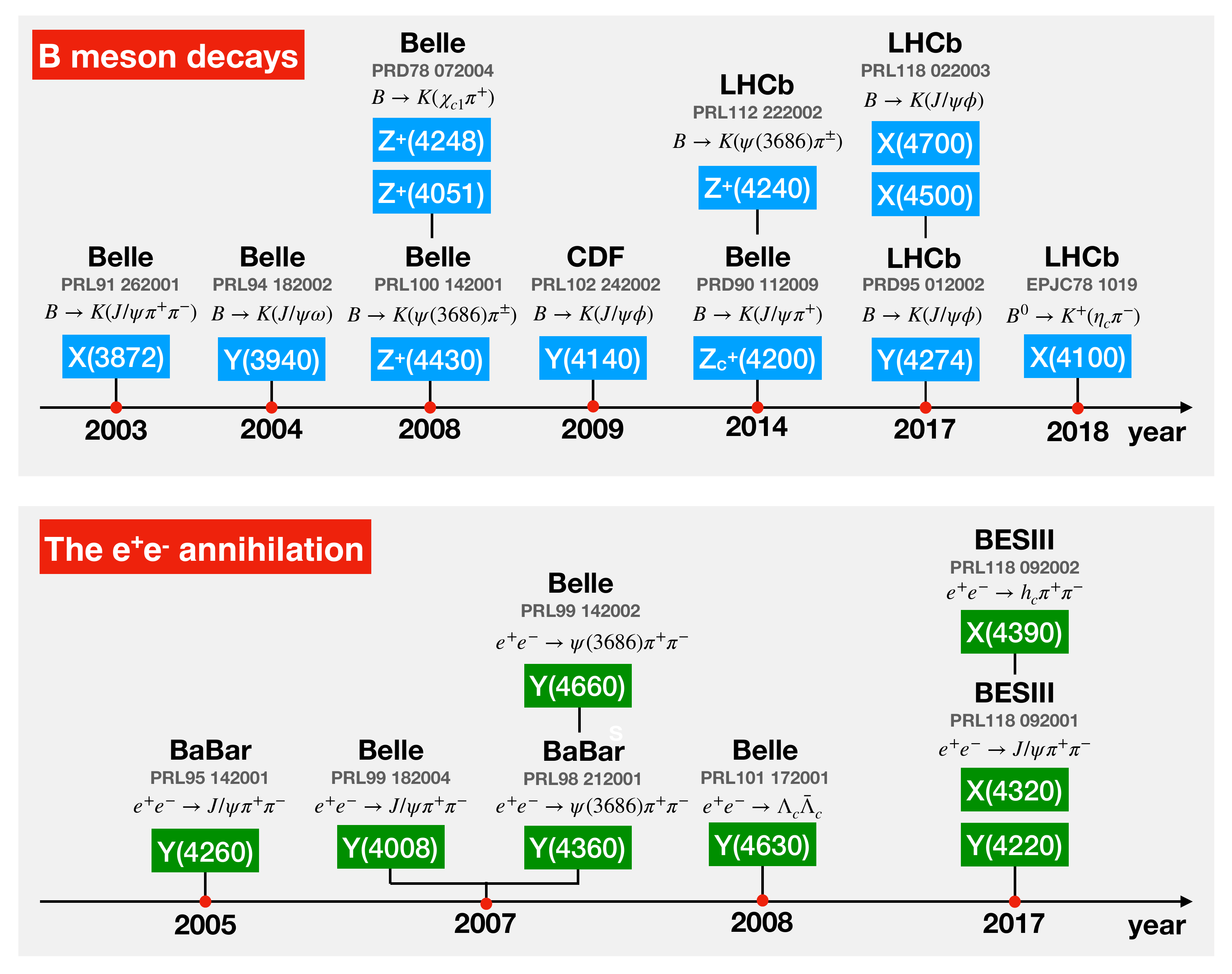}
\caption{Charmonium-like $XYZ$ states from $B$ meson decays and the
$e^+e^-$ annihilation.}\label{fig2}
\end{figure}

\begin{figure}[htbp]
\centering
\includegraphics[width=500pt]{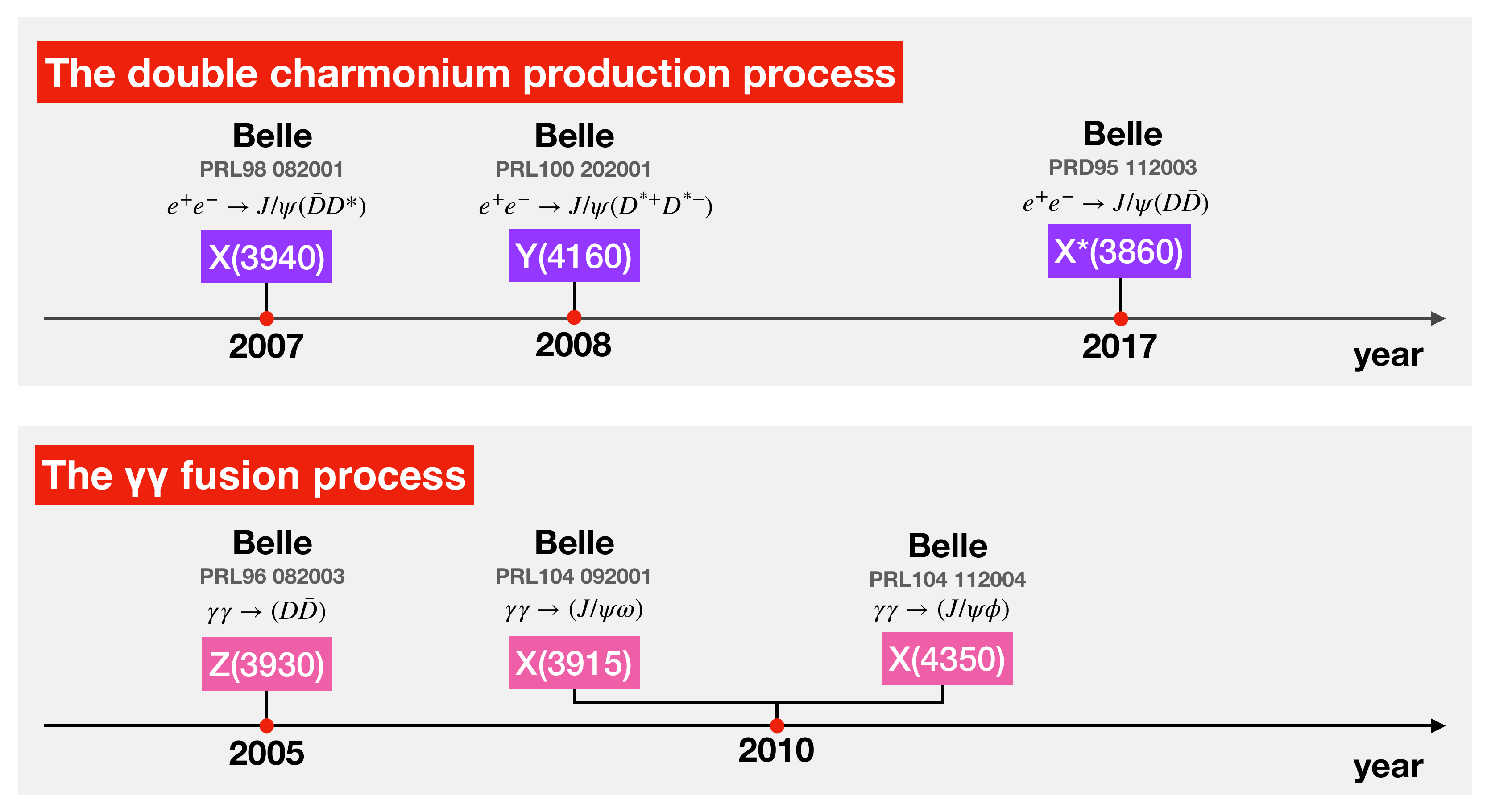}\\
\includegraphics[width=500pt]{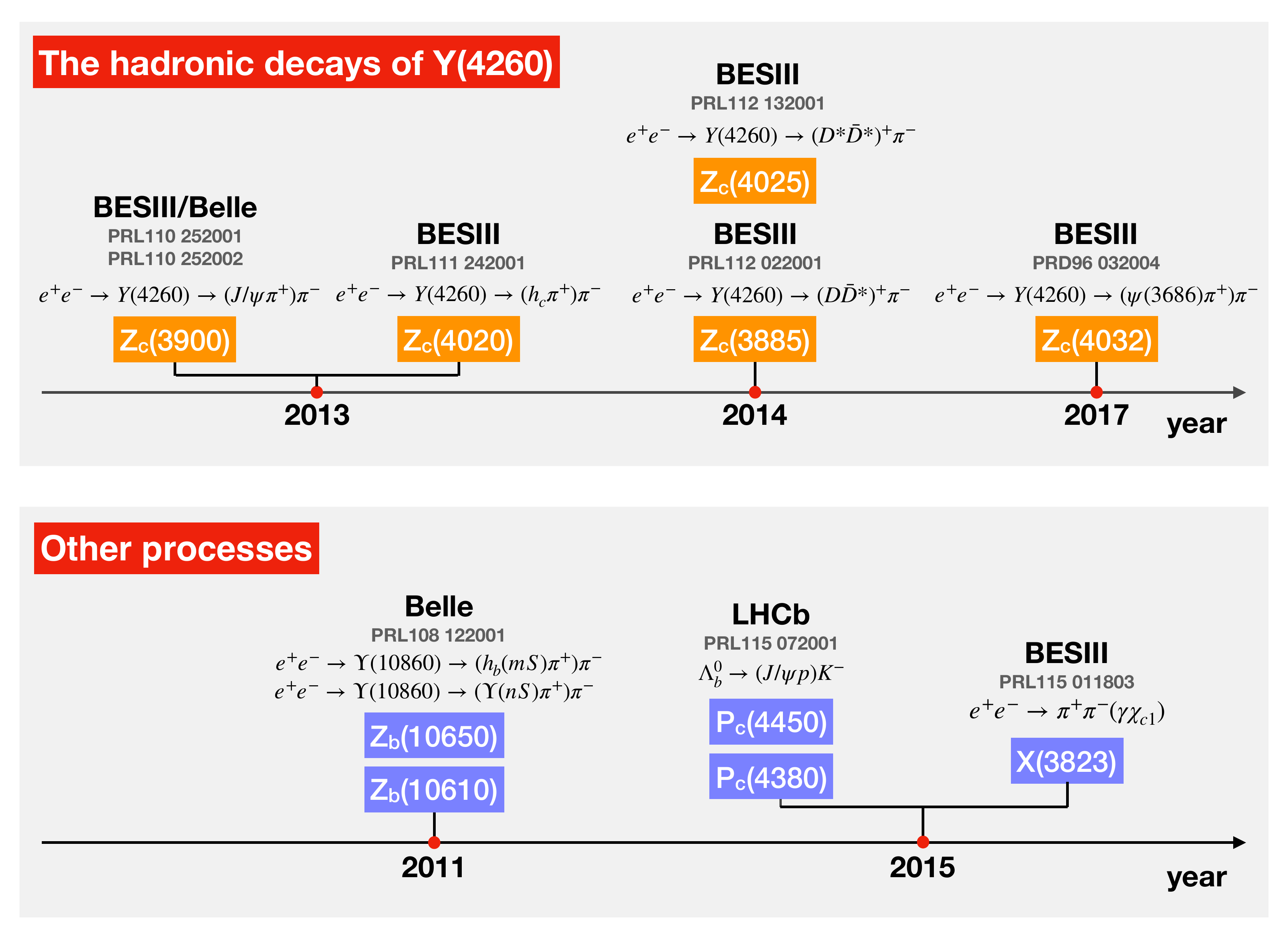}
\caption{Charmonium-like $XYZ$ states from the double charmonium
production process, the $\gamma\gamma$ fusion process, the hadronic
decays of the $Y(4260)$, and some other processes.}\label{fig3}
\end{figure}

In the past 17 years, more and more candidates of the exotic hadrons
were observed in the Belle, BaBar, BESIII, D0, CDF, CMS, and LHCb
experiments, such as a) the charmonium-like $XYZ$ states $Y(4260)$,
$Z_c(4430)$, and $Z_c(3900)$, b) the bottomonium-like states
$Z_b(10610)$ and $Z_b(10650)$, and c) the hidden-charm pentaquark
states $P_c(4380)$ and $P_c(4450)$, etc. In Figs.
\ref{fig2}-\ref{fig3}, we summarize these observations concisely. An
extensive review of the states observed before 2016 can be found in
Ref.~\cite{Chen:2016qju}. Here we briefly introduce the experimental
observations after 2016:
\begin{itemize}
\item In the $J/\psi\phi$ invariant mass spectrum of the
$B\to K~J/\psi~\phi$ decay, LHCb established the existence of the
$Y(4274)$ structure with $6\sigma$ significance
\cite{Aaij:2016nsc,Aaij:2016iza}. This state has a mass
$M=4273.3\pm8.3^{+17.2}_{-3.6}$ MeV, width
$\Gamma=56.2\pm10.9^{+8.4}_{-11.1}$ MeV, and quantum numbers
$J^{PC}=1^{++}$. Before this observation, CDF and CMS had reported
an evidence of a structure around 4274 MeV in the $J/\psi\phi$
invariant mass spectrum \cite{Aaltonen:2011at,Chatrchyan:2013dma}.
Besides the $Y(4274)$ structure, two extra structures with higher
masses were found in the $J/\psi\phi$ invariant mass spectrum, which
are the $X(4500)$ and $X(4700)$ with $J^P = 0^+$. Their resonance
parameters in units of MeV are \cite{Aaij:2016nsc,Aaij:2016iza}
\begin{eqnarray}
M_{X(4500)}&=&4506\pm11^{+12}_{-15},\quad
\Gamma_{X(4500)}=92\pm21^{+21}_{-20},\\
M_{X(4700)}&=&4704\pm10^{+14}_{-24},\quad
\Gamma_{X(4700)}=120\pm31^{+42}_{-33}.
\end{eqnarray}

\item In 2017, BESIII Collaboration
presented more precise measurements of the $e^+e^-\to \pi^+\pi^-
J/\psi$ cross section at the center-of-mass energy from 3.77 to 4.60
GeV \cite{Ablikim:2016qzw} and the $e^+e^-\to \pi^+\pi^- h_c$ cross
section at the center-of-mass energy from 3.896 to 4.600 GeV
\cite{BESIII:2016adj}. Three vector structures $Y(4220)$, $Y(4320)$
and $Y(4390)$ were reported with the resonance parameters in units
of MeV:
\begin{eqnarray*}
\begin{array}{c|ccc} \hline
{\mathrm{states}}& {\mathrm{mass}} &
{\mathrm{width}}&{\mathrm{channel}}\\\hline

Y(4220)&4222.0\pm3.1\pm1.4&44.1\pm4.3\pm2.0&\pi^+\pi^-J/\psi\\

            &4218.4^{+5.5}_{-4.5}\pm0.9&66.0^{+12.3}_{-8.3}\pm0.4&\pi^+\pi^-h_c\\

Y(4320)&4320\pm10.4\pm7.0&101.4^{+25.3}_{-19.7}\pm 10.2&\pi^+\pi^-J/\psi\\

Y(4390)&4391.5^{+6.3}_{-6.8}\pm1.0&139.5^{+16.2}_{-20.6}\pm
0.6&\pi^+\pi^-h_c\\\hline
\end{array}.
\end{eqnarray*}
In fact this updated analysis of the $e^+e^-\to \pi^+\pi^- J/\psi$
cross section \cite{BESIII:2016adj} shows that the $Y(4260)$
contains two substructures, the $Y(4220)$ and $Y(4320)$.

\item The Belle Collaboration carried out an analysis of the $e^+e^-\to J/\psi D\bar{D}$
cross section, and concluded that a broad structure $X^*(3860)$
exists in the $D\bar{D}$ invariant mass spectrum, which has a mass
$M=3862^{+26+40}_{-32-13}$ MeV and width
$201^{+154+88}_{-67\;\;-82}$ MeV \cite{Chilikin:2017evr}. Their
result shows that $J^{PC}=0^{++}$ is favored over $J^{PC}=2^{++}$ at
the level of $2.5\sigma$ \cite{Chilikin:2017evr}.

\item A new charged structure $Z_c(4032)$ was reported by BESIII Collaboration in the
$\psi(3860)\pi^+$ invariant mass spectrum of the $e^+e^-\to
Y(4260)\to \psi(3860)\pi^+\pi^-$ process \cite{Ablikim:2017oaf}.
BESIII confirmed the vector charmonium-like structure $Y(4220)$ in
the $\psi(3860)\pi^+\pi^-$ invariant mass spectrum.

\item  In 2018, LHCb found the evidence of a structure in the
$\eta_c(1S)\pi^-$ invariant mass spectrum of the $B^0\to
\eta_c(1S)K^+\pi^-$ decay, which was named as the $X(4100)$ with a
mass $M=4096\pm20^{+18}_{-22}$ MeV and width
$\Gamma=152\pm58^{+60}_{-35}$ MeV \cite{Aaij:2018bla}. The LHCb
measurement also indicated that the $X(4100)$ structure may be
described under the spin-parity $J^P=0^+$ or $J^P=1^+$ assignments
\cite{Aaij:2018bla}.
\begin{figure}[!th]
\begin{center}
\includegraphics[width=0.8\textwidth,clip]{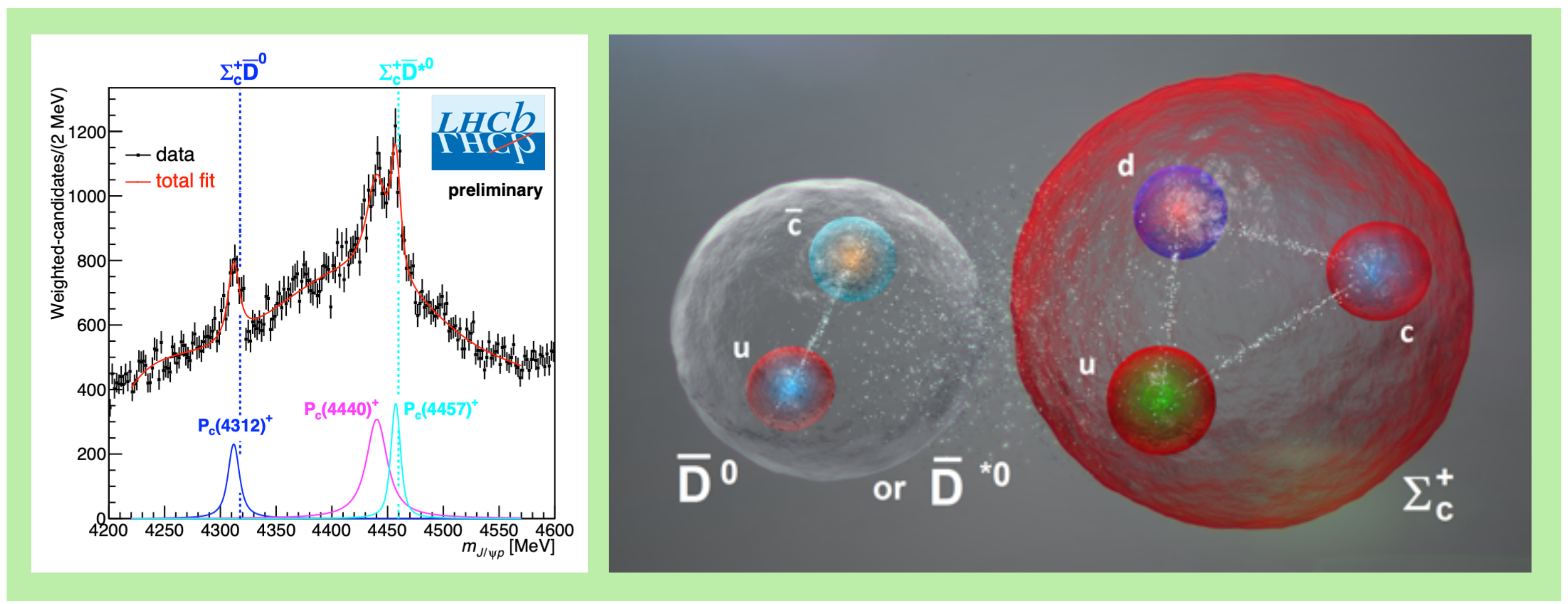}
\end{center}
\caption{ Three new hidden-charm pentaquarks observed by the LHCb Collaboration. Figures were taken
from Ref.~\cite{lhcbnew}.} \label{rspc}
\end{figure}
\item
Very recently, the LHCb Collaboration announced the observation of
three new pentaquarks \cite{lhcbnew} at the Rencontres de Moriond QCD
conference, as shown in Fig.~\ref{rspc}. In the measured $J/\psi p$ invariant mass spectrum, a
new pentaquark $P_c(4312)$ was discovered with a $7.3\sigma$
significance. The new LHCb analysis further found that the
$P_c(4450)$ is composed of two substructures $P_c(4440)$ and
$P_c(4457)$ with 5.4$\sigma$ significance. Their resonance parameters are
collected as following
\begin{eqnarray*}
P_c(4312)^+:\quad m&=&4311.9\pm0.7^{+6.8}_{-0.6}\,{\rm MeV},\\
\Gamma&=&9.8\pm2.7^{+3.7}_{-4.5}\,{\rm MeV},\\
P_c(4440)^+:\quad m&=&4440.3\pm1.3^{+4.1}_{-4.7}\,{\rm MeV},\\
\Gamma&=&20.6\pm4.9^{+8.7}_{-10.1}\,{\rm MeV},\\
P_c(4457)^+:\quad m&=&4457.3\pm1.3^{+0.6}_{-4.1}\,{\rm MeV},\\
\Gamma&=&6.4\pm2.0^{+5.7}_{-1.9}\,{\rm MeV}.
\end{eqnarray*}
The isospin of $P_c(4312)$, $P_c(4440)$ and $P_c(4457)$ is $I=1/2$
since these three pentaquarks were discovered in the $J/\psi p$
channel. The $P_c(4312)^+$ lies below the $\Sigma_c^+\bar{D}^0$
threshold, while the masses of the $P_c(4440)^+$ and $P_c(4457)^+$
are slightly lower than the $\Sigma_c^+\bar{D}^{*0}$ threshold. The
observation of $P_c(4312)$, $P_c(4440)$ and $P_c(4457)$ clearly
confirms the hidden-charm molecular pentaquarks.

\end{itemize}

The present situation of the charmonium-like $XYZ$ and $P_c$ states
is a bit similar to a) that of the ground-state mesons and baryons
in 1960s, and b) that of the charmonia in 1980s. With such abundant
novel phenomenon, the most important task is to identify the genuine
tetraquark and pentaquark signals. There have accumulated hundreds
of investigations of the tetraquark and pentaquark systems with
various phenomenological models. In this review, we will introduce
several typical phenomenological methods/models and their
applications to multiquark states.

Our previous review focused on the hidden-charm and hidden-bottom
multiquark systems ~\cite{Chen:2016qju}. In the present article, we
will summarize the experimental and theoretical progress on the
hidden-charm tetraquark and pentaquark states in the past three
years. Besides the hidden heavy flavor systems, we shall discuss
more interesting tetraquark and pentaquark systems either with open
heavy flavor or with three/four heavy quarks. These new states
include the exotic tetraquarks $QQ\bar Q\bar Q$ and $QQ\bar q \bar
q$ systems where $Q$ is a heavy quark. The $QQ\bar Q\bar Q$ states
may be produced at LHC while the $QQ\bar q \bar q$ system may be
searched for at BelleII. Moreover, we will summarize theoretical
predictions of the multiquark states from various formalisms such as
the chromomagnetic interaction (CMI), constituent quark model, meson
exchange model, heavy quark and heavy diquark symmetry, QCD sum
rules, Faddeev equation for the three body systems, Skyrme model and
the chiral quark-soliton model, and the lattice QCD simulations. We
shall pay special attention to those theoretical schemes which are
not covered or not addressed in depth in our previous review
~\cite{Chen:2016qju}, such as the chromomagnetic interaction, heavy
quark symmetry etc. We shall emphasize the model-independent
predictions of these models which are truly/closely related to
Quantum Chromodynamics (QCD).

Besides the present article and our previous review
~\cite{Chen:2016qju}, there exist many nice reviews of the XYZ
states in literature
\cite{Hosaka:2016pey,Ali:2017jda,Karliner:2017qhf,Guo:2017jvc,Esposito:2016noz,Lebed:2016hpi,Richard:2016eis,Olsen:2017bmm}.
Interested readers are encouraged to consult them for a glimpse of
this extremely active and vast field.

%% file: section2.tex
%
\section{$SU(6)$ symmetry and Chromomagnetic interaction (CMI)}\label{sec3}

The hyperfine structure for atoms is induced by the spin-related
interaction between electrons and nuclei. In a similar situation,
the hyperfine structure in hadron spectroscopy is from the
spin-related interaction between quarks or between quarks and
antiquarks. In the Hamiltonian or Lagrangian formalism, such a term
contains a color factor since quarks interact in the color space.
The simple color-magnetic interaction arises from the
one-gluon-exchange potential and causes the mass splittings of the
conventional hadrons, whose color configuration is unique. After the
inclusion of the quark mass, one obtains an effective description
method for hadron masses, which is the Hamiltonian of the
color-magnetic interaction or chromomagnetic interaction (CMI)
model.

Although the CMI model is simple and the results may sometimes
deviate from reality, it is helpful to understand the basic features
of hadron spectra through the estimated masses, see Ref.
\cite{SilvestreBrac:1992mv} for detailed arguments. This type of
simple models together with symmetry analysis can also be helpful to
confirm dynamical models to some extent. For example, from the
analysis of the mass splittings of the conventional mesons and
baryons, the author of Ref. \cite{KerenZur:2007vp} concluded that
the Cornell potential (Coulomb+linear) is the preferred model.

Within the multiquark systems, there may exist two or more color
configurations. For example, one may decompose the fully heavy
tetraquark $(QQ)_A (\bar Q \bar Q)_B$ into two clusters. There are
two possible color configurations: $6_A \times {\bar 6}_B$ and
${\bar 3}_A \times 3_B$. The
$\vec{\lambda_{i}}\cdot\vec{\lambda_{j}}$ type color-electric
interactions do not induce mixing between different color
configurations. In contrast, the chromomagnetic interaction will not
only mix two different color configurations, but also cause the mass
splitting within the multiplets.

Recent developments in spectroscopy motivated the proposal of
improved CMI models or methods which are becoming powerful tools in
studying interesting multiquark states. We shall discuss recent
developments about the model and application methods in this
section.

\subsection{Hamiltonian for the CMI model}

A realistic quark potential model generally includes the quark
kinetic term, color Coulomb term, color-spin interaction term,
spin-orbital term, tensor term, and color confinement term.
Considering the Hamiltonian in Ref. \cite{DeRujula:1975qlm} for the
$S$-wave hadrons (where the sin-orbit and tensor interactions
vanish) and ignoring the electromagnetic part, one has
\begin{eqnarray}\label{H}
\hat{H}&=&\Big\{L(\vec{r}_{1},\vec{r}_{2},...)+\sum\limits_{i}(m_{0i}+\frac{\vec{p}^2_{i}}{2m_{0i}})+\frac{1}{4}\sum\limits_{i<j}\alpha_{s}
\vec{\lambda_{i}}\cdot\vec{\lambda_{j}}\Big[\frac{1}{|\vec{r}|}-\frac{\pi}{2}\delta^{3}(\vec{r})\Big(\frac{1}{m_{0i}^2}+\frac{1}{m_{0j}^2}\Big)
\nonumber\\
&&-\frac{1}{2m_{0i}m_{0j}}\Big(\frac{\vec{p}_{i}\cdot
\vec{p}_{j}}{|\vec{r}|}+\frac{\vec{r}\cdot(\vec{r}\cdot\vec{p}_{i})
\vec{p}_{j}}{|\vec{r}|^{3}}\Big)\Big]\Big\}
-\frac{\pi}{6}\sum\limits_{i<j}\alpha_{s}
\delta^{3}(\vec{r})\frac{\vec{\lambda_{i}}\cdot\vec{\lambda_{j}}\vec{\sigma}_{i}\cdot\vec{\sigma}_{j}}{m_{0i}m_{0j}}.
\end{eqnarray}
Here, the Gell-Mann matrix $\lambda_i$ should be replaced by
$-\lambda_i^*$ for an antiquark, $L$ is responsible for the quark
binding, $\vec{r}_{i}$, $\vec{p}_{i}$, and $m_{0i}$ are the
position, momentum, and constituent mass of the $i$-th (anti)quark,
respectively, and $\vec{r}=\vec{r}_i-\vec{r}_j$. The average on the
orbital wave function $\Psi_0$ ($L=0$) gives
\begin{eqnarray}\label{mass}
H&=&\langle\Psi_0|\hat{H}|\Psi_0\rangle \equiv
\sum_im_i-\sum\limits_{i<j}C_{ij}\vec{\lambda_{i}}\cdot\vec{\lambda_{j}}\vec{\sigma}_{i}\cdot\vec{\sigma}_{j}
=\sum_im_i+H_{CMI},\nonumber\\
M&=&\sum_{i=1} m_i+\langle {H}_{CMI}\rangle,
\end{eqnarray}
where $m_i$ containing various effects is the effective mass for the
$i$th (anti)quark and $C_{ij}=\langle \alpha_s\delta^3(\vec{r})
\rangle \pi/(6 m_{0i}m_{0j})$. The coupling parameter $C_{ij}$ is
also written as $v/(m_im_j)$ in the literature. Note that $\langle
\alpha_s\delta^3(\vec{r})\rangle$ or $v$ for mesons differs from
that for baryons. This mass expression is also called
Sakharov-Zeldovich formula \cite{Sakharov:1966tua,KerenZur:2007vp}.

Now, the model contains only the color-spin operator and this
Hamiltonian should have an $SU(6)_{cs}$ symmetry. The eigenstate of
$H$ is also that of $SU(6)_{cs}$. One may express the eigenvalue on
a state with Casimir operators. For a state containing only
identical quarks whose flavor-spin-color wave function is $\varphi$,
one has \cite{Oka:2012zz,Maeda:2015hxa}
\begin{eqnarray}
\langle\varphi| -
\sum_{i<j}(\lambda_i\cdot\lambda_j)(\sigma_i\cdot\sigma_j)|\varphi\rangle
&=&8N+\frac43S(S+1)+2C_2[SU(3)_c]-4C_2[SU(6)_{cs}],
\end{eqnarray}
where $N$ is the number of quarks, $S$ is the total spin, and the
definitions for the quadratic Casimir operators are
\begin{eqnarray}
C_2[SU(2)]&=&\frac14\Big(\sum_i^N\vec{\sigma}_i\Big)^2,\qquad C_2[SU(3)]=\frac14\Big(\sum_i^N\vec{\lambda}_i\Big)^2,\nonumber\\
C_2[SU(6)]&=&\frac18\Big[\frac23\Big(\sum_i^N\vec{\sigma}_i\Big)+\Big(\sum_i^N\vec{\lambda}_i\Big)^2+\Big(\sum_i^N\vec{\sigma}_i\vec{\lambda}_i\Big)^2\Big].
\end{eqnarray}
If a representation is specified by the Young diagram
$[f_1,...,f_g]$, one has
\begin{eqnarray}
C_2[SU(g)]=\frac12\left[\sum_i f_i(f_i-2i+g+1)-\frac{N^2}{g}\right].
\end{eqnarray}
For a quark-antiquark pair or two quarks without constraint from the
Pauli principle, the color part and spin part can be calculated
separately and the formula in this special case is
\begin{eqnarray}
\langle\varphi|(\lambda_i\cdot\lambda_j)(\sigma_i\cdot\sigma_j)|\varphi\rangle
=4\Big[C_2[SU(3)_c]-\frac83\Big]\Big[S(S+1)-\frac32\Big].
\end{eqnarray}
If $C_{ij}=C$ and one acts $H_{CMI}$ on a multiquark system whose
quark content is $q^m\bar{q}^n$ ($n+m>3$), the average value can be
expressed as \cite{Jaffe:1976ih,Buccella:2006fn}
\begin{eqnarray}\label{CMI-Casimir}
\langle\varphi| H_{CMI}|\varphi\rangle&=&C\Big\{8N+4C_2[SU(6)_{cs},tot]-\frac43S_{tot}(S_{tot}+1)-2C_2[SU(3)_c,tot]\nonumber\\
&&+4C_2[SU(3)_c,q]+\frac83S_q(S_q+1)-8C_2[SU(6)_{cs},q]\nonumber\\
&&+4C_2[SU(3)_c,\bar{q}]+\frac83S_{\bar{q}}(S_{\bar{q}}+1)-8C_2[SU(6)_{cs},\bar{q}]\Big\}.
\end{eqnarray}
Now, $N=n+m$ is the total number of quarks and antiquarks. One may
also consult Ref. \cite{Aerts:1977rw} for more information.

In principle, the symmetry breaking should be considered. Then the
coupling parameters can be different for different flavors and more
complicated CMI expressions may be obtained (see Eq. (5) of Ref.
\cite{Buccella:2006fn}). In addition, mixing between different
color-spin structures may be induced by the chromomagnetic
interaction. The number of base states for a given $J^{PC}$ is
generally larger than one and we need to know all the relevant CMI
expressions which can be calculated with the knowledge of group
theory \cite{Hogaasen:1978jw,Cui:2005az,Cui:2006mp}. After
diagonalizing the obtained matrices $\langle H_{CMI}\rangle$'s, one
obtains eigenvalues of $H_{CMI}$. Alternatively, one may directly
calculate the matrix elements with the constructed wave functions in
color and spin spaces separately and then combine the results
together \cite{Hogaasen:2004pm}. If the results in the two spaces
have the form
\begin{eqnarray}
\langle\varphi_k| H_C|\varphi_l\rangle&=&\sum_{i<j}\langle\varphi_k| C_{ij}\lambda_i\cdot\lambda_j|\varphi_l\rangle=\sum_{i<j}a_{ij}^{kl}C_{ij},\nonumber\\
\langle\varphi_k| H_S|\varphi_l\rangle&=&\sum_{i<j}\langle\varphi_k|
C_{ij}\sigma_i\cdot\sigma_j|\varphi_l\rangle=\sum_{i<j}b_{ij}^{kl}C_{ij},
\end{eqnarray}
where $k$ or $l$ is the label of a base state, one gets (See also
Ref. \cite{SilvestreBrac:1992mv})
\begin{eqnarray}\label{cmiM.E.}
\langle\varphi_k|
H_{CMI}|\varphi_l\rangle&=&-\sum_{i<j}(a_{ij}^{kl}b_{ij}^{kl})C_{ij}\equiv\sum_{i<j}X_{ij}^{kl}C_{ij}.
\end{eqnarray}

\subsection{CMI models}

In principle, the values of $m_i$ and $C_{ij}$ in the CMI model
\eqref{mass} should be different for various systems, which can be
understood from the reduction procedure in getting the Hamiltonian.
However, it is difficult to exactly calculate these parameters for a
given system without knowing the spatial wave function. Practically,
they are extracted from the masses of conventional hadrons by
assuming that quark-(anti)quark interactions are the same for all
the hadron systems. This assumption certainly leads to uncertainties
on mass estimations. The uncertainty caused by $m_i$ does not allow
us to give accurate multiquark masses while the uncertainty in
coupling parameters has smaller effects and the mass splittings
should be more reliable. To give reasonable description for hadron
spectra in the CMI model, one usually uses modified mass equation
based on the adopted assumptions or chooses refined parameters in
calculations. In the literature, different forms of CMI Hamiltonians
can be found. Here we summarize several versions of the model in
studying heavy quark multiquark states.

\begin{itemize}

\item Models with diquark assumption. The $S$-wave correlated quark-quark
state (diquark) with color=$\bar{3}$ is assumed
\cite{Maiani:2004uc,Lipkin:1977ie}, where the spin of the diquark
can be 1 and 0 (see Ref. \cite{Anselmino:1992vg} for a review). For
the light-light diquark, the spin-0 state is more tightly bound than
the spin-1 state. For the heavy-light diquark, the difference
between the spin-1 and spin-0 states is suppressed by the heavy
quark mass. The formation mechanism of tetraquark states in the
diquark configuration was pictured in Ref. \cite{Brodsky:2014xia}
and the hypothesis that diquarks and antidiquarks in tetraquarks are
separated by a potential barrier was introduced in Ref.
\cite{Maiani:2017kyi}. In this diquark approximation, the
Hamiltonian for $q_1q_2\bar{q}_3\bar{q}_4$ tetraquark states has the
form \cite{Maiani:2004vq,Drenska:2009cd}
    \begin{eqnarray}\label{CMImodel-MaianiI}
    H&=&m_{[q_1q_2]}+m_{[q_3q_4]}+2(\kappa_{q_1q_2})_{\bar{3}}(S_{q_1}\cdot S_{q_2})+2(\kappa_{q_3q_4})_{\bar{3}}(S_{\bar{q}_3}\cdot S_{\bar{q}_4})+2\kappa_{q_1\bar{q}_3}(S_{q_1}\cdot S_{\bar{q}_3})\nonumber\\
    &&+2\kappa_{q_1\bar{q}_4}(S_{q_1}\cdot S_{\bar{q}_4})
    +2\kappa_{q_2\bar{q}_3}(S_{q_2}\cdot S_{\bar{q}_3})+2\kappa_{q_2\bar{q}_4}(S_{q_2}\cdot S_{\bar{q}_4})+2A\vec{L}\cdot(\vec{S}_{q_1q_2}+\vec{S}_{\bar{q}_3\bar{q}_4})+\frac{B}{2}\vec{L}^2,
    \end{eqnarray}
    or the form \cite{Maiani:2014aja}
\begin{eqnarray}\label{CMImodel-MaianiII}
    H&=&M_{00}+\frac12B{\vec L}^2-2a\vec{L}\cdot\vec{S}+2\kappa_{q_1q_2}^\prime(\vec{S}_{q_1}\cdot\vec{S}_{q_2})+2\kappa_{q_3q_4}^\prime(\vec{S}_{{\bar q_3}}\cdot\vec{S}_{{\bar q}_4})].
\end{eqnarray}
The former (latter) Hamiltonian is for the ``type-I'' (``type-II'')
diquark-antidiquark model where orbital excitation between the
diquark and the antidiquark is involved. In the type-II model, the
spin-spin interactions between quark components in different
diquarks are ignored and the coupling coefficients have different
values from those in the type-I model.

\item Model with triquark assumption. To understand the inner structure of
the LHCb pentaquark states, Lebed proposed the diquark-triquark
picture in Ref. \cite{Lebed:2015tna} where the triquark is composed
of a diquark and an antiquark. The diquarks from their generation
are always in the color-$\bar{3}$ representation
\cite{Brodsky:2014xia}. This picture is different from the one
proposed in Ref. \cite{Karliner:2003dt} where the diquark in the
triquark is in the color-$6$ representation. Another difference lies
in the stability. The diquark-triquark state in Ref.
\cite{Karliner:2003dt} looks like a static molecule stabilized by a
$P$-wave barrier, while the state in Ref. \cite{Lebed:2015tna} can
exist in each partial wave and can last as long as the components
continue to separate. Compared with the diquark-diquark-antiquark
configuration, the antiquark belongs to a compact component of the
pentaquark, although the two configurations have the same color
structure. This diquark-triquark picture allows the assignment
consistent with LHCb data and can qualitatively explain the measured
widths. The effective Hamiltonian in this diquark-triquark picture
can be written in a form similar to the model
\eqref{CMImodel-MaianiI} \cite{Zhu:2015bba},
\begin{eqnarray}\label{CMImodel-Qiao}
H=m_\delta+m_\theta+H_{SS}^\delta+H_{SS}^{\bar{\theta}}+H_{SS}^{\delta\bar{\theta}}+H_{SL}+H_{LL},
\end{eqnarray}
where $\delta$ ($\bar{\theta}$) indicates the diquark (triquark) and
the subscripts of $H$ mean the spin-spin, spin-orbital, and
orbital-orbital interactions within or between different clusters.

\item Model with the chromoelectric term. In Ref. \cite{Hogaasen:2013nca},
Hogaasen {\it et al} generalized the CMI model by including the
chromoelectric term,
\begin{eqnarray}\label{CMImodel-Hogaasen}
H=\sum_im_i-\sum_{i,j}A_{ij}\vec{\lambda}_i\cdot\vec{\lambda}_j-\sum_{i,j}C_{ij}\vec{\lambda}_i\cdot\vec{\lambda}_j\vec{\sigma}_i\cdot\vec{\sigma}_j,
\end{eqnarray}
where the effective quark masses have different values from those in
Eq. \eqref{mass} because of the non-vanishing $A_{ij}$. The
contributions from the chromoelectric term were implicitly included
in the model \eqref{mass}.

Ref. \cite{Weng:2018mmf} adopted a modified form of
\eqref{CMImodel-Hogaasen} to discuss the masses of the doubly and
triply heavy conventional baryons. The model Hamiltonian reads
\begin{eqnarray}\label{CMImodel-Deng}
H=-\frac34\sum_{i<j}m_{ij}\vec{F}_i\cdot\vec{F}_j-\sum_{i<j}v_{ij}\vec{S}_i\cdot\vec{S}_j\vec{F}_i\cdot\vec{F}_j,
\end{eqnarray}
where $F_i=\lambda_i/2$, $S_i=\sigma_i/2$, and $m_{ij}$ is the
to-be-fitted mass parameter of the quark pair labeled with $i$ and
$j$.

\item Model motivated with a QCD-string-junction picture \cite{Rossi:2016szw}.
Noting the fact that the effective quark masses and coupling
parameters determined from mesons differ from those from baryons,
Karliner, Rosner, and Nussinov proposed a method to estimate hadron
masses in Refs. \cite{Karliner:2014gca,Karliner:2016zzc}. Since
there is no junction (but one QCD string) for conventional mesons
and one junction (and two additional QCD strings) for conventional
baryons, a universal constant $S$ is added to the baryon mass
expression. For tetraquark states, the number of $S$ added to mass
expressions is 2 because there are two junctions (and two more QCD
strings). Besides, a correction term $B$ representing additional
binding is usually needed. Therefore, the mass formula in this model
generally has the form
    \begin{eqnarray}\label{CMImodel-junction}
    M=M_0+xS+B+M_{HF},
    \end{eqnarray}
where $M_0$ is the effective quark mass term, $x$ is an integer, and
$M_{HF}$ is the hyperfine splitting term.

\item Method with a reference mass scale. From the explicit calculations,
one finds that the model \eqref{mass} usually gives higher masses
for conventional hadrons (see Table \ref{CMI-hadrons} for a
comparison). According to our studies for various multiquark systems
\cite{Wu:2016gas,Chen:2016ont,Wu:2017weo,Luo:2017eub,Wu:2016vtq,Zhou:2018pcv,Li:2018vhp,Wu:2018xdi},
the obtained masses with this model are also the largest values we
can obtain. They should be higher than the realistic masses. Here,
``realistic'' means ``measured'' by assuming that the states do
exist. These observations indicate that the attractions between
quark components in this simple model are not sufficiently
considered. In order to reduce the uncertainties and obtain more
appropriate estimations, one may adopt an alternative form of the
mass formula.

Introducing a reference system and modifying the mass formula, one
gets
\begin{eqnarray}\label{massref}
 M=(M_{ref}-\langle{H}_{CMI}\rangle_{ref})+\langle{H}_{CMI}\rangle,
\end{eqnarray}
where $M_{ref}$ and $\langle{H}_{CMI}\rangle_{ref}$ are the physical
mass of the reference system and the corresponding CMI eigenvalue,
respectively. For $M_{ref}$, one may use the mass of a reference
multiquark state or use the threshold of a reference hadron-hadron
system whose quark content is the same as the considered multiquark
states. With this method, the problem of using extracted quark
masses from conventional hadrons in multiquark systems
\cite{Stancu:2009ka} is evaded and part of missed attraction between
quark components is phenomenologically compensated. Whether this
manipulation gives results close to realistic masses or not can be
tested in more studies. In fact, such an estimation method had been
applied to the multiquark states many years ago
\cite{SilvestreBrac:1992mv,Leandri:1989su,SilvestreBrac:1992yg}.

In calculating the CMI matrix elements for the multiquark states, we
consider all the color-spin configurations and the important color
mixing effects \cite{Vijande:2013qr}. When one adopts this method,
more than one thresholds or scales may exist. Since the size of a
multiquark state is expected to be larger than that of a
conventional hadron and the distance between quark components may be
larger, the attraction between quark components should be weaker.
The highest masses seem to be more reasonable in this method and we
use this assumption in our studies, although we cannot give a
definite answer for the problem which threshold should be adopted.
From the theoretical studies in recent years, it seems that the
predicted masses with such a strategy are still lower than the
realistic values. That is,
\begin{eqnarray}\label{massrange}
max(M_{\eqref{massref}})<M_{realistic}<M_{\eqref{mass}},
\end{eqnarray}
where $M_{\eqref{massref}}$ and $M_{\eqref{mass}}$ are obtained with
Eqs. \eqref{massref} and \eqref{mass}, respectively.

\end{itemize}

\subsection{CMI and Conventional hadrons}\label{CMI:conventionalhadrons}

The parameters in CMI models are effective masses and coupling
coefficients. In the simplest CMI model \eqref{mass}, one may
extract effective quark masses and most of the coupling parameters
with the known hadrons. There are different approaches to determine
them, by fitting the masses of all known hadrons, by fitting the
masses of mesons and baryons differently, or by just extracting
their values with parts of hadron masses. A set of the extracted
effective quark masses is $m_n=361.7$ MeV ($n=u,d$), $m_s=540.3$ MeV,
$m_c=1724.6$ MeV, and $m_{b}=5052.8$ MeV while the obtained coupling
parameters are listed in Table \ref{CMI-parameters}. For the unknown
$C_{c\bar{b}}$, its value was extracted with the meson masses
predicted in the Godfrey-Isgur (GI) model \cite{Godfrey:1985xj}. For
the unknown $C_{s\bar{s}}$, $C_{cc}$, $C_{bb}$, and $C_{cb}$, the
approximation $C_{qq}/C_{q\bar{q}}=C_{nn}/C_{n\bar{n}}\approx 2/3$
\cite{KerenZur:2007vp,Lipkin:1986dx} has been used. This
approximation seems to be applicable for all the coupling
parameters. The parameters presented here are consistent with those
in the literature, e.g. \cite{Maiani:2004vq,Karliner:2014gca}.
\begin{table}[htbp]
\centering
\caption{Coupling parameters in units of MeV. The value of
$C_{c\bar{b}}$ is estimated with the mass splitting in the GI model
\cite{Godfrey:1985xj}. The approximations $C_{cc}=k C_{c\bar{c}}$,
$C_{bb}=k C_{b\bar{b}}$, $C_{cb}=k C_{c\bar{b}}$, and
$C_{s\bar{s}}=C_{ss}/k$ have been adopted, where $k\equiv
C_{nn}/C_{n\bar{n}}\approx2/3$.}\label{CMI-parameters}
\vspace{.3cm}
\scalebox{0.84}{
\renewcommand{\arraystretch}{1.5}
\begin{tabular}{c|c|c|c|c|c|c|c|c|c}\hline
$C_{nn}=18.4$&$C_{ns}=12.1$&$C_{nc}=4.0$&$C_{nb}=1.3$&$C_{ss}=6.5$&$C_{sc}=4.5$&$C_{sb}=1.2$&$C_{cc}=3.3$&$C_{bb}=1.8$&$C_{cb}=2.0$\\
$C_{n\bar{n}}=29.8$&$C_{n\bar{s}}=18.7$&$C_{n\bar{c}}=6.6$&$C_{n\bar{b}}=2.1$&$C_{s\bar{s}}=10.5$&$C_{s\bar{c}}=6.7$&$C_{s\bar{b}}=2.3$&$C_{c\bar{c}}=5.3$&$C_{b\bar{b}}=2.9$&$C_{c\bar{b}}=3.3$\\\hline
\end{tabular}
}
\end{table}

When substituting these parameters into Eq. \eqref{mass} for various
conventional hadrons and comparing the obtained masses with
experiments (Table \ref{CMI-hadrons}), it is obvious that most
theoretical masses are higher than the measured masses. This
observation illustrates that the effective attraction is not
appropriately considered in the model \eqref{mass}. To reproduce the
masses of the conventional hadrons in acceptable uncertainties in
the CMI model, it is necessary to adopt modified Hamiltonian by
including more effects. In the mentioned CMI models, only two
models, \eqref{CMImodel-Hogaasen} or \eqref{CMImodel-Deng} and
\eqref{CMImodel-junction}, are appropriate to achieve the task.
Recently, with the model \eqref{CMImodel-junction}, Karliner and
Rosner obtained reasonable masses for conventional mesons and
baryons in Ref. \cite{Karliner:2014gca}. The predicted mass of
$\Xi_{cc}$, $3627\pm12$ MeV, is very close to that measured by LHCb,
$3621.40\pm0.72\pm0.14$ MeV \cite{Aaij:2017ueg}. With the extended
CMI model whose Hamiltonian is given in Eq. \eqref{CMImodel-Deng},
Weng, Chen, and Deng explored in Ref. \cite{Weng:2018mmf} the masses
of doubly and triply heavy baryons. The parameters were obtained by
fitting masses of known hadrons and they can reproduce the baryon
masses well. The calculated $\Xi_{cc}$ mass ($3633.3\pm9.3$ MeV) is
also close to that measured by LHCb. For other predicted masses, the
values in these two methods are also consistent with each other. All
the predicted $QQq$ and $QQQ$ masses are not far from the lattice
results \cite{Mathur:2018epb,Mathur:2018rwu}, either. The studies
indicate that the CMI model is still powerful once chromoelectric
interactions are appropriately considered, although the model is
very simple.

\begin{table}[htbp]
\centering
\tiny \caption{Comparison for hadron masses measured by experiments
(Ex.)\cite{Tanabashi:2018oca} and calculated by using Eq.
\eqref{mass}, where $X_q\equiv C_{bc}+C_{bq}+C_{cq}$ and
$Y_q\equiv\sqrt{(2C_{bc}-C_{bq}-C_{cq})^2+3(C_{cq}-C_{bq})^2}$ with
$q$ being $n$ or $s$.}\label{CMI-hadrons}
\vspace{.3cm}
\scalebox{1.1}{
\renewcommand{\arraystretch}{1.8}
\begin{tabular}{ccccc||ccccc}\hline
Hadron&CMI&Th.&Ex.&(Th.-Ex.)&Hadron&CMI&Th.&Ex.&(Th.-Ex.)\\\hline
$\pi$&$-16C_{n\bar{n}}$&246.6&139.6&107&$\rho$&$\frac{16}{3}C_{n\bar{n}}$&882.3&775.3&107\\
$K$ &$-16C_{n\bar{s}}$&602.8&493.7&109&$K^*$&$\frac{16}{3}C_{n\bar{s}}$&1001.7&891.8&110\\
&&&&&$\omega$&$\frac{16}{3}C_{n\bar{n}}$&882.3&782.7&100\\
&&&&&$\phi$&$\frac{16}{3}C_{s\bar{s}}$&1136.7&1019.5&117\\
$D$&$-16C_{c\bar{n}}$&1980.7&1869.7&111&$D^{*}$&$\frac{16}{3}C_{c\bar{n}}$&2121.5&2010.3&111\\
$D_s$&$-16C_{c\bar{s}}$&2157.7&1968.3&189&$D_{s}^{*}$&$\frac{16}{3}C_{c\bar{s}}$&2300.6&2112.2&188\\
$B$&$-16C_{b\bar{n}}$&5380.9&5279.5&102&$B^{*}$&$\frac{16}{3}C_{b\bar{n}}$&5425.7&5324.7&101\\
$B_s$&$-16C_{b\bar{s}}$&5556.3&5366.9&189&$B_s^*$&$\frac{16}{3}C_{b\bar{s}}$&5605.4&5415.4&190\\
$\eta_{c}$&$-16C_{c\bar{c}}$&3364.4&2983.9&381&$J/\psi$&$\frac{16}{3}C_{c\bar{c}}$&3477.5&3096.9&381\\
$\eta_{b}$&$-16C_{b\bar{b}}$&10059.2&9399.0&660&$\Upsilon$&$\frac{16}{3}C_{b\bar{b}}$&10121.1&9460.3&661\\
$B_c$&$-16C_{\bar{c}b}$&6724.6&6274.9&450&$B^*_c$ \cite{Godfrey:1985xj}&$\frac{16}{3}C_{\bar{c}b}$&6795.0& \\

$N$&$-8C_{nn}$&937.9&938.3&0&$\Delta$&$8C_{nn}$&1232.3&1232.0&0\\
$\Sigma$&$\frac{8}{3}(C_{nn}-4C_{ns})$&1183.7&1189.4&-6&$\Sigma^*$&$\frac83(C_{nn}+2C_{ns})$&1377.3&1382.8&-6\\
$\Lambda$&$-8C_{nn}$&1116.5&1115.7&1\\
$\Xi$&$\frac{8}{3}(C_{ss}-4C_{ns})$&1330.6&1314.9&16&$\Xi^*$&$\frac{8}{3}(C_{ss}+2C_{ns})$&1524.2&1531.8&-8\\
&&&&&$\Omega$&$8C_{ss}$&1672.9&1672.5&0\\

$\Lambda_{c}$&$-8C_{nn}$&2300.8&2286.5&14\\
$\Sigma_{c}$&$\frac83(C_{nn}-4C_{cn})$&2454.4&2454.0&0&$\Sigma_{c}^*$&$\frac{8}{3}(C_{nn}+2C_{cn})$&2518.4&2518.4&0\\
$\Xi_{c}$&$-8C_{ns}$&2529.8&2467.9&62\\
$\Xi'_{c}$&$\frac{8}{3}(C_{ns}-2C_{cn}-2C_{cs})$&2613.5&2577.4&36&$\Xi_{c}^*$&$\frac{8}{3}(C_{ns}+C_{cn}+C_{cs})$&2681.5&2645.5&36\\
$\Omega_c$&$\frac{8}{3}(C_{ss}-4C_{cs})$&2774.5&2695.2&79&$\Omega_c^*$&$\frac83(C_{ss}+2C_{cs})$&2846.5&2765.9&81\\
$\Lambda_b$&$-8C_{nn}$&5629.0&5619.6&9\\
$\Sigma_b$&$\frac83(C_{nn}-4C_{bn})$&5811.4&5811.3&0&$\Sigma_{b}^{*}$&$\frac83(C_{nn}+2C_{bn})$&5832.2&5832.1&0\\
$\Xi_b$&$-8C_{ns}$&5858.0&5794.5&64\\
$\Xi'_{b}$&$\frac83(C_{ns}-2C_{bn}-2C_{bs})$&5973.7&5935.0&39&$\Xi_b^*$&$\frac83(C_{ns}+C_{bn}+C_{bs})$&5993.7&5955.3&39\\
$\Omega_b$&$\frac{8}{3}(C_{ss}-4C_{bs})$&6137.9&6046.1&92&$\Omega_b^*$&$\frac83(C_{ss}+2C_{bs})$&6157.1&\\
$\Xi_{cc}$&$\frac{8}{3}(C_{cc}-4C_{cn})$&3777.0&3621.4&156&$\Xi_{cc}^*$&$\frac{8}{3}(C_{cc}+2C_{cn})$&3841.0&\\
$\Omega_{cc}$&$\frac{8}{3}(C_{cc}-4C_{cs})$&3950.2&&&$\Omega_{cc}^*$&$\frac{8}{3}(C_{cc}+2C_{cs})$&4022.2&\\
$\Xi_{bb}$&$\frac{8}{3}(C_{bb}-4C_{bn})$&10458.2&&&$\Xi_{bb}^*$&$\frac{8}{3}(C_{bb}+2C_{bn})$&10479.0&\\
$\Omega_{bb}$&$\frac{8}{3}(C_{bb}-4C_{bs})$&10637.9&&&$\Omega_{bb}$&$\frac{8}{3}(C_{bb}+2C_{bs})$&10657.1&\\
$\Xi_{bc}$&$\frac83(-X_n-Y_n)$&7106.6&&&&\\
$\Xi_{bc}^\prime$&$\frac83(-X_n+Y_n)$&7132.4&&&$\Xi_{bc}^*$&$\frac83X_n$&7158.7&&\\
$\Omega_{bc}$&$\frac83(-X_s-Y_s)$&7281.2&&&&\\
$\Omega_{bc}^\prime$&$\frac83(-X_s+Y_s)$&7312.9&&&$\Omega_{bc}^*$&$\frac83X_s$&7338.3&&\\
&&&&&$\Omega_{ccc}$&$8C_{cc}$&5200.0&\\
&&&&&$\Omega_{bbb}$&$8C_{bb}$&15172.7&\\
$\Omega_{ccb}$&$\frac83(C_{cc}-4C_{bc})$&8489.0&&&$\Omega^*_{ccb}$&$\frac83(C_{cc}+2C_{bc})$&8521.6&\\
$\Omega_{bbc}$&$\frac83(C_{bb}-4C_{bc})$&11813.2&&&$\Omega^*_{bbc}$&$\frac83(C_{bb}+2C_{bc})$&11845.8&\\\hline
\end{tabular}
}
\end{table}

\subsection{CMI and $Q\bar{Q}q\bar{q}$}

Most of the observed $XYZ$ states belong to this category. Decades
ago, such type of tetraquark states had been discussed after the
excited charmonium state $\psi(2S)$ was observed. Since the
observation of the $X(3872)$ in 2003, more and more heavy hadrons
with exotic properties were observed and partly confirmed in the
following years. If the states have explicit exotic quantum numbers,
they are good tetraquark candidates. Otherwise, some of these states
may be conventional heavy quarkonia which are affected heavily by
the mixing effects with four-quark components. It's possible some
states are molecules or compact tetraquarks with non-exotic quantum
numbers. In some cases, there exist non-resonant explanations. We
are still far from understanding the nature of the
$Q\bar{Q}q\bar{q}$ states. Here, we first concentrate on recent
theoretical studies and developments within the CMI models,
especially in recent three years. One may consult studies about the
$Q\bar{Q}q\bar{q}$ states before 2016 in our previous review
\cite{Chen:2016qju}.

In Ref. \cite{Maiani:2004vq}, Maiani {\it et al} discussed the
nature of $X(3872)$ in the ``type-I'' diquark-antidiquark model
\eqref{CMImodel-MaianiI}. As the $1^{++}$ tetraquark state, the
properties of $X(3872)$ such as its narrow width and isospin
breaking in decay can be understood. Within this model, they also
discussed the properties of $X(3940)$ [now called $X(3915)$],
$D_s(2317)$, $D_s(2460)$, and $D_{sJ}(2632)$. Latter in Ref.
\cite{Maiani:2005pe}, they interpreted $Y(4260)$ as the first
orbital excitation of a diquark-antidiquark state
($[cs][\bar{c}\bar{s}]$). In Refs.
\cite{Maiani:2007wz,Maiani:2008zz}, the $Z(4430)$ was identified as
the first radial excitation of the tetraquark basic supermultiplet
to which $X(3872)$ belongs. In Ref. \cite{Drenska:2009cd}, the
authors studied the $cs\bar{c}\bar{s}$ states with different
$J^{PC}$ and predicted a $0^{-+}$ state at 4277 MeV decaying into
$J/\psi\phi$.

Hogaasen, Richard, and Sorba discussed the nature of $X(3872)$ in
the tetraquark configuration with the mass equation \eqref{mass} in
Ref. \cite{Hogaasen:2005jv}. By introducing the annihilation term
and different masses for the $u$ and $d$ quarks, they found that the
$X(3872)$ may be interpreted as a $cn\bar{c}\bar{n}$ tetraquark
state where the $c\bar{c}$ pair is mostly a color-octet state. This
configuration can also explain the rough ratio between the
$J/\psi\rho$ and $J/\psi\omega$ decay modes. However, their
parameters could not be applied to the hidden-bottom tetraquark
states.

Cui {\it et al.} considered the partner states of the $X(3872)$ with
the mass equation \eqref{mass} \cite{Cui:2006mp}. They determined
the coupling parameter by assuming the $X(3872)$ as a
$cn\bar{c}\bar{n}$ tetraquark. With the obtained parameters, the
masses of all the ground states with the configurations
$Qq\bar{Q}\bar{q}$, $QQ\bar{q}\bar{q}$, and $QQ\bar{Q}\bar{q}$ were
estimated, where $Q=c,b$ and $q=u,d,s$.

Stancu explored the spectrum of $cs\bar{c}\bar{s}$ tetraquarks with
the CMI model \eqref{mass} after the observation of the $Y(4140)$
\cite{Stancu:2009ka}. The results suggest that the $Y(4140)$ could
be the strange partner of $X(3872)$ with $J^{PC}=1^{++}$ in a
tetraquark interpretation. The later LHCb measurement for the
$X(4274)$ \cite{Aaij:2016iza} indicates that the mass splitting
between the $X(4274)$ and $X(4140)$ is consistent with the CMI model
prediction.

Tetraquark states were also studied with the CMI model \eqref{mass}
together with the possible $qqq\bar{q}\bar{q}\bar{q}$ hexaquark
states \cite{Abud:2009rk}. The $X(4140)$ and $X(4350)$ can be
assigned as $1^{++}$ and $0^{++}$ $cs\bar{c}\bar{s}$ states,
respectively.

In a schematic study of the S-wave $b\bar{b}q\bar{q}$ ($q=u,d,s$)
tetraquarks with the CMI model \eqref{mass}, the authors of Ref.
\cite{Guo:2011gu} obtained the $1^+$ $bu\bar{b}\bar{d}$ or
$bd\bar{b}\bar{u}$ states with masses $10612$ MeV and $10683$ MeV,
which are compatible with the $Z_b(10610)$ and $Z_b(10650)$.

In Refs. \cite{Ali:2011ug,Ali:2014dva}, Ali {\it et al} considered
the tetraquark interpretation for the bottomonium-like states
$Z_b(10610)$ and $Z_b(10650)$ with $I^G(J^P)=1^+(1^+)$. They assumed
that these mesons are $[bu][\bar{b}\bar{d}]$ diquark-antidiquark
states produced in the decay of the $Y(10890)$ which was also
assumed to be a tetraquark state \cite{Ali:2009pi,Ali:2009es}. Using
the Hamiltonian \eqref{CMImodel-MaianiI} and including the meson
loop effects, they were able to reproduce their measured
masses~\cite{Tanabashi:2018oca}.

After the observation of the $Z_c(3900)$, Maiani {\it et al}
interpreted it as the tetraquark state around 3880 MeV
~\cite{Faccini:2013lda}, which was predicted in their previous study
within the diquark-antidiquark model
\eqref{CMImodel-MaianiI}~\cite{Maiani:2004vq}. They also discussed
the possibility of another peak around 3760 MeV.

In Ref. \cite{Hogaasen:2013nca}, Hogaasen {\it et al} used the
Hamiltonian \eqref{CMImodel-Hogaasen} to discuss possible partner
states of the $X(3872)$. Their fitted parameters can reproduce the
mass spectrum of the $c\bar{c}n\bar{n}$ states obtained with Eq.
\eqref{mass} in Ref. \cite{Hogaasen:2005jv}. The $1^+$
$b\bar{b}n\bar{n}$ states ($I=0$ and $I=1$) may lie around 10.62
GeV, and the $Z_b(10610)^\pm$ or $Z_b(10650)^\pm$ is such a
tetraquark candidate, whose $(b\bar{b})$ component is almost a
color-octet state. They also identified the $Z_c(3900)$ implied in
Ref. \cite{Hogaasen:2005jv} as the corresponding $cn\bar{c}\bar{n}$
tetraquark.

After the development of meson spectroscopy for about ten years,
many candidates of exotic hadrons were reported. To understand their
possible natures, the ``type-II'' model \eqref{CMImodel-MaianiII}
was proposed in Ref. \cite{Maiani:2014aja}. According to this model,
the $X(3872)$ is still the state with the flavor wave function
$[cn]_{S=1}[\bar{c}\bar{n}]_{S=0} +
[cn]_{S=0}[\bar{c}\bar{n}]_{S=1}$; the other combination
$[cn]_{S=1}[\bar{c}\bar{n}]_{S=0} -
[cn]_{S=0}[\bar{c}\bar{n}]_{S=1}$ and
$[cn]_{S=1}[\bar{c}\bar{n}]_{S=1}$ can form the $Z_c(3900)$ and
$Z_c(4020)$ with $J^P=1^+$; the $X(3915)$ and $X(3940)$ are
identified to be $[cn]_{S=1}[\bar{c}\bar{n}]_{S=1}$ with $J^P=0^+$
and $[cn]_{S=1}[\bar{c}\bar{n}]_{S=1}$ with $J^P=2^+$, respectively.
Within these assignments, the quark-quark interactions are assumed
to be stronger while the quark-antiquark interactions are
negligible, which implies that the assumed diquarks in tetraquarks
are more compact than those in conventional baryons. In the $P=-$
case, six exotic vector states can be assigned: the $Y(4260)$ has
the same spin structure and spin interactions as the $X(3872)$ but
has an additional $P$-wave excitation between the diquark and the
antidiquark, the $Y(4008)$ is the orbitally excited
$[cn]_{S=0}[\bar{c}\bar{n}]_{S=0}$, the $Y(4630)$ is the $S_{tot}=2$
$P$-wave state, the $Y(4290)$ or $Y(4220)$ is identified as the
orbitally excited $S_{tot}=0$ $[cn]_{S=1}[\bar{c}\bar{n}]_{S=1}$
state with $J^{PC}=1^{--}$, and the $Y(4660)$ and $Y(4360)$ are the
first radial excitations of $Y(4260)$ and $Y(4008)$, respectively.
The observed decay $Y(4260)\to \gamma X(3872)$ was argued to be a
natural consequence of the diquark-antidiquark description for these
two states \cite{Chen:2015dig}.

The authors of Ref. \cite{Anisovich:2015caa} assumed the
$[Qq]_{\bar{3}_c}[\bar{Q}\bar{q}]_{3_c}$ diquark-antidiquark
configuration. Noticing the similar color structure between the
assumed diquark-antidiquark system and the quark-antiquark system,
the authors estimated the masses in a similar way,
\begin{eqnarray}
m^J_{(q\bar{q})}=m_q+m_{\bar{q}}+J(J+1)\Delta,
\end{eqnarray}
where $q$ can also be a scalar or axialvector diquark. After
considering the mass shifts due to the recombined meson-meson
rescattering effects, they presented the masses of
$[cn][\bar{c}\bar{n}]$, $[cs][\bar{c}\bar{s}]$, and
$[cn][\bar{c}\bar{s}]$ tetraquark states.

After assuming the $X(3872)$ as a $1^{++}$ diquark-antidiquark type
tetraquark and using it as an input, the authors of Ref.
\cite{Kim:2016tys} investigated possible assignments for its partner
states through the mass splittings calculated with the CMI method
\eqref{massref}. The masses of two states with $1^{+-}$ and one
state with $2^{++}$ fit very nicely to the $X(3823)$, $X(3900)$, and
$X(3940)$.

In Ref. \cite{Lebed:2016yvr}, Lebed and Polosa proposed that the
$\chi_{c0}(3915)$ [originally called $Y(3940)$, now called
$X(3915)$] is the lightest $0^{++}$ $c\bar{c}s\bar{s}$
diquark-antidiquark state in the ``type-II'' model
\eqref{CMImodel-MaianiII}. This identification may explain its
observed width around 20 MeV. After treating the $Y(4140)$ as a
$1^{++}$ $c\bar{c}s\bar{s}$ state and the $Y(4230)$ and $Y(4360)$ as
$1^{--}$ $c\bar{c}s\bar{s}$ states, they further interpreted the
$Y(4008)$ and $Y(4660)$ as $c\bar{c}s\bar{s}$ states. Moreover, the
$X(4350)$ is interpreted as a $0^{++}$ or $2^{++}$
$c\bar{c}s\bar{s}$ tetraquark state, while the $Y(4274)$ may be an
orbitally excited $c\bar{c}s\bar{s}$ state. In 2006 the LHCb
reported other two exotic structures in the $J/\psi\phi$ channel
\cite{Aaij:2016iza}, the $X(4500)$ and $X(4700)$. In
Ref.~\cite{Maiani:2016wlq} Maiani {\it et al} assigned the $Y(4140)$
as the $1^{++}$ ground tetraquark, and the $X(4500)$ and $X(4700)$
as two radially excited diquark-antidiquark states. They noticed
that the $X(4274)$ corresponds to two almost degenerate states with
$J^{PC}=0^{++}$ and $2^{++}$. Predictions of the other
$[cs][\bar{c}\bar{s}]$ states within this picture will be tested by
upcoming experiments.

Later Zhu {\it et al} presented a study on the hidden charm
tetraquarks in the $[cq]_{\bar{3}_c}[\bar{c}\bar{q}]_{3_c}$
diquark-antidiquark model \eqref{CMImodel-MaianiI}
\cite{Zhu:2016arf}. The authors proposed the following assignments:
the $Z_c(3900)$ with $J^P=1^+$ can be explained as
$[cn]_{\bar{3}_c}[\bar{c}\bar{n}]_{3_c}$, the $Z_c(4020)$ with
$J^P=1^+$ is a companion of $Z_c(3900)$, the $Z(4430)$ is the radial
excitation of $Z_c(3900)$, the $X(4140)$ and $X(4274)$ may be the
$1^+$ tetraquarks with the flavor wave function
$(u\bar{u}+d\bar{d}-2s\bar{s})c\bar{c}$, the $X(4500)$ and $X(4700)$
may be the radial excitations of the $0^+$ tetraquarks with
$(u\bar{u}+d\bar{d}+s\bar{s})c\bar{c}$ and
$(u\bar{u}+d\bar{d}-2s\bar{s})c\bar{c}$ respectively, and the
$Y(4260)$ may be an orbitally excited diquark-antidiquark
tetraquark.

In Ref. \cite{Ali:2017wsf}, Ali {\it et al} studied the
$J^{PC}=1^{--}$ $Y$ tetraquark states with an extended Hamiltonian
of the model \eqref{CMImodel-MaianiII} by including the tensor
coupling contributions. Only two scenarios are consistent with the
Belle, BaBar, and BESIII data in the diquark-antidiquark picture.
Each scenario contains four $Y$ states. Scenario I contains
$Y(4008)$, $Y(4260)$, $Y(4360)$, and $Y(4660)$, while Scenario II
contains $Y(4220)$, $Y(4330)$, $Y(4390)$, and $Y(4660)$.

In Ref. \cite{Wu:2016gas}, we updated estimations for the mass
spectrum of the $cs\bar{c}\bar{s}$ tetraquark states with the CMI
method \eqref{massref}. After assuming the $X(4140)$ to be the
lowest $J^{PC}=1^{++}$ $cs\bar{c}\bar{s}$ state, we confirmed the
mass splittings obtained in Ref. \cite{Stancu:2009ka}. We found that
the $X(4274)$ should be the other $1^{++}$ $cs\bar{c}\bar{s}$
tetraquark, and the $X(4350)$ \cite{Shen:2009vs} is probably a
$0^{++}$ $cs\bar{c}\bar{s}$, while the interpretation of the
$X(4500)$ and $X(4700)$ needs orbital or radial excitations. In this
study we also confirmed the mass inequality \eqref{massrange}, and
found that $M_{realistic}$ is closer to $max(M_{\eqref{massref}})$
than to $M_{\eqref{mass}}$. Similar features may be found in other
systems.

\begin{figure}[htbp]
\begin{center}
\includegraphics[width=0.7\textwidth,clip]{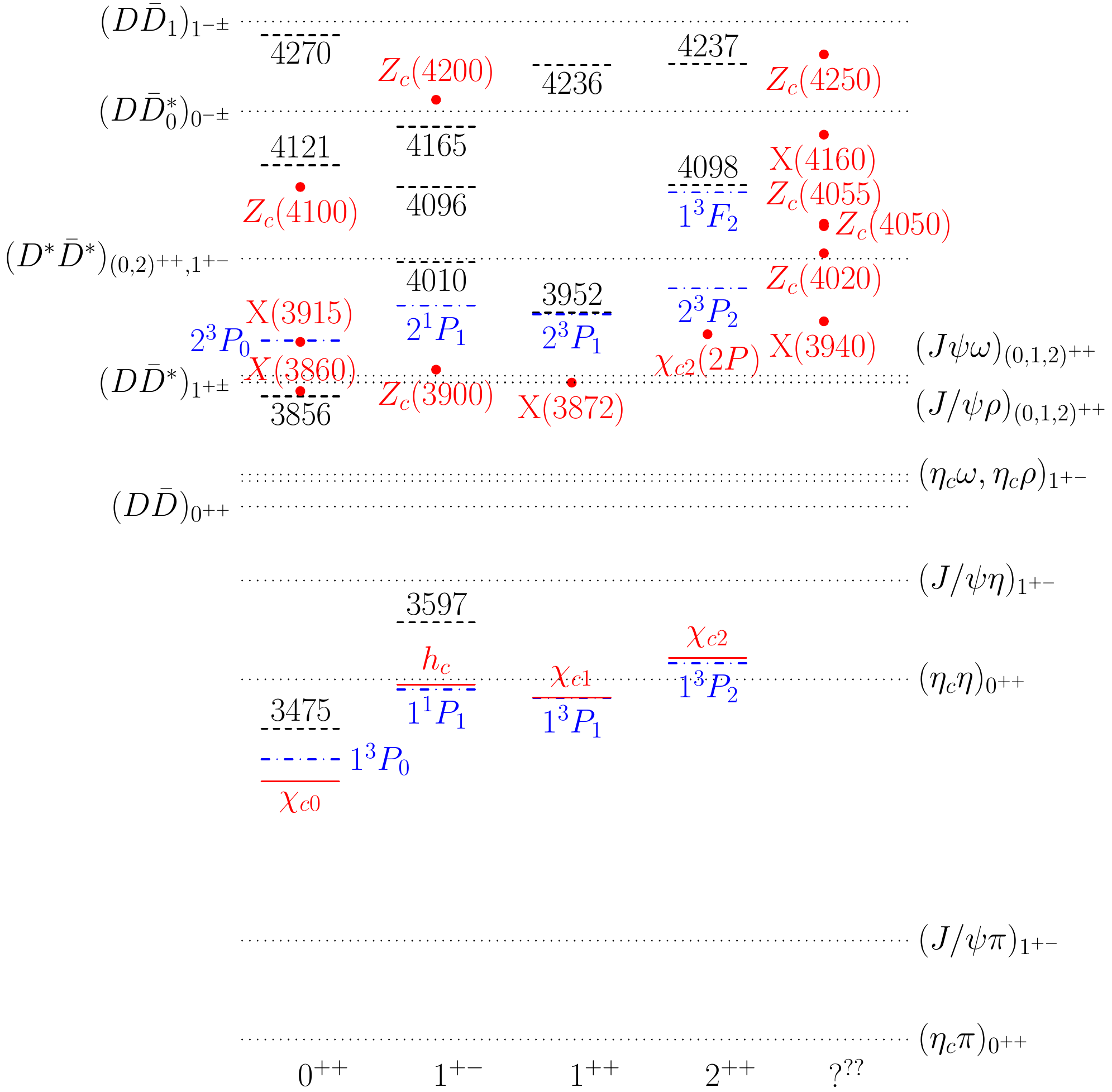}
\end{center}
\caption{Relative positions for the $cn\bar{c}\bar{n}$ tetraquarks
(black dashed lines), quark model predictions for the charmonia
(blue dash-dotted lines), observed charmonia (red solid lines),
states with exotic properties (red solid dots), and various
meson-meson thresholds (black dotted lines). The masses are given in
units of MeV. The subscripts of threshold symbols are $J^{PC}$ in
the $S$-wave case. Note the results are the same for the
$cn\bar{c}\bar{n}$ tetraquark states with $I=0$ and $I=1$. Taken
from Ref.~\cite{Wu:2018xdi}. }\label{fig-cnCN}
\end{figure}

In Ref. \cite{Wu:2018xdi}, we have systematically evaluated the mass
spectra of the $Q_1q_2\bar{Q}_3\bar{q}_4$ ($Q=c,b$ and $q=u,d,s$)
states. Besides the estimation method \eqref{massref}, we also
roughly estimated the tetraquark mass spectra by treating the mass
of the $X(4140)$ as an input. The quark mass differences $m_s-m_n$
and $m_b-m_c$ need to be replaced by other values, since quarks get
different masses in different systems. With this replacement method,
higher masses than $max(M_{\eqref{massref}})$ can be obtained. Such
masses satisfy the inequality \eqref{massrange} and we treat them as
realistic values. We show the results for $cn\bar{c}\bar{n}$ states
in Fig.~\ref{fig-cnCN}. Results for other systems can also be found
in this study~\cite{Wu:2018xdi}. It is possible to assign both the
$Z_c(4100)$ and $X(3860)$ to be the $0^{++}$ $cn\bar{c}\bar{n}$
tetraquark states. With the defined measure reflecting effective
quark interactions (see Sec.~\ref{subsec:effectiveCMI}), we argued
that this assignment is acceptable. Another possible tetraquark
candidate is the $Z_c(4200)$, but we need further studies to answer
whether the assigned structures are correct or not. At present, our
results with this updated method are higher than the masses obtained
in an early paper \cite{SilvestreBrac:1992mv} where the
$cn\bar{c}\bar{n}$, $bn\bar{b}\bar{n}$, and $cn\bar{b}\bar{n}$
states were all found to be stable. The $J=2$ $cn\bar{c}\bar{s}$ and
$bn\bar{b}\bar{s}$ are also weakly bound, although the binding
energy in the $J=2$ sector is very weak.

\subsection{CMI and $QQ\bar{q}\bar{q}$}
\label{sec2:QQqq}

The study of the doubly heavy tetraquark states $QQ\bar{q}\bar{q}$
($Q=c,b$, $q=u,d,s$) has a long history in the literature. Various
approaches have been applied to such systems, and the most
interesting state is the lowest isoscalar $1^+$ $cc\bar{u}\bar{d}$
(or its antiparticle $ud\bar{c}\bar{c}$), named $T_{cc}$. The
$T_{cc}$ and its partner tetraquark states are explicitly exotic.
They do not suffer from possible annihilation effects and their
inner quark interactions may provide enough attractions to stabilize
the four-quark system. One may consult Refs.
\cite{Ballot:1983iv,Lipkin:1986dw,Zouzou:1986qh,Heller:1986bt,Carlson:1987hh,Manohar:1992nd,SilvestreBrac:1992mv,Ericson:1993wy,SilvestreBrac:1993ry,SilvestreBrac:1993ss,Semay:1994ht,Moinester:1995fk,Pepin:1996id,Abud:2009rk,Brink:1998as,SchaffnerBielich:1998ci,Gelman:2002wf,Stewart:1998hk,Michael:1999nq,Barnes:1999hs,Cook:2002am}
for relevant investigations before 2003. Whether they are bound or
not depends on models and adopted approximations. Generally
speaking, the lowest $I(J^P)=0(1^+)$ states are narrow tetraquarks.
Up to now, neither the $T_{cc}$ nor any of its partner states have
been observed in experiments. Recently, the confirmation of the
$\Xi_{cc}$ state reignited interests in these exotic states since
the $QQq$ and $QQ\bar{q}\bar{q}$ systems may be related to each
other with the heavy quark symmetry (see the following Sec.
\ref{sec:symmetry}). In this subsection, we mainly concentrate on
recent developments in studies with CMI models since 2003.

In Ref. \cite{Cui:2006mp}, Cui {\it et al} investigated the mass
spectrum of the $QQ\bar{q}\bar{q}$ tetraquark states with Eq.
\eqref{mass} using the mass of $X(3872)$ as input. The obtained mass
of the most interesting $T_{cc}$ state with $I(J^P)=0(1^+)$ is 3786
MeV. This value is about 100 MeV below the $DD^*$ threshold and thus
this state should be narrow. The $I(J^P)=0(1^+)$ $T_{bb}$ is also a
stable state. The results in Ref. \cite{Abud:2009rk} obtained with
Eq. \eqref{mass} indicate that the lowest $1^+$ $T_{cc}$ state is
also below the $DD^*$ threshold.

In Refs. \cite{Lee:2007tn,Lee:2009rt}, Lee {\it et al} argued that
the $I(J^P)=0(1^+)$ $T_{cc}$, $T_{bb}$, and $T_{cb}$ states may be
stable against rearrangement decays by using a diquark model with
Eq. \eqref{massref}. In Refs. \cite{Hyodo:2012pm,Hyodo:2017hue}, the
masses of two $I(J^P)=0(1^+)$ $T_{cc}$ states were estimated using
the model \eqref{massref}, one with the color structure
$[cc]_{\bar{3}_c}[\bar{u}\bar{d}]_{3_c}$ and the other with
$[cc]_{6_c}[\bar{u}\bar{d}]_{\bar{6}_c}$. The lower state is 71 MeV
below the $DD^*$ threshold and the higher is 54 MeV above the $DD^*$
threshold. Their productions in $e^+e^-$ collisions were also
investigated.

We have performed a systematic investigation of the
$qq\bar{Q}\bar{Q}$ ($Q=c,b$, $q=u,d,s$) states with the CMI model
\eqref{massref} in Ref. \cite{Luo:2017eub}, where the mixing effects
of different color-spin structures were included. Considering the
uncertainties in the estimation method, the mass of the
$I(J^P)=0(1^+)$ $ud\bar{c}\bar{c}$ state should be larger than 3780
MeV but not exceed 4007 MeV, that of $ud\bar{b}\bar{b}$ should be
between 10483 MeV and 10686 MeV, that of the lowest $I(J^P)=0(0^+)$
$ud\bar{c}\bar{b}$ should be between 7041 MeV and 7256 MeV, and that
of the lowest $0(1^+)$ $ud\bar{c}\bar{b}$ should be between 7106 MeV
and 7321 MeV. Most theoretical predictions fell into these ranges,
see comparison tables in Refs.~\cite{Ebert:2007rn,Luo:2017eub}. Our
results indicate that the mixing effects for the $ud\bar{c}\bar{c}$
systems are not so significant. If our lower bounds on the masses
were reasonable, the $T_{cc/bb/cb}^{I=0}$ and
$ns\bar{c}\bar{c}/ns\bar{b}\bar{b}/ns\bar{c}\bar{b}$ states with
$J^P=1^+$ as well as $T_{cb}^{I=0}$ and $ns\bar{c}\bar{b}$ with
$J^P=0^+$ are probably stable. Our results for
$ud\bar{Q}_1\bar{Q}_2$ ($Q=c,b$) are consistent with an early study
performed in Ref. \cite{SilvestreBrac:1992mv}. Namely, such states
with $I(J^P)=0(1^+)$ are bound. The binding results in these
references~\cite{SilvestreBrac:1992mv,Luo:2017eub} for the
$ud\bar{c}\bar{b}$ with $I(J^P)=0(0^+)$ and $us\bar{b}\bar{b}$ with
$I(J^P)=1/2(1^+)$ are also consistent. If the mass of the realistic
$T_{cc}^{I=0}$ satisfies the inequality \eqref{massrange} and is
closer to 3780 MeV than to 4007 MeV, this tetraquark seems to be
around the $DD^*$ threshold, a situation similar to the $X(3872)$.

Two investigations based on heavy quark symmetry analysis also show
the possible existence of the doubly heavy tetraquark states
\cite{Eichten:2017ffp,Mehen:2017nrh}, which are partly motivated by
LHCb observation of the double-charm baryon $\Xi_{cc}(3621)$
\cite{Aaij:2017ueg} since the role of the $cc$ diquark within
$T_{cc}$ is similar to that in $\Xi_{cc}(3621)$. Later, we will
further introduce this issue in Sec. \ref{sec:symmetry}.

Shortly after the LHCb observation of the $\Xi_{cc}(3621)$
\cite{Aaij:2017ueg}, also motivated by the success of the model
\eqref{CMImodel-junction} in predicting the $\Xi_{cc}$ mass
\cite{Karliner:2014gca}, Karliner and Rosner investigated the masses
of the $cc\bar{u}\bar{d}$, $bb\bar{u}\bar{d}$, and
$bc\bar{u}\bar{d}$ in Ref. \cite{Karliner:2017qjm}. They noticed
that the $I(J^P)=0(1^+)$ $cc\bar{u}\bar{d}$ is just about 7 MeV
above the $D^0D^{*+}$ threshold, the $I(J^P)=0(1^+)$
$bb\bar{u}\bar{d}$ is about 215 MeV below the $B^-\bar{B}^{*0}$
threshold, and the lowest $I(J^P)=0(0^+)$ $bc\bar{u}\bar{d}$ is 11
MeV below the $\bar{B}^0D^0$ threshold. Therefore, they concluded
that there are no stable $cc\bar{u}\bar{d}$. But there are
definitely stable $bb\bar{u}\bar{d}$ states. The lifetime of the
$bb\bar{u}\bar{d}$ was estimated to be 367 fs.

In the diquark-antidiquark model \eqref{CMImodel-MaianiI}, the
authors of Ref. \cite{Yan:2018gik} studied the spectra and decay
properties of the $qq\bar{c}\bar{c}$ ($q=u,d,s$) states. For the
lowest $ud\bar{c}\bar{c}$, its quantum numbers are $J^P=1^+$ and its
mass is 3.6 GeV. This state is 140 (270) MeV below the $DD$ ($DD^*$)
threshold. For the other $P=+$ $ud\bar{c}\bar{c}$ states, the masses
are 3.94 GeV, 3.97 GeV, and 4.04 GeV for the $0^+$, $1^+$, and
$2^+$, respectively. The masses of the $1^-$ $ud\bar{c}\bar{c}$
states range between 3.82 GeV and 4.14 GeV. The masses of their
partner states, $ns\bar{c}\bar{c}$ and $ss\bar{c}\bar{c}$, can also
be found in this work. In addition to the study of the mass
spectrum, the authors discussed their $(q\bar{c})$-$(q\bar{c})$
meson-meson and $(\bar{c}\bar{q}\bar{q})$-$(qqq)$ antibaryon-baryon
decay modes, too. However, their subsequent calculation with Eq.
\eqref{mass} gives an above-threshold $T_{cc\bar{u}\bar{d}}$ and a
below-threshold $T_{bb\bar{u}\bar{d}}$ by adopting the
diquark-antidiquark configuration \cite{Xing:2018bqt}.

\subsection{CMI and $QQ\bar{Q}\bar{Q}$}

Generally speaking, many $XYZ$ exotic states have been observed in
particle experiments, some of which are good tetraquark candidates,
as illustrated in the hidden-heavy case. However, it is still
difficult to distinguish the compact multiquark picture from the
molecular picture once two or more light quarks are involved in the
states. For the full-heavy $QQ\bar{Q}\bar{Q}$ ($Q=c,b$) states,
there does not exist appropriate binding mechanism if they are
treated as loosely bound molecules. If such a state is observed, it
can be identified as a compact tetraquark state bound by short-range
gluon-exchange interactions. In the literature, one may also find
lots of theoretical studies on the heavy-full tetraquark states with
various methods, such as MIT bag model and quark potential model
\cite{Iwasaki:1975pv,Iwasaki:1976cn,Iwasaki:1977qw,Chao:1980dv,Ader:1981db,Ballot:1983iv,Heller:1985cb,Badalian:1985es,Lipkin:1986dw,Zouzou:1986qh,Heller:1986bt,Kalashnikova:1988bx,SilvestreBrac:1992mv,SilvestreBrac:1993ss}.
Similar to the $T_{cc}$ studies, controversial conclusions on their
existence were drawn when different models were applied. In recent
years, experimental developments in search of exotic states
reignited our interests in the fully heavy tetraquark states. Here,
we focus on recent theoretical studies on the $QQ\bar{Q}\bar{Q}$
states with CMI models.

In Refs. \cite{Berezhnoy:2011xy,Berezhnoy:2011xn}, the mass spectra
of the diquark-antidiquark $[cc][\bar{c}\bar{c}]$,
$[bb][\bar{b}\bar{b}]$, and $[bc][\bar{b}\bar{c}]$ states were
studied with the Hamiltonian \eqref{CMImodel-MaianiI}. The authors
determined the coupling parameters by solving the nonrelativistic
Schr\"odinger equation \cite{Kiselev:2002iy}. The lowest
$cc\bar{c}\bar{c}$ and $bc\bar{b}\bar{c}$ with $J^{PC}=2^{++}$ were
found to be above the $J/\psi J/\psi$ and $\Upsilon J/\psi$
thresholds, respectively. But the lowest states with other quantum
numbers are all below relevant meson-meson thresholds. These states
may be stable.

In Ref. \cite{Wu:2016vtq}, we have systematically investigated the
mass spectrum of the $QQ\bar{Q}\bar{Q}$ ($Q=c,b$) compact tetraquark
states with the method \eqref{massref}. 
Our estimated values are lower bounds on realistic tetraquark
masses, and it seems that no stable $QQ\bar{Q}\bar{Q}$ states exist.
From our study, the only possible states, which are stable or
relatively narrow, are $bb\bar{b}\bar{c}$ and $bc\bar{b}\bar{c}$.
The unbinding result for the $cc\bar{b}\bar{b}$ states is consistent
with an early CMI calculation in Ref. \cite{SilvestreBrac:1992mv}.
Our conclusions are consistent with those drawn from dynamical
studies within constituent quark models (see Sec. \ref{sec:CQM}).
For example, the authors of Ref. \cite{Liu:2019zuc} concluded that
no narrow bound $QQ\bar{Q}\bar{Q}$ states are expected in
experiments. For comparison, their mass values also satisfy the mass
inequality \eqref{massrange}. Since their masses are closer to our
$max(M_{\eqref{massref}})$ than to $M_{\eqref{mass}}$, which feature
is similar to the $cs\bar{c}\bar{s}$ case, the results in this
potential calculation are probably closer to the realistic case.

In Ref. \cite{Karliner:2016zzc}, Karliner {\it et al} studied the
$QQ\bar{Q}\bar{Q}$ ($Q=c,b$) states with the model
\eqref{CMImodel-junction} motivated by the QCD-string-junction
picture in the diquark-antidiquark configuration. The predicted
lowest-lying $cc\bar{c}\bar{c}$ state with $J^{PC}=0^{++}$ has a
mass $6192\pm 25$ MeV, which is just below the $J/\psi J/\psi$
threshold but can decay to $\eta_c\eta_c$. The predicted lowest
$bb\bar{b}\bar{b}$ has a mass $18826\pm 25$ MeV, which is just 28
MeV above the $\eta_b\eta_b$ threshold and may have a width narrow
enough to be observed.

Last year, LHCb searched for a possible exotic meson
$X_{b\bar{b}b\bar{b}}$ in the $\Upsilon(1S)\mu^+\mu^-$ invariant
mass distribution in Ref. \cite{Aaij:2018zrb}. Such a study was
motivated by theoretical predictions and by the observation of
$\Upsilon\Upsilon$ production by CMS \cite{Khachatryan:2016ydm}.
However, no significant excess was found in the mass range between
17.5 GeV and 20 GeV. The authors of Ref. \cite{Esposito:2018cwh}
studied a possible $bb\bar{b}\bar{b}$ di-bottomonium with the
diquark-antidiquark picture \eqref{CMImodel-MaianiII}. The obtained
mass is approximately 100 MeV below the $\Upsilon\Upsilon$
threshold. They argued that the decay of this state into
$\Upsilon\mu\mu$ is unlikely to be observed at LHC, which is
consistent with the LHCb experiment~\cite{Aaij:2018zrb}.

\subsection{CMI and $QQ\bar{Q}\bar{q}$}

If the $QQ\bar{Q}\bar{Q}$ states do exist, the gluon-exchange
interactions would provide enough binding forces for heavy quarks,
and such interactions could also bind three heavy quarks and one
light quark into compact $QQ\bar{Q}q$ ($Q=c,b$, $q=u,d,s$) states.
In such systems, no long-range particle-exchange force exists,
neither. Among these states, the $cc\bar{b}\bar{q}$ and
$bb\bar{c}\bar{q}$ tetraquarks are explicitly exotic. Although the
$Qc\bar{c}\bar{q}$ or $Qb\bar{b}\bar{q}$ tetraquarks look like the
excited $D$, $B$, $D_s$, or $B_s$ mesons, their excitation energy
may not be explained by simple orbital or radial excitations if such
states were observed in the future experiments. Similar arguments
were used in the prediction of the hidden-charm pentaquarks
\cite{Wu:2010jy}. Therefore, the exotic tetraquark nature of the
$QQ\bar{Q}q$ would be easily identified once observed. Up to now,
such systems are poorly studied.

More than twenty years ago, Silvestre-Brac investigated various
tetraquark systems with the CMI method \eqref{massref}
\cite{SilvestreBrac:1992mv}. For the $QQ\bar{Q}\bar{q}$ systems, no
bound states with $J=0$ but weakly bound $cc\bar{b}\bar{n}$,
$bc\bar{b}\bar{n}$, and $bc\bar{c}\bar{n}$ states with $J=2$ were
found.

In addition to the $Qq\bar{Q}\bar{q}$ and $QQ\bar{q}\bar{q}$ states,
the authors of Ref. \cite{Cui:2006mp} also presented results for the
$QQ\bar{Q}\bar{q}$ states using the mass equation \eqref{mass}. From
the obtained masses, many states, e.g. the $J^P=2^+$
$qc\bar{c}\bar{c}$, $qb\bar{c}\bar{c}$, and $qc\bar{b}\bar{b}$, were
found to be below their respective lowest threshold of rearrangement
decays and should be narrow.

We have performed a systematic study of the $QQ\bar{Q}\bar{q}$
systems in Ref. \cite{Chen:2016ont} with the CMI method
\eqref{massref}. Our results indicate that the lowest tetraquarks
are only near-threshold bound or resonant states. Since our
estimated values are just lower bounds on their realistic masses,
all such tetraquarks are probably not stable. However, there may
exist relatively narrow and thus detectable states such as
$bc\bar{b}\bar{q}$ and $bc\bar{c}\bar{q}$ if the interactions inside
the diquarks are attractive, although they are above the thresholds
for the rearrangement decay channels.

\subsection{CMI and $Q\bar{Q}qqq$}

In 2015, LHCb observed two pentaquark-like resonances $P_c(4380)$
and $P_c(4450)$ with opposite $P$-parities in the decay
$\Lambda_b^0\to J/\psi pK^-$ \cite{Aaij:2015tga}. After that, in
Ref. \cite{Aaij:2016phn}, the LHCb collaboration further applied a
model-independent method to the same sample and the result supports
their previous model-dependent observation. Later in Ref.
\cite{Aaij:2016ymb}, a full amplitude analysis of $\Lambda_b^0 \to
J/\psi p \pi^-$ decays also supports the evidence for exotic hadron
contributions from the $P_c(4380)$, $P_c(4450)$, and $Z_c(4200)$.
Besides the LHCb, the CLAS12 at JLab may be a possible place to
observe the $P_c$ states in $J/\psi$ photoproduction
\cite{Blin:2016dlf,Kubarovsky:2016whd}. Its spin and photocouplings
may be measured with the future data, too. Now, the search for the
LHCb pentaquark in the photo-production process in Hall C at
Jefferson Lab has been approved
\cite{Meziani:2016lhg,Joosten:2018gyo}. In future, these two $P_c$
states and their partner states can also be searched for by
PANDA/FAIR \cite{Lutz:2009ff}, EIC, and JPARC, etc.

Before the observation of the LHCb $P_c$ states, many studies had
been performed on the existence of possible hidden-charm pentaquarks
\cite{Wu:2010jy,Wu:2010vk,Wang:2011rga,Yang:2011wz,Yuan:2012wz,Wu:2012md,Garcia-Recio:2013gaa,Xiao:2013yca,Huang:2013mua,Li:2014gra,Wang:2015xwa,Uchino:2015uha,Garzon:2015zva}.
Here we show several examples. In Refs. \cite{Wu:2010jy,Wu:2010vk},
Wu {\it et al} predicted the hidden-charm $N^*$ and $\Lambda^*$
resonances above 4 GeV in the molecule picture. In studying the
predicted $nnnc\bar{c}$ and $udsc\bar{c}$ in Ref.
\cite{Yuan:2012wz}, the authors adopted three kinds of quark-quark
hyperfine interactions, the flavor-spin interaction based on the
meson exchange, the CMI based on one-gluon exchange, and the
instanton-induced interaction based on the non-perturbative QCD
vacuum structure. In Ref. \cite{Huang:2013mua}, the discovery
potential of the hidden-charm baryon resonances via photoproduction
was investigated. In Ref. \cite{Garzon:2015zva}, the effects of the
hidden charm $N^*$ in the $\pi^-p\to D^-\Sigma_c^+$ reaction near
threshold were discussed. In Ref. \cite{Hofmann:2005sw}, there were
some discussions about meson-baryon type hidden-charm pentaquarks
with the help of flavor $SU(4)$ symmetry. After the LHCb
observation, more and more studies appear on the nature of the $P_c$
states and relevant predictions. In Ref. \cite{Chen:2015loa}, the
$P_c(4380)$ and $P_c(4450)$ were interpreted as the loosely bound
$\Sigma_cD^*$ and $\Sigma_c^*D^*$ molecules, respectively, in the
one-boson-exchange model. In Ref. \cite{Chen:2015moa}, the quantum
numbers of these two states were suggested to be $J^P=3/2^-$ and
$5/2^+$, respectively. In Ref. \cite{Kopeliovich:2015vqa}, the
masses of the pentaquarks with hidden beauty and strangeness were
simply estimated with the heavy quark symmetry. For other studies of
the hidden-charm pentaquark states before 2016, one may consult
Refs. \cite{Chen:2016qju,Chen:2016heh}. Here we concentrate on
recent developments with CMI models.

In Ref. \cite{Maiani:2015vwa}, Maiani {\it et al} argued that the
observed pentaquarks by LHCb are naturally expected
diquark-diquark-antiquark states in an extended picture of Refs.
\cite{Maiani:2004vq,Ali:2009pi}. They assigned the compositions for
the two $P_c$ states as
\begin{eqnarray}
P(3/2^-)&=&\{\bar{c}[cq]_{S=1}[q^\prime q^{\prime\prime}]_{S=1},L=0\},\nonumber\\
P(5/2^+)&=&\{\bar{c}[cq]_{S=1}[q^\prime
q^{\prime\prime}]_{S=0},L=1\}.
\end{eqnarray}

In Ref. \cite{Anisovich:2015cia}, the pentaquarks were assumed as
$\bar{c}[cu]_{\bar{3}_c}[ud]_{\bar{3}_c}$ diquark-diquark-antiquark
states where all possible spins of diquarks were considered. Their
masses were estimated with the formula taking into account the spin
splitting,
\begin{eqnarray}
M_{P_c}\simeq
m_{\bar{c}}+M_{[cu][ud]}=m_{\bar{c}}+m_{[cu]}+m_{[ud]}+J_{[cu][ud]}(J_{[cu][ud]}+1)\Delta,
\end{eqnarray}
where $M_{[cu][ud]}$ and $J_{[cu][ud]}$ are the mass and angular
momentum of the diquark-diquark system. According to the obtained
pentaquark masses in this analysis, the $P_c(4450)$ is a good
candidate of $I(J^P)=1/2(5/2^-)$ diquark-diquark-antidiquark state,
but the $P_c(4380)$ was suggested to be a broad bump in the
$3/2^+$-wave resulting from rescatterings in the $pJ/\psi$ channel.
This scheme differs from the classification in Refs.
\cite{Maiani:2015vwa,Ali:2016dkf}. In Ref. \cite{Anisovich:2015xja},
the contributions from the possible $1^+$ $[cn][\bar{c}\bar{s}]$
diquark-antidiquark states \cite{Anisovich:2015caa} with masses
around 4189$\sim$4300 MeV in the $K^-J/\psi$ channel of the
$\Lambda_b^0\to K^-J/\psi p$ decay were discussed. The authors
emphasised that such exotic mesons can imitate broad bumps in the
$pJ/\psi$ channel. The study of the pentaquark spectrum was extended
to the strange $c\bar{c}$ case in Ref.  \cite{Anisovich:2015zqa}.
Further discussions in this scheme can be found in Refs.
\cite{Anisovich:2017aqa,Anisovich:2017ubd}.

In Ref. \cite{Zhu:2015bba}, Zhu and Qiao studied the pentaquark
states in the diquark-triquark picture where both diquark and
triquark are not compact objects. Their investigated pentaquarks
have the structure
$\delta\bar{\theta}=[Qu]_{\bar{3}_c}[ud\bar{Q}]_{3_c}$ or
$\delta\bar{\theta}=[Qu]_{\bar{3}_c}[us\bar{Q}]_{3_c}$ ($Q=c,b$).
Their masses were calculated with the model Hamiltonian
\eqref{CMImodel-Qiao}, where the diquark masses were extracted from
the $1^{++}$ $X(3872)$ and the $1^{--}$ $Y_b(10890)$ by assuming that
they are diquark-antidiquark states, and the triquark masses were
estimated by summing the quark masses. Their results suggest that
the $P_c(4380)$ is a mixed state of three diquark-triquark
structures with $J^P=3/2^-$, and the $P_c(4450)$ can be interpreted
as the diquark-triquark pentaquark state
$\{[cu]_{S=1}[(ud)_{S=0}\bar{c}]_{S=1/2}\}_{L=1}$ with $J^P=5/2^+$.
The authors also predicted the spectra of $[cu][us\bar{c}]$,
$[bu][us\bar{b}]$, and $[bu][us\bar{b}]$, and discussed possible
production and decay processes for these pentaquark states.

In the $\bar{c}{\cal Q}{\cal Q}^\prime=\bar{c}[cq][q^\prime
q^{\prime\prime}]$ ($q=u,d,s$) diquark-diquark-antiquark picture,
Ali {\it et al} investigated the pentaquark spectrum using the
extended model of \eqref{CMImodel-MaianiII} in Ref.
\cite{Ali:2016dkf},
\begin{eqnarray}
H=H_{[{\cal Q}{\cal Q}^\prime]}+H_{\bar{c}[{\cal Q}{\cal
Q}^\prime]}+H_{S_{\cal P}L_{\cal P}}+H_{L_{\cal P}L_{\cal P}},
\end{eqnarray}
where $L_{\cal P}$ and $S_{\cal P}$ are the orbital and spin angular
momenta of the pentaquark state, respectively. They also adopted the
extended model \eqref{CMImodel-MaianiI} by including additional
spin-spin terms. Identifying the observed $P_c(4450)$ as the state
$P(5/2^+)=\{\bar{c}[cu]_{S=1}[ud]_{S=0},L=1\}$, the masses of the
other $S$- and $P$-wave pentaquark states were predicted. According
to their results, identifying the state
$P(3/2^-)=\{\bar{c}[cu]_{S=1}[ud]_{S=1},L=0\}$ to be the observed
$P_c(4380)$ is problematic because its production from the
$\Lambda_b^0$ decay is suppressed by the heavy quark symmetry. They
suggest the LHCb Collaboration reanalyze their data to search for
the lower $J^P=3/2^-$ pentaquark in the range 4140-4130 MeV. In Ref.
\cite{Ali:2017ebb}, the same authors extended their studies of the
$J^P=3/2^-$ and $J^P=5/2^+$ pentaquark states to the $J^P=1/2^\pm$
cases. The anticipated discovery modes in $b$-baryons decays were
also discussed.

In Ref. \cite{Wu:2017weo}, we have systematically studied the
$Q_1\bar{Q}_2q_3q_4q_5$ ($Q=c,b$, $q=u,d,s$) pentaquark states with
the method \eqref{massref}, by assuming $Q_1\bar{Q}_2$ to be in the
color $8$ representation. The LHCb $P_c$ states fall in the mass
region of the studied $nnnc\bar{c}$ system. Most such pentaquark
states were found to have $S$-wave open-heavy decays and their
widths should be broad. In contrast, the $J^P=5/2^-$ states do not
decay through the $S$-wave, so their widths should not be very
broad. We also found low-mass $\Lambda$-type $udsQ_1\bar{Q}_2$
states with $J^P=1/2^-$ in this model. One may understand the
existence of such a state from the CMI matrix elements. In the
flavor $SU(3)$ symmetric limit, two states in the $1_f$
representation and three states in the $8_f$ representations have
negative CMI matrix elements. Their mixing due to the symmetry
breaking effectively provides additional attraction, which leads to
a low-mass $\Lambda$-type pentaquark finally. This result agrees
with the study in Ref. \cite{Irie:2017qai} where detailed
investigations with one-gluon exchange and instanton-induced
interactions were performed.

In the study of hidden-charm pentaquark states, Buccella discussed
the masses of the compact $c\bar{c}uud$ with the method similar to
\eqref{massref} in Ref. \cite{Buccella:2018jnt}, where both $cc$ and
$uud$ are color octet states. For the $J^P=3/2^-$ case, he predicted
four pentaquarks around 4360 MeV, 4409 MeV, 4491 MeV, and 4560 MeV.
The first two states couple strongly to $J/\psi p$, and they appear
as a resonance with a mass 4380 MeV in this channel, while the third
and forth states couple strongly to $\Lambda_c\bar{D}^*$ and
$\Sigma_c\bar{D}^*$, respectively. For the $J^P=5/2^+$ $P_c(4450)$,
the two color octets were assumed to be in $P$-wave.

In Ref. \cite{Santopinto:2016pkp}, the authors assigned the observed
$P_c(4380)$ as a compact $uudc\bar{c}$ pentaquark with $J^P=3/2^-$
in the lowest $SU_f(3)$ multiplet and considered its flavor-octet
partner states using the extended G\"ursey-Radicati (GR) formula
\cite{Gursey:1992dc},
\begin{eqnarray}
M_{GR}=M_0+AS(S+1)+DY+E[I(I+1)-\frac14Y^2]+GC_2(SU(3))+FN_C,
\end{eqnarray}
where $M_0$ is a scale parameter, $I$ is the isospin, $Y$ is the
hypercharge, $C_2(SU(3))$ is the eigenvalue of the $SU_f(3)$ Casimir
operator, and $N_c$ is a counter of $c$ or $\bar{c}$ quark. With the
obtained parameters which were determined by fitting the masses of
the ground state baryons, the authors found a state with the mass
4377 MeV and the same quantum numbers as the $P_c(4380)$ (charge,
spin, parity). Possible discovery channels in bottom baryon decays
and the partial decay widths for all the pentaquarks were discussed.
In Ref. \cite{Ortiz-Pacheco:2018ccl}, an extended $SU(4)$ quark
model was developed. The mass formula for baryons has the form
\begin{eqnarray}
M^2=\Big(\sum_i m_i\Big)^2+M_{sf}^2+M_{orb}^2,
\end{eqnarray}
where the spin-flavor operator $M_{sf}^2$ has a generalized form of
the GR formula and the orbital contribution term has the form
$M_{orb}^2=\alpha^\prime L$ with $\alpha^\prime$ being the linear
trajectory slope. Applying the formula to the $J^{P}=3/2^-$ $L=0$
$qqqc\bar{c}$ ($q=u,d,s$) states, the authors found that all the
hidden-charm pentaquarks are below the relevant $J/\psi$-baryon
thresholds. Therefore, the LHCb $P_c$ states were proposed to be the
excited pentaquarks within this scheme.

\subsection{CMI and $QQqq\bar{q}$ }

Up to now, the $\Xi_{cc}$ baryon is the unique one of the
experimentally confirmed doubly heavy hadrons. Its observation
motivated heated discussions on the possible $QQ\bar{q}\bar{q}$
($Q=c,b$, $q=u,d,s$) tetraquarks. If one adds one more light quark,
the existence of the $QQqq\bar{q}$ pentaquark baryons might also be
possible, because the light quark interactions inside the
$qq\bar{q}$ cluster may provide stronger attractions and play an
important role in binding the five quarks, especially when the
color-spin mixing effects are considered. Compared with the
conventional $QQq$ baryons, the $QQqq\bar{q}$ masses are probably
not very large if the complicated interactions among quark
components significantly lower their masses. As a result, the
properties of the excited $QQq$ baryons may be affected by the
$QQqq\bar{q}$ states through coupled channel effects. Knowledge
about the basic features for the $QQqq\bar{q}$ spectra may be
helpful for us to understand possible structures of heavy hadrons.
In the literature, there exist some explorations of such systems
which focus on the $(Qqq)$-$(Q\bar{q})$ molecules. However, such
systems have not been paid enough attention to.

In Ref. \cite{Zhou:2018pcv}, we have systematically estimated the
masses of the possible $QQqq\bar{q}$ pentaquark masses in the CMI
method \eqref{massref}. Our results suggest that there exist many
states below relevant open-heavy baryon-meson thresholds, and thus
stable pentaquarks are expected. Since the realistic pentaquarks
probably have higher masses than the values in Ref.
\cite{Zhou:2018pcv}. It is not very clear where the
$(QQq)$-$(q\bar{q})$ type baryon-meson thresholds locate since most
of the $(QQq)$ baryons have not been observed yet. Further study is
needed to verify whether such states are stable or not. Our results
can be a helpful guide for deeper investigations.

To search for stable pentaquark states, the authors of Ref.
\cite{Park:2018oib} presented a systematic analysis on the
flavor-color-spin structures of the heavy quark pentaquark systems
with the method \eqref{massref}. The authors noticed that the
$I(J^P)=0(1/2^-)$ $udcc\bar{s}$ state is perhaps the most stable
one.

\subsection{CMI and $QQQq\bar{q}$ }

The $QQQq\bar{q}$ system is the mirror-type structure of the
$Q\bar{Q}qqq$ by exchanging the heavy and light flavors. Such states
are related with the triply heavy conventional baryons, which have
not been observed so far. This situation is similar to the relation
between $Q\bar{Q}$ mesons and $Q\bar{Q}q\bar{q}$ tetraquark states.
With the potential observations of $QQQ$ baryons in future
measurements, it is also instructive to study the feature of the
$QQQq\bar{q}$ states. Besides, the study of the bound state and
scattering problems induced by the observed $\Xi_{cc}$ baryons can
help experiments to look for exotic phenomena.

In Ref. \cite{Li:2018vhp}, we have discussed whether the compact
$QQQq\bar{q}$ pentaquark states are possible with the CMI method
\eqref{massref} by assuming that the $QQQ$ is always a color-octet
state, similar to Ref. \cite{Wu:2017weo}. Our results indicate that
the compact pentaquarks should not be stable against their
rearrangement decays. On the other hand, the near-threshold
baryon-meson molecules are possible, see Sec.
\ref{sec:mesonexchange}. Therefore, the gap between the lowest
threshold and the lowest compact pentaquark leaves room for the
identification of hadronic molecules. If experiments could observe
an exotic resonance around some low-lying threshold, its nature as a
molecule will be preferred over a compact pentaquark. As a
by-product, we conjectured the maximum values for the realistic
masses of the conventional triply heavy baryons, which are
consistent with most theoretical predictions in the literature.

\subsection{Mass constraints for $QQq$ and $QQQ$ baryons}

With the simplest CMI model \eqref{mass}, one cannot get accurate
masses for conventional hadrons because of the adopted assumption,
which has been illustrated in Sec. \ref{CMI:conventionalhadrons}.
The estimated multiquark masses with the model or its variants might
also be far from the realistic values. Irrespective of the results,
however, we find that some mass constraints for conventional hadrons
$QQq$ and $QQQ$ may be conjectured when estimating multiquark masses
with the CMI method \eqref{massref}.

In some cases, there may be two or more reference thresholds when
one discusses a system with the help of Eq. \eqref{massref}.
Obviously, they lead to different multiquark masses. Then one may
establish an inequality for the thresholds of these reference
systems. If one reference system contains a hadron without measured
mass, the inequality can be used to constrain the mass of that
hadron, e.g. $QQq$ or $QQQ$. For convenience, we use $H_1$ and $H_2$
($H_1^\prime$ and $H_2^\prime$) to denote the two hadrons in the
reference system $R$ ($R^\prime$). What we observed from the results
for various systems is that the multiquark masses estimated with $R$
are usually higher than those with $R^\prime$, if both $H_1$ and
$H_2$ contain heavy quarks while $H_1^\prime$ or $H_2^\prime$ does
not contain heavy quarks, i.e.
\begin{eqnarray}
M_{H_1}+M_{H_2}-\langle H_{CMI}\rangle_{H_1}-\langle
H_{CMI}\rangle_{H_2}> M_{H_1^\prime}+M_{H_2^\prime}-\langle
H_{CMI}\rangle_{H_1^\prime}-\langle H_{CMI}\rangle_{H_2^\prime}.
\end{eqnarray}
If $H_2^\prime$ does not contain heavy quarks, that means
\begin{eqnarray}
M_{H_1^\prime}< \Big[M_{H_1}-\langle
H_{CMI}\rangle_{H_1}\Big]+\Big[M_{H_2}-\langle
H_{CMI}\rangle_{H_2}\Big]-\Big[M_{H_2^\prime}-\langle
H_{CMI}\rangle_{H_2^\prime}\Big]+\langle
H_{CMI}\rangle_{H_1^\prime}.
\end{eqnarray}
For example, in estimating the masses of the $cccs\bar{s}$ states,
the reference systems may be $\Omega_{cc}D_s$ or $\Omega_{ccc}\phi$,
and one expects that the latter system leads to lower pentaquark
masses. Then we get the upper bound for the mass of $\Omega_{ccc}$
(see Table \ref{maxQQQmasses}). If another system is considered,
e.g. $cccs\bar{n}$, a different value of mass can be obtained. The
minimum mass should be the upper bound. In Ref. \cite{Li:2018vhp},
we have obtained mass inequalities between the triply heavy baryons
and the doubly heavy baryons with this feature, which are also
collected in Table \ref{maxQQQmasses}.

With the help of Ref. \cite{Luo:2017eub}, we may similarly constrain
the masses of doubly heavy baryons with $Qqq$ baryons and $Q\bar{q}$
mesons. Such inequalities are also shown in Table
\ref{maxQQQmasses}. Numerically, the mass within this approach is
smaller than the upper limit with Eq. \eqref{mass}, which is
illustrated by comparing masses in Tables \ref{CMI-hadrons} and
\ref{maxQQQmasses}. Up to now, only the doubly charmed baryon
$\Xi_{cc}$ was discovered by LHCb and no other baryons containing
two or more heavy quarks are observed. It is obvious that the
measured mass of $\Xi_{cc}$ is consistent with the upper bound for
$\Xi_{cc}$. For other baryon states, one may compare the constraints
given here with the masses in the literature
\cite{Karliner:2014gca,Wei:2015gsa,Wei:2016jyk,Weng:2018mmf,Li:2018vhp}.
We find most of the theoretical predictions satisfy such
constraints.

Since the masses in the second column of Table \ref{maxQQQmasses}
are physical values, these constraints can be tested only after all
the relevant ground-state baryons are observed. In this sense these
inequalities are just conjectures. With these constraints, we may
judge whether an observed heavy baryon is a ground state or not,
{\it i.e.}, once experiments observed states higher than such
bounds, they should not be ground states.

\begin{table}[htbp]
\centering
\caption{The conjectured constraints on the masses of ground-state
triply and doubly heavy baryons, where the hadron symbols in the
second column represent their masses. The estimations for the upper
limits are given in units of MeV. We have used the estimations for
the upper limits of doubly heavy baryons when obtaining those for
triple heavy baryons.}\label{maxQQQmasses}
\vspace{.3cm}
\scalebox{0.9}{
\renewcommand{\arraystretch}{1.4}
\begin{tabular}{ccc}\hline \hline
State&Upper limit&Estimation\\ \hline
$\Omega_{ccc}$&$({\Omega_{cc}^*}+{D_s}-\phi)+\frac{16}{3}(C_{s\bar{s}}+3C_{c\bar{s}}+C_{cc}-C_{cs})$&4977\\
$\Omega_{bbb}$&$({\Omega_{bb}^*}+{B_s}-\phi)+\frac{16}{3}(C_{s\bar{s}}+3C_{b\bar{s}}+C_{bb}-C_{bs})$&14937\\
$\Omega_{bbc}^*$&$({\Omega_{bb}^*}+{D_s}-\phi)+\frac{16}{3}(C_{s\bar{s}}+3C_{c\bar{s}}+C_{bc}-C_{bs})$&11610\\
$\Omega_{ccb}^*$&$({\Omega_{bc}^*}+{D_s}-\phi)+\frac{8}{3}(2C_{s\bar{s}}+6C_{c\bar{s}}+C_{cc}+C_{bc}-C_{cs}-C_{bs})$&8285\\
$\Omega_{bbc}$&${\Omega_{bbc}^*}-16C_{bc}$&11577\\
$\Omega_{ccb}$&${\Omega_{ccb}^*}-16C_{bc}$&8253\\
\hline
$\Xi_{cc}$&$({\Xi_c^\prime}+{D_s}-{\phi})+\frac83(C_{cc}-C_{ns}-2C_{cn}+2C_{cs}+2C_{s\bar{s}}+6C_{c\bar{s}})$&3669\\
$\Xi_{bb}$&$({\Xi_b^\prime}+{B_s}-{\phi})+\frac83(C_{bb}-C_{ns}-2C_{bn}+2C_{bs}+2C_{s\bar{s}}+6C_{b\bar{s}})$&10347\\
$\Omega_{cc}$&$(\Omega_c+D_s-\phi)+\frac83(C_{cc}-C_{ss}+2C_{s\bar{s}}+6C_{c\bar{s}})$&3799\\
$\Omega_{bb}$&$(\Omega_b+B_s-\phi)+\frac83(C_{bb}-C_{ss}+2C_{s\bar{s}}+6C_{b\bar{s}})$&10474\\
$\Xi_{bc}^*$&$(\Xi_b^\prime+D_s-\phi)+\frac83(C_{bc}-C_{ns}+C_{cn}+3C_{bn}+2C_{bs}+2C_{s\bar{s}}+6C_{c\bar{s}})$&7048\\
$\Omega_{bc}^*$&$(\Omega_b+D_s-\phi)+\frac83(C_{bc}-C_{ss}+C_{cs}+5C_{bs}+2C_{s\bar{s}}+6C_{c\bar{s}})$&7174\\
$\Xi_{cc}^*$&$\Xi_{cc}+16C_{cn}$&3733\\
$\Xi_{bb}^*$&$\Xi_{bb}+16C_{bn}$&10368\\
$\Omega_{cc}^*$&$\Omega_{cc}+16C_{cs}$&3871\\
$\Omega_{bb}^*$&$\Omega_{bb}+16C_{bs}$&10493\\
$\Xi_{bc}$&$\Xi_{bc}^*-\frac83(2X_n+Y_n)$&6996\\
$\Xi_{bc}^\prime$&$\Xi_{bc}^*-\frac83(2X_n-Y_n)$&7022\\
$\Omega_{bc}$&$\Omega_{bc}^*-\frac83(2X_s+Y_s)$&7117\\
$\Omega_{bc}^\prime$&$\Omega_{bc}^*-\frac83(2X_s-Y_s)$&7149\\ \hline
\hline
\end{tabular}
}
\end{table}

\subsection{Effective CMI for multiquark states}\label{subsec:effectiveCMI}

When studying the properties of the $QQ\bar{Q}\bar{Q}$ tetraquark
states \cite{Wu:2016vtq}, we checked the mass shifts due to
variations of coupling parameters. What the mass shift reflects is
the effective color-spin interaction between a pair of quark
components. For a multiquark state, its mass $M$ in the CMI model
can be expressed as a function of the effective quark masses and
coupling constants
\begin{eqnarray}
M = M(m_n,m_s,m_c,m_b,C_{12},C_{13},\cdots) \, .
\end{eqnarray}
In the case that mixing among different color-spin structures is
considered, this mass function cannot be expressed as a simple
formula like Eq. \eqref{CMI-Casimir}. However, we may approximately
use a linear form to show the relation between $M$ and the
parameters. To do that, we consider effective chromomagnetic
interactions by making a change on the coupling constants. When one
reduces a parameter $C_{ij}$, the mass of the state becomes larger
or smaller. If the effective CMI between the $i$th quark component
and the $j$th quark component is attractive (repulsive), the
reduction of $C_{ij}$ will lead to a larger (smaller) $M$. This
effect can be reflected by a dimensionless measure, which was
defined in Ref. \cite{Li:2018vhp},
\begin{eqnarray}
K_{ij}=\frac{\Delta M}{\Delta C_{ij}} \, .
\end{eqnarray}
Here $\Delta C_{ij}$ is the variation of the coupling parameter
$C_{ij}$, and $\Delta M=\Delta\langle H_{CMI}\rangle$ is the
corresponding variation of the multiquark mass. When $\Delta
C_{ij}\to 0$, $K_{ij}\to \frac{\partial M}{\partial C_{ij}}$ becomes
a constant, then we have
\begin{eqnarray}\label{massmix}
M&=& M_0+\sum_{i<j}K_{ij}C_{ij} \, ,
\end{eqnarray}
where $M_0=\sum_{i=1} m_i$ of Eq. \eqref{mass} or
$M_0=M_{ref}-\langle{H}_{CMI}\rangle_{ref}$ of Eq. \eqref{massref}.
The validity of this mass formula in the mixing case is easy to be
checked with numerical results. Obviously, negative (positive)
$K_{ij}$ indicates attractive (repulsive) effective CMI.

In practice, the value of $K_{ij}$ can be extracted by reducing
$C_{ij}$ slightly, while it can also be derived with the matrix
$\langle H_{CMI}\rangle$ and its eigenvalues. 
Consider a mixed state
\begin{eqnarray}
\psi_{mix}=x_k\varphi_k \, ,
\end{eqnarray}
which is a mixture of different bases $\varphi_k$'s with
corresponding coefficients $x_k$'s. If $E_{CMI}$ represents an
eigenvalue of $H_{CMI}$, one has
\begin{eqnarray}
E_{CMI}=\langle\psi_{mix}|{H}_{CMI}|\psi_{mix}\rangle=x_kx_l\langle\varphi_k|{H}_{CMI}|\varphi_l\rangle.
\end{eqnarray}
The Hamiltonian acting on bases can be expressed as (see Eq.
\eqref{cmiM.E.})
\begin{eqnarray}
\langle\varphi_k|{H}_{CMI}|\varphi_l\rangle=\langle
H_{CMI}\rangle_{kl}=\sum_{m<n}X^{kl}_{mn}C_{mn},
\end{eqnarray}
where $X^{kl}_{mn}$ can be read out from the obtained $\langle
H_{CMI}\rangle$. Then
\begin{eqnarray}
E_{CMI}=x_kx_l\langle\varphi_k|{H}_{CMI}|\varphi_l\rangle =x_kx_l
\langle H_{CMI}\rangle_{kl}=\sum_{m<n}(x_k X^{kl}_{mn}
x_l)C_{mn}=\sum_{m<n}K_{mn}C_{mn},
\end{eqnarray}
{\it i.e.}, $K_{mn}=x_kX^{kl}_{mn}x_l$. With this expression, one
can understand how contributions from different structures affect
the multiquark mass.

Although Eq. \eqref{massmix} has a linear form, it does not mean
that the multiquark mass is linearly related with the coupling
parameters, because $K_{ij}$'s also depend on $C_{ij}$'s. The
effects on the multiquark mass due to uncertainties of coupling
parameters can only be seen with the values of $K_{ij}$ roughly. If
the absolute value of a $K_{ij}$ is large, it is necessary to reduce
the uncertainty of the corresponding $C_{ij}$ as much as possible.
This possibility of large $K_{ij}$ can appear in some multiquark
states. An application of the measure is to qualitatively guess the
stability of tetraquark states. For states with the configuration
$q_1q_2\bar{q}_3\bar{q}_4$, their dominant decays are through the
fall-apart or rearrangement mechanism. However, there still exist
some special configurations which lead to relatively stable states
although their masses may be high. For example, if both the $q_1q_2$
and $q_3q_4$ interactions in a state are effectively attractive
while other quark-antiquark interactions are repulsive, such a state
should be relatively difficult to fall apart into meson-meson decay
channels. From the systematic study of the $Qq\bar{Q}\bar{q}$
($Q=c,b$, $q=u,d,s$) systems \cite{Wu:2018xdi}, the second highest
$J^P=0^+$ states usually satisfy this condition, which is helpful
for us to identify the possible nature of the exotic mesons.
However, it is not easy to qualitatively discuss the relative
stability of pentaquark states with $K_{ij}$'s.

\subsection{A short summary for CMI}

\begin{table}[htbp]
\begin{center}
\caption{All types of heavy quark tetraquark and pentaquark systems
($Q=c,b$, $q=u,d,s$). The states which were discussed in literatures
are given with bold fonts and the systems that we investigated and
reviewed here are marked with a dagger.}\label{typesofheavymulti}
\vspace{.3cm}
\scalebox{0.9}{
\renewcommand{\arraystretch}{1.4}
\begin{tabular}{l|l}\hline
Tetraquarks&Pentaquarks\\
\hline
\bm{$Qq\bar{q}\bar{q}$} & \bm{$\bar{Q}qqqq$}, \bm{$Qqqq\bar{q}$}\\
\bm{$Q\bar{Q}q\bar{q}^\dag$}, \bm{$QQ\bar{q}\bar{q}^\dag$} & \bm{$Q\bar{Q}qqq^\dag$}, \bm{$QQqq\bar{q}^\dag$}\\
\bm{$qQ\bar{Q}\bar{Q}^\dag$} & \bm{$q\bar{q}QQQ^\dag$}, \bm{$qqQQ\bar{Q}$}\\
\bm{$QQ\bar{Q}\bar{Q}^\dag$} & $\bar{q}QQQQ$, $qQQQ\bar{Q}$\\
&$QQQQ\bar{Q}$\\\hline
\end{tabular}
}
\end{center}
\end{table}

The CMI models play an important role in understanding the
multiquark systems. Although the Hamiltonian is simple, the models
do catch the basic features of spectra since the mass splittings of
hadrons reply on the basic controlling symmetries of the quark
world. In recent years, developments about the CMI model
investigations lead to series of important results. All types of
heavy quark tetraquark and pentaquark systems that could exist in
nature are listed in Table \ref{typesofheavymulti}. We have reviewed
the CMI studies on four types of tetraquark systems and three types
of pentaquark systems in this section. Some interesting observations
are:

\begin{itemize}
\item The $Q\bar{Q}q\bar{q}$ states. The $X(4140)$, $X(4274)$, and
$X(4350)$ observed in $J/\psi\phi$, the $X(3860)$ observed in
$D\bar{D}$, and the $Z_c(4100)$ observed in $\eta_c\pi$ can be
consistently assigned as the compact tetraquark states with
$J^{PC}=1^{++}$, $1^{++}$, $0^{++}$, $0^{++}$, and $0^{++}$,
respectively within the framework of the chromomagnetic
interactions. Additional high-lying and relatively narrow tetraquark
states are also expected whose effective chromomagnetic interactions
inside the diquarks are attractive while those between quark and
antiquarks are repulsive. Uncovering the inner structures of the
exotic $XYZ$ states needs more experimental investigations and
further theoretical studies from various aspects.

\item
The $QQ\bar{q}\bar{q}$ states. From the quark mass ratio $m_q/m_Q$
dependence of the $QQ\bar{q}\bar{q}$ masses, the existence of such
doubly heavy tetraquarks becomes inevitable when the ratio is small
enough. From the CMI model calculations, it is still unclear whether
the lowest $cc\bar{u}\bar{d}$ state is bound or not depending on the
details of the quark interactions. An interesting possibility would
be that the mass of the $T_{cc}$ is around the $DD^*$ threshold,
similar to the masses of $X(3872)$ and $Z_c(3900)$. On the other
hand, the existence of the bound $T_{bb}$ seems without doubt. The
$T_{cc}$ and $T_{bb}$ states have only repulsive color-spin
interactions inside the heavy diquarks. In contrast, not only the
lowest but also heavier $bc\bar{u}\bar{d}$ states may exist with
relatively narrow widths because the effective color-spin
interaction inside the $bc$ and $ud$ diquarks can be both
attractive. The existence of the doubly heavy tetraquarks is
certainly one of the most important issues in hadron physics.

\item The $QQ\bar{Q}\bar{Q}$ and $QQ\bar{Q}\bar{q}$ states. The lowest
compact $cc\bar{c}\bar{c}$ and $bb\bar{b}\bar{b}$ states with
$J^{PC}=0^{++}$ from the CMI models seem unstable if all the
color-spin structures are considered. Although the compact
$bc\bar{b}\bar{c}$ and $bb\bar{b}\bar{c}$ states may also have
rearrangement decay patterns, the effectively attractive color-spin
interactions inside the diquarks probably lead to relatively narrow
tetraquarks. Searching for $bc\bar{b}\bar{c}$ states in the
$J/\psi\Upsilon$ channel seems to be feasible at LHC. Moreover, the
relatively narrow $bc\bar{c}\bar{q}$ and $bc\bar{b}\bar{q}$
tetraquarks probably exist.

\item The $Q\bar{Q}qqq$ states. Although the $P_c(4380)^+$ and
$P_c(4450)^+$ states can be assigned as compact $c\bar{c}uud$
pentaquark states in the CMI models, their inner structures still
remain unresolved. An interesting observation is that a low mass
isoscalar $c\bar{c}uds$ pentaquark is possible which may be searched
for in the $J/\psi\Lambda$ and $\eta_c\Lambda$ channels. Similarly,
the isoscalar $b\bar{c}uds$, $c\bar{b}uds$, and $b\bar{b}uds$ states
may be searched for in their rearrangement strong decay channels.

\item
The $QQqq\bar{q}$ and $QQQq\bar{q}$ states. All the $\Xi_{cc}$,
$T_{cc}$, and $ccqq\bar{q}$ (denoted as $P_{cc}$) states contain the
$cc$ diquark. If the $T_{cc}$ were bound, the lowest $QQqq\bar{q}$
pentaquarks might also be bound states. Since the other $QQq$
baryons except $\Xi_{cc}$ have not been observed, it is not clear
whether low mass structures such as isoscalar $bcns\bar{n}$,
$bcnn\bar{s}$, and $bcnn\bar{n}$ have rearrangement decay channels
or not yet. The isoscalar $ccnn\bar{n}$, $ccnn\bar{s}$, and
$ccns\bar{n}$ states are probably the most stable doubly charmed
pentaquark. The triply heavy $QQQq\bar{q}$ ($P_{QQQ}$) structures
seem unstable.

\end{itemize}

%% file: section3.tex
\section{Constituent quark models}\label{sec:CQM}

Various versions of nonrelativistic and relativistic constituent
quark models can be found in the literature, which were proposed to
understand hadron properties. Almost all of them incorporate both
the short-range one-gluon-exchange (OGE) force and the term
representing the color confinement, in either the coordinate or
momentum space. Some of them include the additional flavor-dependent
Goldstone-boson-exchange (GBE) force from the spontaneously broken
chiral symmetry and/or the screening effects from the
quark-antiquark pair creation. There are also models containing only
GBE and confinement potentials. In some approaches, one has to adopt
the diquark and triquark approximations in order to reduce the
few-body problem to a two-body problem. In this section, we focus on
recent studies of compact multiquark states within constituent quark
models. We will also discuss some $XYZ$ states which are candidates of
the conventional or hybrid charmonium.

One of the widely adopted potentials was proposed by Godfrey and
Isgur~\cite{Godfrey:1985xj,Capstick:1986bm}, which was very
successful in explaining conventional meson and baryon spectra. Now
we denote it as the GI model. In studying multiquark states, there
also exist many types of potentials, two of which were very popular.
They were proposed by Bhaduri, Cohler, and Nogami in Ref.
\cite{Bhaduri:1981pn}, and by Semay and Silvestre-Brac in Refs.
\cite{Semay:1994ht,SilvestreBrac:1996bg}. The latter references
provided four potentials corresponding to different fitted
parameters, which we shall call as BCN and Grenoble (AL1, AP1, AL2,
and AP2) in the present review. The potential between $i$th and
$j$th quark components in the BCN model reads
\begin{eqnarray}
V_{ij}(r)=-\frac{3}{16}\vec{\lambda}_i\cdot\vec{\lambda}_j\Big[\frac{\hbar^2\kappa_\sigma}{m_im_jc^2r_0^2}\frac{e^{-r/r_0}}{r}\vec{\sigma}_i\cdot\vec{\sigma}_j-\frac{\kappa}{r}+\frac{r}{a^2}-\Delta\Big],
\end{eqnarray}
while that in the Grenoble model has the form
\begin{eqnarray}
V_{ij}=-\frac{3}{16}\vec{\lambda}_i\cdot\vec{\lambda}_j\Big[\frac{2\pi\kappa^\prime}{3m_im_j}\frac{exp(-r^2/r_0^2)}{\pi^{3/2}r_0^3}\vec{\sigma}_i\cdot\vec{\sigma}_j
-\frac{\kappa}{r}+\lambda r^p-\Lambda\Big]
\end{eqnarray}
with $r_0(m_i,m_j)=A(2m_im_j/(m_i+m_j))^{-B}$.

In the following subsections, we shall separately review the
theoretical progress of various constituent quark models on the
$Q\bar{Q}q\bar{q}$, $QQ\bar{q}\bar{q}$, $QQ\bar{Q}\bar{Q}$,
$QQ\bar{Q}\bar{q}$, and $Q\bar{Q}qqq$ multiquark states.

\subsection{$Q\bar{Q}q\bar{q}$}

Some neutral $XYZ$ states are probably the conventional quarkonium. In
some cases, the bare quarkonium states are strongly affected by the
couple-channel effects. A typical example is the $X(3872)$
\cite{Kalashnikova:2018vkv} which is affected strongly by the
$D\bar{D}^*$ scattering states. In fact, the mixing between the
$Q\bar{Q}$ and $Q\bar{Q}q\bar{q}$ may be accounted for through the
$^3P_0$ quark pair creation in the vacuum, which could be
equivalently incorporated by the screened potential. The $Q\bar{Q}$
core contributions could be calculated even at the hadron level
through Weinberg's compositeness theorem \cite{Weinberg:1965zz}. On
the other hand, the $Z$ states do not have a quarkonium core.

\subsubsection{Possible quarkonium or hybrid assignments}

The authors of Ref. \cite{Deng:2016stx} studied the charmonium
spectrum and electromagnetic transitions between charmonium states
in a nonrelativistic quark potential model where both the linear and
screened confinement potentials are considered. Their results
indicate that the possibility of the $X(3872)$ being a
$\chi_{c1}(2P)$ dominant state cannot be excluded from its radiative
decay properties. See also discussions in Ref.~\cite{Achasov:2016vxb}. 
The $X(3823)$ as the $\psi_2(1D)$ is supported by
its radiative decay properties. The mass splitting between the
$2^3P_2$ state and the $2^3P_1$ state does not support the
assignment of the $X(3915)$ as the $\chi_{c0}(2P)$ state. In the
screened potential model, the $Y(4260)$ and $Y(4360)$ may be good
candidates of the $\psi(4S)$ and $\psi_1(3D)$, respectively. The
$X(4140)$ or $X(4274)$ might be identified as the $\chi_{c1}(3P)$
state. From the subsequent study on the open-charm strong decays of
the higher charmonium states in Ref. \cite{Gui:2018rvv}, the authors
found that it is possible to assign the $Y(4660)$, $X(4500)$,
$X(3940)$, and $X(3860)$ as $\psi(5S)$, $\chi_{c0}(4P)$,
$\eta_c(3S)$, and $\chi_{c0}(2P)$, respectively. The assignment for
the $X(4140)$ and $X(4274)$ in the same model as the excited
$\chi_{c1}$ states is difficult. The adopted model does not lead to
a self-consistent description for the vector $Y(4230/4260,4360)$ and
the scalar $X(4700)$.

In Ref. \cite{Ortega:2016hde}, the authors investigated possible
assignments for the four $J/\psi\phi$ structures in a coupled
channel scheme by using a nonrelativistic constituent quark model
\cite{Vijande:2004he,Segovia:2008zz}. They found that the $X(4140)$
seems to be a cusp because of the near coincidence of the
$D_sD_s^*$ and $J/\psi\phi$ thresholds, while the $X(4274)$,
$X(4500)$, and $X(4700)$ appear as conventional $3^3P_1$, $4^3P_0$,
and $5^3P_0$ charmonia, respectively.

In Ref. \cite{Molina:2017iaa}, a relativistic Dirac potential model
inspired by the OGE interaction was developed to study the
charmonium spectrum in the momentum space. Due to the relativistic
nature of the model, the spin-orbit, spin-spin, and tensor effects
were all automatically included in the calculation. Two types of the
scalar potentials with screening factors were considered. The
authors determined the model parameters by fitting the masses of the
eight resonances below the $D\bar{D}$ threshold and reproduced the
overall structure of the charmonium spectrum. Analysing the
predicted masses for high-lying resonances, they identified the
$X(3915)$, $X(3872)$, and $\chi_{c2}(2P)$ as $2^3P_J$ states with
$J$=0, 1, and 2, respectively, while the $X(3915)$ is not well
described and the $X(3872)$ is slightly higher than the measured
mass. The $\psi(3823)$ was well reproduced, a result consistent with
Ref. \cite{Deng:2016stx}, but it is difficult to accommodate the
$Y(4260)$, $Y(4360)$, $Y(4660)$, $X(3940)$, $X(4160)$, and $X(4140)$
from their results. The authors also suggested to perform deeper
dynamical studies by considering effects such as molecule/tetraquark
admixture and threshold effects.

In studying the properties of the exotic states, the authors of Ref.
\cite{Bhavsar:2018umj} discussed their possible heavy quarkonium
assignments by solving the relativistic Dirac equation with a linear
confinement potential. Their results indicate that the $X(4140)$ is
a mixture of two $P$-wave charmonia, and the $Y(4630)$ and $Y(4660)$
are admixed states of the $S$-$D$ waves. In Ref.
\cite{Kher:2018wtv}, the spectroscopy of the charmonium states was
investigated with a Coulomb plus linear potential. From their
calculations, the authors assigned the $Y(4660)$, $X(3872)$,
$X(3915)$, and $X(4274)$ to be the $5^3S_1$, $2^3P_1$, $2^3P_0$, and
$3^3P_1$, respectively.

In Ref. \cite{Fu:2018yxq}, the masses and decay widths of the vector
charmonium states from $J/\psi$ to $\psi(4160)$ were calculated by
using the instantaneous Bethe-Salpeter equation with a screened
Cornell potential and the $^3P_0$ model by considering the mixing
effects between the $(n+1)S$ and $nD$ states. Most of the results
were consistent with experimental data. The authors further
investigated the mixing between $4S$ and $3D$ states, and found that
it is still possible to assign the $Y(4260)$ and $Y(4360)$ to be the
$4S$-$3D$ mixed charmonium states. The branching ratios of their
decays into the $D\bar{D}$ were predicted to be small.

In a study of the heavy quarkonium hybrids \cite{Oncala:2017hop}
based on the strong coupling regime of pNRQCD (potential
nonrelativistic QCD), the authors found that most of the isospin
zero $XYZ$ states fit well either as the hybrid or standard
quarkonium candidates. The $X(3823)$ is compatible with the
charmonium $\psi_2(1D)$. The $X(3872)$ is compatible with the
$\chi_{c1}(2P)$, but mixing with $D^0\bar{D}^{0*}$ may have large
contributions. The $X(3915)$ and $X(3940)$ are compatible with the
charmonium states $\chi_{c0}(2P)$ and $h_c(2P)$, respectively, but
$D_s^+D_s^-$ may contribute here. The $Y(4008)$ is compatible with a
$1^{--}$ hybrid state, which mixes with a spin-1 charmonium. The
$X(4140)$ and $X(4160)$ are compatible with the $1^{++}$ hybrid
states. These states may be affected by the $D_s^{*}\bar{D}_s$
threshold. The $Y(4230)$ and $Y(4260)$ are compatible with the
$1^{--}$ $2D$ charmonium state having a dominant spin-0 hybrid
component. The $X(4274)$ is compatible with the charmonium
$\chi_{c1}(3P)$, which may be affected by the $D_s^{*+}D_s^{*-}$
threshold. The $X(4350)$ is a hybrid or a conventional $3P$
charmonium. The $Y(4320)$, $Y(4360)$, and $Y(4390)$ are compatible
with the spin-0 $1^{--}$ hybrid. The $X(4500)$ is compatible with a
$0^{++}$ hybrid state, but its mixing with the spin-1 charmonium is
little and it is difficult to understand its observation in the
$J\psi\phi$ channel. The $Y(4630)$ is compatible with the charmonium
$\psi(3D)$. The $Y(4660)$ is compatible with a spin-0 $1^{--}$
hybrid state. The $X(4700)$ is compatible with the charmonium
$\chi_{c0}(4P)$. In the bottom case, the $\Upsilon(10860)$ is
compatible with the bottomonium $\Upsilon(5S)$, the $Y_b(10890)$
\cite{Chen:2008xia} is compatible with the spin-0 $1^{--}$ hybrid,
and the $\Upsilon(11020)$ with the bottomonium $\psi(4D)$.

In Ref. \cite{Berwein:2015vca}, a nonrelativistic effective field
theory describing heavy quarkonium hybrids was constructed. The
authors discussed possible hybrid assignments for the exotic $X$ and
$Y$ states, such as the $Y(4220)$, $X(4350)$, and $Y(4260)$. In Ref.
\cite{Miyamoto:2018zfr}, Miyamoto and Yasui studied the spectra and
decay widths of the hybrid quarkonia in a hyperspherical coordinate
approach, and discussed the possibility of the $Y(4260)$, $Y(4360)$,
$\psi(4415)$, $Y(4660)$, and $\Upsilon(10860)$ being the hybrid
states. According to their results, in the charmonium sector, the
ground state of a magnetic gluon hybrid and the first excited state
of an electric gluon hybrid lie close to the $\psi(4415)$. In the
bottomonium sector, the first excited electric hybrid and the ground
magnetic hybrid appear a few hundred MeV above the
$\Upsilon(10860)$. If one of the exotic meson candidates is a
hybrid, its constituent gluon was expected to be magnetic. In a
nonrelativistic quark model with the Cornell like potential, the
authors of Ref. \cite{Bruschini:2018lse} found that assigning the
$\Upsilon(10860)$ as a mixture of the conventional $\Upsilon(5S)$
and the lowest hybrid state may give a plausible explanation for its
$\pi\pi\Upsilon(nS)$ ($n=1,2,3$) production rates.

\subsubsection{Schemes of the $Q\bar{Q}$ core admixed with coupled channels}

Very recently, BESIII observed a new decay mode for the $X(3872)$,
$\pi^0\chi_{c1}$, in the processes $e^+e^-\to\gamma \pi^0\chi_{cJ}$
($J=0,1,2$) and found
$Br(X(3872)\to\pi^0\chi_{c1})/Br(X(3872)\to\pi^+\pi^-J/\psi)=0.88^{+0.33}_{-0.27}\pm0.10$
\cite{Ablikim:2019soz}. This new measurement disfavors the
$\chi_{c1}(2P)$ interpretation for the $X(3872)$ presented in Ref.
\cite{Dubynskiy:2007tj}. An early measured ratio $Br(X(3872)\to
\psi(2S)\gamma)/Br(X(3872)\to J/\psi\gamma)=3.4\pm1.4$ by BaBar
\cite{Aubert:2008ae} did not support a pure $\bar{D}^0D^*$ molecule
interpretation, while Belle's result of $Br(X(3872)\to
\psi(2S)\gamma)/Br(X(3872)\to J/\psi\gamma)<2.1$
\cite{Bhardwaj:2011dj} indicates that the $c\bar{c}$ contribution in
$X(3872)$ is probably not large. These measurements strongly
indicate that the $X(3872)$ probably contains a larger $D\bar{D}^*$
component than the $c\bar{c}$. Inspired by these experimental
measurements, there has been continuous progress of theoretical
studies on the coupled nature of the $X(3872)$ and other $X$ and $Y$
states in recent years.

In Ref. \cite{Cincioglu:2016fkm}, the authors investigated the
quarkonium contributions to the meson molecules like the $X(3872)$
and its heavy-quark spin-flavor partners. They found that the
$c\bar{c}$ core produces an extra attraction, and thus less
attractive $D\bar{D}^*$ interaction is needed to form a bound
$X(3872)$. But such an attraction does not appear in the $2^{++}$
sector. The $c\bar{c}$ content in the $X(3872)$ was found to be
10$\sim$30\%.

With the extended Friedrichs scheme, Zhou and Xiao of Ref.
\cite{Zhou:2017dwj} studied the radially excited $P$-wave charmonium
spectrum. According to their analyses, the $X(3872)$ is dynamically
generated by the coupling of the bare $\chi_{c1}(2P)$ and continuum
states, with large molecule components (64-85\%). The observed
$X(3860)$ has a much larger width than the $\chi_{c0}(2P)$
charmonium although their masses are close. The state $h_{c}(2P)$
was predicted to be around 3890 MeV with a pole width about 44 MeV.
Later in Ref. \cite{Zhou:2017txt}, they considered the isospin
breaking effects of the $X(3872)$ with the same scheme, and obtained
the reasonable ratio 0.58$\sim$0.92 between the $J/\psi \pi\pi\pi$
and $J/\psi\pi\pi$ channels. In Ref. \cite{Zhou:2018hlv}, the
bottomonium counterpart of the $X(3872)$ around 10615 MeV and the
$\chi_{b1}(4P)$ around 10771 MeV were predicted. The reason for the
non-observation of the $X_b$ signal in the $\Upsilon\pi\pi$
\cite{Chatrchyan:2013mea} and $\Upsilon\pi\pi\pi$ \cite{He:2014sqj}
channels was also discussed.

In Ref. \cite{Ferretti:2018tco}, the coupled channel effects on the
heavy quarkonium states like $\chi_{\rm c}(2P)$ and $\chi_{\rm
b}(3P)$ were studied by considering the meson loop corrections. The
authors found that the $X(3872)$ is a superposition of a $c\bar{c}$
core and meson-meson continuum components, while the $\chi_b(3P)$
should be a pure bottomonium. The authors of Ref.
\cite{Ortega:2019tby} presented an unquenched method to consider the
coupling between the quark-antiquark states and meson-meson states
for a given $J^{PC}$. After the investigation on channels related
with the $X(3872)$, the authors got three states. The state that can
be realized as the $X(3872)$ has a dominant $D\bar{D}^*$ component
and a posterior $2^3P_1$ $c\bar{c}$. The state identified as the
$X(3940)$ has a predominant $2^3P_1$ $c\bar{c}$ component. The third
state is an almost pure $1^3P_1$ $c\bar{c}$ state. The coupled channel 
effects for the bottomonium with realistic wave were also studied under 
the framework of $^3P_0$ model in Ref. \cite{Lu:2016mbb}.

By assuming the $X(3872)$ as a mixing state of the 2P charmonium and
$\bar{D}D^*$ molecule, the authors of Ref. \cite{Cincioglu:2019gzd}
analyzed its radiative decays into $J/\psi\gamma$ and
$\psi(2S)\gamma$ in an effective field theory. They found that a
wide range of charmonium probability in the $X(3872)$ is consistent
with the experimental ratio $Br(X\to\psi(2S)\gamma)/Br(X\to
J\psi\gamma)$. In the case of the destructive interferences between
the long-range meson loops and the short-range counter-term, a
strong constraint on the $c\bar{c}$ admixture is found ($<15\%$ is
expected).

To understand the nature of $X(3872)$, the spectral function of the
charmonium state $\chi_{c1}(2P)$ coupled to $DD^*$ mesons was
studied in Ref. \cite{Giacosa:2019zxw}. The authors found two poles
in the complex plane, one pole corresponding to the $c\bar{c}$
object $\chi_{c1}(2P)$ and the other virtual pole just below the
$D^0D^{*0}$ threshold. The $X(3872)$ as a mixed object emerges from
an interplay between a $c\bar{c}$ state and the thresholds. With
their approach, the existence of $X(3872)$ is not possible without
the seed charmonium state.

In Ref. \cite{Qin:2016spb}, the $Y(4260)$ was treated as an $S$-wave
$\bar{D}D_1 +c.c.$ molecule containing a small charmonium component.
Extracting parameters from experimental results, the authors found
that the physical wave function of the $Y(4260)$ is
$0.363|c\bar{c}\rangle+0.932|\bar{D}D_1+c.c.\rangle$. In Ref.
\cite{Lu:2017yhl}, Lu, Anwar, and Zou studied the coupling between
the $1^{--}$ charmonium structures and relevant meson-antimeson
structures with the $^3P_0$ pair creation model. Their results
support the picture that the $Y(4260)$ is a $D_1\bar{D}$ molecule
having a non-negligible $\psi(nD)$ core.

\subsubsection{Tetraquark states}

In the chiral SU(3) quark model \cite{Zhang:2007xa}, the masses of
the $Qq\bar{Q}\bar{q}^\prime$ ($Q=c,b$, $q=u,d,s$) states with
$J^{PC}=(0,1,2)^{++}$ and $1^{+-}$ were calculated. It was
impossible to assign the $X(3872)$ and $Y(3940)$ [now called
$X(3915)$] as pure $cn\bar{c}\bar{n}$ tetraquark states with
$J^{PC}=1^{++}$ and $2^{++}$, respectively. From this study, narrow
tetraquarks in the hidden-bottom sector are possible. The study of
the spectrum in the framework of the constituent quark model in Ref.
\cite{Vijande:2007fc} did not show any bound four-quark
$c\bar{c}n\bar{n}$ states for low-lying $J^{PC}$, ruling out the
possibility that the $X(3872)$ is a compact tetraquark system, if no
additional correlations are considered in simple quark models.

In  Ref. \cite{Coito:2016ads}, the $Z_c(3900)$, $Z_c(4020)$,
$Z_c(4050)$, $Z_b(10610)$, and $Z_b(10650)$ were studied within a
coupled-channel Schr\"odinger model by coupling the
$I^G(J^{PC})=1^-(1^{++})$ and $1^+(1^{+-})$ excited quark-antiquark
pairs and their OZI-allowed decay channels $D^{(*)}\bar{D}^{(*)}$ or
$B^{(*)}\bar{B}^{(*)}$. Poles matching experimental data for these
mesons were all found. In Ref. \cite{Patel:2016otd}, the nature of
the $\Upsilon(10890)$ and other exotic states in the bottom sector
was discussed in the diquark-antidiquark configuration in a quark
potential model. The authors found that the $Z_b(10650)$ is probably
a radially excited diquark-antidiquark state, and that the
$\Upsilon(10890)$ might be the tetraquark $Y_b(10890)$ rather than a
conventional $b\bar{b}$ state.

In Ref. \cite{Lu:2016cwr}, Lu and Dong presented a study for the
$[cs]_{\bar{3}_c}[\bar{c}\bar{s}]_{3_c}$ diquark-antidiquark states
in the GI model by including the color screening effects. In this
model, the $X(4140)$ can be regarded as the $A\bar{S}$ type
tetraquark state, and the $X(4700)$ as the first radially excited
$A\bar{A}$ or $S\bar{S}$ state, where $S$ ($A$) denotes the scalar
(axial-vector) diquark. They also studied the case that the internal
orbital or radial excitation is allowed for diquarks
\cite{Chen:2016oma,Wang:2016gxp}. Their results indicate that the
$X(4500)$ can be explained as the tetraquark composed of one
$2^1S_0$ diquark and one $1^1S_0$ antidiquark, while the $X(4350)$
is a tetraquark composed of one $2^3S_1$ and one $1^3S_1$. However,
there is no assignment for the $X(4274)$ in this model, which is
proposed to be the candidate of the charmonium $\chi_{c1}(3P)$,
supported by the decay width calculation in Ref.
\cite{Barnes:2005pb} and the arguments in Ref. \cite{Liu:2016onn}.
Their results also indicate that it is  possible to assign the
$X(3915)$ as the lightest $cs\bar{c}\bar{s}$ state, which supports
the proposal given in Ref. \cite{Lebed:2016yvr}. Besides, some
tetraquark masses are close to those of the $Y(4630)$ and $Y(4660)$.

In Ref. \cite{Yang:2017prf}, possible $D^{(*)}\bar{D}^{*}$ molecular
states were studied in the BCN model and a chiral constituent quark
model by including the $s$-channel one gluon exchange. The authors
found the $J^{PC}=(1,2)^{++}$ bound states in the charm sector and
$(0,1,2)^{++}, 1^{+-}$ bound states in the bottom sector. Good
candidates for the $X(3872)$ and $Z_b(10610)$ were then obtained. In
another work \cite{Yang:2017rmm}, the same authors considered the
$S$-wave $\bar{Q}Q\bar{q}q$ ($Q=c,b$, $q=u,d,s$) systems with two
chiral constituent quark models in the meson-antimeson picture. From
their results, several $B\bar{B}^*$ bound states were found. If the
hidden-color channels are also included, a bound state
$[c\bar{q}]^*[q\bar{c}]^*$ with $I(J^{PC})=1(0^{++})$ was also
found. The obtained $B\bar{B}^*$ and $B^*\bar{B}^*$ states with
$I(J^{PC})=1(1^{+-})$ can be related to the $Z_b(10610)$ and
$Z_b(10650)$, respectively.

In Ref. \cite{Anwar:2018sol}, the investigation for the spectroscopy
of the hidden-charm $[qc]_{\bar{3}_c}[\bar q \bar c]_{3_c}$ and
$[sc]_{\bar{3}_c}[\bar s \bar c]_{3_c}$ tetraquarks was performed in
a relativized diquark model. The authors found possible assignments
for the $X(3872)$, $Z_c(3900)$, $Z_c(4020)$, $Y(4008)$, $Z_c(4240)$,
$Y(4260)$, $Y(4360)$, $Y(4630)$, and $Y(4660)$ in the
$c\bar{c}n\bar{n}$ sector, and for the $X(4140)$, $X(4500)$, and
$X(4700)$ in the $c\bar{c}s\bar{s}$ sector, but did not for the
$X(4274)$ and $Z_c(4430)$.

In the dynamical diquark model, the authors of Ref.
\cite{Giron:2019bcs} presented a study on the spectrum of
charmoniumlike tetraquarks by solving the Schr\"odinger equation
with the Born-Oppenheimer potentials calculated numerically on the
lattice. Choosing the $X(3872)$ or $Z_c(4430)$ as a reference
diquark-antidiquark state, the authors obtained a spectrum that
agrees well with the observed charmoniumlike states.

Yang and Ping adopted a chiral quark model with the exchange of
$\pi$, $\kappa$, and $\eta$ and considered the diquark-antidiquark
and meson-meson configurations in Ref. \cite{Yang:2019dxd}. They
investigated the spectrum of $cs\bar c\bar s$ states and found that
the $X(4274)$ and $X(4350)$ are good candidates of the compact
tetraquark states with $J^{PC}=1^{++}$ and $0^{++}$, respectively.
The $X(4700)$ can be explained as the $2S$ radial excited tetraquark
state with $J^{PC}=0^{++}$. However, no appropriate matching states
for $X(4140)$ and $X(4500)$ were found.

\subsection{$QQ\bar{q}\bar{q}$}

The doubly heavy $QQ\bar{q}\bar{q}$ tetraquark systems are
particularly interesting depending on the spatial configurations of
two heavy quarks. In the extreme case that the $QQ$ pair is in the
color anti-triplet and stays very close to each other, the compact
heavy quark pair acts like one static color source. Two light
anti-quarks circle around this point-like color source. This
configuration is very similar to the Helium atom in QED. We denote
it as the ``QCD Helium atom''. On the other hand, if the heavy quarks
are well separated from each other, the $\bar{q}\bar{q}$ pair is
shared by the two heavy quarks. This is the QCD valence bond. Such a
configuration is the QCD analogue of the hydrogen molecule in QED.
We denote it as the "QCD Hydrogen molecule". In general, the
$QQ\bar{q}\bar{q}$ tetraquark system may be the superposition of the
{\it atomic}  and {\it molecular} structures.

In Ref. \cite{Janc:2004qn}, using the BCN potential and the Grenoble
AL1 potential, the authors studied the binding problem of the
$T_{cc}$ (the lowest isoscalar $1^+$ $cc\bar{u}\bar{d}$, see
Sec.~\ref{sec2:QQqq}). According to their calculation, the $T_{cc}$
with the {\it molecular} structure is weakly bound against the
$DD^*$ threshold, but it can become {\it atomic} with the inclusion
of the three-body force.

In Ref. \cite{Ebert:2007rn}, the masses of the tetraquark states
$QQ\bar{q}\bar{q}$ ($Q=c,b$, $q=u,d,s$) were calculated in the
diquark-antidiquark picture with a relativistic quark model. All the
$(cc)(\bar{q}\bar{q}^\prime)$ states are not bound, and only the
$I(J^P)=0(1^+)$ $(bb)(\bar{n}\bar{n})$ state lies below the $BB^*$
threshold. The authors of Ref. \cite{Zhang:2007mu} studied the
$QQ\bar{q}\bar{q}$ four-quark bound states in chiral $SU(3)$ quark
model. Their calculation indicates that a $bb\bar{n}\bar{n}$ state
with $I(J^P)=0(1^+)$ is bound, but there is no bound state in the
$cc\bar{q}\bar{q}$ system. With three different quark models one of
which is chiral quark model, the authors of Ref. \cite{Yang:2009zzp}
studied the $QQ\bar{n}\bar{n}$ ($Q=s,c,b$, $n=u,d$) spectrum. Their
results show that only the $bb\bar{n}\bar{n}$ state with
$I(J^P)=0(1^+)$ is bound within these models.

The authors of Ref. \cite{Vijande:2009kj} studied possible compact
four-quark states $QQ\bar{n}\bar{n}$ with two different constituent
quark models, one of which contains Goldstone boson exchanges
between quarks. They noticed that the $I(J^P)=0(1^+)$
$cc\bar{n}\bar{n}$ and $bb\bar{n}\bar{n}$ are bound and should be
narrow, which was consistent with their previous studies
\cite{Vijande:2003ki,Vijande:2006jf,Vijande:2007rf}. Further study
of the exotic $bc\bar n\bar n$ four-quark states in Ref.
\cite{Caramees:2018oue} showed two isoscalar bound states with
$J^P=0^+$ and $1^+$. In Ref. \cite{Feng:2013kea}, the Bethe-Salpeter
equations for the ground-state $QQ\bar{u}\bar{d}$ were established
in the diquark-antidiquark picture. According to their numerical
results, the $cc\bar{u}\bar{d}$ and $bb\bar{u}\bar{d}$ bound states
with $I(J^P)=0(1^+)$ and the $bc\bar{u}\bar{d}$ bound states with
$I(J^P)=0(0^+)$ and $0(1^+)$ should all be stable.

In Ref. \cite{Park:2013fda}, Park and Lee investigated the $1^+$
$T_{bb}$ and $T_{cc}$ states with the BCN potential by including the
contribution from the $6_c$-$\bar{6}_c$ diquark-antidiquark color
structure, which was neglected in Ref. \cite{Brink:1998as}. The
$T_{cc}$ was found to be 100 MeV above the $DD^*$ threshold, while
the $T_{bb}$ was about 100 MeV below the $BB^*$ threshold. The
$6_c$-$\bar{6}_c$ contribution was found to be negligible. Later in
Ref. \cite{Park:2018wjk}, the masses of the doubly heavy tetraquarks
were updated after the observation of the $\Xi_{cc}$. Now the
$ud\bar{b}\bar{b}$ tetraquark is bound by 121 MeV and the
$us\bar{b}\bar{b}$ is bound by 7 MeV. In an investigation with one
gluon exchange potential, the authors of Ref.
\cite{Czarnecki:2017vco} studied the existence of the bound
$Q_1\bar{q}_2Q_3\bar{q}_4$ states. They found that there are no
stable $cc\bar{n}\bar{n}$ and $bc\bar{n}\bar{n}$ states, but stable
$bb\bar{u}\bar{d}$ states are possible.

The authors of Ref. \cite{Richard:2018yrm} discussed approximations
used in the multiquark studies, {\it i.e.}, diquark,
Born-Oppenheimer, Hall-Post inequalities, color-mixing, and
spin-dependent corrections. They pointed out that the $1^+$
$cc\bar{u}\bar{d}$ is at the edge of binding, and more delicate
studies are still needed. For the $1^+$ $bb\bar{u}\bar{d}$, the spin
effects or color mixing effects are needed to achieve a binding
state.

\subsection{$QQ\bar{Q}\bar{Q}$ and $QQ\bar{Q}\bar{q}$}

The study in Ref. \cite{Lloyd:2003yc} with a nonrelativistic
potential model indicates that the lowest $0^{++}$ all-charm
tetraquark state is below the $\eta_c\eta_c$ threshold. A study for
the fully-heavy tetraquarks was performed in Ref.
\cite{Anwar:2017toa} with a nonrelativistic effective field theory
(NREFT) at the leading order and a relativized diquark-antidiquark
model. Both approaches give a $bb\bar{b}\bar{b}$ ground state around
18.72 GeV with $J^{PC}=0^{++}$, which is below the $\eta_b\eta_b$
threshold. Mass inequalities for all type $QQ\bar{Q}\bar{Q}$
($Q=c,b$) were also investigated.

The authors of Ref. \cite{Bai:2016int} presented a calculation for
the mass spectrum of the ground-state $0^{++}$ $bb\bar{b}\bar{b}$
tetraquark states using a diffusion Monte-Carlo method to solve the
non-relativistic many-body problem, whose potential is based on the
flux-tube model. A state around 18.69 GeV was found to be about 100
MeV below the $\eta_b\eta_b$ threshold.

In Ref. \cite{Barnea:2006sd}, the masses of the $cc\bar{c}\bar{c}$
tetraquark states were calculated in a constituent quark model with
the hyperspherical formalism. The authors noted that the $0^{++}$
state is 58 MeV above the $\eta_c\eta_c$ threshold, $1^{+-}$ state
is 14 MeV above the $J/\psi\eta_c$ threshold, and $2^{++}$ state is
22 MeV below the $J/\psi J/\psi$ threshold. In Ref.
\cite{Debastiani:2017msn}, the authors investigated the spectroscopy
of $[cc][\bar{c}\bar{c}]$ in a diquark-antidiquark configuration
using a non-relativistic model, whose potential is a
Cornell-inspired type. Their results show very compact tetraquarks,
which are below their thresholds of spontaneous dissociation into
low-lying charmonium pairs.

In Ref. \cite{Richard:2017vry}, Richard {\it et al} found that
full-charm and full-beauty tetraquarks are unbound in the
chromoelectric model with additive potentials. If the naive
color-additive model of confinement is replaced by a string-inspired
interaction, however, the $bc\bar{b}\bar{c}$ case might be
favorable. In additional to studies about the $QQ\bar{q}\bar{q}$
states, the authors also discussed the existence of
$cc\bar{c}\bar{c}$, $bb\bar{b}\bar{b}$, and $bb\bar{c}\bar{c}$
tetraquarks in Ref. \cite{Czarnecki:2017vco}. Because of lack of
strongly separated mass scales, they concluded that no such bound
states exist.

In a dynamical study with a nonrelativistic potential model in Ref.
\cite{Liu:2019zuc}, the mass spectra of all types of full-heavy
$Q_1Q_2\bar{Q}_3\bar{Q}_4$ ($Q=c,b$) tetraquark states were
obtained. No bound states can be formed below the thresholds of any
meson pairs $(Q_1\bar{Q}_3)-(Q_2\bar{Q}_4)$ or
$(Q_1\bar{Q}_4)-(Q_2\bar{Q}_3)$, and thus such states with narrow
widths are not expected in experiments. This conclusion about
spectrum confirmed the arguments from the CMI estimations given in
Ref. \cite{Wu:2016vtq}. In a study with a nonrelativistic chiral
quark model~\cite{Chen:2019dvd}, the analysis of $bb\bar{b}\bar{b}$
states indicates that the $J^P=(0,1,2)^+$ states are all higher than
the corresponding thresholds, in both meson-meson and
diquark-antidiquark configurations. The results also confirmed the
conclusions from the CMI calculation with the method \eqref{massref}
\cite{Wu:2016vtq}.

In Ref. \cite{SilvestreBrac:1993ss}, Silvestre-Brac and Semay did
not find any bound $QQ\bar{Q}\bar{q}$ states with a nonrelativistic
quark model, whose potential was proposed by Bhaduri {\it et al}
\cite{Bhaduri:1981pn}. This result is roughly consistent with the
estimations with the CMI model \eqref{massref}
\cite{SilvestreBrac:1992mv}.

\subsection{$Q\bar{Q}qqq$ and $QQqq\bar{q}$}

Considering the quark delocalization color screening effects in a
chiral quark model, Huang {\it et al} investigated the $P_c$-like
hidden-charm pentaquarks in the molecule configuration in Ref.
\cite{Huang:2015uda}. According to their results, the $P_c(4380)$
can be explained as the $\Sigma_c^*\bar{D}$ molecule with
$J^P=3/2^-$. Other hidden-charm and hidden-bottom bound states were
also investigated. But all the positive-parity states were found to
be unbound. They studied the hidden-strangeness pentaquarks in Ref.
\cite{Huang:2018ehi}. In Ref. \cite{Yang:2015bmv}, the same group
performed a dynamical calculation of the five-quark systems in the
framework of a chiral quark model. The $P_c(4380)$ was again
suggested to be a $\Sigma_c^*\bar{D}$ molecule. All the obtained
positive-parity states are unbound in their calculation, unless the
effective $\sigma$ meson exchange is employed. The authors did not
adopt the assignment for the $P_c(4450)$ as the $\Sigma_c\bar{D}^*$
molecule because of the inconsistent parity, although their masses
are close to each other. They presented a similar study of the
hidden-bottom pentaquarks in Ref. \cite{Yang:2018oqd}.

In Ref. \cite{Gerasyuta:2015djk}, the solution for the relativistic
five-quark equations favors the assignment for the quantum numbers
of the pentaquarks $P_c(4380)$ and $P_c(4450)$ to be $J^P=5/2^+$ and
$3/2^-$, respectively.

In the molecule picture, the authors of Ref. \cite{Ortega:2016syt}
studied the LHCb $P_c$ states with a chiral constituent quark model
\cite{Vijande:2004he,Segovia:2008zz} and confirmed the existence of
several $\bar{D}^{(*)}\Sigma_c^{(*)}$ structures near the $P_c$
masses. In Ref. \cite{Takeuchi:2016ejt}, the isospin-half
$J^P=(1/2,3/2,5/2)^-$ $uudc\bar{c}$ pentaquarks with the color-octet
$uds$ component were investigated with the quark cluster model.
Three structures were found to be around the
$\Sigma_c^{(*)}\bar{D}^{(*)}$ thresholds: one bound state, two
resonances, and one large cusp. The authors argued that the
$P_c(4450)$ may arise from these structures.

In Ref. \cite{Park:2017jbn}, the authors studied the possible
pentaquark nature of the $P_c(4380)$ with $I(J^P)=1/2(3/2^-)$ using
a hyperfine+confinement potential. With the variational method, they
found that the ground state is the isolated $p$ and $J/\psi$ state,
and the excited state lies far above the observed $P_c(4380)$. This
observation led to the conclusion that the observed $P_c(4380)$
cannot be a compact $J^P=3/2^-$ pentaquark generated by the
conventional two-body quark interactions. In Ref.
\cite{Stancu:2017azl}, Stancu studied the effects of three-body
chromoelectric interaction in pentaquark states and found that the
three-body confining interaction can stabilize the pentaquark states
with an appropriate sign. She suggested that perhaps the fall apart
decay mode of the ground-state $nnnc\bar{c}$ pentaquarks in Ref.
\cite{Park:2017jbn} may be avoided if such an interaction is
included.

In Ref. \cite{Richard:2017una}, Richard, Valcarce, and Vijande
studied the hidden-charm pentaquarks in a constituent model with the
Grenoble AL1 potential by solving the five-body problem. They found
that the lowest states with $(I,J)=(3/2,1/2)$ and $(3/2,3/2)$ are
below their lowest $S$- and $D$-wave thresholds, $\eta_c\Delta$ and
$\Sigma_c\bar{D}$, and thus are expected to be stable. In a recent
work \cite{Hiyama:2018ukv}, Hiyama, Hosaka, Oka, and Richard studied
the five-body scattering problem with a nonrelativistic quark model
by including explicitly open channels such as $J/\psi N$, $\eta_cN$,
and $\Lambda_c\bar{D}$, etc. They also adopted the Grenoble
potentials (AP1 and AL1). Their results are compatible with those
obtained in Ref. \cite{Richard:2017una}. According to their
analysis, there does not exist any resonance at the energies of the
LHCb $P_c$ states, while two narrow states are possible lying at
4690 MeV and 4920 MeV with $J^P=1/2^-$ and $3/2^-$, respectively.

In Ref. \cite{Stancu:2019qga}, Stancu investigated the spectrum of
the $uudc\bar{c}$ hidden-charm pentaquark states with an $SU(4)$
Goldstone-boson-exchange model, in which the mass splittings are
accounted for by the flavor-spin hyperfine interaction. She found
that the lowest positive-parity states with $J^P=1/2^+$ and $3/2^+$
are below the lowest negative-parity state with $J^P=1/2^-$. The
results accommodate a $3/2^+$ $P_c(4380)$ and a $5/2^+$ $P_c(4450)$.

With the Bethe-Salpeter equation, the spectrum of the heavy
pentaquark states $Q\bar{Q}qqq$ ($Q=c,b$, $q=u,d,s$) were studied in
the diquark-diquark-antiquark configuration in Ref.
\cite{Giannuzzi:2019esi}. The masses of the $P_c$ like hidden-charm
pentaquarks $J=3/2$ and $5/2$ states are $\sim 300$ MeV higher than
the experimental values. The authors also calculated the masses of
$QQqq\bar{q}$ states. The mass of the lowest $[cn][cn]\bar{n}$ state
is around 4.54 GeV which is larger than our result obtained in the
CMI model \cite{Zhou:2018pcv}.

In Ref. \cite{Giron:2019bcs}, the charmoniumlike pentaquark states
were investigated in the dynamical diquark model by assuming that
they are diquark-triquark states. Using the observed $P_c(4380)$ and
$P_c(4450)$ as inputs, the authors predicted a number of other
unobserved states. Especially the ground hidden-charm pentaquarks
were below the $J/\psi N$ even $\eta_cN$ threshold.

The authors of Ref. \cite{Wang:2016dzu} presented a study of the
magnetic moments of the hidden-charm pentaquark states with the
isospin $(I,I_3)=(1/2,1/2)$ and $J^P=(1/2\sim 7/2)^\pm$ in the
molecular, diquark-triquark, and diquark-diquark-antiquark models.
The obtained magnetic moments are different, although the
description for pentaquark masses and decay patterns can be achieved
in all these models. Their results may be used to understand the
inner structures and distinguish phenomenological models. In Ref.
\cite{Wang:2016wxg}, the axial charges of the hidden-charm
pentaquarks were studied in the chiral quark model. Their results
can also be used to distinguish the underlying structures of the
$P_c$ states.

\subsection{Color flux-tube model}

The flux-tube model~\cite{Isgur:1983wj,Isgur:1984bm} was first extracted from the strong-coupling Hamiltonian
lattice formulation~\cite{Kogut:1974ag} and the early descriptions of flux tubes as
cylindrical bags of colored fields~\cite{Gnadig:1976pn}. In the flux-tube model, the gluonic
excitation was considered as the transverse vibration of the string-like
flux tube between a quark-antiquark pair. The ordinary quark model was
contained under the zero angular momentum approximation, while exotica
can also be achieved for the non-zero angular momentum~\cite{Isgur:1984bm,Barnes:1995hc}.

The string potential in Hamiltonian was usually in a standard linear
form to investigate the confinement phenomenon~\cite{Barnes:1995hc}.
Such a flux-tube model has been extensively used to study the hadron
masses, decays and production characteristics~\cite{Barnes:1995hc,Merlin:1986tz,Page:1998gz}.
Recently, a naive color flux-tube model was developed by considering a harmonic-like confinement
potential rather than a linear one to study the tetraquark states~\cite{Wang:1984gg,Ping:2006nc}.

In Ref.~\cite{Deng:2014gqa}, Deng  {\it et. al.} studied the lowest
charged tetraquark states $Qq\bar Q^\prime\bar q^\prime$ ($Q=c, b$
and $q=u, d, s$) in the framework of the color flux-tube model with
a four-body confinement potential. Within the color flux-tube model,
the total hamiltonian of the diquark-antidiquark can be expressed
as,
\begin{eqnarray}
\nonumber
H_f&=&\sum_{i=1}^f\left(m_i+\frac{p_i^2}{2m_i}\right)-T_C+\sum_{i>j}^fV_{ij}+V_{min}^C(f)\,
,
\\ V_{ij}&=&V_{ij}^B+V_{ij}^\sigma+V_{ij}^G\, ,
\end{eqnarray}
in which $f=2$ or $f=4$, $T_C$ is the center-of-mass kinetic energy,
$p_i$ is the momentum of the $i$-th quark, $V_{min}^C(f)$ is the
quadratic confinement potential. The interaction between two
different quarks is described by $V_{ij}$, which contains the
one-boson-exchange potential $V_{ij}^B$, the $\sigma$-meson exchange
potential $V_{ij}^\sigma$, the one-gluon-exchange potential
$V_{ij}^G$. In the diquark-antidiquark configuration, the energy of
the tetraquark states $Qq\bar Q^\prime\bar q^\prime$ was then
calculated by solving the four-body Schr\"odinger equation. The
numerical results indicated that some compact resonance states were
formed with the four-body quadratic potential. These resonances can
not decay into $Q\bar q^\prime$ and $\bar Q^\prime q$ but into
$Q\bar Q^\prime$ and $q\bar q^\prime$ through the breakdown and
recombination of the flux tubes. The authors interpreted the charged
states $Z_c(3900)$ and $Z_c(4025)/Z_c(4020)$ as the $S$-wave
tetraquarks $[cu][\bar c\bar d]$ with quantum numbers $I=1$ and
$J=1$ and 2, respectively.

Later in Ref.~\cite{Deng:2015lca}, they systematically studied the
charged $Z_c^+$ family in the color flux-tube model and considered
the spin-orbit interactions between quarks. Their investigation
indicated that the $Z_c(3900)/Z_c(3885)$, $Z_c(3930)$,
$Z_c(4025)/Z_c(4020)$, $Z_c(4050)$, $Z_c(4250)$, and $Z_c(4200)$ can
be described as the $[cu][\bar c\bar d]$ tetraquark states with
$n^{2S+1}L_J$ and $J^P$ of $1^3S_1$ and $1^+$, $2^3S_1$ and $1^+$,
$1^5S_2$ and $2^+$, $1^3P_1$ and $1^-$, $1^5D_1$ and $1^+$, $1^3D_1$
and $1^+$, respectively. However, the two heavier charged states
$Z_c^+(4430)$ and $Z_c^+(4475)$ can not be explained as tetraquark
states in this model. In Ref.~\cite{Zhou:2015frp}, Zhou, Deng and
Ping also employed the flux-tube model to investigate the tetraquark
states $cq\bar c\bar q$ with $IJ^{PC}=01^{--}$. The vector states
$Y(4008)$, $Y(4140)$, $Y(4260)$, and $Y(4360)$ were described as the
$1^1P_1$, $1^5P_1$, $2^5P_1$ and $1^5F_1$ hidden-charm tetraquarks,
respectively.

In Ref.~\cite{Deng:2016rus}, the hidden-charm pentaquark states were investigated
within the framework of the color flux-tube model including a five-body confinement potential.
The results showed that the main component of the $P_c^+(4380)$ can be described as a
compact $uudc\bar c$ pentaquark state with the pentagonal structure and $J^P=\frac{3}{2}^-$.
However, the $P_c^+(4450)$ state can not be interpreted as a compact pentaquark, since the
masses of the positive parity $uudc\bar c$ states were extracted much higher than those of the
$P_c^+(4380)$ and $P_c^+(4450)$ in the color flux-tube model.

Recently, they also studied the stability of the doubly heavy tetraquark states $[QQ][\bar q\bar q]$
in the color flux-tube model~\cite{Deng:2018kly}. They solved the four-body Schr\"odinger equation
to calculate the energies of all doubly heavy tetraquark states with both the color configurations
$[[QQ]_{\bf{\bar 3}_c}[\bar q\bar q]_{\bf 3_c}]_{\bf 1}$ and $[[QQ]_{\bf{6}_c}[\bar q\bar q]_{\bf{\bar 6}_c}]_{\bf 1}$.
Their numerical results showed that the states $[bb][\bar u\bar d]$ with $IJ^P=01^+$ and $[bb][\bar q^\prime\bar s]$ with $IJ^P=\frac{1}{2}1^+$ were the most promising stable doubly heavy tetraquark states against strong interaction.
The doubly heavy tetraquark states $qq\bar Q\bar Q$ have also been addressed by a fully unitary and
microscopic quark model with a triple string flip-flop potential~\cite{Bicudo:2015bra}. They found several tetraquark bound states and resonances and calculated their masses and decay widths by computing the $T$ matrix and finding the
pole positions in the complex energy plane.

\subsection{A short summary}

We summarize some interesting observations for the multiquark states
from quark model investigations in recent years:

\begin{itemize}

\item
The physical $X(3872)$ should be a mixed state due to the strong
channel coupling between the bare quark model state $\chi_{c1}(2P)$
and the $D\bar{D}^*$ scattering state. Assigning the $X(3915)$ as
the $\chi_{c0}(2P)$ charmonium is not supported by all quark model
calculations. The $X(3940)$ can be assigned as the $\eta_c(3S)$ or
$h_c(2P)$ (affected by $D_s\bar{D}_s$).

\item
In the charmonium or hybrid picture, the $X(4140)$ is a mixed
$c\bar{c}$, a hybrid affected by $D_s\bar{D}_s^*$, or a cusp. The
$X(4274)$ is the $\chi_{c1}(3P)$ affected by $D_s^*\bar{D}_s^*$. The
$X(4350)$ is the $\chi_{c0}(3P)$ or a hybrid. The $X(4500)$ is the
$\chi_{c0}(4P)$ or a hybrid. The $X(4700)$ is the $\chi_{c0}(4P)$ or
$\chi_{c0}(5P)$.

\item
Perhaps both the mixed $4S$-$3D$ $c\bar{c}$ states and the
$D_1\bar{D}^{(*)}$ states contribute to the observed
$Y(4260)/Y(4230)$ and $Y(4360)$, while the $Y(4630)$ and $Y(4660)$
are dominantly mixed $5S$-$4D$ even $3D$ charmonia. Their masses can
also be reproduced in tetraquark models.

\item
The masses of $Z_c(3900)$ and $Z_c(4020)$ may be obtained in both
molecule picture and diquark-antidiquark picture. The $Z_b(10610)$
and $Z_b(10650)$ should be $B^{(*)}\bar{B}^{(*)}$ molecules,
although the diquark-antidiquark interpretation for $Z_b(10650)$ is
also possible. The nature of $Z_c(4430)$ is still unclear.

\item
Whether the lowest $cc\bar{u}\bar{d}$ tetraquark state is above or
below the $DD^*$ threshold is still an open question, but the lowest
$bb\bar{u}\bar{d}$ state should lie below the $B\bar{B}^*$
threshold.

\item
Few models favor the bound $cc\bar{c}\bar{c}$ and $bb\bar{b}\bar{b}$
states. A bound state in the $bc\bar{b}\bar{c}$ case might be
favorable. Bound $QQ\bar{Q}\bar{q}$ states should not exist.

\item
The quark model calculations favor the molecule interpretations for
the $P_c(4380)$ and $P_c(4450)$.

\end{itemize}

%% file: section4.tex
\section{Meson exchange and scattering methods}\label{sec:mesonexchange}

In the study of the hadronic molecules, almost all the
meson-exchange models are constructed at the hadron level. One first
derives the effective boson-exchange potentials in coordinate or
momentum space, and then solves the bound state problem or
scattering problem of two hadrons. From the obtained binding energy
or scattering phase shifts, one extracts the resonance information.
Such a formalism is a straightforward extension of the traditional
meson exchange models in nuclear force.

There are meson-exchange models based on the quark-level
interactions, e.g. based on the chiral quark model. After
integrating out the inner degrees of freedom inside the hadrons,
hadron-level effective meson-exchange potentials can be obtained. An
example is illustrated in Ref. \cite{Li:2010cy}. Of course, the
molecule problems can also be dealt with at the quark level
directly.

The color force of the van de Waals type is not described by the
meson-exchange but by two-gluon-exchange interactions. This type
interaction was applied to understand the nature of the exotic
states in the hadroquarkonium picture where a heavy quarkonium is
embedded in a light hadronic matter. This hypothesis dated back to
early nineties \cite{Anwar:2018bpu}. Although the interaction
between ground-state hadrons may be weak and no bound state can be
formed, probably such a force may result in bound states or
resonances for radially and orbitally excited hadrons.

In the chiral limit, the eight pseudoscalar meson masses vanish as
the up, down and strangeness current quark masses go to zero. In
contrast, the masses of the nucleon, rho meson or heavy
mesons/baryons etc remain finite or even large in the chiral limit.
They are sometimes denoted as the matter fields. If they contain the
light quark degree of freedom, the matter fields will interact with
the pseudoscalar mesons. The rigorous framework is the well-known
chiral perturbation theory.

In order to ensure the chiral invariance, one should replace the
simple derivative operator by the chirally covariant derivative
operator in the construction of the kinetic energy of the matter
fields at the leading order Lagrangian. The chirally covariant
derivative operator contains the chiral connection. In other words,
the kinetic term of the matter fields always induce the contact
seagull term in the form of $M^\ast {\pi\partial_\mu \pi \over
F_\pi^2} M$, where the $M$ is the matter field, $\pi$ is the
pseudoscalar meson, and $F_\pi$ is the pion decay constant. The
model-independent seagull term is a direct manifestation of the
chiral symmetry, which does not introduce new coupling constants.

Now we move on to the scattering process of the pseudoscalar meson
and the matter fields. The physical $\pi M$ scattering amplitude is
composed of the seagull amplitude, the corrections from higher-order
tree-level diagrams and loop corrections. In other words, the
seagull amplitude does not saturate the whole amplitude. Sometimes
the seagull amplitude may be not dominant.

In the chiral unitary approach, only the seagull amplitude is
iterated (unitarized) through the Bethe-Salpeter or
Lippmann-Schwinger equation because of its easy-be-dealt-with and
model-independent expression. From the pole positions of the
unitarized amplitudes, one can extract information of bound states
or resonances. Such states are called dynamically generated states
in the literature. Since the seagull amplitude may miss some
important physics as illustrated above, the chiral unitary approach
sometimes generates spurious signals.

In the scattering of the light vector meson and the matter fields,
one may treat the vector meson as the ``gauge" field and introduce
the hidden local symmetry into the Lagrangian. Then the chiral
unitary formalism can be extended to the processes involving light
vector mesons in a similar way.

If only the pions are exchanged between two heavy hadrons, the
potentials can be organized order by order in the framework of the
chiral perturbation theory ($\chi$PT). One first constructs the
tree-level Lagrangians order by order. Then one computes the higher
order chiral corrections from the loop diagrams such as the triangle
diagram, the box diagrams etc where only the two-particle
irreducible contributions are kept. From the above scattering
amplitudes, one derives the effective potentials at the given chiral
order which enters the Schr\"odinger equation or Lippmann-Schwinger
equation in the iteration. Again, one can extract the binding energy
after solving the Schr\"odinger equation. One may also extract the
resonance parameters either from the phase shifts or from the poles
of the scattering amplitudes. This approach mimics the modern
nuclear force in terms of $\chi$PT which is pioneered by Weinberg
\cite{Weinberg:1990rz,Weinberg:1991um}.

\subsection{$Q\bar{Q}q\bar{q}$}

In Ref. \cite{Chen:2016qju}, we performed a comprehensive review of
the hidden-charm tetraquark states. In this section, we mainly
introduce the theoretical progress on the systems with the
$Q\bar{Q}q\bar{q}$ configuration after 2016. Different theoretical
groups still focused on the interaction between a charmed meson and
an anticharmed meson. The central issue remains to answer whether
these observed $XYZ$ charmoniumlike states can be understood in the
hidden-charm molecular scenario.

In Ref. \cite{Rathaud:2016tys}, the mass spectra for the heavy
meson-antimeson bound states were discussed with the Hellmann plus
one-pion-exchange potential. The results indicate that the $X(3872)$
is not a pure molecule. The authors of Ref. \cite{Liu:2016kqx}
discussed possible hadronic bound states formed by heavy-light
mesons and heavy-light antimesons in the one-pion-exchange approach,
where a heavy-light meson belongs to the $(0^-,1^-)$ multiplet or
its chiral partner multiplet $(0^+,1^+)$. The coupled channel
results show that only the charm isosinglet molecules with
$J^{PC}=1^{++}$ are bound, which can be identified as the $X(3872)$.
The analysis indicates that the reported $1^{+-}$ isotriplet
$Z_c(3900)^\pm$ is at best a near threshold resonance, and the
$Z_c(4020)^\pm$ at best also a threshold effect. The study for the
hidden-bottom meson-antimeson systems leads to the identification of
the isotriplet exotic molecule with $J^{PC}=1^{+-}$ as a mixture of
the $Z_b^+(10610)$ and $Z_b^+(10650)$. A bound isosinglet
hidden-bottom molecule $X_b(10532)$ with $J^{PC}=1^{++}$ analogous
to the $X(3872)$ was proposed.

A recent systematic study of the possible hadronic molecules
composed of the $S$-wave heavy-light mesons can be found in Ref.
\cite{Liu:2017mrh}. The assignment of the $X(3872)$ as a shallow
$D\bar{D}^*$ molecule is supported, while assigning the $Z_c(3900)$,
$Z_c(4020)$, $Z_b(10610)$, and $Z_b(10650)$ as molecules is not
favorable. Moreover, the $Y(4140)$ cannot be assigned as a molecule
composed of $D_s^{*+}D_s^{*-}$, neither.

With the Bethe-Salpeter equation, the authors of Ref.
\cite{Chen:2015fdn} calculated the mass of the $X(4140)$ in the
molecule picture. They assigned this meson to be a mixed state of
three pure molecule states $D^{*0}\bar{D}^{*0}$, $D^{*+}D^{*-}$, and
$D_s^{*+}D_s^{*-}$. In Ref. \cite{Wang:2017dcq}, the $D\bar{D}^*$
molecule in the Bethe-Salpeter equation approach was discussed with
the ladder and instantaneous approximations by considering $\sigma$,
$\pi$, $\eta$, $\rho$, and $\omega$ exchanges. The $X(3872)$ can be
a molecular bound state.

The possibility of the $X(4274)$ as a $P$-wave
$D_s\bar{D}_{s0}(2317)$ state in a quasipotential Bethe-Salpeter
equation approach was discussed in Ref. \cite{He:2016pfa}. A pole at
$4275\pm11i$ MeV was produced through the $P$-wave interaction with
the $J/\psi\phi$ channel. If this state can be interpreted as a
hadronic molecule, an $S$-wave $D_s\bar{D}_{s0}(2317)$ bound state
below the $J/\psi\phi$ threshold should also exist.

In Ref. \cite{He:2017mbh}, He and Chen investigated the
$D^*\bar{D}_1$ interaction in a quasipotential Bethe-Salpeter
equation approach by considering one-pion-exchange interaction. They
obtained a bound state at 4384 MeV with $I^G(J^{PC})=0^-(1^{--})$
and a resonance at 4.461+$i$39 MeV with $I^G(J^{P})=1^+(1^+)$. The
former (latter) state was assigned to be the $Y(4390)$ (Z(4430)).
Later in \cite{Chen:2017abq}, Chen, Xiao, and He found that the
molecule assignment for the $Y(4390)$ can naturally explain its
observation in $e^+e^-\to\pi\pi h_c$ and its absence in
$e^+e^-\to\pi\pi J/\psi$.

In a study with chiral unitary method in Ref. \cite{Sakai:2017avl},
Sakai, Roca, and Oset investigated the $(\bar{b}n)$-$(c\bar{n})$ and
$(\bar{b}n)$-$(\bar{c}n)$ interactions. They found isoscalar bound
states above 7 GeV for the $0^+$ $BD$, $1^+$ $B^*D$ and $BD^*$, and
$(0,1,2)^+$ $B^*D^*$ systems. In Ref. \cite{Sun:2017wgf}, the
$D\bar{D}^*$ interaction was investigated with an extended hidden
gauge symmetry in the chiral unitary model. A bound state slightly
lower than the $D\bar{D}^*$ threshold was dynamically generated in
the isoscalar channel, which was related to the $X(3872)$. An
isoscalar $B\bar{B}^*$ bound state was also found.

These authors in Ref. \cite{Ortega:2018cnm} studied the $Z_c(3900)$
and $Z_c(4020)$ structures in a constituent quark model by
considering the coupled-channels $D^{(*)}\bar{D}^*+h.c.$, $\pi
J/\psi$, and $\rho\eta_c$. They found that these two structures are
virtual states, whose effects as shown in the production line shapes
can be seen as the $D^{(*)}\bar{D}^{(*)}$ threshold cusps.

In the hadrocharmonium picture, the authors of Ref.
\cite{Panteleeva:2018ijz} studied possible $\psi(2S)\phi$ bound
states. They authors found that the obtained $S$-wave vector-vector
bound state corresponds to a mass-degenerate multiples, and the
$X(4274)$ can be such a state. The degeneracy indicates that two
more structures around the $X(4274)$ should exist if this picture is
correct. In Ref. \cite{Ferretti:2018kzy}, the bound states formed by
$\eta_c$ or $J/\psi$ and isoscalar mesons in the hadrocharmonium
picture were investigated. From their results, the $X(3915)$,
$X(3940)$, $X(4160)$, $Y(4260)$, and $Y(4360)$ are
$\eta_c\eta^\prime$, $\eta_cf_0$, $\eta_cf_1$, $J/\psi f_1$, and
$J/\psi f_2$ states, respectively. A different assignment was
proposed in Ref.~\cite{Molina:2009ct} to interpret the $X(3915)$, $X(3940)$, and $X(4160)$ as
dynamically generated resonances in the coupled channels of $D^* \bar D^*$ and $D_s^* \bar D_s^*$,
with some relevant ones.

Voloshin interpreted the $Z_c(4100)$ and $Z_c(4200)$ as two states
of hadrocharmonium related by the charm quark spin symmetry
\cite{Voloshin:2018vym}. Later in Ref. \cite{Voloshin:2019ilw}, he
predicted the existence of the strange hadrocharmonium resonances
$Z_{cs}(4250)$ and $Z_{cs}(4350)$ which decay dominantly into
$\eta_cK$ and $J/\psi K$, respectively.

In a study of the scattering problem related with near-threshold
heavy-flavor resonances \cite{Gao:2018jhk}, the authors found that
both the $D\bar{D}^*$ and other hadronic degrees of freedom are
equally important inside the $Z_c(3900)$. The $D^*\bar{D}^*$,
$D_s\bar{D}^*_s$, and $\Lambda_c\bar{\Lambda}_c$ components inside
the $X(4020)$, $X(4140)$, and $Y(4660)$, respectively, are not so
important. For the $Y(4260)$, this study favors its interpretation
as the $D_1\bar{D}$ molecule. A previous study with the same
approach~\cite{Kang:2016ezb} indicates that both the $Z_b(10610)$
and $Z_b(10650)$ are dominated by the $B^{(*)}\bar{B}^*$ component,
76\% and 68 \%, respectively. See also discussions in Ref.~\cite{Dias:2014pva}.

\subsection{$QQ\bar{q}\bar{q}$}

In Ref. \cite{Molina:2010tx}, besides the $D_{s2}^*(2573)$, the
authors also studied the doubly charmed vector-vector meson states
within the hidden gauge formalism in a coupled channel unitary
approach. They obtained a pole around 3970 MeV in the
$I(J^P)=0(1^+)$ $D^*D^*$ channel, which is 100 MeV above the $DD^*$
threshold. Later in an extended work \cite{Sakai:2017avl}, the
$(\bar{b}n)$-$(c\bar{n})$ and $(\bar{b}n)$-$(\bar{c}n)$ interactions
were systematically investigated in order to see whether there are
dynamically generated states. The authors noted that the
$I(J^P)=0(1^+)$ bound states in $B^*\bar{D}$, $B\bar{D}^*$, and
$B^*\bar{D}^*$ systems are all possible. In Ref.
\cite{Carames:2011zz}, the doubly charmed exotic states as
meson-meson molecules were investigated. The authors studied the
scattering problem by solving the Lippmann-Schwinger equation, where
the effective potentials were derived from the chiral constituent
quark model \cite{Valcarce:2005em}. Their results suggest the
existence of a stable $QQ\bar{q}\bar{q}$ state with $I(J^P)=0(1^+)$.

The authors of Ref. \cite{Ohkoda:2012hv} presented a study of
possible doubly charm and doubly bottom molecular states composed of
a pair of heavy mesons, by solving the coupled Schr\"odinger
equations through the one-boson-exchange (OBEP) method. The authors
got bound and/or resonant states of various quantum numbers up to
$J\leq2$. In the $I(J^P)=0(1^+)$ case, they obtained the $DD^*$ and
$\bar{B}\bar{B}^*$ bound states with binding energies about tens of
MeV. In another coupled channel calculation for the molecule problem
of $(Q\bar{q})$-$(Q\bar{q})$ ($Q=c,b$, $q=u,d,s$) systems in the
OBEP framework \cite{Li:2012ss}, the $I(J^P)=0(1^+)$ $DD^*$ and
$\bar{B}\bar{B}^*$ bound states were also obtained. In addition, the
$I(J^P)=0(1^+)$ $D\bar{B}^*$ molecule were found to be bound
together with several other molecules of various configurations.
Studies of their partner molecules can be found in Ref.
\cite{Sun:2012sy}.

Using the lattice potentials from
Refs.~\cite{Bicudo:2012qt,Bicudo:2015kna,Bicudo:2015vta} and the
Born-Oppenheimer approximation, Bicudo {\it et al} studied the
$ud\bar{b}\bar{b}$ tetraquark states and found an $I(J^P)=0(1^+)$
state about 90 MeV below the $BB^*$ threshold. Later in Ref.
\cite{Bicudo:2017szl}, an $I(J^P)=0(1^-)$ resonance with
$m=10576^{+4}_{-4}$ MeV and $\Gamma=112^{+90}_{-103}$ MeV was found.
The authors of Ref. \cite{SanchezSanchez:2017xtl} discussed the
possible states with the exotic doubly charmed configurations
${D}_{s0}^{*}(2317)D$ and ${D}_{s1}^{*}(2460){D}^{*}$ based on the
kaon-exchange interactions. The authors obtained bound states in the
$J^P=0^-$ and $2^-$ channels.

\begin{figure}[htpb]
\begin{center}
\includegraphics[scale=0.45]{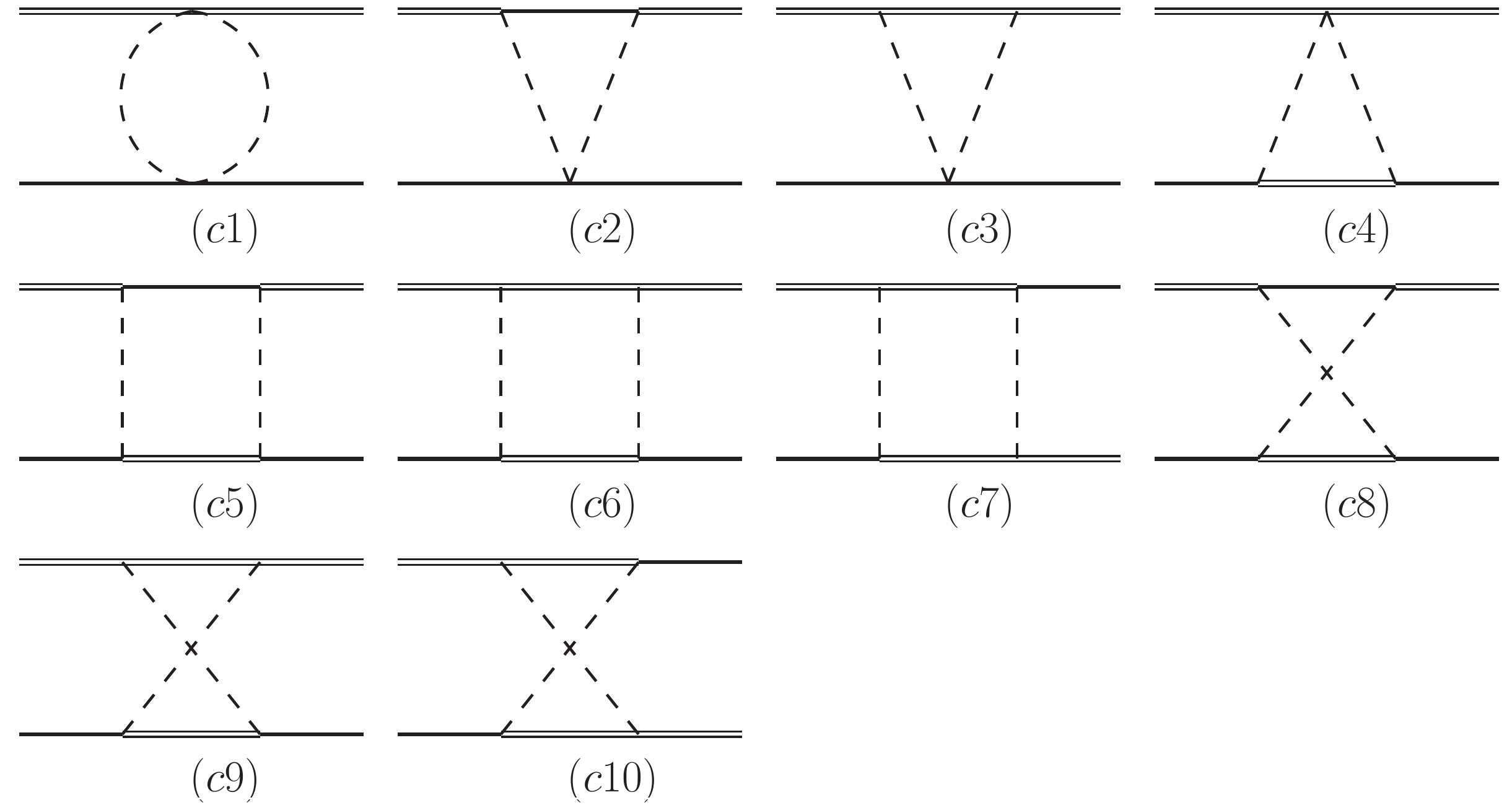}
\caption{Two-pion-exchange diagrams at $O(\epsilon^2)$ for the
$DD^*$ interaction. The solid, double-solid, and dashed lines stand
for $D$, $D^{*}$, pion, respectively. Figure was taken from Ref.
\cite{Xu:2017tsr}}\label{O22piDiagram}
\end{center}
\end{figure}

In the framework of the chiral effective field theory, the authors
of Ref. \cite{Xu:2017tsr} derived the effective $DD^*$ potentials up
to the order ${\cal O}(\epsilon^2)$, where $\epsilon$ can be the
momentum of the pion, residual momentum of heavy mesons, or the
$D$-$D^*$ mass splitting. As shown in Fig. \ref{O22piDiagram}, the
two pion exchange contribution which is equivalent to
phenomenological $\rho$ and $\sigma$ exchange contribution, was
considered in their calculation. With the obtained potentials in the
coordinate space, the authors found an $I(J^P)=0(1^+)$ bound state
by solving the Schr\"odinger equation. Further study in the same
framework \cite{Wang:2018atz} showed that two $I(J^P)=0(1^+)$ bound
states, $\bar{B}\bar{B}^*$ and $\bar{B}^*\bar{B}^*$, are possible.

\subsection{$Q\bar{Q}qqq$}

In 2011, the authors of Ref. \cite{Yang:2011wz} predicted the
existence of the hidden-charm molecular pentaquarks in the framework
of one boson exchange model. Later, the hidden-charm $P_c(4380)$ and
$P_c(4450)$ states were announced by LHCb  in the $\Lambda_b\to
J/\psi p K$ process \cite{Aaij:2015tga}, which inspired extensive
discussions of the molecular pentaquarks with the $Q\bar{Q}qqq$
content. The research status of the hidden-charm molecular
pentaquarks composed of an S-wave charmed baryon and an S-wave
anti-charmed meson was covered in great details in Ref.
\cite{Chen:2016qju}. Since the investigation along the same line
continues, we discuss the progress in the past three years in this
section. The coupled channel bound states problem of
$\Sigma_c^{(*)}\bar{D}^{(*)}$ was studied with the one-pion-exchange
model in Ref. \cite{Shimizu:2016rrd}. The authors found the
existence of one or two bound states with the binding energy of
several MeV below the $\Sigma_c^*\bar{D}$ threshold. This result
indicates that the $P_c(4380)$ can be understood as a loosely bound
molecule. The extended analysis to the $b\bar{b}$, $b\bar{c}$, and
$c\bar{b}$ cases also shows pentaquarks lying about 10 MeV below the
corresponding thresholds. For the $P_c(4450)$, it was proposed to be
the Feshbach resonance. In Ref. \cite{Shimizu:2017xrg}, an extended
study for the coupled system of $I(J^P)=\frac12(\frac32^-)$ $J/\psi
N$-$\Lambda_c\bar{D}^{(*)}$-$\Sigma_c^{(*)}\bar{D}^{(*)}$ with the
complex scaling method was performed. The potential included both
the pion exchange and $D^{(*)}$ meson exchange force. The
contribution of the $J/\psi N$ channel was very small, and a
pentaquark candidate with a value close to the $P_c(4380)$ was
obtained.

In Ref. \cite{Yamaguchi:2016ote}, the coupled $(\bar{c}n)$-$(cnn)$
meson-baryon systems were investigated in a one-meson-exchange
potential model. From the obtained bound and resonant states with
$I(J^P)=1/2(3/2,5/2)^\pm$, the authors concluded that the $J^P$
assignments for the $P_c(4380)$ and $P_c(4450)$ are $3/2^+$ and
$5/2^-$, respectively, in agreement with the LHCb results. In Ref.
\cite{Yamaguchi:2017zmn}, the study was extended to include coupling
with compact five-quark channels and to the hidden-bottom case. The
short-range effective potential from the compact five-quark channel
was found to be attractive, which plays an important role in
producing the $P_c$ states.

The author of Ref. \cite{He:2016pfa} explored the
$\bar{D}^*\Sigma_c$ interactions with a Bethe-Saltpeter equation
method in the hadronic molecule picture by considering the channel
coupling to $J/\psi p$. A pole near the $\bar{D}^*\Sigma_c$
threshold with $J^P=5/2^+$ can be found. When this $P$-wave state is
produced near threshold, the $S$-wave pole corresponding to
$J^P=3/2^-$ locates around 4390 MeV. These two poles are suggested
to be related with the $P_c(4450)$ and $P_c(4380)$, respectively.
Later in Ref.~\cite{He:2017aps}, by replacing the $\bar{c}$ quark
with the $\bar{s}$ quark, the author interpreted the $N(1875)$ and
$N(2100)$ as the strange molecular partners of the $P_c(4380)$ and
$P_c(4450)$, respectively.

In Ref. \cite{Chen:2016ryt}, possible strange hidden-charm
pentaquarks in the $\Sigma_c^{(*)}\bar{D}_s^*$ and
$\Xi_c^{(\prime,*)}\bar{D}^*$ systems were investigated in a
one-boson-exchange model. Promising candidates including
$I(J^P)=0(1/2^-)$ $\Xi^\prime_c\bar{D}^*$ and $0(1/2,3/2)^-$
$\Xi_c^*\bar{D}^*$ states were predicted. This hidden-charm
pentaquark with strangeness is accessible at LHCb.

The authors of Ref. \cite{Geng:2017hxc} discussed the scale
invariance in hadron molecules like the
$\Lambda_{c1}\bar{D}$-$\Sigma_c\bar{D}^*$ ($\Lambda_{c1}$ denotes
$\Lambda_c(2595)$) coupled state, which is related to the
$P_c(4450)$. The nearly on-shell pion exchange transition can
generate a long-range $1/r^2$ potential which can lead to
approximate scale invariance for the equation describing the
molecule if the attraction is strong enough. As a result, an
Efimov-like geometrical spectrum in two-hadron systems is possible.
The molecules, $\Lambda_{c1}\bar{D}$-$\Sigma_c\bar{D}^*$ of $1/2^+$,
$\Lambda_{c1}D$-$\Sigma_cD^*$ of $1/2^+$,
$\Lambda_{c1}\bar{\Xi}_b$-$\Sigma_c\bar{\Xi}_b^\prime$ of $0^+/1^-$,
and $\Lambda_{c1}\Xi_b$-$\Sigma_c\Xi_b^\prime$ of $1^+$, seem to
satisfy the condition for the approximate scale invariance.

In Ref. \cite{Chen:2017vai}, Chen, Hosaka, and Liu explored the
intermediate- and short-range forces in the framework of
one-boson-exchange model and studied various hadron-hadron bound
state problems. They found $S$-wave $\Lambda_c\bar{D}$,
$\Lambda_cB$, $\Lambda_b\bar{D}$, and $\Lambda_bB$ molecules. In the
study of bound state problems in Ref. \cite{Rathaud:2018fna}, the
authors proposed a one-boson-exchange potential model by adding a
screen Yukawa-like potential. From their results, the $P_c(4450)$
was interpreted as an $I(J^P)=1/2(3/2^-)$ $\Sigma_c\bar{D}^*$
molecule.

The authors of Ref. \cite{Shen:2017ayv} presented an exploratory
study of possible $\bar{D}\Lambda_c$-$\bar{D}\Sigma_c$ and
$B\Lambda_c$-$B\Sigma_c$ resonances by extending the J\"ulich-Bonn
dynamical coupled-channel framework, which is a unitary meson-baryon
exchange model. The authors found one pole in each partial wave up
to $J^P=5/2^\pm$ in the hidden-charm case and several poles in
partial waves up to $G_{17}$ in the hidden-bottom case. The very
narrow pole with $J^P=1/2^-$ in the hidden-charm case was
interpreted as a $\bar{D}\Sigma_c$ bound state, which was predicted
in Refs. \cite{Wu:2010jy,Wu:2010vk}.

In Ref. \cite{Huang:2018wed}, Huang and Ping investigated the
hidden-charm and hidden-bottom pentaquark resonances in
hadron-hadron scattering processes in the framework of the quark
delocalization color screening model. They found a few narrow
hidden-charm resonances above 4.2 GeV and some narrow hidden-bottom
states above 11 GeV. Besides, they also noticed bound
$(Q\bar{Q})$-$N$ states from the behavior of the low-energy phase
shifts obtained with coupled channel calculations.

In Ref. \cite{Eides:2015dtr}, the authors interpreted the
$P_c(4450)$ as a $\psi(2S)N$ bound state generated by the
charmonium-nucleon interaction in terms of charmonium chromoelectric
polarizabilities and the nucleon energy-momentum distribution. They
obtained two almost degenerate narrow states at the position of the
$P_c(4450)$ with $J^P=1/2^-$ and $3/2^-$. The authors of Ref.
\cite{Perevalova:2016dln} confirmed the results and predicted the
isospin-3/2 $\psi(2S)\Delta$ narrow bound states around 4.5 GeV and
broader resonances around 4.9 GeV in the framework of the Skyrme
model. Later in Ref. \cite{Eides:2017xnt}, with a QCD inspired
approach, the authors interpreted the $P_c(4450)$ as a bound
$\psi(2S)N$ state with $J^P=3/2^-$, which is a member of one of two
almost degenerate hidden-charm baryon octets. According to their
study, one has to assign the $P_c(4380)$ as a $J^P=5/2^+$ state but
no natural interpretation was found in the hadroquarkonium picture.
The authors also studied the other hadroquarkonia and compared their
results with those in the one-pion exchange approach. They did not
find any $\Upsilon(1S)N$ bound state but found an inconclusive
$\Upsilon(2S)N$ state.

A study of the LHCb $P_c$ states in the hadroquarkonium picture was
carried out in Ref. \cite{Anwar:2018bpu}. The authors solved the
Schr\"odinger equation with the potential from QCD multiple expansion
and got spectrum. The $P_c(4380)$ and $P_c(4450)$ can be interpreted
as the $\psi(2S)N$ and $\chi_{c2}(1P)N$ bound states, respectively.
They also predicted the hadroquarkonium states in the hidden-bottom
sector.

Unfortunately, no evidence of the exotic $\psi(2S)p$ state was
observed by LHCb in the $\Lambda_b^0$ decay into $\psi(2S)p\pi^-$
\cite{Aaij:2018jlf}. The hadrocharmonium picture is being
challenged.

After the observation of $P_c(4380)$ and $P_c(4450)$, their possible
inner structures were proposed either as tightly bound pentaquark
states
\cite{Maiani:2015vwa,Anisovich:2015cia,Li:2015gta,Li:2015gta,Ghosh:2015ksa,Lebed:2015tna,Zhu:2015bba}
or loosely bound molecular states \cite{Chen:2015loa,Chen:2015moa}.
However, the experimental data in Ref. \cite{Aaij:2015tga} was
unable to distinguish these configurations. In 26 March 2019, at the
Rencontres de Moriond QCD conference, the LHCb Collaboration
reported the observation of three new pentaquarks \cite{lhcbnew}. The
observed $P_{c}(4312)^+$ may correspond to the $\Sigma_c\bar{D}$
molecule with $I(J^{P})=1/2(1/2^-)$, while the $P_c(4440)^+$ and
$P_c(4457)^+$ can be identified as the $\Sigma_c\bar{D}$ molecular
states with $I(J^{P})=1/2(1/2^-)$ and $1/2(3/2^-)$. The current
measurement strongly supports the molecular hidden-charm pentaquarks
predicted in Refs. \cite{Yang:2011wz,Wu:2010jy,Wu:2010vk}, which
shall become a milestone in the exploration of the multiquark
matter. After their announcement, these three pentaquark states
were immediately studied in the QCD sum rules~\cite{Chen:2019bip},
the one-boson-exchange(OBE) model~\cite{Chen:2019asm},
a contact-range effective field theory and a contact-range effective field theory~\cite{Liu:2019Pc}. The isospin breaking decay pattern was studied in~\cite{Guo:2019Pc}.
In Ref.~\cite{Chen:2019asm}, the authors studied these new
pentaquarks in a direct calculation with the OBE model. Their result
supported that the $P_{c}(4312)^+$, $P_c(4440)^+$
and $P_c(4457)^+$ may correspond to the loosely bound $\Sigma_c\bar{D}$ molecule
with ($I=1/2, J^P=1/2^-$),  $\Sigma_c\bar{D}^\ast$ with ($I=1/2, J^P=1/2^-$)
and $\Sigma_c\bar{D}^\ast$ with ($I=1/2, J^P=3/2^-$), respectively.

\subsection{$QQqq\bar{q}$}

In Refs. \cite{Hofmann:2005sw,Hofmann:2006qx}, various baryon-meson
molecular states were studied by considering vector-meson exchange
forces as well as the coupled channel effects with the broken
$SU(4)$ flavor symmetry. In the charm number $C=2$ sector, the
authors predicted more than 10 states with various quantum numbers.
The authors of Ref. \cite{Guo:2011dd} predicted a possible
$\Xi_{cc}\bar{K}$ molecule, and the authors of Ref.
\cite{Romanets:2012hm} systematically studied dynamically generated
double-charm baryons in a coupled channel unitary model. In a chiral
unitary approach, the authors of Ref. \cite{Dias:2018qhp} studied
various dynamically generated double-charm meson-baryon states
through vector meson exchanges interactions. They found ten
molecular pentaquark states.

In Ref. \cite{Xu:2010fc}, the $\Lambda_cD$ and $\Lambda_b\bar{B}$
interactions and related bound state problems were studied. The
authors derived two-pion-exchange potentials and regularized the
divergence with a phenomenological cutoff. With solutions from the
Schr\"odinger equation, they concluded that the $\Lambda_b\bar{B}$
bound state is possible, but the $\Lambda_cD$ bound state can only
be prudently expected. In Ref. \cite{Chen:2017vai}, Chen, Hosaka,
and Liu explored the intermediate- and short-range forces in the
framework of one-boson-exchange model and studied various
hadron-hadron bound state problems. They found $S$-wave
$\Lambda_c{D}$, $\Lambda_c\bar{B}$, $\Lambda_bD$, and
$\Lambda_b\bar{B}$ molecules.

In addition to the study of the $I(J^P)=1/2(3/2^-)$ coupled system
$J/\psi N$-$\Lambda_c\bar{D}^{(*)}$-$\Sigma_c^{(*)}\bar{D}^{(*)}$ in
Ref. \cite{Shimizu:2017xrg}, Shimizu and Harada also considered the
coupled system $\Lambda_cD^{(*)}$-$\Sigma_c^{(*)}D^{(*)}$ in the
$I(J^P)=1/2(3/2^-)$ channel. The authors used the same
one-pion-exchange potential after the antimesons are replaced by
mesons. They noted that there exists a doubly charmed baryon of the
$ccnn\bar{n}$ type as a hadronic molecule, named $\Xi_{cc}^*(4380)$,
whose mass and width are close to those of $P_c(4380)$.

The observation of the double-charm baryon $\Xi_{cc}(3621)$ by LHCb
stimulated the investigations of the molecule-type $QQqq\bar{q}$
states in the literature. In Ref. \cite{Guo:2017vcf}, Guo considered
the scattering of the $QQq$ baryons and Goldstone bosons in a chiral
effective theory. According to their calculation, there are a pair
of bound and virtual states near the $\Xi_{cc}K$ threshold,
indicating that there exists an interesting exotic state with the
quark component $ccud\bar{s}$. A resonance pole around the
$\Xi_{cc}\bar{K}$ threshold, a bound state below the
$\Xi_{cc}\bar{K}$ threshold, and two resonances around the
$\Xi_{cc}\pi$ and $\Omega_{cc}K$ thresholds were also found. The
authors of Ref. \cite{Rathaud:2018fna} presented a study of various
meson-baryon and baryon-baryon/antibaryon states in a
one-boson-exchange potential model compensated by a screened
Yukawa-like potential. A series of double-heavy molecules were
obtained.

\subsection{$QQQq\bar{q}$}

In addition to the molecular bound or resonant states with charm
number $C=0$, $C=-1$, $C=1$, and $C=2$, the states with $C=3$ were
also studied in Refs. \cite{Hofmann:2005sw,Hofmann:2006qx}. A flavor
singlet bound state with $J^P=1/2^-$ ($3/2^-$) with mass around 4.3
(4.3) GeV is possible, which results from the scattering of a $QQq$
baryon and a $Q\bar{q}$ meson as well as a $QQQ$ baryon and a
$q\bar{q}$ meson. In the systematic study of triply charmed
dynamically generated baryons in Ref. \cite{Romanets:2012hm}, the
authors found one $J^P=1/2^-$ bound state around 4.4 GeV and one
$J^P=3/2^-$ bound state around 4.5 GeV by coupling
$(ccc)$-$(q\bar{q})$ and $(ccq)$-$(c\bar{q})$ channels.

\begin{figure}[!htbp]
\center
\includegraphics[width=4.4in]{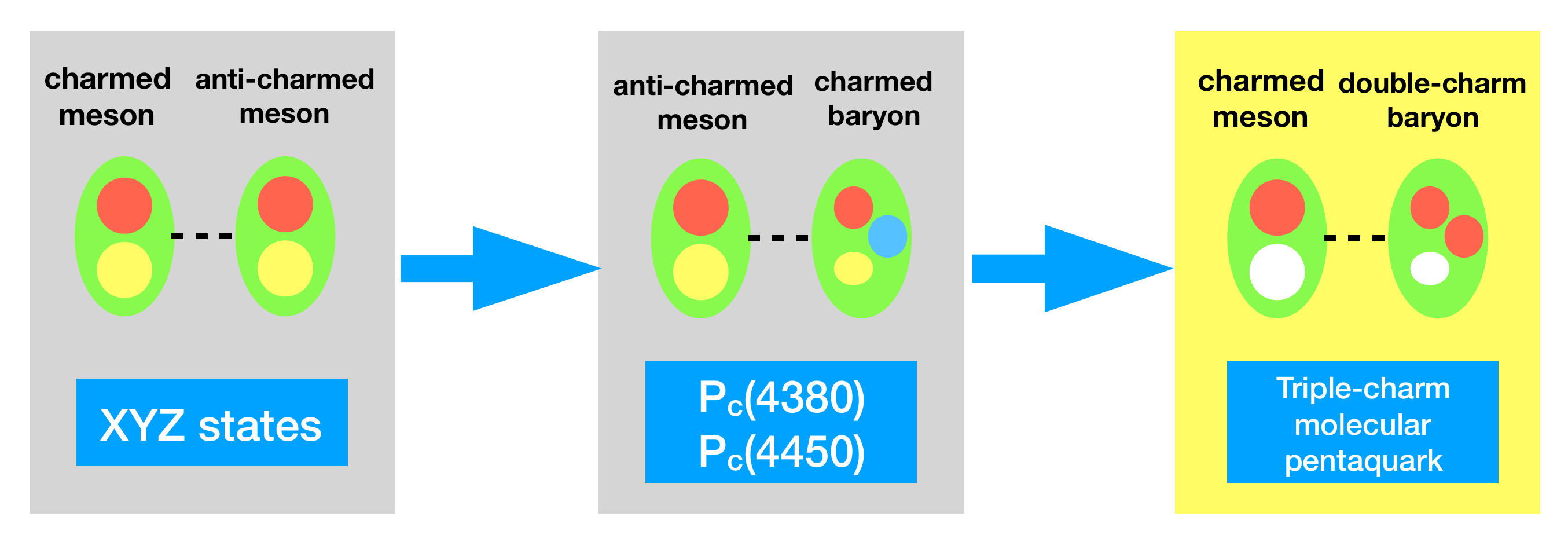}
\caption{Evolution of interaction of hadrons and the corresponding
connections with charmonium-like $XYZ$ states,
$P_c(4380)/P_{c}(4450)$ and triple-charm molecular pentaquark.
Figure was taken from Ref. \cite{Chen:2017jjn}.}\label{fig:1}
\end{figure}

The discovery of the charmonium-like $XYZ$ states and $P_c(4380)$
and $P_c(4450)$ pushed the exploration of the hadronic molecular
states in the past decade, which was based on the interaction
between charmed hadron and anti-charmed hadron. The observation of
the double-charm baryon $\Xi_{cc}^{++}(3621)$ drives us to further
explore the interaction between the double-charm baryon and charmed
meson, which is a natural extension of the one boson exchange model
as shown in Fig. \ref{fig:1}. The same idea was further extended to
the investigation of the interaction between the double-charm baryon
and charm baryon in Ref. \cite{Chen:2018pzd}.

Due to the above motivation, in a one-boson-exchange potential
model, possible molecular states formed by the $\Xi_{cc}(3621)$ and
a charmed state $D$ and $D^*$ were discussed in Ref.
\cite{Chen:2017jjn}. Two isoscalar bound states, $\Xi_{cc}D$ of
$1/2^-$ and $\Xi_{cc}D^*$ of $3/2^-$, were found. Similar bound
states with the same quantum numbers were found by replacing
$D^{(*)}$ with $\bar{B}^{(*)}$. In Ref. \cite{Wang:2019aoc},
possible triple-charm molecular pentaquarks $\Xi_{cc}D_1$ and
$\Xi_{cc}D_2^*$ were further discussed. The $I(J^P)=0(1/2^+,3/2^+)$
$\Xi_{cc}D_1$ and $I(J^P)=0(3/2^+,5/2^+)$ $\Xi_{cc}D_2^*$ loosely
bound molecules are possible.

\subsection{A short summary}

Since 2003, the observations of a series of charmonium-like
near-threshold $XYZ$ states have stimulated strong interest in the
interactions between two heavy mesons. Later, the coupled-channel
effect, various hyperfine interactions and recoil corrections were
introduced into the one-boson-exchange model step by step. Within
this simple framework, the hidden-charm molecular type pentaquarks
were predicted. In 2015, LHCb collaboration did surprise us with two
$P_c$ states. Very recently, LHCb collaboration updated their
analysis with a ten times larger data sample, which strongly
supports the molecular pentaquarks. At present, the one boson
exchange model remains an effective and popular tool to deal with
the hadronic interactions.

However, this framework sometimes lacks the definite predictive
power because of too many unknown parameters such as the coupling
constants and cutoff parameter. The original one-boson-exchange
model was proposed for the nuclear force where there exists plenty
of experimental data such as the deuteron binding energy and enough
nucleon nucleon scattering data, which can be used to fix all the
unknown model parameters. In contrast, except the pionic couplings,
most of the light meson and heavy hadron interaction vertices remain
unknown. Especially, the bound state or resonance solution is very
sensitive to the cutoff parameter in the form factor which is
introduced to suppress the ultraviolet contribution.

One may borrow the same formalism from the modern nuclear force and
construct the interaction potential between two heavy hadrons in the
framework of chiral perturbation theory. The pion exchange is still
responsible for the long-range force. The two-pion exchange mimics
the scalar meson and vector meson exchange to some extent. The
contact heavy hadron interaction contributes to the short-range
interaction in terms of the low energy constants which may be
extracted through fitting to the lattice simulation data on the
heavy hadron scattering if possible. One may expect significant
progress along this direction in the future.

In short summary, many loosely bound molecular states or resonances
such as the hidden-charm and open-strange pentaquark states and
triple-charm pentaquark states have been proposed either within the
one-boson-exchange model or with various unitary schemes, which may
be searched for at LHCb and BelleII.

%% file: section5.tex
\section{Heavy quark symmetry and multiquark states}\label{sec:symmetry}

QCD exhibits the chiral symmetry in the limit when quarks are
massless, and the heavy quark symmetry \cite{Manohar:2000dt} in the
limit when quarks have infinitely large masses, both of which play
important roles in understanding properties of hadrons and their
interactions. The latter symmetry indicates both the heavy quark
flavor symmetry (HQSS) which means that the dynamics is not affected by the
exchange of heavy quark flavors, and the heavy quark spin symmetry (HQSS)
which means that the dynamics is irrelevant of the heavy quark spin.
For hadrons containing a single heavy quark, there exist degenerate
multiplets which can be classified by the angular momentum of the
light degree of freedom inside the hadrons. Recently, discussions
about the application of the heavy quark symmetry to hadronic
systems containing one heavy quark were performed in
Ref.~\cite{Hosaka:2016ypm}.

In the heavy quark limit, the heavy quark core of a hadron looks
like a static color source, and the hadron resembles an atomic
system. For conventional mesons with one heavy quark and
conventional baryons with two heavy quarks, the heavy color source
is of course in the representation $3_c/\bar 3_c$. They can be
related with the heavy antiquark-diquark symmetry (HADS)
\cite{Savage:1990di,Brambilla:2005yk,Fleming:2005pd,Cohen:2006jg}
since the motion and spin interactions of the color source are
negligible in this limit. An example of such relations is given in
Ref.~\cite{Cohen:2006jg}
\begin{eqnarray}
M_{\Xi_{cc}^*}-M_{\Xi_{cc}}=\frac34(M_{D^*}-M_D).
\end{eqnarray}
For multiquark hadrons with two heavy quarks, the heavy quark core
can also be in the representation $6_c$. One may naively expect that
the lowest states contain the color-$\bar{3}_c$ heavy diquarks only,
and then the doubly heavy tetraquarks can be related to the doubly
heavy baryons. In fact, mass relations between $QQ\bar{q}\bar{q}$,
$QQq$, $Qqq$, and $Q\bar{q}$ states \cite{Eichten:2017ffp},
\begin{eqnarray}
M(QQ\bar{q}\bar{q})-M(QQq)=M(Qqq)-M(Q\bar{q}),
\end{eqnarray}
can be found according to the heavy quark symmetry. In Ref.
\cite{Cohen:2006jg}, other mass relations were given. Similarly, one
may understand that relations exist between $QQqq\bar{q}$
pentaquarks and $Qq\bar{q}\bar{q}$ tetraquarks. With these heavy
quark relations and heavy quark mass corrections, one may predict
relevant multiquark masses.

The heavy quark symmetry may also be used to discuss hadronic molecule
problem, where no additional light particles are created and the
heavy hadron components remain distinct and move
non-relativistically \cite{Cai:2019orb}. From the existence of one
molecule which consists of two or more heavy hadrons, one can expect
possible existence of its partner molecules according to the heavy
quark symmetry. The respective binding energies can be obtained  by
solving the Schr\"odinger equation.

\subsection{$Q\bar{Q}q\bar{q}$}

In Ref. \cite{Cleven:2015era}, the exotic charmonium spectrum in
hadron-charmonium, tetraquark, and hadronic molecule scenarios were
investigated by employing heavy quark spin symmetry. The spectrum
patterns from different models turn out to be quite distinct. For
the $Y(4260)$, a lighter $0^{-+}$ partner state around 4140 MeV
should exist only within the hadrocharmonium picture.

Three hadronic partner states of the $X(3872)$ with $J^{PC}=0^{++}$,
$1^{+-}$, and $2^{++}$ should exist in the strict heavy-quark limit,
if it is a $1^{++}$ $D\bar{D}^*$ molecular state
\cite{Baru:2016iwj}. Once the one-pion-exchange interaction was
included, this result was found to be robust only if all relevant
channels and partial waves are considered. The heavy-quark spin
symmetry implied spin partners of the $Z_b(10610)$ and $Z_b(10650)$,
which were studied by assuming they are $B\bar{B}^*$ and
$B^*\bar{B}^*$ molecules, respectively in Ref. \cite{Baru:2017gwo}.
With the force from short-range contact terms and the
one-pion-exchange potential, the authors predicted the existence of
an isovector $J^{PC}=2^{++}$ tensor state lying a few MeV below the
$B^*\bar{B}^*$ threshold.

In Ref. \cite{Liu:2019stu}, the heavy quark symmetry partners of the
$X(3872)$ were discussed by assuming that this exotic meson is a
$D\bar{D}^*$ molecule. With the one-boson-exchange
($\pi,\sigma,\rho,\omega$) potential model in which the cutoff
parameter was extracted from $X(3872)$, the bound state problems of
these meson-antimeson systems were discussed. The location and
quantum numbers of the $Z_b(10610)$ and $Z_b(10650)$ were correctly
reproduced. Other bound states such as $I(J^{PC})=0^+(2^{++})$
$D^*\bar{D}^*$, $0^+(1^{++})$ $B\bar{B}^*$, and $0^+(2^{++})$
$B^*\bar{B}^*$ were predicted. We refer interested readers to
Ref.~\cite{AlFiky:2005jd}, which introduced an effective Lagrangian implementing the heavy 
quark symmetry to describe those molecular states. See also Refs.~\cite{Chiladze:1998ti,Petrov:2005tp}
where the non-relativistic effective theory was applied to describe heavy quarkonium hybrids .

In Ref. \cite{Voloshin:2016cgm}, Voloshin proposed the existence of
light quark spin symmetry (LQSS) in molecular $Z_b$ resonances,
$Z_b(10610)\sim |B^*\bar{B},B\bar{B}^*\rangle$ and $Z_b(10650)\sim
|B^*\bar{B}^*\rangle$, by noticing the weak coupling of the
$Z_b(10650)$ to the $B^*\bar{B}+B\bar{B}^*$ channel. In the case of
free heavy meson pairs, the spin structure of the two $Z_b$ states
reads
\begin{eqnarray}
Z_b(10610)\sim\frac{1}{\sqrt2}(1_{b\bar{b}}^-\otimes
0_{q\bar{q}}^-+0_{b\bar{b}}^-\otimes 1_{q\bar{q}}^-),\qquad
Z_b(10650)\sim\frac{1}{\sqrt2}(1_{b\bar{b}}^-\otimes
0_{q\bar{q}}^--0_{b\bar{b}}^-\otimes 1_{q\bar{q}}^-).
\end{eqnarray}
In the case of interacting heavy meson pairs, if the interaction in
the state $1_{q\bar{q}}^-$ differed from that in $0_{q\bar{q}}^-$,
the two $Z_b$ states would not be the eigenstates for interaction
mesons. Then, one has to assume that the interaction between the
mesons does not depend on the total spin of the light
quark-antiquark pair, which is the proposed LQSS. Unlike the HQSS,
this approximate symmetry is not a symmetry from the underlying QCD,
and is manifestly broken by the pion exchange force. Although the
LQSS is unexpected, it was used to predict the existence of four new
isovector negative $G$-parity resonances, $0^+$ $B\bar{B}$, $1^+$
$B\bar{B}^*$, $0^+$ $B^*\bar{B}^*$, and $2^+$ $B^*\bar{B}^*$,
together with the HQSS. The application of this symmetry to the
charm quark case would be complicated, because of the enhanced
violation of HQSS and the breaking of isospin symmetry.

\subsection{$QQ\bar{q}\bar{q}$}

In Ref. \cite{Eichten:2017ffp}, Eichten and Quigg calculated the
masses of the $QQ\bar{q}\bar{q}$ ($Q=c,b$, $q=u,d,s$) tetraquark
states based on mass relations obtained from the heavy quark
symmetry and finite-mass corrections. According to their  results,
the $I(J^P)=0(1^+)$ $cc\bar{u}\bar{d}$ state is 102 MeV above the
$D^+D^{*0}$ threshold, while the $bb\bar{u}\bar{d}$ state is 121 MeV
below the $B^-\bar{B}^{*0}$ threshold. The finding is consistent
with that by Karliner and Rosner \cite{Karliner:2017qjm}. Besides,
this calculation indicated a stable $bb\bar{n}\bar{s}$ bound state
with $I(J^P)=0(0^+)$ but no stable $bc\bar{u}\bar{d}$.

The chiral Lagrangian that uses HADS to relate doubly heavy
tetraquarks to singly heavy baryons was given in Ref.
\cite{Mehen:2017nrh}, which is an extension of Ref.
\cite{Hu:2005gf}. The author constructed the chiral Lagrangian for
multiplets containing singly heavy mesons and doubly heavy baryons.
With these Lagrangians and the measured mass of the $\Xi_{cc}$,
Mehen studied properties of the five lowest-lying excited $ccn$
states together with doubly heavy tetraquarks. He noted that the
lowest $I(J^P)=0(1^+)$ $bb\bar{u}\bar{d}$ state should be stable. If
its mass is below 10405 MeV, he argued that the $I(J^P)=1(1^+)$
$bb\bar{u}\bar{d}$ state should also be stable.

In Ref. \cite{Cai:2019orb}, the authors presented a
model-independent argument about the existence of near threshold
exotic mesons containing two heavy quarks. Based on Born-Oppenheimer
and semi-classical considerations, they found the existence of
parametrically narrow tetraquarks, which are close to the threshold
of two heavy mesons.

\subsection{$Q\bar{Q}qqq$ and $QQQq\bar{q}$}

In Ref. \cite{Shimizu:2018ran}, the structure of heavy quark spin
multiplets for the $S$-wave $\bar{P}^{(*)}\Sigma_Q^{(*)}$ molecular
states was studied. The authors introduced the light cloud spin
(LCS, the allowed angular momentum of the light degree of freedom)
basis, and constructed the unitary transformation matrices in
relating the LCS basis and the molecule basis. They found four types
of multiplets classified by the structure of heavy quark spin and
LCS, and discussed their restrictions on decay widths of the
molecules. In Ref. \cite{Shimizu:2019jfy}, they extended the
formalism to study the heavy quark spin multiplet structures of the
$P_c$-like pentaquarks as $P$-wave hadronic molecules. Solving the
coupled-channel Schr\"odinger equation in the OPEP model and comparing
results with the previous study \cite{Shimizu:2018ran}, they pointed
out that it seems difficult to explain the masses and decay widths
of the $P_c(4380)$ and $P_c(4450)$ simultaneously.

In Ref. \cite{Liu:2018zzu}, the authors discussed the possible
partner state of the $P_c(4450)$ according to the heavy quark spin
symmetry assuming it to be a $J^P=3/2^-$ $\bar{D}^*\Sigma_c$ bound
state. They predicted the existence of a $5/2^-$
$\bar{D}^*\Sigma_c^*$ molecule with a binding energy similar to the
$P_c(4450)$. In fact, before the observation of $P_c$ states by
LHCb, the authors of Ref. \cite{Garcia-Recio:2013gaa} had discussed
the heavy-quark spin multiplet structures in hidden-charm molecules.

Besides the hidden-charm partner state of the $P_c(4450)$, the
authors of Ref. \cite{Liu:2018zzu} also discussed the possible
open-charm molecule partners of the $P_c(4450)$ using HADS. They
predicted the $\Xi_{cc}\Sigma_c$ molecule with $J^P=0^+$,
$\Xi_{cc}\Sigma_c^*$ molecule with $J^P=1^+$, $\Xi_{cc}^*\Sigma_c$
molecule with $J^P=2^+$, and $\Xi_{cc}^*\Sigma_c^*$ molecule with
$J^P=3^+$. The binding energies of all these states are in the
20$\sim$30 MeV range.

%% file: section6.tex
\section{QCD sum rules}

The formalism of QCD sum rules is a powerful and successful
non-perturbative method~\cite{Shifman:1978bx,Reinders:1984sr}, which
has been widely applied to study the mass spectra and decay
properties of various exotic hadrons. Since we have thoroughly
reviewed its applications to the hidden-charm pentaquark and
tetraquark states in Ref.~\cite{Chen:2016qju} and its applications
to the open-charm tetraquark states in Ref.~\cite{Chen:2016spr}, we
shall only briefly introduce the recent progress of this method in
the present view, and we refer interested readers to
Refs.~\cite{Chen:2016qju,Chen:2016spr,Shifman:1978bx,Reinders:1984sr,Colangelo:2000dp,Narison:2002pw,Nielsen:2009uh,Albuquerque:2018jkn}
for more discussions.

%
\subsection{A short induction to QCD sum rules}
%

A key idea of QCD sum rules is the quark-hadron duality, i.e., the
equivalence of the (integrated) correlation functions at both the
hadron level and the quark-gluon level.

When studying the mass spectrum of the hadron $H$, one calculates
the two-point correlation function
%
\begin{equation}
\Pi(q^2) \, \equiv \, i \int d^4x e^{iqx} \langle 0 | {\mathbb T}
\eta(x) { \eta^\dagger } (0) | 0 \rangle \, , \label{def:pi}
\end{equation}
%
at both the hadron and quark-gluon levels. Here $\eta(x)$ is an
interpolating current which has the same quantum numbers as the
hadron $H$, and their coupling is defined to be
\begin{eqnarray}
\langle 0 | \eta(0) | H \rangle \equiv f_H \, .
\end{eqnarray}
The interpolating current $\eta(x)$ can partly reflect the internal
structure of the hadron $H$, but we still do not fully understand
this relation~\cite{Chen:2018kuu,Cui:2019roq}. We shall discuss this
in details in the next subsection.

At the hadron level one expresses $\Pi(q^2)$ in the form of the
dispersion relation:
%
\begin{eqnarray}
\Pi(q^2) &=& {1\over\pi}\int^\infty_{s_<}\frac{{\rm
Im}\Pi(s)}{s-q^2-i\varepsilon}ds \label{eq:disper}
\\ \nonumber &\equiv& \int^\infty_{s_<}\frac{\rho_{\rm phen}(s)}{s-q^2-i\varepsilon}ds \, ,
\end{eqnarray}
%
where $\rho_{\rm phen}(s) \equiv {\rm Im}\Pi(s)/\pi$ is the spectral
density, for which one usually adopts a parametrization of one pole
dominance for the ground state $H$ and a continuum contribution:
%
\begin{eqnarray}
\rho_{\rm phen}(s) & \equiv & \sum_n\delta(s-M^2_n)\langle
0|\eta|n\rangle\langle n|{\eta^\dagger}|0\rangle \ \nonumber\\ &=&
f^2_H \delta(s-M^2_H)+ \rm{continuum}\, . \label{eq:rho}
\end{eqnarray}
%

At the quark-gluon level, one calculates $\Pi(q^2)$ using the method
of operator product expansion (OPE), and calculates the spectral
density $\rho_{\rm OPE}(s)$ up to certain order in the expansion.
After performing the Borel transformation at both the hadron and
quark-gluon levels, and approximating the continuum using $\rho_{\rm
OPE}(s)$ above a threshold value $s_0$, one obtains the sum rule
equation
%
\begin{equation}
\Pi(s_0, M_B^2) \equiv f^2_H e^{-M_H^2/M_B^2} = \int^{s_0}_{s_<}
e^{-s/M_B^2} \rho_{\rm OPE}(s) ds \label{eq_fin} \, ,
\end{equation}
%
which can be used to calculate $M_H$ through
%
\begin{equation}
M^2_H = {1\over{\Pi(s_0, M_B^2)}}{\frac{\partial \Pi(s_0,
M_B^2)}{\partial(-1/M_B^2)}} = \frac{\int^{s_0}_{s_<}
e^{-s/M_B^2}s\rho_{\rm OPE}(s)ds}{\int^{s_0}_{s_<}
e^{-s/M_B^2}\rho_{\rm OPE}(s)ds} \, . \label{eq_LSR}
\end{equation}
%

The above two-point correlation function is investigated when
extracting hadron masses, while one can also consider the
three-point correlation function to study their decay properties:
\begin{eqnarray}
T_{A \rightarrow BC}(p, p^\prime, q) = \int d^4x d^4y e^{ip^\prime
x} e ^{i q y} \langle0|{\mathbb T} \eta^{B}(x) \eta^C(y)
\eta^{A\dagger}(0) |0\rangle \, .
\end{eqnarray}
Here $p$, $p^\prime$, and $q$ are the momenta of $A$, $B$, and $C$,
respectively. The above correlation function can be used to study
the $A \rightarrow BC$ decay. One still calculates it at both the
hadron and quark-gluon levels.

At the hadron level one expresses $T_{A \rightarrow BC}(p, p^\prime,
q)$ as:
\begin{eqnarray}\label{eq:rhopiPH}
T_{A \rightarrow BC}(p, p^\prime, q) &=& g_{A \rightarrow BC} \times
{f_A f_B f_C \over (m_A^2 - p^2) (m_B^2 - p^{\prime2})(m_C^2 - q^2)}
\, ,
\end{eqnarray}
where $g_{A \rightarrow BC}$ is the coupling constant, and $f_A$,
$f_B$, and $f_C$ are the relevant decay constants.

At the quark-gluon level one calculates $T_{A \rightarrow BC}(p,
p^\prime, q)$ using the method of operator product expansion (OPE).
Again, by using the quark-hadron duality to relate the expressions
of $T_{A \rightarrow BC}(p, p^\prime, q)$ at the hadron and
quark-gluon levels, one can calculate the coupling constant $g_{A
\rightarrow BC}$, and further evaluate the decay width of the $A
\rightarrow BC$ process.

%
\subsection{Interpolating currents and their relations to physical states}
%

When applying the method of QCD sum rules to investigate a physical
state, one always needs to construct the relevant interpolating
current. However, we still do not fully understand their relations:
the interpolating current sees only the quantum numbers of the
physical state, so it can also couple to other physical states and
thresholds having the same quantum numbers; while one can sometimes
construct more than one interpolating currents, all of which couple
to the same physical state.

In 2006 we first systematically constructed all the $u d \bar s \bar
s$ interpolating currents of $J^{PC} = 0^{++}$ in a local product
form, and found that there are five independent currents for this
channel~\cite{Chen:2006hy}. They can be either in the
diquark-antidiquark form ($[qq][\bar q \bar q]$)
%
\begin{eqnarray}
\nonumber\label{define_diquark_current} && S_6 = (\bar{s}_a \gamma_5
C \bar{s}_b^T)(u_a^T C \gamma_5 d_b)\, ,
\\ \nonumber &&
V_6 = (\bar{s}_a \gamma_\mu \gamma_5 C \bar{s}_b^T)(u_a^T C
\gamma^\mu \gamma_5 d_b)\, ,
\\ &&
T_3 = (\bar{s}_a \sigma_{\mu\nu} C \bar{s}_b^T)(u_a^T C
\sigma^{\mu\nu} d_b)\, ,
\\ \nonumber &&
A_3 = (\bar{s}_a \gamma_\mu C \bar{s}_b^T)(u_a^T C \gamma^\mu d_b)\,
,
\\ \nonumber &&
P_6 = (\bar{s}_a C \bar{s}_b^T)(u_a^T C d_b)\, .
\end{eqnarray}
%
or in the meson-meson form ($[\bar q q][\bar q q]$)
%
\begin{eqnarray}
\nonumber\label{define_meson_current} && S_1 =
(\bar{s}_au_a)(\bar{s}_bd_b)\, ,~~~~~~~~~~~~~~~~~~~~~~~~~~~~ S_8 =
(\bar{s}_a{\lambda^n_{ab}}u_b)(\bar{s}_c{\lambda^n_{cd}}d_d)\, ,
\\ \nonumber &&
V_1 = (\bar{s}_a\gamma_\mu u_a)(\bar{s}_b\gamma^\mu d_b)\,
,~~~~~~~~~~~~~~~~~~~~~~ V_8 = (\bar{s}_a\gamma_\mu {\lambda^n_{ab}}
u_b)(\bar{s}_c\gamma^\mu {\lambda^n_{cd}} d_d)\, ,
\\ &&
T_1 = (\bar{s}_a\sigma_{\mu\nu}u_a)(\bar{s}_b\sigma^{\mu\nu}d_b)\,
,~~~~~~~~~~~~~~~~~~~ T_8 = (\bar{s}_a\sigma_{\mu\nu}
{\lambda^n_{ab}} u_b)(\bar{s}_c\sigma^{\mu\nu} {\lambda^n_{cd}}
d_d)\, ,
\\ \nonumber &&
A_1 =
(\bar{s}_a\gamma_\mu\gamma_5u_a)(\bar{s}_b\gamma^\mu\gamma_5d_b )\,
,~~~~~~~~~~~~~~~ A_8 = (\bar{s}_a\gamma_\mu\gamma_5 {\lambda^n_{ab}}
u_b)(\bar{s}_c\gamma^\mu\gamma_5 {\lambda^n_{cd}} d_d)\, ,
\\ \nonumber &&
P_1 = (\bar{s}_a\gamma_5u_a)(\bar{s}_b\gamma_5d_b)\,
,~~~~~~~~~~~~~~~~~~~~~~ P_8 = (\bar{s}_a\gamma_5 {\lambda^n_{ab}}
u_b)(\bar{s}_c\gamma_5 {\lambda^n_{cd}} d_d) \, ,
\end{eqnarray}
%
and can be related to each other through the Fierz
transformation~\cite{Fierz,Zong:1994ww,Maruhn:2000af,Chen:2006hy}.
In the above expressions, $a$ and $b$ are color indices, $C =
i\gamma_2 \gamma_0$ is the charge conjugation operator, and the
superscript $T$ represents the transpose of Dirac indices. Note that
only five of the ten currents in Eqs.~(\ref{define_meson_current})
are independent, and all of them can be written as combinations of
Eqs.~(\ref{define_diquark_current}). We refer to
Ref.~\cite{Chen:2006hy} for detailed discussion.

Later we applied the same method to systematically study various
tetraquark, pentaquark, dibaryon and baryonium
states~\cite{Chen:2006hy,Chen:2006zh,Chen:2007xr,Chen:2007mp,Chen:2008qw,Chen:2008iw,Chen:2008ne,Jiao:2009ra,Chen:2009gs,Chen:2010jd,Chen:2011qu,Du:2012pn,Chen:2012ut,Chen:2013jra,Chen:2013gnu,Chen:2014vha,Chen:2015fwa,Chen:2016ymy,Chen:2016oma}.
Since the internal structure of multiquark states is quite
complicated, we also applied this method to systemically
conventional mesons and
baryons~\cite{Chen:2008qv,Chen:2009sf,Chen:2010ba,Dmitrasinovic:2011yf,Chen:2011rh,Chen:2012ex,Chen:2012vs,Chen:2013efa,Dmitrasinovic:2016hup,Chen:2017sbg,Cui:2017udv}.
We clearly verified in many cases that the local diquark-antidiquark
($[qq][\bar q \bar q]$) and meson-meson ($[\bar q q][\bar q q]$)
currents can be related to each other, so a straightforward
conclusion is that the method of QCD sum rules can not actually
differentiate the compact tetraquark and the hadronic molecule, when
local four-quark currents are taken into account.

After so many QCD sum rule studies, we have obtained some
experiences on hadronic interpolating currents and their relations
to hadronic states, for examples,
\begin{enumerate}

\item
The singly heavy baryons were systematically studied in
Refs.~\cite{Liu:2007fg,Huang:2009is,Chen:2015kpa,Mao:2015gya,Chen:2016phw,Mao:2017wbz,Chen:2017sci,Zhou:2014ytp,Zhou:2015ywa},
where we found that they have rich internal structures (see
references of the review~\cite{Chen:2016spr} for more relevant
discussions). There can be as many as three $P$-wave excited
$\Omega_c$ states of $J^P =1/2^-$, three of $J^P =3/2^-$, and one of
$J^P =5/2^-$. For each state we constructed one relevant
interpolating current. Maybe not all of them exist in nature, but
the LHCb experiment~\cite{Aaij:2017nav} did observe as many as five
excited $\Omega_c$ states at the same time, all of which are
candidates of $P$-wave charmed baryons.

\item
The mass spectra of vector and axial-vector hidden-charm tetraquark
states were systematically investigated in Ref.~\cite{Chen:2010ze},
where we found as many as eight $q c \bar q \bar c$ ($q=u/d$)
interpolating currents of $J^{PC} = 1^{--}$. Comparably, there were
also many vector charmonium-like states observed in particle
experiments~\cite{Tanabashi:2018oca}, including the
$Y(4008)$~\cite{Yuan:2007sj}, $Y(4220)$~\cite{Aubert:2005rm},
$Y(4320)$~\cite{Ablikim:2016qzw}, $Y(4360)$~\cite{Aubert:2007zz},
$Y(4630)$~\cite{Pakhlova:2008vn}, and $Y(4660)$~\cite{Wang:2007ea},
etc.

\item
The hidden-charm pentaquark states having spin
$J={1\over2}/{3\over2}/{5\over2}$ were systematically studied in
Refs.~\cite{Chen:2015moa,Chen:2016otp,Xiang:2017byz}, where we
constructed hundreds of hidden-charm pentaquark interpolating
currents. In 2015 the first two hidden-charm pentaquark states,
$P_c(4380)$ and $P_c(4450)$, were discovered by LHCb~\cite{Aaij:2015tga}.
In this year LHCb discovered a new candidate $P_c(4312)$, and at the same time
they separated the $P_c(4450)$ into two structures, $P_c(4440)$ and $P_c(4457)$~\cite{lhcbnew}.
One may expect there would exist more hidden-charm pentaquark states to be discovered in future experiments.

\end{enumerate}
The internal structures of (exotic) hadrons are not so simple. In
QCD sum rule studies, we can construct relevant interpolating
currents to partly reflect such internal structures, but their
relations are even more complicated.

To clarify this problem, we found a good subject in
Ref.~\cite{Chen:2018kuu} that there are only two independent $ss\bar
s \bar s$ tetraquark currents of $J^{PC} =
1^{--}$~\cite{Chen:2008ej}:
\begin{eqnarray}
&& \eta_{1\mu} = (s_a^T C \gamma_5 s_b) (\bar{s}_a \gamma_\mu
\gamma_5 C \bar{s}_b^T) - (s_a^T C \gamma_\mu \gamma_5 s_b)
(\bar{s}_a \gamma_5 C \bar{s}_b^T) \label{def:eta1} \, ,
\\ && \eta_{2\mu} = (s_a^T C \gamma^\nu s_b) (\bar{s}_a \sigma_{\mu\nu} C \bar{s}_b^T)
- (s_a^T C \sigma_{\mu\nu} s_b) (\bar{s}_a \gamma^\nu C \bar{s}_b^T)
\label{def:eta2} \, ,
\end{eqnarray}
We can also construct four non-vanishing meson-meson $(\bar ss)
(\bar s s)$ interpolating currents with $J^{PC} = 1^{--}$, but they
all depend on the above two diquark-antidiquark $(ss) (\bar s \bar
s)$ ones. Their relations can be derived by using the Fierz
transformation~\cite{Fierz,Zong:1994ww,Maruhn:2000af,Chen:2006hy}.

We have separately used $\eta_{1\mu}$ and $\eta_{2\mu}$ to perform
QCD sum rule analyses in Ref.~\cite{Chen:2008ej}, by calculating the
diagonal terms:
\begin{eqnarray}
\langle 0 | T \eta_{1\mu}(x) { \eta_{1\nu}^\dagger } (0) | 0
\rangle~~~{\rm and }~~~\langle 0 | T \eta_{2\mu}(x) {
\eta_{2\nu}^\dagger } (0) | 0 \rangle \, . \label{eq:diagonal}
\end{eqnarray}
The masses extracted are both around 2.3 GeV, which can be
used to explain the $Y(2175)$ of $J^{PC} = 1^{--}$.

To study the relations between interpolating currents and their
relevant physical states, given that we do not know how to
diagonalize physical states, in Ref.~\cite{Chen:2018kuu} we further
calculated the following off-diagonal term
\begin{eqnarray}
\langle 0 | T \eta_{1\mu}(x) { \eta_{2\nu}^\dagger } (0) | 0 \rangle
\, , \label{eq:offdiagonal}
\end{eqnarray}
which were found to be non-zero, suggesting that $\eta_{1\mu}$ and
$\eta_{2\mu}$ can couple to the same physical state. Based on
Eqs.~(\ref{eq:diagonal}) and (\ref{eq:offdiagonal}), two new
currents were obtained in Ref.~\cite{Chen:2018kuu}
\begin{eqnarray}
J_{1\mu} &=& \cos\theta~\eta_{1\mu} + \sin\theta~i~\eta_{2\mu} \, ,
\\ \nonumber J_{2\mu} &=& \sin\theta~\eta_{1\mu} + \cos\theta~i~\eta_{2\mu} \, ,
\end{eqnarray}
satisfying
\begin{eqnarray}
&& \langle 0 | T J_{1\mu}(x) { J_{2\nu}^\dagger } (0) | 0 \rangle
\ll \sqrt{\langle 0 | T J_{1\mu}(x) { J_{1\nu}^\dagger } (0) | 0
\rangle \langle 0 | T J_{2\mu}(x) { J_{2\nu}^\dagger } (0) | 0
\rangle} \, .
\end{eqnarray}
Hence, these two new currents are non-correlated, and should not
strongly couple to the same physical state. In
Ref.~\cite{Chen:2018kuu} we assumed they couple to two different
states, whose masses were extracted to be
\begin{eqnarray}
M_{J_1} &=& 2.41 \pm 0.25  {\rm~GeV} \, ,
\\ M_{J_2} &=& 2.34 \pm 0.17  {\rm~GeV} \, ,
\end{eqnarray}
with the mass splitting
\begin{equation}
\Delta M = 71 ^{+172}_{-~48}  {\rm~MeV} \, .
\end{equation}
The mass extracted using $J_{2\mu}$ is consistent with the
experimental mass of the $Y(2175)$, suggesting that $J_{2\mu}$ may
couple to the $Y(2175)$; while the mass extracted using $J_{1\mu}$
is a bit larger, suggesting that the $Y(2175)$ may have a partner
state, whose mass is around $71 ^{+172}_{-~48}$ MeV larger. The
latter may be used to explain the structure in the $\phi f_0(980)$
invariant mass spectrum at around 2.4
GeV~\cite{Aubert:2007ur,Ablikim:2007ab,Shen:2009zze,Ablikim:2014pfc,Shen:2009mr}.

Later this method was applied in Ref.~\cite{Cui:2019roq} to study
the $s s \bar s \bar s$ tetraquark states with $J^{PC} = 1^{+-}$.
Again two independent interpolating currents were found, but any one
of them was found to give reliable QCD sum rule results. The mass
was extracted to be $2.00^{+0.10}_{-0.09}$~GeV, which can be used to
explain the $X(2063)$ observed by BESIII
recently~\cite{Ablikim:2018xuz}.

\begin{figure*}[hbt]
\begin{center}
\includegraphics[width=1\textwidth]{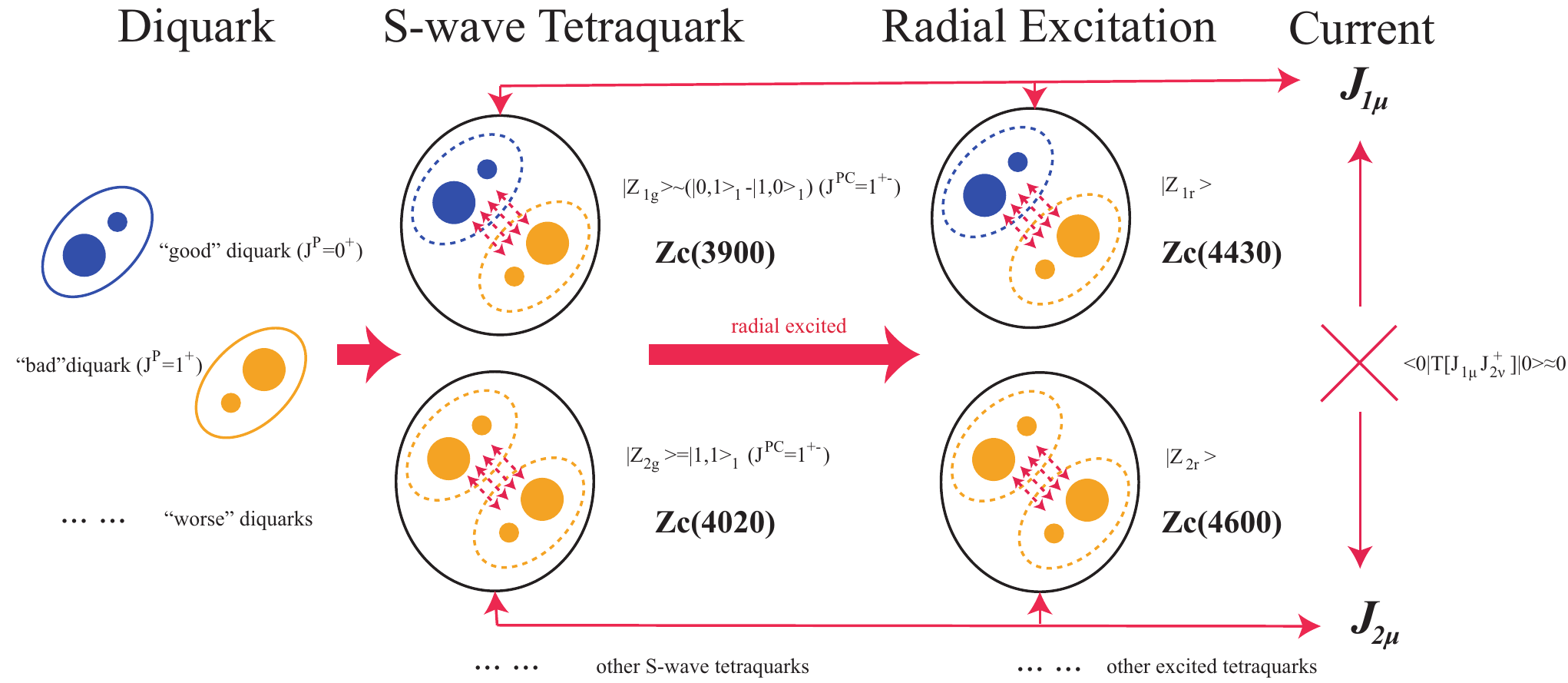}
\caption{Possible interpretations of the $Z_c(3900)$, $Z_c(4020)$,
$Z_c(4430)$, and $Z_c(4600)$, supported by a) the phenomenological
analyses within the type II diquark-antidiquark
model~\cite{Maiani:2014aja}, and b) the QCD sum rule analyses
performed in Ref.~\cite{Chen:2019osl}. Taken from
Ref.~\cite{Chen:2019osl}.} \label{fig:zc4600}
\end{center}
\end{figure*}

In Ref.~\cite{Chen:2019osl}, the authors applied the same method to
investigate four charged charmonium-like states at the same time,
including the $Z_c(3900)$, $Z_c(4020)$, $Z_c(4430)$, and
$Z_c(4600)$~\cite{Aaij:2019ipm}. Their results suggest that the
$Z_c(3900)$ and $Z_c(4020)$ are two $S$-wave tetraquark states with
$J^{PC} = 1^{+-}$, and the other two higher states can be
established as their first radial excitations; the $Z_c(3900)$ and
$Z_c(4430)$ both contains one ``good'' diquark with $J^P = 0^+$ and
one ``bad'' diquark with $J^P = 1^+$, while the $Z_c(4020)$ and
$Z_c(4600)$ both contains two ``bad'' diquarks with $J^P = 1^+$; an
illustration is given in Fig.~\ref{fig:zc4600}.

%
%

\subsection{$Q\bar Q q \bar q$ and $QQ \bar Q \bar Q$}

Recently, the QCD sum rule method has been extensively used to study
the tetraquark
systems~\cite{Matheus:2006xi,Navarra:2006nd,Lee:2007gs,Bracco:2008jj,Albuquerque:2008up,Matheus:2009vq,Albuquerque:2009ak,Wang:2009wk,Wang:2009ue,Zhang:2009st,Zhang:2010mv,Zhang:2010mw,Zhang:2011jja,Chen:2012pe,Chen:2013omd,Chen:2013wva,Wang:2013exa,Wang:2013daa,Chen:2014fza,Chen:2015fsa,Chen:2015ata,Wang:2016mmg,Huang:2016rro,Chen:2016mqt,Chen:2017rhl,Fu:2018ngx,Zhang:2018mnm}.
Most of these works have been already reviewed in
Refs.~\cite{Nielsen:2009uh,Chen:2016qju}. In this paper, we shall
only introduce some very recent studies on the hidden-charm
tetraquarks and pentaquarks. We will also introduce some
investigations on the doubly, triply and fully charmed and bottom
tetraquark states. These tetraquark systems are probably very stable
and narrow against the strong decays, and thus motivated lots of
theoretical interests.
In this subsection we review the QCD sum rule studies on hidden-charm tetraquarks states in recent years. 

\begin{figure}[htbp]
\centering
\includegraphics[width=4in]{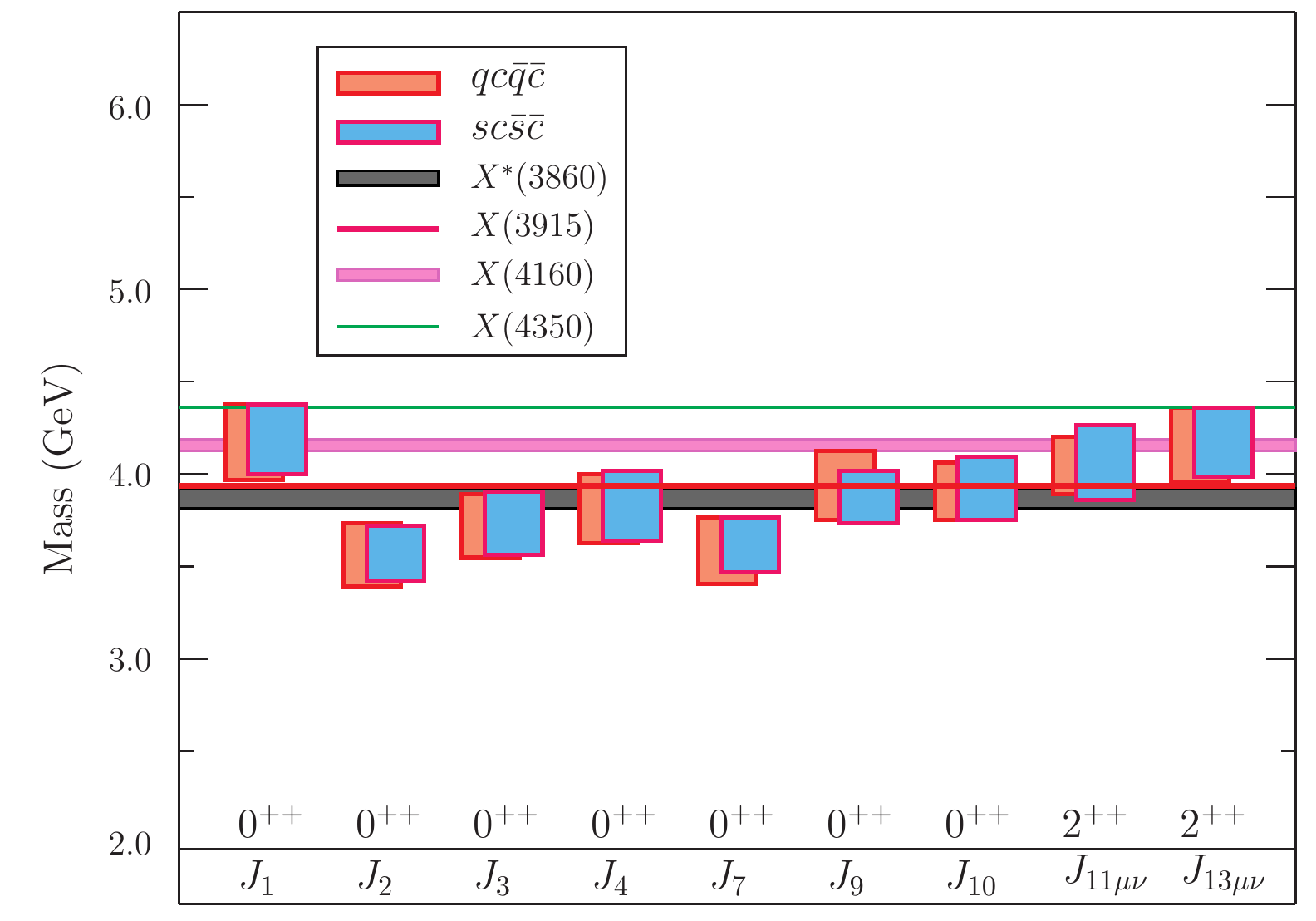}\\
\caption{The mass spectra for the hidden-charm $qc\bar q\bar c$ and
$sc\bar s\bar c$ tetraquark states with $J^{PC}=0^{++}$ and
$2^{++}$, taken from Ref.~\cite{Chen:2017dpy}} \label{Spectra0++2++}
\end{figure}

To study the structures of the $X(3915), X(4350), X(4160)$ and
$X^\ast(3860)$ in the diquark-antidiquark model, the authors of
Ref.~\cite{Chen:2017dpy} investigated the hidden-charm and
hidden-bottom $qc\bar q\bar c$, $sc\bar s\bar c$, $qb\bar q\bar b$,
$sb\bar s\bar b$ tetraquark states with $J^{PC}=0^{++}$ and $2^{++}$
in the framework of QCD sum rules. They constructed ten scalar and
four tensor interpolating currents to calculate the two-point
correlation functions. After performing the numerical analyses, they
obtained the mass spectra for the hidden-charm $qc\bar q\bar c$ and
$sc\bar s\bar c$ tetraquark. As shown in Fig.~\ref{Spectra0++2++},
the mass of the $0^{++}$ hidden-charm $qc\bar q\bar c$ tetraquark
extracted from the currents $J_4(x)$, $J_9(x)$ and $J_{10}(x)$ was
about $3.8-3.9$ GeV, which was consistent with the mass of the
$X^\ast(3860)$ state. However, the tensor $qc\bar q\bar c$
tetraquark state was calculated to be around $4.06-4.16$, which was
a bit higher than the mass of $X^\ast(3860)$. This implies that the
assignment $J^{PC}=0^{++}$ is favored for $X^\ast(3860)$ than
$2^{++}$, if it is a tetraquark. The same conclusion can be obtained
for $X(3915)$. For $X(4160)$, their results cannot distinguish the
scalar and tensor configurations, due to the nearly degenerated
masses for them. It is also shown that the $X(4350)$ cannot be
interpreted as a $sc\bar s\bar c$ tetraquark with either
$J^{PC}=0^{++}$ or $2^{++}$. The $X^\ast(3860)$ was also studied as
a tetraquark state in Ref.~\cite{Wang:2017lbl}.

In Ref.~\cite{Chen:2016oma}, the newly observed $X(4500)$ and
$X(4700)$ were studied together with the $X(4140)$ and $X(4274)$
states, based on the diquark-antidiquark configuration in the
framework of QCD sum rules. The authors found that the $X(4500)$ and
$X(4700)$ can be both interpreted as $D$-wave tetraquark states with
the quark content $cs\bar{c}\bar{s}$ and quantum numbers $J^P =
0^+$: the $X(4500)$ consists of one $D$-wave ``bad'' diquark and one
$S$-wave ``bad'' antidiquark, with the antisymmetric color structure
$[\mathbf{\bar 3_c}]_{cs} \otimes [\mathbf{3_c}]_{\bar{c}\bar s}$;
the $X(4700)$ consists of similar diquarks, but with the symmetric
color structure $[\mathbf{6_c}]_{cs} \otimes [\mathbf{ \bar
6_c}]_{\bar{c}\bar s}$. These two interpretations were remarkably
similar to those obtained in Ref.~\cite{Chen:2010ze}, in which the
$X(4140)$ and $X(4274)$ can be both interpreted as $S$-wave
$cs\bar{c}\bar{s}$ tetraquark states of $J^P = 1^+$, but with
distinct color structures. They also calculated the masses of the
hidden-bottom partner states of the $X(4500)$ and $X(4700)$ and
suggested to search for them in the $\Upsilon \phi$ invariant mass
distribution. The masses and decay widths for the $X(4500)$,
$X(4700)$, $X(4140)$ and $X(4274)$ mesons were also studied as the
$cs\bar{c}\bar{s}$ tetraquark states in
Refs.~\cite{Wang:2016ujn,Wang:2016tzr,Agaev:2017foq} and molecule
state in Ref.~\cite{Wang:2016dcb} by using QCD sum rules.

The QCD sum rule has also been performed to study the doubly
hidden-charm/bottom tetraquark states in Ref.~\cite{Chen:2016jxd},
in which the moment method was adopted instead of the Borel
transformation. The moment was defined as the $n$-th derivative of
the correlation function $\Pi(Q^2)$ in the Euclidean region
\begin{align}
\nonumber
M_n(Q^2_0)&=\frac{1}{n!}\bigg(-\frac{d}{dQ^2}\bigg)^n\Pi(Q^2)|_{Q^2=Q_0^2}
\\
&=\int_{16m_Q^2}^{\infty}\frac{\rho(s)}{(s+Q^2_0)^{n+1}}ds
=\frac{f_X^2}{(m_X^2+Q_0^2)^{n+1}}\big[1+\delta_n(Q_0^2)\big]\, ,
\label{moment}
\end{align}
where $\rho(s)$ is the spectral function and the narrow resonance
approximation was adopted in the last step. The $\delta_n(Q_0^2)$ in
Eq.~\eqref{moment} contains the contributions of higher states and
the continuum. It tends to zero as $n$ goes to infinity for a
certain value of $Q_0^2$. Then the hadron mass of the lowest lying
resonance can be obtained as
\begin{align}
m_X=\sqrt{\frac{M_{n}(Q_0^2)}{M_{n+1}(Q_0^2)}-Q_0^2}\, .
\end{align}

\begin{figure}[htbp]
\centering
\includegraphics[width=4.2in]{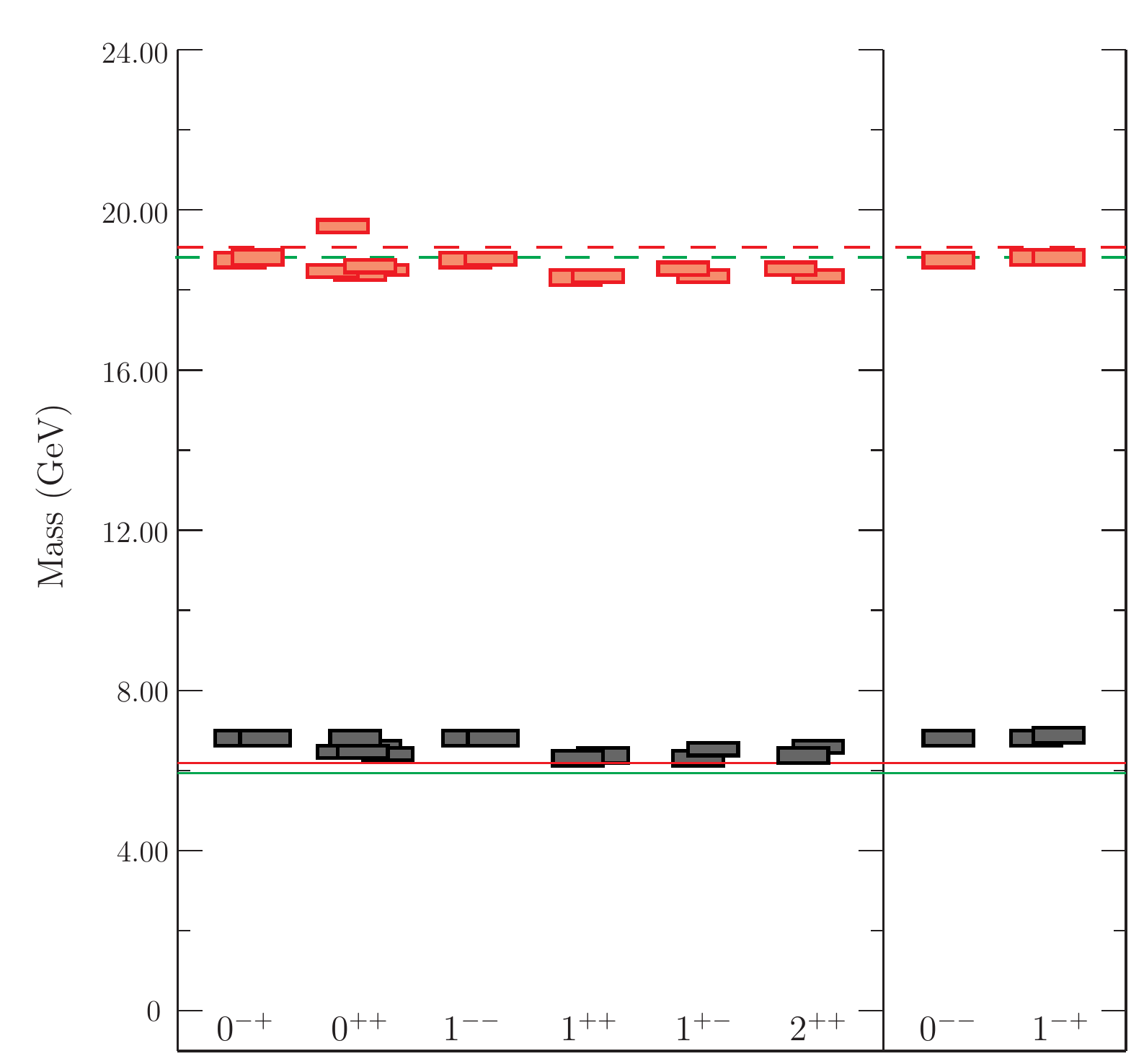}\\
\caption{The tetraquark mass spectra for the $cc\bar c\bar c$ and
$bb\bar b\bar b$ systems, taken from Ref.~\cite{Chen:2016jxd}. The
green and red solid (dashed) lines indicate the two-charmonium
(bottomonium) thresholds $\eta_c(1S)\eta_c(1S)$
($\eta_b(1S)\eta_b(1S)$) and $J/\psi J/\psi$
($\Upsilon(1S)\Upsilon(1S)$), respectively.} \label{Spectraccccbbbb}
\end{figure}

In Refs.~\cite{Chen:2016jxd,Chen:2018cqz}, the authors studied both
the $cc\bar c\bar c$ and $bb\bar b\bar b$ tetraquark systems with
various quantum numbers. As shown in Fig.~\ref{Spectraccccbbbb}, all
the $bb\bar b\bar b$ tetraquark states were predicted to be slightly
below than the mass thresholds of $\eta_b(1S)\eta_b(1S)$ and
$\Upsilon(1S)\Upsilon(1S)$, while the $cc\bar c\bar c$ tetraquarks
lied above the two-charmonium thresholds $\eta_c(1S)\eta_c(1S)$ and
$J/\psi J/\psi$. Moreover, the positive parity states with
$J^{PC}=0^{++}, 1^{++}, 1^{+-}, 2^{++}$ were lighter than the
negative parity states.

In Refs.~\cite{Wang:2017jtz,Wang:2018poa}, the $cc\bar c\bar c$ and
$bb\bar b\bar b$ tetraquark states with $J^{PC}=0^{++}, 1^{+-},
1^{--}, 2^{++}$ were also studied in the Borel sum rule method. The
extracted masses for these tetraquarks were slightly different with
those obtained in Ref.~\cite{Chen:2016jxd}.

\subsection{$QQ\bar q\bar q$ and $QQ\bar Q \bar q$}

The triply heavy tetraquark states $QQ\bar{Q}\bar{q}$ with $J^P=0^+$
and $1^+$ were also studied in Ref.~\cite{Jiang:2017tdc}, including
the explicitly exotic states $cc\bar{b}\bar{q}$ and the hidden
exotic states $cc\bar{c}\bar{q}$, $bb\bar{b}\bar{q}$,
$cb\bar{b}\bar{q}$. The masses are lower than those estimated with
the CMI model~\cite{Chen:2016ont}. Especially, the
$bb\bar{b}\bar{q}$ states were predicted to lie below the
bottomonium plus $B^{(\ast)}$ thresholds and their
OZI(Okubo-Zweig-Iizuka)-allowed strong decays are kinematically
forbidden. These tetraquarks were expected to be very narrow and the
authors suggested to search for them in the final states with a $B$
meson plus a light meson or photon processes. More studies are
needed to understand such triple-heavy tetraquark states.

In Ref.~\cite{Du:2012wp}, the mass spectra for the doubly
charmed/bottom $QQ\bar q\bar q$, $QQ\bar q\bar s$, $QQ\bar s\bar s$
tetraquark states with quantum numbers $J^P=0^-, 0^+, 1^-$ and $1^+$
have been calculated. Their numerical results indicated that the
masses of the doubly bottom tetraquarks $bb\bar q\bar q$, $bb\bar
q\bar s$, $bb\bar s\bar s$ were below the corresponding thresholds
of the two-bottom mesons, two-bottom baryons, and one doubly bottom
baryon plus one anti-nucleon. These doubly bottom tetraquarks are
thus predicted to be stable against the strong decays. These results
were consistent with the earlier study in
Ref.~\cite{Navarra:2007yw}. Further in Ref.~\cite{Chen:2013aba}, the
open-flavor scalar and axial-vector $bc\bar q\bar q$ and $bc\bar
s\bar s$ tetraquark states were also studied and their masses were
below the $D^{(\ast)}\bar B^{(\ast)}$ and $D^{(\ast)}_s\bar
B^{(\ast)}_s$ thresholds respectively, suggesting dominantly weak
decay mechanisms.

Later, the doubly heavy tetraquark states have attracted lots of
attention~\cite{Wang:2017dtg,Agaev:2018vag,Agaev:2018khe,Sundu:2019feu},
motivated by the discovery of doubly charmed baryon $\Xi_{cc}^{++}$.
In Ref.~\cite{Agaev:2018vag}, the authors calculated both the masses
and two-body strong decay width for the $cc\bar s\bar s$ and $cc\bar
d\bar s$ tetraquarks with $J^P=0^-$. They obtained
$m_T=(4390\pm150)$ MeV and $\Gamma=(302\pm113)$ MeV for the $cc\bar
s\bar s$ tetraquark, while $\tilde m_T=(4265\pm140)$ MeV and
$\tilde\Gamma=(171\pm52)$ MeV for the $cc\bar d\bar s$ tetraquark.
Since the masses of the doubly bottomed tetraquark were predicted
below the two-meson thresholds, the weak decays of the axial-vector
$T_{bb\bar u\bar d}^-$ state to the scalar state $Z_{bc\bar u\bar
d}^0$ were investigated using the three-point sum rule
approach~\cite{Agaev:2018khe}. They recalculated the spectroscopic
parameters of the $T_{bb\bar u\bar d}^-$ and $Z_{bc\bar u\bar d}^0$
states in the framework of two-point QCD sum rules. The semileptonic
decay channels $T_{bb\bar u\bar d}^-\to Z_{bc\bar u\bar d}^0
l\bar\nu_l (l=e,\, \nu,\, \tau)$ were then investigated with the
width and lifetime $\Gamma=(7.17\pm1.23)\times 10^{-8}$ MeV and
$\tau=9.18^{+1.90}_{-1.34}$ fs. Later in Ref.~\cite{Sundu:2019feu},
they also studied the semileptonic decays of the scalar tetraquark
$Z_{bc\bar u\bar d}^0$ to final states $T_{bs\bar u\bar
d}^-e^+\nu_e$ and $T_{bs\bar u\bar d}^-\mu^+\nu_\mu$. After
reanalysing the spectroscopic parameters of $T_{bs\bar u\bar
d}^-$, they used the three-point sum rules to evaluate the weak form
factors and extrapolated them to the whole momentum region. The
total semileptonic decay width and lifetime of the scalar tetraquark
$Z_{bc\bar u\bar d}^0$ were finally obtained as
$\Gamma=(2.37\pm0.36)\times 10^{-11}$ MeV and
$\tau=27.8^{+4.9}_{-3.7}$ ps.

\subsection{$Q\bar Q q q q$}

Since the discovery of the $P_c(4380)$ and $P_c(4450)$ by LHCb in
2015~\cite{Aaij:2015tga}, there have been lots of theoretical
studies using various methods, and the QCD sum rule approach is one
of the most popular one.

In Ref.~\cite{Chen:2015moa}, we applied the method of QCD sum rules
and studied the $P_c(4380)$ and $P_c(4450)$ as exotic hidden-charm
pentaquarks composed of an anti-charmed meson and a charmed baryon.
Our results suggest that the $P_c(4380)$ and $P_c(4450)$ have
quantum numbers $J^P = 3/2^-$ and $5/2^+$, respectively. We also
predicted the masses of their hidden-bottom partners to be
$11.55^{+0.23}_{-0.14}$~GeV and $11.66^{+0.28}_{-0.27}$~GeV.
This study was later expanded in Refs.~\cite{Chen:2016otp,Xiang:2017byz}, where we found two mixing currents, and
again our results suggest that they can be identified as
hidden-charm pentaquark states having $J^P = 3/2^-$ and $5/2^+$,
while there still exist other possible spin-parity assignments, such
as $J^P = 3/2^+$ and $5/2^-$, which still needs to be clarified in
further theoretical and experimental studies.

Especially, in Ref.~\cite{Chen:2016otp} we systematically constructed all the possible local hidden-charm pentaquark
currents with spin $J = {1\over2}/{3\over2}/{5\over2}$ and quark contents $uud c \bar c$, through which we found that the internal structure of hidden-charm pentaquark states is quite complicated.
We derived some mass predictions, and we summarized in Table~\ref{tab:sumruleresult} all the mass predictions that are extracted from single currents and less than 4.5~GeV.
The coincidence of these mass predictions with the masses of $P_c(4312)$, $P_c(4440)$, and $P_c(4457)$ measured in the recent LHCb experiment~\cite{lhcbnew} drives us to the ``molecular'' picture~\cite{Chen:2019bip} that the $P_c(4312)$ can be well explained as the $[\Sigma_c^{++} \bar D^-]$ bound state with $J^P = 1/2^-$; the $P_c(4440)$ and $P_c(4457)$ can be explained as the $[\Sigma_c^{*++} \bar D^{-}]$ and $[\Sigma_c^{+} \bar D^{*0}]$ bound states with $J^P = 3/2^-$, respectively, while one of them may also be explained as the $[\Sigma_c^{+} \bar D^0]$ bound state with $J^P = 1/2^-$ or the $[\Sigma_c^{*+} \bar D^{*0}]$ bound state with $J^P = 5/2^-$; there is still a place for the $P_c(4380)$, {\it i.e.}, to be explained as the $[\Sigma_c^{++} \bar D^{*-}]$ bound state with $J^P = 3/2^-$. The $[\Sigma_c \bar D^*]$ bound state with $J^P = 1/2^-$ was not investigated in Ref.~\cite{Chen:2016otp}. See Ref.~\cite{Chen:2019bip} for detailed discussions of these possible interpretations.

\renewcommand{\arraystretch}{1.5}
\begin{table*}[]
\begin{center}
\caption{Mass predictions for the hidden-charm pentaquark states with spin $J = {1\over2}/{3\over2}/{5\over2}$ and quark contents $uud c \bar c$~\cite{Chen:2016otp}. They are extracted using hidden-charm pentaquark currents composed of color-singlet charmed baryon fields and anti-charmed meson fields, {\it i.e.}, $[\epsilon^{abc} u_a d_b c_c] [\bar c_d u_d]$ or $[\epsilon^{abc} u_a u_b c_c] [\bar c_d d_d]$, where the subscripts $a,b,c,d$ are color indices. We summarize here all the mass predictions that are extracted from single currents and less than 4.5~GeV.}
\vspace{.3cm}
\begin{tabular}{cc|cc|cc}
\hline\hline
~~\mbox{Current}~~ & ~~\mbox{Structure}~~ & \mbox{$s_0$ [GeV$^2$]} & \mbox{Borel Mass [GeV$^2$]} & ~~\mbox{Mass [GeV]}~~ & ~~\mbox{($J$, $P$)}~~
\\ \hline
$\xi_{14}$         & $[\Sigma_c^{+} \bar D^0]$      & $20 - 24$ & $4.12 - 4.52$ & $4.45^{+0.17}_{-0.13}$ & ($1/2,-$)
\\
$\psi_2$           & $[\Sigma_c^{++} \bar D^-]$    & $19 - 23$ & $3.95 - 4.47$ & $4.33^{+0.17}_{-0.13}$ & ($1/2,-$)
\\ \hline
$\xi_{33\mu}$      & $[\Sigma_c^+ \bar D^{*0}]$    & $20 - 24$ & $3.97 - 4.41$ & $4.46^{+0.18}_{-0.13}$ & ($3/2,-$)
\\
$\psi_{2\mu}$      & $[\Sigma_c^{*++} \bar D^-]$    & $20 - 24$ & $3.88 - 4.41$ & $4.45^{+0.16}_{-0.13}$ & ($3/2,-$)
\\
$\psi_{9\mu}$      & $[\Sigma_c^{++} \bar D^{*-}]$    & $19 - 23$ & $3.94 - 4.27$ & $4.37^{+0.18}_{-0.13}$ & ($3/2,-$)
\\ \hline
$\xi_{13\mu\nu}$   & $[\Sigma_c^{*+} \bar D^{*0}]$  & $20 - 24$ & $3.51 - 4.00$ & $4.50^{+0.18}_{-0.12}$ & ($5/2,-$)
\\ \hline\hline
\end{tabular}
\label{tab:sumruleresult}
\end{center}
\end{table*}

In Ref.~\cite{Azizi:2016dhy}, Azizi {\it et al} performed QCD sum
rules analyses on the hidden-charm pentaquark states with $J^P =
3/2^\pm$ and $5/2^\pm$. They adopted a molecular picture for $J^P =
3/2^\pm$ states and a mixed current in a molecular form for
$5/2^\pm$, and their analyses suggested that the $P_c(4380)$ and
$P_c(4450)$ can be considered as hidden-charm pentaquark states with
$J^P = 3/2^-$ and $5/2^+$, respectively. Later in
Ref.~\cite{Azizi:2017bgs}, they applied the same method to study the
hidden-bottom pentaquark states with spin $J = 3/2$ and $5/2$, and
predicted their masses. The strong decays and electromagnetic
multipole moments of the $P_c^+(4380)$ were later studied in
Refs.~\cite{Azizi:2018bdv,Ozdem:2018qeh,Azizi:2018jky}.

In Ref.~\cite{Wang:2015epa}, Wang constructed the
diquark-diquark-antiquark type interpolating currents, and his QCD
sum rule studies also supported to assign the $P_c(4380)$ and
$P_c(4450)$ as $J^P = 3/2^-$ and $5/2^+$ pentaquark states,
respectively. This study was later expanded in
Refs.~\cite{Wang:2015ava,Wang:2018waa,Wang:2018lhz}.

%% file: section7.tex
%
\section{Three-body system}
\label{sect:6}

There are various two-body interpretations of the exotic hadrons, as
we have reviewed in previous sections. There are also some
three-body interpretations for exotic hadrons. For example,
combining the Faddeev equations with the chiral unitary
model~\cite{MartinezTorres:2007sr,MartinezTorres:2008gy}, the
authors interpreted the $Y(2175)$ as a dynamically generated state
in the $\phi K \bar K$ system. In this section, we review some of
these studies. We note that the three-body system have also been
extensively studied within the chiral effective field theory~\cite{Epelbaum:2002vt,Epelbaum:2000mx,Bernard:2007sp,Bernard:2011zr,Petschauer:2015elq}
and by using Lattice QCD~\cite{Meissner:2014dea,Mai:2017bge,Meng:2017jgx,Doring:2018xxx,Mai:2018djl,Briceno:2018mlh,Briceno:2017tce,Briceno:2012rv,Polejaeva:2012ut,Guo:2017crd,Guo:2017ism,Kreuzer:2008bi,Jansen:2015lha,Bour:2012hn,Hammer:2017kms,Meng:2017jgx,Romero-Lopez:2018rcb,Pang:2019dfe} in recent years, but we shall
not review them in the present paper, and just recommend interested
readers to the above references
for detailed discussions.

\subsection{A short introduction to Faddeev equations}

\begin{figure}[tb]
\begin{center}
\epsfig{file=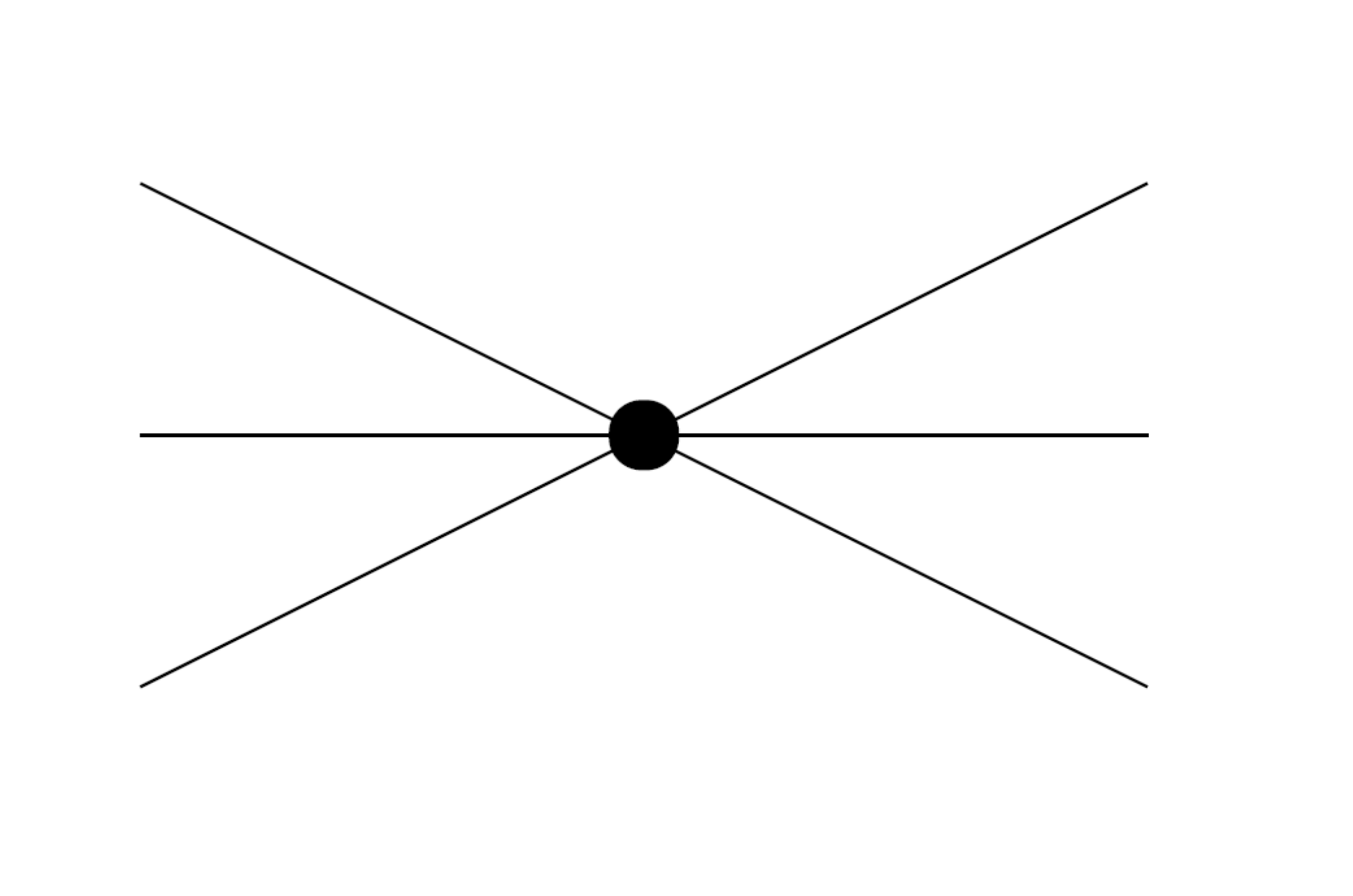,scale=0.4} \caption{``Direct''
three-body interactions. \label{sec6:3body}}
\end{center}
\end{figure}

When studying a three-body system, usually one does not consider the
``direct'' three-body interactions, as shown in
Fig.~\ref{sec6:3body}. Instead, one needs to consider all the
possible two-body exchanges/interactions in a system of three
particles. One can use the Faddeev equations, named after their
inventor Ludvig Faddeev~\cite{Faddeev:1960su}. We refer to
Ref.~\cite{Glockle} for detailed instructions.

\begin{figure}[tb]
\begin{center}
\epsfig{file=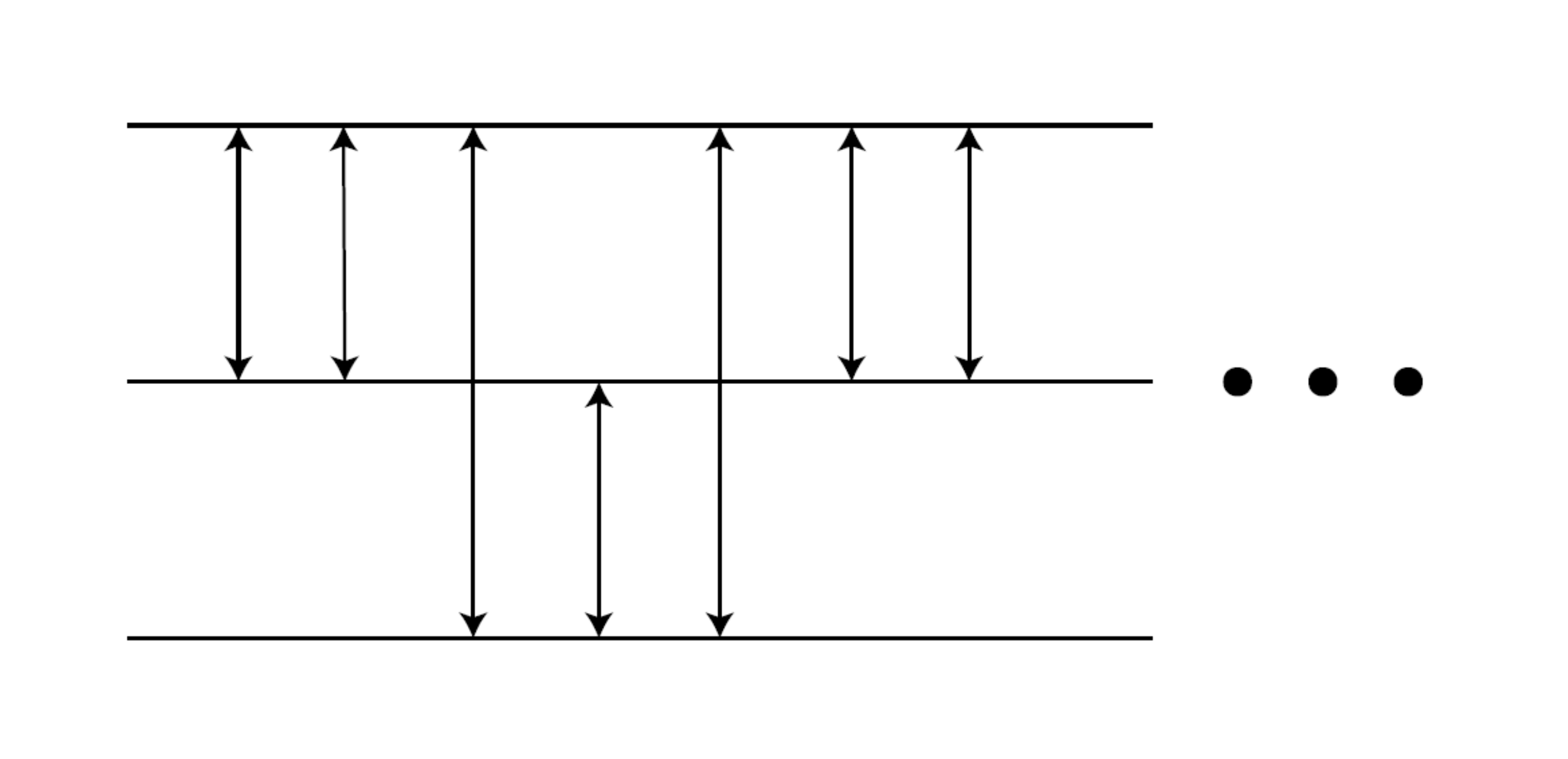,scale=0.4} \caption{Diagrammatic
representation of Eq.~(\ref{sec6:iterate}), where all the pair
forces in each two-body subsystem have been summed up iteratedly to
infinite order. \label{sec6:2body}}
\end{center}
\end{figure}

In a three-body system there are three different two-body
subsystems. The idea of the Faddeev equations is to sum up all the
pair forces in each two-body subsystem to infinite order among all
three particles. First we write the Schr{\" o}dinger equation for a
three-body system as
\begin{equation}
\left ( H_0 + \sum_{i=1}^3 V_i \right) \Psi = E \Psi \, ,
\label{sec6:schrodinger}
\end{equation}
where $H_0$ is the kinetic energy of the relative motion for three
particles; $V_1 = V_{23}$, $V_2 = V_{31}$, and $V_3 = V_{12}$ are
the three interactions in the two-body subsystems. Then we transfer
Eq.~(\ref{sec6:schrodinger}) into its integral form:
\begin{equation}
\Psi = {1 \over E - H_0} \sum_{i=1}^3 V_i \Psi \equiv G_0 \sum_{i}
V_i \Psi \, ,
\end{equation}
and further iterate it many times to be:
\begin{equation}
\Psi = G_0 \sum_{i} V_i~G_0 \sum_{j} V_j~G_0 \sum_{k}
V_k~\cdots~\Psi \, . \label{sec6:iterate}
\end{equation}
This can be shown graphically in Fig.~\ref{sec6:2body}, where all
the pair forces in each two-body subsystem have been summed up
iteratedly to infinite order.

\begin{figure}[tb]
\begin{center}
\epsfig{file=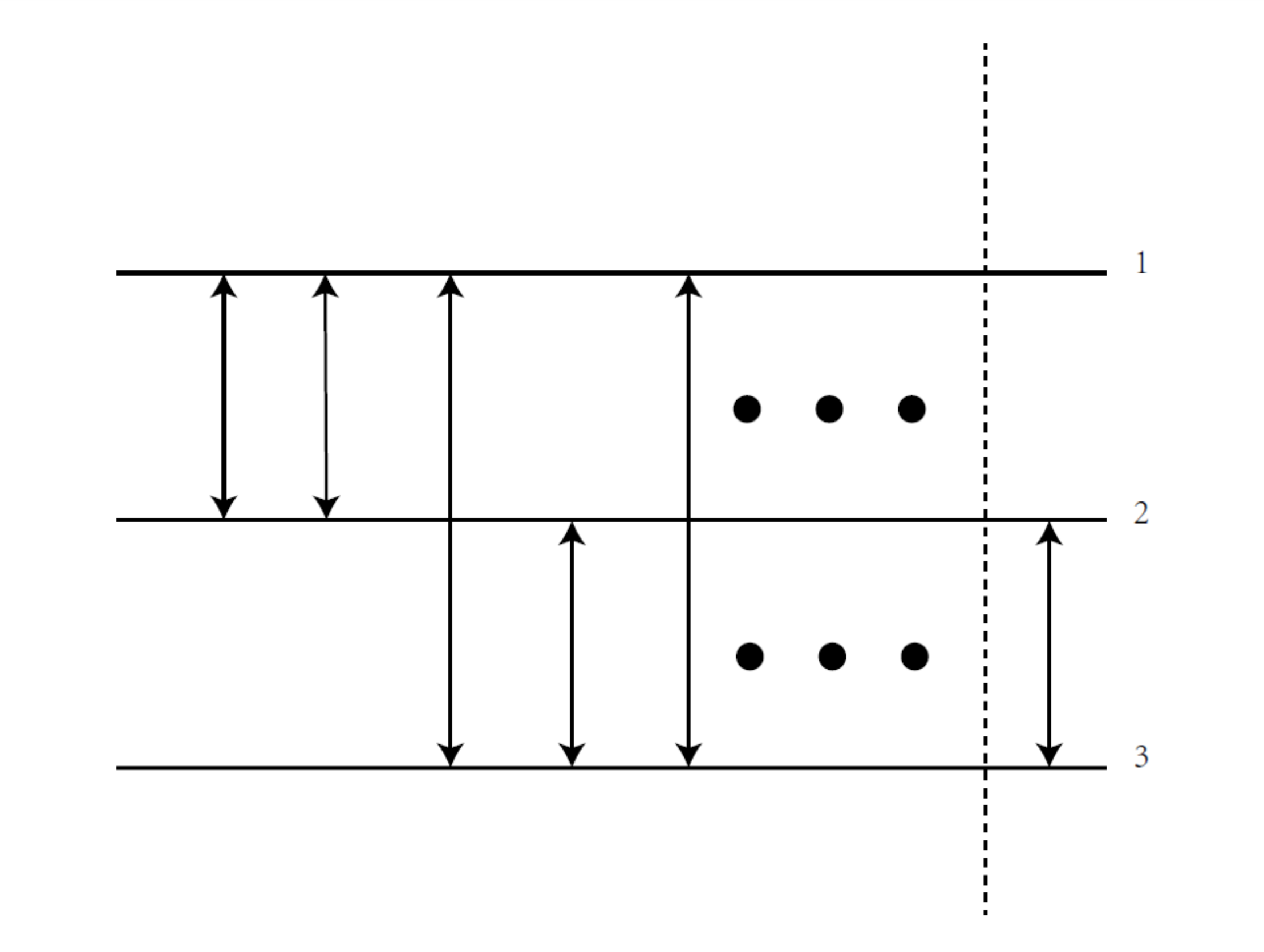,scale=0.4} \caption{Diagrammatic
representation of the Faddeev component $\psi_1$.
\label{sec6:2body1}}
\end{center}
\end{figure}

To solve Eq.~(\ref{sec6:iterate}), we decompose $\Psi$ into three
Faddeev components:
\begin{equation}
\Psi = \sum_i \psi_i = \sum_i G_0 V_i \Psi \, ,
\end{equation}
where $\psi_i = G_0 V_i \Psi$ is that part of $\Psi$ having $V_i$ as
the last interaction to the left. As an example, we show $\psi_1$
graphically in Fig.~\ref{sec6:2body1}.

We can further iterate $\psi_i$ to be:
\begin{equation}
\psi_i = G_0 V_i \Psi = G_0 V_i \sum_j \psi_j = G_0 V_i \psi_i + G_0
V_i \sum_{j \neq i} \psi_j \, ,
\end{equation}
so that
\begin{equation}
\psi_i = ({\bf 1} - G_0 V_i)^{-1} G_0 V_i \sum_{j \neq i} \psi_j \,
.
\end{equation}
The kernel can be simplified to be
\begin{eqnarray}
( {\bf 1} - G_0 V_i)^{-1} G_0 V_i &=& ( {\bf 1} + G_0 V_i + G_0 V_i
G_0 V_i + \cdots) G_0 V_i
\\ \nonumber &=& G_0( V_i + V_i G_0 V_i + V_i G_0 V_i G_0 V_i + \cdots)
\\ \nonumber &\equiv& G_0 t_i \, ,
\end{eqnarray}
where $t_i = ( {\bf 1} - V_i G_0)^{-1} V_i $ sums up $V_i$ to
infinite order.

Finally, we arrive at three coupled equations
\begin{equation}
\psi_i = G_0 t_i \sum_{j \neq i} \psi_j \, ,
\end{equation}
which are called as the Faddeev equations. For identical particles,
these equations can be simplified. Again, we refer
Refs.~\cite{Faddeev:1960su,Glockle} for detailed discussions.

\subsection{Applications of Faddeev equations to exotic hadrons}

The Faddeev equations have been widely applied to study the
quantum-mechanical three-body problem, such as three-nucleon bound
states ($NNN$~\cite{Stadler:1991zz}, $\Xi
NN$~\cite{Garcilazo:2016ylj} and $\Omega
NN$~\cite{Garcilazo:2019igo}) and $\bar K NN$ quasi-bound
state~\cite{Shevchenko:2007ke} in nuclear physics. Here we review
its applications to exotic hadrons in hadron physics.

In Refs.~\cite{MartinezTorres:2007sr,MartinezTorres:2008gy} the
authors developed a method to study the meson-meson-baryon system by
solving the Faddeev equations together with the chiral unitary
model. Here we briefly review its application to the
$Y(2175)$~\cite{MartinezTorres:2008gy}. There the $Y(2175)$ was
interpreted as a $\phi f_0(980)$ resonant state, with the $f_0(980)$
dynamically generated in the $K \bar K$ channel. Hence, to study the
$Y(2175)$, the authors solved the Faddeev equations in the
three-body $\phi K \bar K$ system.

\begin{figure}[tb]
\begin{center}
\epsfig{file=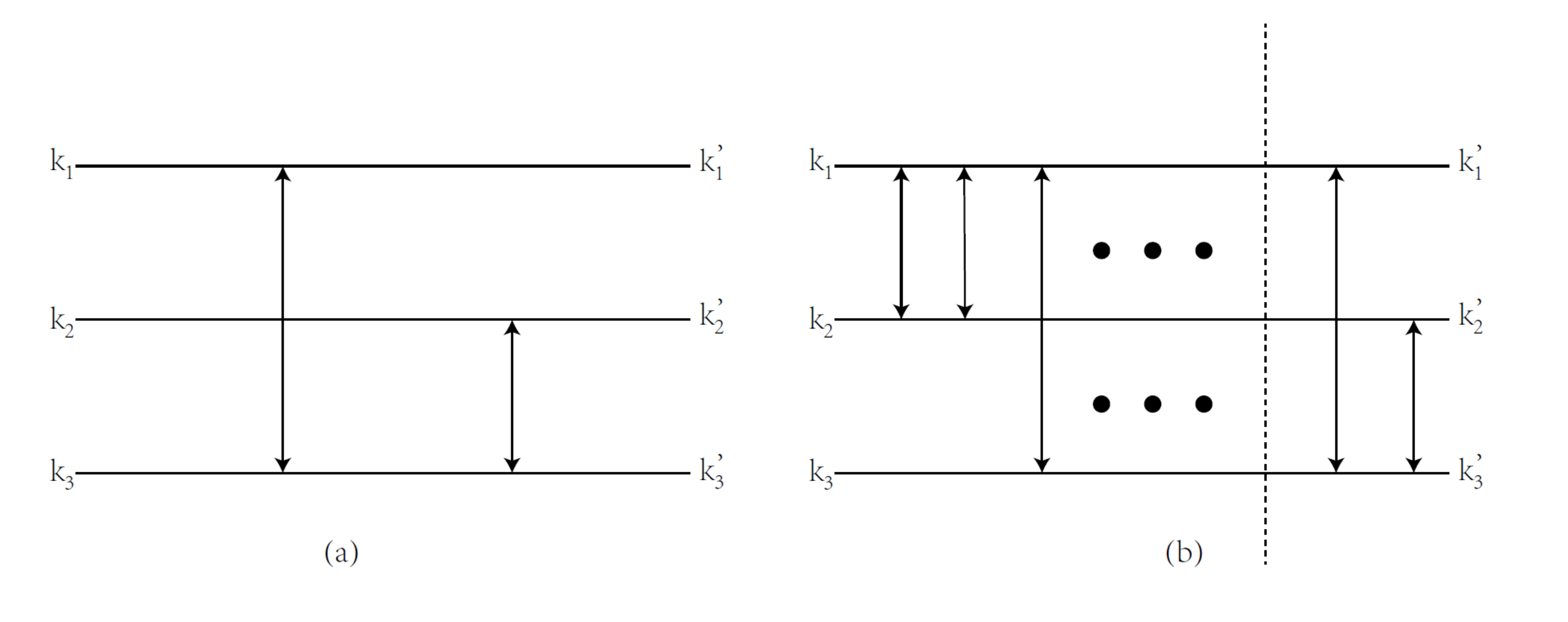,scale=0.4} \caption{Diagrammatic
representations of the terms (a) $t^1 g^{12} t^2$ and (b)
$T_R^{12}$. \label{sec6:unitary2body}}
\end{center}
\end{figure}

The full three-body scattering matrix is
\begin{equation}
T = T^1 + T^2 + T^3 \, ,
\end{equation}
where $T^i$ is the Faddeev partition. Note that $\Psi$ and $\psi_i$
are used in the previous subsection. They further write $T^i$ as
\begin{eqnarray}
T^i &=& t^i \delta^3(\vec k_i^\prime - \vec k_i) + T_R^{ij} +
T_R^{ik} \, ,  ~~~~~ i \neq j \neq k = 1, 2, 3 \, ,
\end{eqnarray}
where $t^i$ is the two-body scattering matrix where the particle $i$
is a spectator; the term $t^i \delta^3(\vec k_i^\prime - \vec k_i)$
corresponds to a ``disconnected'' diagram and should be removed;
$T^{ij}_R$ sums up all the diagrams with the last two $t$ matrices
being $t_j$ and $t_i$:
\begin{eqnarray}
T^{ij}_R &=& t^i g^{ij} t^j + t^i g^{ij} t^j g^{jk} t^k + t^i g^{ij}
t^j g^{ji} t^i + \cdots
\\ \nonumber &\rightarrow& t^i g^{ij} t^j + t^i \left[ G^{iji} T^{ji}_R + G^{ijk} T^{jk}_R \right] \, , ~~~~~~~~~~~~~~~ i \neq j \neq k = 1, 2, 3 \, .
\label{sec:unitary}
\end{eqnarray}
In the above expression $g^{ij}$ is the three-body propagator:
\begin{equation}
g^{ij} = \left( \prod_{r=1}^D { 1 \over 2 E_r} \right) { 1 \over
\sqrt{s} - E_i(\vec k_i) - E_j(\vec k_j) - E_k(\vec k_i + \vec k_j)
+ i \epsilon} \, .
\end{equation}
All the terms $t^ig^{ij}t^jg^{jk}t^k$ are written as
$t^i~G^{ijk}\left( t^j g^{jk} t^k \right)\Big |_{\rm on-shell}$:
\begin{equation}
G^{ijk} = \int {d^3 k^{''} \over (2\pi)^3} { N_l \over 2E_l} {N_m
\over 2 E_m} {F^{ijk}(\sqrt{s},\vec k^{''}) \over \sqrt{s_{lm}} -
E_l(\vec k^{''}) - E_m( \vec k^{''}) + i \epsilon} \, , ~~~~~ i \neq
j , \, j \neq k , \, i \neq l \neq m \, ,
\end{equation}
where
\begin{equation}
F^{ijk} = t^j(\sqrt{s_{int}}(\vec k^{''})) \left( g^{jk}_{\rm
off-shell} \over g^{jk}_{\rm on-shell} \right)
[t^j(\sqrt{s_{int}}(\vec k_j^{'}))]^{-1} \, .
\end{equation}
The detailed explanations of the above equations can be found in
Refs.~\cite{MartinezTorres:2007sr,MartinezTorres:2008gy}. As
examples, we show $t^1 g^{12} t^2$ and $T_R^{12}$ graphically in
Fig.~\ref{sec6:unitary2body}.

\begin{figure}[tb]
\begin{center}
\epsfig{file=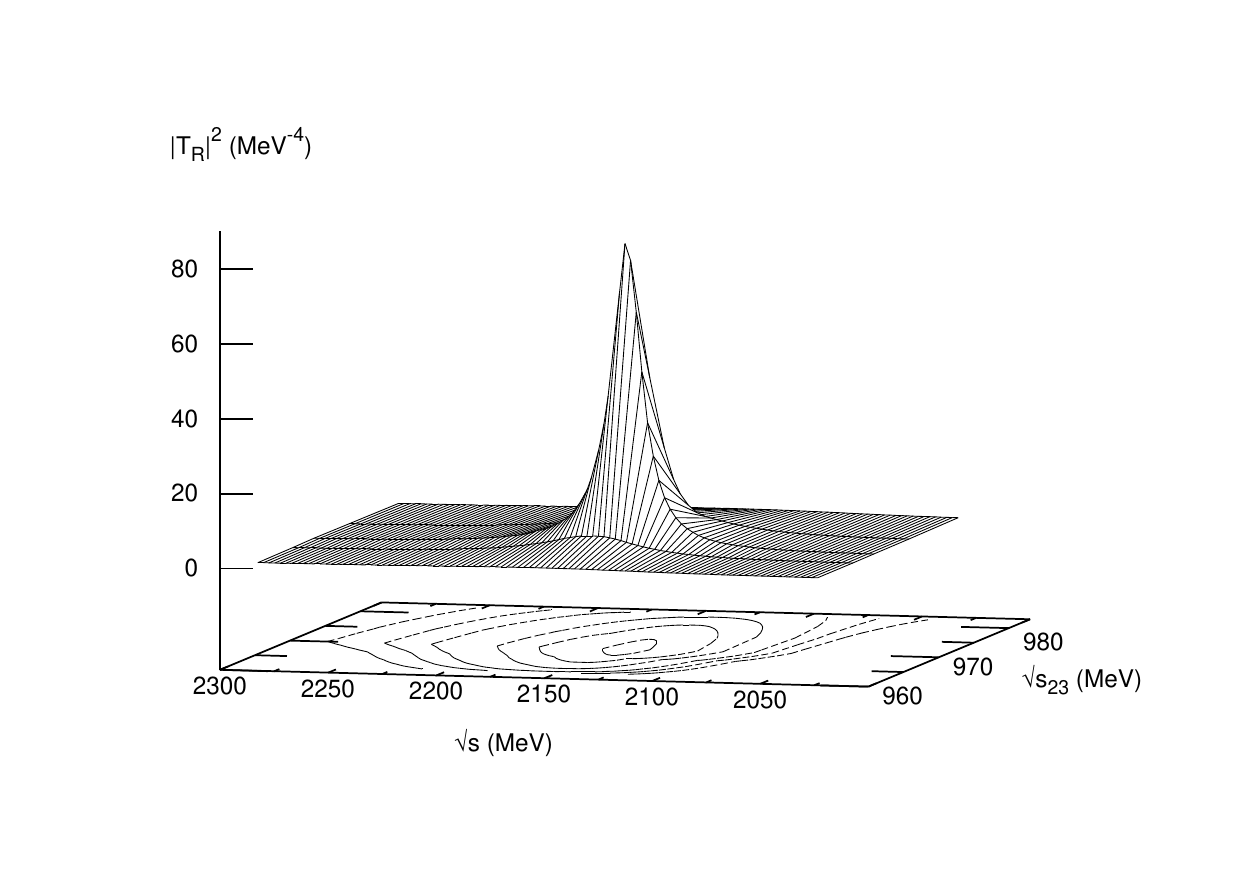,scale=1} \caption{The $\phi K \bar
K$ squared amplitude $|T_R|^2 \equiv |T^{12}_R + T^{13}_R + T^{21}_R
+ T^{23}_R + T^{31}_R + T^{32}_R|^2$. The pole at $\sqrt{s} \sim
2150$ MeV is associated to the $Y(2175)$, and the pole at
$\sqrt{s_{23}} \sim 970$ MeV is associated to the $f_0(980)$. Taken
from Ref.~\cite{MartinezTorres:2008gy}. \label{sec6:Y2175phif02}}
\end{center}
\end{figure}

The two-body scattering matrices $t^i$ for the interactions of
pseudoscalar and pseudoscalar mesons can be found in
Ref.~\cite{Oller:1997ti}, while those for the interactions of
pseudoscalar and vector mesons can be found in
Ref.~\cite{Roca:2005nm}. Using them as inputs, the authors solved
the Faddeev equations, Eqs.~(\ref{sec:unitary}), in the chiral
unitary approach. They found a clear sharp peak of $|T_R|^2 \equiv
|T^{12}_R + T^{13}_R + T^{21}_R + T^{23}_R + T^{31}_R + T^{32}_R|^2$
around 2150 MeV with a narrow width about tens of MeV, as shown in
Fig.~\ref{sec6:Y2175phif02}. This structure can be associated to the
$Y(2175)$, so their results indicate that the $Y(2175)$ is a
dynamically generated resonant state in the $\phi K \bar K$ system.
We refer to Ref.~\cite{MartinezTorres:2008gy} for more discussions,
where the coupled $\phi \pi \pi$ channel has also been taken into
account.

The above method has been systematically applied to study the
meson-meson-baryon (two pseudoscalar mesons and one octet baryon)
system, and many dynamically generated resonant states are proposed. For
examples, the $N^*(1710)$ is interpreted as a dynamically generated resonance in the $\pi \pi N$
system with $I = 1/2$ and $J^P = 1/2^+$~\cite{Khemchandani:2008rk}.
The three mesons system has also been systematically studied within
the same approach in
Refs.~\cite{MartinezTorres:2009xb,MartinezTorres:2018zbl}.
For examples, the $Y(4260)$ is interpreted as a dynamically
generated resonance in the $J/\psi K \bar K$
system~\cite{MartinezTorres:2009xb}; an exotic state with
mass around 4140 MeV and $I = 1/2$ was predicted in the $D D K$
system~\cite{MartinezTorres:2018zbl}.

A similar approach was applied to study the $\bar K N \pi$ system by
solving the Faddeev equations with relativistic kinematics in
Ref.~\cite{Gal:2011yp}, and a $\bar K N\pi$ resonance with $I=1$ and
$J^P = {3/2}^-$ was predicted to be around $1570\pm10$ MeV. The
$P$-wave three-body $B^*B^* \bar K$ system is investigated in
Ref.~\cite{Valderrama:2018knt}, and the results suggest that there
exist $P$-wave $B^*B^* \bar K$ bound states with quantum numbers
$J^P = 0^+/1^+/2^+$, all of which locate at 30-40 MeV below the $B^*
B_{s1}$ threshold. The $B^{(*)}B^*B^*$ systems are investigated in
Ref.~\cite{Garcilazo:2018rwu}, and a unique bound state of three B
mesons with $I = 1/2$ and $J^P = 2^-$ was predicted to be around 90
MeV below the threshold of three $B$-mesons.

\subsection{Fixed center approximation}

The Faddeev equations are sometimes not easy to be solved, so the
fixed center approximation was proposed to deal with
them~\cite{Chand:1962ec,Barrett:1999cw,Deloff:1999gc,Kamalov:2000iy}.
In Ref.~\cite{Roca:2010tf} this approach within the chiral unitary
model was applied to study the $f_2(1270)$, $\rho_3(1690)$,
$f_4(2050)$, $\rho_5(2350)$ and $f_6(2510)$ resonances as
multi-$\rho(770)$ states, and later on to many other exotic
structures. In this paper we briefly review its application to the
three-body $N \rho \rho$ system~\cite{Sun:2011fr}.

\begin{figure}[tb]
\begin{center}
\epsfig{file=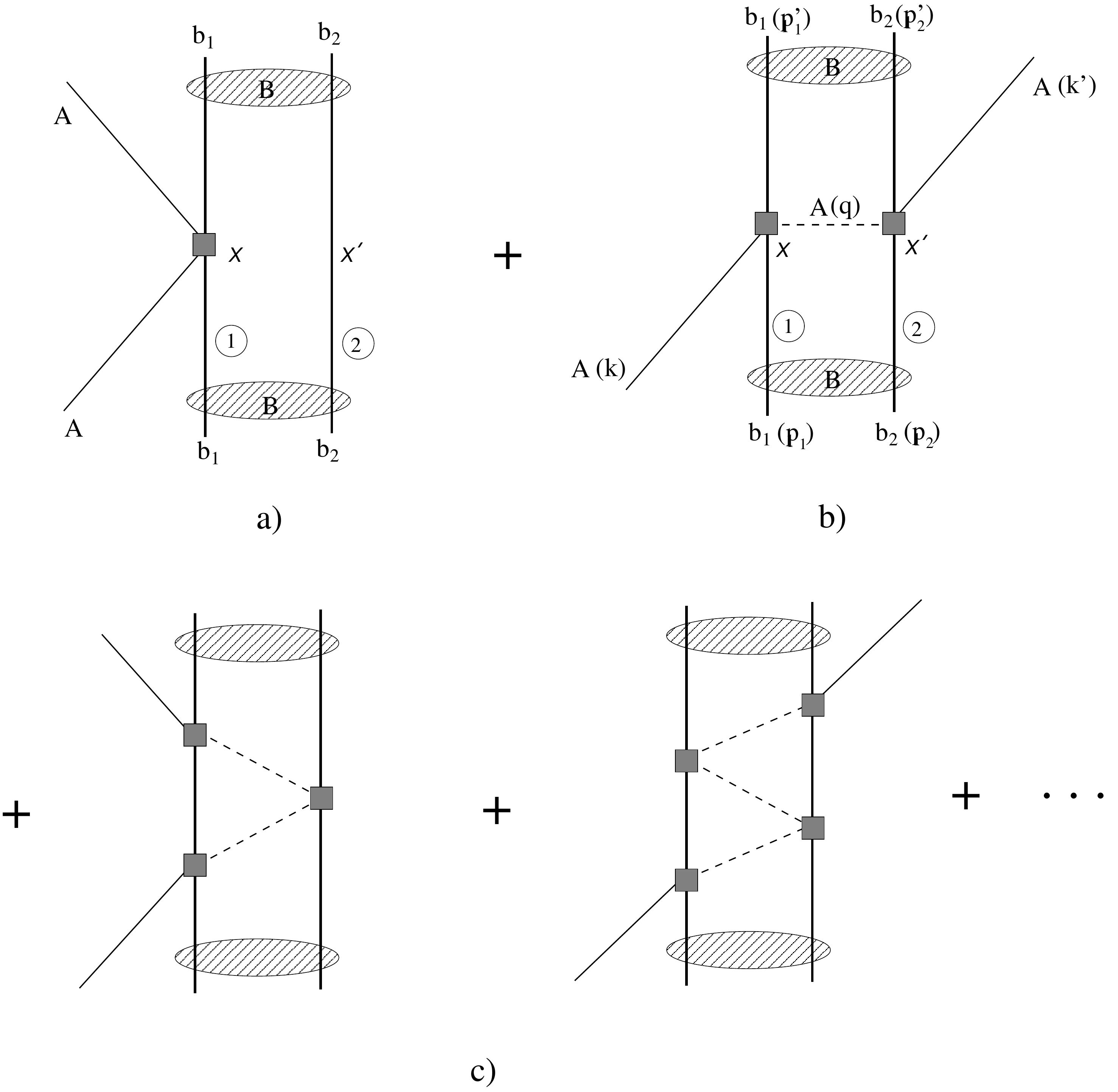,scale=0.4} \caption{Diagrammatic
representation of the fixed center approximation to the Faddeev
equations. $a)$, $b)$ and $c)$ represent single, double, and
multiple scattering contributions, respectively. Taken from
Ref.~\cite{Sun:2011fr}. \label{sec6:FCAdiagram}}
\end{center}
\end{figure}

In Ref.~\cite{Sun:2011fr} the authors investigated the three-body $N
\rho \rho$ system by considering that two of the $\rho$ mesons are
clusterized forming an strongly binding $f_2(1270)$ resonance, i.e.,
$N$-$(\rho \rho)_{f_2(1270)}$. In such a situation one can use the
fixed center approximation to the Faddeev equations. We show this
diagrammatically in Fig.~\ref{sec6:FCAdiagram}, where the external
particle $N$ interacts successively with the other two $\rho$ mesons
forming the $f_2(1270)$. The total scattering matrix is
\begin{equation}
T = T_1 + T_2 \, , \label{sec6:FCA1}
\end{equation}
where $T_i$ ($i=1,2$) are the two partition functions, accounting
for all the diagrams starting with the interaction of $N$ and the
$i$-th $\rho$ meson of the compound system:
\begin{eqnarray}
T_1 &=& t_1 + t_1 G_0 T_2 \, , \label{sec6:FCA2}
\\ T_2 &=& t_2 + t_2 G_0 T_1 \, .
\label{sec6:FCA3}
\end{eqnarray}
Here $t_i$ is the $N \rho$ scattering amplitude, which has been
systematically investigated in Ref.~\cite{Oset:2009vf} and the
results therein can be used here as inputs; $G_0$ is the loop
function for the particle $N$ propagating inside the compound system
\begin{eqnarray}
G_0(s) &=& \frac{1}{\sqrt{2\omega_{f_2(1270)} 2\omega_{f_2(1270)}'}}
\displaystyle{ \int \frac{d^3\vec{q}}{(2\pi)^3} \times
F_{f_2(1270)}(q) \frac{M_N}{E_N(\vec{q})}\frac{1}{q^0-E_N(\vec{q}) +
i\epsilon}} \, ,
\end{eqnarray}
where the form factor $F_{f_2(1270)}(q)$ is
\begin{eqnarray}
F_{f_2(1270)}(q) &=& \frac{1}{\cal N} \int_{p<\Lambda,|\vec p-\vec
q|<\Lambda} d^3p \times \frac{1}{M_{f_2}-2\omega_\rho(\vec p)}
\frac{1}{M_{f_2}-2\omega_\rho(\vec p-\vec q)} \, ,
\end{eqnarray}
and the normalization factor $\cal N$ is
\begin{eqnarray}
{\cal N} = \int_{p<\Lambda}d^3p \frac{1}{\left(
M_{f_2}-2\omega_\rho(\vec p) \right)^2} \, .
\end{eqnarray}
We recommend interested readers to
Refs.~\cite{Roca:2010tf,Sun:2011fr} for detailed explanations of the
above equations, where the wave function of the $f_2(1270)$ as well
as its width are discussed.

\begin{figure}[tb]
\begin{center}
\epsfig{file=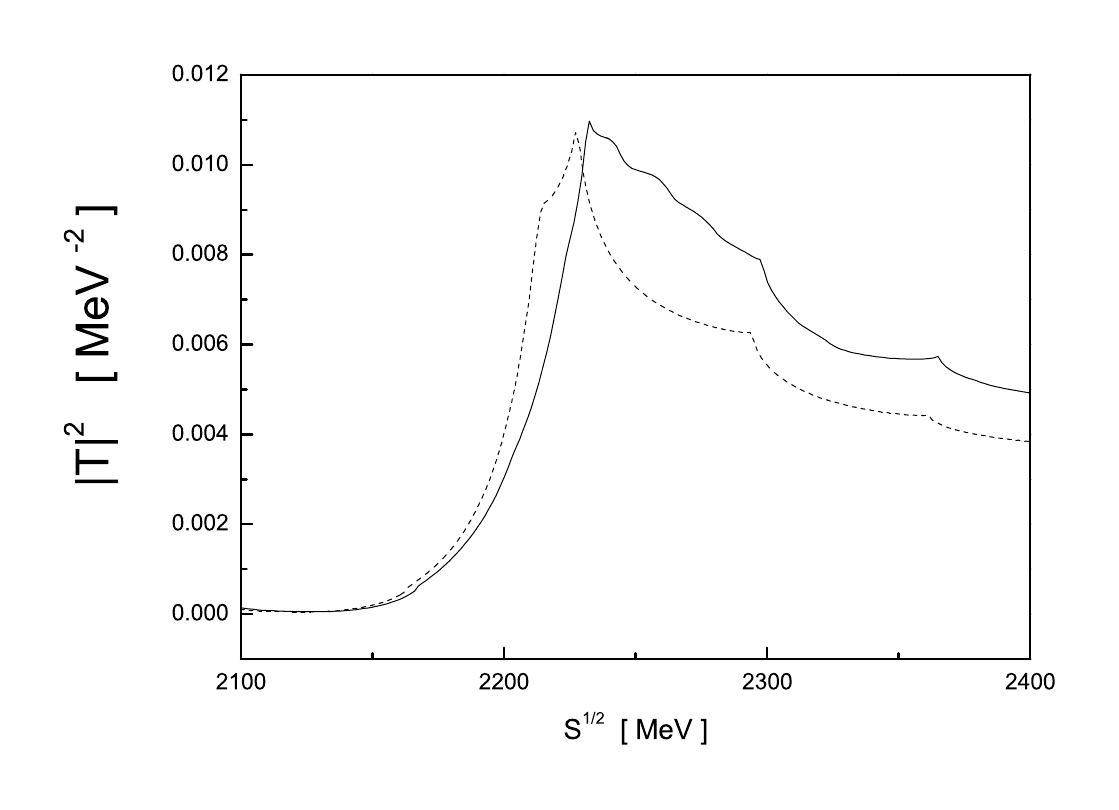,scale=0.8}
\epsfig{file=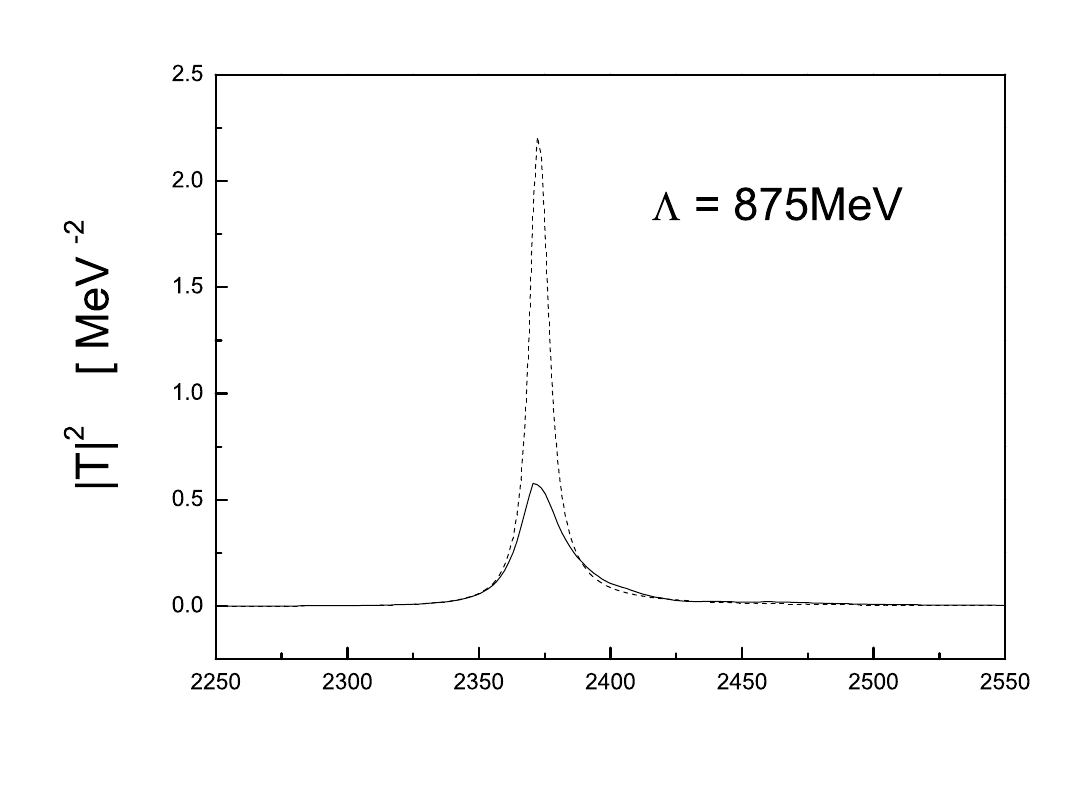,scale=0.8} \caption{We show the
unitarized $N \rho \rho$ squared amplitude $|T|^2$ in the left
panel, and the unitarized $\Delta \rho \rho$ squared amplitude in
the right panel. In both panels the solid and dashed lines denote
the cases with and without the $f_2(1270)$ decay width,
respectively. Taken from Ref.~\cite{Sun:2011fr}. \label{sec6:tnf2}}
\end{center}
\end{figure}

Because there are two identical $\rho$ mesons, one has $T_1 = T_2$,
and Eqs.~(\ref{sec6:FCA1}), (\ref{sec6:FCA2}) and (\ref{sec6:FCA3})
can be simplified to be:
\begin{eqnarray}
T &=& 2 T_1 = 2 t_1 + 2 t_1 G_0 T_1 = {2 t_1 \over 1 - G_0 t_1} \, .
\end{eqnarray}
The results are shown in the left panel of Fig.~\ref{sec6:tnf2},
where there is a peak at around $2227$ MeV with a width of $100$
MeV. This peak does not have a standard Breit-Wigner form, so it
could be due to the cusp effect.

In Ref.~\cite{Sun:2011fr} the authors also applied the same approach
to investigate the three-body $\Delta \rho \rho$ system. The results
are shown in the right panel of Fig.~\ref{sec6:tnf2}, where there is
a peak around 2372 MeV with an approximate Breit-Wigner shape. This
peak can be associated with the $\Delta(2390)$ of $J^P =
7/2^+$~\cite{Cutkosky:1979fy}.

The fixed center approximation to the Faddeev equations has been
systematically applied to study the $X$-multi-$\rho$
systems~\cite{YamagataSekihara:2010qk,Xiao:2012dw}. For
examples, in Ref.~\cite{YamagataSekihara:2010qk} the $K^*_2(1430)$,
$K^*_3 (1780)$, $K^*_4(2045)$ and $K^*_5(2380)$ are interpreted as
molecules made of an increasing number of $\rho(770)$ and one
$K^*(892)$ mesons; in Ref.~\cite{Xiao:2012dw} the $D^*$-multi-$\rho$
system was investigated and several charmed resonances, $D^*_3$,
$D^*_4$, $D^*_5$ and $D^*_6$, were predicted to be around 2800-2850
MeV, 3075-3200 MeV, 3360-3375 MeV and 3775 MeV, respectively. The
above method has also been applied to study the $NK\bar K$
system~\cite{Xie:2010ig}, the $\eta K \bar K$
system~\cite{Liang:2013yta}, the $D K \bar K$
system~\cite{Debastiani:2017vhv}, the $\bar K N N$
system~\cite{Bayar:2011qj}, the $BD\bar D$
system~\cite{Dias:2017miz}, and the $K D \bar D^*$
system~\cite{Ren:2018pcd}, etc.

%% file: section8.tex
%
\section{The Skyrme model and the chiral quark-soliton model}
\label{sect:5}

The Skyrme model was proposed by T.~H.~R.~Skyrme in 1961 by
introducing the Skyrme term to the nonlinear sigma
model~\cite{Skyrme:1961vq}. Within this model the baryons appear as
collective excitations of the meson fields. In fact, the non-trivial
topological field configuration, which is called as soliton, is
identified with the baryon. Later in 1979, E. Witten developed this
model by arguing that baryons indeed emerge as solitons in the large
$N_C$ generalization of QCD~\cite{Witten:1979kh}. He also studied
static properties of nucleons in the Skyrme model with two light
flavors~\cite{Adkins:1983ya}. The Skyrme model was widely applied in
hadron physics. Based on the Skyrme model one might qualitatively
understand the EMC observation that the quark spin contribution to
the total nucleon spin was unexpectedly small~\cite{Brodsky:1988ip}.
This feature is intensively studied in the Skyrme model with three
light flavors in order to understand the nucleon's strangeness
content~\cite{Weigel:1995cz}.

In the large $N_C$-limit the chiral quark-soliton model
($\chi$QSM)~\cite{Diakonov:1986yh,Reinhardt:1989fv,Meissner:1989kq,Wakamatsu:1990ud}.
is quite similar to the Skyrme model, and sometimes they are taken
as the same method/approach. In the Skyrme model one has integrated
out the quark fields, and so directly operates with the action
involving the pseudoscalar mesons without the quark fields.
Actually, one can obtain the effective action of the Skyrme type
from the one of the chiral quark-soliton model by a gradient
expansion.

The Skyrme model and the chiral quark-soliton model with three light
flavors can be naturally applied to study exotic baryons, which
emerge as rotational excitations of nucleons~\cite{Diakonov:1997mm}.
In this paper we briefly review some of these applications,
separately in the following subsections. We note that there are lots
of investigations within this approach, and we refer interested
readers to the lecture note by ~Diakonov~\cite{Diakonov:1997sj}, the
one by Weigel~\cite{Weigel:2008zz}, and the one by Ma and
~Harada~\cite{Ma:2016npf} for detailed instructions.

\subsection{A short introduction to the Skyrme model}

The Lagrangian of the nonlinear sigma model is
\begin{equation}
\mathcal{L}_{NL\sigma} = {f_\pi^2 \over 4} {\rm Tr} \left(
\partial_\mu U(x) \partial^\mu U^\dagger(x) \right) \, .
\label{lag:nlsigma}
\end{equation}
Here $f_\pi$ is the decay constant of the Nambu-Goldstone boson
$\pi$, and the field $U(x)$ is defined as
\begin{eqnarray}
U(x) \equiv {\rm exp} \left(i {\tau_i \phi_i \over f_\pi}\right) &=&
{\rm cos} \left({\tau_i \phi_i \over f_\pi}\right) + i {\rm sin}
\left({\tau_i \phi_i \over f_\pi}\right)
\\ \nonumber &=& {1\over f_\pi} \left (\sigma + i \tau_i \pi_i \right) = { M \over f_\pi} \, ,
\end{eqnarray}
where $M \equiv \sigma + i \tau_i \pi_i$ is the meson field, and the
field variables $\phi_i$ are related to $\pi_i$ through
\begin{eqnarray}
{\tau_i \pi_i \over f_\pi} = {\rm sin} \left({\tau_i \phi_i \over
f_\pi}\right) \, .
\end{eqnarray}

Under the chiral transformation, the field $U(x)$ transforms as
$U(x) \rightarrow g_L U(x) g_R^\dagger$, so the
Lagrangian~(\ref{lag:nlsigma}) is chirally invariant. With the
nonlinear realization of the chiral symmetry, the Hamiltonian
density of the nonlinear sigma model can be obtained as
\begin{equation}
\mathcal{H}_{NL\sigma} = {f_\pi^2 \over 4} {\rm Tr} \left(
\partial_0 U^\dagger \partial_0 U \right) + {f_\pi^2 \over 4} {\rm
Tr} \left( \partial_i U(x) \partial_i U^\dagger(x) \right) \, ,
\label{ham:nlsigma}
\end{equation}
and its energy can be expressed as
\begin{eqnarray}
E_{NL\sigma} &=& {f_\pi^2 \over 4} \int d^3 x {\rm Tr} \left(
\partial_0 U^\dagger \partial_0 U \right) + {f_\pi^2 \over 4} \int
d^3 x {\rm Tr} \left( \partial_i U(x) \partial_i U^\dagger(x)
\right) \label{energy:nlsigma}
\\ \nonumber &\equiv& E_{NL\sigma}^{\rm rotation} + E_{NL\sigma}^{\rm static} \, .
\end{eqnarray}
Here the first term $E_{NL\sigma}^{\rm rotation}$ is the rotational
energy, and the second term $E_{NL\sigma}^{\rm static}$ is the
static energy. The static energy $E_{NL\sigma}^{\rm static}$ is not
stable: under the rescaling of the space coordinates $U({\bf x})
\rightarrow U(\lambda {\bf x})$, one has $E_{NL\sigma}^{\rm static}
\rightarrow  E_{NL\sigma}^{\rm static} / \lambda$.

To avoid the above stability problem of the static energy, Skyrme
introduced the so-called Skyrme term, and extended the nonlinear
sigma model Lagrangian into the Skyrme model:
\begin{equation}
\mathcal{L}_{\rm Skyrme} = {f_\pi^2 \over 4} {\rm Tr} \left(
\partial_\mu U \partial^\mu U^\dagger \right) + {1 \over 32 e^2}
{\rm Tr} \left\{ \left[ U^\dagger \partial_\mu U , U^\dagger
\partial_\nu U \right] \left[ U^\dagger \partial^\mu U , U^\dagger
\partial^\nu U \right] \right\} \, , \label{lag:skyrme}
\end{equation}
where $e$ is a dimensionless parameter, indicating the magnitude of
the soliton. Its energy can be obtained as
\begin{eqnarray}
E_{\rm Skyrme} &=& - \int d^3 x {\rm Tr} \left[ {f_\pi^2 \over 4}
L_0 L_0  + {1 \over 16 e^2} [L_0 , L_i]^2 \right] - \int d^3 x {\rm
Tr} \left[ {f_\pi^2 \over 4} L_i L_i  + {1 \over 32 e^2} [L_i ,
L_j]^2 \right] \label{energy:nlsigma}
\\ \nonumber &\equiv& E_{\rm Skyrme}^{\rm rotation} + E_{\rm Skyrme}^{\rm static} \, .
\end{eqnarray}
One can solve the following extremum stable condition
\begin{eqnarray}
{d E_{\rm Skyrme}^{\rm static}(\lambda) \over d \lambda}
\Bigg|_{\lambda = 1} = 0 \, ,
\end{eqnarray}
and the obtained stabilized solutions are called as Skyrme solitons
or skyrmions.

Based on Eq.~(\ref{lag:skyrme}), the Witten-Wess-Zumino term was
added at the level of the action to reproduce QCD's low energy
anomalous structure~\cite{Witten:1979kh,Witten:1983tw,Wess:1971yu}:
\begin{eqnarray}
\Gamma^{\rm WZ}_{\rm Skyrme} &=& - {i N_c \over 240 \pi^2} \int d^5x
\epsilon^{\mu \nu \rho \sigma \tau} {\rm Tr}\left[ L_\mu L_\nu
L_\rho L_\sigma L_\tau \right] \, ,
\end{eqnarray}
where $N_c = 3$ is the number of colors. Sometimes the symmetry
breaking terms are also necessary:
\begin{eqnarray}
\mathcal{L}_{\rm Skyrme}^{\rm SB} &=& {f_\pi^2 m_\pi^2 + 2 f_K^2
m_K^2 \over 12} {\rm Tr}\left[ U + U^\dagger - 2 \right] + \sqrt3
\times {f_\pi^2 m_\pi^2 - f_K^2 m_K^2 \over 6} {\rm Tr}\left[
\lambda_8 (U + U^\dagger) \right]
\\ \nonumber && + {f_K^2 - f_\pi^2 \over 12} {\rm Tr}\left[ (1 - \sqrt3 \lambda_8) (U (\partial_\mu U)^\dagger \partial^\mu U +  U^\dagger \partial_\mu U (\partial^\mu U)^\dagger ) \right]  \, .
\end{eqnarray}

The Skyrme model can be used to derive a series of
states~\cite{Manohar:1984ys,Chemtob:1985ar,Mazur:1984yf,Jain:1984gp}.
In the three-flavor case, the lowest states are the $(\mathbf{8},
1/2)$ and $(\mathbf{10}, 3/2)$, which can be used to describe the
ground-state octet baryons of $J^P = 1/2^+$ and decuplet baryons of
$J^P = 3/2^+$. The remaining states are all exotic baryons, and the
states at the next level are the $(\mathbf{\overline{10}}, 1/2)$,
$(\mathbf{27}, 1/2)$ and $(\mathbf{27}, 3/2)$.

\begin{figure}[tb]
\begin{center}
\epsfig{file=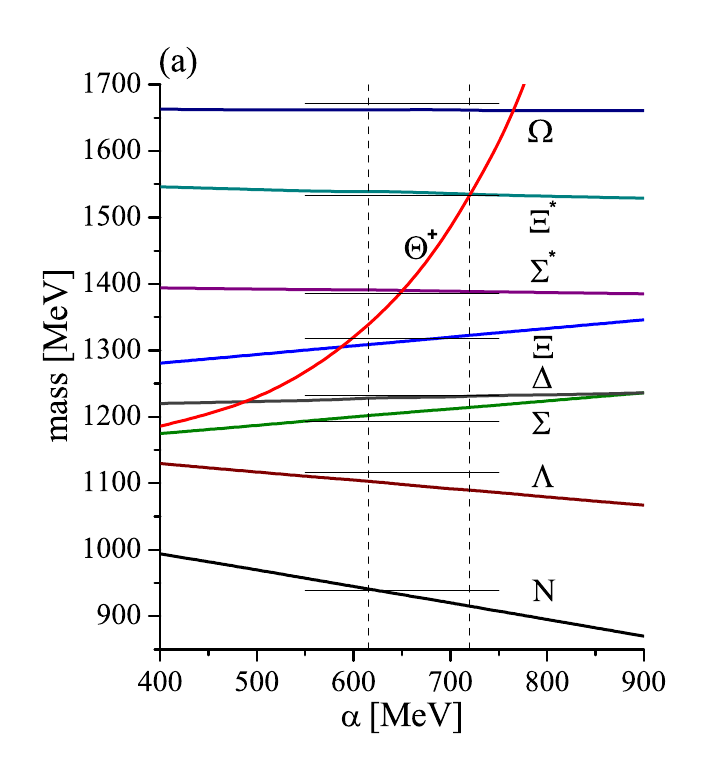,scale=1.5} \caption{Masses of the
ordinary baryons together with the mass of the $\Theta^+$, as
functions of the parameter $\alpha$. Taken from
Ref.~\cite{Praszalowicz:2003ik}. \label{sec6:skyrme}}
\end{center}
\end{figure}

In the chiral symmetry limit one can write the effective Hamiltonian
as:
\begin{eqnarray}
\mathcal{H}_{\rm Skyrme} = M_{cl} + {1 \over 2 I_1} J(J+1) + {1
\over 2 I_2} \left( C_2(\mathcal{R}) - J(J+1) - {N_c^2 \over 12}
\right) + \mathcal{H}^\prime_{\rm Skyrme} \, ,
\label{eq:hamiltonskyrme}
\end{eqnarray}
where $J$ denotes the baryon spin, $C_2(\mathcal{R})$ denotes the
Casimir operator for the $SU(3)$ representation $\mathcal{R}$, and
the other three parameters $M_{cl}$ and $I_{1,2}$ are treated as
free parameters. In the Skyrme model the symmetry breaking
Hamiltonian is
\begin{eqnarray}
\mathcal{H}^\prime_{\rm Skyrme} = - \alpha D^{(8)}_{88}(A)  \, ,
\end{eqnarray}
which leads to both the first order correction to the baryon mass
\begin{eqnarray}
M^{(1)}_{B(\mathcal{R})} = - \alpha
\delta^{\mathcal{R}}_{B(\mathcal{R})}  \, ,
\end{eqnarray}
and the second order correction
\begin{eqnarray}
M^{(2)}_{B(\mathcal{R})} = - 2 I_2 \alpha^2 \sum_{\mathcal{R}^\prime
\neq \mathcal{R}} { (\delta^{\mathcal{R}^\prime}_{B(\mathcal{R})})^2
\over C_2(\mathcal{R}^\prime) - C_2(\mathcal{R}) }  \, ,
\end{eqnarray}
We refer interested readers to
Refs.~\cite{Praszalowicz,Praszalowicz:2003ik} for detailed
discussions on the above equations. Especially, the mass of the $u u
d d \bar s$ pentaquark belonging to $(\mathbf{\overline{10}}, 1/2)$
was predicted to be~\cite{Praszalowicz,Praszalowicz:2003ik}
\begin{eqnarray}
M_{\Theta^+} = M_8 + {3\over2} {\alpha^2 \over \epsilon} - {2\over8}
\alpha - {3 \over 112} \epsilon \approx 1530 ~{\rm MeV}  \, .
\end{eqnarray}
However, this mass value is quite sensitive to the choice of
parameters. For example, the authors of
Ref.~\cite{Praszalowicz:2003ik} plot in Fig.~\ref{sec6:skyrme} the
results of the constrained fits, and one can find a rather steep
rise of $M_{\Theta^+}$ with $\alpha$. The above mass value 1530~{\rm
MeV} is very close to the mass of the $\Theta^+$ observed by
LEPS~\cite{Nakano:2003qx}, whose observation quickly led to renewed
interest in the Skyrme model descriptions of baryons. Unfortunately,
the $\Theta^+$ was not confirmed in a series of high precise
particle experiments~\cite{Liu:2014yva}.

The Skyrme model has many advantages in describing hadron physics,
and we refer to
Refs.~\cite{Schechter:1999hg,Weigel:2008zz,Ma:2016npf} for detailed
discussions. Especially, the Skyrme model can be applied to study
both baryon and meson physics in free space~\cite{Adkins:1983ya},
while it can also be applied to study nuclear matter and the medium
modified hadron
properties~\cite{Wuest:1987rc,Klebanov:1985qi,Lee:2003aq}. In recent
years, the Skyrme model is extended so that a) the effects of the
higher resonances and higher chiral order terms can be
self-consistently
analysed~\cite{Ma:2012kb,Ma:2012zm,Ma:2013ela,Ma:2013ooa}, and b)
properties of heavy baryons containing a heavy quark can be
systematically
investigated~\cite{Callan:1985hy,Nowak:1992um,Bardeen:1993ae,Nowak:2004jg,Harada:2012dm}.

\subsection{A short introduction to the chiral quark-soliton model}

In the previous subsection we have briefly introduced the Skyrme
model. In this subsection we shall briefly introduce the chiral
quark-soliton model~\cite{Diakonov:1987ty,Christov:1995vm}. Both of
them have similar group structure, and can be applied to study the
$\Theta^+$ pentaquark belonging to the exotic
$(\mathbf{\overline{10}}, 1/2)$ flavor representation. In fact, the
prediction of the narrow width of the $\Theta^+$ in the chiral
quark-soliton model~\cite{Diakonov:1997mm} stimulated the
experimental search by LEPS~\cite{Nakano:2003qx}. We refer to
Refs.~\cite{Diakonov:1997sj,Goeke:2004ht} for detailed discussions.

According to the chiral symmetry, the interaction of pseudoscalar
mesons with constituent quarks can be written as:
\begin{eqnarray}
\mathcal{L}_{eff} = \bar q \left[ i \partial^\mu \gamma_\mu - M {\rm
exp}(i \gamma_5 \pi^A \lambda^A / F_\pi) \right] q \, ,
\end{eqnarray}
where the $\pi^A$ fields are the ordinary pseudoscalar mesons $\pi$,
$K$, and $\eta$. In the large $N_c$ limit baryons emerge as solitons
of this chiral action~\cite{Witten:1979kh}. The model of baryons
based on the large $N_c$ limit is called the chiral quark-soliton
model~\cite{Goeke:2004ht}.

Actually, we can also use the word ``soliton'' for the
self-consistent pion mean field in the nucleon. Because these
Hartree-Fock states composed of valence- and sea-quarks are
degenerate with respect to rotations in both space and flavor-space,
we can use the following collective Hamiltonian to describe them in
the rigid rotor approximation~\cite{Blotz:1992br,Blotz:1992pw}:
\begin{eqnarray}
\mathcal{H}_{\rm \chi QSM} = M_{cl} + {J(J + 1) \over 2I_1} + {1
\over 2I_2}\left(\mathcal{C}_2(\mathcal{R}) - J(J + 1) - {N^2_c
\over 12}\right) + H^\prime_{\rm \chi QSM} \, ,
\end{eqnarray}
where $M_{cl}$ is the classical mean-field energy of the quark
system, $J$ is the baryon spin, and $C_2(\mathcal{R})$ is the
Casimir operator for the $SU(3)$ representation $\mathcal{R}$. This
Hamiltonian is very similar to the one of the Skyrme model given in
Eq.~(\ref{eq:hamiltonskyrme}) except a different symmetry breaking
term, {\it i.e.},
\begin{eqnarray}
H^\prime_{\rm \chi QSM} = \alpha D^{(8)}_{88} + \beta Y + {\gamma
\over \sqrt3} D^{(8)}_{8i} J_i \, ,
\end{eqnarray}
where $\alpha$, $\beta$, and $\gamma$ are parameters of order
$\mathcal{O}(m_s)$, $D^{(\mathcal{R})}_{ab}$ are $SU(3)$ Wigner
rotation matrices, and $J_i$ are collective spin operators.

\begin{figure}[tb]
\begin{center}
\epsfig{file=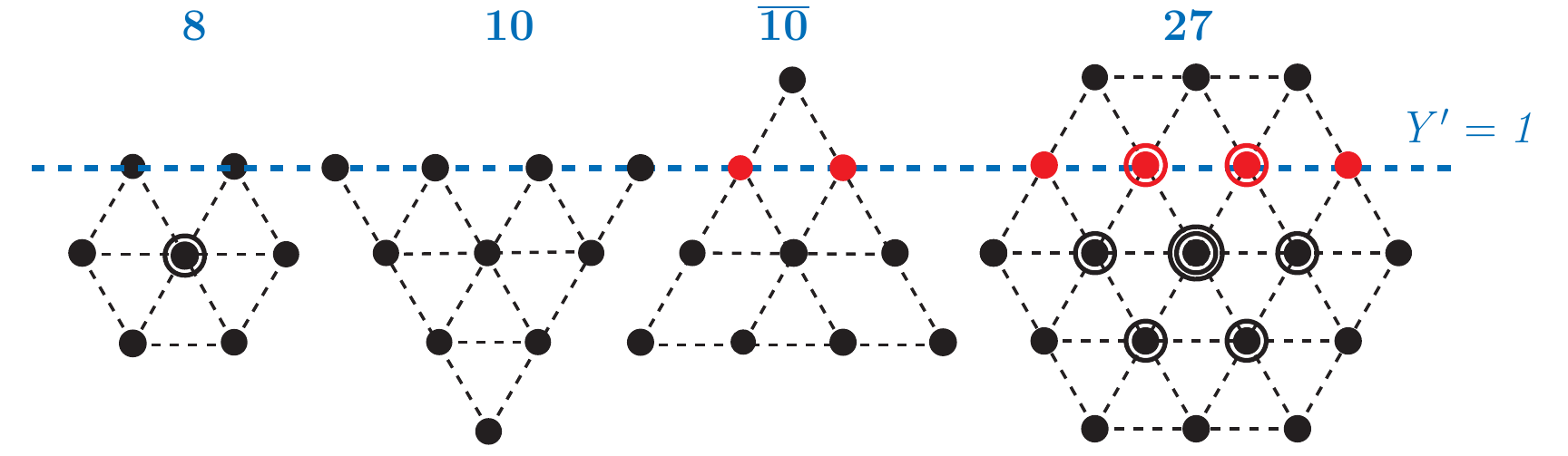,scale=1} \caption{Weight
diagrams for the lowest-lying baryon multiplets, including the
flavor representations $\mathbf{8}$, $\mathbf{10}$,
$\mathbf{\overline{10}}$, and $\mathbf{27}$. Taken from
Ref.~\cite{Yang:2018gju}. \label{sec6:chiQSM}}
\end{center}
\end{figure}

The symmetry-breaking term $H^\prime_{\rm \chi QSM}$ mixes different
$SU(3)$ representations, so the collective wave functions can be
written as the following linear combinations:
\begin{eqnarray}
| B_\mathbf{8} \rangle &=& | \mathbf{8}_{1/2} , B \rangle +
c^B_{\overline{10}} | \mathbf{\overline{10}}_{1/2}, B \rangle +
c^B_{27} | \mathbf{27}_{1/2}, B \rangle \, ,
\\ | B_\mathbf{10} \rangle &=& | \mathbf{10}_{3/2} , B \rangle + a^B_{27} | \mathbf{27}_{3/2}, B \rangle + a^B_{35} | \mathbf{35}_{3/2}, B \rangle \, ,
\\ | B_\mathbf{\overline{10}} \rangle &=& | \mathbf{\overline{10}}_{1/2} , B \rangle + d^B_{8} | \mathbf{8}_{1/2}, B \rangle + d^B_{27} | \mathbf{27}_{1/2}, B \rangle + d^B_{\overline{35}} | \mathbf{\overline{35}}_{1/2}, B \rangle \, ,
\\ \nonumber && \cdots
\end{eqnarray}
where $|B_{\mathcal{R}}\rangle$ are the states which reduce to the
$SU(3)$ representation $\mathcal{R}$ when $m_s \rightarrow 0$ and
the spin index $J_3$ is suppressed. $c^B_{\mathcal{R}}$,
$a^B_{\mathcal{R}}$, and $d^B_{\mathcal{R}}$ are the $m_s$-dependent
coefficients. For example, we show in Fig.~\ref{sec6:chiQSM} the
weight diagrams for the lowest-lying baryon multiplets, which
contains the baryons with the hypercharge $Y^\prime = 1$.

Assuming the state $N^*(1710)$ to be a member of the
$(\mathbf{\overline{10}}, 1/2)$ flavor representation, Diakonov,
Petrov and Polyakov investigated an exotic $Z^+$ baryon with spin
1/2, isospin 0 and strangeness +1, and predicted its mass to be
about 1530 MeV and its total width to be less than 15
MeV~\cite{Diakonov:1997mm}. Now this state was known as the
$\Theta^+$, which was first observed by LEPS~\cite{Nakano:2003qx}.
However, we note that this small width actually resulted 
from an arithmetic error, as explained by Jaffe in Ref.~\cite{Jaffe:2004qj},
and correcting this error gives a width twice as large at least.
See Ref.~\cite{Weigel:1995cz} for detailed discussions.

Later in Refs.~\cite{Yang:2010fm,Yang:2011qe} the masses of the
lowest-lying baryons were calculated by including the effects of
isospin symmetry breaking arising from both the quark masses and
electromagnetic self-energies. The authors used the following
collective Hamiltonian in the $SU(3)$ chiral soliton
model~\cite{Blotz:1992br,Blotz:1992pw}:
\begin{equation}
\mathcal{H} = M_{cl} + \mathcal{H}_{rot} + \mathcal{H}_{SB} \, ,
\end{equation}
where $\mathcal{H}_{rot}$ is the $1/N_c$ rotational energy, and
$\mathcal{H}_{SB}$ are the symmetry-breaking terms including both
isospin and $SU(3)$ flavor symmetry breaking effects:
\begin{eqnarray}
\mathcal{H}_{rot} &=& {1 \over 2 I_1} \sum_{i=1}^3 J_i^2 + {1 \over
2 I_2} \sum_{p=4}^7 J_p^2 \, ,
\\ \mathcal{H}_{SB} &=& (m_d - m_u) \left( {\sqrt3\over2} \alpha D^{(8)}_{38}(\mathcal{R}) + \beta T_3 + {\gamma \over 2} \sum_{i=1}^3 D^{(8)}_{3i}(\mathcal{R}) J_i \right)
\\ \nonumber && + (m_s - {m_u + m_d \over 2}) \left( \alpha D^{(8)}_{88}(\mathcal{R}) + \beta Y + {\gamma \over \sqrt3} \sum_{i=1}^3 D^{(8)}_{8i}(\mathcal{R}) J_i \right)
\\ \nonumber && + (m_u + m_d + m_s) \sigma \, .
\end{eqnarray}
We refer interested readers to Refs.~\cite{Yang:2010fm,Yang:2011qe}
for detailed discussions on the above equations.

The baryon mass splittings due to the isospin and $SU(3)$ flavor
symmetry breaking up to the first order were investigated in
Ref.~\cite{Yang:2010fm}, and those due to the second order flavor
symmetry breaking were investigated in Ref.~\cite{Yang:2011qe}. The
authors estimated the mass of the $N^*$ as a member of the baryon
antidecuplet to be $1687$~MeV, which can be used to explain the
$N^*(1685)$, a narrow bump-like structure observed by GRAAL in the
$\gamma n \rightarrow \eta n$ quasi-free cross section in
2006~\cite{Kuznetsov:2006kt}. The authors also estimated the
pion-nucleon sigma term to be $\sigma_{\pi N} = (50.5 \pm 5.4)$~MeV.

In 2015, another narrow resonant structure was observed by the GRAAL
Collaboration in real Compton scattering off the proton. Its mass
and width were determined to be $M = 1.726 \pm 0.002 \pm 0.005$~GeV
and $\Gamma = 21 \pm 7$~MeV, respectively. In the chiral
quark-soliton model, this structure can be identified as a member of
the eikosiheptaplet ($\mathbf{27}$) with spin
3/2~\cite{Praszalowicz:2007zza}. Later in Ref.~\cite{Yang:2018gju}
the authors investigated the strong and radiative decay widths of
the narrow nucleon resonances $N^*(1685)$ and $N^*(1726)$ together
within the $SU(3)$ chiral quark-soliton model~\cite{Yang:2018gju}.
Especially, they found $\Gamma_{p^*(1726)\rightarrow p \gamma} /
\Gamma_{n^*(1726)\rightarrow n \gamma} = 3.76 \pm 0.64$, indicating
that the production of $N^*(1726)$ is more likely to be observed in
the proton channel.

%% file: section9.tex
\section{Progresses from Lattice QCD}

Lattice QCD is the unique non-perturbative theoretical framework to
study the hadron spectroscopy starting from the first principle QCD
Lagrangian. In a lattice QCD simulation, the hadron mass $E_n$ can
be extracted from the $N\times N$ time-dependence correlation
function
\begin{eqnarray}
\nonumber C_{ij}(t)&=&\langle
\Omega|\mathcal{O}_i(t)\mathcal{O}_j^\dagger(0)|\Omega\rangle
\\ &=&\sum_{n=1}^Ne^{-E_nt}\langle \Omega|\mathcal{O}_i|n\rangle\langle n|\mathcal{O}_j^\dagger(0)|\Omega\rangle\, ,
\end{eqnarray}
where the interpolating operator $\mathcal{O}_j^\dagger(0) (j=1, 2,
\cdots, N)$ creates the state of interest from the vacuum
$|\Omega\rangle$ at time $t=0$ and $\mathcal{O}_i(t)$ annihilates
the state at a later Euclidean time $t$. $E_n$ is the eigenvalue of
the Hamiltonian for the system. These creation and annihilation
operators should have the same $J^{PC}$ quantum numbers, which are
called interpolators. In principle they can couple to all physical
eigenstates $|n \rangle $ with the same given quantum numbers but
different magnitudes of overlaps
\begin{equation}
\langle n|\mathcal{O}_i^\dagger|\Omega\rangle\equiv Z_i^n\, .
\end{equation}

In lattice QCD, the correlation functions are usually evaluated by
using a large basis of interpolators constructing of quark and gluon
fields, the Dirac gamma matrices and the gauge-covariant derivative
operator. To search for the tetraquark and pentaquark signatures,
the two-particle interpolators are usually adopted to deduce the
two-point correlation functions. If the interpolator basis is
complete enough, one can extract and identify all two-particle
discrete energy levels with the given quantum numbers from the full
correlation functions. The exotic multiquark states are related to
the extra energy levels in addition to the expected non-interacting
two-particle states. The two-particle energies are obtained from the
correlation function by solving the generalized eigenvalue problem
(GEVP)~\cite{Luscher:1990ck,Blossier:2009kd}:
\begin{equation}
C_{ij}(t)v_j^{n}(t\, , t_0)=\lambda_n(t\, ,
t_0)C_{ij}(t_0)v_j^{n}(t\, , t_0)\, ,
\end{equation}
where $v_j^{n}(t\, , t_0)$ is an eigenvector and $t_0$ a suitably
chosen reference time $t_0<t$. The eigenvalues feature the relevant
state energies~\cite{Luscher:1990ck}
\begin{equation}
\lambda_n(t\, , t_0)\approx A_ne^{-E_n(t-t_0)}+\cdots\, .
\end{equation}
Fitting this exponential decay behavior of the time dependence for
each of these quantities, the exact discrete spectrum $E_n$ can be
extracted from the effective mass plateau of the eigenvalue
$\lambda_n$~\cite{Bulava:2009jb,Dudek:2007wv,Mahbub:2013ala,Kiratidis:2016hda,Kiratidis:2015vpa}
\begin{equation}
E_n^{eff}(t)= \ln\frac{\lambda_n(t)}{\lambda_n(t+1)}\, .
\end{equation}

In the infinite volume, a bound state of two particles can be
defined as a discrete energy eigenstate of the Hamiltonian with
energy level below the two-particle threshold within the L\"uscher's
method~\cite{Luscher:1986pf,Luscher:1985dn,Luscher:1990ck,Luscher:1991cf,Luscher:1990ux}.
In a two-particle scattering process, the resonant character on the
lattice can be observed as the negative center of mass (c.m.)
momentum $q^2$, which corresponds to the attractive interaction
between the two particles and renders the energy level lower than
the non-interacting threshold. The L\"uscher's formula provides a
direct relation of $q^2$ and the elastic scattering phase shift
$\delta(q)$ in the infinite volume, e. g. in the case of the
$s$-wave elastic scattering
\begin{equation}
q\cot\delta(q)=\frac{1}{\pi^3/2}\mathcal{Z}_{00}(1; q^2)\, ,
\end{equation}
where $\mathcal{Z}_{00}(1; q^2)$ is the zeta-function. In the limit
of $q^2\to-\infty$, the phase shift will approximate
$\cot\delta(q)\approx-1$ if there exists a true bound state at the
particular energy in the infinite volume. Therefore, one usually
needs to study the quantities $q^2$ and $\cot\delta(q)$ to really
identify a bound state in a lattice simulation. We refer to a recent
review article~\cite{Briceno:2017max} for detailed discussion, and
refer to Refs.~\cite{Doring:2011vk,Doring:2011ip} from
the viewpoints of the chiral perturbation theory.

In the last decade, there are some remarkable progresses in the
study of new hadron states by lattice QCD. In this section, we try
to introduce these investigations, such as the studies on the
$Y(4260)$, $X(3872)$, the charged $Z_c$ states, the doubly heavy
tetraquark states and the hidden-charm pentaquark states.

\subsection{$Y(4260)$}

The underlying structure of the $Y(4260)$ meson was studied by TWQCD
Collaboration~\cite{Chiu:2005ey}. They calculated the mass spectra
of the hidden-charm hybrid mesons ($\bar cgc$), molecules and
diquark-antidiquark tetraquarks with $J^{PC}=1^{--}$, in quenched
lattice QCD with exact chiral symmetry. The authors computed the
time correlation functions for all operators in the above various
schemes and detected all these exotic resonances. However, the
lowest-lying hybrid charmonium lies about 250 MeV above the mass of
$Y(4260)$. In contrast, the $1^{--}$ hidden-charm molecule ($\bar
uc\bar cu$) and tetraquark ($uc\bar u\bar c$) states were detected
with masses 4238(31)(57) MeV and 4267(68)(83) MeV respectively,
which are in good agreement with the mass of $Y(4260)$. Finally, the
authors suggested the $Y(4260)$ to be a $D_1\bar D$ molecule since
it has a better overlap with the molecular operator than any other
ones. Besides, they also detected resonances of molecular and
tetraquark states with quark contents $cs\bar c\bar s$ around
$4450\pm100$ MeV, and with quark contents $cc\bar c\bar c$  around
$6400\pm50$ MeV.

Before the observation of $Y(4260)$, the mass spectrum of the
charmonium hybrids were predicted by the UKQCD quenched lattice
NRQCD~\cite{Lacock:1996ny,Manke:1998yg,Juge:1999aw} and anisotropic
lattice simulations~\cite{Drummond:1999db}. In
Ref.~\cite{Liu:2012ze}, the Hadron Spectrum Collaboration performed
dynamical lattice QCD calculations to study the mass spectra of
highly excited charmonium and charmonium hybrid mesons. To provide a
large basis of operators with a range of spatial structures in each
channel, the authors used the derivative-based construction for
operators of the general form
$\bar{\psi}\Gamma\stackrel{\leftrightarrow}{D}_i\stackrel{\leftrightarrow}{D}_j\cdots\psi$
to calculate the two-point correlation functions. Their computations
were presented for up and down quarks corresponding to a pion mass
of 400 MeV at a single lattice spacing. They calculated the
dynamical spectrum of charmonium hybrids with exhaustive quantum
numbers including the exotic ones ($0^{+-}, 1^{-+}, 2^{+-}$) up to
4.5 GeV. Their results identified the scheme that the hybrid meson
can be interpreted as a colour-octet quark-antiquark pair coupled to
a $J_g^{P_gC_g}=1^{+-}$ chromomagnetic gluonic excitation. In this
scheme, the lightest hybrid supermultiplet consists of
negative-parity states $J^{PC}=(0, 1, 2)^{-+}, 1^{--}$, in which the
colour-octet quark-antiquark pair is in S-wave. Such a pattern
suggested that the lightest gluonic excitation has an energy scale
of 1.2-1.3 GeV, compared with the conventional charmonium bound
states. The first excited hybrid supermultiplet composed of a P-wave
colour-octet quark-antiquark pair coupled to an excited gluonic
field would contain positive-parity states with $J^{PC}=0^{+-},
(1^{+-})^3, (2^{+-})^2, 3^{+-}, (0, 1, 2)^{++}$. The vector hybrid
charmonium with $J^{PC}=1^{--}$ was extracted at
$M-M_{\eta_c}\approx1.3$ GeV, which agrees well with the mass of
$Y(4260)$ and supports the hybrid interpretation of this state. The
charmonium hybrid interpretation for $Y(4260)$ was also performed in
Ref.~\cite{Luo:2005zg}, in which the authors calculated the masses
of the ground and first excited states for the $\bar cgc$ hybrid
mesons with $J^{PC}=0^{-+}, 1^{--}$ and $1^{++}$. They finally
concluded that the $Y(4260)$ meson can not be identified as the
ground state of the $1^{--}$ hybrid, but most probably the first
excited state of the $1^{--}$ hybrid charmonium state. However,
the QCD sum rule calculation gave much lower mass prediction
for the vector hybrid charmonium, which was around 3.4 GeV~\cite{Chen:2013zia}.
The heavy quarkonium hybrid states with various flavor structures
and quantum numbers have been extensively studied in
Refs.~\cite{Chen:2013zia,Chen:2013eha,Chen:2014fza}.

\begin{figure}[htbp]
\centering
\includegraphics[width=6in]{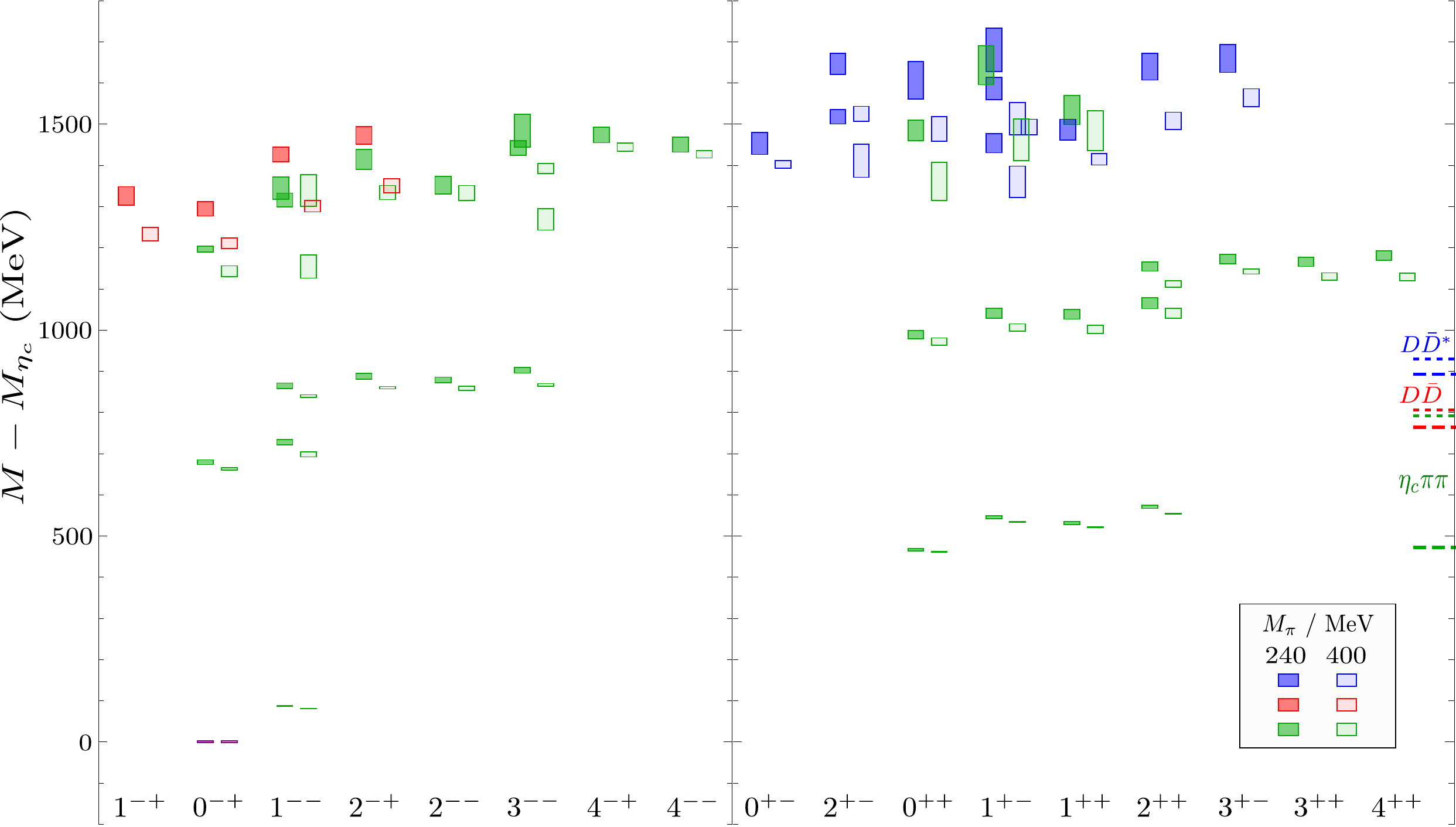}\\
\caption{The lattice calculations of charmonium spectra with
$m_\pi\sim240$ MeV from Ref.~\cite{Cheung:2016bym}, comparing to the
spectra with $m_\pi\sim400$ MeV in Ref.~\cite{Liu:2012ze}. The green
boxes are normal $c\bar c$ charmonium states, while red and dark
blue boxes are $\bar cgc$ charmonium hybrid candidates of the
lightest and first-excited supermultiplet respectively. The dashed
lines indicate the corresponding non-interacting hadron thresholds.}
\label{44charmonium240}
\end{figure}

Since the lattice investigations are always performed with
unphysical-heavy light quarks, it is significant to explore the
changes in the pattern of states with light-quark masses closer to
their physical values~\cite{Guo:2012tg}. For the charmonium systems,
the light-quark mass dependence enters through the sea quark content
in the dynamical gauge field ensembles. In the case of the
open-charm charmed mesons, the existence of a valence light quark
affects the spectrum directly. To investigate the light quark
dependence effect on lattice QCD simulation, the Hadron Spectrum
Collaboration presented spectra of the excited charmonium,
open-charm $D$ and $D_s$ mesons from dynamical lattice QCD
calculations with a pion mass $m_\pi\sim240$ MeV in
Ref.~\cite{Cheung:2016bym}. The charmonium spectra labelled by
$J^{PC}$ were shown in Fig.~\ref{44charmonium240} together with
their previous results with $m_\pi\sim400$ MeV in
Ref.~\cite{Liu:2012ze}. The masses of the low-lying states are
generally consistent between these two ensembles within statistical
uncertainties. Only the hyperfine splitting $M_{J/\psi}-M_{\eta_c}$
has a small but statistically significant increase when the pion
mass decreased. For the higher states in the spectrum, the masses
are generally larger with the $m_\pi\sim240$ MeV ensemble.
Especially for the hybrids, there is a small but significant
increase in their masses as the pion mass is reduced, which will
increase the mass splitting between the hybrids and low-lying
conventional charmonium states. However, the statistical
uncertainties at higher energies are larger. The unstable nature of
these states above threshold may be
important~\cite{Dudek:2010wm,Moir:2013ub,Moir:2016srx}. The authors
thus concluded that the hybrid mesons appear to show a mild increase
in mass as the light quark mass is decreased, but the pattern of
states and supermultiplet structure are unchanged.

\begin{figure}[htbp]
\centering
\includegraphics[width=4in]{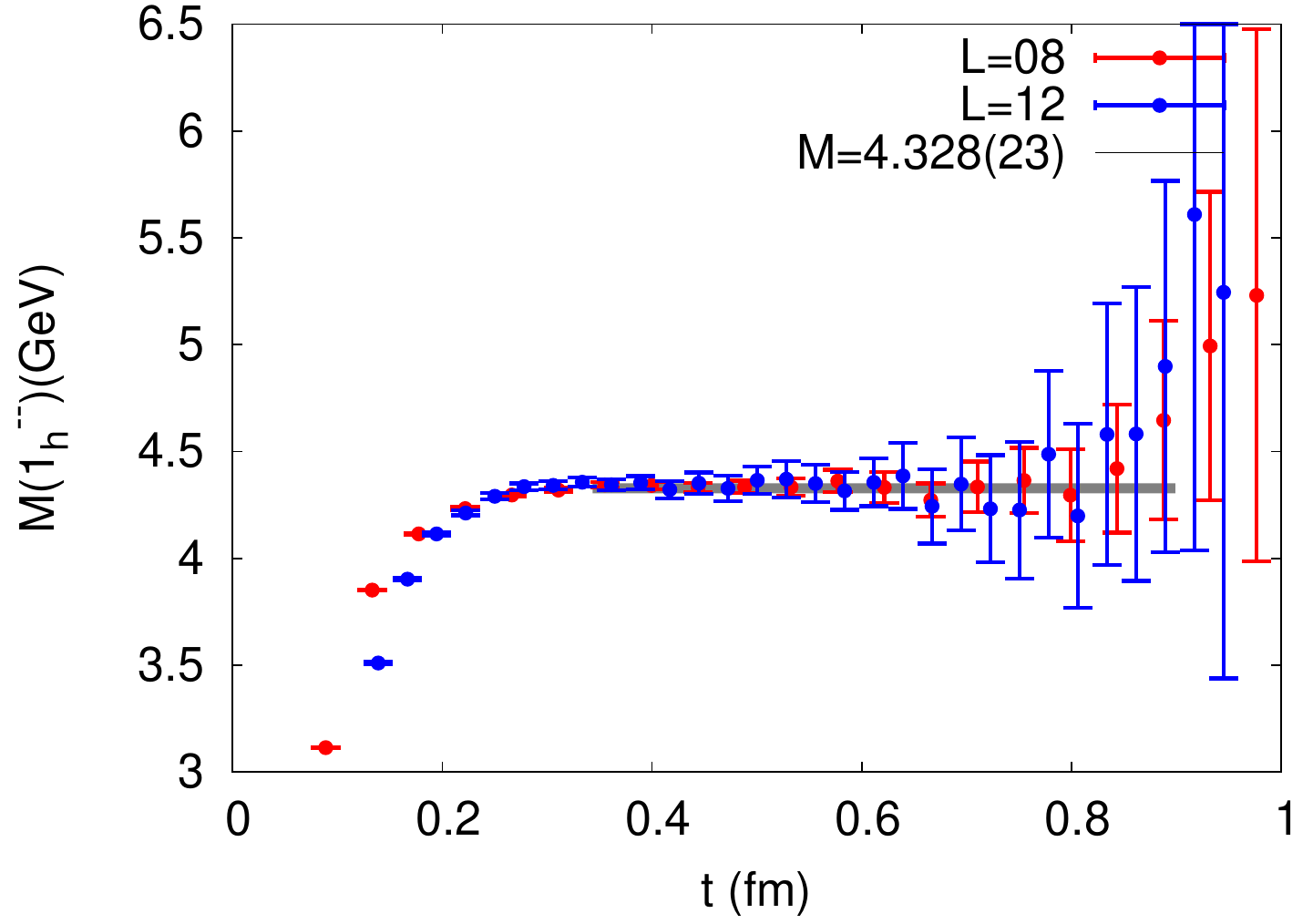}\\
\caption{The effective mass plateaus of $C(\omega,\, t)$ for
$\beta=2.4$ (red points) and $\beta=2.8$ (blue points), taken from
Ref.~\cite{Chen:2016ejo}} \label{hybrid_chen}
\end{figure}

In Ref.~\cite{Chen:2016ejo}, Chen {\it et al.} investigated the
existence of the exotic vector charmonia by constructing the
interpolating operators $O^{(H)}_i\left(x,\, t;\,
r\right)=\left(\bar c^a\gamma_5c^b\right)\left(x,\,
t\right)B^{ab}_i\left(x+r,\, t\right)$, where $i$ is the spatial
index and $B^{ab}_i(x)=\frac{1}{2}\epsilon_{ijk}F^{ab}_{jk}$ the
chromomagnetic field tensor. Compared to the usual hybrid operator,
this is a new type of the hybrid-like operator that the charm
quark-antiquark pair component $\bar c^a\gamma_5c^b$ and the gluon
field $B^{ab}_i$ are split into two parts by an explicit spatial
displacement $r$. Such a configuration can resemble the
center-of-mass motion of the $\bar cc$ recoiling against an
additional degree of freedom, and thus is expected to suppress the
coupling to the conventional charmonia. They accordingly calculated
the two-point correlation functions $C^{H}(r,\, t)$ by using these
operators. To eliminate the contribution from the conventional
charmonium states, they combined linearly the correlation functions
at two specific different $r$ as $C(\omega,\, t)=C^{H}(r_1,\,
t)-\omega C^{H}(r_2,\, t)$, where $\omega$ is a tunable parameter.
For the numerical analyses, they used the tadpole-improved gauge
action \cite{Morningstar:1997ff,Morningstar:1999rf,Chen:2005mg} to
generate gauge configurations on anisotropic lattices. Two lattices
$L^3\times T=8^3\times 96(\beta=2.4)$ and $12^3\times
144(\beta=2.8)$ with different lattice spacings were used to check
the discretization artifacts. After fitting the effective mass
plateaus as shown in Fig.~\ref{hybrid_chen}, they observed a vector
charmonium-like state $X$ with a mass of $4.33(2)$ GeV with exotic
nature. In addition, they calculated the leptonic decay constant of
this exotic state to be $f_X<40$ MeV, which accordingly led to a
very small leptonic decay width $\Gamma(X\to e^+e^-)<40$ eV. The
mass and leptonic decay width of this signal are consistent with the
production and decay properties of $Y(4260)$.

\subsection{$X(3872)$}

The TWQCD Collaboration also studied the $X(3872)$ state as a
hidden-charm four-quark $qc\bar q\bar c$ meson with $J^{PC}=1^{++}$
in quenched lattice QCD with exact chiral
symmetry~\cite{Chiu:2006hd}. They used both molecular operator
$(\bar q \gamma_i c)(\bar c \gamma_5 q) - (\bar c \gamma_i q)(\bar q
\gamma_5 c)$ and tetraquark operator $(q^T C \gamma_i c)(\bar q C
\gamma_5\bar c^T) - (\bar q^T C \gamma_i \bar c)(q C \gamma_5 c^T)$
to compute the time-correlation functions for two lattice volumes
$24^3\times 48$ and $20^3\times 40$. For both the molecular and
tetraquark operators, they detected a resonance with mass around
$3890\pm30$ MeV in the limit $m_q\to m_u$, which is identified as
$X(3872)$. Their results showed that $X(3872)$ has good overlap with
the molecular operator as well as the tetraquark operator in the
quenched approximation. Comparing with their previous study for
$Y(4260)$, they concluded that $X(3872)$ may be more tightly bound
than $Y(4260)$, which had better overlap with the molecular operator
than any tetraquark ones~\cite{Chiu:2005ey}. Since only one energy
level was extracted near the $D\bar D^\ast$ threshold, the result
could not support or disfavor the existence of $X(3872)$. Letting
$m_q\to m_s$, they predicted a heavier $1^{++}$ $cs\bar c\bar s$
exotic meson around $4100\pm50$ MeV. By using the same method and
interpolating currents as above, they investigated the mass spectrum
of the $(cs\bar c\bar q)/(cq\bar c\bar s)$ tetraquark states with
$J^P=1^+$. They also detected an axial-vector $(cs\bar c\bar
q)/(cq\bar c\bar s)$ resonance around $4010 \pm 50$
MeV~\cite{Chiu:2006us}. In these investigations of TWQCD
Collaboration, the authors neglected the disconnected quark loop
diagrams for annihilation channels. However, these diagrams provide
the mixing effects between states created by $\bar cc$ and $\bar
cq\bar qc$ operators, which have been proven to be important for the
lattice studies of light scalar
tetraquarks~\cite{Prelovsek:2010kg,Guo:2013nja}, and the string
breaking in the static limit~\cite{Bali:2005fu}.

As indicated above, the mass spectra of the highly excited
charmonium states have been calculated up to around 4.5 GeV in
dynamical lattice QCD by the Hadron Spectrum Collaboration in
Ref.~\cite{Liu:2012ze}. The first radial excitation of the P-wave
$1^{++}$ charmonium state is extracted about 110 MeV above the mass
of $X(3872)$, where the mass of $\eta_c$ is subtracted from the
calculated mass in order to reduce the systematic error from the
tuning of the bare charm quark mass. However, the results presented
in Ref.~\cite{Liu:2012ze} are for unphysically-heavy up and down
quarks corresponding to $m_\pi\approx400$ MeV. The dynamical lattice
simulation with lighter up and down quarks $m_\pi\approx240$ MeV
showed that only a very small mass increase for the low-lying and
even higher charmonium states~\cite{Cheung:2016bym}. Moreover,
several other lattice simulations found one excited $1^{++}$
charmonium state near the mass of
$X(3872)$~\cite{Bali:2011dc,Bali:2012ua,Mohler:2012na}. However,
none of these simulations can unambiguously determine whether this
state is $X(3872)$ or the scattering $D\bar D^\ast$ state.

In Ref.~\cite{Bali:2011rd}, the authors calculated the charmonium
spectrum including higher spin and gluonic excitations. In
particular, they studied the mixing of charmonia with states created
by the hidden-charm molecular operators by using the variational
generalized eigenvalue method as well as improved stochastic
all-to-all propagator methods. Their simulations were employed with
light pseudoscalar mass down to $280$ MeV and $n_F = 2$ sea quarks.
They investigated the binding between pairs of $D$ and anti-$D$
mesons in the pseudoscalar, vector and axialvector sectors. Only the
axialvector channel was clearly attractive. Moreover, the mixing
effects between the isoscalar radial charmonium states and $\bar
cq\bar qc$ molecular states were very small for the pseudoscalar and
vector channels. But for the $1^{++}$ channel, they found a
significant binding $m_{D^\ast\bar D}-(m_{D^\ast}+m_{\bar
D})=88(26)$ MeV. This binding is much bigger than the mass
difference $m_{X(3872)}-(m_{D^\ast}+m_{\bar D})$.

For the molecular interpretation of $X(3872)$, many dynamical
studies supported the attractive channel of $D$ and $\bar{D}^*$
mesons. One can consult Refs.~\cite{Chen:2016qju,Guo:2017jvc} and
references therein for these discussions. Due to the coupled channel
effects, the $S$-wave $D\bar{D}^*$ scattering state with
$J^{PC}=1^{++}$ can easily couple to the $\chi_{c1}^\prime(2P)$
state~\cite{Suzuki:2005ha,Meng:2005er} and significantly lower the
mass of the pure $\chi_{c1}^\prime(2P)$ state predicted in the GI
model~\cite{Godfrey:1985xj}.

\begin{figure*}[tb]
\begin{center}
\includegraphics*[width=0.7\textwidth,clip]{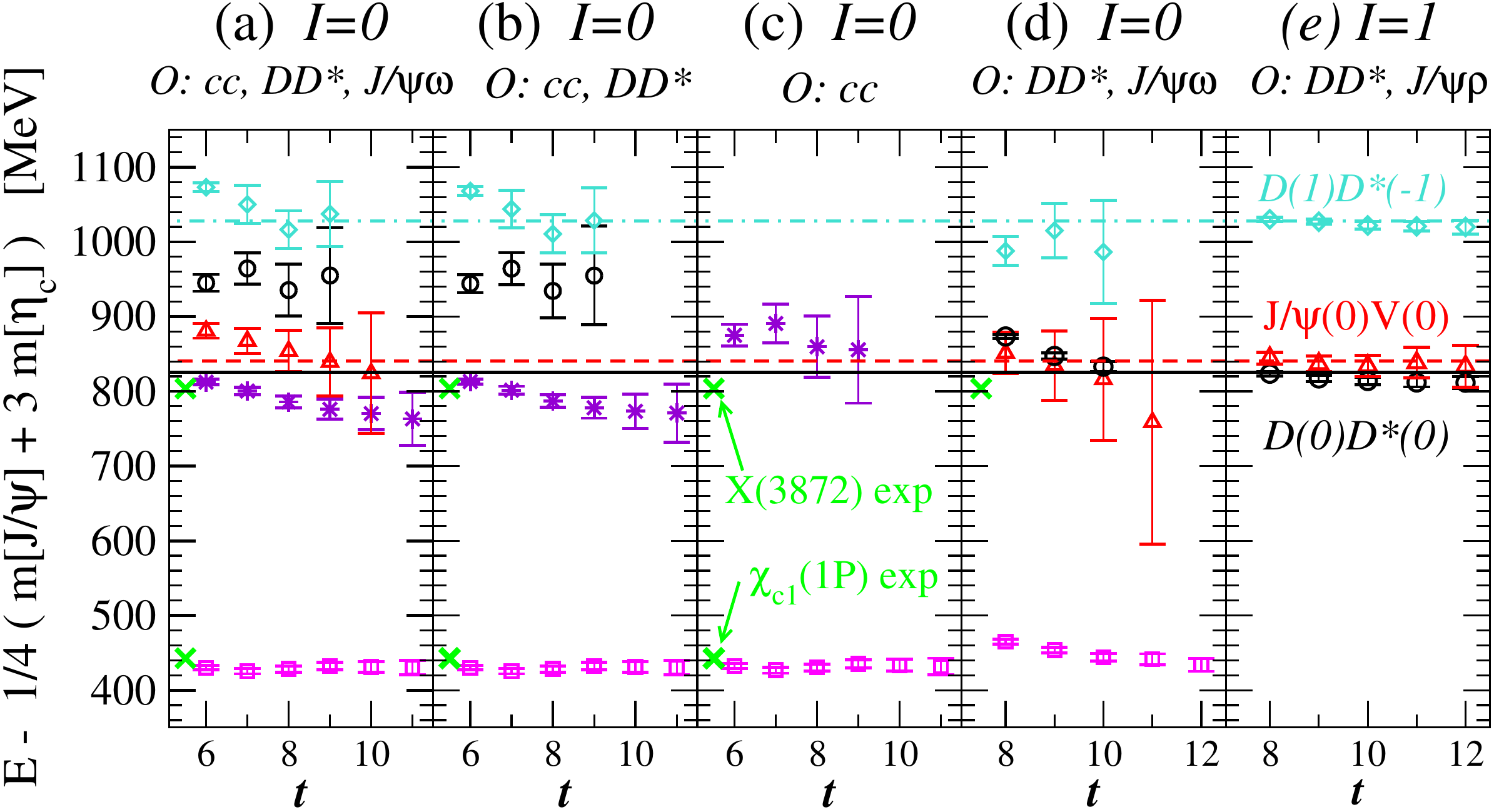}
\end{center}
\caption{The spectrum $E_n-\tfrac{1}{4}(m_{\eta_c}+3m_{J/\psi})$ in
the effective mass plateau region for the $J^{PC}=1^{++}$ channel
with $I=0$ and $I=1$, from Ref.~\cite{Prelovsek:2013cra}. Different
interpolator basis is chosen for each plot. Dashed lines represent
energies $E^{n.i.}$ of the non-interacting scattering
states.}\label{fig:44X3872}
\end{figure*}

In lattice QCD, this picture was supported by the dynamical $N_f =
2$ lattice simulation in Ref.~\cite{Prelovsek:2013cra}, which was
based on one ensemble of Clover-Wilson dynamical configurations with
$m_\pi=266$ MeV. The authors identified the low-lying
$\chi_{c1}(1P)$ and $X(3872)$ as well as the nearby discrete
scattering levels $D\bar{D}^*$ and $J/\psi\omega$ for both $I=0$ and
$I=1$ channels. They counted the number of lattice states near the
$D\bar{D}^*$ threshold in order to establish the existence of the
$X(3872)$. In their simulation, they chose the interpolating fields
$\mathcal{O}_i$ that couple to $\bar cc$ as well as the scattering
states, i.e., $\mathcal{O}^{\bar c c}_{1-8}$, $\mathcal{O}^{D
D^*}_i(i=1,2,3)$, $\mathcal{O}^{J/\psi \omega}$ (for $I=0$), and
$\mathcal{O}^{J/\psi\rho}$ (for $I=1$). Finally, they found a
candidate for the charmoniumlike $X(3872)$ state $11\pm 7$ MeV below
the $D\bar D^\ast$ with $J^{PC} = 1^{++}$ and $I = 0$, in addition
to the nearby $D \bar D^*$ and $J/\psi \omega$ discrete scattering
states, as shown in Fig.~\ref{fig:44X3872}. They extracted large and
negative $D \bar D^*$ scattering length, $a^{DD^*}_0 = -1.7 \pm 0.4$
fm, and the effective range, $r^{DD^*}_0 = 0.5\pm0.1$ fm. Moreover,
they found the $\chi_{c1}(1P)$ state but no candidate for the
$X(3872)$ in the $I=0$ channel if they used only five scattering
interpolators but no $\mathcal{O}^{\bar c c}_{1-8}$ interpolating
fields. In the $I =1$ channel, they did not find a candidate for
$X(3872)$, which is consistent with the assignment of the state in
PDG~\cite{Tanabashi:2018oca}.

Later, the isoscalar $J^{PC} = 1^{++}$ charmonium state $X(3872)$
was also investigated by the Fermilab Lattice and MILC
Collaborations on gauge field configurations with $2+1+1$ flavors of
highly improved staggered sea quarks (HISQ) with clover (Fermilab interpretation) charm quarks and HISQ light valence
quarks~\cite{Lee:2014uta}. They used a combination of the $c\bar c$
and $D\bar D^\ast+\bar DD^\ast$ interpolating operators. They found
an isosinglet candidate of $X(3872)$ with an energy level 13(6) MeV
below the $D\bar D^\ast$ threshold, which agrees with the result in
Ref.~\cite{Prelovsek:2013cra}.

These studies were then extended by Padmanath, Lang, and Prelovsek
by utilizing a large basis of the $\bar cc$, two-meson and
diquark-antidiquark interpolating fields (altogether 22
interpolators)~\cite{Padmanath:2015era}. The Wick contractions were
considered including both the connected contraction diagrams and the
diagrams for annihilation channels in which the light quarks do not
propagate from source to sink. Their simulation was performed with
$N_f=2$ and $m_\pi=266$ MeV, which aims at the possible
charmonium-like $J^{PC} = 1^{++}$ signatures for both $I=0$ and
$I=1$ channels. They found a lattice candidate for $X(3872)$ with
$I=0$ close to the experimental data only if both the $\bar cc$ and
$D \bar D^*$ interpolators are included. However, they found no
candidate if the diquark-antidiquark and $D \bar D^*$ interpolators
were used in the absence of $\bar cc$, which indicates that the
$\bar cc$ Fock component is crucial for the $X(3872)$ while the
four-quark component $O^{4q}$ alone does not produce the signal.
Their simulation also implies a combined dominance of the $\bar c c$
and $D \bar D^*$ operators in determining the position of the energy
levels, while the tetraquark $O^{4q}$ operators do not significantly
affect the results. Moreover, their results do not support the
neutral or charged $X(3872)$ in the $I = 1$ channel. And no
signature of the exotic $\bar cc\bar ss$ state (candidate for the
$Y(4140)$) with $I = 0$ are found below 4.2 GeV. We refer readers to
Refs.~\cite{Baru:2013rta,Jansen:2013cba,Garzon:2013uwa,Albaladejo:2013aka}
for the analytic investigations considering the quark mass
dependence, the volume dependence and the effect from the isospin
breaking.

The unphysical pion mass $m_\pi=266$ MeV on the lattice is still
much larger than its physical mass 140 MeV. Within the molecular
scheme, the long range one-pion-exchange force plays a dominant role
in the formation of the loosely bound molecular state
\cite{Li:2012cs,Zhao:2014gqa}, which decays exponentially as the
pion mass increases. The present lattice simulation with the pion
mass $m_\pi=266$ MeV was obtained from a rather small lattice
volume, which has already indicated the important role of both the
$\bar c c$ and $D \bar D^*$ components in the formation of the
$X(3872)$ signal on the lattice. As the pion mass approaches the
physical value 140 MeV, one shall expect a larger $D \bar D^*$
component within the physical $X(3872)$ state.

\subsection{The charged $Z_c$ states}

The observations of the charged charmonium-like states $Z_c(3900)$
and $Z_c(4020)$ trigged lots of investigations of the molecular
interpretations for their inner
structures~\cite{Zhao:2014gqa,Liu:2008tn,Sun:2012zzd,He:2014nya}. In
Ref.~\cite{Prelovsek:2013xba}, Prelovsek and Leskovec searched for
the $Z_c(3900)$ in the $J^{PC} = 1^{+-}$ and $I = 1$ channel in
their lattice simulation with degenerate dynamical $u/d$ quarks and
$m_\pi\approx266$ MeV. They used six meson-meson type of
interpolators $O_i^{DD^\ast}$ and $O_i^{J/\psi\pi} (i=1, 2, 3)$.
They calculated the time-dependence of the $6\times6$ correlation
matrix by considering the connected diagrams of Wick contraction.
They omitted the diagrams involving charm annihilation, which effect
was proven to be suppressed due to the OZI rule for the conventional
charmonium~\cite{Levkova:2010ft}. They observed four energy levels
which almost exactly coincide with the energies of four
non-interacting $D \bar D^*$ and $J/\psi \pi$ scattering states.
They did not find any additional energy level relating to the
$Z_c(3900)$ state. They suggested to perform simulations including
meson-meson interpolators as well as some different type of
interpolators, such as diquark-antidiquark interpolators with the
same quantum numbers.

Such a simulation was performed in Ref.~\cite{Prelovsek:2014swa} by
using 18 meson-meson $O^{M_1M_2}$ interpolating operators and 4
diquark-antidiquark $O^{4q}$ interpolators with structure $[\bar
c\bar d]_{3_c}[cu]_{\bar 3_c}$ in the $I^GJ^{PC}=1^+1^{+-}$ channel.
The two-meson operators $O^{M_1M_2}$ included interpolators such as
$J/\psi\pi$, $\eta_c\rho$, $D\bar D^\ast$ and $D^\ast\bar D^\ast$.
They aimed to extract and identify all 13 two-meson energy levels
from the corresponding $22\times22$ correlation functions and
examine the possible additional states related to the exotic $Z_c^+$
hadron. As a result, they found all expected lowest thirteen levels,
appearing near the corresponding non-interacting energies of the
two-particle states. A two-particle level will disappear from the
spectrum or become very noisy when the corresponding $O^{M_1M_2}$
operator is absent from the correlator matrix. However, the energy
spectrum was unaffected after the $O^{4q}_{1-4}$ operators were
omitted. Besides these two-particle levels, they did not find any
additional state below 4.2 GeV relating to an exotic $Z_c^+$
candidate. They listed some possible reasons for the absence of the
$Z_c^+$ candidate: a) the experimental resonance $Z_c^+(3900)$ might
not be of dynamical origin since it has not been seen in the $B$
meson decay process; b) it might be a couple-channel threshold
effect~\cite{Chen:2013coa,Swanson:2014tra}; c) their simulation was
performed at a large unphysical $m_\pi\approx266$ MeV; d) their
interpolator basis may not be complete enough to render a $Z_c$
state in addition to thirteen two-meson states; e) the S-D mixing
effect may be important for creating the experimental $Z_c$ states.
In Ref.~\cite{Lee:2014uta}, the authors used the Highly Improved
Staggered Quark action to search for the $Z_c(3900)$ with a
combination of the $J/\psi\pi$ and $D\bar D^\ast+\bar DD^\ast$
channels. Again, they found no evidence for the existence of
$Z_c(3900)$. A similar analysis was performed in
Ref.~\cite{Guerrieri:2014nxa} by considering a basis of five $I=0$
$D^{(\ast)}\bar D^\ast$ interpolating operators and three $I=1$
$D\bar D^\ast$ operators with the same $J^P=1^+$. Their results
didn't show any unknown energy level.

These negative results of searching for the exotic charged $Z_c$ in
lattice QCD simulations were also supported by the investigations of
the resonance-like $Z_c$ structures in the low-energy scattering of
the $(D^{(\ast)}\bar D^\ast)^\pm$ systems by the CLQCD
Collaboration~\cite{Chen:2014afa,Chen:2015jwa}. In
Ref.~\cite{Chen:2014afa}, Chen {\it et al.} presented an exploratory
lattice study of the low-energy scattering of the $(D\bar
D^\ast)^\pm$ two-meson system by using single-channel L\"uscher's
finite-size technique. The calculation was based on $N_f=2$ twisted
mass fermion configuration of size $32^3\times64$ with a lattice
spacing of about $0.067$ fm with three pion mass values $m_\pi=300$
MeV, 420 MeV and 485 MeV. They constructed the $S$-wave two-meson
operators with $I^GJ^{PC}=1^+1^{+-}$ to calculate the correlation
functions. Since the energy being considered is very close to the
$(D\bar D^\ast)^\pm$ threshold, they computed the scattering length
$a_0$ and effective range $r_0$ to search for the evidence of the
existence of $Z_c(3900)$. They found negative values of the
scattering length $a_0$ for all three pion masses, which indicated a
weak repulsive interaction between the two mesons $D$ and $\bar
D^*$. They also checked the possibility of the bound state for the
negative energy shifts. None of them was consistent with the signal
of a bound state. Based on their results, they concluded that their
lattice QCD simulation did not support a bound state in the
$I^GJ^{PC}=1^+1^{+-}$ channel corresponding to $Z_c(3900)$, at least
for the pion mass values being studied. Later in
Ref.~\cite{Chen:2015jwa}, they performed a similar investigation for
the low-energy scattering of the $D^*\bar D^*$ two-meson system in
the same channel with $I^GJ^{PC}=1^+1^{+-}$. Their results gave
negative values of the scattering lengths for all three pion masses,
which again indicated a weak repulsive interaction between the two
vector charmed mesons, and did not support a bound state of the two
mesons in the $J^{P}=1^+$ channel corresponding to the
$Z_c(4020)/Z_c(4025)$ state.

\begin{figure}[!th]
\begin{center}
\includegraphics[width=0.6\textwidth,clip]{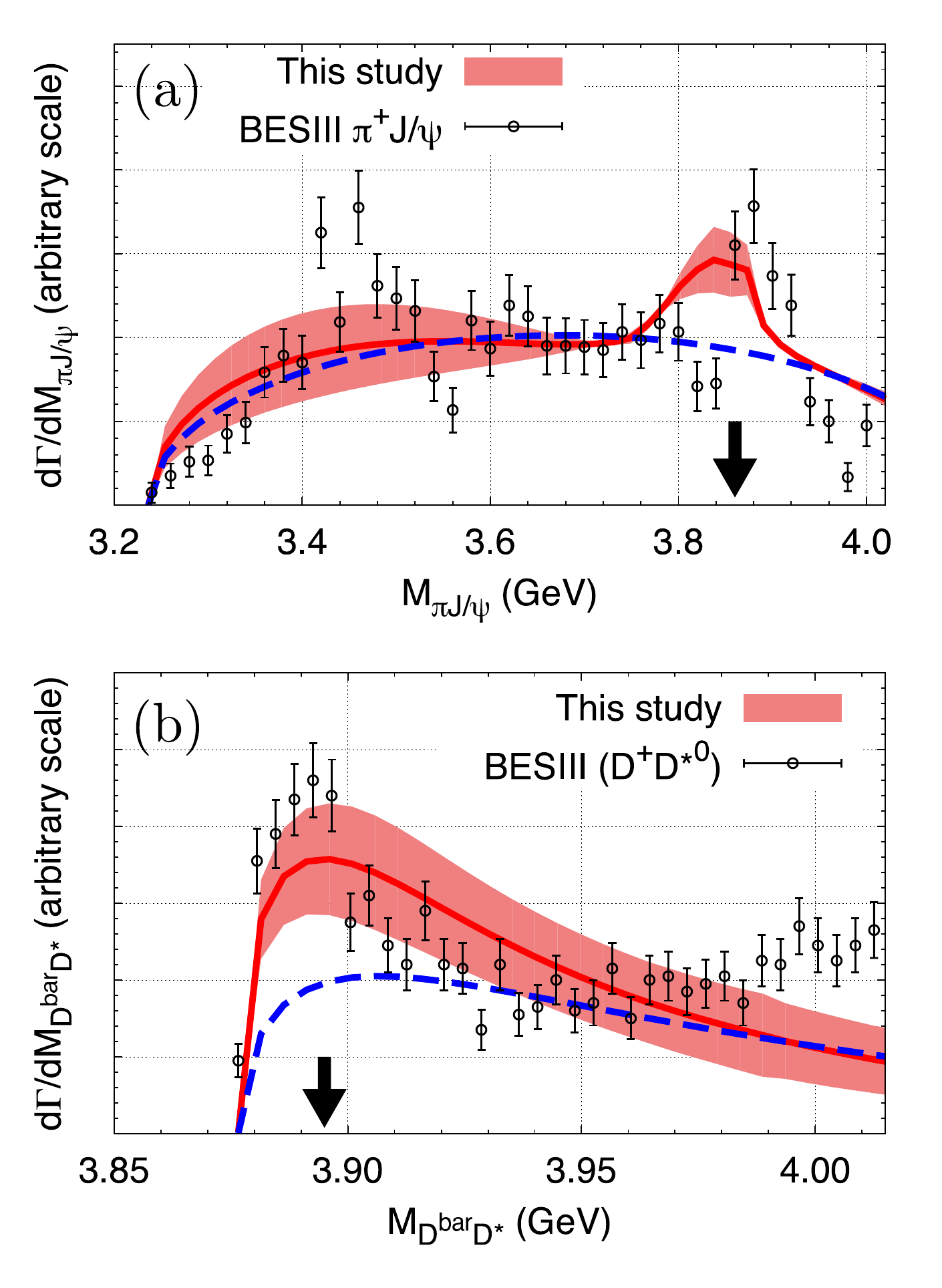}
\end{center}
\caption{ The invariant mass spectrum of (a) $Y(4260) \to \pi \pi
J/\psi$ and (b)  $Y(4260) \to \pi \bar{D}D^{*}$ established with the
coupled-channel potential $V^{\alpha \beta}$ from the lattice QCD
simulation at $m_\pi=411$ MeV, taken from Ref.~\cite{Ikeda:2016zwx}.
The vertical black arrows show the reproduced peak positions from
the lattice QCD calculations.The blue dashed lines show the
invariant mass spectra without the off-diagonal components of
$V^{\alpha \beta}$.} \label{44Zc3900}
\end{figure}

The lattice QCD studies of the spectrum and the low-energy
scattering parameters with the standard L\"uscher's method gave no
candidates for the $Z_c(3900)$ and $Z_c(4025)$ in all $J/\psi\pi$,
$\eta_c\rho$, $D\bar D^\ast$ and $D^\ast\bar D^\ast$ two-meson
channels. These results indicated that the $Z_c(3900)$ and
$Z_c(4025)$ may not be the conventional resonance states. In
Refs.~\cite{Ikeda:2016zwx,Ikeda:2017mee}, the HAL QCD Collaboration
studied the $\pi J/\psi-\rho\eta_c-\bar DD^\ast$ coupled-channel
interactions from $(2+1)$-flavor full QCD simulations at three pion
masses $m_\pi=411$ MeV, 570 MeV, 701 MeV in order to explore the
structure of $Z_c(3900)$. To consider the interactions among the
$\pi J/\psi$, $\rho\eta_c$ and $\bar DD^\ast$ channels, the authors
calculated the $S$ matrix from the equal-time Nambu-Bethe-Salpeter
(NBS) wave functions by using the coupled-channel HAL QCD
method~\cite{Aoki:2011gt,Aoki:2012bb,Sasaki:2015ifa}. They extracted
the $s$-wave coupled-channel potential $V^{\alpha\beta}$ and found
that all diagonal potentials $V^{\bar DD^\ast, \, \bar DD^\ast}$,
$V^{\rho\eta_c, \, \rho\eta_c}$ and $V^{\pi J/\psi, \, \pi J/\psi}$
were very weak, which indicated that the $Z_c(3900)$ was neither a
simple $\pi J/\psi$ hadrocharmonium nor $DD^\ast$ molecule state.
Instead, they found the $\rho\eta_c-\bar DD^\ast$ coupling and the
$\pi J/\psi-\bar DD^\ast$ coupling from the off-diagonal potentials
were both strong, which implied that the structure of $Z_c(3900)$ can
be explained as a threshold cusp. They further calculated the
invariant mass spectra and pole positions associated with the
coupled-channel two-body $T$ matrix on the basis of
$V^{\alpha\beta}$. Their results supported the $Z_c(3900)$ resonance
as a threshold cusp induced by the strong $\pi J/\psi-\bar DD^\ast$
coupling. They further made a semiphenomenological analysis of the
three-body decays $Y(4260)\to\pi\pi J/\psi, \pi\bar DD^\ast$
processes. As shown in Fig.~\ref{44Zc3900}, they found that the
coupled-channel potential $V^{\alpha\beta}$ can well reproduce the
experimental peak structures in the $Y(4260)\to\pi\pi J/\psi$ and
$Y(4260)\to\pi\bar DD^\ast$ decays. When the off-diagonal components
of $V^{\alpha\beta}$ were turned off, the peak structures at 3.9 GeV
disappeared in both cases. Accordingly, the authors concluded that
the $Z_c(3900)$ was not a conventional resonance but a threshold
cusp~\cite{Ikeda:2016zwx,Ikeda:2017mee}.

\begin{figure}[!th]
\begin{center}
\includegraphics[width=0.8\textwidth,clip]{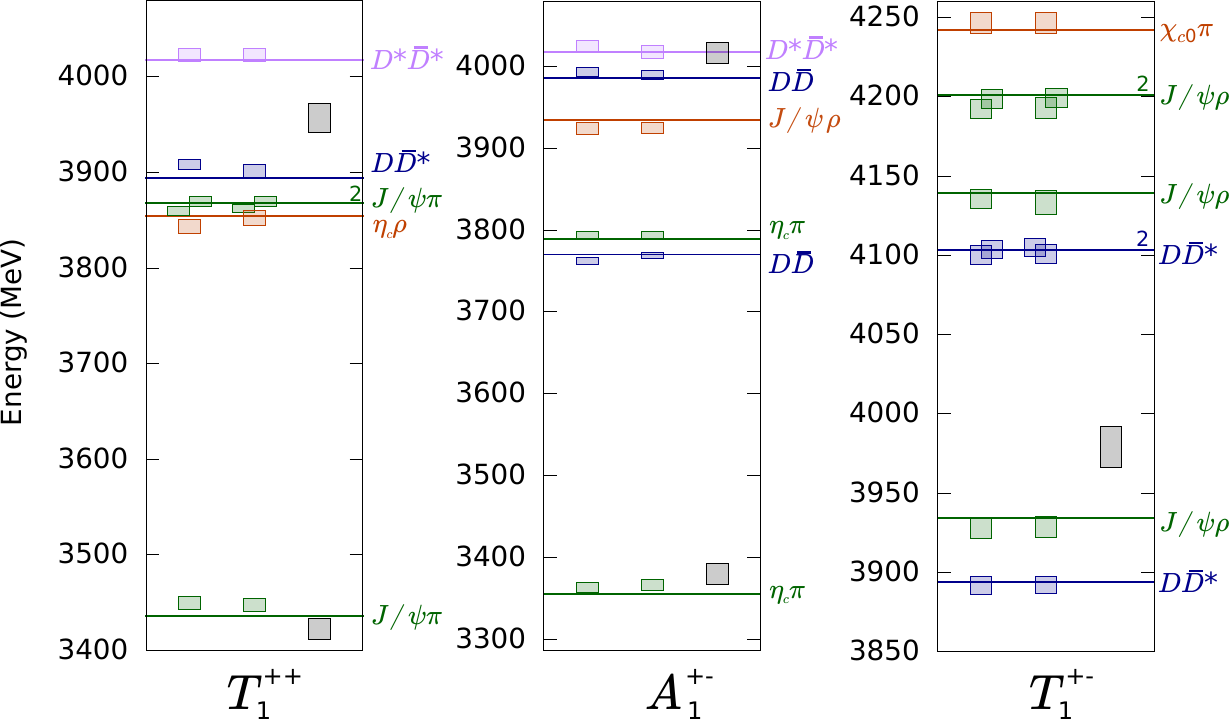}
\end{center}
\caption{ Figure was taken from Ref.~\cite{Cheung:2017tnt}. The
spectrum plot in the hidden-charm isospin-1 $\Lambda^{PG}=T_1^{++},
A_1^{+-}, T_1^{+-}$ lattice irreps calculated using the full basis
of meson-meson and tetraquark operators (left column), only
meson-meson operators (middle column) and only tetraquark operators
(right column). Horizontal lines denote the non-interacting
meson-meson energy levels, and boxes give the lattice QCD computed
energies.} \label{44cqcbarqbarplots}
\end{figure}

As indicated in Refs.~\cite{Prelovsek:2013xba,Prelovsek:2014swa},
including a large basis of four-quark interpolating operators in the
calculations could lead to more reliable determination of
finite-volume mass spectra and scattering amplitudes. In the very
recent lattice QCD study~\cite{Cheung:2017tnt}, the Hadron Spectrum
Collaboration constructed 29 hidden-charm interpolating currents and
48 doubly-charmed interpolating currents with different
color-flavor-spatial-spin structures. Using these diverse bases of
meson-meson operators as well as compact tetraquark operators, they
computed the two-point correlation functions in the isospin-1
hidden-charm and doubly-charmed sectors using the distillation
framework~\cite{Peardon:2009gh}. Their calculations were performed
on an anisotropic $16^3\times128$ lattice volume using 478
configurations and a Clover fermion action with $N_f=2+1$ flavors of
dynamical quarks. The mass of the two degenerate light quarks
corresponds to $m_\pi=391$ MeV while the strange quark was tuned so
that its mass approximated the physical value~\cite{Edwards:2008ja}.
The multiple energy levels associated with the meson-meson levels
which could be degenerate in the non-interacting limit were
extracted reliably. In Fig.~\ref{44cqcbarqbarplots}, the mass
spectra in the isospin-1 hidden-charm sector were shown for the
lattice irreps $\Lambda^{PG}=T_1^{++}, A_1^{+-}, T_1^{+-}$
corresponding to the $I=1$ and $J^{PC}=1^{+-}, 0^{++}, 1^{++}$ for
the charge neutral channels, respectively. One found that every
energy level had a dominant overlap onto one meson-meson operator,
while the addition of tetraquark operators to a basis of meson-meson
operators did not significantly affect the finite-volume spectrum.
For all these channels, the number of energy levels in the computed
spectra was equal to the number of non-interacting two-meson levels
expected in the considered energy region and they all lie close to
the non-interacting levels. The authors concluded that there were no
strong indications for any bound state or narrow resonance in these
channels~\cite{Cheung:2017tnt}. Especially for the isovector
$T_1^{++}$ irreps with $J^{PC}=1^{+-}$, they didn't find any
candidate for the $Z_c(3900)$ and $Z_c(4025)$ states. Since the
$X(3872)$ has been admitted as an
isosinglet~\cite{Tanabashi:2018oca}, this study gave no hint for its
existence. However, if the $X(3872)$ exists as a tetraquark
~\cite{Prelovsek:2013cra,Lee:2014uta,Padmanath:2015era}, its
isospin-1 partner would appear in this channel with
$I^GJ^{PC}=1^+1^{+-}$. However, no clear signal for such an isovector
hidden-charm tetraquark state was found in
Ref.~\cite{Cheung:2017tnt}.

Besides the studies of $Z_c(3900)$ and $Z_c(4025)$ in low-energy
scattering of the $(D^{(\ast)}\bar D^\ast)^\pm$ systems, the CLQCD
Collaboration also investigated the low-energy near threshold
scattering of the $(\bar D_1D^\ast)^\pm$ system in both the $S$-wave
($A_1$ with $I^GJ^{PC}=1^+0^{--}$) and $P$-wave ($T_1$ with
$I^GJ^{PC}=1^+1^{+-}$) channels~\cite{Meng:2009qt,Chen:2016lkl}. In
Ref.~\cite{Chen:2016lkl}, the authors performed their calculations
by using lattice QCD with $N_f=2$ twisted mass fermion
configurations of size $32^3\time64$ with lattice spacing
$a\approx0.067$ fm at three pion masses $m_\pi=307.0$ MeV, 423.6
MeV, 488.4 MeV. For the $\bar D_1D^\ast$ scattering in the $S$-wave
$A_1$ channel, they obtained the values of the lowest $q^2$ to be in
the range $[-0.7,\, -0.5]$, which increased by an order of magnitude
compared with their previous quenched lattice QCD result in
Ref.~\cite{Meng:2009qt}. This resulted in stronger attractive
interaction between the two charmed mesons. The phase shift was also
checked to satisfy $\cot\delta(q^2)\approx-1$, which confirmed the
existence of a shallow bound state in this channel. In the $P$-wave
$T_1$ channel, similar conclusions were reached by inspecting the
lowest values of $q^2$ and the quantity $\cot\delta(q^2)$. Based on
these results, the author concluded that the interactions between a
$(\bar D_1D^\ast)^\pm$ system are attractive and a possible bound
state below the threshold may exist for both pseudoscalar and
axial-vector channels. Especially, the resonance candidate in the
$I^GJ^{PC}=1^+1^{+-}$ channel may provide some hints on the nature
of the $Z^+(4430)$ state~\cite{Chen:2016lkl}.

\subsection{The doubly-charmed/bottom tetraquark states}

The idea of searching for stable tetraquark states from the first
principle lattice QCD has a long history and is still attractive
now. In 1990's, there were some efforts to calculate the
interactions and potential between two heavy-light mesons in lattice
QCD~\cite{Richards:1990xf,Mihaly:1996ue,Stewart:1998hk,Michael:1999nq,Pennanen:1999xi,Green:1999mf},
inspired by many phenomenological studies of the stability for these
systems~\cite{Ader:1981db,Zouzou:1986qh,Lipkin:1986dw,Carlson:1987hh,Richard:1990wf,Manohar:1992nd,SilvestreBrac:1993ry,Bander:1994sp,Moinester:1995fk,Pepin:1996id,Brink:1998as}.

In Ref.~\cite{Mihaly:1996ue}, the authors extracted an effective
potential between the heavy-light mesons (HLHL systems) from the
quark correlation functions in the framework of quenched lattice QCD
with Kogut-Susskind fermions. The correlation functions were
calculated by including the two kinds of diagrams corresponding to
the pure gluon-exchange part and the quark-exchange part of the
mesonic interactions. The gauge field configurations were generated
on a periodic $8^3\times16$ lattice with inverse gauge coupling
$\beta=5.6$ corresponding to a lattice spacing $a\approx0.19$ fm.
Finally, they were able to extract the $MM$ potential from the
Euclidean time behavior of these correlators. The result showed that
the resulting potential was attractive at short heavy-quark
distances and the interaction was stronger for smaller light quark
masses. They also found that the quark-exchange diagram played a
significant role only for distance $r$ much less than $2a$. This
result was supported by the investigation in
Ref.~\cite{Stewart:1998hk}, where an adiabatic approximation was
used to derive the binding energy potential between two heavy-light
mesons in quenched SU(2)-color lattice QCD. The derived binding
potential was attractive at short and medium range. The UKQCD
Collaboration studied the $BB$ system at fixed heavy quark
separation $R$ in quenched~\cite{Michael:1999nq} and
unquenched~\cite{Pennanen:1999xi} SU(3) lattice QCD. They found
evidence for deep binding at small $R$ in the $I_q, \, S_q=(0, \,
0)$ and $(1, \, 1)$ (where $I_q$ and $S_q$ are total isospin and
spin for the two light quarks, respectively) cases, for both
quenched and unquenched results. The binding energy was about
$200-400$ MeV at $R=0$ and very short-ranged. It was essentially
contributed by the gluon-exchange diagram and insensitive to the
light quark mass. The $(0, \, 1)$ channel at $R=0$ was attractive
for unquenched and repulsive for quenched, which was the only
difference between the quenched and unquenched results. The
quark-exchange diagram contributed at larger separations
$R\approx0.5$ fm, where the evidence for weak binding was found. The
authors concluded that the exotic $bb\bar q\bar q$-mesons exist as
states stable under strong interactions.

However, the extracted potentials remain largely unexplored in the
above lattice calculations due to the large statistical
uncertainties. In Ref.~\cite{Detmold:2007wk}, the NPLQCD
Collaboration studied the potentials between two $B$ mesons in the
heavy-quark limit by choosing a relatively small lattice volume in
order to explore the intermediate and short-distance component of
the potential. Their calculations were performed on $16^3\times32$
quenched lattices with a spatial length of $\sim1.6$ fm at a pion
mass $m_\pi\approx400$ MeV. They found nonzero central potentials in
all four $(I,\, s_l)=(0, \, 0), \, (0, \, 1), \, (1, \, 0), \, (1,
\, 1)$ spin-isospin channels, where $s_l$ is the total spin of the
light quarks. At short distance, they found clear evidence of
repulsion between the $B$ mesons in the $I\neq s_l$ channels and
attraction in the $I=s_l$ channels. Later, the possible tetraquark
bound states in HLHL system were suggested in
Refs.~\cite{Bali:2010xa,Brown:2012tm}, based on their lattice QCD
calculations of the interaction potentials for various channels in
the heavy-quark limit.

\begin{figure}[!th]
\begin{center}
\includegraphics[width=0.6\textwidth,clip]{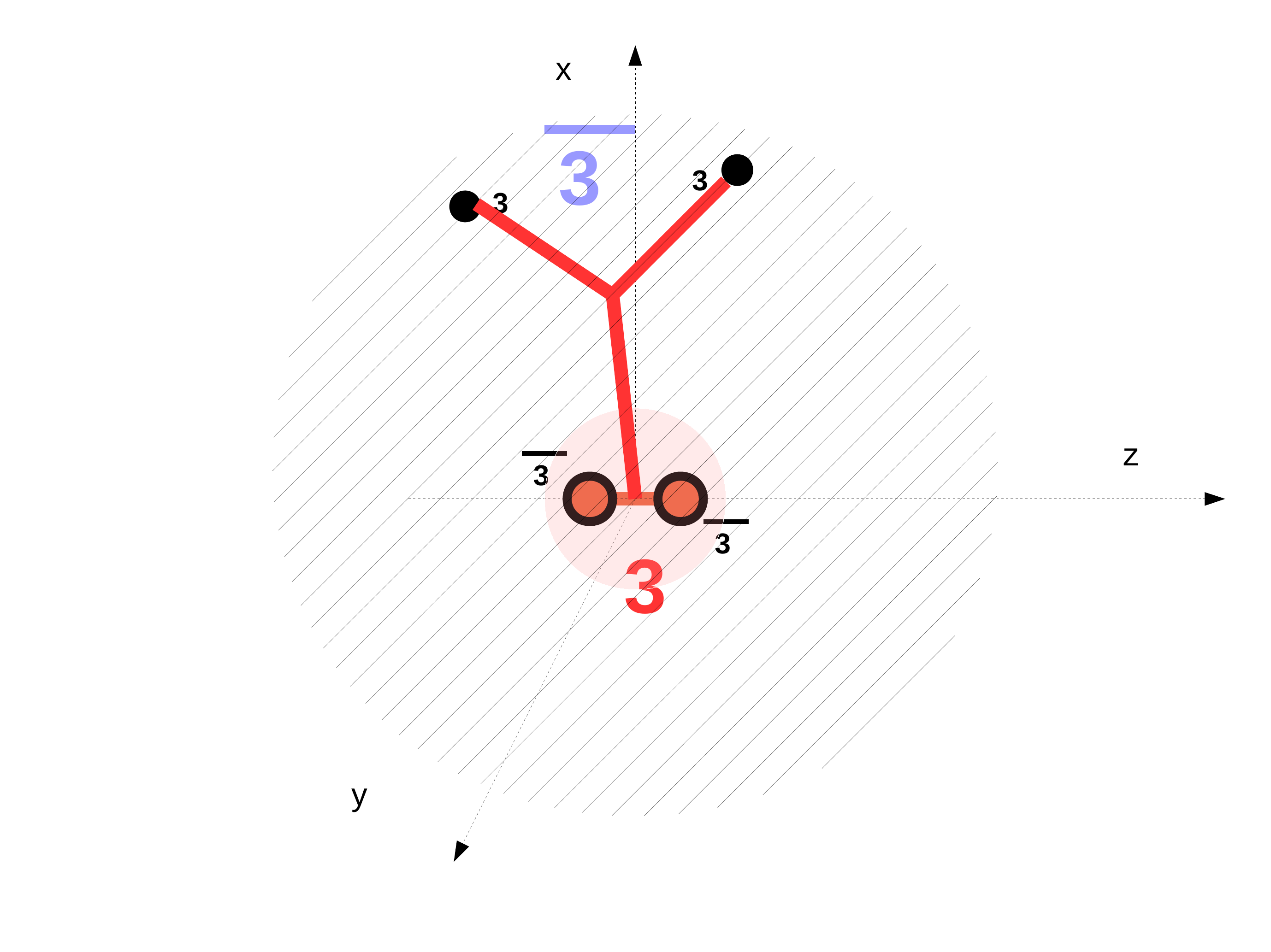}
\end{center}
\caption{The $\bar b\bar b$ pair forms a color triplet antidiquark
at small separation $r$. The screening of the light quarks has
essentially little effect on the $\bar b\bar b$ interaction, due to
the much farther separated light quarks $qq$. Figure was taken from
Ref.~\cite{Bicudo:2015vta}.} \label{44TbbScreening}
\end{figure}

Besides, the $BB$ potential was also computed by using the Wilson
twisted mass lattice QCD with two flavors of degenerate dynamical
quarks ($m_{\pi}\approx340$ MeV)~\cite{Wagner:2010ad,Wagner:2011ev}.
The authors calculated 36 independent trial state potentials using
$24^3\times48$ gauge field configurations. They established a simple
rule to judge whether a $BB$ potential is attractive or repulsive: a
$BB$ potential is attractive if the trial state is symmetric under
the meson exchange, while the potential is repulsive if the trial
state is antisymmetric under the meson exchange, where the meson
exchange means the combined exchange of flavor, spin and parity.
Using the dynamical results obtained in
Refs.~\cite{Wagner:2010ad,Wagner:2011ev}, Bicudo and Wagner found
strong indication for the existence of a $\bar b\bar bqq$ tetraquark
bound state~\cite{Bicudo:2012qt}. Assuming that the pair of
antibottom quarks was immersed in a cloud of two light quarks whose
screening has little effect on the $\bar b\bar b$ interaction, the
authors suggested the following ansatz to model the heavy
antiquark-antiquark potential:
\begin{equation}
V(r)=-\frac{\alpha}{r}e^{-(r/d)^p}\, , \label{Sec44:Screeningansatz}
\end{equation}
where $r$ is the separation of two antibottom quarks, $d$
characterizes the size of the $B$ meson and $p$ is an exponent. They
focused on a scalar isosinglet and a vector isotriplet tetraquark
channels, which were found to have attractive interactions between
two static-light mesons in the dynamical lattice
simulations~\cite{Wagner:2010ad,Wagner:2011ev}. Fitting the ansatz
Eq.~\eqref{Sec44:Screeningansatz} to the lattice results for the
heavy antiquark-antiquark interactions, the authors were able to
determine all three parameters $\alpha, d$ and $p$ for the scalar
isosinglet while only two parameters $\alpha$ and $d$ for the vector
isotriplet. They derived an analytical rule for the existence or
non-existence of a bound tetraquark state for the radial equation in
three dimensions. Applying this rule to angular momentum $l=0$, they
found the condition for having at least one bound state
\begin{equation}
\mu\alpha d\geq\frac{9\pi^2}{128\times2^{1/p}\Gamma^2(1+1/2p)}\, ,
\label{Sec44:condition}
\end{equation}
where $\mu$ is the reduced antibottom quark mass. According to this
condition, the authors only found a strong indication for the
existence of a bound state in the isoscalar channel. To investigate
the existence of this bound state rigorously, they numerically
solved the Schr\"odinger equation to calculate the binding for heavy
$\bar b\bar bqq$ tetraquarks. They confirmed the existence of this
$\bar b\bar bqq$ tetraquark state with a confidence level of around
$1.8\sigma$...$3.0\sigma$ and binding energy of approximately 30
MeV...57 MeV. For the vector isotriplet channel, however, both the
analytical rule and binding energy obtained from solving
Schr\"odinger equation indicated that the existence of a bound state
in this channel is rather questionable. Nevertheless, possible
sources of systematic errors should be considered, including the
choice of the heavy quark mass and the lattice spacing, the lattice
QCD finite volume effects, the unphysically heavy light quark
masses.

In Ref.~\cite{Bicudo:2015vta}, they extended their computations for
the light quark combinations $qq \in \{ (ud-du)/\sqrt{2} \ , \ uu ,
(ud+du)/\sqrt{2} , dd \}$~\cite{Bicudo:2012qt} to similar systems
with heavier strange and charm quarks, i.e.\ $qq=ss$ and $qq=cc$, by
investigating and quantifying systematic uncertainties in detail.
They confirmed the existence of the isoscalar $ud\bar b\bar b$
tetraquark state with quantum numbers $I(J^P)=0(1^+)$ and no bound
tetraquark state in the isotriplet channel with $I(J^P)=1(0^+),
1(1^+), 1(2^+)$. They found no $ss\bar b\bar b$ and $cc\bar b\bar b$
tetraquarks exist in any channel. Later they considered the
light-quark mass dependence and computed the $BB$ potential as a
function of their spatial separation $r$, by using the twisted mass
lattice QCD (tmQCD) with a lattice spacing around 0.079 fm and three
infinite-volume pion mass $m_\pi\approx340$ MeV, 480 MeV, 650
MeV~\cite{Bicudo:2015kna}. This allowed for an extrapolation of the
potentials to the physical pion mass, showing the tendency that the
binding in the scalar isosinglet channel became stronger towards the
physical pion mass, while the pion mass dependence was relatively
mild close to the physical point. Accordingly, the authors concluded
that the two $B$ mesons could form a tetraquark state in the scalar
isosinglet channel with the binding energy $E_B=-90^{+43}_{-36}$
MeV. For the vector isotriplet channel, there was no binding and the
results were essentially independent on the pion mass.

Including the effects of the heavy $\bar b$ quark spins, Bicudo {\it
et. al.} reanalyzed the $ud\bar b\bar b$ tetraquark binding energy
by solving a coupled-channel Schr\"odinger equation with the
Born-Oppenheimer approximation~\cite{Bicudo:2016ooe}. They found
that the spin of heavy $\bar b$ quarks decreased the binding energy.
However, the attraction was sufficiently strong that the previously
predicted $ud\bar b\bar b$ bound tetraquark state persisted in
$I(J^P)=0(1^+)$ channel. Later in Ref.~\cite{Bicudo:2017szl}, a new
tetraquark $ud\bar b\bar b$ resonance with quantum numbers
$I(J^P)=0(1^-)$ was predicted.

However, the above lattice simulations for the $ud\bar b\bar b$
tetraquarks were all performed with $m_\pi>340$ MeV in the static
heavy quark limit. Recently, Francis {\it et. al.} investigated the
possibility of the $ud\bar b\bar b$ and $ls\bar b\bar b$ ($l=u, d$)
tetraquark bound states using $n_f=2+1$ lattice QCD ensembles with
sufficiently low pion masses $m_\pi\approx164, 299$ and 415 MeV and
the near-physical $m_K$~\cite{Francis:2016hui}. The authors
considered one diquark-antidiquark interpolating operator and a
meson-meson operator to calculate the correlation functions. They
used the NRQCD lattice
action~\cite{Manohar:1997qy,Lepage:1992tx,Thacker:1990bm} to
calculate the bottom quark propagators, which avoided the static
approximation for the heavy $\bar b$ quarks. Finally, they found
unambiguous signals for the $J^P=1^+$ tetraquark states in the
$ud\bar b\bar b$ and $ls\bar b\bar b$ channels, which lied 189(10)
and 98(7) MeV below the corresponding free two-meson thresholds.
These $ud\bar b\bar b$ and $ls\bar b\bar b$ tetraquarks can only
decay via the weak interaction, implying that these states are
rather stable. Later in Ref.~\cite{Francis:2018jyb}, they extended
their investigation to study the $qq^\prime\bar b\bar c$ tetraquark
states with the same method. They found the evidence of a stable
$I(J^P)=0(1^+)$ $ud\bar b\bar c$ tetraquark, which lies about
$15-61$ MeV below the corresponding $(\bar DB^\ast)$ threshold.

There were also some studies on the doubly-charmed tetraquark states
in lattice QCD. In Ref.~\cite{Ikeda:2013vwa}, Ikeda {\it et. al.}
investigated the S-wave meson-meson interactions in the $D-D, \bar
K-D, D-D^\ast$ and $\bar K-D^\ast$ systems with both $I=0$ and
$I=1$, by using $(2+1)$-flavor full QCD gauge configurations
generated at $m_\pi=410\sim700$ MeV. They employed the relativistic
heavy-quark action to treat the charm quark dynamics on the lattice.
Using the HAL QCD method, the authors extracted the S-wave
potentials in lattice QCD simulations and then calculated the
meson-meson scattering phase shifts and scattering lengths. Their
results showed that the interactions were repulsive and insensitive
to the pion mass for the $I=1$ channels while they were attractive
and growing with $m_\pi$ for the $I=0$ channels. However, the S-wave
scattering phase shifts in these attractive channels indicated that
no bound states or resonances were formed at the pion mass
$m_\pi=410\sim700$ MeV, particularly in the $I=0$  $D-D^\ast$
channel corresponding to the doubly-charmed tetraquarks $T_{cc}$
with quantum numbers $I(J^P)=0(1^+)$.

\begin{figure}[!th]
\begin{center}
\includegraphics[width=0.8\textwidth,clip]{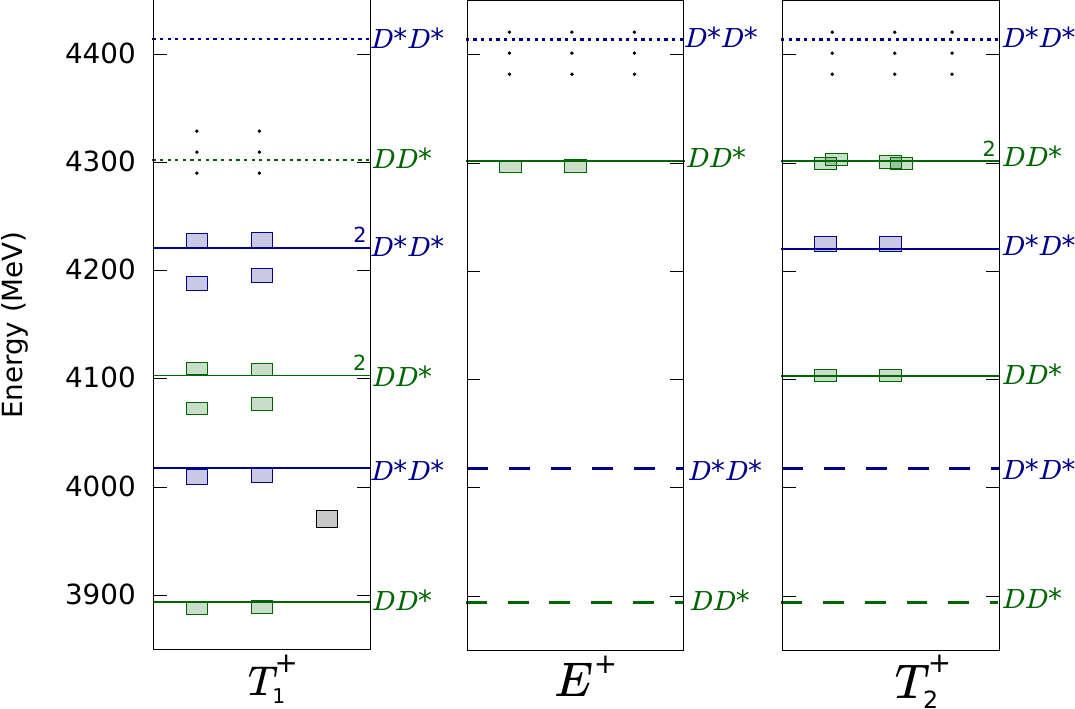}
\end{center}
\caption{ Figure was taken from Ref.~\cite{Cheung:2017tnt}. The
spectrum plot for the isospin-0 doubly-charmed sector with quark
flavor content $cc\bar l\bar l$. The three columns of boxes and the
horizontal lines are the same as in Fig.~\ref{44cqcbarqbarplots}.}
\label{44ccqqplots}
\end{figure}

\begin{figure}[!th]
\begin{center}
\includegraphics[width=0.7\textwidth,clip]{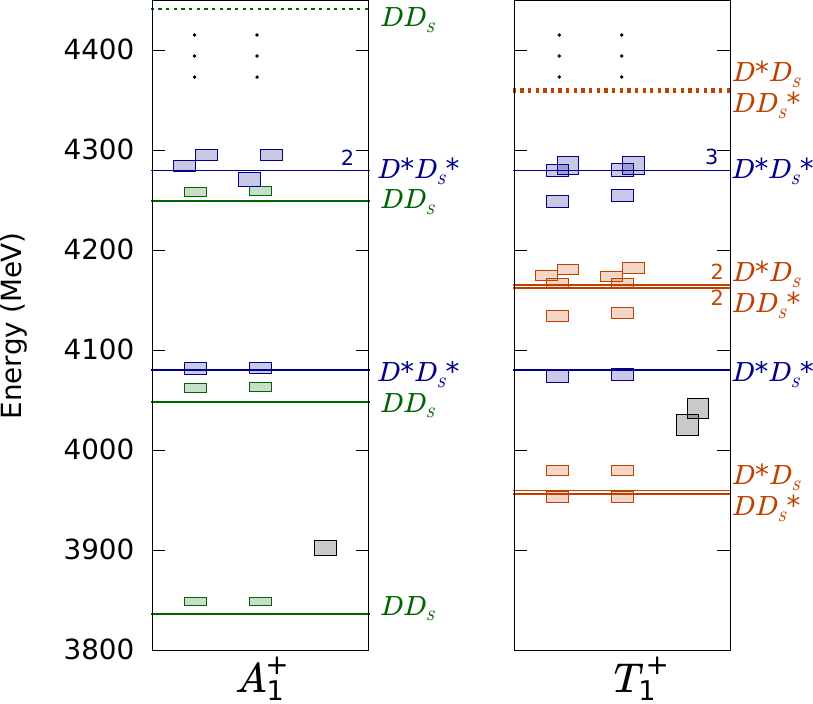}
\end{center}
\caption{ As Fig.~\ref{44ccqqplots} but for the
isospin-$\frac{1}{2}$ doubly-charmed sector with quark flavor
content $cc\bar l\bar s$. Figure was taken from
Ref.~\cite{Cheung:2017tnt}.} \label{44ccqsplots}
\end{figure}

The mass spectra for the doubly-charmed tetraquark states were also
studied by the Hadron Spectrum Collaboration, using 48 interpolating
currents~\cite{Cheung:2017tnt}. Using these diverse bases of the
meson-meson operators as well as compact tetraquark operators, they
calculated the mass spectra for the flavor content $cc\bar l\bar l$
with isospin-0 in the  lattice irreps $\Lambda^{P}=T_1^{+}, E^{+},
T_2^{+}$ corresponding to $J^{P}=1^{+}, 2^{+}, 2^{+}$ respectively
as shown in Fig.~\ref{44ccqqplots}, and mass spectra for the flavor
$cc\bar l\bar s$ with isospin-$\frac{1}{2}$ in the irreps
$\Lambda^{P}=A_1^{+}, T_1^{+}$ corresponding to $J^{P}=0^{+}, 1^{+}$
respectively as shown in Fig.~\ref{44ccqsplots}. Similar to the
results for the isospin-1 hidden-charm sector, the number of energy
levels in the computed spectra for the doubly-charmed sector was
equal to the number of non-interacting two-meson levels and the
majority of energies were at most slightly shifted from the
non-interacting levels. There were no strong indications that there
was any bound state or narrow resonance for the $T_{cc}$ tetraquarks
in these channels~\cite{Cheung:2017tnt}.

A comprehensive lattice QCD calculation of the tetraquark states
with quark contents $q_1q_2\bar{Q}\bar{Q}, \, q_1,q_2 \subset
u,d,s,c$ and $Q \equiv b,c$ in both spin zero ($J=0$) and spin one
($J=1$) sectors was presented on three dynamical $N_f=2+1+1$ highly
improved staggered quark ensembles at lattice spacings of about
$0.12, 0.09$, and $0.06$ fm~\cite{Junnarkar:2018twb}. The authors
employed the overlap fermion action for the light, strange, and charm
quarks~\cite{Neuberger:1997fp,Neuberger:1998wv} while a NRQCD
formulation for the bottom quark~\cite{Lepage:1992tx}. They
performed the standard method for GEVP to obtain the effective
masses of the ground state energy levels. The results of the spin
one tetraquark states showed the presence of the ground state energy
levels which were below their respective thresholds for all $ud\bar
b\bar b, us\bar b\bar b$, $uc\bar b\bar b, sc\bar b\bar b$ and
$ud\bar c\bar c, us\bar c\bar c$ light flavor combinations. The mass
spectra for all spin zero tetraquark ground state were found to lie
above their respective thresholds, strongly disfavor the existence
of any heavy bound tetraquark state with spin zero.

\subsection{Other tetraquark configurations}

After the $X(5568)$ was observed by the D0 Collaboration, Lang {\it
et. al.} investigated the $S$-wave $B_s\pi^+$ scattering using
lattice QCD to search for its existence with the flavor $\bar bs\bar
du$ and spin-parity $J^P=0^+$~\cite{Lang:2016jpk}. They employed
gauge configurations with $N_f=2+1$ dynamical quarks and lattice
spacing $a=0.0907(13)$ fm at rather low pion mass
$m_\pi=162.6(2.2)(2.3)$ MeV. For the strange quark, they used a
partially quenched setup with the valence mass close to the physical
point leading to $m_K=504(1)(7)$ MeV~\cite{Lang:2014yfa}. The bottom
quark was treated as a valence quark using the Fermilab
method~\cite{ElKhadra:1996mp,Oktay:2008ex}. They took into account
the $B^+\bar K^0$ channel for completeness. They didn't find a
candidate for $X(5568)$ with $J^P=0^+$.

In Ref.~\cite{Hughes:2017xie}, the authors studied the low-lying
spectrum of the $bb\bar b\bar b$ system using the lattice
nonrelativistic QCD methodology to search for a stable tetraquark
state below the corresponding lowest noninteracting bottomonium-pair
threshold. They constructed a full $S$-wave basis for the
$1_c\times1_c, 8_c\times8_c$ meson-meson type of interpolating
operators and the $\bar 3_c\times 3_c, 6_c\times\bar 6_c$
diquark-antidiquark type of operators with quantum numbers
$J^{PC}=0^{++}$ coupling to $2\eta_b$ and $2\Upsilon$, the
$J^{PC}=1^{+-}$ coupling to the $\eta_b\Upsilon$ and $J^{PC}=2^{++}$
coupling to $2\Upsilon$. They employed four gluon field ensembles at
lattice spacings ranging from $a=0.06-0.12$ fm, and one ensemble
which had physical light-quark masses. All ensembles had  $u, d, s$
and $c$ quarks in the sea. The results showed no evidence of a
stable tetraquark candidate below the noninteracting
bottomonium-pair thresholds in any channel by studying a full
$S$-wave color-spin basis of QCD operators. To ensure the robustness
of this conclusion, they added an auxiliary scalar potential into
the QCD interactions with the objective of pushing a near threshold
tetraquark increasingly lower than the threshold. As a result, they
found no indication of any state below the noninteracting $2\eta_b$
threshold in the $0^{++}, 1^{+-}$ or $2^{++}$ channel.

\subsection{Pentaquarks}

The charmonium-nucleon scattering at low energy has been studied in
lattice QCD before the LHCb's observation of the hidden-charm
pentaquark states. In Ref.~\cite{Kawanai:2010ev}, Kawanai and Sasaki
studied the charmonium-nucleon potentials for both the $\eta_c-N$
and $J/\psi-N$ systems in quenched lattice QCD. The central and
spin-independent potentials were calculated from the equal-time BS
amplitude through the effective Schr\"odinger equation. They
performed the quenched lattice QCD simulations on two different
lattice sizes with a lattice cutoff of $a^{-1}\approx2.1$ GeV. They
used nonperturbatively $\mathcal{O}(a)$ improved Wilson fermions for
light quarks~\cite{Luscher:1996ug} and a relativistic heavy quark
(RHQ) action for the charm quark~\cite{Aoki:2001ra}. As a result,
the authors found that the all charmonium-nucleon potentials were
weakly attractive at short distances at $m_\pi=640$ MeV, while the
attractive interaction in the $J/\psi-N$ systems were rather
stronger than that of the $\eta_c-N$ system. However, these
potentials were still not strong enough to form a bound state in the
$J/\psi-N$ system. Moreover, the attractive interaction in the
$\eta_c-N$ system tends to get slightly weaker as the light quark
mass decreases, which may imply that the strength of the $\eta_c-N$
potential at the physical point becomes much weaker than that
measured at $m_\pi=640$ MeV. Similar conclusions were also obtained
previously in the preliminary quenched lattice QCD study of the
$\eta_c-N$ and $J/\psi-N$ scattering for $m_\pi=293-598$
MeV~\cite{Yokokawa:2006td}, and in the computation of the scattering
lengths of the charmonium scattering with light hadrons in full QCD
at $m_\pi\approx197$ MeV~\cite{Liu:2008rza}. However, the NPL
Collaboration reported a contradictory result in
Ref.~\cite{Beane:2014sda} that the $\eta_c-N$ system with
$J^P=\frac{1}{2}^-$ has a deeply bound state with binding energy of
19.8(2.6) MeV, by using lattice QCD calculations at very heavy pion
mass $m_\pi=805$ MeV and a single lattice spacing $b=0.145(2)$ fm.

To further discriminate the conflicting results between
Refs.~\cite{Kawanai:2010ev} and~\cite{Beane:2014sda}, Sugiura {\it
et. al.} performed the time-dependent HAL QCD method to study the
charmonium-nucleon effective central interactions in
Ref.~\cite{Sugiura:2017vks}. The authors focused on three $S$-wave
channels $\eta_cN (J^P=\frac{1}{2}^-)$, $J/\psi N
(J^P=\frac{1}{2}^-)$, and $J/\psi N (J^P=\frac{3}{2}^-)$, which were
a part of the coupled channels that can couple to the hidden-charm
pentaquark states. The calculations were employed with $2+1$ flavor
full QCD gauge configurations at the lattice spacing $a=0.1209$ fm.
The clover action was used for all quarks including charm quarks.
They calculated the effective central potentials in the two $J/\psi
N$ and one $\eta_cN$ channels for different time slices. The
scattering phase shift for $J/\psi N$ was also obtained by solving
the $S$-wave radial Schr\"odinger equation. They found the
attractive interactions for all channels, however, not strong enough
to form bound states. This result confirmed the time-independent
method observation in Ref.~\cite{Kawanai:2010ev}.

\begin{figure}[!th]
\begin{center}
\includegraphics[width=0.9\textwidth,clip]{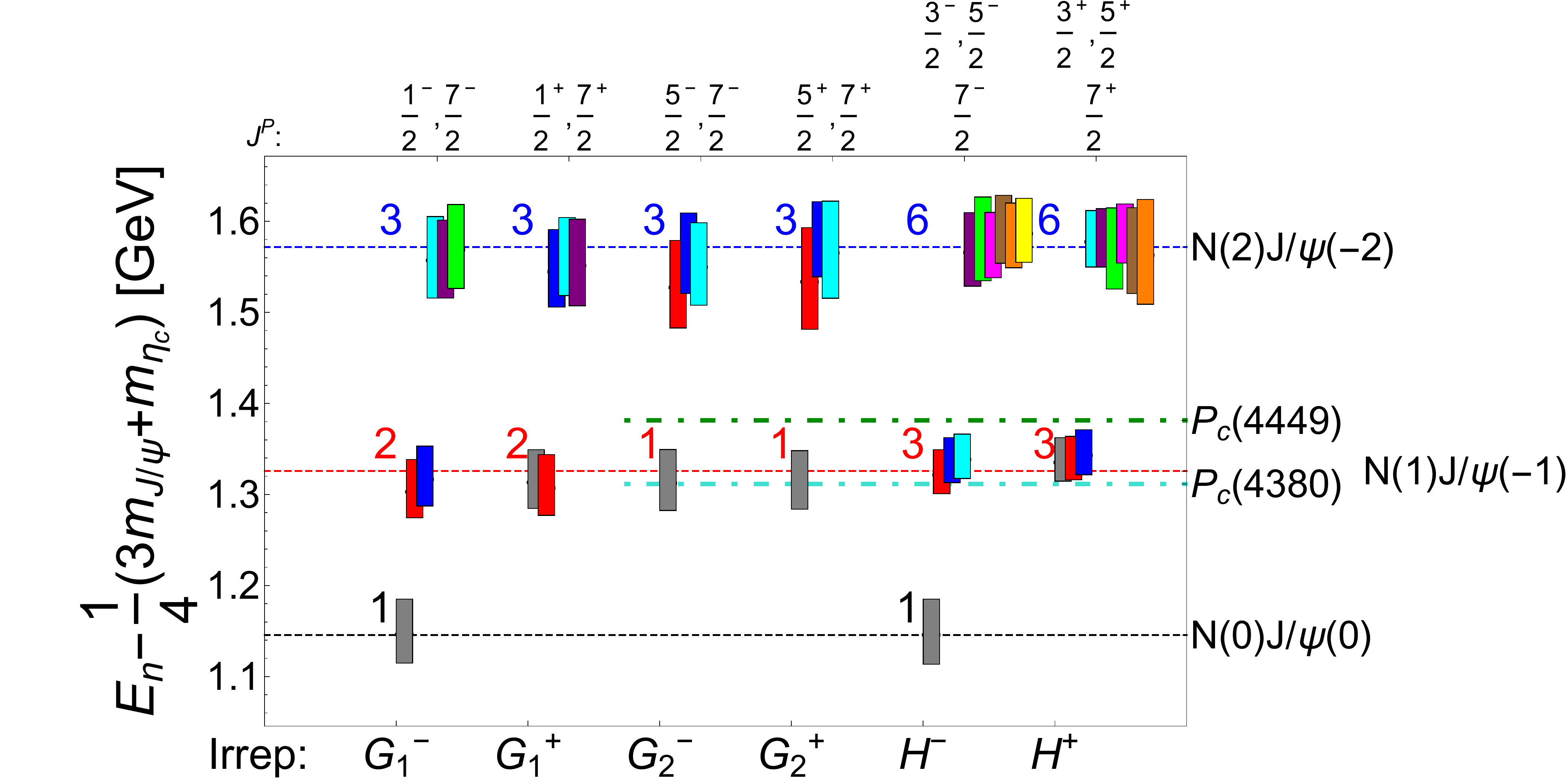}
\end{center}
\caption{Eigenenergies of $J/\psi N$ system in one-channel
approximation for all lattice irreps, denoted as the boxes. The
numbers denote the degeneracy of the expected states in the
non-interacting limit. Dashed lines represent the non-interacting
energies $E_N(p)+E_{J/\psi}(-p)$ for different value of momentum
$p$. The green and turquoise dash-dotted lines correspond to the
experimental masses of $P_c(4380)$ and $P_c(4450)$. Figure was taken
from Ref.~\cite{Skerbis:2018lew}.} \label{44JpsiN}
\end{figure}

\begin{figure}[!th]
\begin{center}
\includegraphics[width=0.9\textwidth,clip]{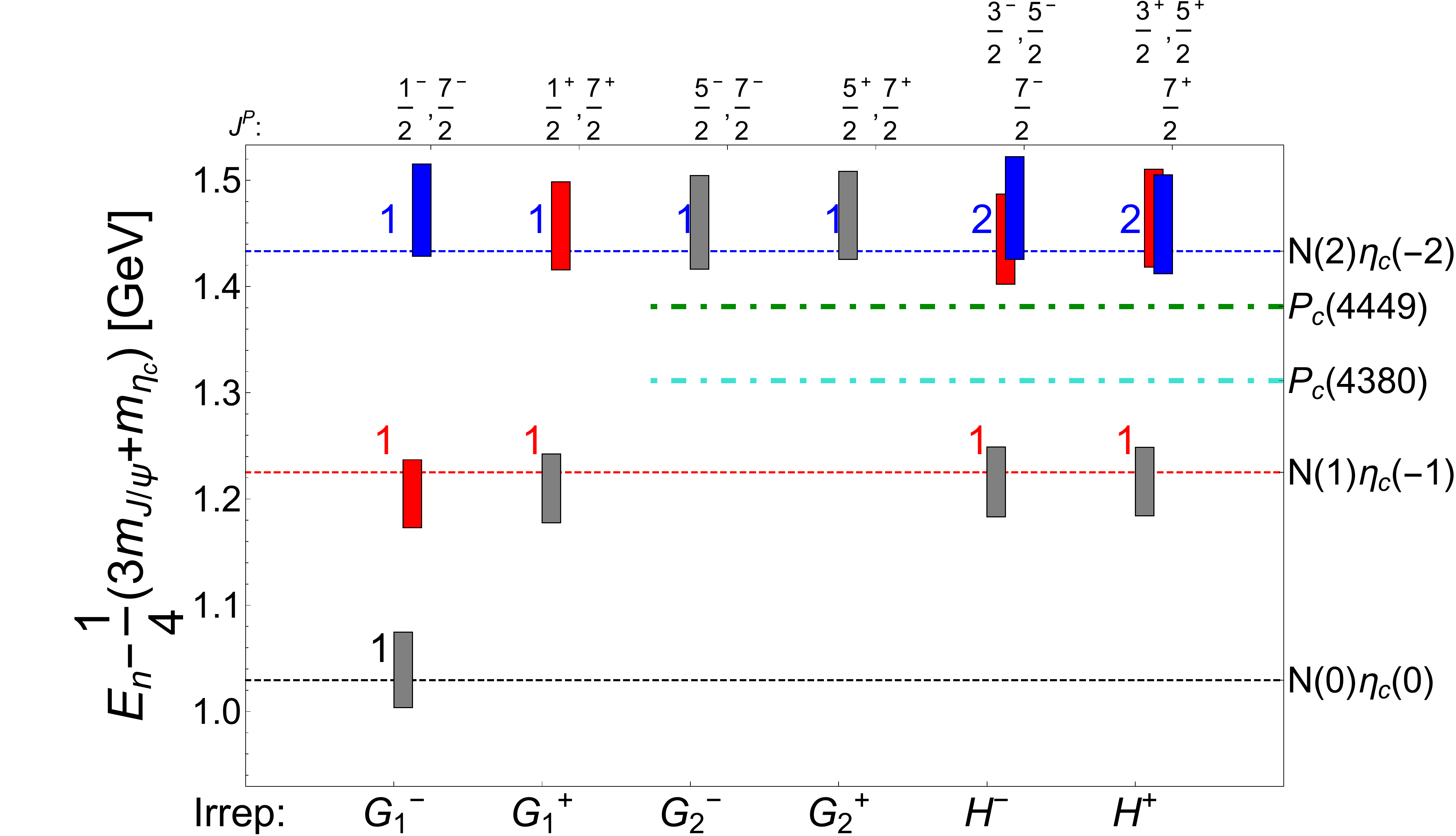}
\end{center}
\caption{Eigenenergies of $\eta_c N$ system in one-channel
approximation for all lattice irreps. Notations are the same with
those in Fig.~\ref{44JpsiN}. Figure was taken from
Ref.~\cite{Skerbis:2018lew}.} \label{44EtacN}
\end{figure}

The first lattice simulation reaching the energy region of the
hidden-charm pentaquark states $P_c(4380)$ and $P_c(4450)$ was
performed in Ref.~\cite{Skerbis:2018lew}, in which both the $S$-wave
and $P$-wave, both $\eta_cN$ and $J/\psi N$ systems were studied including
all possible $J^P$ assignments. The authors considered the
nucleon-charmonium systems with total momentum zero of the form
$O\sim N(p)M(-p)$, where $N$ is nucleon and $M$ denotes $J/\psi$ or
$\eta_c$. They performed the simulations on the $N_f=2$ ensemble
with a lattice spacing $a=0.1239$ fm at the pion mass $m_\pi=266$
MeV. The Wilson Clover action was adopted for all light quarks and
charm quarks. The energy spectra were calculated in the one-channel
approximation for all six irreps of the discrete lattice group
$O_h$: irrep $G_1^\pm$ with $J^P=\frac{1}{2}^\pm, \,
\frac{7}{2}^\pm$, irrep $G_2^\pm$ with $J^P=\frac{5}{2}^\pm, \,
\frac{7}{2}^\pm$, irrep $H^\pm$ with $J^P=\frac{3}{2}^\pm, \,
\frac{5}{2}^\pm, \, \frac{7}{2}^\pm$. The final eigenenergies were
presented in Fig.~\ref{44JpsiN} for the $J/\psi N$ system and in
Fig.~\ref{44EtacN} for the $\eta_cN$ system. As seen in these plots,
the extracted lattice energy spectra were exactly consistent with
the prediction of the non-interacting nucleon-charmonium systems. No
additional eigenstates were found, which implied that there was no
strong indication for a hidden-charm pentaquark resonance in the
one-channel approximation for $\eta_cN$ and $J/\psi N$ scatterings.

The hadroquarkonium picture~\cite{Dubynskiy:2008mq} was proposed for
the modification of the potential between a static $c\bar c$ pair
(in the heavy quark limit) induced by the presence of octet and
decuplet light baryons. In Ref.~\cite{Alberti:2016dru}, the authors
performed lattice QCD simulations on a CLS ensemble with $N_f=2+1$
flavors of nonperturbatively improved Wilson quarks at a pion mass
$m_\pi\approx 223$ MeV at a lattice spacing $a=0.0854$ fm. The shift
of the $c\bar c$ binding energies due to the presence of the nucleon
was extracted to be down only by a few MeV, similar in strength to
the deuterium binding. It was dubious whether such a small
attraction survives in the infinite volume limit and supports the
existences of bound states or resonances.

\subsection{A short summary for lattice QCD}

We summarize some important conclusions from the present lattice QCD
investigations as follows:
\begin{itemize}

\item Lattice QCD simulations found an isosinglet $I=0$ candidate for the
$X(3872)$ state close to the experimental data, only if both the
conventional $c\bar c$ operators and the two-meson $D\bar D^\ast$
operators were used. However, they found no candidate if the
diquark-antidiquark and $D \bar D^*$ interpolators were used in the
absence of $c\bar c$, implying the $c\bar c$ Fock component is
crucial for the $X(3872)$. The inclusion or exclusion of the
diquark-antidiquark operators does not affect the lattice energy
levels significantly.

\item The lattice QCD simulations for the mass spectra of the
hidden-charm four-quark systems didn't find any additional energy
level in the $I^G(J^{PC})=1^+(1^{+-})$ channel associated with the
$Z_c(3900)$ and $Z_c(4020)$ states. The lattice investigations of
the multiple channel scattering indicated that the $Z_c(3900)$ is
not a conventional resonance but a threshold cusp.

\item A quenched lattice QCD calculation showed that the interactions
between the $(\bar D_1D^\ast)^\pm$ system are attractive for both
the pseudoscalar ($S$-wave) and axial-vector ($P$-wave) channels.
The $Z^+(4430)$ might be a candidate of the $P$-wave $\bar
D_1D^\ast$ molecular state.

\item The $1^{--}$ hybrid charmonium mass from Lattice calculations
is consistent with the experimental mass of the $Y(4260)$, which
supported the hybrid interpretation for this state.

\item Lots of lattice simulations showed that the $ud\bar
b\bar b$ tetraquark state with quantum numbers $I(J^P)=0(1^+)$ lies
below the $BB^{(\ast)}$ thresholds, implying that this state is
stable against the strong and electromagnetic interaction. This
observation has been confirmed at the sufficiently low pion masses
$m_\pi\approx164$ MeV.

\item There is no evidence of a $bb\bar b\bar b$ tetraquark state with a mass
below the lowest noninteracting $2\eta_b$ threshold in the $0^{++},
1^{+-}$ or $2^{++}$ channel in the lattice nonrelativistic QCD
studies.

\item To date, there are few lattice studies for many other $XYZ$
structures reported in the last 16 years. Future investigations are
badly needed to study these claimed signals in various channels.

\item Most of the present lattice simulations were performed with a
rather large pion mass. The pion exchange force plays a pivotal role
in the formation of the loosely bound molecular states, which decays
exponentially with $m_\pi$ and is very sensitive to the pion mass.
Full QCD simulations near the physical point are crucial to explore
the underlying structures of the near-threshold $XYZ$ states.

\item The lattice studies of the $\eta_cN$ and $J/\psi N$ scatterings in
the one-channel approximation found no strong indication for the
existence of $P_c(4380)$ and $P_c(4450)$ pentaquark states. The
strong coupling between the $J/\psi N$ with other two-hadron
channels might be important for the formation of the $P_c$ states.
The multiple channel scattering need to be investigated in the
future lattice simulations to confirm or refute the existence of the
$P_c$ states.

\end{itemize}

%% file: section10.tex
\section{Production and decay properties}

In this subsection, we shall discuss recent theoretical progress on
the productions and decay properties of the multiquark states, which
are crucial to understand the nature of exotic states. In
particular, the non-resonant explanations for some $XYZ$ states depend
on their production processes.

The production of the multiquark systems at the quark
level remains very challenging since nonperturbative effects are
usually involved. One has to turn to effective formalisms such as
the nonrelativistic QCD (NRQCD) framework
\cite{Ma:2003zk,Hyodo:2017hue,Hyodo:2012pm,Jin:2014nva,Li:2017ghe}
and quark combination models \cite{Han:2009jw}. At the hadron level,
one may check whether it is necessary to introduce new resonances
through refitting the production cross sections and line shapes.
There are investigations of the interference effects on line shapes,
$R$ values, and resonance properties due to the nearby hadrons.
There are also explorations of amplitude pole structures in
understanding the nature of resonances by considering the
rescattering or meson loop mechanism. An interesting observation is
that the anomalous triangle singularity (ATS) in final state
interactions through the triangle diagram may mimic resonance-like
structures in the mass invariant distributions of measured channels,
when all the three particles in the loop approach their on-shell
conditions simultaneously. If this mechanism were confirmed, the
number of genuine resonances might be reduced. In the literature,
there are also discussions about productions of exotic hadrons in
heavy ion collisions \cite{Cho:2010db,Cho:2011ew,Cho:2017dcy}.

One may adopt the widely used Cornell model or $^3P_0$ pair creation
model to study the decays of higher charmonium(like) states
\cite{Liu:2016sip,Xiao:2018iez}. The quark-interchange model was
proposed for the rearrangement decays of the multiquark states.
\cite{Ackleh:1991dy,Wong:2001td,Barnes:2003dg,Wang:2018pwi}. The
rescattering mechanism may be used to study the decays at the hadron
level \cite{Guo:2016iej}.

\subsection{$Q\bar{Q}q\bar{q}$}

\subsubsection{The $X$ states}

In Ref. \cite{Zhou:2015uva}, Zhou, Xiao, and Zhou reanalyzed the
$\gamma\gamma\to D\bar{D}$ and $\gamma\gamma\to J/\psi\omega$
processes where the $X(3930)$ [now called $\chi_{c2}(3930)$] and the
$X(3915)$ were observed, respectively. Their results indicate that
these two mesons are the same $2^{++}$ state. In Ref.
\cite{Baru:2017fgv}, Baru, Hanhart, and Nefediev investigated
whether the $X(3915)$ is indeed the tensor partner of the $X(3872)$
in the molecule picture~\cite{Albaladejo:2017blx} by analysing the production amplitudes. 
They found that current data favor a scalar assignment, if the $X(3915)$
is a $D^*\bar{D}^*$ molecule. Its nature would require some exotic
interpretation (i.e., neither a regular quarkonium state nor the
$D^*\bar{D}^*$ spin partner of the $X(3872)$), if it is indeed
dominated by the helicity-0 contribution of the nearby tensor state.

The hadronic effects on the $X(3872)$ meson abundance in heavy 
ion collisions were studied in Ref. \cite{Cho:2013rpa}, where the 
authors found that the absorption cross sections and the time evolution 
of the $X(3872)$ meson abundance were strongly dependent on its 
structure and quantum numbers. 
Combining the double parton scattering and basic ideas of the color
evaporation model, the authors of Ref. \cite{Carvalho:2015nqf}
developed a model for the tetraquark production. They predicted the
$X(3872)$ production cross section at LHC $\sqrt{s}=14$ TeV to be in
the order of tens of nb. In Ref. \cite{Wang:2015rcz}, Wang and Zhao
discussed dependence of the ratio between the $B_c$ semileptonic and
nonleptonic decays into the $X(3872)$. They demonstrated that the
ratios would be universal and predictable if the $X(3872)$
production mechanism is through the $c\bar{c}$ component, while
significant deviations from the predicted results would be a clear
signal for the non-standard charmonium structure at the short
distance.

In Ref. \cite{Hsiao:2016vck}, the authors emphasized the connection
between the decays $\Lambda_b\to X_c^0\Lambda$ and $B^-\to
X_c^0K^-$, with $X_c^0=c\bar{c}q\bar{q}$ ($q=u,d,s$) since these
processes together with the case $X_c^0\to J/\psi$ arise from the
$b\to c\bar{c}s$ transition at the quark level. Assuming the
$X(3872)$ and $X(4140)$ to be tetraquark states, Hsiao and Geng
predicted the branching ratios of $\Lambda_b\to \Lambda
X(3872)\to\Lambda J/\psi\pi^+\pi^-$ and $\Lambda_b\to \Lambda
X(4140)\to\Lambda J/\psi\phi$ are $(5.2\pm1.8)\times10^{-6}$ and
$(4.7\pm2.6)\times10^{-6}$, respectively, which were expected to be
accessible at LHCb. In Ref. \cite{Hsiao:2016pml}, they predicted the
branching fractions for $B^-\to X_c^0\pi^-$, $\bar{B}^0\to
X_c^0\bar{K}^0$, $\bar{B}_s^0\to X_c^0\bar{K}^0$, $B_c^-\to
X_c^0\pi^-$, $B_c^-\to X_c^0K^-$, $B_c^-\to J/\psi
\mu^-\bar{\nu}_\mu$, and $B_c^-\to X_c^0\mu^-\bar{\nu}_\mu$.

In Ref. \cite{Wang:2016ydp}, productions of the $X(3940)$ and
$X(4160)$ in exclusive weak decays of $B_c$ were studied with the
improved Bethe-Salpeter method by assigning these two mesons as
$\eta_c(3S)$ and $\eta_c(4S)$, respectively. The predicted branching
ratios are Br$(B_c^+\to X(3940)e^+\nu_e)=1.0\times10^{-4}$ and
Br$(B_c^+\to X(4160)e^+\nu_e)=2.4\times10^{-5}$. Nonleptonic $B_c$
decays to these states were also investigated. In Ref.
\cite{Wang:2016mqb}, the same group investigated the two-body
open-charm OZI-allowed strong decays of these two mesons with the
$^3P_0$ model. The decay width of $\eta_c(3S)$ is around 34 MeV and
close to the width of $X(3940)$. Therefore, the $\eta_c(3S)$ is a
good candidate of $X(3940)$. Although the width of $\eta_c(4S)$ is
around 70 MeV and close to the lower limit of the width of
$X(4160)$, the ratio $\Gamma(D\bar{D}^*)/\Gamma(D^*\bar{D}^*)$ of
$\eta_c(4S)$ is larger than the experimental data of $X(4160)$. The
authors concluded that $\eta_c(4S)$ is not the candidate of
$X(4160)$.

In Ref. \cite{Liu:2016onn}, Liu tried to understand the nature of
$X(4140)$, $X(4274)$, $X(4500)$, and $X(4700)$ in the process
$B^+\to J/\psi\phi K^+$ with the rescattering mechanism. The
$X(4700)$ and $X(4140)$ may be simulated by the rescattering
effects, but the $X(4274)$ and $X(4500)$ could not if the quantum
numbers of $X(4274)$ and $X(4500)$ are $1^{++}$ and $0^{++}$,
respectively, which indicates that the $X(4274)$ and $X(4500)$ could
be genuine resonances. The arguments from the mass, decay, and
production
\cite{Godfrey:1985xj,Barnes:2005pb,delAmoSanchez:2010jr,Aaij:2016iza,Aaij:2016nsc}
lead to the suggestion that the $X(4274)$ may be the conventional
$\chi_{c1}(3P)$. In Ref. \cite{Liu:2015cah}, Liu and Oka
investigated the radiative transition processes $e^+e^-\to \gamma
J/\psi\phi$, $\gamma J/\psi\omega$, and $\pi^0 J/\psi\eta$ to search
for the $c\bar{c}s\bar{s}$ states, such as the $X(4140)$, $X(4274)$,
$X(4350)$, and $X(3915)$, by considering the rescattering processes
via the charmed-strange meson loops. They found that the anomalous
triangle singularity (ATS) may appear as narrow peaks, when some
special conditions are satisfied, and thus nonresonance explanation
is possible for their structures. Contrary to genuine resonances,
the location of the ATS peaks depends on kinematic configurations,
and their movements may be used to distinguish the ATS peaks from
genuine resonances, once high resolution experiments are available.

In Ref. \cite{Abreu:2016dfe}, the production of $X(3872)$ as a
$D^*\bar{D}$ molecule was considered in the process
$\bar{D}^{(*)}D^{(*)}\to \pi X(3872)$ by using the heavy meson
effective theory. In Ref. \cite{Abreu:2016qci}, the $X(3872)$
production and absorption in the hot hadron gas produced in heavy
ion collisions were studied. The authors found that the average
$X(3872)$ number as a $D\bar{D}^*$ molecule,
$N_{X(mol)}\approx7.8\times10^{-4}$, is about 80 times larger than
that as a tetraquark state, since the production mechanisms are
quite different. Similar investigations about the $Z_b(10610)$ and
$Z_b(10650)$ cases can be found in Refs.
\cite{Abreu:2016xlr,Abreu:2017nuc,Abreu:2017pos}.

The authors of Ref. \cite{Jin:2016vjn} studied the production and
decay of the exotic hadrons in multiproduction processes at high
energy hadron colliders in the non-relativistic wave function
framework by treating the exotic states as hadronic molecules. The
rapidity and transverse momentum distributions of the  $X(3872)$,
$Y(4260)$, and $P_c(4380)$ at $\sqrt{s}=8$ TeV in $pp$ collisions
were investigated.

The authors of Ref. \cite{Chen:2016iua} discussed the possibility of
the $X(4140)$ being the $\chi_{c1}(3P)$ state through the
$\chi_{c1}\pi^+\pi^-$ invariant mass spectrum in the $B\to
K\chi_{c1}\pi^+\pi^-$ process. The analysis for the $D\bar{D}$
invariant mass in the $B\to KD\bar{D}$ decay results in a
$\chi_{c0}(3P)$ candidate with a mass around 4080 MeV. From the
decay behaviors of the $\chi_{cJ}(3P)$, the assignments of the
$X(4140)$ and $X(4080)$ as the $\chi_{c1}(3P)$ and $\chi_{c0}(3P)$
are reinforced respectively.

In Ref. \cite{Kang:2016jxw}, Kang and Oller introduced a
near-threshold parameterization to analyze the $X(3872)$ production
processes $p\bar{p}\to J/\psi\pi\pi+...$, $B\to KJ/\psi\pi\pi$, and
$B\to KD\bar{D}^{*0}$. The data can be reproduced with similar
quality for $X(3872)$ being a bound and/or a virtual state. The
authors of Ref. \cite{Wang:2017mrt} studied the $J/\psi\phi$ mass
distribution of the $B^+\to J/\psi\phi K^+$ reaction, and found that the
contributions of the narrow $X(4140)$ and the $X(4160)$ are needed, 
which provided an explanation for the large width of $X(4140)$ 
observed by LHCb \cite{Aaij:2016nsc,Aaij:2016iza}.
The $X(4160)$ state was found to be strongly tied to the
$D_s^{*+}D_s^{*-}$ channel. 

In Ref. \cite{Dai:2018tgo}, Dai, Dias, and Oset studied the
$B_c^-\to \pi^-J/\psi\omega$ and $B_c^-\to\pi^-D^*\bar{D}^*$
reactions. They found that the molecular states $X(3940)$ and
$X(3930)$ couple mostly to $D^*\bar{D}^*$ in the $2^{++}$ and
$0^{++}$ channels respectively, and have big influence on the
$J/\psi\omega$ mass distribution. Later in Ref.
\cite{Ikeno:2018ugx}, the authors studied the semileptonic decay of
the $B_c^-$ meson into the $2^{++}$ $X(3930)$, $0^{++}$ $X(3940)$,
and $2^{++}$ $X(4160)$ resonances by treating them as dynamically
generated states. 

In Ref. \cite{Goncalves:2018hiw}, the productions of the $X(4350)\to
\phi J/\psi$ and $X(3915)\to\omega J/\psi$ in $\gamma\gamma$
interactions in $pp/pPb/PbPb$ collisions at the LHC were
investigated. The results with both $J^{PC}=0^{++}$ and $2^{++}$
indicate that the experimental study is feasible and can be used to
check the existence of these two states. In Refs.
\cite{Braaten:2018eov,Braaten:2019sxh}, Braaten, He, and Ingles
pointed out that the production rate of $X(3872)$ accompanied by a
pion at a high energy hadron collider should be larger than that of
$X(3872)$ without a pion, if the $X(3872)$ is a weakly bound
charm-meson molecule. This type of production may happen through the
creation of the $J=1$ $D^*\bar{D}^*$ pair at short distances
followed by a rescattering process. In Ref. \cite{Braaten:2019yua},
the production of $X(3872)$ accompanied by a pion in $B$ meson
decays was discussed. Such a production should be observable, which
can be used to support for the interpretation of $X(3872)$ as a
molecule.

In Ref. \cite{Andronic:2019wva}, the transverse momentum spectra and
yields of the $X(3872)$, $J/\psi$, and $\psi(2S)$ for heavy-ion
collisions at the LHC energies are predicted within the framework of
the statistical hadronization model. The production yield of
$X(3872)$ was found to be about 1\% relative to that for $J/\psi$.
In Ref. \cite{Guo:2019qcn}, Guo proposed a method to study whether
the $X(3872)$ is above or below the $D^0\bar{D}^{*0}$ threshold. He
found that the line shape of $X(3872)\gamma$ is very sensitive to
the deviation of the $X(3872)$ mass from the threshold due to a
triangle singularity. This indirect method can be applied to
experiments producing copious $D^{*0}\bar{D}^{*0}$ pairs.

In Ref. \cite{Wu:2016dws}, the production of $1^{++}$
$X_b=B\bar{B}^*$ molecule in the process $\Upsilon(5S,6S)\to\gamma
X_b$ was investigated by using the rescattering mechanism. The
production ratios are orders of $10^{-5}$. In Ref.
\cite{Gonzalez:2016fsr}, a comparative study for the strong decays
of $X(3915)$ as a $0^{++}$ charmonium state into $D\bar{D}$ and
$J/\psi\omega$ channels were performed with the Cornell potential
model, and a generalized screen potential model. The suppression of
the $D\bar{D}$ mode can be understood with both models, while the
significant $J/\psi\omega$ mode favors the latter model.

In Ref. \cite{Wang:2017dcq}, the decay $X(3872)\to J/\psi\gamma$ was
investigated in the Bethe-Salpeter equation approach if it is a
$D\bar{D}^*$ molecule. The decay width was found to be in the range
8.7$\sim$49.5 keV. The authors of Ref. \cite{Chen:2018xok} studied
the two-photon decay of the $X(3915)$ state if it as a
$D^{*0}\bar{D}^{*0}$ molecule. The obtained decay width is 36 keV.
The decay width into $J/\psi\omega$ is 66 MeV in the same molecular
picture \cite{Chen:2018fbs}.

\subsubsection{The $Y$ states}

In Ref. \cite{Chen:2015bft}, in the Fano-like interference scenario,
Chen, Liu, Li, and Ke studied the asymmetric line shapes of the
$Y(4260)$ and $Y(4360)$ structures in their discovery processes
$e^+e^-\to\pi\pi J/\psi$ and $e^+e^-\to\pi\pi\psi(3686)$ by
considering the interference between continuum and resonance
contributions. They found that the broad structure $Y(4008)$ can be
induced by the Fano-like interference. Their numerical results
indicate that the $Y(4008)$, $Y(4260)$, and $Y(4360)$ may be not
genuine resonances, which may explain the absence of these states in
the $R$ value scan and the non-observation of their open-charm decay
modes. The Fano-like phenomena were also expected in processes such
as $e^+e^-\to\pi\pi\psi(3770)$ and $e^+e^-\to K\bar{K}J/\psi$. In
Ref. \cite{Chen:2017uof}, Chen, Liu, and Matsuki studied the
interference effects in understanding the $Y(4320)$ and $Y(4390)$.
They found that the signals of these two structures can be reproduced
by introducing the interference effects between $\psi(4160)$,
$\psi(4415)$, and background contributions. Therefore, the $Y(4320)$
and $Y(4390)$ may be not genuine resonances. A better description of
experimental results needs the introduction of $Y(4220)$, which was
proposed to be the charmonium $\psi(4S)$. In Ref.
\cite{Wang:2019mhs}, the updated data of $Y(4220)$ were used to
categorize this state into the $J/\psi$ family in a $4S$-$3D$ mixing
scheme. The present experimental data seem to support the
charmonium assignment. The assignments of higher charmonia states
and properties of newly predicted states were also discussed.

In Ref. \cite{Liu:2016sip}, the open-charm decay mode $\Lambda
_c\bar{\Lambda }_c$ of $Y(4630)$ was explored in the $^3P_0$ pair
creation model by assuming this exotic meson to be a $P$-wave
diquark-antidiquark state \cite{Maiani:2014aja}. The decay
mechanisms of $Y(4630)$ in both tetraquark and molecule pictures
were studied in Ref. \cite{Liu:2016nbm}. The $p\bar{p}$, $\pi\pi$,
and $K\bar{K}$ decay modes of $Y(4630)$ were discussed in the
$\Lambda_c\bar{\Lambda}_c$ rescattering mechanism in Ref.
\cite{Guo:2016iej}.

In Ref. \cite{Wang:2016fhj}, the role of the s-channel $Y(4630)$ in
the $p\bar{p}\rightarrow\Lambda_c\bar{\Lambda}_c$ reaction near
threshold was investigated. The authors found that this state gives
a clear bump structure with the magnitude of 10 $\mu$b, which can be
tested by the PANDA experiment. In Ref. \cite{Wang:2017sxq},
productions of $Y(4220)$ in the processes $e^+e^-\to Y(4220)\to
p\bar{p}\pi^0$ and $p\bar{p} \to Y(4220) \pi^0$ were studied
simultaneously, where the nucleon and its excitations play an
important role in the reactions. Therefore, the PANDA experiment can
also be used to study the properties of $Y(4220)$.

In Ref. \cite{Qin:2016spb}, Qin, Xue, and Zhao investigated the
production mechanism of $Y(4260)$ in $e^+e^-$ annihilation by
assuming it to be a $\bar{D}D_1 +c.c.$ molecule admixed with a
compact $c\bar{c}$ component. They found that the heavy quark spin
symmetry breaking plays a crucial role for the $Y(4260)$ production,
and the molecule picture can describe the available observables
simultaneously. They also found that $\bar{D}D^*\pi+c.c.$ is one of
the important decay channels of $Y(4260)$. In Ref.
\cite{Cleven:2016qbn}, Cleven and Zhao demonstrated that the
molecule picture of $Y(4260)$ can explain the cross section line
shape of $e^+e^-\to \chi_{c0}\omega$ around the $Y(4260)$ mass
region. Further in Ref. \cite{Xue:2017xpu}, Xue, Jing, Guo, and Zhao
showed that the partial widths of the molecular $Y(4260)$ to
$D^*\bar{D}^*$ and $D_s^*\bar{D}_s^*$ are much smaller than the
width to $\bar{D}D^*\pi+c.c.$, and the $Y(4260)$ contributions to
the cross section line shapes of the $e^+e^-\to D^*\bar{D}^*$ and
$D_s^*\bar{D}_s^*$ are rather small in most of the energy region
from thresholds to about 4.6 GeV. They also found that the
interferences of the $Y(4260)$ with nearby charmonium states
$\psi(4040)$, $\psi(4160)$, and $\psi(4415)$ produce a dip around
4.22 GeV in the $e^+e^-\to D^*\bar{D}^*$ cross section line shape.

In Ref. \cite{Dai:2017fwx}, Dai, Haidenbauer, and Meissner
investigated the reaction $e^+e^-\rightarrow
\Lambda^+_c\bar\Lambda^-_c$ at energies close to the threshold after
considering the effects from the $Y(4630)$ resonance and the final
state interactions. The mass and width of this state were found to
be around 4652 MeV and 63 MeV, respectively, and thus the $Y(4630)$
and $Y(4660)$ could be the same state. In Ref. \cite{Zhang:2018zog},
a combined fit was performed to the cross sections of
$e^+e^-\to\omega\chi_{c0}$, $\pi\pi h_c$, $\pi\pi J/\psi$,
$\pi\pi\psi(3686)$, and $\pi^+D^0D^{*-}+c.c.$ with three resonances
$Y(4220)$, $Y(4390)$, and $Y(4660)$. The authors found that the
first two resonances are sufficient to explain these cross sections
below 4.6 GeV, so the $Y(4320)$, $Y(4360)$, and $Y(4390)$ should be
one state.

In Ref. \cite{Piotrowska:2018rzl}, the $D\bar{D}^*$ loop
contributions to the $e^+e^-$ reaction around 4.0 GeV were
considered with effective Lagrangians. The authors noticed a second
pole dynamically generated by the meson-meson quantum fluctuations
in addition to the charmonium $\psi(4040)$. This pole leads to the
controversial state $Y(4008)$ observed in $e^+e^-\to\pi\pi J/\psi$.
The authors also expect that the pole could be visible in the
process $e^+e^-\to \psi(4040)\to D\bar{D}^*$. The same mechanism may
be applied to the $Y(4260)$ and $\psi(4160)$.

In Ref. \cite{Coito:2019cts}, the relation between the $\psi(4160)$
and $Y(4260)$ within a unitarized effective Lagrangian approach was
studied. The authors found that the $Y(4260)$ is not an independent
resonance but a manifestation of the $\psi(4160)$ due to the final
state interactions between the $D_s^*\bar{D}_s^*$ and the $J/\psi
f_0(980)$. In Ref. \cite{Chen:2019mgp}, Chen, Dai, Guo, and Kubis
studied the processes $e^-e^+\to Y(4260)\to J/\psi\pi\pi(K\bar{K})$
in the dispersion theory, and analyzed the roles of the light-quark
$SU(3)$ singlet and octet states in the transitions. According to
their investigations, the $Y(4260)$ contains a large light-quark
component, and it is neither a hybrid nor a conventional charmonium
state. The analysis of the ratio of the octet and singlet components
indicates that the $Y(4260)$ has a sizeable $\bar{D}D_1$ component,
although it is not a pure $\bar{D}D_1$ molecule.

Very recently, BESIII studied the reaction $e^+e^-\to \pi^+\pi^-
D\bar{D}$ at center-of-mass energies above 4.08 GeV, and observed
$\psi(3770)$ and $D_1(2420)\bar{D}+c.c.$ in this process. Neither
fast rise of the cross section above the $D_1(2420)\bar{D}$
threshold nor obvious structure is visible from the measurements,
which is different from the expectation of Ref. \cite{Wang:2013cya}.

In Ref. \cite{Chen:2016byt}, Chen {\it et al} studied the decay of
$Y(4260)$ into $Z_c(3900)\pi$ by assuming that they are $D\bar{D}_1$
and $D\bar{D}^*$ molecules, respectively. The partial width was
found to be around 1.2$\sim$3.0 MeV. In Ref. \cite{Chen:2018fsi},
the open-charm decays of the charmonium $\psi(3^3D_1)$ were studied
in the $^3P_0$ pair creation model. The total width is consistent
with those of both $Y(4320)$ and $Y(4390)$. One of the dominant
decays of $Y(4390)$ is $Y(4390)\to D_1\bar{D}\to \pi^+D^0D^{*-}$.

In Ref. \cite{Xiao:2018iez}, Xiao et al. studied the
$\Lambda_c\bar{\Lambda}_c$ decay mode of the higher vector
charmonium states around 4.6 GeV in the $^3P_0$ pair creation model.
They found that the partial decay widths for $S$-wave charmonia are
about a few MeV, and those for $D$-wave charmonia are less than one
MeV. The $Y(4660)$ is very likely to be an $S$-wave charmonium, if
the $Y(4630)$ and $Y(4660)$ are the same state.

\subsubsection{The $Z_c$ states}

In Ref. \cite{Gong:2016hlt}, the final state interaction effects on
the $Z_c(3900)$ was studied. In the analysis, the authors calculated
the amplitudes of the $e^+e^-\to J/\psi\pi\pi$, $h_c\pi\pi$, and
$D\bar{D}^*\pi$ processes, and searched poles in the complex energy
plane. They found only one pole in the nearby energy region in
different Riemann sheets and concluded that the $Z_c(3900)$ is a
molecular state according to the pole counting rule
\cite{Morgan:1992ge,Au:1986vs}. In Ref. \cite{Gong:2016jzb}, the
same group noted that the triangle singularity mechanism did not
produce the observed $Z_c(3900)$ peak. Their study again favors the
molecular assignment for the $Z_c(3900)$.

In Ref. \cite{Albaladejo:2016jsg}, to understand the nature of
$Z_c(3900)$, the authors considered the $J/\psi\pi$ and $D^*\bar{D}$
coupled-channel T-matrix analysis in a finite volume. Comparing
their calculations with the lattice results
\cite{Prelovsek:2014swa}, they concluded that it is difficult to
distinguish whether the $Z_c(3900)$ is a resonance originating from
a pole above the $D^*\bar{D}$ threshold or a virtual state below the
threshold. In Ref. \cite{Pilloni:2016obd}, the authors carried out a
thorough amplitude analysis of the $Z_c(3900)$ data and considered
four different scenarios corresponding to the pure QCD states,
virtual states, or purely kinematical enhancements. They concluded
that it is not possible to distinguish between these scenarios at
that time, which contradicted the conclusion of Ref.
\cite{Gong:2016jzb}.

Both the $X(3872)$ and $Z_c(3900)$ were observed in the $e^+e^-$
annihilations (into $\gamma X(3872)$ and $\pi Z_c(3900)$,
respectively), but only the $X(3872)$ was observed in $B$ decay. In
order to understand this puzzle, Yang, Wang, and Meissner studied
the isospin amplitudes in the exclusive $B$ decay into $D^*\bar{D}K$
in Ref. \cite{Yang:2017nde}. Their analysis indicates that the
production of the isovector $C=-$ $D^*\bar{D}$ state is highly
suppressed compared to the isospin singlet one, if both states are
$D\bar{D}^*$ molecules. Their findings seem to support the molecular
nature of $Z_c(3900)$.

In Ref. \cite{He:2017lhy}, He and Chen investigated the $\pi
J/\psi$-$\bar{D}^*D$ interaction in the $Y(4260)$ decay in a
coupled-channel quasipotential Bethe-Salpeter equation approach in
order to understand the origin of the $Z_c(3900)$ and $Z_c(3885)$.
They found a virtual state around 3870 MeV, and concluded that the
two $Z_c$ states are from the same virtual state. In Ref.
\cite{Zhao:2018xrd}, Zhao analyzed possible quantum numbers and
various explanations of the $Z_c(4100)$ state via its production in
the $B$ decay process. He found that the state can be either induced
by final state interaction effects arising from an $S$-wave
$D^*\bar{D}^*$ rescattering with $I^G(J^{PC})=1^-(0^{++})$, or it is
a $P$-wave resonance with $I^G(J^{PC})=1^-(1^{-+})$ arising from the
$D^*\bar{D}^*$ interaction. The latter structure can be regarded as
an excited state of $Z_c(4020)$.

In Ref. \cite{Cao:2018vmv}, Cao and Dai discussed possible $J^P$
assignments for the $Z_c(4100)$, $Z_1(4050)$ [called $Z_c(4050)$ in
PDG \cite{Tanabashi:2018oca}], and $Z_2(4250)$ and proposed that the
$Z_c(4100)$ observed in $\eta_c\pi$ by LHCb \cite{Aaij:2018bla} and
the $Z_1(4050)$ observed in $\chi_{c1}\pi$ by Belle
\cite{Mizuk:2008me} correspond to the same state. They suggested
that the $Z_2(4250)$ should be a $1^+$ or $1^-$ state, while the
$Z_c(4100)/Z_1(4050)$ could be a $0^+$ or $1^-$ state.

In Ref. \cite{Nakamura:2019btl}, the authors demonstrated that the
$Z_c(4430)$ and $Z_c(4200)$ can be consistently interpreted as
kinematical singularities from the triangle diagrams. They explained
why the $Z_c(4200)$ contribution was observed in $\Lambda_b\to
J/\psi p\pi$ but $Z_c(4430)$ was not. In Ref.
\cite{Nakamura:2019emd}, the $Z_1(4050)$ and $Z_2(4250)$ structures
observed in $\bar{B}^0\to \chi_{c1}K^-\pi^+$ were also explained
with the triangle singularities, and the $J^P$ was predicted to be
$1^-$ ($1^+$ or $1^-$) for $Z_1(4050)$ ($Z_2(4250)$).

In Ref. \cite{Wu:2019vbk}, the production of $Z_c(3900)$ and
$Z_c(4020)$ in $B_c$ decays was studied with an effective Lagrangian
approach by considering meson loop contributions. The branching
ratios of $B_c$ decays to $Z_c(3900)\pi$ and $Z_c(4020)\pi$ are
around $10^{-4}$ and $10^{-7}$, respectively. In a light-front
model, Ke and Li studied the decays of $Z_c(3900)$ and $Z_c(4020)$
[called $X(4020)$ in \cite{Tanabashi:2018oca}] into $h_c\pi$ by
assuming them as $D\bar{D}^*$ and $D^*\bar{D}^*$ molecules,
respectively \cite{Ke:2016owt}. They found that the $Z_c(4020)$
seems to be a molecular state. If the $Z_c(3900)$ is also a
molecule, it would be observable in $e^+e^-\to h_c\pi$, since the
partial width of $Z_c(3900)$ is only three times smaller than that
of $Z_c(4020)$. The nonobservation of $Z_c(3900)$ in this channel
with precise measurements would be helpful to rule out its molecule
interpretation.

The role of the charged exotic states in the reaction $e^+e^- \to
\psi(2S)\pi^+\pi^-$ was considered with a dispersive approach in
Ref. \cite{Molnar:2019uos}. The authors found that the $Z_c(3900)$
state plays an important role in explaining the BESIII's invariant
mass distribution at both $\sqrt{s}=4.226$ and 4.258 GeV
\cite{Ablikim:2017oaf}. The sharp narrow structure at
$\sqrt{s}=4.416$ GeV can be explained with a heavier charged state
whose mass and width are about $4.016$ GeV and 52 MeV, respectively.
At $\sqrt{s}=4.358$ GeV, no intermediate $Z_c$ state is necessary.
The authors also concluded that the $\pi\pi$ final state interaction
leads to the $\pi\pi$ invariant mass distribution for these four
energies.

In Ref. \cite{Agaev:2016dev}, the strong decay widths of $Z_c(3900)$
into $J/\psi\pi$ and $\eta_c\rho$ were calculated with the coupling
constants in the QCD sum rule method with the assumption that the
$Z_c(3900)$ is a diquark-antidiquark state with $J^{PC}=1^{+-}$. The
resulting decay width of these two modes is around 66 MeV, larger
than the PDG value $28.2\pm2.6$ MeV \cite{Tanabashi:2018oca}. In
Ref. \cite{Wu:2016ypc}, the charmless decays of the $Z_c(3900)$ and
$Z_c(4025)$ [called $X(4020)$ in \cite{Tanabashi:2018oca}] into
vector-pseudoscalar (VP) type light mesons were investigated with
the rescattering mechanism by assuming these mesons to be
$I^G(J^{PC})=1^+(1^{+-})$ $D\bar{D}^*$ and $D^*\bar{D}^*$ molecules,
respectively. The branching ratios of $Z_c(3900)\to VP$
($Z_c(4025)\to VP$) are typically of the order $10^{-3}$
($10^{-5}$).

In Ref. \cite{Goerke:2016hxf}, the strong decay widths $Z_c(3900)\to
J/\psi\pi$ ($\eta_c\rho$, $\bar{D}^0D^{*+}$, $\bar{D}^{*0}D^+$) and
$Z(4430)\to J/\psi$ ($\psi(2S)\pi$) were calculated within a
covariant quark model. The authors found that the tetraquark-type
(molecule-type) current is in disaccord (accordance) with the
experimental observation that $Z_c(3900)$ has a much stronger
coupling to $D\bar{D}^*/D^*\bar{D}$ than to $J/\psi\pi$. They found
that the $Z(4430)$ state is a good candidate for the compact
tetraquark state, and predicted the partial decay width
$\Gamma(Z_{4430}^+\to D^{*+}+\bar{D}^{*0})$ to be around 24 MeV.

In Ref. \cite{Wang:2018pwi}, Wang {\it et al} investigated the
strong decays of $Z_c(3900)\sim D^*\bar{D}$, $Z_c(4020)\sim
D^*\bar{D}^*$, $Z_c(4430)\sim \bar{D}D^*(2S)$ or $\bar{D}^*D(2S)$,
$Z_b(10610)\sim B^*\bar{B}$, and $Z_b(10650)\sim B^*\bar{B}^*$ into
a heavy quarkonium plus a pion in a relativized quark model in the
molecule picture. The decay proceeds through the interchange of a
heavy quark and a light quark as well as the exchange of a gluon.
Therefore, the exchanged $Q\bar{q}$ or $\bar{Q}q$ is a color-octet
state. According to their results, the $Z_c(3900)$ and $Z_c(4020)$
have a larger coupling with $\psi(2S)\pi$ than $J/\psi\pi$, but the
partial width $\Gamma(Z_c(3900)\to J/\psi\pi)$ is much larger than
$\Gamma(Z_c(3900)\to \psi(2S)\pi)$. This feature is consistent with
the experimental observations. The obtained decay ratios also favor
the molecule assignments for the $Z_b$ states. However, the
calculation does not favor the pure molecule assignment for the
$Z_c(4430)$, which is different from the study with the naive
nonrelativistic quark model \cite{Liu:2014eka}.

In Ref. \cite{Voloshin:2019ivc}, Voloshin considered the decays
$Z_c(4020)\to X(3872)\gamma$ and $Z_c(4020)\to X(3872)\pi$. He
assumed that $X(3872)$ is dominantly an $S$-wave $D^0\bar{D}^{*0}$
molecule and that the $Z_c(4020)$ is an $S$-wave $D^*\bar{D}^*$
resonance, where the transitions occur through $D^*\to
D\gamma,D\pi$. The rates were found to be at most several tenths
percent in terms of the branching fraction for the $Z_c(4020)$.

\subsubsection{The $Z_b$ states}

In Ref. \cite{Chen:2015jgl}, the role of the $Z_b$ states in the
$\Upsilon(nS)\to \Upsilon(mS)\pi\pi$ ($m<n\le 3$) transitions was
analyzed by studying the final $\pi\pi$ rescattering effects with
the dispersion theory. The $Z_b$ effects in $\Upsilon(2S)\to
\Upsilon(1S)\pi\pi$ and $\Upsilon(3S)\to \Upsilon(2S)\pi^0\pi^0$ are
very small, but those in $\Upsilon(3S)\to \Upsilon(1S)\pi\pi$ are
significant. The inclusion of the $Z_b$ states and $\pi\pi$
interactions can explain the observed anomaly of
$\Upsilon(3S)\to\Upsilon\pi\pi$. Further in Ref.
\cite{Chen:2016mjn}, the effects of the two intermediate $Z_b$
states and bottom meson loops on the dipion transition processes
$\Upsilon(4S)\to \Upsilon(1S,2S)\pi\pi$ were studied. The authors
found that the contribution from meson loops is comparable to those
from the chiral contact terms and the $Z_b$-exchange terms. With the
inclusion of $Z_b$'s and meson loops, the authors can reproduce the
experimental two-hump behavior of the $\pi\pi$ spectra in
$\Upsilon(4S)\to \Upsilon(2S)\pi\pi$. In the process
$\Upsilon(4S)\to \Upsilon(1S)\pi\pi$, they expect a narrow dip
around 1 GeV in the $\pi\pi$ invariant mass distribution.

In Ref. \cite{Guo:2016bjq}, Guo et al studied the line shapes of the
near-threshold states $Z_b(10610)$ and $Z_b(10650)$ with
$I^G(J^{PC})=1^+(1^{+-})$ using their parametrization method. They
found that the $Z_b(10610)$ is a virtual state on the second Riemann
sheet near the $B\bar{B}^*$ threshold, while the $Z_b(10650)$ is a
resonance on the third or fourth Riemann sheet near the
$B^*\bar{B}^*$ threshold.

In Ref. \cite{Bondar:2016pox}, Bondar and Voloshin discussed the
possibility that the production channels in the $\Upsilon(6S)$ mass
region are dominated by the $Z_b(10610)$ (not $Z_b(10650)$) due to
the triangle singularity in $e^+e^-\to B_1(5721) \bar{B}\to
Z_b(10610)\pi$. If this mechanism is dominant, any nonresonant
background not associated with the $Z_b(10610)$ should be
suppressed. A similar structure near 11.06 GeV in the $e^+e^-$
annihilations into $\Upsilon(ns)\pi\pi$ and $h_b(kP)\pi\pi$ was also
expected.

In Ref. \cite{Xiao:2017uve}, the hidden-bottom decays of the
$Z_b(10610)$ and $Z_b(10650)$ into $\Upsilon(nS)\pi$ ($n=1,2,3$),
$\eta_b(mS)\rho$ ($m=1,2$), and $\eta_b(mS)\gamma$ were analyzed
with the effective Lagrangian approach, via final state interactions
or meson loop contributions. With this mechanism, the final state
interactions were found to be important, and the branching ratios
into $\eta_b(mS)\rho$ and $\eta_b(mS)\gamma$ were predicted. In the
same framework, the hidden-charm decays of the $Z_c(3900)$ and
$Z_c(4020)$ states in Ref. \cite{Xiao:2018kfx} were consistent with
the recent BESIII measurement \cite{Yuan:2018inv}.

In Ref. \cite{Voloshin:2017gnc}, the bottomoniumlike $Z_b(10610)$
and $Z_b(10650)$ resonances were proposed to be a mixture of the
molecular $B\bar{B}^*/B^*\bar{B}$ and $B^*\bar{B}^*$ states with one
mixing angle $\theta$. With $\theta\approx 0.2$, the ratio
$\Gamma[Z_b(10650)\to B\bar{B}^*/B^*\bar{B}]/\Gamma[Z_b(10650)\to
B^*\bar{B}^*]\sim 0.04$ and a definite interference pattern in the
process $\Upsilon(5S)\to B\bar{B}^*/B^*\bar{B} \pi$ were predicted.

In Ref. \cite{Wang:2018jlv}, the experimental data for the two $Z_b$
states were analyzed simultaneously in the molecule picture using
the Lippmann-Schwinger equations. The long-range pion interaction
does not affect the line shapes if only $S$-waves are considered,
but the situation is slightly different once $D$-waves are included.
The study also indicates that the two $Z_b$ states can be described
by poles on the unphysical Riemann sheets, where the $Z_b(10610)$ is
associated with a virtual state just below the $B\bar{B}^*$
threshold, while the $Z_b(10650)$ is likely an above-threshold
shallow state. In Ref. \cite{Baru:2019xnh}, the same method was
applied to the spin partner states $W_b$'s with
$J^{PC}=(0,1,2)^{++}$. The pionful (pionless) approach leads to the
threshold cusp (above-threshold hump) in the line shapes.

In Ref. \cite{Wu:2018xaa}, the production of $Z_b(10610)$ and
$Z_b(10650)$ from the $\Upsilon(5S,6S)$ decays through bottom-meson
loops was investigated. The contributions from the
$B_1^\prime\bar{B}^{(*)}/B_0^*\bar{B}^*$ loops dominate those from
the $B^{(*)}\bar{B}^{(*)}$ loops due to the larger pionic couplings
in the triangle diagrams. The branching ratios of $\Upsilon(6S)\to
Z_b(10610)\pi$ and $Z_b(10650)\pi$ were predicted to be around 4\%.

In Ref. \cite{Goerke:2017svb}, the strong decays of $Z_b(10610)$ and
$Z_b(10650)$ were studied with molecular type currents using the same
method in studying the decay widths of $Z_c(3900)$ and $Z(4430)$
\cite{Goerke:2016hxf}. The $\Upsilon(1S)\pi$ and $\eta_b\rho$ decay
modes are suppressed but not as much as the Belle data.

Inspired by BESIII's new measurement
$\Gamma(Z_c(3900)\to\rho\eta_c)/\Gamma(Z_c\to\pi J/\psi)=2.1\pm0.8$
 \cite{Yuan:2018inv}, Voloshin estimated the cross section
of the processes $e^+e^-\to\gamma W_{bJ}$ at the maximum of the
$\Upsilon(5S)$ resonance to be about 0.1 pb \cite{Voloshin:2018pqn},
where $W_{bJ}$ are the predicted isovector $B^{(*)}\bar{B}^{(*)}$
meson-antimeson states from the considerations of the heavy quark
spin symmetry and light quark spin symmetry \cite{Voloshin:2016cgm}.

\subsection{$QQ\bar{q}\bar{q}$ and $QQ\bar{Q}\bar{Q}$}

In Ref. \cite{DelFabbro:2004ta}, the feasibility of the production
and detection of $T_{cc}$ at various facilities was discussed. Its
production is comparable to the doubly charmed baryon production and
prompt $J/\psi c\bar{c}$ production. Thus, its observations at
SELEX, LHC, and Belle should be possible.

The production of the $T_{cc}$ states in the electron-positron
process has been discussed in Refs.
\cite{Hyodo:2012pm,Hyodo:2017hue,Jin:2014nva}. To search for these
exotic states at $B$ factories, the authors of Ref.
\cite{Jin:2014nva} proposed a more efficient analysis method. If a
three-jet event may be identified by their energy and angular
distributions, the doubly charmed states can be searched for in the
most energetic jet, which may suppress the combination background
fluctuations of the reconstructed hadron mass spectra.

In \cite{Esposito:2013fma}, the authors argued that the doubly
charmed tetraquarks are a straightforward consequence of the
diquark-antidiquark model \cite{Maiani:2004vq,Maiani:2014aja}.
Assuming that these tetraquarks are above-threshold particles, the
authors explored their decay modes and rates as well as their
production mechanisms. They showed that the doubly charmed
tetraquarks could be produced at LHC from the decays of $B_c$ and
$\Xi_{bc}$.

In Ref. \cite{Hong:2018mpk}, the authors noted that the hadronic
effects on the $T_{cc}$ tetraquark state are insignificant in
relativistic heavy ion collisions. The authors also noticed that the
$D^+D^+\pi^-$ and $D^0D^0\pi^+$ decay modes of $T_{cc}$ should be
the most probable channels to reconstruct it from heavy ion
collisions. Besides, the order of yields at LHC and RHIC reads,
$D>\Xi_{cc}^{(*)}>T_{cc}$(molecule)$>T_{cc}$(compact).

In Ref. \cite{Ali:2018ifm}, Ali {\it et al} explored the production
of $T^{\{bb\}}_{[\bar{u}\bar{d}]}$, $T^{\{bb\}}_{[\bar{u}\bar{s}]}$,
$T^{\{bb\}}_{[\bar{d}\bar{s}]}$, and $\Xi_{bbq}$ ($q=u,d,s$), in
$Z$-boson decays. The lifetimes of these tetraquarks were estimated
to be around 0.5 ps. Besides, their signature decay modes were
studied. In Ref. \cite{Ali:2018xfq}, the discovery potential of
these tetraquarks at LHC was discussed. According to the estimated
cross section $\sigma(pp\to T^{\{bb\}}+X)$, the prospects of
discovering these states are excellent.

In Ref. \cite{Xing:2018bqt}, the semileptonic 2-, 3-, and 4-body
weak decays of $T_{QQ\bar{q}\bar{q}}$ ($Q=c,b$, $q=u,d,s$) were
systematically investigated. The nonleptonic decays and possible
golden decay channels at LHCb and Belle II experiments were also
discussed. The weak decays of the $I(J^P)=0(1^+)$ $T_{bb;\bar{u}
\bar{d}}^{-}$ tetraquark state to the scalar diquark-antidiquark
$Z_{bc;\bar{u}\bar{d}}=[bc][\bar{u}\bar{d}]$ were investigated in
\cite{Agaev:2018khe} using the QCD sum rule approach, and the
lifetime of the $T_{bb;\bar{u} \bar{d}}^{-}$ was found to be around
9.2 fs, which is shorter than that obtained in Ref.
\cite{Ali:2018ifm}.

In Refs. \cite{Gershon:2018gda,Ridgway:2019zks}, the inclusive decay
mode $\Xi_{bbq}\to B_c^{(*)-}+X_{c,s,q}$ was proposed as a potential
discovery channel for the doubly bottom baryon at LHC, which is
based on the fact that $B_c^{(*)}$ can only arise from the weak
decay of a hadron with two bottom quarks. If the $\Xi_{bbq}$ states
could be observed, such a decay mode may also be used to search for
the bound $T_{bb}$ states.

The possible observation of the $[cc][\bar{c}\bar{c}]$ and $[bc][\bar{b}\bar{c}]$ states with
$J^{PC}=2^{++}$ at LHC was discussed in Refs. \cite{Berezhnoy:2011xy,Berezhnoy:2011xn}.
The color connections of $QQ\bar{Q}\bar{Q}$ states and the induced
hadronization effects on $\Xi_{cc}$ and $T_{cc}$ in $e^+e^-$
annihilation processes were studied in \cite{Jin:2013bra}. A study
of $T_{4c}=cc\bar{c}\bar{c}$ production in double parton scattering
was given in Ref. \cite{Carvalho:2015nqf}. Possible channels to
search for such a $bb\bar{b}\bar{b}$ state were discussed in Refs.
\cite{Eichten:2017ual,Vega-Morales:2017pmm}.

In Ref. \cite{Li:2019uch}, the authors investigated the lifetime and
weak decays of the fully heavy tetraquark state $bb\bar{c}\bar{c}$
with $J^P=0^+$. They obtained a lifetime value around 0.1$\sim$0.3
ps and discussed the golden channels to search for this state.

\subsection{$Q\bar{Q}qqq$}

The analysis of the process $\Lambda_b\to J/\psi K^-p$ in Ref.
\cite{Roca:2015dva} by Roca, Nieves, and Oset supports the
$P_c(4450)$ as a $3/2^-$ molecule of mostly $\bar{D}^*\Sigma_c$ and
$\bar{D}^*\Sigma_c^*$. In Ref. \cite{Roca:2016tdh}, an improved
analysis by Roca and Oset indicates that the existence of the
$P_c(4380)$ state cannot be undoubtedly claimed with only the fit to
the $K^-p$ and $J/\psi p$ mass distributions. 
The investigation of $\Lambda_b^0\to J/\psi K^-p$ and $J/\psi
p\pi^-$ in Ref. \cite{Xiao:2016ogq} indicates that the s-/u-channel
contributions are not negligible. The obtained bound states and line
shape of the $J/\psi N$ mass distribution favor the assignment of
the $P_c(4450)$ as a $\bar{D}^*\Sigma_c$ molecule.

In Ref. \cite{Lu:2016roh}, the process $\Lambda_b\to J/\psi
K^0\Lambda$ was studied with the contributions of a hidden-charm
pentaquark-like molecule state \cite{Wu:2010jy,Wu:2010vk}. A clear
peak in the $J/\psi\Lambda$ mass distribution was found. In Ref.
\cite{Huang:2016tcr}, the contributions of the hidden-charm $N^*$
states with $J^P=1/2^-$ and $3/2^-$ to the reaction $\gamma p\to
\bar{D}^{*0}\Lambda_c^+$ were discussed. These states lead to clear
peak structures in the total cross sections. The contributions of a
hidden-beauty $N^*(11052)$ to the $\pi^-p$ scattering were discussed
similarly in Ref. \cite{Cheng:2016ddp}. The authors of Ref.
\cite{Xie:2017gwc} suggested that a clean peak should be seen for a
hidden-charm resonance (mass$\sim$4265 MeV), which couples to the
$\eta_c p$ channel in the decay $\Lambda_b\to \eta_c K^-p$.

In Ref. \cite{Wang:2016vxa}, the production of the $P_c$ states and
their yields in ultrarelativistic heavy ion collisions were
discussed in both the compact pentaquark picture and the molecule
picture. In Ref. \cite{Schmidt:2016cmd}, the production cross
section for a hidden-charm pentaquark in proton-nucleus collisions
was found to be sizable by assuming if it is a molecule of a
charmonium and a light baryon.

In Ref. \cite{Liu:2016dli}, Liu and Oka discussed the rescattering
effects in the reaction $\pi^-p\to\pi^-J/\psi p$ due to the
intermediate open-charm hadrons. They showed that the triangle
singularity peaks can simulate the pentaquark-like resonances in the
$J/\psi p$ invariant mass distributions. In Ref. \cite{Guo:2016bkl},
Guo {\it et al} proposed that the $P_c(4450)$ observed in the
process $\Lambda_b\to J/\psi pK$ might be due to a triangle
singularity around the $\chi_{c1}p$ threshold. A discussion on
triangle singularities in Ref. \cite{Bayar:2016ftu} indicates that
the narrow $P_c(4450)$ would have an origin other than the triangle
singularity from the $\Lambda^*$-charmonium-proton intermediate
states, if this baryon has quantum numbers $J^P=3/2^-,5/2^+$.

In studying the nature of $Z_c(3900)$ with the pole counting rule in
Ref. \cite{Gong:2016hlt}, the authors suggested that the $P_c(4450)$
pentaquark can be related to $Z_c(3900)$ by replacing a $ud$ pair by
a $\bar{d}$. They speculated that the $P_c(4450)$ should be a
$\Sigma_c\bar{D}^*$ molecule with $J=3/2$ rather than $J=5/2$.

In Ref. \cite{Kim:2016cxr}, Kim {\it et al} considered the reaction
$\pi^-p\to J/\psi n$ based on the hybridized Regge model and
included the contributions of the LHCb $P_c$ states. The total cross
sections were found to be about 1 nb in the vicinity of the heavy
pentaquark masses. In Ref. \cite{Karliner:2017qje}, Karliner and
Rosner discussed the $J/\psi N$ photoproduction on deuterium,
through which the simultaneous investigation of the $c\bar{c}uud$
and $c\bar{c}ddu$ isospin partner states can be achieved. The
photoproduction of $P_c(4450)$ on a deuteron target may be used to
check whether it is a genuine isospin-half resonance.

The authors of Ref. \cite{Zhou:2017bhq} studied the pentaquark
production in the semileptonic decays $\Lambda_b\to J/\psi
p\ell^-\bar{\nu}_\ell$ where $\ell=e,\mu$. The branching ratios are
about two orders of magnitude smaller than the decay $\Lambda_b\to
p\mu\bar{\nu}_\mu$, which is accessible to the ongoing experiments
at the LHCb.

In Ref. \cite{Li:2017ghe}, the hidden-charm pentaquark production at
the $e^+e^-$ colliders was discussed in the NRQCD factorization
framework, by treating the pentaquark as a $c\bar{c}nnn$ state with
a color-octet $c\bar{c}$ configuration. It is possible to search
for the direct pentaquark production signal at $e^+e^-$ colliders.
The authors of Ref. \cite{Buccella:2018jnt} discussed the production
mechanism of the $P_c$ pentaquarks in $\Lambda_b$ decays and assumed
that they are $(c\bar{c})_{8_c}(qqq)_{8_c}$ states. The decay
happens through $b\to c+s+\bar{c}$ followed by a gluon emission
($\to u\bar{u}$) from $s$. Then $s$ and $\bar{u}$ form the kaon
meson while the other quarks form the pentaquarks. The two color
octet $(c\bar{c})_{8_c}$ and $(qqq)_{8_c}$ components in $S$-wave
($P$-wave) lead to the $J^P=3/2^-$ ($5/2^+$) state. This mechanism
can explain why only a few states with nonminimal constituents were
discovered so far.

In Ref. \cite{Pilloni:2018kwm}, the difference between partial-wave
analysis formalisms used in the construction of three-body
amplitudes involving fermions was discussed, particularly in the
decay $\Lambda_b\to J/\psi K^-p$.  The authors found meaningful
effects on the resonance pole position extraction, which is
particularly relevant when several resonances overlap and the
quantum number assignment is not stable. In Ref.
\cite{Paryev:2018fyv}, the authors discussed the role of the
hidden-charm pentaquark resonance $P_c(4450)$ in $J/\psi$
photoproduction off $^{12}C$ and $^{208}Pb$ target nuclei near
threshold. The presence of $P_c(4450)$ produces additional
enhancements above threshold in the total $J/\psi$ creation cross
section on nuclei, which can be tested in future JLab experiments.

In Ref. \cite{Voloshin:2019wxx}, Voloshin discussed the hidden-charm
pentaquark production in $\bar{p}d$ collisions, which is possible
due to the coupling of charmonia to the $p\bar{p}$ channel. He found
that the pentaquark formation may happen with the nucleons moving
slowly inside the deuteron due to the masses of the pentaquark,
charmonium, and the nucleon being close to a special kinetic
relation. The formation cross section of hypothetical pentaquark
states decaying to $\eta_cN$ was much larger than that of
pentaquarks into $J/\psi N$, because of a much larger $\eta_c\to
p\bar{p}$ decay width.

The authors of Ref. \cite{Lu:2016nnt} studied the strong decay mode
$J/\psi p$ of the $P_c$ states in the molecule scenario. The partial
decay widths are significantly different for various $J^P$
assignments. The $S$-wave $\Sigma_c\bar{D}^*$ pictures for the
$P_c(4380)$ and $P_c(4450)$ and the $\Sigma_c^*\bar{D}$ assignment
for the $P_c(4380)$ with $J^P=3/2^-$ are all allowed by the present
experimental data.

In Ref. \cite{Shen:2016tzq}, the different decay properties of the
$\bar{D}\Sigma_c^*$ and $\bar{D}^*\Sigma_c$ molecules were
discussed. The $P_c(4380)$ was proposed as a $\bar{D}\Sigma_c^*$
molecule and the decay channel $\bar{D}^*\Lambda_c$ was suggested to
be used to disentangle its nature. In Ref. \cite{Lin:2017mtz}, the
decay properties of the $P_c(4380)$ and $P_c(4450)$ were studied
with the effective Lagrangian framework by treating them as
meson-baryon molecules. The $\bar{D}\Sigma_c^*$ and
$\bar{D}^*\Sigma_c$ molecules have different decay branching ratios.
The $P_c(4380)$ ($P_c(4450)$) was proposed to be a $3/2^-$
$\bar{D}\Sigma_c^*$ ($5/2^+$ $\bar{D}^*\Sigma_c$) molecule. The
cross sections for the processes $\gamma p\to J/\psi p$ and $\pi
p\to J/\psi p$ through the $S$-channel $P_c$ states were also
calculated. The same framework was used to discuss the decay
behaviors of the strange and beauty partners of the $P_c$ hadronic
molecules in Ref. \cite{Lin:2018kcc}.

Besides the spectrum, the authors of Ref. \cite{Eides:2017xnt} also
calculated the partial decay width of the $J^P=3/2^-$ $P_c(4450)$
into $J/\psi N$ to be around 11 MeV by treating it as a
hadroquarkonium state. Later in Ref. \cite{Eides:2018lqg}, the
decays of the pentaquark $P_c(4450)$ were further investigated in
both hadrocharmonium and molecular pictures. The authors found that
the decay patterns are vastly different. The decays of the $P_c$
molecule into $J/\psi$ are strongly suppressed, while the opposite
happens in the hadrochamonium case.

In Ref. \cite{Azizi:2018bdv}, the authors studied the strong decay
width of $P_c(4380)$ into $J/\psi N$ by treating it as a
$\bar{D}^*\Sigma_c$ molecule with $J^P=3/2^+$ or $3/2^-$. The
coupling constants were calculated with QCD sum rules, and the
obtained decay width is around 187 (213) MeV for the case $P=+$
($P=-$).

After the new pentaquark states $P_c(4312)$, $P_c(4440)$, and $P_c(4457)$ were announced~\cite{lhcbnew}, Guo, Jing, Meissner and Sakai suggested to search for the $P_c(4457)$ in the $J/\psi\Delta$ mode in order to test its nature \cite{Guo:2019Pc}. If the state is a $\Sigma_c\bar{D}^*$ molecule, the authors pointed out that the isospin breaking decay ratio ${\cal B}(P_c(4457)\to J/\psi\Delta^+)/{\cal B}(P_c(4457)\to J/\psi p)$ would be at the level ranging from a few percent to about $30\%$.

\subsection{A short summary}

We give a short summary for the production and decay properties of
the heavy tetraquark and pentaquark states based on the recent
studies in the literature.

\begin{itemize}

\item The $X$ states. In the conventional charmonium picture, the
$X(3940)$ could be the $\eta_c(3S)$ while the $X(4140)$ or $X(4274)$
could be the $\chi_{c1}(3P)$. In the molecule picture, the $X(4160)$
could be a $D_s^*\bar{D}_s^*$ molecule.

\item
The $Y$ states. From the theoretical studies of production and
decay, it seems that not all $Y$ states ($Y(4008)$,
$Y(4220)/Y(4230)$, $Y(4260)$, $Y(4320)$, $Y(4360)$, $Y(4390)$,
$Y(4630)$, and $Y(4660)$) shall exist. The interference effects,
coupled channel effects, or final state interactions may make
several genuine states behave like more resonance signals. Up to
now, we have much deeper understandings about the nature of
$X(3872)$ than 15 years ago, although more experimental measurements
are still needed. Different from the $X(3872)$, the $J^{PC}=1^{--}$
structures in the mass region $4.0\sim4.6$ GeV seem more complicated
and challenging. The conventional charmonium states, theoretically
expected but elusive hybrid charmonium mesons, and molecular states
coincide here. Do they mix? The nature of $Y(4260)$ is still far
from being understood, although it was observed almost 15 years ago.

\item
The $Z_c$ and $Z_b$ states. The studies of productions and decays
favor the molecule assignments for the $Z_c(3900)$ and $Z_c(4020)$.
Whether the $Z_c(4430)$ state is a molecule, a compact tetraquark,
or a structure induced by triangle singularity is still unclear. One
cannot distinguish the nature of $Z_c(4100)$ with the present
experimental $B$ decay data, either. Is it a P-wave $D^*\bar{D}^*$
resonance or a state caused by final state interaction effects? The
$Z_b(10610)$ and $Z_b(10650)$ should be the $B\bar{B}^*$ and
$B^*\bar{B}^*$ molecular states, respectively.

\item
The $Q\bar{Q}qqq$ states. The $\Sigma_c^*\bar{D}$ and
$\Sigma_c\bar{D}^*$ molecule assignments for $P_c(4380)$ and
$P_c(4450)$, respectively, can explain experimental measurements for
their production and decay properties, although their nonresonance
interpretations cannot be excluded.

\item
So far, a large number of investigations on exotic productions have
been presented in the literature while the investigations on decays
are still lacking. Further studies on decay properties of the exotic
hadrons, especially at the quark level, are necessary, which play a
key role in understanding the inner structures of the exotic states.
On the experimental side, searching for expected multiquarks,
checking predicted features, and measuring more physical quantities
are definitely needed.

\item
At present, all the multiquark candidates are produced at $e^+e^-$,
$e^+p$, $pp$, $pA$, and $AA$ colliders. Future $p\bar{p}$,
$\bar{p}A$, and $eA$ colliders are still under construction. Besides
these collision processes, the $\gamma\gamma,\gamma A$ interactions,
charmonia decays, bottomonia decays, $B$ decays, and $\Lambda_b$
decays are also used to study the multiquark states. From the
experimental fact that all the confirmed exotic candidates were
observed in the $e^+e^-$ ($\gamma^*$), $\gamma\gamma$, or hadron
decay processes, it seems that producing multiquark states in
multiproduction processes is very difficult and undetectable
\cite{Jin:2016cpv,Han:2009jw}. If this is true, searching for the
conventional $bbb$ baryons should be more feasible and important
than searching for $bb\bar{b}\bar{b}$ compact states. The exotic
hadrons we may observe in the near future would be dominantly states
in the charm sector. They are indeed the structures being studied at
LHC (see Table 26 of Ref. \cite{Cerri:2018ypt}).

\end{itemize}

%% file: section11.tex
\section{Summary and perspective}

\begin{figure}[htbp]
\centering
\includegraphics[width=6in]{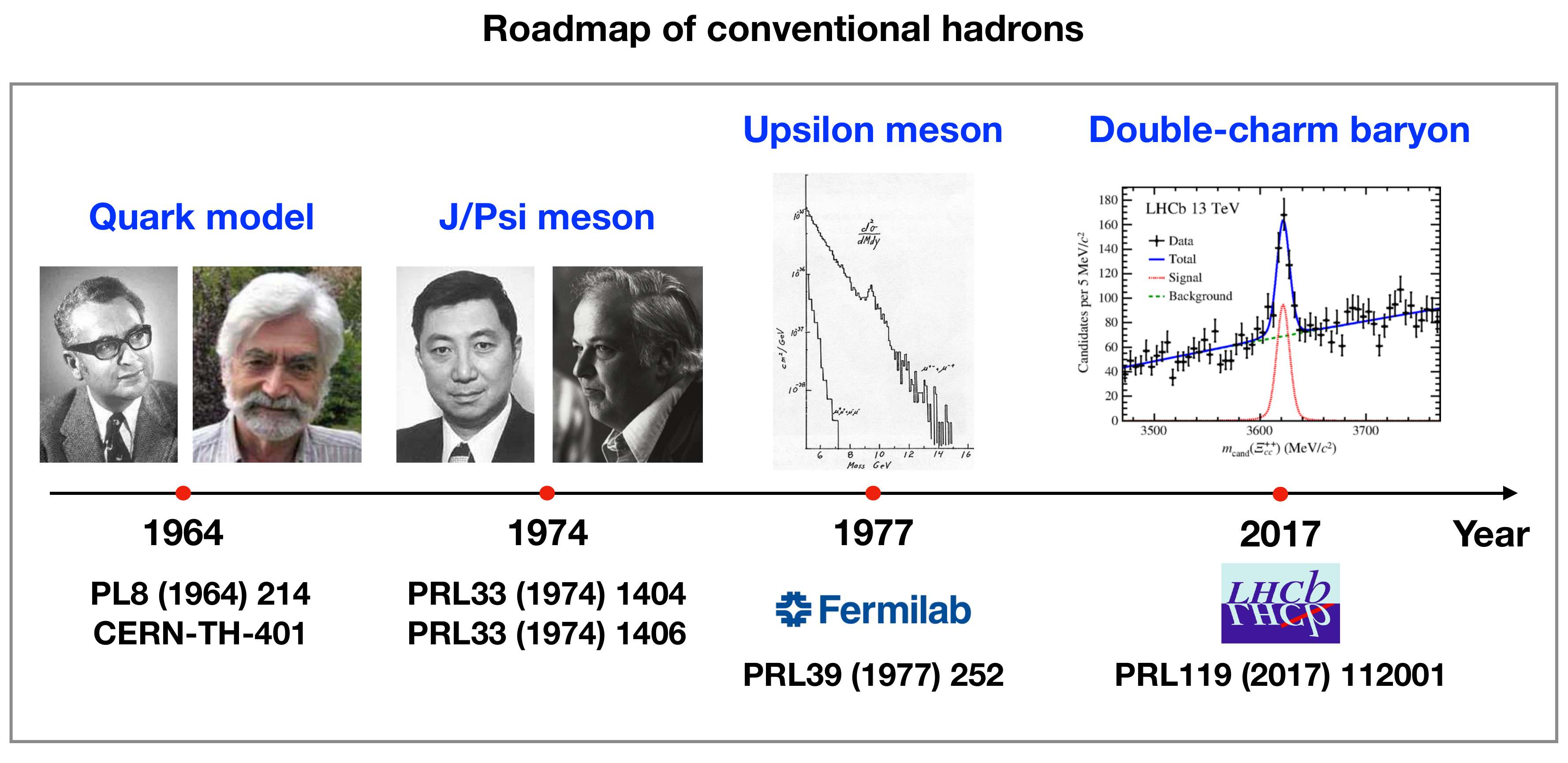}
\caption{The roadmap of the conventional hadrons.}
\label{RoadmapofQM}
\end{figure}

At the end of this review, we would give a brief summary and
perspective for the progress on the multiquark physics. It is useful
and instructive to recall the establishment of the quark model
before we discuss the multiquark systems. The quark model was first
proposed in 1964 to deal with the classification of hadrons. In
1974, the discovery of the $J/\psi$ meson gave the direct evidence
for the existence of the charm quark, which was known as ``November
Revolution". Actually, the existence of the charm quark was already
predicted in the GIM mechanism before its observation
~\cite{Glashow:1970gm}. The bottom quark was discovered by E288
experiment in 1977 at Fermilab. After that, plenty of bound states
and excitations of the heavy mesons and baryons have been observed
in the following decades ~\cite{Tanabashi:2018oca}. We show the
roadmap of the conventional hadrons in Fig.~\ref{RoadmapofQM}.

Although we still can't understand QCD confinement from the first
principle, the quark model has achieved indubitable success in
systematizing the properties of mesons and baryons. For the exotic
hadron states such as the multiquark systems, one may turn to those
phenomenological models for help which are closely related to QCD
since the rigorous solutions of soft QCD are still unavailable. The
interplay between the phenomenological models and experimental
observations will also improve their prediction capabilities. The
discovery of the $\Xi_{cc}(3621)$ in 2017 was a successful example
for the collaboration between experimenters and theorists. Inspired
by this observation, one will further expect the future discovery of
the $\Omega_{ccc}$ baryon.

Searching for the multiquark states is another central issue of the
hadron physics. In the past several decades, there accumulated
abundant investigations on tetraquarks and pentaquarks based on
various phenomenological models. However, the progress was still
rudimentary and far away from our expectation due to the lack of the
sufficiently accurate experimental results. The year of 2003 is very
important in the history of the exotic hadrons. Since 2003, lots of
quarkonium-like $XYZ$ states and pentaquark $P_c$ states have been
observed in the modern facilities, such as BaBar, Belle, CLEO-c, D0,
CDF, BESIII, and LHCb etc.

The desires for the nature of these new hadron states have promoted
the model building. Some model predictions agree with the
experimental data. For example, the masses of some $XYZ$ states can
be reproduced in the tetraquark configurations. However, there are
still many observables which can not be computed accurately enough
to be compared with the experimental measurements, e. g., the
branching fractions and decay widths for some important channels.
Moreover, some new resonances can usually be explained in different
models, implying the present phenomenological investigations are unable to
distinguish their underlying structures. More efforts are needed to
bring down the error bars in future.

Nevertheless, we can still find some critical breakthroughs in the
phenomenological aspect. The investigations for the doubly heavy
tetraquarks with various methods such as the QCD sum
rules~\cite{Du:2012wp}, CMI
model~\cite{Luo:2017eub,Karliner:2017qjm},
LQCD~\cite{Junnarkar:2018twb}, and the heavy-quark symmetry
prediction~\cite{Eichten:2017ffp} lead to consistent predictions
that the doubly bottom $ud\bar b\bar b$ tetraquark states would
probably be stable against the strong interaction. Such an agreement
may imply critical interaction mechanisms in these tetraquark
systems, which can be a breaking point for our understanding of the
multiquark states. To date, most of the multiquark candidates were
observed in the charmonium sector, such as the hidden-charm XYZ
states and $P_c$ pentaquark states. From the theoretical point of
view, more tetraquark and pentaquark states with different flavor
and color structures and different quantum numbers should also be
possible. In the past several years, the authors have made some
efforts along this direction. We predicted the pentaquark states
with hidden-charm and open-strange flavors. Based on the discovery
of $\Xi_{cc}(3621)$, we studied the interactions between the
$\Xi_{cc}(3621)$ baryon and $S$-wave charmed mesons and predicted
the existence of the triply charmed pentaquark states.

\begin{figure}[htbp]
\centering
\includegraphics[width=3in]{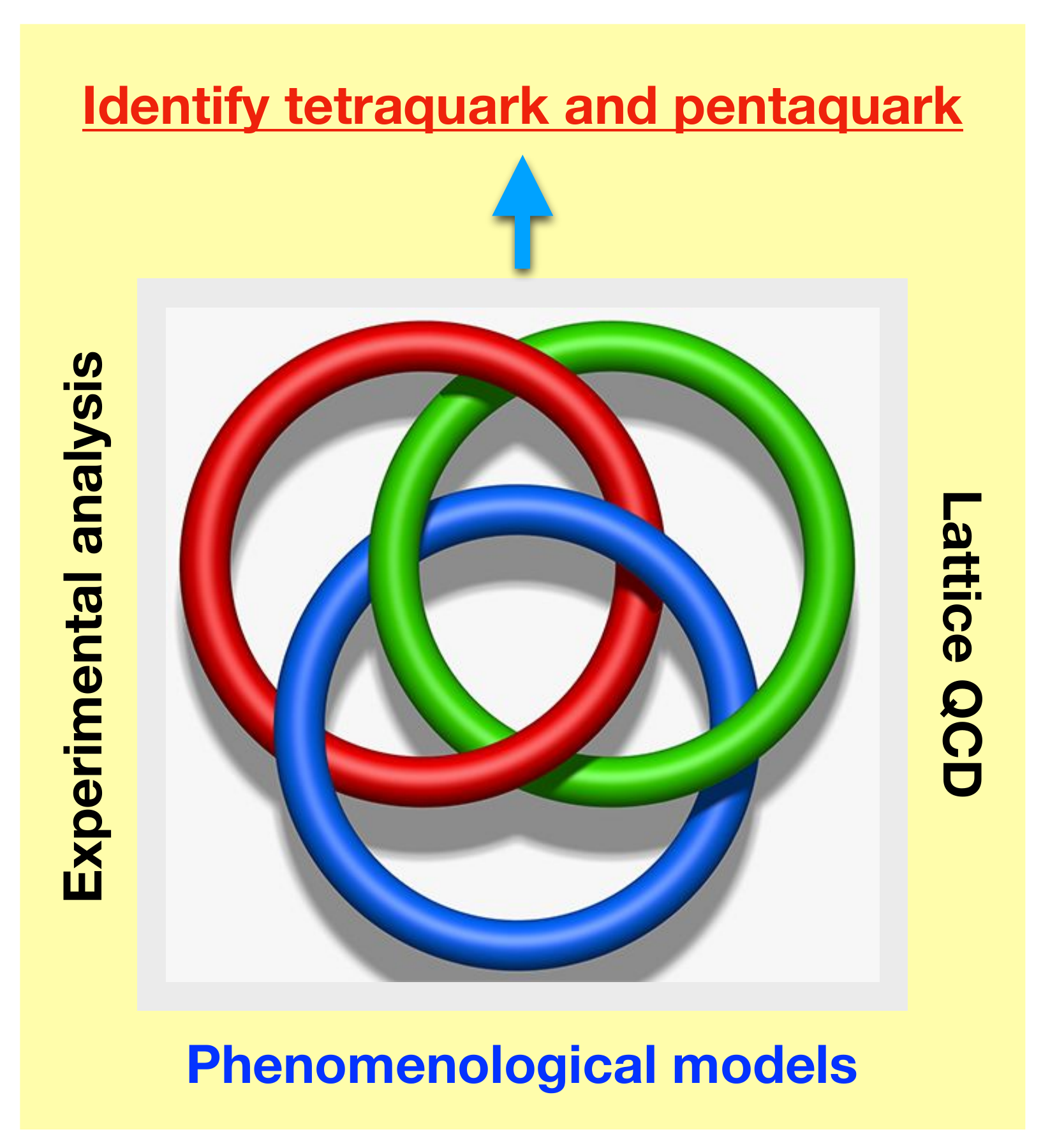}\\
\caption{The interplay of the experimental analyses,
phenomenological models and lattice QCD on the understanding of the
new hadron spectroscopy.} \label{borromean}
\end{figure}

Last but not the least, we shall emphasize the unique and pivotal
role of the lattice QCD in the search of the multiquark states. The
lattice QCD simulations start from the first principle and can
investigate the few-hadron reaction processes and extract important
information such as the resonance properties, the hadronic
interactions, scattering amplitudes, phase shifts, binding energies,
mass spectra, decay widths and so on. In the past few years,
tremendous lattice results from quenched to unquenched calculations
become accurate enough to be compared with both the experimental
data and phenomenological results. Especially the simulations using
a large basis of operators including the multi-hadron operators have
become a unique, vital and powerful tool to explore the
near-threshold exotic states. Some very successful examples include
the lattice QCD investigations of the $X(3872)$ and $Z_c(3900)$
states, the doubly bottom tetraquark states, which provide very
clear and intuitive picture of the underlying structures of the
exotic states. However, only the $\eta_c-N$ and $J/\psi-N$
scattering processes were studied in lattice QCD for the pentaquark
systems, which is far away from pinning down the existence of the
hidden-charm pentaquark states. In terms of LHCb's recent discovery
of the three hidden-charm pentaquark states, the lattice QCD
simulations of the charmed baryon and anti-charmed meson are badly
needed. At the present stage, many lattice simulations were
performed at the unphysically large pion mass due to the expensive
computational cost. However, the existence of the molecular type of
exotic hadrons is very sensitive to the pion mass since the pion
exchange force decays exponentially with the pion mass. Further
dynamical lattice QCD simulations using ensembles with near-physical
or even the physical pion mass will be extremely valuable for the
$P_c$ systems. As shown in Fig.~\ref{borromean}, the interplay among
experiments, phenomenological models and lattice QCD and their joint
progress will definitely sharpen our understanding of the new hadron
spectroscopy and QCD itself.

When we were preparing this review, the LHCb Collaboration reported
a new charmonium state with $J^{PC}=3^{--}$ via $pp$ collision in
the PHI-PSI 2019 workshop. The BESIII Collaboration has observed the
processes $e^+e^-\to\pi^+\pi^-\psi(3770)$ and $D_1(2420)^0\bar
D^0+c.c.$ for the first time~\cite{Ablikim:2019faj} and measured the
cross section of the process
$e^+e^-\to\omega\chi_{c0}$~\cite{Ablikim:2019apl}, which provided
new useful information to study the $XYZ$ states. Very recently, LHCb
has observed CP violation in the charmed meson
decays~\cite{Aaij:2019kcg}, which is definitely an important
milestone in the history of particle physics. These excellent
observations at LHCb demonstrated its distinguished performances and
capabilities to study the exotic hadron states. The Belle II
Collaboration released the Belle II Physics Book~\cite{Kou:2018nap}
in 2018 and started to take data with a fully instrumented detector
in March, 2019. Although the main aim of Belle II experiment is
searching for CP-violation and revealing violations of the symmetry
between particles and anti-particles, it will also search for the
exotic hadrons and make precision measurements of their properties.
To some extent, BelleII shall become a factory of the
charmonium-like $XYZ$ states. The BESIII experiment at BEPCII will
continue to contribute to the field of the $XYZ$ states in the near
future. They will perform unique investigations on the $XYZ$ states
around $\sqrt{s}=4.2$ GeV, 4.38 GeV and 4.6 GeV. Moreover, the
upcoming PANDA experiment will also study the multiquark and other
exotic hadrons in future. The coming decade shall witness a new
landscape of the hadron spectroscopy.

%% file: multiquark.bbl
\begin{thebibliography}{100}
\expandafter\ifx\csname url\endcsname\relax
  \def\url#1{\texttt{#1}}\fi
\expandafter\ifx\csname urlprefix\endcsname\relax\def\urlprefix{URL }\fi
\expandafter\ifx\csname href\endcsname\relax
  \def\href#1#2{#2} \def\path#1{#1}\fi

\bibitem{Chen:2016qju}
H.-X. Chen, W.~Chen, X.~Liu, S.-L. Zhu, {The hidden-charm pentaquark and
  tetraquark states}, Phys. Rept. 639 (2016) 1--121.
\newblock \href {http://arxiv.org/abs/1601.02092} {\path{arXiv:1601.02092}},
  \href {http://dx.doi.org/10.1016/j.physrep.2016.05.004}
  {\path{doi:10.1016/j.physrep.2016.05.004}}.

\bibitem{Okun:1962kca}
L.~B. Okun, {The theory of weak interaction}, in: {High-energy physics.
  Proceedings, 11th International Conference, ICHEP'62, Geneva, Switzerland,
  Jul 4-11, 1962}, 1962, pp. 845--866.

\bibitem{GellMann:1964nj}
M.~Gell-Mann, {A Schematic Model of Baryons and Mesons}, Phys. Lett. 8 (1964)
  214--215.
\newblock \href {http://dx.doi.org/10.1016/S0031-9163(64)92001-3}
  {\path{doi:10.1016/S0031-9163(64)92001-3}}.

\bibitem{Zweig:1964jf}
G.~Zweig, {An SU(3) model for strong interaction symmetry and its breaking.
  Version 2}, in: D.~Lichtenberg, S.~P. Rosen (Eds.), DEVELOPMENTS IN THE QUARK
  THEORY OF HADRONS. VOL. 1. 1964 - 1978, 1964, pp. 22--101.

\bibitem{Zweig:1981pd}
G.~Zweig, {An $SU(3)$ model for strong interaction symmetry and its breaking.
  Version 1}, 1964.

\bibitem{Barnes:1964pd}
V.~E. Barnes, et~al., {Observation of a Hyperon with Strangeness -3}, Phys.
  Rev. Lett. 12 (1964) 204--206.
\newblock \href {http://dx.doi.org/10.1103/PhysRevLett.12.204}
  {\path{doi:10.1103/PhysRevLett.12.204}}.

\bibitem{GellMann:1961ky}
M.~Gell-Mann, {The Eightfold Way: A Theory of strong interaction symmetry},
  1961.

\bibitem{Neeman:1961jhl}
Y.~Ne'eman, {Derivation of strong interactions from a gauge invariance}, Nucl.
  Phys. 26 (1961) 222--229.
\newblock \href {http://dx.doi.org/10.1016/0029-5582(61)90134-1}
  {\path{doi:10.1016/0029-5582(61)90134-1}}.

\bibitem{Aubert:1974js}
J.~J. Aubert, et~al., {Experimental Observation of a Heavy Particle $J$}, Phys.
  Rev. Lett. 33 (1974) 1404--1406.
\newblock \href {http://dx.doi.org/10.1103/PhysRevLett.33.1404}
  {\path{doi:10.1103/PhysRevLett.33.1404}}.

\bibitem{Augustin:1974xw}
J.~E. Augustin, et~al., {Discovery of a Narrow Resonance in $e^+e^-$
  Annihilation}, Phys. Rev. Lett. 33 (1974) 1406--1408, [Adv. Exp. Phys. 5, 141
  (1976)].
\newblock \href {http://dx.doi.org/10.1103/PhysRevLett.33.1406}
  {\path{doi:10.1103/PhysRevLett.33.1406}}.

\bibitem{Glashow:1970gm}
S.~L. Glashow, J.~Iliopoulos, L.~Maiani, {Weak Interactions with Lepton-Hadron
  Symmetry}, Phys. Rev. D2 (1970) 1285--1292.
\newblock \href {http://dx.doi.org/10.1103/PhysRevD.2.1285}
  {\path{doi:10.1103/PhysRevD.2.1285}}.

\bibitem{Abrams:1974yy}
G.~S. Abrams, et~al., {The Discovery of a Second Narrow Resonance in e+ e-
  Annihilation}, Phys. Rev. Lett. 33 (1974) 1453--1455, [Adv. Exp.
  Phys.5,150(1976)].
\newblock \href {http://dx.doi.org/10.1103/PhysRevLett.33.1453}
  {\path{doi:10.1103/PhysRevLett.33.1453}}.

\bibitem{Rapidis:1977cv}
P.~A. Rapidis, et~al., {Observation of a Resonance in $e^+ e^-$ Annihilation
  Just Above Charm Threshold}, Phys. Rev. Lett. 39 (1977) 526, [Erratum: Phys.
  Rev. Lett.39,974(1977)].
\newblock \href {http://dx.doi.org/10.1103/PhysRevLett.39.526,
  10.1103/PhysRevLett.39.974} {\path{doi:10.1103/PhysRevLett.39.526,
  10.1103/PhysRevLett.39.974}}.

\bibitem{Goldhaber:1977qn}
G.~Goldhaber, et~al., {$D$ and $D^*$ Meson Production Near 4-GeV in $e^+ e^-$
  Annihilation}, Phys. Lett. 69B (1977) 503--507.
\newblock \href {http://dx.doi.org/10.1016/0370-2693(77)90855-3}
  {\path{doi:10.1016/0370-2693(77)90855-3}}.

\bibitem{Brandelik:1978ei}
R.~Brandelik, et~al., {Total Cross-section for Hadron Production by $e^+ e^-$
  Annihilation at Center-of-mass Energies Between 3.6 {GeV} and 5.2 {GeV}},
  Phys. Lett. B76 (1978) 361.
\newblock \href {http://dx.doi.org/10.1016/0370-2693(78)90807-9}
  {\path{doi:10.1016/0370-2693(78)90807-9}}.

\bibitem{Siegrist:1976br}
J.~Siegrist, et~al., {Observation of a Resonance at 4.4-GeV and Additional
  Structure Near 4.1-GeV in $e^+ e^-$ Annihilation}, Phys. Rev. Lett. 36 (1976)
  700.
\newblock \href {http://dx.doi.org/10.1103/PhysRevLett.36.700}
  {\path{doi:10.1103/PhysRevLett.36.700}}.

\bibitem{Eichten:1974af}
E.~Eichten, K.~Gottfried, T.~Kinoshita, J.~B. Kogut, K.~D. Lane, T.-M. Yan,
  {The Spectrum of Charmonium}, Phys. Rev. Lett. 34 (1975) 369--372, [Erratum:
  Phys. Rev. Lett. 36, 1276 (1976)].
\newblock \href {http://dx.doi.org/10.1103/PhysRevLett.34.369}
  {\path{doi:10.1103/PhysRevLett.34.369}}.

\bibitem{Eichten:1978tg}
E.~Eichten, K.~Gottfried, T.~Kinoshita, K.~D. Lane, T.-M. Yan, {Charmonium: The
  Model}, Phys. Rev. D17 (1978) 3090, [Erratum: Phys. Rev. D21, 313 (1980)].
\newblock \href {http://dx.doi.org/10.1103/PhysRevD.17.3090,
  10.1103/PhysRevD.21.313} {\path{doi:10.1103/PhysRevD.17.3090,
  10.1103/PhysRevD.21.313}}.

\bibitem{Eichten:1979ms}
E.~Eichten, K.~Gottfried, T.~Kinoshita, K.~D. Lane, T.-M. Yan, {Charmonium:
  Comparison with Experiment}, Phys. Rev. D21 (1980) 203.
\newblock \href {http://dx.doi.org/10.1103/PhysRevD.21.203}
  {\path{doi:10.1103/PhysRevD.21.203}}.

\bibitem{Barbieri:1975jd}
R.~Barbieri, R.~Kogerler, Z.~Kunszt, R.~Gatto, {Meson Masses and Widths in a
  Gauge Theory with Linear Binding Potential}, Nucl. Phys. B105 (1976)
  125--138.
\newblock \href {http://dx.doi.org/10.1016/0550-3213(76)90064-X}
  {\path{doi:10.1016/0550-3213(76)90064-X}}.

\bibitem{Stanley:1980zm}
D.~P. Stanley, D.~Robson, {Nonperturbative Potential Model for Light and Heavy
  Quark anti-Quark Systems}, Phys. Rev. D21 (1980) 3180--3196.
\newblock \href {http://dx.doi.org/10.1103/PhysRevD.21.3180}
  {\path{doi:10.1103/PhysRevD.21.3180}}.

\bibitem{Carlson:1983rw}
J.~Carlson, J.~B. Kogut, V.~R. Pandharipande, {Hadron Spectroscopy in a Flux
  Tube Quark Model}, Phys. Rev. D28 (1983) 2807.
\newblock \href {http://dx.doi.org/10.1103/PhysRevD.28.2807}
  {\path{doi:10.1103/PhysRevD.28.2807}}.

\bibitem{Richardson:1978bt}
J.~L. Richardson, {The Heavy Quark Potential and the Upsilon, J/psi Systems},
  Phys. Lett. 82B (1979) 272--274.
\newblock \href {http://dx.doi.org/10.1016/0370-2693(79)90753-6}
  {\path{doi:10.1016/0370-2693(79)90753-6}}.

\bibitem{Buchmuller:1980bm}
W.~Buchmuller, G.~Grunberg, S.~H.~H. Tye, {The Regge Slope and the Lambda
  Parameter in QCD: An Empirical Approach via Quarkonia}, Phys. Rev. Lett. 45
  (1980) 103, [Erratum: Phys. Rev. Lett.45,587(1980)].
\newblock \href {http://dx.doi.org/10.1103/PhysRevLett.45.103,
  10.1103/PhysRevLett.45.587} {\path{doi:10.1103/PhysRevLett.45.103,
  10.1103/PhysRevLett.45.587}}.

\bibitem{Buchmuller:1980su}
W.~Buchmuller, S.~H.~H. Tye, {Quarkonia and Quantum Chromodynamics}, Phys. Rev.
  D24 (1981) 132.
\newblock \href {http://dx.doi.org/10.1103/PhysRevD.24.132}
  {\path{doi:10.1103/PhysRevD.24.132}}.

\bibitem{Martin:1980rm}
A.~Martin, {A Simultaneous FIT of B anti-B, C anti-C, S anti-S, (BCS Pairs) and
  C anti-S Spectra}, Phys. Lett. 100B (1981) 511--514.
\newblock \href {http://dx.doi.org/10.1016/0370-2693(81)90617-1}
  {\path{doi:10.1016/0370-2693(81)90617-1}}.

\bibitem{Bhanot:1978mj}
G.~Bhanot, S.~Rudaz, {A New Potential for Quarkonium}, Phys. Lett. 78B (1978)
  119--124.
\newblock \href {http://dx.doi.org/10.1016/0370-2693(78)90362-3}
  {\path{doi:10.1016/0370-2693(78)90362-3}}.

\bibitem{Quigg:1977dd}
C.~Quigg, J.~L. Rosner, {Quarkonium Level Spacings}, Phys. Lett. 71B (1977)
  153--157.
\newblock \href {http://dx.doi.org/10.1016/0370-2693(77)90765-1}
  {\path{doi:10.1016/0370-2693(77)90765-1}}.

\bibitem{Fulcher:1991dm}
L.~P. Fulcher, {Perturbative QCD, a universal QCD scale, long range spin orbit
  potential, and the properties of heavy quarkonia}, Phys. Rev. D44 (1991)
  2079--2084.
\newblock \href {http://dx.doi.org/10.1103/PhysRevD.44.2079}
  {\path{doi:10.1103/PhysRevD.44.2079}}.

\bibitem{Gupta:1993pd}
S.~N. Gupta, J.~M. Johnson, W.~W. Repko, C.~J. Suchyta, III, {Heavy quarkonium
  potential model and the p wave singlet state of charmonium}, Phys. Rev. D49
  (1994) 1551--1555.
\newblock \href {http://arxiv.org/abs/hep-ph/9312205}
  {\path{arXiv:hep-ph/9312205}}, \href
  {http://dx.doi.org/10.1103/PhysRevD.49.1551}
  {\path{doi:10.1103/PhysRevD.49.1551}}.

\bibitem{Zeng:1994vj}
J.~Zeng, J.~W. Van~Orden, W.~Roberts, {Heavy mesons in a relativistic model},
  Phys. Rev. D52 (1995) 5229--5241.
\newblock \href {http://arxiv.org/abs/hep-ph/9412269}
  {\path{arXiv:hep-ph/9412269}}, \href
  {http://dx.doi.org/10.1103/PhysRevD.52.5229}
  {\path{doi:10.1103/PhysRevD.52.5229}}.

\bibitem{Ebert:2002pp}
D.~Ebert, R.~N. Faustov, V.~O. Galkin, {Properties of heavy quarkonia and $B_c$
  mesons in the relativistic quark model}, Phys. Rev. D67 (2003) 014027.
\newblock \href {http://arxiv.org/abs/hep-ph/0210381}
  {\path{arXiv:hep-ph/0210381}}, \href
  {http://dx.doi.org/10.1103/PhysRevD.67.014027}
  {\path{doi:10.1103/PhysRevD.67.014027}}.

\bibitem{Godfrey:1985xj}
S.~Godfrey, N.~Isgur, {Mesons in a Relativized Quark Model with
  Chromodynamics}, Phys. Rev. D32 (1985) 189--231.
\newblock \href {http://dx.doi.org/10.1103/PhysRevD.32.189}
  {\path{doi:10.1103/PhysRevD.32.189}}.

\bibitem{Capstick:1986bm}
S.~Capstick, N.~Isgur, {Baryons in a Relativized Quark Model with
  Chromodynamics}, Phys. Rev. D34 (1986) 2809, [AIP Conf. Proc.132,267(1985)].
\newblock \href {http://dx.doi.org/10.1103/PhysRevD.34.2809, 10.1063/1.35361}
  {\path{doi:10.1103/PhysRevD.34.2809, 10.1063/1.35361}}.

\bibitem{Tanabashi:2018oca}
M.~Tanabashi, et~al., {Review of Particle Physics}, Phys. Rev. D98~(3) (2018)
  030001.
\newblock \href {http://dx.doi.org/10.1103/PhysRevD.98.030001}
  {\path{doi:10.1103/PhysRevD.98.030001}}.

\bibitem{Aaij:2017ueg}
R.~Aaij, et~al., {Observation of the doubly charmed baryon $\Xi_{cc}^{++}$},
  Phys. Rev. Lett. 119~(11) (2017) 112001.
\newblock \href {http://arxiv.org/abs/1707.01621} {\path{arXiv:1707.01621}},
  \href {http://dx.doi.org/10.1103/PhysRevLett.119.112001}
  {\path{doi:10.1103/PhysRevLett.119.112001}}.

\bibitem{Klempt:2009pi}
E.~Klempt, J.-M. Richard, {Baryon spectroscopy}, Rev. Mod. Phys. 82 (2010)
  1095--1153.
\newblock \href {http://arxiv.org/abs/0901.2055} {\path{arXiv:0901.2055}},
  \href {http://dx.doi.org/10.1103/RevModPhys.82.1095}
  {\path{doi:10.1103/RevModPhys.82.1095}}.

\bibitem{Klempt:2007cp}
E.~Klempt, A.~Zaitsev, {Glueballs, Hybrids, Multiquarks. Experimental facts
  versus QCD inspired concepts}, Phys. Rept. 454 (2007) 1--202.
\newblock \href {http://arxiv.org/abs/0708.4016} {\path{arXiv:0708.4016}},
  \href {http://dx.doi.org/10.1016/j.physrep.2007.07.006}
  {\path{doi:10.1016/j.physrep.2007.07.006}}.

\bibitem{Bai:2003sw}
J.~Z. Bai, et~al., {Observation of a near threshold enhancement in th $p
  \bar{p}$ mass spectrum from radiative $J/\psi\to \gamma p \bar{p}$ decays},
  Phys. Rev. Lett. 91 (2003) 022001.
\newblock \href {http://arxiv.org/abs/hep-ex/0303006}
  {\path{arXiv:hep-ex/0303006}}, \href
  {http://dx.doi.org/10.1103/PhysRevLett.91.022001}
  {\path{doi:10.1103/PhysRevLett.91.022001}}.

\bibitem{Ablikim:2005um}
M.~Ablikim, et~al., {Observation of a resonance X(1835) in $J /\psi \to \gamma
  \pi^+ \pi^- \eta^\prime$}, Phys. Rev. Lett. 95 (2005) 262001.
\newblock \href {http://arxiv.org/abs/hep-ex/0508025}
  {\path{arXiv:hep-ex/0508025}}, \href
  {http://dx.doi.org/10.1103/PhysRevLett.95.262001}
  {\path{doi:10.1103/PhysRevLett.95.262001}}.

\bibitem{Ablikim:2010au}
M.~Ablikim, et~al., {Confirmation of the $X(1835)$ and observation of the
  resonances $X(2120)$ and $X(2370)$ in $J/\psi\to \gamma
  \pi^+\pi^-\eta^\prime$}, Phys. Rev. Lett. 106 (2011) 072002.
\newblock \href {http://arxiv.org/abs/1012.3510} {\path{arXiv:1012.3510}},
  \href {http://dx.doi.org/10.1103/PhysRevLett.106.072002}
  {\path{doi:10.1103/PhysRevLett.106.072002}}.

\bibitem{Ablikim:2006dw}
M.~Ablikim, et~al., {Observation of a near-threshold enhancement in the omega
  phi mass spectrum from the doubly OZI suppressed decay J / psi --> gamma
  omega phi}, Phys. Rev. Lett. 96 (2006) 162002.
\newblock \href {http://arxiv.org/abs/hep-ex/0602031}
  {\path{arXiv:hep-ex/0602031}}, \href
  {http://dx.doi.org/10.1103/PhysRevLett.96.162002}
  {\path{doi:10.1103/PhysRevLett.96.162002}}.

\bibitem{Ablikim:2007ab}
M.~Ablikim, et~al., {Observation of Y(2175) in J / psi ---> eta phi f(0)(980)},
  Phys. Rev. Lett. 100 (2008) 102003.
\newblock \href {http://arxiv.org/abs/0712.1143} {\path{arXiv:0712.1143}},
  \href {http://dx.doi.org/10.1103/PhysRevLett.100.102003}
  {\path{doi:10.1103/PhysRevLett.100.102003}}.

\bibitem{Ablikim:2014pfc}
M.~Ablikim, et~al., {Study of $J/\psi \to \eta \phi \pi^+ \pi^-$ at BESIII},
  Phys. Rev. D91~(5) (2015) 052017.
\newblock \href {http://arxiv.org/abs/1412.5258} {\path{arXiv:1412.5258}},
  \href {http://dx.doi.org/10.1103/PhysRevD.91.052017}
  {\path{doi:10.1103/PhysRevD.91.052017}}.

\bibitem{Diakonov:1997mm}
D.~Diakonov, V.~Petrov, M.~V. Polyakov, {Exotic anti-decuplet of baryons:
  Prediction from chiral solitons}, Z. Phys. A359 (1997) 305--314.
\newblock \href {http://arxiv.org/abs/hep-ph/9703373}
  {\path{arXiv:hep-ph/9703373}}, \href
  {http://dx.doi.org/10.1007/s002180050406} {\path{doi:10.1007/s002180050406}}.

\bibitem{Nakano:2003qx}
T.~Nakano, et~al., {Evidence for a narrow $S = +1$ baryon resonance in
  photoproduction from the neutron}, Phys. Rev. Lett. 91 (2003) 012002.
\newblock \href {http://arxiv.org/abs/hep-ex/0301020}
  {\path{arXiv:hep-ex/0301020}}, \href
  {http://dx.doi.org/10.1103/PhysRevLett.91.012002}
  {\path{doi:10.1103/PhysRevLett.91.012002}}.

\bibitem{Zhu:2004xa}
S.-L. Zhu, {Pentaquarks}, Int. J. Mod. Phys. A19 (2004) 3439--3469.
\newblock \href {http://arxiv.org/abs/hep-ph/0406204}
  {\path{arXiv:hep-ph/0406204}}, \href
  {http://dx.doi.org/10.1142/S0217751X04019676}
  {\path{doi:10.1142/S0217751X04019676}}.

\bibitem{Liu:2014yva}
T.~Liu, Y.~Mao, B.-Q. Ma, {Present status on experimental search for
  pentaquarks}, Int. J. Mod. Phys. A29~(13) (2014) 1430020.
\newblock \href {http://arxiv.org/abs/1403.4455} {\path{arXiv:1403.4455}},
  \href {http://dx.doi.org/10.1142/S0217751X14300208}
  {\path{doi:10.1142/S0217751X14300208}}.

\bibitem{MartinezTorres:2010zzb}
A.~Martinez~Torres, E.~Oset, {A novel interpretation of the '$\Theta^{+}(1540)$
  pentaquark' peak}, Phys. Rev. Lett. 105 (2010) 092001.
\newblock \href {http://arxiv.org/abs/1008.4978} {\path{arXiv:1008.4978}},
  \href {http://dx.doi.org/10.1103/PhysRevLett.105.092001}
  {\path{doi:10.1103/PhysRevLett.105.092001}}.

\bibitem{Aubert:2003fg}
B.~Aubert, et~al., {Observation of a narrow meson decaying to $D_s^+ \pi^0$ at
  a mass of 2.32 GeV/c$^2$}, Phys. Rev. Lett. 90 (2003) 242001.
\newblock \href {http://arxiv.org/abs/hep-ex/0304021}
  {\path{arXiv:hep-ex/0304021}}, \href
  {http://dx.doi.org/10.1103/PhysRevLett.90.242001}
  {\path{doi:10.1103/PhysRevLett.90.242001}}.

\bibitem{Barnes:2003dj}
T.~Barnes, F.~E. Close, H.~J. Lipkin, {Implications of a $DK$ molecule at
  $2.32$ GeV}, Phys. Rev. D68 (2003) 054006.
\newblock \href {http://arxiv.org/abs/hep-ph/0305025}
  {\path{arXiv:hep-ph/0305025}}, \href
  {http://dx.doi.org/10.1103/PhysRevD.68.054006}
  {\path{doi:10.1103/PhysRevD.68.054006}}.

\bibitem{Cheng:2003kg}
H.-Y. Cheng, W.-S. Hou, {B decays as spectroscope for charmed four quark
  states}, Phys. Lett. B566 (2003) 193--200.
\newblock \href {http://arxiv.org/abs/hep-ph/0305038}
  {\path{arXiv:hep-ph/0305038}}, \href
  {http://dx.doi.org/10.1016/S0370-2693(03)00834-7}
  {\path{doi:10.1016/S0370-2693(03)00834-7}}.

\bibitem{Szczepaniak:2003vy}
A.~P. Szczepaniak, {Description of the D*(s)(2320) resonance as the D pi atom},
  Phys. Lett. B567 (2003) 23--26.
\newblock \href {http://arxiv.org/abs/hep-ph/0305060}
  {\path{arXiv:hep-ph/0305060}}, \href
  {http://dx.doi.org/10.1016/S0370-2693(03)00865-7}
  {\path{doi:10.1016/S0370-2693(03)00865-7}}.

\bibitem{Besson:2003cp}
D.~Besson, et~al., {Observation of a narrow resonance of mass $2.46$ GeV/c$^2$
  decaying to $D^{*}_s\pi^0$ and confirmation of the $D^*_{sJ}(2317)$ state},
  Phys. Rev. D68 (2003) 032002, [Erratum: Phys. Rev. D75, 119908 (2007)].
\newblock \href {http://arxiv.org/abs/hep-ex/0305100}
  {\path{arXiv:hep-ex/0305100}}, \href
  {http://dx.doi.org/10.1103/PhysRevD.68.032002, 10.1103/PhysRevD.75.119908}
  {\path{doi:10.1103/PhysRevD.68.032002, 10.1103/PhysRevD.75.119908}}.

\bibitem{vanBeveren:2003kd}
E.~van Beveren, G.~Rupp, {Observed D(s)(2317) and tentative D(2030) as the
  charmed cousins of the light scalar nonet}, Phys. Rev. Lett. 91 (2003)
  012003.
\newblock \href {http://arxiv.org/abs/hep-ph/0305035}
  {\path{arXiv:hep-ph/0305035}}, \href
  {http://dx.doi.org/10.1103/PhysRevLett.91.012003}
  {\path{doi:10.1103/PhysRevLett.91.012003}}.

\bibitem{Dai:2006uz}
Y.-B. Dai, X.-Q. Li, S.-L. Zhu, Y.-B. Zuo, {Contribution of DK continuum in the
  QCD sum rule for D(sJ) (2317)}, Eur. Phys. J. C55 (2008) 249--258.
\newblock \href {http://arxiv.org/abs/hep-ph/0610327}
  {\path{arXiv:hep-ph/0610327}}, \href
  {http://dx.doi.org/10.1140/epjc/s10052-008-0591-9}
  {\path{doi:10.1140/epjc/s10052-008-0591-9}}.

\bibitem{Choi:2003ue}
S.~K. Choi, et~al., {Observation of a narrow charmonium-like state in exclusive
  $B^{\pm} \to K^{\pm} \pi^+ \pi^- J/\psi$ decays}, Phys. Rev. Lett. 91 (2003)
  262001.
\newblock \href {http://arxiv.org/abs/hep-ex/0309032}
  {\path{arXiv:hep-ex/0309032}}, \href
  {http://dx.doi.org/10.1103/PhysRevLett.91.262001}
  {\path{doi:10.1103/PhysRevLett.91.262001}}.

\bibitem{Meng:2007cx}
C.~Meng, K.-T. Chao, {Decays of the $X(3872)$ and $\chi_{c1}(2P)$ charmonium},
  Phys. Rev. D75 (2007) 114002.
\newblock \href {http://arxiv.org/abs/hep-ph/0703205}
  {\path{arXiv:hep-ph/0703205}}, \href
  {http://dx.doi.org/10.1103/PhysRevD.75.114002}
  {\path{doi:10.1103/PhysRevD.75.114002}}.

\bibitem{Aaij:2016nsc}
R.~Aaij, et~al., {Amplitude analysis of $B^+\to J/\psi \phi K^+$ decays}, Phys.
  Rev. D95~(1) (2017) 012002.
\newblock \href {http://arxiv.org/abs/1606.07898} {\path{arXiv:1606.07898}},
  \href {http://dx.doi.org/10.1103/PhysRevD.95.012002}
  {\path{doi:10.1103/PhysRevD.95.012002}}.

\bibitem{Aaij:2016iza}
R.~Aaij, et~al., {Observation of $J/\psi\phi$ structures consistent with exotic
  states from amplitude analysis of $B^+\to J/\psi \phi K^+$ decays}, Phys.
  Rev. Lett. 118~(2) (2017) 022003.
\newblock \href {http://arxiv.org/abs/1606.07895} {\path{arXiv:1606.07895}},
  \href {http://dx.doi.org/10.1103/PhysRevLett.118.022003}
  {\path{doi:10.1103/PhysRevLett.118.022003}}.

\bibitem{Aaltonen:2011at}
T.~Aaltonen, et~al., {Observation of the $Y(4140)$ structure in the
  $J/\psi\phi$ mass spectrum in $B^\pm\to J/\psi\phi K^\pm$ decays}, Mod. Phys.
  Lett. A32~(26) (2017) 1750139.
\newblock \href {http://arxiv.org/abs/1101.6058} {\path{arXiv:1101.6058}},
  \href {http://dx.doi.org/10.1142/S0217732317501395}
  {\path{doi:10.1142/S0217732317501395}}.

\bibitem{Chatrchyan:2013dma}
S.~Chatrchyan, et~al., {Observation of a peaking structure in the $J/\psi \phi$
  mass spectrum from $B^{\pm} \to J/\psi \phi K^{\pm}$ decays}, Phys. Lett.
  B734 (2014) 261--281.
\newblock \href {http://arxiv.org/abs/1309.6920} {\path{arXiv:1309.6920}},
  \href {http://dx.doi.org/10.1016/j.physletb.2014.05.055}
  {\path{doi:10.1016/j.physletb.2014.05.055}}.

\bibitem{Ablikim:2016qzw}
M.~Ablikim, et~al., {Precise measurement of the $e^+e^-\to \pi^+\pi^-J/\psi$
  cross section at center-of-mass energies from 3.77 to 4.60 GeV}, Phys. Rev.
  Lett. 118~(9) (2017) 092001.
\newblock \href {http://arxiv.org/abs/1611.01317} {\path{arXiv:1611.01317}},
  \href {http://dx.doi.org/10.1103/PhysRevLett.118.092001}
  {\path{doi:10.1103/PhysRevLett.118.092001}}.

\bibitem{BESIII:2016adj}
M.~Ablikim, et~al., {Evidence of Two Resonant Structures in $e^+ e^- \to \pi^+
  \pi^- h_c$}, Phys. Rev. Lett. 118~(9) (2017) 092002.
\newblock \href {http://arxiv.org/abs/1610.07044} {\path{arXiv:1610.07044}},
  \href {http://dx.doi.org/10.1103/PhysRevLett.118.092002}
  {\path{doi:10.1103/PhysRevLett.118.092002}}.

\bibitem{Chilikin:2017evr}
K.~Chilikin, et~al., {Observation of an alternative $\chi_{c0}(2P)$ candidate
  in $e^+ e^- \rightarrow J/\psi D \bar{D}$}, Phys. Rev. D95 (2017) 112003.
\newblock \href {http://arxiv.org/abs/1704.01872} {\path{arXiv:1704.01872}},
  \href {http://dx.doi.org/10.1103/PhysRevD.95.112003}
  {\path{doi:10.1103/PhysRevD.95.112003}}.

\bibitem{Ablikim:2017oaf}
M.~Ablikim, et~al., {Measurement of $e^{+}e^{-}\rightarrow
  \pi^{+}\pi^{-}\psi(3686)$ from 4.008 to 4.600~GeV and observation of a
  charged structure in the $\pi^{\pm}\psi(3686)$ mass spectrum}, Phys. Rev.
  D96~(3) (2017) 032004, [erratum: Phys. Rev.D99,no.1,019903(2019)].
\newblock \href {http://arxiv.org/abs/1703.08787} {\path{arXiv:1703.08787}},
  \href {http://dx.doi.org/10.1103/PhysRevD.96.032004,
  10.1103/PhysRevD.99.019903} {\path{doi:10.1103/PhysRevD.96.032004,
  10.1103/PhysRevD.99.019903}}.

\bibitem{Aaij:2018bla}
R.~Aaij, et~al., {Evidence for an $\eta _c(1S) \pi ^-$ resonance in $B^0
  \rightarrow \eta _c(1S) K^+\pi ^-$ decays}, Eur. Phys. J. C78~(12) (2018)
  1019.
\newblock \href {http://arxiv.org/abs/1809.07416} {\path{arXiv:1809.07416}},
  \href {http://dx.doi.org/10.1140/epjc/s10052-018-6447-z}
  {\path{doi:10.1140/epjc/s10052-018-6447-z}}.

\bibitem{lhcbnew}
\mbox{Talk given by T.~Skwarnicki, on behalf of the LHCb Collaboration at
  Moriond2019}.

\bibitem{Hosaka:2016pey}
A.~Hosaka, T.~Iijima, K.~Miyabayashi, Y.~Sakai, S.~Yasui, {Exotic hadrons with
  heavy flavors: X, Y, Z, and related states}, PTEP 2016~(6) (2016) 062C01.
\newblock \href {http://arxiv.org/abs/1603.09229} {\path{arXiv:1603.09229}},
  \href {http://dx.doi.org/10.1093/ptep/ptw045}
  {\path{doi:10.1093/ptep/ptw045}}.

\bibitem{Ali:2017jda}
A.~Ali, J.~S. Lange, S.~Stone, {Exotics: Heavy Pentaquarks and Tetraquarks},
  Prog. Part. Nucl. Phys. 97 (2017) 123--198.
\newblock \href {http://arxiv.org/abs/1706.00610} {\path{arXiv:1706.00610}},
  \href {http://dx.doi.org/10.1016/j.ppnp.2017.08.003}
  {\path{doi:10.1016/j.ppnp.2017.08.003}}.

\bibitem{Karliner:2017qhf}
M.~Karliner, J.~L. Rosner, T.~Skwarnicki, {Multiquark States}, Ann. Rev. Nucl.
  Part. Sci. 68 (2018) 17--44.
\newblock \href {http://arxiv.org/abs/1711.10626} {\path{arXiv:1711.10626}},
  \href {http://dx.doi.org/10.1146/annurev-nucl-101917-020902}
  {\path{doi:10.1146/annurev-nucl-101917-020902}}.

\bibitem{Guo:2017jvc}
F.-K. Guo, C.~Hanhart, U.-G. Meissner, Q.~Wang, Q.~Zhao, B.-S. Zou, {Hadronic
  molecules}, Rev. Mod. Phys. 90~(1) (2018) 015004.
\newblock \href {http://arxiv.org/abs/1705.00141} {\path{arXiv:1705.00141}},
  \href {http://dx.doi.org/10.1103/RevModPhys.90.015004}
  {\path{doi:10.1103/RevModPhys.90.015004}}.

\bibitem{Esposito:2016noz}
A.~Esposito, A.~Pilloni, A.~D. Polosa, {Multiquark Resonances}, Phys. Rept. 668
  (2016) 1--97.
\newblock \href {http://arxiv.org/abs/1611.07920} {\path{arXiv:1611.07920}},
  \href {http://dx.doi.org/10.1016/j.physrep.2016.11.002}
  {\path{doi:10.1016/j.physrep.2016.11.002}}.

\bibitem{Lebed:2016hpi}
R.~F. Lebed, R.~E. Mitchell, E.~S. Swanson, {Heavy-Quark QCD Exotica}, Prog.
  Part. Nucl. Phys. 93 (2017) 143--194.
\newblock \href {http://arxiv.org/abs/1610.04528} {\path{arXiv:1610.04528}},
  \href {http://dx.doi.org/10.1016/j.ppnp.2016.11.003}
  {\path{doi:10.1016/j.ppnp.2016.11.003}}.

\bibitem{Richard:2016eis}
J.-M. Richard, {Exotic hadrons: review and perspectives}, Few Body Syst.
  57~(12) (2016) 1185--1212.
\newblock \href {http://arxiv.org/abs/1606.08593} {\path{arXiv:1606.08593}},
  \href {http://dx.doi.org/10.1007/s00601-016-1159-0}
  {\path{doi:10.1007/s00601-016-1159-0}}.

\bibitem{Olsen:2017bmm}
S.~L. Olsen, T.~Skwarnicki, D.~Zieminska, {Nonstandard heavy mesons and
  baryons: Experimental evidence}, Rev. Mod. Phys. 90~(1) (2018) 015003.
\newblock \href {http://arxiv.org/abs/1708.04012} {\path{arXiv:1708.04012}},
  \href {http://dx.doi.org/10.1103/RevModPhys.90.015003}
  {\path{doi:10.1103/RevModPhys.90.015003}}.

\bibitem{SilvestreBrac:1992mv}
B.~Silvestre-Brac, {Systematics of Q**2 (anti-Q**2) systems with a
  chromomagnetic interaction}, Phys. Rev. D46 (1992) 2179--2189.
\newblock \href {http://dx.doi.org/10.1103/PhysRevD.46.2179}
  {\path{doi:10.1103/PhysRevD.46.2179}}.

\bibitem{KerenZur:2007vp}
B.~Keren-Zur, {Testing confining potentials through meson/baryon hyperfine
  splitting ratio}, Annals Phys. 323 (2008) 631--642.
\newblock \href {http://arxiv.org/abs/hep-ph/0703011}
  {\path{arXiv:hep-ph/0703011}}, \href
  {http://dx.doi.org/10.1016/j.aop.2007.04.010}
  {\path{doi:10.1016/j.aop.2007.04.010}}.

\bibitem{DeRujula:1975qlm}
A.~De~Rujula, H.~Georgi, S.~L. Glashow, {Hadron Masses in a Gauge Theory},
  Phys. Rev. D12 (1975) 147--162.
\newblock \href {http://dx.doi.org/10.1103/PhysRevD.12.147}
  {\path{doi:10.1103/PhysRevD.12.147}}.

\bibitem{Sakharov:1966tua}
A.~D. Sakharov, {\relax Ya}.~B. Zel'dovich, {Kvarkovaia struktura i massy
  sil'novzaimodeistvuyushchikh chastits}, Yad. Fiz. 4 (1966) 395--400.

\bibitem{Oka:2012zz}
M.~Oka, {Origin of the short-range part of generalized two- and three-body
  nuclear force}, Nucl. Phys. A881 (2012) 6--13.
\newblock \href {http://dx.doi.org/10.1016/j.nuclphysa.2012.01.007}
  {\path{doi:10.1016/j.nuclphysa.2012.01.007}}.

\bibitem{Maeda:2015hxa}
S.~Maeda, M.~Oka, A.~Yokota, E.~Hiyama, Y.-R. Liu, {A model of charmed
  baryon¨Cnucleon potential and two- and three-body bound states with charmed
  baryon}, PTEP 2016~(2) (2016) 023D02.
\newblock \href {http://arxiv.org/abs/1509.02445} {\path{arXiv:1509.02445}},
  \href {http://dx.doi.org/10.1093/ptep/ptv194}
  {\path{doi:10.1093/ptep/ptv194}}.

\bibitem{Jaffe:1976ih}
R.~L. Jaffe, {Multi-Quark Hadrons. 2. Methods}, Phys. Rev. D15 (1977) 281.
\newblock \href {http://dx.doi.org/10.1103/PhysRevD.15.281}
  {\path{doi:10.1103/PhysRevD.15.281}}.

\bibitem{Buccella:2006fn}
F.~Buccella, H.~Hogaasen, J.-M. Richard, P.~Sorba, {Chromomagnetism, flavour
  symmetry breaking and S-wave tetraquarks}, Eur. Phys. J. C49 (2007) 743--754.
\newblock \href {http://arxiv.org/abs/hep-ph/0608001}
  {\path{arXiv:hep-ph/0608001}}, \href
  {http://dx.doi.org/10.1140/epjc/s10052-006-0142-1}
  {\path{doi:10.1140/epjc/s10052-006-0142-1}}.

\bibitem{Aerts:1977rw}
A.~T.~M. Aerts, P.~J.~G. Mulders, J.~J. de~Swart, {Multi-Baryon States in the
  Bag Model}, Phys. Rev. D17 (1978) 260.
\newblock \href {http://dx.doi.org/10.1103/PhysRevD.17.260}
  {\path{doi:10.1103/PhysRevD.17.260}}.

\bibitem{Hogaasen:1978jw}
H.~Hogaasen, P.~Sorba, {The Systematics of Possibly Narrow Quark States with
  Baryon Number One}, Nucl. Phys. B145 (1978) 119--140.
\newblock \href {http://dx.doi.org/10.1016/0550-3213(78)90417-0}
  {\path{doi:10.1016/0550-3213(78)90417-0}}.

\bibitem{Cui:2005az}
Y.~Cui, X.-L. Chen, W.-Z. Deng, S.-L. Zhu, {The J**P = 1+ ud anti-s anti-s
  tetraquark}, Phys. Rev. D73 (2006) 014018.
\newblock \href {http://arxiv.org/abs/hep-ph/0511150}
  {\path{arXiv:hep-ph/0511150}}, \href
  {http://dx.doi.org/10.1103/PhysRevD.73.014018}
  {\path{doi:10.1103/PhysRevD.73.014018}}.

\bibitem{Cui:2006mp}
Y.~Cui, X.-L. Chen, W.-Z. Deng, S.-L. Zhu, {The Possible Heavy Tetraquarks qQ
  anti-q anti-Q, qq anti-Q anti-Q and qQ anti-Q anti-Q}, HEPNP 31 (2007) 7--13.
\newblock \href {http://arxiv.org/abs/hep-ph/0607226}
  {\path{arXiv:hep-ph/0607226}}.

\bibitem{Hogaasen:2004pm}
H.~Hogaasen, P.~Sorba, {The Colour triplet qq anti-q cluster and pentaquark
  models}, Mod. Phys. Lett. A19 (2004) 2403--2410.
\newblock \href {http://arxiv.org/abs/hep-ph/0406078}
  {\path{arXiv:hep-ph/0406078}}, \href
  {http://dx.doi.org/10.1142/S0217732304015592}
  {\path{doi:10.1142/S0217732304015592}}.

\bibitem{Maiani:2004uc}
L.~Maiani, F.~Piccinini, A.~D. Polosa, V.~Riquer, {A New look at scalar
  mesons}, Phys. Rev. Lett. 93 (2004) 212002.
\newblock \href {http://arxiv.org/abs/hep-ph/0407017}
  {\path{arXiv:hep-ph/0407017}}, \href
  {http://dx.doi.org/10.1103/PhysRevLett.93.212002}
  {\path{doi:10.1103/PhysRevLett.93.212002}}.

\bibitem{Lipkin:1977ie}
H.~J. Lipkin, {Are There Charmed - Strange Exotic Mesons?}, Phys. Lett. 70B
  (1977) 113--116.
\newblock \href {http://dx.doi.org/10.1016/0370-2693(77)90357-4}
  {\path{doi:10.1016/0370-2693(77)90357-4}}.

\bibitem{Anselmino:1992vg}
M.~Anselmino, E.~Predazzi, S.~Ekelin, S.~Fredriksson, D.~B. Lichtenberg,
  {Diquarks}, Rev. Mod. Phys. 65 (1993) 1199--1234.
\newblock \href {http://dx.doi.org/10.1103/RevModPhys.65.1199}
  {\path{doi:10.1103/RevModPhys.65.1199}}.

\bibitem{Brodsky:2014xia}
S.~J. Brodsky, D.~S. Hwang, R.~F. Lebed, {Dynamical Picture for the Formation
  and Decay of the Exotic XYZ Mesons}, Phys. Rev. Lett. 113~(11) (2014) 112001.
\newblock \href {http://arxiv.org/abs/1406.7281} {\path{arXiv:1406.7281}},
  \href {http://dx.doi.org/10.1103/PhysRevLett.113.112001}
  {\path{doi:10.1103/PhysRevLett.113.112001}}.

\bibitem{Maiani:2017kyi}
L.~Maiani, A.~D. Polosa, V.~Riquer, {A Theory of X and Z Multiquark
  Resonances}, Phys. Lett. B778 (2018) 247--251.
\newblock \href {http://arxiv.org/abs/1712.05296} {\path{arXiv:1712.05296}},
  \href {http://dx.doi.org/10.1016/j.physletb.2018.01.039}
  {\path{doi:10.1016/j.physletb.2018.01.039}}.

\bibitem{Maiani:2004vq}
L.~Maiani, F.~Piccinini, A.~D. Polosa, V.~Riquer, {Diquark-antidiquarks with
  hidden or open charm and the nature of $X(3872)$}, Phys. Rev. D71 (2005)
  014028.
\newblock \href {http://arxiv.org/abs/hep-ph/0412098}
  {\path{arXiv:hep-ph/0412098}}, \href
  {http://dx.doi.org/10.1103/PhysRevD.71.014028}
  {\path{doi:10.1103/PhysRevD.71.014028}}.

\bibitem{Drenska:2009cd}
N.~V. Drenska, R.~Faccini, A.~D. Polosa, {Exotic Hadrons with Hidden Charm and
  Strangeness}, Phys. Rev. D79 (2009) 077502.
\newblock \href {http://arxiv.org/abs/0902.2803} {\path{arXiv:0902.2803}},
  \href {http://dx.doi.org/10.1103/PhysRevD.79.077502}
  {\path{doi:10.1103/PhysRevD.79.077502}}.

\bibitem{Maiani:2014aja}
L.~Maiani, F.~Piccinini, A.~D. Polosa, V.~Riquer, {The $Z(4430)$ and a New
  Paradigm for Spin Interactions in Tetraquarks}, Phys. Rev. D89 (2014) 114010.
\newblock \href {http://arxiv.org/abs/1405.1551} {\path{arXiv:1405.1551}},
  \href {http://dx.doi.org/10.1103/PhysRevD.89.114010}
  {\path{doi:10.1103/PhysRevD.89.114010}}.

\bibitem{Lebed:2015tna}
R.~F. Lebed, {The Pentaquark Candidates in the Dynamical Diquark Picture},
  Phys. Lett. B749 (2015) 454--457.
\newblock \href {http://arxiv.org/abs/1507.05867} {\path{arXiv:1507.05867}},
  \href {http://dx.doi.org/10.1016/j.physletb.2015.08.032}
  {\path{doi:10.1016/j.physletb.2015.08.032}}.

\bibitem{Karliner:2003dt}
M.~Karliner, H.~J. Lipkin, {A Diquark - triquark model for the K N pentaquark},
  Phys. Lett. B575 (2003) 249--255.
\newblock \href {http://arxiv.org/abs/hep-ph/0402260}
  {\path{arXiv:hep-ph/0402260}}, \href
  {http://dx.doi.org/10.1016/j.physletb.2003.09.062}
  {\path{doi:10.1016/j.physletb.2003.09.062}}.

\bibitem{Zhu:2015bba}
R.~Zhu, C.-F. Qiao, {Pentaquark states in a diquark¨Ctriquark model}, Phys.
  Lett. B756 (2016) 259--264.
\newblock \href {http://arxiv.org/abs/1510.08693} {\path{arXiv:1510.08693}},
  \href {http://dx.doi.org/10.1016/j.physletb.2016.03.022}
  {\path{doi:10.1016/j.physletb.2016.03.022}}.

\bibitem{Hogaasen:2013nca}
H.~H{\o}gaasen, E.~Kou, J.-M. Richard, P.~Sorba, {Isovector and hidden-beauty
  partners of the X(3872)}, Phys. Lett. B732 (2014) 97--100.
\newblock \href {http://arxiv.org/abs/1309.2049} {\path{arXiv:1309.2049}},
  \href {http://dx.doi.org/10.1016/j.physletb.2014.03.027}
  {\path{doi:10.1016/j.physletb.2014.03.027}}.

\bibitem{Weng:2018mmf}
X.-Z. Weng, X.-L. Chen, W.-Z. Deng, {Masses of doubly heavy-quark baryons in an
  extended chromomagnetic model}, Phys. Rev. D97~(5) (2018) 054008.
\newblock \href {http://arxiv.org/abs/1801.08644} {\path{arXiv:1801.08644}},
  \href {http://dx.doi.org/10.1103/PhysRevD.97.054008}
  {\path{doi:10.1103/PhysRevD.97.054008}}.

\bibitem{Rossi:2016szw}
G.~Rossi, G.~Veneziano, {The string-junction picture of multiquark states: an
  update}, JHEP 06 (2016) 041.
\newblock \href {http://arxiv.org/abs/1603.05830} {\path{arXiv:1603.05830}},
  \href {http://dx.doi.org/10.1007/JHEP06(2016)041}
  {\path{doi:10.1007/JHEP06(2016)041}}.

\bibitem{Karliner:2014gca}
M.~Karliner, J.~L. Rosner, {Baryons with two heavy quarks: Masses, production,
  decays, and detection}, Phys. Rev. D90~(9) (2014) 094007.
\newblock \href {http://arxiv.org/abs/1408.5877} {\path{arXiv:1408.5877}},
  \href {http://dx.doi.org/10.1103/PhysRevD.90.094007}
  {\path{doi:10.1103/PhysRevD.90.094007}}.

\bibitem{Karliner:2016zzc}
M.~Karliner, S.~Nussinov, J.~L. Rosner, {$Q Q \bar Q \bar Q$ states: masses,
  production, and decays}, Phys. Rev. D95~(3) (2017) 034011.
\newblock \href {http://arxiv.org/abs/1611.00348} {\path{arXiv:1611.00348}},
  \href {http://dx.doi.org/10.1103/PhysRevD.95.034011}
  {\path{doi:10.1103/PhysRevD.95.034011}}.

\bibitem{Wu:2016gas}
J.~Wu, Y.-R. Liu, K.~Chen, X.~Liu, S.-L. Zhu, {$X(4140)$, $X(4270)$, $X(4500)$
  and $X(4700)$ and their $cs\bar{c}\bar{s}$ tetraquark partners}, Phys. Rev.
  D94~(9) (2016) 094031.
\newblock \href {http://arxiv.org/abs/1608.07900} {\path{arXiv:1608.07900}},
  \href {http://dx.doi.org/10.1103/PhysRevD.94.094031}
  {\path{doi:10.1103/PhysRevD.94.094031}}.

\bibitem{Chen:2016ont}
K.~Chen, X.~Liu, J.~Wu, Y.-R. Liu, S.-L. Zhu, {Triply heavy tetraquark states
  with the $QQ\bar{Q}\bar{q}$ configuration}, Eur. Phys. J. A53~(1) (2017) 5.
\newblock \href {http://arxiv.org/abs/1609.06117} {\path{arXiv:1609.06117}},
  \href {http://dx.doi.org/10.1140/epja/i2017-12199-3}
  {\path{doi:10.1140/epja/i2017-12199-3}}.

\bibitem{Wu:2017weo}
J.~Wu, Y.-R. Liu, K.~Chen, X.~Liu, S.-L. Zhu, {Hidden-charm pentaquarks and
  their hidden-bottom and $B_c$-like partner states}, Phys. Rev. D95~(3) (2017)
  034002.
\newblock \href {http://arxiv.org/abs/1701.03873} {\path{arXiv:1701.03873}},
  \href {http://dx.doi.org/10.1103/PhysRevD.95.034002}
  {\path{doi:10.1103/PhysRevD.95.034002}}.

\bibitem{Luo:2017eub}
S.-Q. Luo, K.~Chen, X.~Liu, Y.-R. Liu, S.-L. Zhu, {Exotic tetraquark states
  with the $qq\bar{Q}\bar{Q}$ configuration}, Eur. Phys. J. C77~(10) (2017)
  709.
\newblock \href {http://arxiv.org/abs/1707.01180} {\path{arXiv:1707.01180}},
  \href {http://dx.doi.org/10.1140/epjc/s10052-017-5297-4}
  {\path{doi:10.1140/epjc/s10052-017-5297-4}}.

\bibitem{Wu:2016vtq}
J.~Wu, Y.-R. Liu, K.~Chen, X.~Liu, S.-L. Zhu, {Heavy-flavored tetraquark states
  with the $QQ\bar{Q}\bar{Q}$ configuration}, Phys. Rev. D97~(9) (2018) 094015.
\newblock \href {http://arxiv.org/abs/1605.01134} {\path{arXiv:1605.01134}},
  \href {http://dx.doi.org/10.1103/PhysRevD.97.094015}
  {\path{doi:10.1103/PhysRevD.97.094015}}.

\bibitem{Zhou:2018pcv}
Q.-S. Zhou, K.~Chen, X.~Liu, Y.-R. Liu, S.-L. Zhu, {Surveying exotic
  pentaquarks with the typical $QQqq\bar{q}$ configuration}, Phys. Rev. C98~(4)
  (2018) 045204.
\newblock \href {http://arxiv.org/abs/1801.04557} {\path{arXiv:1801.04557}},
  \href {http://dx.doi.org/10.1103/PhysRevC.98.045204}
  {\path{doi:10.1103/PhysRevC.98.045204}}.

\bibitem{Li:2018vhp}
S.-Y. Li, Y.-R. Liu, Y.-N. Liu, Z.-G. Si, J.~Wu, {Pentaquark states with the
  $QQQq\bar{q}$ configuration in a simple model}, Eur. Phys. J. C79~(1) (2019)
  87.
\newblock \href {http://arxiv.org/abs/1809.08072} {\path{arXiv:1809.08072}},
  \href {http://dx.doi.org/10.1140/epjc/s10052-019-6589-7}
  {\path{doi:10.1140/epjc/s10052-019-6589-7}}.

\bibitem{Wu:2018xdi}
J.~Wu, X.~Liu, Y.-R. Liu, S.-L. Zhu, {Systematic studies of charmonium-,
  bottomonium-, and $B_c$-like tetraquark states}, Phys. Rev. D99~(1) (2019)
  014037.
\newblock \href {http://arxiv.org/abs/1810.06886} {\path{arXiv:1810.06886}},
  \href {http://dx.doi.org/10.1103/PhysRevD.99.014037}
  {\path{doi:10.1103/PhysRevD.99.014037}}.

\bibitem{Stancu:2009ka}
F.~Stancu, {Can Y(4140) be a c anti-c s anti-s tetraquark?}, J. Phys. G37
  (2010) 075017.
\newblock \href {http://arxiv.org/abs/0906.2485} {\path{arXiv:0906.2485}},
  \href {http://dx.doi.org/10.1088/0954-3899/37/7/075017}
  {\path{doi:10.1088/0954-3899/37/7/075017}}.

\bibitem{Leandri:1989su}
J.~Leandri, B.~Silvestre-Brac, {Systematics of $\bar{Q} Q^-$4 Systems With a
  Pure Chromomagnetic Interaction}, Phys. Rev. D40 (1989) 2340--2352.
\newblock \href {http://dx.doi.org/10.1103/PhysRevD.40.2340}
  {\path{doi:10.1103/PhysRevD.40.2340}}.

\bibitem{SilvestreBrac:1992yg}
B.~Silvestre-Brac, J.~Leandri, {Systematics of q-6 systems in a simple
  chromomagnetic model}, Phys. Rev. D45 (1992) 4221--4239.
\newblock \href {http://dx.doi.org/10.1103/PhysRevD.45.4221}
  {\path{doi:10.1103/PhysRevD.45.4221}}.

\bibitem{Vijande:2013qr}
J.~Vijande, A.~Valcarce, J.-M. Richard, {Adiabaticity and color mixing in
  tetraquark spectroscopy}, Phys. Rev. D87~(3) (2013) 034040.
\newblock \href {http://arxiv.org/abs/1301.6212} {\path{arXiv:1301.6212}},
  \href {http://dx.doi.org/10.1103/PhysRevD.87.034040}
  {\path{doi:10.1103/PhysRevD.87.034040}}.

\bibitem{Lipkin:1986dx}
H.~J. Lipkin, {Relations Between Meson and Baryon Hyperfine Splittings}, Phys.
  Lett. B171 (1986) 293--294.
\newblock \href {http://dx.doi.org/10.1016/0370-2693(86)91551-0}
  {\path{doi:10.1016/0370-2693(86)91551-0}}.

\bibitem{Mathur:2018epb}
N.~Mathur, M.~Padmanath, S.~Mondal, {Precise predictions of charmed-bottom
  hadrons from lattice QCD}, Phys. Rev. Lett. 121~(20) (2018) 202002.
\newblock \href {http://arxiv.org/abs/1806.04151} {\path{arXiv:1806.04151}},
  \href {http://dx.doi.org/10.1103/PhysRevLett.121.202002}
  {\path{doi:10.1103/PhysRevLett.121.202002}}.

\bibitem{Mathur:2018rwu}
N.~Mathur, M.~Padmanath, {Lattice QCD study of doubly-charmed strange baryons},
  Phys. Rev. D99~(3) (2019) 031501.
\newblock \href {http://arxiv.org/abs/1807.00174} {\path{arXiv:1807.00174}},
  \href {http://dx.doi.org/10.1103/PhysRevD.99.031501}
  {\path{doi:10.1103/PhysRevD.99.031501}}.

\bibitem{Maiani:2005pe}
L.~Maiani, V.~Riquer, F.~Piccinini, A.~D. Polosa, {Four quark interpretation of
  Y(4260)}, Phys. Rev. D72 (2005) 031502.
\newblock \href {http://arxiv.org/abs/hep-ph/0507062}
  {\path{arXiv:hep-ph/0507062}}, \href
  {http://dx.doi.org/10.1103/PhysRevD.72.031502}
  {\path{doi:10.1103/PhysRevD.72.031502}}.

\bibitem{Maiani:2007wz}
L.~Maiani, A.~D. Polosa, V.~Riquer, {The Charged $Z(4433)$: Towards a new
  spectroscopy,~}\href {http://arxiv.org/abs/0708.3997}
  {\path{arXiv:0708.3997}}.

\bibitem{Maiani:2008zz}
L.~Maiani, A.~D. Polosa, V.~Riquer, {The charged Z(4430) in the
  diquark-antidiquark picture}, New J. Phys. 10 (2008) 073004.
\newblock \href {http://dx.doi.org/10.1088/1367-2630/10/7/073004}
  {\path{doi:10.1088/1367-2630/10/7/073004}}.

\bibitem{Hogaasen:2005jv}
H.~Hogaasen, J.~M. Richard, P.~Sorba, {A Chromomagnetic mechanism for the
  $X(3872)$ resonance}, Phys. Rev. D73 (2006) 054013.
\newblock \href {http://arxiv.org/abs/hep-ph/0511039}
  {\path{arXiv:hep-ph/0511039}}, \href
  {http://dx.doi.org/10.1103/PhysRevD.73.054013}
  {\path{doi:10.1103/PhysRevD.73.054013}}.

\bibitem{Abud:2009rk}
M.~Abud, F.~Buccella, F.~Tramontano, {Hints for the existence of hexaquark
  states in the baryon-antibaryon sector}, Phys. Rev. D81 (2010) 074018.
\newblock \href {http://arxiv.org/abs/0912.4299} {\path{arXiv:0912.4299}},
  \href {http://dx.doi.org/10.1103/PhysRevD.81.074018}
  {\path{doi:10.1103/PhysRevD.81.074018}}.

\bibitem{Guo:2011gu}
T.~Guo, L.~Cao, M.-Z. Zhou, H.~Chen, {The Possible candidates of tetraquark :
  $Z_b(10610)$ and $Z_b(10650)$}\href {http://arxiv.org/abs/1106.2284}
  {\path{arXiv:1106.2284}}.

\bibitem{Ali:2011ug}
A.~Ali, C.~Hambrock, W.~Wang, {Tetraquark Interpretation of the Charged
  Bottomonium-like states $Z_b^{+-}(10610)$ and $Z_b^{+-}(10650)$ and
  Implications}, Phys. Rev. D85 (2012) 054011.
\newblock \href {http://arxiv.org/abs/1110.1333} {\path{arXiv:1110.1333}},
  \href {http://dx.doi.org/10.1103/PhysRevD.85.054011}
  {\path{doi:10.1103/PhysRevD.85.054011}}.

\bibitem{Ali:2014dva}
A.~Ali, L.~Maiani, A.~D. Polosa, V.~Riquer, {Hidden-Beauty Charged Tetraquarks
  and Heavy Quark Spin Conservation}, Phys. Rev. D91~(1) (2015) 017502.
\newblock \href {http://arxiv.org/abs/1412.2049} {\path{arXiv:1412.2049}},
  \href {http://dx.doi.org/10.1103/PhysRevD.91.017502}
  {\path{doi:10.1103/PhysRevD.91.017502}}.

\bibitem{Ali:2009pi}
A.~Ali, C.~Hambrock, I.~Ahmed, M.~J. Aslam, {A case for hidden $b\bar{b}$
  tetraquarks based on $e^+e^- \to b\bar{b}$ cross section between
  $\sqrt{s}=10.54$ and 11.20 GeV}, Phys. Lett. B684 (2010) 28--39.
\newblock \href {http://arxiv.org/abs/0911.2787} {\path{arXiv:0911.2787}},
  \href {http://dx.doi.org/10.1016/j.physletb.2009.12.053}
  {\path{doi:10.1016/j.physletb.2009.12.053}}.

\bibitem{Ali:2009es}
A.~Ali, C.~Hambrock, M.~J. Aslam, {A Tetraquark interpretation of the BELLE
  data on the anomalous $\Upsilon(1S) \pi^+\pi^-$ and $\Upsilon(2S) \pi^+\pi^-$
  production near the $\Upsilon(5S)$ resonance}, Phys. Rev. Lett. 104 (2010)
  162001, [Erratum: Phys. Rev. Lett. 107, 049903 (2011)].
\newblock \href {http://arxiv.org/abs/0912.5016} {\path{arXiv:0912.5016}},
  \href {http://dx.doi.org/10.1103/PhysRevLett.104.162001,
  10.1103/PhysRevLett.107.049903} {\path{doi:10.1103/PhysRevLett.104.162001,
  10.1103/PhysRevLett.107.049903}}.

\bibitem{Faccini:2013lda}
L.~Maiani, V.~Riquer, R.~Faccini, F.~Piccinini, A.~Pilloni, A.~D. Polosa, {A
  $J^{PG}=1^{++}$ Charged Resonance in the $Y(4260) \to \pi^+ \pi^- J/\psi$
  Decay?}, Phys. Rev. D87~(11) (2013) 111102.
\newblock \href {http://arxiv.org/abs/1303.6857} {\path{arXiv:1303.6857}},
  \href {http://dx.doi.org/10.1103/PhysRevD.87.111102}
  {\path{doi:10.1103/PhysRevD.87.111102}}.

\bibitem{Chen:2015dig}
H.~X. Chen, L.~Maiani, A.~D. Polosa, V.~Riquer, {$Y(4260)\rightarrow \gamma +
  X(3872)$ in the diquarkonium picture}, Eur. Phys. J. C75~(11) (2015) 550.
\newblock \href {http://arxiv.org/abs/1510.03626} {\path{arXiv:1510.03626}},
  \href {http://dx.doi.org/10.1140/epjc/s10052-015-3781-2}
  {\path{doi:10.1140/epjc/s10052-015-3781-2}}.

\bibitem{Anisovich:2015caa}
V.~V. Anisovich, M.~A. Matveev, A.~V. Sarantsev, A.~N. Semenova, {Exotic mesons
  with hidden charm as diquark-antidiquark states}, Int. J. Mod. Phys. A30
  (2015) 1550186.
\newblock \href {http://arxiv.org/abs/1507.07232} {\path{arXiv:1507.07232}},
  \href {http://dx.doi.org/10.1142/S0217751X15501869}
  {\path{doi:10.1142/S0217751X15501869}}.

\bibitem{Kim:2016tys}
H.~Kim, K.~S. Kim, M.-K. Cheoun, D.~Jido, M.~Oka, {Testing the tetraquark
  structure for the X resonances in the low-lying region}, Eur. Phys. J.
  A52~(7) (2016) 184.
\newblock \href {http://arxiv.org/abs/1602.07540} {\path{arXiv:1602.07540}},
  \href {http://dx.doi.org/10.1140/epja/i2016-16184-0}
  {\path{doi:10.1140/epja/i2016-16184-0}}.

\bibitem{Lebed:2016yvr}
R.~F. Lebed, A.~D. Polosa, {$\chi^{\vphantom\dagger}_{c0}(3915)$ As the
  Lightest $c\bar c s \bar s$ State}, Phys. Rev. D93~(9) (2016) 094024.
\newblock \href {http://arxiv.org/abs/1602.08421} {\path{arXiv:1602.08421}},
  \href {http://dx.doi.org/10.1103/PhysRevD.93.094024}
  {\path{doi:10.1103/PhysRevD.93.094024}}.

\bibitem{Maiani:2016wlq}
L.~Maiani, A.~D. Polosa, V.~Riquer, {Interpretation of Axial Resonances in
  J/psi-phi at LHCb}, Phys. Rev. D94~(5) (2016) 054026.
\newblock \href {http://arxiv.org/abs/1607.02405} {\path{arXiv:1607.02405}},
  \href {http://dx.doi.org/10.1103/PhysRevD.94.054026}
  {\path{doi:10.1103/PhysRevD.94.054026}}.

\bibitem{Zhu:2016arf}
R.~Zhu, {Hidden charm octet tetraquarks from a diquark-antidiquark model},
  Phys. Rev. D94~(5) (2016) 054009.
\newblock \href {http://arxiv.org/abs/1607.02799} {\path{arXiv:1607.02799}},
  \href {http://dx.doi.org/10.1103/PhysRevD.94.054009}
  {\path{doi:10.1103/PhysRevD.94.054009}}.

\bibitem{Ali:2017wsf}
A.~Ali, L.~Maiani, A.~V. Borisov, I.~Ahmed, M.~Jamil~Aslam, A.~{\relax Ya}.
  Parkhomenko, A.~D. Polosa, A.~Rehman, {A new look at the Y tetraquarks and
  $\Omega _c$ baryons in the diquark model}, Eur. Phys. J. C78~(1) (2018) 29.
\newblock \href {http://arxiv.org/abs/1708.04650} {\path{arXiv:1708.04650}},
  \href {http://dx.doi.org/10.1140/epjc/s10052-017-5501-6}
  {\path{doi:10.1140/epjc/s10052-017-5501-6}}.

\bibitem{Shen:2009vs}
C.~P. Shen, et~al., {Evidence for a new resonance and search for the $Y(4140)$
  in the $\gamma \gamma \to \phi J/\psi$ process}, Phys. Rev. Lett. 104 (2010)
  112004.
\newblock \href {http://arxiv.org/abs/0912.2383} {\path{arXiv:0912.2383}},
  \href {http://dx.doi.org/10.1103/PhysRevLett.104.112004}
  {\path{doi:10.1103/PhysRevLett.104.112004}}.

\bibitem{Ballot:1983iv}
J.~l. Ballot, J.~M. Richard, {FOUR QUARK STATES IN ADDITIVE POTENTIALS}, Phys.
  Lett. 123B (1983) 449--451.
\newblock \href {http://dx.doi.org/10.1016/0370-2693(83)90991-7}
  {\path{doi:10.1016/0370-2693(83)90991-7}}.

\bibitem{Lipkin:1986dw}
H.~J. Lipkin, {A MODEL INDEPENDENT APPROACH TO MULTI - QUARK BOUND STATES},
  Phys. Lett. B172 (1986) 242--247.
\newblock \href {http://dx.doi.org/10.1016/0370-2693(86)90843-9}
  {\path{doi:10.1016/0370-2693(86)90843-9}}.

\bibitem{Zouzou:1986qh}
S.~Zouzou, B.~Silvestre-Brac, C.~Gignoux, J.~M. Richard, {FOUR QUARK BOUND
  STATES}, Z. Phys. C30 (1986) 457.
\newblock \href {http://dx.doi.org/10.1007/BF01557611}
  {\path{doi:10.1007/BF01557611}}.

\bibitem{Heller:1986bt}
L.~Heller, J.~A. Tjon, {On the Existence of Stable Dimesons}, Phys. Rev. D35
  (1987) 969.
\newblock \href {http://dx.doi.org/10.1103/PhysRevD.35.969}
  {\path{doi:10.1103/PhysRevD.35.969}}.

\bibitem{Carlson:1987hh}
J.~Carlson, L.~Heller, J.~A. Tjon, {Stability of Dimesons}, Phys. Rev. D37
  (1988) 744.
\newblock \href {http://dx.doi.org/10.1103/PhysRevD.37.744}
  {\path{doi:10.1103/PhysRevD.37.744}}.

\bibitem{Manohar:1992nd}
A.~V. Manohar, M.~B. Wise, {Exotic Q Q anti-q anti-q states in QCD}, Nucl.
  Phys. B399 (1993) 17--33.
\newblock \href {http://arxiv.org/abs/hep-ph/9212236}
  {\path{arXiv:hep-ph/9212236}}, \href
  {http://dx.doi.org/10.1016/0550-3213(93)90614-U}
  {\path{doi:10.1016/0550-3213(93)90614-U}}.

\bibitem{Ericson:1993wy}
T.~E.~O. Ericson, G.~Karl, {Strength of pion exchange in hadronic molecules},
  Phys. Lett. B309 (1993) 426--430.
\newblock \href {http://dx.doi.org/10.1016/0370-2693(93)90957-J}
  {\path{doi:10.1016/0370-2693(93)90957-J}}.

\bibitem{SilvestreBrac:1993ry}
B.~Silvestre-Brac, C.~Semay, {Spectrum and decay properties of diquonia}, Z.
  Phys. C59 (1993) 457--470.
\newblock \href {http://dx.doi.org/10.1007/BF01498626}
  {\path{doi:10.1007/BF01498626}}.

\bibitem{SilvestreBrac:1993ss}
B.~Silvestre-Brac, C.~Semay, {Systematics of L = 0 q-2 anti-q-2 systems}, Z.
  Phys. C57 (1993) 273--282.
\newblock \href {http://dx.doi.org/10.1007/BF01565058}
  {\path{doi:10.1007/BF01565058}}.

\bibitem{Semay:1994ht}
C.~Semay, B.~Silvestre-Brac, {Diquonia and potential models}, Z. Phys. C61
  (1994) 271--275.
\newblock \href {http://dx.doi.org/10.1007/BF01413104}
  {\path{doi:10.1007/BF01413104}}.

\bibitem{Moinester:1995fk}
M.~A. Moinester, {How to search for doubly charmed baryons and tetraquarks}, Z.
  Phys. A355 (1996) 349--362.
\newblock \href {http://arxiv.org/abs/hep-ph/9506405}
  {\path{arXiv:hep-ph/9506405}}, \href
  {http://dx.doi.org/10.1007/s002180050123} {\path{doi:10.1007/s002180050123}}.

\bibitem{Pepin:1996id}
S.~Pepin, F.~Stancu, M.~Genovese, J.~M. Richard, {Tetraquarks with color blind
  forces in chiral quark models}, Phys. Lett. B393 (1997) 119--123.
\newblock \href {http://arxiv.org/abs/hep-ph/9609348}
  {\path{arXiv:hep-ph/9609348}}, \href
  {http://dx.doi.org/10.1016/S0370-2693(96)01597-3}
  {\path{doi:10.1016/S0370-2693(96)01597-3}}.

\bibitem{Brink:1998as}
D.~M. Brink, F.~Stancu, {Tetraquarks with heavy flavors}, Phys. Rev. D57 (1998)
  6778--6787.
\newblock \href {http://dx.doi.org/10.1103/PhysRevD.57.6778}
  {\path{doi:10.1103/PhysRevD.57.6778}}.

\bibitem{SchaffnerBielich:1998ci}
J.~Schaffner-Bielich, A.~P. Vischer, {Charmlets}, Phys. Rev. D57 (1998)
  4142--4153.
\newblock \href {http://arxiv.org/abs/nucl-th/9710064}
  {\path{arXiv:nucl-th/9710064}}, \href
  {http://dx.doi.org/10.1103/PhysRevD.57.4142}
  {\path{doi:10.1103/PhysRevD.57.4142}}.

\bibitem{Gelman:2002wf}
B.~A. Gelman, S.~Nussinov, {Does a narrow tetraquark cc anti-u anti-d state
  exist?}, Phys. Lett. B551 (2003) 296--304.
\newblock \href {http://arxiv.org/abs/hep-ph/0209095}
  {\path{arXiv:hep-ph/0209095}}, \href
  {http://dx.doi.org/10.1016/S0370-2693(02)03069-1}
  {\path{doi:10.1016/S0370-2693(02)03069-1}}.

\bibitem{Stewart:1998hk}
C.~Stewart, R.~Koniuk, {Hadronic molecules in lattice QCD}, Phys. Rev. D57
  (1998) 5581--5585.
\newblock \href {http://arxiv.org/abs/hep-lat/9803003}
  {\path{arXiv:hep-lat/9803003}}, \href
  {http://dx.doi.org/10.1103/PhysRevD.57.5581}
  {\path{doi:10.1103/PhysRevD.57.5581}}.

\bibitem{Michael:1999nq}
C.~Michael, P.~Pennanen, {Two heavy - light mesons on a lattice}, Phys. Rev.
  D60 (1999) 054012.
\newblock \href {http://arxiv.org/abs/hep-lat/9901007}
  {\path{arXiv:hep-lat/9901007}}, \href
  {http://dx.doi.org/10.1103/PhysRevD.60.054012}
  {\path{doi:10.1103/PhysRevD.60.054012}}.

\bibitem{Barnes:1999hs}
T.~Barnes, N.~Black, D.~J. Dean, E.~S. Swanson, {B B intermeson potentials in
  the quark model}, Phys. Rev. C60 (1999) 045202.
\newblock \href {http://arxiv.org/abs/nucl-th/9902068}
  {\path{arXiv:nucl-th/9902068}}, \href
  {http://dx.doi.org/10.1103/PhysRevC.60.045202}
  {\path{doi:10.1103/PhysRevC.60.045202}}.

\bibitem{Cook:2002am}
M.~S. Cook, H.~R. Fiebig, {A Lattice study of interaction mechanisms in a heavy
  light meson meson system}\href {http://arxiv.org/abs/hep-lat/0210054}
  {\path{arXiv:hep-lat/0210054}}.

\bibitem{Lee:2007tn}
S.~H. Lee, S.~Yasui, W.~Liu, C.~M. Ko, {Charmed exotics in Heavy Ion
  Collisions}, Eur. Phys. J. C54 (2008) 259--265.
\newblock \href {http://arxiv.org/abs/0707.1747} {\path{arXiv:0707.1747}},
  \href {http://dx.doi.org/10.1140/epjc/s10052-007-0516-z}
  {\path{doi:10.1140/epjc/s10052-007-0516-z}}.

\bibitem{Lee:2009rt}
S.~H. Lee, S.~Yasui, {Stable multiquark states with heavy quarks in a diquark
  model}, Eur. Phys. J. C64 (2009) 283--295.
\newblock \href {http://arxiv.org/abs/0901.2977} {\path{arXiv:0901.2977}},
  \href {http://dx.doi.org/10.1140/epjc/s10052-009-1140-x}
  {\path{doi:10.1140/epjc/s10052-009-1140-x}}.

\bibitem{Hyodo:2012pm}
T.~Hyodo, Y.-R. Liu, M.~Oka, K.~Sudoh, S.~Yasui, {Production of doubly charmed
  tetraquarks with exotic color configurations in electron-positron
  collisions}, Phys. Lett. B721 (2013) 56--60.
\newblock \href {http://arxiv.org/abs/1209.6207} {\path{arXiv:1209.6207}},
  \href {http://dx.doi.org/10.1016/j.physletb.2013.02.045}
  {\path{doi:10.1016/j.physletb.2013.02.045}}.

\bibitem{Hyodo:2017hue}
T.~Hyodo, Y.-R. Liu, M.~Oka, S.~Yasui, {Spectroscopy and production of doubly
  charmed tetraquarks}\href {http://arxiv.org/abs/1708.05169}
  {\path{arXiv:1708.05169}}.

\bibitem{Ebert:2007rn}
D.~Ebert, R.~N. Faustov, V.~O. Galkin, W.~Lucha, {Masses of tetraquarks with
  two heavy quarks in the relativistic quark model}, Phys. Rev. D76 (2007)
  114015.
\newblock \href {http://arxiv.org/abs/0706.3853} {\path{arXiv:0706.3853}},
  \href {http://dx.doi.org/10.1103/PhysRevD.76.114015}
  {\path{doi:10.1103/PhysRevD.76.114015}}.

\bibitem{Eichten:2017ffp}
E.~J. Eichten, C.~Quigg, {Heavy-quark symmetry implies stable heavy tetraquark
  mesons $Q_iQ_j \bar q_k \bar q_l$}, Phys. Rev. Lett. 119~(20) (2017) 202002.
\newblock \href {http://arxiv.org/abs/1707.09575} {\path{arXiv:1707.09575}},
  \href {http://dx.doi.org/10.1103/PhysRevLett.119.202002}
  {\path{doi:10.1103/PhysRevLett.119.202002}}.

\bibitem{Mehen:2017nrh}
T.~Mehen, {Implications of Heavy Quark-Diquark Symmetry for Excited Doubly
  Heavy Baryons and Tetraquarks}, Phys. Rev. D96~(9) (2017) 094028.
\newblock \href {http://arxiv.org/abs/1708.05020} {\path{arXiv:1708.05020}},
  \href {http://dx.doi.org/10.1103/PhysRevD.96.094028}
  {\path{doi:10.1103/PhysRevD.96.094028}}.

\bibitem{Karliner:2017qjm}
M.~Karliner, J.~L. Rosner, {Discovery of doubly-charmed $\Xi_{cc}$ baryon
  implies a stable ($b b \bar{u} \bar{d}$) tetraquark}, Phys. Rev. Lett.
  119~(20) (2017) 202001.
\newblock \href {http://arxiv.org/abs/1707.07666} {\path{arXiv:1707.07666}},
  \href {http://dx.doi.org/10.1103/PhysRevLett.119.202001}
  {\path{doi:10.1103/PhysRevLett.119.202001}}.

\bibitem{Yan:2018gik}
X.~Yan, B.~Zhong, R.~Zhu, {Doubly charmed tetraquarks in a diquark¨Cantidiquark
  model}, Int. J. Mod. Phys. A33~(16) (2018) 1850096.
\newblock \href {http://arxiv.org/abs/1804.06761} {\path{arXiv:1804.06761}},
  \href {http://dx.doi.org/10.1142/S0217751X18500963}
  {\path{doi:10.1142/S0217751X18500963}}.

\bibitem{Xing:2018bqt}
Y.~Xing, R.~Zhu, {Weak Decays of Stable Doubly Heavy Tetraquark States}, Phys.
  Rev. D98~(5) (2018) 053005.
\newblock \href {http://arxiv.org/abs/1806.01659} {\path{arXiv:1806.01659}},
  \href {http://dx.doi.org/10.1103/PhysRevD.98.053005}
  {\path{doi:10.1103/PhysRevD.98.053005}}.

\bibitem{Iwasaki:1975pv}
Y.~Iwasaki, {A Possible Model for New Resonances-Exotics and Hidden Charm},
  Prog. Theor. Phys. 54 (1975) 492.
\newblock \href {http://dx.doi.org/10.1143/PTP.54.492}
  {\path{doi:10.1143/PTP.54.492}}.

\bibitem{Iwasaki:1976cn}
Y.~Iwasaki, {Is a State c anti-c c anti-c Found at 6.0-GeV?}, Phys. Rev. Lett.
  36 (1976) 1266.
\newblock \href {http://dx.doi.org/10.1103/PhysRevLett.36.1266}
  {\path{doi:10.1103/PhysRevLett.36.1266}}.

\bibitem{Iwasaki:1977qw}
Y.~Iwasaki, {How to Find eta(C) and a Possible State Charm anti-Charm Charm
  anti-Charm}, Phys. Rev. D16 (1977) 220.
\newblock \href {http://dx.doi.org/10.1103/PhysRevD.16.220}
  {\path{doi:10.1103/PhysRevD.16.220}}.

\bibitem{Chao:1980dv}
K.-T. Chao, {The (c c) - (anti-c anti-c) (Diquark - anti-Diquark) States in e+
  e- Annihilation}, Z. Phys. C7 (1981) 317.
\newblock \href {http://dx.doi.org/10.1007/BF01431564}
  {\path{doi:10.1007/BF01431564}}.

\bibitem{Ader:1981db}
J.~P. Ader, J.~M. Richard, P.~Taxil, {DO NARROW HEAVY MULTI - QUARK STATES
  EXIST?}, Phys. Rev. D25 (1982) 2370.
\newblock \href {http://dx.doi.org/10.1103/PhysRevD.25.2370}
  {\path{doi:10.1103/PhysRevD.25.2370}}.

\bibitem{Heller:1985cb}
L.~Heller, J.~A. Tjon, {On Bound States of Heavy $Q^2 \bar{Q}^2$ Systems},
  Phys. Rev. D32 (1985) 755.
\newblock \href {http://dx.doi.org/10.1103/PhysRevD.32.755}
  {\path{doi:10.1103/PhysRevD.32.755}}.

\bibitem{Badalian:1985es}
A.~M. Badalian, B.~L. Ioffe, A.~V. Smilga, {FOUR QUARK STATES IN THE HEAVY
  QUARK SYSTEM}, Nucl. Phys. B281 (1987) 85.
\newblock \href {http://dx.doi.org/10.1016/0550-3213(87)90248-3}
  {\path{doi:10.1016/0550-3213(87)90248-3}}.

\bibitem{Kalashnikova:1988bx}
{\relax Yu}.~S. Kalashnikova, I.~M. Narodetsky, {DISSOCIATION CALCULATIONS FOR
  S WAVE Q**2 anti-Q**2 SYSTEMS}, Z. Phys. C43 (1989) 273.
\newblock \href {http://dx.doi.org/10.1007/BF01588215}
  {\path{doi:10.1007/BF01588215}}.

\bibitem{Berezhnoy:2011xy}
A.~V. Berezhnoy, A.~K. Likhoded, A.~V. Luchinsky, A.~A. Novoselov, {Double
  J/psi-meson Production at LHC and 4c-tetraquark state}, Phys. Rev. D84 (2011)
  094023.
\newblock \href {http://arxiv.org/abs/1101.5881} {\path{arXiv:1101.5881}},
  \href {http://dx.doi.org/10.1103/PhysRevD.84.094023}
  {\path{doi:10.1103/PhysRevD.84.094023}}.

\bibitem{Berezhnoy:2011xn}
A.~V. Berezhnoy, A.~V. Luchinsky, A.~A. Novoselov, {Tetraquarks Composed of 4
  Heavy Quarks}, Phys. Rev. D86 (2012) 034004.
\newblock \href {http://arxiv.org/abs/1111.1867} {\path{arXiv:1111.1867}},
  \href {http://dx.doi.org/10.1103/PhysRevD.86.034004}
  {\path{doi:10.1103/PhysRevD.86.034004}}.

\bibitem{Kiselev:2002iy}
V.~V. Kiselev, A.~K. Likhoded, O.~N. Pakhomova, V.~A. Saleev, {Mass spectra of
  doubly heavy Omega $Q Q^\prime$ baryons}, Phys. Rev. D66 (2002) 034030.
\newblock \href {http://arxiv.org/abs/hep-ph/0206140}
  {\path{arXiv:hep-ph/0206140}}, \href
  {http://dx.doi.org/10.1103/PhysRevD.66.034030}
  {\path{doi:10.1103/PhysRevD.66.034030}}.

\bibitem{Liu:2019zuc}
M.-S. Liu, Q.-F. L¨¹, X.-H. Zhong, Q.~Zhao, {Fully-heavy tetraquarks}\href
  {http://arxiv.org/abs/1901.02564} {\path{arXiv:1901.02564}}.

\bibitem{Aaij:2018zrb}
R.~Aaij, et~al., {Search for beautiful tetraquarks in the
  $\Upsilon(1S)\mu^+\mu^-$ invariant-mass spectrum}, JHEP 10 (2018) 086.
\newblock \href {http://arxiv.org/abs/1806.09707} {\path{arXiv:1806.09707}},
  \href {http://dx.doi.org/10.1007/JHEP10(2018)086}
  {\path{doi:10.1007/JHEP10(2018)086}}.

\bibitem{Khachatryan:2016ydm}
V.~Khachatryan, et~al., {Observation of $\Upsilon$(1S) pair production in
  proton-proton collisions at $ \sqrt{s}=8 $ TeV}, JHEP 05 (2017) 013.
\newblock \href {http://arxiv.org/abs/1610.07095} {\path{arXiv:1610.07095}},
  \href {http://dx.doi.org/10.1007/JHEP05(2017)013}
  {\path{doi:10.1007/JHEP05(2017)013}}.

\bibitem{Esposito:2018cwh}
A.~Esposito, A.~D. Polosa, {A $bb\bar b\bar b$di-bottomonium at the LHC?}, Eur.
  Phys. J. C78~(9) (2018) 782.
\newblock \href {http://arxiv.org/abs/1807.06040} {\path{arXiv:1807.06040}},
  \href {http://dx.doi.org/10.1140/epjc/s10052-018-6269-z}
  {\path{doi:10.1140/epjc/s10052-018-6269-z}}.

\bibitem{Wu:2010jy}
J.-J. Wu, R.~Molina, E.~Oset, B.~S. Zou, {Prediction of narrow $N^*$ and
  $\Lambda^*$ resonances with hidden charm above 4 GeV}, Phys. Rev. Lett. 105
  (2010) 232001.
\newblock \href {http://arxiv.org/abs/1007.0573} {\path{arXiv:1007.0573}},
  \href {http://dx.doi.org/10.1103/PhysRevLett.105.232001}
  {\path{doi:10.1103/PhysRevLett.105.232001}}.

\bibitem{Aaij:2015tga}
R.~Aaij, et~al., {Observation of $J/\psi p$ Resonances Consistent with
  Pentaquark States in $\Lambda_b^0 \to J/\psi K^-p$ Decays}, Phys. Rev. Lett.
  115 (2015) 072001.
\newblock \href {http://arxiv.org/abs/1507.03414} {\path{arXiv:1507.03414}},
  \href {http://dx.doi.org/10.1103/PhysRevLett.115.072001}
  {\path{doi:10.1103/PhysRevLett.115.072001}}.

\bibitem{Aaij:2016phn}
R.~Aaij, et~al., {Model-independent evidence for $J/\psi p$ contributions to
  $\Lambda_b^0\to J/\psi p K^-$ decays}, Phys. Rev. Lett. 117~(8) (2016)
  082002.
\newblock \href {http://arxiv.org/abs/1604.05708} {\path{arXiv:1604.05708}},
  \href {http://dx.doi.org/10.1103/PhysRevLett.117.082002}
  {\path{doi:10.1103/PhysRevLett.117.082002}}.

\bibitem{Aaij:2016ymb}
R.~Aaij, et~al., {Evidence for exotic hadron contributions to $\Lambda_b^0 \to
  J/\psi p \pi^-$ decays}, Phys. Rev. Lett. 117~(8) (2016) 082003, [Addendum:
  Phys. Rev. Lett.118,119901(2017)].
\newblock \href {http://arxiv.org/abs/1606.06999} {\path{arXiv:1606.06999}},
  \href {http://dx.doi.org/10.1103/PhysRevLett.118.119901,
  10.1103/PhysRevLett.117.082003, 10.1103/PhysRevLett.117.109902}
  {\path{doi:10.1103/PhysRevLett.118.119901, 10.1103/PhysRevLett.117.082003,
  10.1103/PhysRevLett.117.109902}}.

\bibitem{Blin:2016dlf}
A.~N. Hiller~Blin, C.~Fern¨¢ndez-Ram¨ªrez, A.~Jackura, V.~Mathieu, V.~I.
  Mokeev, A.~Pilloni, A.~P. Szczepaniak, {Studying the P$_c$(4450) resonance in
  J/$\psi$ photoproduction off protons}, Phys. Rev. D94~(3) (2016) 034002.
\newblock \href {http://arxiv.org/abs/1606.08912} {\path{arXiv:1606.08912}},
  \href {http://dx.doi.org/10.1103/PhysRevD.94.034002}
  {\path{doi:10.1103/PhysRevD.94.034002}}.

\bibitem{Kubarovsky:2016whd}
V.~Kubarovsky, M.~B. Voloshin, {Search for Hidden-Charm Pentaquark with
  CLAS12}\href {http://arxiv.org/abs/1609.00050} {\path{arXiv:1609.00050}}.

\bibitem{Meziani:2016lhg}
Z.~E. Meziani, et~al., {A Search for the LHCb Charmed 'Pentaquark' using
  Photo-Production of $J/{\psi}$ at Threshold in Hall C at Jefferson Lab}\href
  {http://arxiv.org/abs/1609.00676} {\path{arXiv:1609.00676}}.

\bibitem{Joosten:2018gyo}
S.~Joosten, Z.~E. Meziani, {Heavy Quarkonium Production at Threshold: from JLab
  to EIC}, PoS QCDEV2017 (2018) 017.
\newblock \href {http://arxiv.org/abs/1802.02616} {\path{arXiv:1802.02616}},
  \href {http://dx.doi.org/10.22323/1.308.0017}
  {\path{doi:10.22323/1.308.0017}}.

\bibitem{Lutz:2009ff}
M.~F.~M. Lutz, et~al., {Physics Performance Report for PANDA: Strong
  Interaction Studies with Antiprotons,~}\href {http://arxiv.org/abs/0903.3905}
  {\path{arXiv:0903.3905}}.

\bibitem{Wu:2010vk}
J.-J. Wu, R.~Molina, E.~Oset, B.~S. Zou, {Dynamically generated $N^{*}$ and
  $\Lambda^*$ resonances in the hidden charm sector around 4.3 GeV}, Phys. Rev.
  C84 (2011) 015202.
\newblock \href {http://arxiv.org/abs/1011.2399} {\path{arXiv:1011.2399}},
  \href {http://dx.doi.org/10.1103/PhysRevC.84.015202}
  {\path{doi:10.1103/PhysRevC.84.015202}}.

\bibitem{Wang:2011rga}
W.~L. Wang, F.~Huang, Z.~Y. Zhang, B.~S. Zou, {$\Sigma_c \bar{D}$ and
  $\Lambda_c \bar{D}$ states in a chiral quark model}, Phys. Rev. C84 (2011)
  015203.
\newblock \href {http://arxiv.org/abs/1101.0453} {\path{arXiv:1101.0453}},
  \href {http://dx.doi.org/10.1103/PhysRevC.84.015203}
  {\path{doi:10.1103/PhysRevC.84.015203}}.

\bibitem{Yang:2011wz}
Z.-C. Yang, Z.-F. Sun, J.~He, X.~Liu, S.-L. Zhu, {The possible hidden-charm
  molecular baryons composed of anti-charmed meson and charmed baryon}, Chin.
  Phys. C36 (2012) 6--13.
\newblock \href {http://arxiv.org/abs/1105.2901} {\path{arXiv:1105.2901}},
  \href {http://dx.doi.org/10.1088/1674-1137/36/1/002}
  {\path{doi:10.1088/1674-1137/36/1/002}}.

\bibitem{Yuan:2012wz}
S.~G. Yuan, K.~W. Wei, J.~He, H.~S. Xu, B.~S. Zou, {Study of $qqqc\bar{c}$ five
  quark system with three kinds of quark-quark hyperfine interaction}, Eur.
  Phys. J. A48 (2012) 61.
\newblock \href {http://arxiv.org/abs/1201.0807} {\path{arXiv:1201.0807}},
  \href {http://dx.doi.org/10.1140/epja/i2012-12061-2}
  {\path{doi:10.1140/epja/i2012-12061-2}}.

\bibitem{Wu:2012md}
J.-J. Wu, T.~S.~H. Lee, B.~S. Zou, {Nucleon Resonances with Hidden Charm in
  Coupled-Channel Models}, Phys. Rev. C85 (2012) 044002.
\newblock \href {http://arxiv.org/abs/1202.1036} {\path{arXiv:1202.1036}},
  \href {http://dx.doi.org/10.1103/PhysRevC.85.044002}
  {\path{doi:10.1103/PhysRevC.85.044002}}.

\bibitem{Garcia-Recio:2013gaa}
C.~Garcia-Recio, J.~Nieves, O.~Romanets, L.~L. Salcedo, L.~Tolos, {Hidden charm
  $N$ and $\Delta$ resonances with heavy-quark symmetry}, Phys. Rev. D87 (2013)
  074034.
\newblock \href {http://arxiv.org/abs/1302.6938} {\path{arXiv:1302.6938}},
  \href {http://dx.doi.org/10.1103/PhysRevD.87.074034}
  {\path{doi:10.1103/PhysRevD.87.074034}}.

\bibitem{Xiao:2013yca}
C.~W. Xiao, J.~Nieves, E.~Oset, {Combining heavy quark spin and local hidden
  gauge symmetries in the dynamical generation of hidden charm baryons}, Phys.
  Rev. D88 (2013) 056012.
\newblock \href {http://arxiv.org/abs/1304.5368} {\path{arXiv:1304.5368}},
  \href {http://dx.doi.org/10.1103/PhysRevD.88.056012}
  {\path{doi:10.1103/PhysRevD.88.056012}}.

\bibitem{Huang:2013mua}
Y.~Huang, J.~He, H.-F. Zhang, X.-R. Chen, {Discovery potential of hidden charm
  baryon resonances via photoproduction}, J. Phys. G41~(11) (2014) 115004.
\newblock \href {http://arxiv.org/abs/1305.4434} {\path{arXiv:1305.4434}},
  \href {http://dx.doi.org/10.1088/0954-3899/41/11/115004}
  {\path{doi:10.1088/0954-3899/41/11/115004}}.

\bibitem{Li:2014gra}
X.-Q. Li, X.~Liu, {A possible global group structure for exotic states}, Eur.
  Phys. J. C74~(12) (2014) 3198.
\newblock \href {http://arxiv.org/abs/1409.3332} {\path{arXiv:1409.3332}},
  \href {http://dx.doi.org/10.1140/epjc/s10052-014-3198-3}
  {\path{doi:10.1140/epjc/s10052-014-3198-3}}.

\bibitem{Wang:2015xwa}
X.-Y. Wang, X.-R. Chen, {Production of the superheavy baryon
  $\Lambda_{c\bar{c}}^{*}$ (4209) in kaon-induced reaction}, Eur. Phys. J.
  A51~(7) (2015) 85.
\newblock \href {http://arxiv.org/abs/1504.01075} {\path{arXiv:1504.01075}},
  \href {http://dx.doi.org/10.1140/epja/i2015-15085-0}
  {\path{doi:10.1140/epja/i2015-15085-0}}.

\bibitem{Uchino:2015uha}
T.~Uchino, W.-H. Liang, E.~Oset, {Baryon states with hidden charm in the
  extended local hidden gauge approach}, Eur. Phys. J. A52~(3) (2016) 43.
\newblock \href {http://arxiv.org/abs/1504.05726} {\path{arXiv:1504.05726}},
  \href {http://dx.doi.org/10.1140/epja/i2016-16043-0}
  {\path{doi:10.1140/epja/i2016-16043-0}}.

\bibitem{Garzon:2015zva}
E.~J. Garzon, J.-J. Xie, {Effects of a $N^*_{c\overline{c}}$ resonance with
  hidden charm in the $\pi^-p \to D^-\Sigma_c^+$ reaction near threshold},
  Phys. Rev. C92~(3) (2015) 035201.
\newblock \href {http://arxiv.org/abs/1506.06834} {\path{arXiv:1506.06834}},
  \href {http://dx.doi.org/10.1103/PhysRevC.92.035201}
  {\path{doi:10.1103/PhysRevC.92.035201}}.

\bibitem{Hofmann:2005sw}
J.~Hofmann, M.~F.~M. Lutz, {Coupled-channel study of crypto-exotic baryons with
  charm}, Nucl. Phys. A763 (2005) 90--139.
\newblock \href {http://arxiv.org/abs/hep-ph/0507071}
  {\path{arXiv:hep-ph/0507071}}, \href
  {http://dx.doi.org/10.1016/j.nuclphysa.2005.08.022}
  {\path{doi:10.1016/j.nuclphysa.2005.08.022}}.

\bibitem{Chen:2015loa}
R.~Chen, X.~Liu, X.-Q. Li, S.-L. Zhu, {Identifying exotic hidden-charm
  pentaquarks}, Phys. Rev. Lett. 115~(13) (2015) 132002.
\newblock \href {http://arxiv.org/abs/1507.03704} {\path{arXiv:1507.03704}},
  \href {http://dx.doi.org/10.1103/PhysRevLett.115.132002}
  {\path{doi:10.1103/PhysRevLett.115.132002}}.

\bibitem{Chen:2015moa}
H.-X. Chen, W.~Chen, X.~Liu, T.~G. Steele, S.-L. Zhu, {Towards exotic
  hidden-charm pentaquarks in QCD}, Phys. Rev. Lett. 115~(17) (2015) 172001.
\newblock \href {http://arxiv.org/abs/1507.03717} {\path{arXiv:1507.03717}},
  \href {http://dx.doi.org/10.1103/PhysRevLett.115.172001}
  {\path{doi:10.1103/PhysRevLett.115.172001}}.

\bibitem{Kopeliovich:2015vqa}
V.~Kopeliovich, I.~Potashnikova, {Simple estimates of the masses of pentaquarks
  with hidden beauty or strangeness}, Phys. Rev. D93 (2016) 074012.
\newblock \href {http://arxiv.org/abs/1510.05958} {\path{arXiv:1510.05958}},
  \href {http://dx.doi.org/10.1103/PhysRevD.93.074012}
  {\path{doi:10.1103/PhysRevD.93.074012}}.

\bibitem{Chen:2016heh}
R.~Chen, X.~Liu, S.-L. Zhu, {Hidden-charm molecular pentaquarks and their
  charm-strange partners}, Nucl. Phys. A954 (2016) 406--421.
\newblock \href {http://arxiv.org/abs/1601.03233} {\path{arXiv:1601.03233}},
  \href {http://dx.doi.org/10.1016/j.nuclphysa.2016.04.012}
  {\path{doi:10.1016/j.nuclphysa.2016.04.012}}.

\bibitem{Maiani:2015vwa}
L.~Maiani, A.~D. Polosa, V.~Riquer, {The New Pentaquarks in the Diquark Model},
  Phys. Lett. B749 (2015) 289--291.
\newblock \href {http://arxiv.org/abs/1507.04980} {\path{arXiv:1507.04980}},
  \href {http://dx.doi.org/10.1016/j.physletb.2015.08.008}
  {\path{doi:10.1016/j.physletb.2015.08.008}}.

\bibitem{Anisovich:2015cia}
V.~V. Anisovich, M.~A. Matveev, J.~Nyiri, A.~V. Sarantsev, A.~N. Semenova,
  {Pentaquarks and resonances in the $pJ/\psi$ spectrum,~}\href
  {http://arxiv.org/abs/1507.07652} {\path{arXiv:1507.07652}}.

\bibitem{Ali:2016dkf}
A.~Ali, I.~Ahmed, M.~J. Aslam, A.~Rehman, {Heavy quark symmetry and weak decays
  of the $b$-baryons in pentaquarks with a $c\bar{c}$ component}, Phys. Rev.
  D94~(5) (2016) 054001.
\newblock \href {http://arxiv.org/abs/1607.00987} {\path{arXiv:1607.00987}},
  \href {http://dx.doi.org/10.1103/PhysRevD.94.054001}
  {\path{doi:10.1103/PhysRevD.94.054001}}.

\bibitem{Anisovich:2015xja}
V.~V. Anisovich, M.~A. Matveev, A.~V. Sarantsev, A.~N. Semenova, {Pentaquarks
  and strange tetraquark mesons}, Mod. Phys. Lett. A30~(38) (2015) 1550212.
\newblock \href {http://arxiv.org/abs/1509.03028} {\path{arXiv:1509.03028}},
  \href {http://dx.doi.org/10.1142/S0217732315502120}
  {\path{doi:10.1142/S0217732315502120}}.

\bibitem{Anisovich:2015zqa}
V.~V. Anisovich, M.~A. Matveev, J.~Nyiri, A.~V. Sarantsev, A.~N. Semenova,
  {Non-strange and strange pentaquarks with hidden charm}, Int. J. Mod. Phys.
  A30 (2015) 1550190.
\newblock \href {http://arxiv.org/abs/1509.04898} {\path{arXiv:1509.04898}},
  \href {http://dx.doi.org/10.1142/S0217751X15501900}
  {\path{doi:10.1142/S0217751X15501900}}.

\bibitem{Anisovich:2017aqa}
V.~V. Anisovich, M.~A. Matveev, J.~Nyiri, A.~N. Semenova, {Narrow pentaquarks
  as diquark¨Cdiquark¨Cantiquark systems}, Mod. Phys. Lett. A32~(29) (2017)
  1750154.
\newblock \href {http://arxiv.org/abs/1706.01336} {\path{arXiv:1706.01336}},
  \href {http://dx.doi.org/10.1142/S0217732317501541}
  {\path{doi:10.1142/S0217732317501541}}.

\bibitem{Anisovich:2017ubd}
V.~V. Anisovich, M.~A. Matveev, J.~Nyiri, A.~V. Sarantsev, A.~N. Semenova,
  {Diquark-diquark-antiquark model for pentaquarks with hidden charm: current
  status and problems}\href {http://arxiv.org/abs/1711.10736}
  {\path{arXiv:1711.10736}}.

\bibitem{Ali:2017ebb}
A.~Ali, I.~Ahmed, M.~J. Aslam, A.~Rehman, {Mass spectrum of spin-1/2
  pentaquarks with a $c\bar{c}$ component and their anticipated discovery modes
  in $b$-baryon decays}\href {http://arxiv.org/abs/1704.05419}
  {\path{arXiv:1704.05419}}.

\bibitem{Irie:2017qai}
Y.~Irie, M.~Oka, S.~Yasui, {Flavor-singlet hidden charm pentaquark}, Phys. Rev.
  D97~(3) (2018) 034006.
\newblock \href {http://arxiv.org/abs/1707.04544} {\path{arXiv:1707.04544}},
  \href {http://dx.doi.org/10.1103/PhysRevD.97.034006}
  {\path{doi:10.1103/PhysRevD.97.034006}}.

\bibitem{Buccella:2018jnt}
F.~Buccella, {b decays: a factory for hidden charm multiquark states}, Phys.
  Rev. D98~(11) (2018) 114011.
\newblock \href {http://arxiv.org/abs/1801.03723} {\path{arXiv:1801.03723}},
  \href {http://dx.doi.org/10.1103/PhysRevD.98.114011}
  {\path{doi:10.1103/PhysRevD.98.114011}}.

\bibitem{Santopinto:2016pkp}
E.~Santopinto, A.~Giachino, {Compact pentaquark structures}, Phys. Rev. D96~(1)
  (2017) 014014.
\newblock \href {http://arxiv.org/abs/1604.03769} {\path{arXiv:1604.03769}},
  \href {http://dx.doi.org/10.1103/PhysRevD.96.014014}
  {\path{doi:10.1103/PhysRevD.96.014014}}.

\bibitem{Gursey:1992dc}
F.~Gursey, L.~A. Radicati, {Spin and unitary spin independence of strong
  interactions}, Phys. Rev. Lett. 13 (1964) 173--175.
\newblock \href {http://dx.doi.org/10.1103/PhysRevLett.13.173}
  {\path{doi:10.1103/PhysRevLett.13.173}}.

\bibitem{Ortiz-Pacheco:2018ccl}
E.~Ortiz-Pacheco, R.~Bijker, C.~Fern{\'u}ndez-Ram{\'i}rez, {Hidden charm
  pentaquarks: mass spectrum, magnetic moments, and photocouplings}\href
  {http://arxiv.org/abs/1808.10512} {\path{arXiv:1808.10512}}.

\bibitem{Park:2018oib}
W.~Park, S.~Cho, S.~H. Lee, {Where is the stable Pentaquark}\href
  {http://arxiv.org/abs/1811.10911} {\path{arXiv:1811.10911}}.

\bibitem{Wei:2015gsa}
K.-W. Wei, B.~Chen, X.-H. Guo, {Masses of doubly and triply charmed baryons},
  Phys. Rev. D92~(7) (2015) 076008.
\newblock \href {http://arxiv.org/abs/1503.05184} {\path{arXiv:1503.05184}},
  \href {http://dx.doi.org/10.1103/PhysRevD.92.076008}
  {\path{doi:10.1103/PhysRevD.92.076008}}.

\bibitem{Wei:2016jyk}
K.-W. Wei, B.~Chen, N.~Liu, Q.-Q. Wang, X.-H. Guo, {Spectroscopy of singly,
  doubly, and triply bottom baryons}, Phys. Rev. D95~(11) (2017) 116005.
\newblock \href {http://arxiv.org/abs/1609.02512} {\path{arXiv:1609.02512}},
  \href {http://dx.doi.org/10.1103/PhysRevD.95.116005}
  {\path{doi:10.1103/PhysRevD.95.116005}}.

\bibitem{Bhaduri:1981pn}
R.~K. Bhaduri, L.~E. Cohler, Y.~Nogami, {A Unified Potential for Mesons and
  Baryons}, Nuovo Cim. A65 (1981) 376--390.
\newblock \href {http://dx.doi.org/10.1007/BF02827441}
  {\path{doi:10.1007/BF02827441}}.

\bibitem{SilvestreBrac:1996bg}
B.~Silvestre-Brac, {Spectrum and static properties of heavy baryons}, Few Body
  Syst. 20 (1996) 1--25.
\newblock \href {http://dx.doi.org/10.1007/s006010050028}
  {\path{doi:10.1007/s006010050028}}.

\bibitem{Kalashnikova:2018vkv}
Y.~S. Kalashnikova, A.~V. Nefediev, {$X(3872)$ in the molecular model}\href
  {http://arxiv.org/abs/1811.01324} {\path{arXiv:1811.01324}}, \href
  {http://dx.doi.org/10.3367/UFNe.2018.08.038411}
  {\path{doi:10.3367/UFNe.2018.08.038411}}.

\bibitem{Weinberg:1965zz}
S.~Weinberg, {Evidence That the Deuteron Is Not an Elementary Particle}, Phys.
  Rev. 137 (1965) B672--B678.
\newblock \href {http://dx.doi.org/10.1103/PhysRev.137.B672}
  {\path{doi:10.1103/PhysRev.137.B672}}.

\bibitem{Deng:2016stx}
W.-J. Deng, H.~Liu, L.-C. Gui, X.-H. Zhong, {Charmonium spectrum and their
  electromagnetic transitions with higher multipole contributions}, Phys. Rev.
  D95~(3) (2017) 034026.
\newblock \href {http://arxiv.org/abs/1608.00287} {\path{arXiv:1608.00287}},
  \href {http://dx.doi.org/10.1103/PhysRevD.95.034026}
  {\path{doi:10.1103/PhysRevD.95.034026}}.

\bibitem{Achasov:2016vxb}
N.~Achasov, {Physics of the charmonium-like state X(3872)}, EPJ Web Conf. 125
  (2016) 04002.
\newblock \href {http://dx.doi.org/10.1051/epjconf/201612504002}
  {\path{doi:10.1051/epjconf/201612504002}}.

\bibitem{Gui:2018rvv}
L.-C. Gui, L.-S. Lu, Q.-F. L¨¹, X.-H. Zhong, Q.~Zhao, {Strong decays of higher
  charmonium states into open-charm meson pairs}, Phys. Rev. D98~(1) (2018)
  016010.
\newblock \href {http://arxiv.org/abs/1801.08791} {\path{arXiv:1801.08791}},
  \href {http://dx.doi.org/10.1103/PhysRevD.98.016010}
  {\path{doi:10.1103/PhysRevD.98.016010}}.

\bibitem{Ortega:2016hde}
P.~G. Ortega, J.~Segovia, D.~R. Entem, F.~Fern¨¢ndez, {Canonical description of
  the new LHCb resonances}, Phys. Rev. D94~(11) (2016) 114018.
\newblock \href {http://arxiv.org/abs/1608.01325} {\path{arXiv:1608.01325}},
  \href {http://dx.doi.org/10.1103/PhysRevD.94.114018}
  {\path{doi:10.1103/PhysRevD.94.114018}}.

\bibitem{Vijande:2004he}
J.~Vijande, F.~Fernandez, A.~Valcarce, {Constituent quark model study of the
  meson spectra}, J. Phys. G31 (2005) 481.
\newblock \href {http://arxiv.org/abs/hep-ph/0411299}
  {\path{arXiv:hep-ph/0411299}}, \href
  {http://dx.doi.org/10.1088/0954-3899/31/5/017}
  {\path{doi:10.1088/0954-3899/31/5/017}}.

\bibitem{Segovia:2008zz}
J.~Segovia, A.~M. Yasser, D.~R. Entem, F.~Fernandez, {$J^{PC}=1^{--}$ hidden
  charm resonances}, Phys. Rev. D78 (2008) 114033.
\newblock \href {http://dx.doi.org/10.1103/PhysRevD.78.114033}
  {\path{doi:10.1103/PhysRevD.78.114033}}.

\bibitem{Molina:2017iaa}
D.~Molina, M.~De~Sanctis, C.~Fernandez-Ramirez, {Charmonium spectrum with a
  Dirac potential model in the momentum space}, Phys. Rev. D95~(9) (2017)
  094021.
\newblock \href {http://arxiv.org/abs/1703.08097} {\path{arXiv:1703.08097}},
  \href {http://dx.doi.org/10.1103/PhysRevD.95.094021}
  {\path{doi:10.1103/PhysRevD.95.094021}}.

\bibitem{Bhavsar:2018umj}
T.~Bhavsar, M.~Shah, P.~C. Vinodkumar, {Status of quarkonia-like negative and
  positive parity states in a relativistic confinement scheme}, Eur. Phys. J.
  C78~(3) (2018) 227.
\newblock \href {http://arxiv.org/abs/1803.07249} {\path{arXiv:1803.07249}},
  \href {http://dx.doi.org/10.1140/epjc/s10052-018-5694-3}
  {\path{doi:10.1140/epjc/s10052-018-5694-3}}.

\bibitem{Kher:2018wtv}
V.~Kher, A.~K. Rai, {Spectroscopy and decay properties of charmonium}, Chin.
  Phys. C42~(8) (2018) 083101.
\newblock \href {http://arxiv.org/abs/1805.02534} {\path{arXiv:1805.02534}},
  \href {http://dx.doi.org/10.1088/1674-1137/42/8/083101}
  {\path{doi:10.1088/1674-1137/42/8/083101}}.

\bibitem{Fu:2018yxq}
H.-F. Fu, L.~Jiang, {Coupled-Channel-Induced $S-D$ mixing of Charmonia and
  Possible Assignments for $Y(4260)$ and $Y(4360)$}\href
  {http://arxiv.org/abs/1812.00179} {\path{arXiv:1812.00179}}.

\bibitem{Oncala:2017hop}
R.~Oncala, J.~Soto, {Heavy Quarkonium Hybrids: Spectrum, Decay and Mixing},
  Phys. Rev. D96~(1) (2017) 014004.
\newblock \href {http://arxiv.org/abs/1702.03900} {\path{arXiv:1702.03900}},
  \href {http://dx.doi.org/10.1103/PhysRevD.96.014004}
  {\path{doi:10.1103/PhysRevD.96.014004}}.

\bibitem{Chen:2008xia}
K.~F. Chen, et~al., {Observation of an enhancement in $e^+e^- \to
  \Upsilon(1S)\pi^+ \pi^-$, $\Upsilon(2S)\pi^+ \pi^-$, and $\Upsilon(3S)\pi^+
  \pi^-$ production around $\sqrt{s}=10.89$ GeV at Belle}, Phys. Rev. D82
  (2010) 091106.
\newblock \href {http://arxiv.org/abs/0810.3829} {\path{arXiv:0810.3829}},
  \href {http://dx.doi.org/10.1103/PhysRevD.82.091106}
  {\path{doi:10.1103/PhysRevD.82.091106}}.

\bibitem{Berwein:2015vca}
M.~Berwein, N.~Brambilla, J.~Tarr{\'u}s~Castell{\`a}, A.~Vairo, {Quarkonium
  Hybrids with Nonrelativistic Effective Field Theories}, Phys. Rev. D92~(11)
  (2015) 114019.
\newblock \href {http://arxiv.org/abs/1510.04299} {\path{arXiv:1510.04299}},
  \href {http://dx.doi.org/10.1103/PhysRevD.92.114019}
  {\path{doi:10.1103/PhysRevD.92.114019}}.

\bibitem{Miyamoto:2018zfr}
T.~Miyamoto, S.~Yasui, {Hyperspherical-coordinate approach to the spectra and
  decay widths of hybrid quarkoniaA hyperspherical-coordinate approach to the
  spectra and decay widths of hybrid quarkonia}, Phys. Rev. D98~(9) (2018)
  094027.
\newblock \href {http://arxiv.org/abs/1806.07970} {\path{arXiv:1806.07970}},
  \href {http://dx.doi.org/10.1103/PhysRevD.98.094027}
  {\path{doi:10.1103/PhysRevD.98.094027}}.

\bibitem{Bruschini:2018lse}
R.~Bruschini, P.~Gonz{\'a}lez, {A plausible explanation of $\Upsilon(10860)$},
  Phys. Lett. B791 (2019) 409--413.
\newblock \href {http://arxiv.org/abs/1811.08236} {\path{arXiv:1811.08236}},
  \href {http://dx.doi.org/10.1016/j.physletb.2019.03.017}
  {\path{doi:10.1016/j.physletb.2019.03.017}}.

\bibitem{Ablikim:2019soz}
M.~Ablikim, et~al., {Observation of the decay $X(3872) \to \pi^0
  \chi_{c1}(1P)$}\href {http://arxiv.org/abs/1901.03992}
  {\path{arXiv:1901.03992}}.

\bibitem{Dubynskiy:2007tj}
S.~Dubynskiy, M.~B. Voloshin, {Pionic transitions from $X(3872)$ to
  $\chi_{cJ}$}, Phys. Rev. D77 (2008) 014013.
\newblock \href {http://arxiv.org/abs/0709.4474} {\path{arXiv:0709.4474}},
  \href {http://dx.doi.org/10.1103/PhysRevD.77.014013}
  {\path{doi:10.1103/PhysRevD.77.014013}}.

\bibitem{Aubert:2008ae}
B.~Aubert, et~al., {Evidence for $X(3872) \to \psi(2S) \gamma$ in $B^\pm \to
  X(3872) K^\pm$ decays, and a study of $B \to c \bar{c} \gamma K$}, Phys. Rev.
  Lett. 102 (2009) 132001.
\newblock \href {http://arxiv.org/abs/0809.0042} {\path{arXiv:0809.0042}},
  \href {http://dx.doi.org/10.1103/PhysRevLett.102.132001}
  {\path{doi:10.1103/PhysRevLett.102.132001}}.

\bibitem{Bhardwaj:2011dj}
V.~Bhardwaj, et~al., {Observation of $X(3872)\to J/\psi \gamma$ and search for
  $X(3872)\to\psi'\gamma$ in $B$ decays}, Phys. Rev. Lett. 107 (2011) 091803.
\newblock \href {http://arxiv.org/abs/1105.0177} {\path{arXiv:1105.0177}},
  \href {http://dx.doi.org/10.1103/PhysRevLett.107.091803}
  {\path{doi:10.1103/PhysRevLett.107.091803}}.

\bibitem{Cincioglu:2016fkm}
E.~Cincioglu, J.~Nieves, A.~Ozpineci, A.~U. Yilmazer, {Quarkonium Contribution
  to Meson Molecules}, Eur. Phys. J. C76~(10) (2016) 576.
\newblock \href {http://arxiv.org/abs/1606.03239} {\path{arXiv:1606.03239}},
  \href {http://dx.doi.org/10.1140/epjc/s10052-016-4413-1}
  {\path{doi:10.1140/epjc/s10052-016-4413-1}}.

\bibitem{Zhou:2017dwj}
Z.-Y. Zhou, Z.~Xiao, {Understanding $X(3862)$, $X(3872)$, and $X(3930)$ in a
  Friedrichs-model-like scheme}, Phys. Rev. D96~(5) (2017) 054031, [Erratum:
  Phys. Rev.D96,no.9,099905(2017)].
\newblock \href {http://arxiv.org/abs/1704.04438} {\path{arXiv:1704.04438}},
  \href {http://dx.doi.org/10.1103/PhysRevD.96.099905,
  10.1103/PhysRevD.96.054031} {\path{doi:10.1103/PhysRevD.96.099905,
  10.1103/PhysRevD.96.054031}}.

\bibitem{Zhou:2017txt}
Z.-Y. Zhou, Z.~Xiao, {Comprehending Isospin breaking effects of $X(3872)$ in a
  Friedrichs-model-like scheme}, Phys. Rev. D97~(3) (2018) 034011.
\newblock \href {http://arxiv.org/abs/1711.01930} {\path{arXiv:1711.01930}},
  \href {http://dx.doi.org/10.1103/PhysRevD.97.034011}
  {\path{doi:10.1103/PhysRevD.97.034011}}.

\bibitem{Zhou:2018hlv}
Z.-Y. Zhou, D.-Y. Chen, Z.~Xiao, {Does the bottomonium counterpart of $X(3872)$
  exist?}, Phys. Rev. D99~(3) (2019) 034005.
\newblock \href {http://arxiv.org/abs/1810.03452} {\path{arXiv:1810.03452}},
  \href {http://dx.doi.org/10.1103/PhysRevD.99.034005}
  {\path{doi:10.1103/PhysRevD.99.034005}}.

\bibitem{Chatrchyan:2013mea}
S.~Chatrchyan, et~al., {Search for a new bottomonium state decaying to
  $\Upsilon(1S)\pi^+\pi^-$ in pp collisions at $\sqrt{s}$ = 8 TeV}, Phys. Lett.
  B727 (2013) 57--76.
\newblock \href {http://arxiv.org/abs/1309.0250} {\path{arXiv:1309.0250}},
  \href {http://dx.doi.org/10.1016/j.physletb.2013.10.016}
  {\path{doi:10.1016/j.physletb.2013.10.016}}.

\bibitem{He:2014sqj}
X.~H. He, et~al., {Observation of $e^+e^- \to \pi^+ \pi^- \pi^0 \chi_{bJ}$ and
  Search for $X_b \to \omega \Upsilon(1S)$ at $\sqrt{s}=10.867$ GeV}, Phys.
  Rev. Lett. 113~(14) (2014) 142001.
\newblock \href {http://arxiv.org/abs/1408.0504} {\path{arXiv:1408.0504}},
  \href {http://dx.doi.org/10.1103/PhysRevLett.113.142001}
  {\path{doi:10.1103/PhysRevLett.113.142001}}.

\bibitem{Ferretti:2018tco}
J.~Ferretti, E.~Santopinto, {Threshold corrections of $\chi_c$ (2 P ) and
  $\chi_b$ (3 P ) states and J /$\psi \rho$ and J /$\psi \omega$ transitions of
  the $\chi$ (3872) in a coupled-channel model}, Phys. Lett. B789 (2019)
  550--555.
\newblock \href {http://arxiv.org/abs/1806.02489} {\path{arXiv:1806.02489}},
  \href {http://dx.doi.org/10.1016/j.physletb.2018.12.052}
  {\path{doi:10.1016/j.physletb.2018.12.052}}.

\bibitem{Ortega:2019tby}
P.~G. Ortega, D.~R. Entem, F.~Fern¨¢ndez, {Unquenching the quark model in a
  non-perturbative scheme}\href {http://arxiv.org/abs/1901.02484}
  {\path{arXiv:1901.02484}}.

\bibitem{Lu:2016mbb}
Y.~Lu, M.~N. Anwar, B.-S. Zou, {Coupled-Channel Effects for the Bottomonium
  with Realistic Wave Functions}, Phys. Rev. D94~(3) (2016) 034021.
\newblock \href {http://arxiv.org/abs/1606.06927} {\path{arXiv:1606.06927}},
  \href {http://dx.doi.org/10.1103/PhysRevD.94.034021}
  {\path{doi:10.1103/PhysRevD.94.034021}}.

\bibitem{Cincioglu:2019gzd}
E.~Cincioglu, A.~Ozpineci, {Radiative decay of the $X(3872)$ as a mixed
  molecule-charmonium state in effective field theory}\href
  {http://arxiv.org/abs/1901.03138} {\path{arXiv:1901.03138}}.

\bibitem{Giacosa:2019zxw}
F.~Giacosa, M.~Piotrowska, S.~Coito, {$X(3872)$ as virtual companion pole of
  the charm-anticharm state $\chi_{c1}(2P)$}\href
  {http://arxiv.org/abs/1903.06926} {\path{arXiv:1903.06926}}.

\bibitem{Qin:2016spb}
W.~Qin, S.-R. Xue, Q.~Zhao, {Production of $Y(4260)$ as a hadronic molecule
  state of $\bar{D}D_1 +c.c.$ in $e^+e^-$ annihilations}, Phys. Rev. D94~(5)
  (2016) 054035.
\newblock \href {http://arxiv.org/abs/1605.02407} {\path{arXiv:1605.02407}},
  \href {http://dx.doi.org/10.1103/PhysRevD.94.054035}
  {\path{doi:10.1103/PhysRevD.94.054035}}.

\bibitem{Lu:2017yhl}
Y.~Lu, M.~N. Anwar, B.-S. Zou, {$X(4260)$ Revisited: A Coupled Channel
  Perspective}, Phys. Rev. D96~(11) (2017) 114022.
\newblock \href {http://arxiv.org/abs/1705.00449} {\path{arXiv:1705.00449}},
  \href {http://dx.doi.org/10.1103/PhysRevD.96.114022}
  {\path{doi:10.1103/PhysRevD.96.114022}}.

\bibitem{Zhang:2007xa}
H.~X. Zhang, M.~Zhang, Z.~Y. Zhang, {Study of Qq anti-Q anti-q-prime states in
  chiral SU(3) quark model}, Chin. Phys. Lett. 24 (2007) 2533--2536.
\newblock \href {http://arxiv.org/abs/0705.2470} {\path{arXiv:0705.2470}},
  \href {http://dx.doi.org/10.1088/0256-307X/24/9/019}
  {\path{doi:10.1088/0256-307X/24/9/019}}.

\bibitem{Vijande:2007fc}
J.~Vijande, E.~Weissman, N.~Barnea, A.~Valcarce, {Do $c \bar{c} n \bar{n}$
  bound states exist?}, Phys. Rev. D76 (2007) 094022.
\newblock \href {http://arxiv.org/abs/0708.3285} {\path{arXiv:0708.3285}},
  \href {http://dx.doi.org/10.1103/PhysRevD.76.094022}
  {\path{doi:10.1103/PhysRevD.76.094022}}.

\bibitem{Coito:2016ads}
S.~Coito, {Radially excited axial mesons and the enigmatic $Z_c$ and $Z_b$ in a
  coupled-channel model}, Phys. Rev. D94~(1) (2016) 014016.
\newblock \href {http://arxiv.org/abs/1602.07821} {\path{arXiv:1602.07821}},
  \href {http://dx.doi.org/10.1103/PhysRevD.94.014016}
  {\path{doi:10.1103/PhysRevD.94.014016}}.

\bibitem{Patel:2016otd}
S.~Patel, P.~C. Vinodkumar, {Tetraquark states in the bottom sector and the
  status of the $Y_b$ (10890) state}, Eur. Phys. J. C76~(7) (2016) 356.
\newblock \href {http://arxiv.org/abs/1606.01047} {\path{arXiv:1606.01047}},
  \href {http://dx.doi.org/10.1140/epjc/s10052-016-4186-6}
  {\path{doi:10.1140/epjc/s10052-016-4186-6}}.

\bibitem{Lu:2016cwr}
Q.-F. L¨¹, Y.-B. Dong, {X(4140) , X(4274) , X(4500) , and X(4700) in the
  relativized quark model}, Phys. Rev. D94~(7) (2016) 074007.
\newblock \href {http://arxiv.org/abs/1607.05570} {\path{arXiv:1607.05570}},
  \href {http://dx.doi.org/10.1103/PhysRevD.94.074007}
  {\path{doi:10.1103/PhysRevD.94.074007}}.

\bibitem{Chen:2016oma}
H.-X. Chen, E.-L. Cui, W.~Chen, X.~Liu, S.-L. Zhu, {Understanding the internal
  structures of the $X(4140)$, $X(4274)$, $X(4500)$ and $X(4700)$}, Eur. Phys.
  J. C77~(3) (2017) 160.
\newblock \href {http://arxiv.org/abs/1606.03179} {\path{arXiv:1606.03179}},
  \href {http://dx.doi.org/10.1140/epjc/s10052-017-4737-5}
  {\path{doi:10.1140/epjc/s10052-017-4737-5}}.

\bibitem{Wang:2016gxp}
Z.-G. Wang, {Scalar tetraquark state candidates: $X(3915)$, $X(4500)$ and
  $X(4700)$}, Eur. Phys. J. C77~(2) (2017) 78.
\newblock \href {http://arxiv.org/abs/1606.05872} {\path{arXiv:1606.05872}},
  \href {http://dx.doi.org/10.1140/epjc/s10052-017-4640-0}
  {\path{doi:10.1140/epjc/s10052-017-4640-0}}.

\bibitem{Barnes:2005pb}
T.~Barnes, S.~Godfrey, E.~S. Swanson, {Higher charmonia}, Phys. Rev. D72 (2005)
  054026.
\newblock \href {http://arxiv.org/abs/hep-ph/0505002}
  {\path{arXiv:hep-ph/0505002}}, \href
  {http://dx.doi.org/10.1103/PhysRevD.72.054026}
  {\path{doi:10.1103/PhysRevD.72.054026}}.

\bibitem{Liu:2016onn}
X.-H. Liu, {How to understand the underlying structures of $X(4140)$,
  $X(4274)$, $X(4500)$ and $X(4700)$}, Phys. Lett. B766 (2017) 117--124.
\newblock \href {http://arxiv.org/abs/1607.01385} {\path{arXiv:1607.01385}},
  \href {http://dx.doi.org/10.1016/j.physletb.2017.01.008}
  {\path{doi:10.1016/j.physletb.2017.01.008}}.

\bibitem{Yang:2017prf}
Y.-C. Yang, Z.-Y. Tan, J.~Ping, H.-S. Zong, {Possible $D^{(*)}\bar{D}^{(*)}$
  and $B^{(*)}\bar{B}^{(*)}$ molecular states in the extended constituent quark
  models}, Eur. Phys. J. C77~(9) (2017) 575.
\newblock \href {http://arxiv.org/abs/1703.09718} {\path{arXiv:1703.09718}},
  \href {http://dx.doi.org/10.1140/epjc/s10052-017-5137-6}
  {\path{doi:10.1140/epjc/s10052-017-5137-6}}.

\bibitem{Yang:2017rmm}
Y.-C. Yang, Z.-Y. Tan, H.-S. Zong, J.~Ping, {Dynamical study of $S$-wave
  $\bar{Q}Q\bar{q}q$ system}, Few Body Syst. 60~(1) (2019) 9.
\newblock \href {http://arxiv.org/abs/1712.09285} {\path{arXiv:1712.09285}},
  \href {http://dx.doi.org/10.1007/s00601-018-1477-5}
  {\path{doi:10.1007/s00601-018-1477-5}}.

\bibitem{Anwar:2018sol}
M.~N. Anwar, J.~Ferretti, E.~Santopinto, {Spectroscopy of the hidden-charm
  $[qc][\bar q \bar c]$ and $[sc][\bar s \bar c]$ tetraquarks in the
  relativized diquark model}, Phys. Rev. D98~(9) (2018) 094015.
\newblock \href {http://arxiv.org/abs/1805.06276} {\path{arXiv:1805.06276}},
  \href {http://dx.doi.org/10.1103/PhysRevD.98.094015}
  {\path{doi:10.1103/PhysRevD.98.094015}}.

\bibitem{Giron:2019bcs}
J.~F. Giron, R.~F. Lebed, C.~T. Peterson, {The Dynamical Diquark Model: First
  Numerical Results}\href {http://arxiv.org/abs/1903.04551}
  {\path{arXiv:1903.04551}}.

\bibitem{Yang:2019dxd}
Y.~Yang, J.~Ping, {Investigation of $cs\bar c\bar s$ tetraquark in the chiral
  quark model}\href {http://arxiv.org/abs/1903.08505}
  {\path{arXiv:1903.08505}}.

\bibitem{Janc:2004qn}
D.~Janc, M.~Rosina, {The $T_{cc} = DD^*$ molecular state}, Few Body Syst. 35
  (2004) 175--196.
\newblock \href {http://arxiv.org/abs/hep-ph/0405208}
  {\path{arXiv:hep-ph/0405208}}, \href
  {http://dx.doi.org/10.1007/s00601-004-0068-9}
  {\path{doi:10.1007/s00601-004-0068-9}}.

\bibitem{Zhang:2007mu}
M.~Zhang, H.~X. Zhang, Z.~Y. Zhang, {$QQ \bar{q}\bar{q}$ four-quark bound
  states in chiral SU(3) quark model}, Commun. Theor. Phys. 50 (2008) 437--440.
\newblock \href {http://arxiv.org/abs/0711.1029} {\path{arXiv:0711.1029}},
  \href {http://dx.doi.org/10.1088/0253-6102/50/2/31}
  {\path{doi:10.1088/0253-6102/50/2/31}}.

\bibitem{Yang:2009zzp}
Y.~Yang, C.~Deng, J.~Ping, T.~Goldman, {S-wave $Q Q \bar q \bar q$ state in the
  constituent quark model}, Phys. Rev. D80 (2009) 114023.
\newblock \href {http://dx.doi.org/10.1103/PhysRevD.80.114023}
  {\path{doi:10.1103/PhysRevD.80.114023}}.

\bibitem{Vijande:2009kj}
J.~Vijande, A.~Valcarce, N.~Barnea, {Exotic meson-meson molecules and compact
  four-quark states}, Phys. Rev. D79 (2009) 074010.
\newblock \href {http://arxiv.org/abs/0903.2949} {\path{arXiv:0903.2949}},
  \href {http://dx.doi.org/10.1103/PhysRevD.79.074010}
  {\path{doi:10.1103/PhysRevD.79.074010}}.

\bibitem{Vijande:2003ki}
J.~Vijande, F.~Fernandez, A.~Valcarce, B.~Silvestre-Brac, {Tetraquarks in a
  chiral constituent quark model}, Eur. Phys. J. A19 (2004) 383.
\newblock \href {http://arxiv.org/abs/hep-ph/0310007}
  {\path{arXiv:hep-ph/0310007}}, \href
  {http://dx.doi.org/10.1140/epja/i2003-10128-9}
  {\path{doi:10.1140/epja/i2003-10128-9}}.

\bibitem{Vijande:2006jf}
J.~Vijande, A.~Valcarce, K.~Tsushima, {Dynamical study of bf QQ - anti-u anti-d
  mesons}, Phys. Rev. D74 (2006) 054018.
\newblock \href {http://arxiv.org/abs/hep-ph/0608316}
  {\path{arXiv:hep-ph/0608316}}, \href
  {http://dx.doi.org/10.1103/PhysRevD.74.054018}
  {\path{doi:10.1103/PhysRevD.74.054018}}.

\bibitem{Vijande:2007rf}
J.~Vijande, E.~Weissman, A.~Valcarce, N.~Barnea, {Are there compact heavy
  four-quark bound states?}, Phys. Rev. D76 (2007) 094027.
\newblock \href {http://arxiv.org/abs/0710.2516} {\path{arXiv:0710.2516}},
  \href {http://dx.doi.org/10.1103/PhysRevD.76.094027}
  {\path{doi:10.1103/PhysRevD.76.094027}}.

\bibitem{Caramees:2018oue}
T.~F. Caram¨¦s, J.~Vijande, A.~Valcarce, {Exotic $bc\bar q\bar q$ four-quark
  states}, Phys. Rev. D99~(1) (2019) 014006.
\newblock \href {http://arxiv.org/abs/1812.08991} {\path{arXiv:1812.08991}},
  \href {http://dx.doi.org/10.1103/PhysRevD.99.014006}
  {\path{doi:10.1103/PhysRevD.99.014006}}.

\bibitem{Feng:2013kea}
G.~Q. Feng, X.~H. Guo, B.~S. Zou, {$QQ^{\prime} \bar u \bar d$ bound state in
  the Bethe-Salpeter equation approach}\href {http://arxiv.org/abs/1309.7813}
  {\path{arXiv:1309.7813}}.

\bibitem{Park:2013fda}
W.~Park, S.~H. Lee, {Color spin wave functions of heavy tetraquark states},
  Nucl. Phys. A925 (2014) 161--184.
\newblock \href {http://arxiv.org/abs/1311.5330} {\path{arXiv:1311.5330}},
  \href {http://dx.doi.org/10.1016/j.nuclphysa.2014.02.008}
  {\path{doi:10.1016/j.nuclphysa.2014.02.008}}.

\bibitem{Park:2018wjk}
W.~Park, S.~Noh, S.~H. Lee, {Masses of the doubly heavy tetraquarks in a
  constituent quark model}, Nucl. Phys. A983 (2019) 1--19.
\newblock \href {http://arxiv.org/abs/1809.05257} {\path{arXiv:1809.05257}},
  \href {http://dx.doi.org/10.1016/j.nuclphysa.2018.12.019}
  {\path{doi:10.1016/j.nuclphysa.2018.12.019}}.

\bibitem{Czarnecki:2017vco}
A.~Czarnecki, B.~Leng, M.~B. Voloshin, {Stability of tetrons}, Phys. Lett. B778
  (2018) 233--238.
\newblock \href {http://arxiv.org/abs/1708.04594} {\path{arXiv:1708.04594}},
  \href {http://dx.doi.org/10.1016/j.physletb.2018.01.034}
  {\path{doi:10.1016/j.physletb.2018.01.034}}.

\bibitem{Richard:2018yrm}
J.-M. Richard, A.~Valcarce, J.~Vijande, {Few-body quark dynamics for doubly
  heavy baryons and tetraquarks}, Phys. Rev. C97~(3) (2018) 035211.
\newblock \href {http://arxiv.org/abs/1803.06155} {\path{arXiv:1803.06155}},
  \href {http://dx.doi.org/10.1103/PhysRevC.97.035211}
  {\path{doi:10.1103/PhysRevC.97.035211}}.

\bibitem{Lloyd:2003yc}
R.~J. Lloyd, J.~P. Vary, {All charm tetraquarks}, Phys. Rev. D70 (2004) 014009.
\newblock \href {http://arxiv.org/abs/hep-ph/0311179}
  {\path{arXiv:hep-ph/0311179}}, \href
  {http://dx.doi.org/10.1103/PhysRevD.70.014009}
  {\path{doi:10.1103/PhysRevD.70.014009}}.

\bibitem{Anwar:2017toa}
M.~N. Anwar, J.~Ferretti, F.-K. Guo, E.~Santopinto, B.-S. Zou, {Spectroscopy
  and decays of the fully-heavy tetraquarks}, Eur. Phys. J. C78~(8) (2018) 647.
\newblock \href {http://arxiv.org/abs/1710.02540} {\path{arXiv:1710.02540}},
  \href {http://dx.doi.org/10.1140/epjc/s10052-018-6073-9}
  {\path{doi:10.1140/epjc/s10052-018-6073-9}}.

\bibitem{Bai:2016int}
Y.~Bai, S.~Lu, J.~Osborne, {Beauty-full Tetraquarks}\href
  {http://arxiv.org/abs/1612.00012} {\path{arXiv:1612.00012}}.

\bibitem{Barnea:2006sd}
N.~Barnea, J.~Vijande, A.~Valcarce, {Four-quark spectroscopy within the
  hyperspherical formalism}, Phys. Rev. D73 (2006) 054004.
\newblock \href {http://arxiv.org/abs/hep-ph/0604010}
  {\path{arXiv:hep-ph/0604010}}, \href
  {http://dx.doi.org/10.1103/PhysRevD.73.054004}
  {\path{doi:10.1103/PhysRevD.73.054004}}.

\bibitem{Debastiani:2017msn}
V.~R. Debastiani, F.~S. Navarra, {A non-relativistic model for the
  $[cc][\bar{c}\bar{c}]$ tetraquark}, Chin. Phys. C43~(1) (2019) 013105.
\newblock \href {http://arxiv.org/abs/1706.07553} {\path{arXiv:1706.07553}},
  \href {http://dx.doi.org/10.1088/1674-1137/43/1/013105}
  {\path{doi:10.1088/1674-1137/43/1/013105}}.

\bibitem{Richard:2017vry}
J.-M. Richard, A.~Valcarce, J.~Vijande, {String dynamics and metastability of
  all-heavy tetraquarks}, Phys. Rev. D95~(5) (2017) 054019.
\newblock \href {http://arxiv.org/abs/1703.00783} {\path{arXiv:1703.00783}},
  \href {http://dx.doi.org/10.1103/PhysRevD.95.054019}
  {\path{doi:10.1103/PhysRevD.95.054019}}.

\bibitem{Chen:2019dvd}
X.~Chen, {Analysis of beauty-full $bb\bar{b}\bar{b}$ state}\href
  {http://arxiv.org/abs/1902.00008} {\path{arXiv:1902.00008}}.

\bibitem{Huang:2015uda}
H.~Huang, C.~Deng, J.~Ping, F.~Wang, {Possible pentaquarks with heavy quarks},
  Eur. Phys. J. C76~(11) (2016) 624.
\newblock \href {http://arxiv.org/abs/1510.04648} {\path{arXiv:1510.04648}},
  \href {http://dx.doi.org/10.1140/epjc/s10052-016-4476-z}
  {\path{doi:10.1140/epjc/s10052-016-4476-z}}.

\bibitem{Huang:2018ehi}
H.~Huang, X.~Zhu, J.~Ping, {$P_{c}$-like pentaquarks in hidden strange sector},
  Phys. Rev. D97~(9) (2018) 094019.
\newblock \href {http://arxiv.org/abs/1803.05267} {\path{arXiv:1803.05267}},
  \href {http://dx.doi.org/10.1103/PhysRevD.97.094019}
  {\path{doi:10.1103/PhysRevD.97.094019}}.

\bibitem{Yang:2015bmv}
G.~Yang, J.~Ping, {The structure of pentaquarks $P_c^+$ in the chiral quark
  model}, Phys. Rev. D95~(1) (2017) 014010.
\newblock \href {http://arxiv.org/abs/1511.09053} {\path{arXiv:1511.09053}},
  \href {http://dx.doi.org/10.1103/PhysRevD.95.014010}
  {\path{doi:10.1103/PhysRevD.95.014010}}.

\bibitem{Yang:2018oqd}
G.~Yang, J.~Ping, J.~Segovia, {Hidden-bottom pentaquarks}, Phys. Rev. D99~(1)
  (2019) 014035.
\newblock \href {http://arxiv.org/abs/1809.06193} {\path{arXiv:1809.06193}},
  \href {http://dx.doi.org/10.1103/PhysRevD.99.014035}
  {\path{doi:10.1103/PhysRevD.99.014035}}.

\bibitem{Gerasyuta:2015djk}
S.~M. Gerasyuta, V.~I. Kochkin, {Relativistic five-quark equations and the LHCb
  pentaquarks}\href {http://arxiv.org/abs/1512.04040}
  {\path{arXiv:1512.04040}}.

\bibitem{Ortega:2016syt}
P.~G. Ortega, D.~R. Entem, F.~Fern¨¢ndez, {LHCb pentaquarks in constituent
  quark models}, Phys. Lett. B764 (2017) 207--211.
\newblock \href {http://arxiv.org/abs/1606.06148} {\path{arXiv:1606.06148}},
  \href {http://dx.doi.org/10.1016/j.physletb.2016.11.008}
  {\path{doi:10.1016/j.physletb.2016.11.008}}.

\bibitem{Takeuchi:2016ejt}
S.~Takeuchi, M.~Takizawa, {The hidden charm pentaquarks are the hidden
  color-octet $uud$ baryons?}, Phys. Lett. B764 (2017) 254--259.
\newblock \href {http://arxiv.org/abs/1608.05475} {\path{arXiv:1608.05475}},
  \href {http://dx.doi.org/10.1016/j.physletb.2016.11.034}
  {\path{doi:10.1016/j.physletb.2016.11.034}}.

\bibitem{Park:2017jbn}
W.~Park, A.~Park, S.~Cho, S.~H. Lee, {$P_c(4380)$ in a constituent quark
  model}, Phys. Rev. D95~(5) (2017) 054027.
\newblock \href {http://arxiv.org/abs/1702.00381} {\path{arXiv:1702.00381}},
  \href {http://dx.doi.org/10.1103/PhysRevD.95.054027}
  {\path{doi:10.1103/PhysRevD.95.054027}}.

\bibitem{Stancu:2017azl}
F.~Stancu, {Stability of pentaquarks with a two- plus three-body chromoelectric
  interaction}, Phys. Rev. D96~(1) (2017) 014007.
\newblock \href {http://arxiv.org/abs/1705.02490} {\path{arXiv:1705.02490}},
  \href {http://dx.doi.org/10.1103/PhysRevD.96.014007}
  {\path{doi:10.1103/PhysRevD.96.014007}}.

\bibitem{Richard:2017una}
J.~M. Richard, A.~Valcarce, J.~Vijande, {Stable heavy pentaquarks in
  constituent models}, Phys. Lett. B774 (2017) 710--714.
\newblock \href {http://arxiv.org/abs/1710.08239} {\path{arXiv:1710.08239}},
  \href {http://dx.doi.org/10.1016/j.physletb.2017.10.036}
  {\path{doi:10.1016/j.physletb.2017.10.036}}.

\bibitem{Hiyama:2018ukv}
E.~Hiyama, A.~Hosaka, M.~Oka, J.-M. Richard, {Quark model estimate of
  hidden-charm pentaquark resonances}, Phys. Rev. C98~(4) (2018) 045208.
\newblock \href {http://arxiv.org/abs/1803.11369} {\path{arXiv:1803.11369}},
  \href {http://dx.doi.org/10.1103/PhysRevC.98.045208}
  {\path{doi:10.1103/PhysRevC.98.045208}}.

\bibitem{Stancu:2019qga}
F.~Stancu, {Spectrum of the $uudc \bar c$ hidden charm pentaquark with an SU(4)
  flavor-spin hyperfine interaction}\href {http://arxiv.org/abs/1902.07101}
  {\path{arXiv:1902.07101}}.

\bibitem{Giannuzzi:2019esi}
F.~Giannuzzi, {Heavy pentaquark spectroscopy in the diquark model}\href
  {http://arxiv.org/abs/1903.04430} {\path{arXiv:1903.04430}}.

\bibitem{Wang:2016dzu}
G.-J. Wang, R.~Chen, L.~Ma, X.~Liu, S.-L. Zhu, {Magnetic moments of the
  hidden-charm pentaquark states}, Phys. Rev. D94~(9) (2016) 094018.
\newblock \href {http://arxiv.org/abs/1605.01337} {\path{arXiv:1605.01337}},
  \href {http://dx.doi.org/10.1103/PhysRevD.94.094018}
  {\path{doi:10.1103/PhysRevD.94.094018}}.

\bibitem{Wang:2016wxg}
G.-J. Wang, Z.-W. Liu, S.-L. Zhu, {Axial charges of the hidden-charm pentaquark
  states}, Phys. Rev. C94~(6) (2016) 065202.
\newblock \href {http://arxiv.org/abs/1608.07824} {\path{arXiv:1608.07824}},
  \href {http://dx.doi.org/10.1103/PhysRevC.94.065202}
  {\path{doi:10.1103/PhysRevC.94.065202}}.

\bibitem{Isgur:1983wj}
N.~Isgur, J.~E. Paton, {A Flux Tube Model for Hadrons}, Phys. Lett. 124B (1983)
  247--251.
\newblock \href {http://dx.doi.org/10.1016/0370-2693(83)91445-4}
  {\path{doi:10.1016/0370-2693(83)91445-4}}.

\bibitem{Isgur:1984bm}
N.~Isgur, J.~E. Paton, {A Flux Tube Model for Hadrons in QCD}, Phys. Rev. D31
  (1985) 2910.
\newblock \href {http://dx.doi.org/10.1103/PhysRevD.31.2910}
  {\path{doi:10.1103/PhysRevD.31.2910}}.

\bibitem{Kogut:1974ag}
J.~B. Kogut, L.~Susskind, {Hamiltonian Formulation of Wilson's Lattice Gauge
  Theories}, Phys. Rev. D11 (1975) 395--408.
\newblock \href {http://dx.doi.org/10.1103/PhysRevD.11.395}
  {\path{doi:10.1103/PhysRevD.11.395}}.

\bibitem{Gnadig:1976pn}
P.~Gnadig, P.~Hasenfratz, J.~Kuti, A.~S. Szalay, {The Quark Bag Model with
  Surface Tension}, Phys. Lett. 64B (1976) 62--66.
\newblock \href {http://dx.doi.org/10.1016/0370-2693(76)90358-0}
  {\path{doi:10.1016/0370-2693(76)90358-0}}.

\bibitem{Barnes:1995hc}
T.~Barnes, F.~E. Close, E.~S. Swanson, {Hybrid and conventional mesons in the
  flux tube model: Numerical studies and their phenomenological implications},
  Phys. Rev. D52 (1995) 5242--5256.
\newblock \href {http://arxiv.org/abs/hep-ph/9501405}
  {\path{arXiv:hep-ph/9501405}}, \href
  {http://dx.doi.org/10.1103/PhysRevD.52.5242}
  {\path{doi:10.1103/PhysRevD.52.5242}}.

\bibitem{Merlin:1986tz}
J.~Merlin, J.~E. Paton, {Spin Interactions in the Flux Tube Model and Hybrid
  Meson Masses}, Phys. Rev. D35 (1987) 1668.
\newblock \href {http://dx.doi.org/10.1103/PhysRevD.35.1668}
  {\path{doi:10.1103/PhysRevD.35.1668}}.

\bibitem{Page:1998gz}
P.~R. Page, E.~S. Swanson, A.~P. Szczepaniak, {Hybrid meson decay
  phenomenology}, Phys. Rev. D59 (1999) 034016.
\newblock \href {http://arxiv.org/abs/hep-ph/9808346}
  {\path{arXiv:hep-ph/9808346}}, \href
  {http://dx.doi.org/10.1103/PhysRevD.59.034016}
  {\path{doi:10.1103/PhysRevD.59.034016}}.

\bibitem{Wang:1984gg}
F.~Wang, C.~W. Wong, {MULTI-QUARK STRINGS}, Nuovo Cim. A86 (1985) 283,
  [Erratum: Nuovo Cim.A90,324(1985)].
\newblock \href {http://dx.doi.org/10.1007/BF02812694}
  {\path{doi:10.1007/BF02812694}}.

\bibitem{Ping:2006nc}
J.-l. Ping, C.-r. Deng, F.~Wang, {Quantum chromodynamics quark benzene}, Phys.
  Lett. B659 (2008) 607--611.
\newblock \href {http://arxiv.org/abs/hep-ph/0610390}
  {\path{arXiv:hep-ph/0610390}}, \href
  {http://dx.doi.org/10.1016/j.physletb.2007.11.051}
  {\path{doi:10.1016/j.physletb.2007.11.051}}.

\bibitem{Deng:2014gqa}
C.~Deng, J.~Ping, F.~Wang, {Interpreting $Z_c(3900)$ and $Z_c(4025)/Z_c(4020)$
  as charged tetraquark states}, Phys. Rev. D90~(5) (2014) 054009.
\newblock \href {http://arxiv.org/abs/1402.0777} {\path{arXiv:1402.0777}},
  \href {http://dx.doi.org/10.1103/PhysRevD.90.054009}
  {\path{doi:10.1103/PhysRevD.90.054009}}.

\bibitem{Deng:2015lca}
C.~Deng, J.~Ping, H.~Huang, F.~Wang, {Systematic study of $Z_c^+$ family from a
  multiquark color flux-tube model}, Phys. Rev. D92~(3) (2015) 034027.
\newblock \href {http://arxiv.org/abs/1507.06408} {\path{arXiv:1507.06408}},
  \href {http://dx.doi.org/10.1103/PhysRevD.92.034027}
  {\path{doi:10.1103/PhysRevD.92.034027}}.

\bibitem{Zhou:2015frp}
P.~Zhou, C.-R. Deng, J.-L. Ping, {Identification of $Y(4008)$, $Y(4140)$,
  $Y(4260)$, and $Y(4360)$ as Tetraquark States}, Chin. Phys. Lett. 32~(10)
  (2015) 101201.
\newblock \href {http://dx.doi.org/10.1088/0256-307X/32/10/101201}
  {\path{doi:10.1088/0256-307X/32/10/101201}}.

\bibitem{Deng:2016rus}
C.~Deng, J.~Ping, H.~Huang, F.~Wang, {Heavy pentaquark states and a novel color
  structure}, Phys. Rev. D95~(1) (2017) 014031.
\newblock \href {http://arxiv.org/abs/1608.03940} {\path{arXiv:1608.03940}},
  \href {http://dx.doi.org/10.1103/PhysRevD.95.014031}
  {\path{doi:10.1103/PhysRevD.95.014031}}.

\bibitem{Deng:2018kly}
C.~Deng, H.~Chen, J.~Ping, {Systematical investigation on the stability of
  doubly heavy tetraquark states}\href {http://arxiv.org/abs/1811.06462}
  {\path{arXiv:1811.06462}}.

\bibitem{Bicudo:2015bra}
P.~Bicudo, M.~Cardoso, {Tetraquark bound states and resonances in the unitary
  and microscopic triple string flip-flop quark model, the
  light-light-antiheavy-antiheavy $q q \bar Q\bar Q$ case study}\href
  {http://arxiv.org/abs/1509.04943} {\path{arXiv:1509.04943}}.

\bibitem{Li:2010cy}
M.-T. Li, Y.-B. Dong, Z.-Y. Zhang, {A study of meson-meson potential in the
  chiral quark model}, Chin. Phys. C35 (2011) 622--628.
\newblock \href {http://arxiv.org/abs/1010.2283} {\path{arXiv:1010.2283}},
  \href {http://dx.doi.org/10.1088/1674-1137/35/7/005}
  {\path{doi:10.1088/1674-1137/35/7/005}}.

\bibitem{Anwar:2018bpu}
J.~Ferretti, E.~Santopinto, M.~Naeem~Anwar, M.~A. Bedolla, {The
  baryo-quarkonium picture for hidden-charm and bottom pentaquarks and LHCb
  $P_{\rm c}(4380)$ and $P_{\rm c}(4450)$ states}, Phys. Lett. B789 (2019)
  562--567.
\newblock \href {http://arxiv.org/abs/1807.01207} {\path{arXiv:1807.01207}},
  \href {http://dx.doi.org/10.1016/j.physletb.2018.09.047}
  {\path{doi:10.1016/j.physletb.2018.09.047}}.

\bibitem{Weinberg:1990rz}
S.~Weinberg, {Nuclear forces from chiral Lagrangians}, Phys. Lett. B251 (1990)
  288--292.
\newblock \href {http://dx.doi.org/10.1016/0370-2693(90)90938-3}
  {\path{doi:10.1016/0370-2693(90)90938-3}}.

\bibitem{Weinberg:1991um}
S.~Weinberg, {Effective chiral Lagrangians for nucleon-pion interactions and
  nuclear forces}, Nucl. Phys. B363 (1991) 3--18.
\newblock \href {http://dx.doi.org/10.1016/0550-3213(91)90231-L}
  {\path{doi:10.1016/0550-3213(91)90231-L}}.

\bibitem{Rathaud:2016tys}
D.~P. Rathaud, A.~K. Rai, {Dimesonic states with the heavy-light flavour
  mesons}, Eur. Phys. J. Plus 132~(8) (2017) 370.
\newblock \href {http://arxiv.org/abs/1608.03781} {\path{arXiv:1608.03781}},
  \href {http://dx.doi.org/10.1140/epjp/i2017-11641-3}
  {\path{doi:10.1140/epjp/i2017-11641-3}}.

\bibitem{Liu:2016kqx}
Y.~Liu, I.~Zahed, {Heavy Exotic Molecules with Charm and Bottom}, Phys. Lett.
  B762 (2016) 362--370.
\newblock \href {http://arxiv.org/abs/1608.06535} {\path{arXiv:1608.06535}},
  \href {http://dx.doi.org/10.1016/j.physletb.2016.09.045}
  {\path{doi:10.1016/j.physletb.2016.09.045}}.

\bibitem{Liu:2017mrh}
M.-Z. Liu, D.-J. Jia, D.-Y. Chen, {Possible hadronic molecular states composed
  of $S$-wave heavy-light mesons}, Chin. Phys. C41~(5) (2017) 053105.
\newblock \href {http://arxiv.org/abs/1702.04440} {\path{arXiv:1702.04440}},
  \href {http://dx.doi.org/10.1088/1674-1137/41/5/053105}
  {\path{doi:10.1088/1674-1137/41/5/053105}}.

\bibitem{Chen:2015fdn}
X.~Chen, X.~Lu, R.~Shi, X.~Guo, {Mass of \emph{Y}(4140) in Bethe-Salpeter
  equation for quarks}\href {http://arxiv.org/abs/1512.06483}
  {\path{arXiv:1512.06483}}.

\bibitem{Wang:2017dcq}
Z.-Y. Wang, J.-J. Qi, X.-H. Guo, C.~Wang, {X(3872) as a molecular $D\bar{D}^*$
  state in the Bethe-Salpeter equation approach}, Phys. Rev. D97~(1) (2018)
  016015.
\newblock \href {http://arxiv.org/abs/1710.07424} {\path{arXiv:1710.07424}},
  \href {http://dx.doi.org/10.1103/PhysRevD.97.016015}
  {\path{doi:10.1103/PhysRevD.97.016015}}.

\bibitem{He:2016pfa}
J.~He, {Understanding spin parity of $P_c(4450)$ and $Y(4274)$ in a hadronic
  molecular state picture}, Phys. Rev. D95~(7) (2017) 074004.
\newblock \href {http://arxiv.org/abs/1607.03223} {\path{arXiv:1607.03223}},
  \href {http://dx.doi.org/10.1103/PhysRevD.95.074004}
  {\path{doi:10.1103/PhysRevD.95.074004}}.

\bibitem{He:2017mbh}
J.~He, D.-Y. Chen, {Interpretation of $Y(4390)$ as an isoscalar partner of
  $Z(4430)$ from $D^*(2010)\bar{D}_1(2420)$ interaction}, Eur. Phys. J. C77~(6)
  (2017) 398.
\newblock \href {http://arxiv.org/abs/1704.08776} {\path{arXiv:1704.08776}},
  \href {http://dx.doi.org/10.1140/epjc/s10052-017-4973-8}
  {\path{doi:10.1140/epjc/s10052-017-4973-8}}.

\bibitem{Chen:2017abq}
D.-Y. Chen, C.-J. Xiao, J.~He, {Hidden-charm decays of Y(4390) in a hadronic
  molecular scenario}, Phys. Rev. D96~(5) (2017) 054017.
\newblock \href {http://dx.doi.org/10.1103/PhysRevD.96.054017}
  {\path{doi:10.1103/PhysRevD.96.054017}}.

\bibitem{Sakai:2017avl}
S.~Sakai, L.~Roca, E.~Oset, {Charm-beauty meson bound states from
  $B(B^*)D(D^*)$ and $B(B^*)\bar D(\bar D^*)$ interaction}, Phys. Rev. D96~(5)
  (2017) 054023.
\newblock \href {http://arxiv.org/abs/1704.02196} {\path{arXiv:1704.02196}},
  \href {http://dx.doi.org/10.1103/PhysRevD.96.054023}
  {\path{doi:10.1103/PhysRevD.96.054023}}.

\bibitem{Sun:2017wgf}
B.-X. Sun, D.-M. Wan, S.-Y. Zhao, {The $D\bar{D}^*$ interaction with isospin
  zero in an extended hidden gauge symmetry approach}, Chin. Phys. C42~(5)
  (2018) 053105.
\newblock \href {http://arxiv.org/abs/1709.07263} {\path{arXiv:1709.07263}},
  \href {http://dx.doi.org/10.1088/1674-1137/42/5/053105}
  {\path{doi:10.1088/1674-1137/42/5/053105}}.

\bibitem{Ortega:2018cnm}
P.~G. Ortega, J.~Segovia, D.~R. Entem, F.~Fern{\'a}ndez, {The $Z_c$ structures
  in a coupled-channels model}, Eur. Phys. J. C79~(1) (2019) 78.
\newblock \href {http://arxiv.org/abs/1808.00914} {\path{arXiv:1808.00914}},
  \href {http://dx.doi.org/10.1140/epjc/s10052-019-6552-7}
  {\path{doi:10.1140/epjc/s10052-019-6552-7}}.

\bibitem{Panteleeva:2018ijz}
J.~Y. Panteleeva, I.~A. Perevalova, M.~V. Polyakov, P.~Schweitzer, {On
  tetraquarks with hidden charm and strangeness as phi-psi(2S)
  hadrocharmonium}\href {http://arxiv.org/abs/1802.09029}
  {\path{arXiv:1802.09029}}.

\bibitem{Ferretti:2018kzy}
J.~Ferretti, {$\eta_{\rm c}$- and $J/\psi$-isoscalar meson bound states in the
  hadro-charmonium picture}, Phys. Lett. B782 (2018) 702--706.
\newblock \href {http://arxiv.org/abs/1805.04717} {\path{arXiv:1805.04717}},
  \href {http://dx.doi.org/10.1016/j.physletb.2018.06.032}
  {\path{doi:10.1016/j.physletb.2018.06.032}}.

\bibitem{Molina:2009ct}
R.~Molina, E.~Oset, {The $Y(3940)$, $Z(3930)$ and the $X(4160)$ as dynamically
  generated resonances from the vector-vector interaction}, Phys. Rev. D80
  (2009) 114013.
\newblock \href {http://arxiv.org/abs/0907.3043} {\path{arXiv:0907.3043}},
  \href {http://dx.doi.org/10.1103/PhysRevD.80.114013}
  {\path{doi:10.1103/PhysRevD.80.114013}}.

\bibitem{Voloshin:2018vym}
M.~B. Voloshin, {$Z_c(4100)$ and $Z_c(4200)$ as hadrocharmonium}, Phys. Rev.
  D98~(9) (2018) 094028.
\newblock \href {http://arxiv.org/abs/1810.08146} {\path{arXiv:1810.08146}},
  \href {http://dx.doi.org/10.1103/PhysRevD.98.094028}
  {\path{doi:10.1103/PhysRevD.98.094028}}.

\bibitem{Voloshin:2019ilw}
M.~B. Voloshin, {Strange hadrocharmonium}\href
  {http://arxiv.org/abs/1901.01936} {\path{arXiv:1901.01936}}.

\bibitem{Gao:2018jhk}
R.~Gao, Z.-H. Guo, X.-W. Kang, J.~A. Oller, {Effective-range-expansion study of
  near threshold heavy-flavor resonances}\href
  {http://arxiv.org/abs/1812.07323} {\path{arXiv:1812.07323}}.

\bibitem{Kang:2016ezb}
X.-W. Kang, Z.-H. Guo, J.~A. Oller, {General considerations on the nature of
  $Z_b(10610)$ and $Z_b(10650)$ from their pole positions}, Phys. Rev. D94~(1)
  (2016) 014012.
\newblock \href {http://arxiv.org/abs/1603.05546} {\path{arXiv:1603.05546}},
  \href {http://dx.doi.org/10.1103/PhysRevD.94.014012}
  {\path{doi:10.1103/PhysRevD.94.014012}}.

\bibitem{Dias:2014pva}
J.~M. Dias, F.~Aceti, E.~Oset, {Study of $B\bar{B}^*$ and $B^*\bar{B}^*$
  interactions in $I=1$ and relationship to the $Z_b(10610)$, $Z_b(10650)$
  states}, Phys. Rev. D91~(7) (2015) 076001.
\newblock \href {http://arxiv.org/abs/1410.1785} {\path{arXiv:1410.1785}},
  \href {http://dx.doi.org/10.1103/PhysRevD.91.076001}
  {\path{doi:10.1103/PhysRevD.91.076001}}.

\bibitem{Molina:2010tx}
R.~Molina, T.~Branz, E.~Oset, {A new interpretation for the $D^*_{s2}(2573)$
  and the prediction of novel exotic charmed mesons}, Phys. Rev. D82 (2010)
  014010.
\newblock \href {http://arxiv.org/abs/1005.0335} {\path{arXiv:1005.0335}},
  \href {http://dx.doi.org/10.1103/PhysRevD.82.014010}
  {\path{doi:10.1103/PhysRevD.82.014010}}.

\bibitem{Carames:2011zz}
T.~F. Carames, A.~Valcarce, J.~Vijande, {Doubly charmed exotic mesons: A gift
  of nature?}, Phys. Lett. B699 (2011) 291--295.
\newblock \href {http://dx.doi.org/10.1016/j.physletb.2011.04.023}
  {\path{doi:10.1016/j.physletb.2011.04.023}}.

\bibitem{Valcarce:2005em}
A.~Valcarce, H.~Garcilazo, F.~Fernandez, P.~Gonzalez, {Quark-model study of
  few-baryon systems}, Rept. Prog. Phys. 68 (2005) 965--1042.
\newblock \href {http://arxiv.org/abs/hep-ph/0502173}
  {\path{arXiv:hep-ph/0502173}}, \href
  {http://dx.doi.org/10.1088/0034-4885/68/5/R01}
  {\path{doi:10.1088/0034-4885/68/5/R01}}.

\bibitem{Ohkoda:2012hv}
S.~Ohkoda, Y.~Yamaguchi, S.~Yasui, K.~Sudoh, A.~Hosaka, {Exotic mesons with
  double charm and bottom flavor}, Phys. Rev. D86 (2012) 034019.
\newblock \href {http://arxiv.org/abs/1202.0760} {\path{arXiv:1202.0760}},
  \href {http://dx.doi.org/10.1103/PhysRevD.86.034019}
  {\path{doi:10.1103/PhysRevD.86.034019}}.

\bibitem{Li:2012ss}
N.~Li, Z.-F. Sun, X.~Liu, S.-L. Zhu, {Coupled-channel analysis of the possible
  $D^{(*)}D^{(*)}, \bar{B}^{(*)}\bar{B}^{(*)}$ and $D^{(*)}\bar{B}^{(*)}$
  molecular states}, Phys. Rev. D88~(11) (2013) 114008.
\newblock \href {http://arxiv.org/abs/1211.5007} {\path{arXiv:1211.5007}},
  \href {http://dx.doi.org/10.1103/PhysRevD.88.114008}
  {\path{doi:10.1103/PhysRevD.88.114008}}.

\bibitem{Sun:2012sy}
Z.-F. Sun, X.~Liu, M.~Nielsen, S.-L. Zhu, {Hadronic molecules with both open
  charm and bottom}, Phys. Rev. D85 (2012) 094008.
\newblock \href {http://arxiv.org/abs/1203.1090} {\path{arXiv:1203.1090}},
  \href {http://dx.doi.org/10.1103/PhysRevD.85.094008}
  {\path{doi:10.1103/PhysRevD.85.094008}}.

\bibitem{Bicudo:2012qt}
P.~Bicudo, M.~Wagner, {Lattice QCD signal for a bottom-bottom tetraquark},
  Phys. Rev. D87~(11) (2013) 114511.
\newblock \href {http://arxiv.org/abs/1209.6274} {\path{arXiv:1209.6274}},
  \href {http://dx.doi.org/10.1103/PhysRevD.87.114511}
  {\path{doi:10.1103/PhysRevD.87.114511}}.

\bibitem{Bicudo:2015kna}
P.~Bicudo, K.~Cichy, A.~Peters, M.~Wagner, {BB interactions with static bottom
  quarks from Lattice QCD}, Phys. Rev. D93~(3) (2016) 034501.
\newblock \href {http://arxiv.org/abs/1510.03441} {\path{arXiv:1510.03441}},
  \href {http://dx.doi.org/10.1103/PhysRevD.93.034501}
  {\path{doi:10.1103/PhysRevD.93.034501}}.

\bibitem{Bicudo:2015vta}
P.~Bicudo, K.~Cichy, A.~Peters, B.~Wagenbach, M.~Wagner, {Evidence for the
  existence of $u d \bar{b} \bar{b}$ and the non-existence of $s s \bar{b}
  \bar{b}$ and $c c \bar{b} \bar{b}$ tetraquarks from lattice QCD}, Phys. Rev.
  D92~(1) (2015) 014507.
\newblock \href {http://arxiv.org/abs/1505.00613} {\path{arXiv:1505.00613}},
  \href {http://dx.doi.org/10.1103/PhysRevD.92.014507}
  {\path{doi:10.1103/PhysRevD.92.014507}}.

\bibitem{Bicudo:2017szl}
P.~Bicudo, M.~Cardoso, A.~Peters, M.~Pflaumer, M.~Wagner, {$u d \bar{b}
  \bar{b}$ tetraquark resonances with lattice QCD potentials and the
  Born-Oppenheimer approximation}, Phys. Rev. D96~(5) (2017) 054510.
\newblock \href {http://arxiv.org/abs/1704.02383} {\path{arXiv:1704.02383}},
  \href {http://dx.doi.org/10.1103/PhysRevD.96.054510}
  {\path{doi:10.1103/PhysRevD.96.054510}}.

\bibitem{SanchezSanchez:2017xtl}
M.~Sanchez~Sanchez, L.-S. Geng, J.-X. Lu, T.~Hyodo, M.~P. Valderrama, {Exotic
  doubly charmed ${D}_{s0}^{*}(2317)D$ and ${D}_{s1}^{*}(2460){D}^{*}$
  molecules}, Phys. Rev. D98~(5) (2018) 054001.
\newblock \href {http://arxiv.org/abs/1707.03802} {\path{arXiv:1707.03802}},
  \href {http://dx.doi.org/10.1103/PhysRevD.98.054001}
  {\path{doi:10.1103/PhysRevD.98.054001}}.

\bibitem{Xu:2017tsr}
H.~Xu, B.~Wang, Z.-W. Liu, X.~Liu, {$D D^{*}$ potentials in chiral perturbation
  theory and possible molecular states}, Phys. Rev. D99~(1) (2019) 014027.
\newblock \href {http://arxiv.org/abs/1708.06918} {\path{arXiv:1708.06918}},
  \href {http://dx.doi.org/10.1103/PhysRevD.99.014027}
  {\path{doi:10.1103/PhysRevD.99.014027}}.

\bibitem{Wang:2018atz}
B.~Wang, Z.-W. Liu, X.~Liu, {$\bar{B}^{(\ast)} \bar{B}^{(\ast)}$ interactions
  in chiral effective field theory}, Phys. Rev. D99~(3) (2019) 036007.
\newblock \href {http://arxiv.org/abs/1812.04457} {\path{arXiv:1812.04457}},
  \href {http://dx.doi.org/10.1103/PhysRevD.99.036007}
  {\path{doi:10.1103/PhysRevD.99.036007}}.

\bibitem{Shimizu:2016rrd}
Y.~Shimizu, D.~Suenaga, M.~Harada, {Coupled channel analysis of molecule
  picture of $P_{c}(4380)$}, Phys. Rev. D93~(11) (2016) 114003.
\newblock \href {http://arxiv.org/abs/1603.02376} {\path{arXiv:1603.02376}},
  \href {http://dx.doi.org/10.1103/PhysRevD.93.114003}
  {\path{doi:10.1103/PhysRevD.93.114003}}.

\bibitem{Shimizu:2017xrg}
Y.~Shimizu, M.~Harada, {Hidden Charm Pentaquark $P_c(4380)$ and Doubly Charmed
  Baryon $\Xi_{cc}^*(4380)$ as Hadronic Molecule States}, Phys. Rev. D96~(9)
  (2017) 094012.
\newblock \href {http://arxiv.org/abs/1708.04743} {\path{arXiv:1708.04743}},
  \href {http://dx.doi.org/10.1103/PhysRevD.96.094012}
  {\path{doi:10.1103/PhysRevD.96.094012}}.

\bibitem{Yamaguchi:2016ote}
Y.~Yamaguchi, E.~Santopinto, {Hidden-charm pentaquarks as a meson-baryon
  molecule with coupled channels for $\bar{D}^{(\ast)}\Lambda_{\rm c}$ and
  $\bar{D}^{(\ast)}\Sigma^{(\ast)}_{\rm c}$}, Phys. Rev. D96~(1) (2017) 014018.
\newblock \href {http://arxiv.org/abs/1606.08330} {\path{arXiv:1606.08330}},
  \href {http://dx.doi.org/10.1103/PhysRevD.96.014018}
  {\path{doi:10.1103/PhysRevD.96.014018}}.

\bibitem{Yamaguchi:2017zmn}
Y.~Yamaguchi, A.~Giachino, A.~Hosaka, E.~Santopinto, S.~Takeuchi, M.~Takizawa,
  {Hidden-charm and bottom meson-baryon molecules coupled with five-quark
  states}, Phys. Rev. D96~(11) (2017) 114031.
\newblock \href {http://arxiv.org/abs/1709.00819} {\path{arXiv:1709.00819}},
  \href {http://dx.doi.org/10.1103/PhysRevD.96.114031}
  {\path{doi:10.1103/PhysRevD.96.114031}}.

\bibitem{He:2017aps}
J.~He, {Nucleon resonances $N(1875)$ and $N(2100)$ as strange partners of LHCb
  pentaquarks}, Phys. Rev. D95~(7) (2017) 074031.
\newblock \href {http://arxiv.org/abs/1701.03738} {\path{arXiv:1701.03738}},
  \href {http://dx.doi.org/10.1103/PhysRevD.95.074031}
  {\path{doi:10.1103/PhysRevD.95.074031}}.

\bibitem{Chen:2016ryt}
R.~Chen, J.~He, X.~Liu, {Possible strange hidden-charm pentaquarks from
  $\Sigma_c^{(*)}\bar{D}_s^*$ and $\Xi^{(',*)}_c\bar{D}^*$ interactions}, Chin.
  Phys. C41~(10) (2017) 103105.
\newblock \href {http://arxiv.org/abs/1609.03235} {\path{arXiv:1609.03235}},
  \href {http://dx.doi.org/10.1088/1674-1137/41/10/103105}
  {\path{doi:10.1088/1674-1137/41/10/103105}}.

\bibitem{Geng:2017hxc}
L.~Geng, J.~Lu, M.~P. Valderrama, {Scale Invariance in Heavy Hadron Molecules},
  Phys. Rev. D97~(9) (2018) 094036.
\newblock \href {http://arxiv.org/abs/1704.06123} {\path{arXiv:1704.06123}},
  \href {http://dx.doi.org/10.1103/PhysRevD.97.094036}
  {\path{doi:10.1103/PhysRevD.97.094036}}.

\bibitem{Chen:2017vai}
R.~Chen, A.~Hosaka, X.~Liu, {Heavy molecules and one-$\sigma/\omega$-exchange
  model}, Phys. Rev. D96~(11) (2017) 116012.
\newblock \href {http://arxiv.org/abs/1707.08306} {\path{arXiv:1707.08306}},
  \href {http://dx.doi.org/10.1103/PhysRevD.96.116012}
  {\path{doi:10.1103/PhysRevD.96.116012}}.

\bibitem{Rathaud:2018fna}
D.~P. Rathaud, A.~K. Rai, {Interaction and Identification of the Meson-Baryon
  molecules}\href {http://arxiv.org/abs/1808.05815} {\path{arXiv:1808.05815}}.

\bibitem{Shen:2017ayv}
C.-W. Shen, D.~R{\"o}nchen, U.-G. Meissner, B.-S. Zou, {Exploratory study of
  possible resonances in heavy meson - heavy baryon coupled-channel
  interactions}, Chin. Phys. C42~(2) (2018) 023106.
\newblock \href {http://arxiv.org/abs/1710.03885} {\path{arXiv:1710.03885}},
  \href {http://dx.doi.org/10.1088/1674-1137/42/2/023106}
  {\path{doi:10.1088/1674-1137/42/2/023106}}.

\bibitem{Huang:2018wed}
H.~Huang, J.~Ping, {Investigating the hidden-charm and hidden-bottom pentaquark
  resonances in scattering process}, Phys. Rev. D99~(1) (2019) 014010.
\newblock \href {http://arxiv.org/abs/1811.04260} {\path{arXiv:1811.04260}},
  \href {http://dx.doi.org/10.1103/PhysRevD.99.014010}
  {\path{doi:10.1103/PhysRevD.99.014010}}.

\bibitem{Eides:2015dtr}
M.~I. Eides, V.~{\relax Yu}. Petrov, M.~V. Polyakov, {Narrow Nucleon-$\psi(2S)$
  Bound State and LHCb Pentaquarks}, Phys. Rev. D93~(5) (2016) 054039.
\newblock \href {http://arxiv.org/abs/1512.00426} {\path{arXiv:1512.00426}},
  \href {http://dx.doi.org/10.1103/PhysRevD.93.054039}
  {\path{doi:10.1103/PhysRevD.93.054039}}.

\bibitem{Perevalova:2016dln}
I.~A. Perevalova, M.~V. Polyakov, P.~Schweitzer, {On LHCb pentaquarks as a
  baryon-$\psi$(2S) bound state: prediction of isospin-$\frac3{2}$ pentaquarks
  with hidden charm}, Phys. Rev. D94~(5) (2016) 054024.
\newblock \href {http://arxiv.org/abs/1607.07008} {\path{arXiv:1607.07008}},
  \href {http://dx.doi.org/10.1103/PhysRevD.94.054024}
  {\path{doi:10.1103/PhysRevD.94.054024}}.

\bibitem{Eides:2017xnt}
M.~I. Eides, V.~{\relax Yu}. Petrov, M.~V. Polyakov, {Pentaquarks with hidden
  charm as hadroquarkonia}, Eur. Phys. J. C78~(1) (2018) 36.
\newblock \href {http://arxiv.org/abs/1709.09523} {\path{arXiv:1709.09523}},
  \href {http://dx.doi.org/10.1140/epjc/s10052-018-5530-9}
  {\path{doi:10.1140/epjc/s10052-018-5530-9}}.

\bibitem{Aaij:2018jlf}
R.~Aaij, et~al., {Observation of the decay
  $\Lambda^0_b\rightarrow\psi(2S)p\pi^-$}, JHEP 08 (2018) 131.
\newblock \href {http://arxiv.org/abs/1806.08084} {\path{arXiv:1806.08084}},
  \href {http://dx.doi.org/10.1007/JHEP08(2018)131}
  {\path{doi:10.1007/JHEP08(2018)131}}.

\bibitem{Li:2015gta}
G.-N. Li, X.-G. He, M.~He, {Some Predictions of Diquark Model for Hidden Charm
  Pentaquark Discovered at the LHCb}, JHEP 12 (2015) 128.
\newblock \href {http://arxiv.org/abs/1507.08252} {\path{arXiv:1507.08252}},
  \href {http://dx.doi.org/10.1007/JHEP12(2015)128}
  {\path{doi:10.1007/JHEP12(2015)128}}.

\bibitem{Ghosh:2015ksa}
R.~Ghosh, A.~Bhattacharya, B.~Chakrabarti, {A study on P$_{c}^{*}$ (4380) and
  P$_{c}^{*}$ in the quasi particle diquark model}[Phys. Part. Nucl.
  Lett.14,no.4,550(2017)].
\newblock \href {http://arxiv.org/abs/1508.00356} {\path{arXiv:1508.00356}},
  \href {http://dx.doi.org/10.1134/S1547477117040100}
  {\path{doi:10.1134/S1547477117040100}}.

\bibitem{Chen:2019bip}
H.-X. Chen, W.~Chen, S.-L. Zhu, {Possible interpretations of the $P_c(4312)$,
  $P_c(4440)$, and $P_c(4457)$}\href {http://arxiv.org/abs/1903.11001}
  {\path{arXiv:1903.11001}}.

\bibitem{Chen:2019asm}
R.~Chen, X.~Liu, Z.-F. Sun, S.-L. Zhu, {Strong LHCb evidence for supporting the
  existence of hidden-charm molecular pentaquarks}\href
  {http://arxiv.org/abs/1903.11013} {\path{arXiv:1903.11013}}.

\bibitem{Liu:2019Pc}
M.-Z. Liu, Y.-W. Pan, F.-Z. Peng, M.~S. Sanchez, L.-S. Geng, A.~Hosaka, M.~P.
  Valderrama, {Emergence of a complete heavy-quark spin symmetry multiplet:
  seven molecular pentaquarks in light of the latest LHCb analysis}\href
  {http://arxiv.org/abs/1903.11560} {\path{arXiv:1903.11560}}.

\bibitem{Guo:2019Pc}
F.-K. Guo, H.-J. Jing, U.-G. Meisser, S.~Sakai, {Isospin breaking decays as a
  diagnosis of the hadronic molecular structure of the $P_c(4457)$}\href
  {http://arxiv.org/abs/1903.11503} {\path{arXiv:1903.11503}}.

\bibitem{Hofmann:2006qx}
J.~Hofmann, M.~F.~M. Lutz, {D-wave baryon resonances with charm from
  coupled-channel dynamics}, Nucl. Phys. A776 (2006) 17--51.
\newblock \href {http://arxiv.org/abs/hep-ph/0601249}
  {\path{arXiv:hep-ph/0601249}}, \href
  {http://dx.doi.org/10.1016/j.nuclphysa.2006.07.004}
  {\path{doi:10.1016/j.nuclphysa.2006.07.004}}.

\bibitem{Guo:2011dd}
F.-K. Guo, U.-G. Meissner, {More kaonic bound states and a comprehensive
  interpretation of the $D_{sJ}$ states}, Phys. Rev. D84 (2011) 014013.
\newblock \href {http://arxiv.org/abs/1102.3536} {\path{arXiv:1102.3536}},
  \href {http://dx.doi.org/10.1103/PhysRevD.84.014013}
  {\path{doi:10.1103/PhysRevD.84.014013}}.

\bibitem{Romanets:2012hm}
O.~Romanets, L.~Tolos, C.~Garcia-Recio, J.~Nieves, L.~L. Salcedo, R.~G.~E.
  Timmermans, {Charmed and strange baryon resonances with heavy-quark spin
  symmetry}, Phys. Rev. D85 (2012) 114032.
\newblock \href {http://arxiv.org/abs/1202.2239} {\path{arXiv:1202.2239}},
  \href {http://dx.doi.org/10.1103/PhysRevD.85.114032}
  {\path{doi:10.1103/PhysRevD.85.114032}}.

\bibitem{Dias:2018qhp}
J.~M. Dias, V.~R. Debastiani, J.~J. Xie, E.~Oset, {Doubly charmed $\Xi_{cc}$
  molecular states from meson-baryon interaction}, Phys. Rev. D98~(9) (2018)
  094017.
\newblock \href {http://arxiv.org/abs/1805.03286} {\path{arXiv:1805.03286}},
  \href {http://dx.doi.org/10.1103/PhysRevD.98.094017}
  {\path{doi:10.1103/PhysRevD.98.094017}}.

\bibitem{Xu:2010fc}
Q.~Xu, G.~Liu, H.~Jin, {Possible bound state of the double heavy meson-baryon
  system}, Phys. Rev. D86 (2012) 114032.
\newblock \href {http://arxiv.org/abs/1012.5949} {\path{arXiv:1012.5949}},
  \href {http://dx.doi.org/10.1103/PhysRevD.86.114032}
  {\path{doi:10.1103/PhysRevD.86.114032}}.

\bibitem{Guo:2017vcf}
Z.-H. Guo, {Prediction of exotic doubly charmed baryons within chiral effective
  field theory}, Phys. Rev. D96~(7) (2017) 074004.
\newblock \href {http://arxiv.org/abs/1708.04145} {\path{arXiv:1708.04145}},
  \href {http://dx.doi.org/10.1103/PhysRevD.96.074004}
  {\path{doi:10.1103/PhysRevD.96.074004}}.

\bibitem{Chen:2018pzd}
R.~Chen, F.-L. Wang, A.~Hosaka, X.~Liu, {Exotic triple-charm deuteronlike
  hexaquarks}, Phys. Rev. D97~(11) (2018) 114011.
\newblock \href {http://arxiv.org/abs/1804.02961} {\path{arXiv:1804.02961}},
  \href {http://dx.doi.org/10.1103/PhysRevD.97.114011}
  {\path{doi:10.1103/PhysRevD.97.114011}}.

\bibitem{Chen:2017jjn}
R.~Chen, A.~Hosaka, X.~Liu, {Prediction of triple-charm molecular pentaquarks},
  Phys. Rev. D96~(11) (2017) 114030.
\newblock \href {http://arxiv.org/abs/1711.09579} {\path{arXiv:1711.09579}},
  \href {http://dx.doi.org/10.1103/PhysRevD.96.114030}
  {\path{doi:10.1103/PhysRevD.96.114030}}.

\bibitem{Wang:2019aoc}
F.-L. Wang, R.~Chen, Z.-W. Liu, X.~Liu, {Possible triple-charm molecular
  pentaquarks from $\Xi_{cc}D_1/\Xi_{cc}D_2^*$ interactions}, Phys. Rev.
  D99~(5) (2019) 054021.
\newblock \href {http://arxiv.org/abs/1901.01542} {\path{arXiv:1901.01542}},
  \href {http://dx.doi.org/10.1103/PhysRevD.99.054021}
  {\path{doi:10.1103/PhysRevD.99.054021}}.

\bibitem{Manohar:2000dt}
A.~V. Manohar, M.~B. Wise, {Heavy quark physics}, Camb. Monogr. Part. Phys.
  Nucl. Phys. Cosmol. 10 (2000) 1--191.

\bibitem{Hosaka:2016ypm}
A.~Hosaka, T.~Hyodo, K.~Sudoh, Y.~Yamaguchi, S.~Yasui, {Heavy Hadrons in
  Nuclear Matter}, Prog. Part. Nucl. Phys. 96 (2017) 88--153.
\newblock \href {http://arxiv.org/abs/1606.08685} {\path{arXiv:1606.08685}},
  \href {http://dx.doi.org/10.1016/j.ppnp.2017.04.003}
  {\path{doi:10.1016/j.ppnp.2017.04.003}}.

\bibitem{Savage:1990di}
M.~J. Savage, M.~B. Wise, {Spectrum of baryons with two heavy quarks}, Phys.
  Lett. B248 (1990) 177--180.
\newblock \href {http://dx.doi.org/10.1016/0370-2693(90)90035-5}
  {\path{doi:10.1016/0370-2693(90)90035-5}}.

\bibitem{Brambilla:2005yk}
N.~Brambilla, A.~Vairo, T.~Rosch, {Effective field theory Lagrangians for
  baryons with two and three heavy quarks}, Phys. Rev. D72 (2005) 034021.
\newblock \href {http://arxiv.org/abs/hep-ph/0506065}
  {\path{arXiv:hep-ph/0506065}}, \href
  {http://dx.doi.org/10.1103/PhysRevD.72.034021}
  {\path{doi:10.1103/PhysRevD.72.034021}}.

\bibitem{Fleming:2005pd}
S.~Fleming, T.~Mehen, {Doubly heavy baryons, heavy quark-diquark symmetry and
  NRQCD}, Phys. Rev. D73 (2006) 034502.
\newblock \href {http://arxiv.org/abs/hep-ph/0509313}
  {\path{arXiv:hep-ph/0509313}}, \href
  {http://dx.doi.org/10.1103/PhysRevD.73.034502}
  {\path{doi:10.1103/PhysRevD.73.034502}}.

\bibitem{Cohen:2006jg}
T.~D. Cohen, P.~M. Hohler, {Doubly heavy hadrons and the domain of validity of
  doubly heavy diquark-anti-quark symmetry}, Phys. Rev. D74 (2006) 094003.
\newblock \href {http://arxiv.org/abs/hep-ph/0606084}
  {\path{arXiv:hep-ph/0606084}}, \href
  {http://dx.doi.org/10.1103/PhysRevD.74.094003}
  {\path{doi:10.1103/PhysRevD.74.094003}}.

\bibitem{Cai:2019orb}
Y.~Cai, T.~Cohen, {On the existence of near threshold exotic hadrons containing
  two heavy quarks}\href {http://arxiv.org/abs/1901.05473}
  {\path{arXiv:1901.05473}}.

\bibitem{Cleven:2015era}
M.~Cleven, F.-K. Guo, C.~Hanhart, Q.~Wang, Q.~Zhao, {Employing spin symmetry to
  disentangle different models for the XYZ states}, Phys. Rev. D92~(1) (2015)
  014005.
\newblock \href {http://arxiv.org/abs/1505.01771} {\path{arXiv:1505.01771}},
  \href {http://dx.doi.org/10.1103/PhysRevD.92.014005}
  {\path{doi:10.1103/PhysRevD.92.014005}}.

\bibitem{Baru:2016iwj}
V.~Baru, E.~Epelbaum, A.~A. Filin, C.~Hanhart, U.-G. Meissner, A.~V. Nefediev,
  {Heavy-quark spin symmetry partners of the X (3872) revisited}, Phys. Lett.
  B763 (2016) 20--28.
\newblock \href {http://arxiv.org/abs/1605.09649} {\path{arXiv:1605.09649}},
  \href {http://dx.doi.org/10.1016/j.physletb.2016.10.008}
  {\path{doi:10.1016/j.physletb.2016.10.008}}.

\bibitem{Baru:2017gwo}
V.~Baru, E.~Epelbaum, A.~A. Filin, C.~Hanhart, A.~V. Nefediev, {Spin partners
  of the Z$_{b}$ (10610) and Z$_{b}$ (10650) revisited}, JHEP 06 (2017) 158.
\newblock \href {http://arxiv.org/abs/1704.07332} {\path{arXiv:1704.07332}},
  \href {http://dx.doi.org/10.1007/JHEP06(2017)158}
  {\path{doi:10.1007/JHEP06(2017)158}}.

\bibitem{Liu:2019stu}
M.-Z. Liu, T.-W. Wu, M.~P. Valderrama, J.-J. Xie, L.-S. Geng, {Heavy-quark spin
  and flavour symmetry partners of the X(3872) revisited: what can we learn
  from the one boson exchange model?}\href {http://arxiv.org/abs/1902.03044}
  {\path{arXiv:1902.03044}}.

\bibitem{AlFiky:2005jd}
M.~T. AlFiky, F.~Gabbiani, A.~A. Petrov, {$X(3872)$: Hadronic molecules in
  effective field theory}, Phys. Lett. B640 (2006) 238--245.
\newblock \href {http://arxiv.org/abs/hep-ph/0506141}
  {\path{arXiv:hep-ph/0506141}}, \href
  {http://dx.doi.org/10.1016/j.physletb.2006.07.069}
  {\path{doi:10.1016/j.physletb.2006.07.069}}.

\bibitem{Chiladze:1998ti}
G.~Chiladze, A.~F. Falk, A.~A. Petrov, {Hybrid charmonium production in B
  decays}, Phys. Rev. D58 (1998) 034013.
\newblock \href {http://arxiv.org/abs/hep-ph/9804248}
  {\path{arXiv:hep-ph/9804248}}, \href
  {http://dx.doi.org/10.1103/PhysRevD.58.034013}
  {\path{doi:10.1103/PhysRevD.58.034013}}.

\bibitem{Petrov:2005tp}
A.~A. Petrov, {X(3872), X(3940) as hybrid charmonium states?}, J. Phys. Conf.
  Ser. 9 (2005) 83--86.
\newblock \href {http://dx.doi.org/10.1088/1742-6596/9/1/013}
  {\path{doi:10.1088/1742-6596/9/1/013}}.

\bibitem{Voloshin:2016cgm}
M.~B. Voloshin, {Light Quark Spin Symmetry in $Z_b$ Resonances?}, Phys. Rev.
  D93~(7) (2016) 074011.
\newblock \href {http://arxiv.org/abs/1601.02540} {\path{arXiv:1601.02540}},
  \href {http://dx.doi.org/10.1103/PhysRevD.93.074011}
  {\path{doi:10.1103/PhysRevD.93.074011}}.

\bibitem{Hu:2005gf}
J.~Hu, T.~Mehen, {Chiral Lagrangian with heavy quark-diquark symmetry}, Phys.
  Rev. D73 (2006) 054003.
\newblock \href {http://arxiv.org/abs/hep-ph/0511321}
  {\path{arXiv:hep-ph/0511321}}, \href
  {http://dx.doi.org/10.1103/PhysRevD.73.054003}
  {\path{doi:10.1103/PhysRevD.73.054003}}.

\bibitem{Shimizu:2018ran}
Y.~Shimizu, Y.~Yamaguchi, M.~Harada, {Heavy quark spin multiplet structure of
  $\bar{P}^{(*)}\Sigma_Q^{(*)}$ molecular states}, Phys. Rev. D98~(1) (2018)
  014021.
\newblock \href {http://arxiv.org/abs/1805.05740} {\path{arXiv:1805.05740}},
  \href {http://dx.doi.org/10.1103/PhysRevD.98.014021}
  {\path{doi:10.1103/PhysRevD.98.014021}}.

\bibitem{Shimizu:2019jfy}
Y.~Shimizu, Y.~Yamaguchi, M.~Harada, {Heavy quark spin multiplet structure of
  $P_c$-like pentaquark as P-wave hadronic molecular state}\href
  {http://arxiv.org/abs/1901.09215} {\path{arXiv:1901.09215}}.

\bibitem{Liu:2018zzu}
M.-Z. Liu, F.-Z. Peng, M.~S¨¢nchez~S¨¢nchez, M.~P. Valderrama, {Heavy-quark
  symmetry partners of the $P_c(4450)$ pentaquark}, Phys. Rev. D98~(11) (2018)
  114030.
\newblock \href {http://arxiv.org/abs/1811.03992} {\path{arXiv:1811.03992}},
  \href {http://dx.doi.org/10.1103/PhysRevD.98.114030}
  {\path{doi:10.1103/PhysRevD.98.114030}}.

\bibitem{Shifman:1978bx}
M.~A. Shifman, A.~I. Vainshtein, V.~I. Zakharov, {QCD and Resonance Physics.
  Theoretical Foundations}, Nucl. Phys. B147 (1979) 385--447.
\newblock \href {http://dx.doi.org/10.1016/0550-3213(79)90022-1}
  {\path{doi:10.1016/0550-3213(79)90022-1}}.

\bibitem{Reinders:1984sr}
L.~J. Reinders, H.~Rubinstein, S.~Yazaki, {Hadron Properties from QCD Sum
  Rules}, Phys. Rept. 127 (1985) 1.
\newblock \href {http://dx.doi.org/10.1016/0370-1573(85)90065-1}
  {\path{doi:10.1016/0370-1573(85)90065-1}}.

\bibitem{Chen:2016spr}
H.-X. Chen, W.~Chen, X.~Liu, Y.-R. Liu, S.-L. Zhu, {A review of the open charm
  and open bottom systems}, Rept. Prog. Phys. 80~(7) (2017) 076201.
\newblock \href {http://arxiv.org/abs/1609.08928} {\path{arXiv:1609.08928}},
  \href {http://dx.doi.org/10.1088/1361-6633/aa6420}
  {\path{doi:10.1088/1361-6633/aa6420}}.

\bibitem{Colangelo:2000dp}
P.~Colangelo, A.~Khodjamirian, {QCD sum rules, a modern perspective,~}\href
  {http://arxiv.org/abs/hep-ph/0010175} {\path{arXiv:hep-ph/0010175}}.

\bibitem{Narison:2002pw}
S.~Narison, {QCD as a theory of hadrons from partons to confinement,~}\href
  {http://arxiv.org/abs/hep-ph/0205006} {\path{arXiv:hep-ph/0205006}}.

\bibitem{Nielsen:2009uh}
M.~Nielsen, F.~S. Navarra, S.~H. Lee, {New Charmonium States in QCD Sum Rules:
  A Concise Review}, Phys. Rept. 497 (2010) 41--83.
\newblock \href {http://arxiv.org/abs/0911.1958} {\path{arXiv:0911.1958}},
  \href {http://dx.doi.org/10.1016/j.physrep.2010.07.005}
  {\path{doi:10.1016/j.physrep.2010.07.005}}.

\bibitem{Albuquerque:2018jkn}
R.~M. Albuquerque, J.~M. Dias, K.~P. Khemchandani, A.~Martinez~Torres, F.~S.
  Navarra, M.~Nielsen, C.~M. Zanetti, {QCD Sum Rules Approach to the $X,~Y$ and
  $Z$ States}\href {http://arxiv.org/abs/1812.08207} {\path{arXiv:1812.08207}}.

\bibitem{Chen:2018kuu}
H.-X. Chen, C.-P. Shen, S.-L. Zhu, {A possible partner state of the $Y(2175)$},
  Phys. Rev. D98~(1) (2018) 014011.
\newblock \href {http://arxiv.org/abs/1805.06100} {\path{arXiv:1805.06100}},
  \href {http://dx.doi.org/10.1103/PhysRevD.98.014011}
  {\path{doi:10.1103/PhysRevD.98.014011}}.

\bibitem{Cui:2019roq}
E.-L. Cui, H.-M. Yang, H.-X. Chen, W.~Chen, C.-P. Shen, {QCD sum rule studies
  of $s s {\bar{s}} {\bar{s}}$ tetraquark states with $J^{PC} = 1^{+-}$}, Eur.
  Phys. J. C79~(3) (2019) 232.
\newblock \href {http://arxiv.org/abs/1901.01724} {\path{arXiv:1901.01724}},
  \href {http://dx.doi.org/10.1140/epjc/s10052-019-6755-y}
  {\path{doi:10.1140/epjc/s10052-019-6755-y}}.

\bibitem{Chen:2006hy}
H.-X. Chen, A.~Hosaka, S.-L. Zhu, {Exotic Tetraquark $ud \bar s \bar s$ of $J^P
  = 0^+$ in the QCD Sum Rule}, Phys. Rev. D74 (2006) 054001.
\newblock \href {http://arxiv.org/abs/hep-ph/0604049}
  {\path{arXiv:hep-ph/0604049}}, \href
  {http://dx.doi.org/10.1103/PhysRevD.74.054001}
  {\path{doi:10.1103/PhysRevD.74.054001}}.

\bibitem{Fierz}
M.~Fierz, Z. Physik 104 (1937) 553.

\bibitem{Zong:1994ww}
H.-S. Zong, F.~Wang, J.-L. Ping, {The Generalized Fierz transformation and its
  application}, Commun. Theor. Phys. 22 (1994) 479--482.

\bibitem{Maruhn:2000af}
J.~A. Maruhn, T.~Buervenich, D.~G. Madland, {Calculating the Fierz
  transformation for higher orders}, Chin. J. Comput. Phys. 169 (2001) 238.
\newblock \href {http://arxiv.org/abs/nucl-th/0007010}
  {\path{arXiv:nucl-th/0007010}}, \href
  {http://dx.doi.org/10.1006/jcph.2001.6727}
  {\path{doi:10.1006/jcph.2001.6727}}.

\bibitem{Chen:2006zh}
H.-X. Chen, A.~Hosaka, S.-L. Zhu, {QCD sum rule study of the masses of light
  tetraquark scalar mesons}, Phys. Lett. B650 (2007) 369--372.
\newblock \href {http://arxiv.org/abs/hep-ph/0609163}
  {\path{arXiv:hep-ph/0609163}}, \href
  {http://dx.doi.org/10.1016/j.physletb.2007.05.031}
  {\path{doi:10.1016/j.physletb.2007.05.031}}.

\bibitem{Chen:2007xr}
H.-X. Chen, A.~Hosaka, S.-L. Zhu, {Light Scalar Tetraquark Mesons in the QCD
  Sum Rule}, Phys. Rev. D76 (2007) 094025.
\newblock \href {http://arxiv.org/abs/0707.4586} {\path{arXiv:0707.4586}},
  \href {http://dx.doi.org/10.1103/PhysRevD.76.094025}
  {\path{doi:10.1103/PhysRevD.76.094025}}.

\bibitem{Chen:2007mp}
H.-X. Chen, A.~Hosaka, S.-L. Zhu, {Light Scalar Mesons in the QCD Sum Rule},
  Prog. Theor. Phys. Suppl. 168 (2007) 186--189.
\newblock \href {http://arxiv.org/abs/0711.1007} {\path{arXiv:0711.1007}},
  \href {http://dx.doi.org/10.1143/PTPS.168.186}
  {\path{doi:10.1143/PTPS.168.186}}.

\bibitem{Chen:2008qw}
H.-X. Chen, A.~Hosaka, S.-L. Zhu, {The $I^G J^{PC} = 1^- 1^{-+}$ Tetraquark
  States}, Phys. Rev. D78 (2008) 054017.
\newblock \href {http://arxiv.org/abs/0806.1998} {\path{arXiv:0806.1998}},
  \href {http://dx.doi.org/10.1103/PhysRevD.78.054017}
  {\path{doi:10.1103/PhysRevD.78.054017}}.

\bibitem{Chen:2008iw}
H.-X. Chen, A.~Hosaka, S.-L. Zhu, {Scalar Tetraquark Currents With Application
  to the QCD Sum Rule}, Mod. Phys. Lett. A23 (2008) 2234--2237.
\newblock \href {http://arxiv.org/abs/0811.1514} {\path{arXiv:0811.1514}},
  \href {http://dx.doi.org/10.1142/S0217732308029095}
  {\path{doi:10.1142/S0217732308029095}}.

\bibitem{Chen:2008ne}
H.-X. Chen, A.~Hosaka, S.-L. Zhu, {The $I^G J^{PC} = 0^+ 1^{-+}$ Tetraquark
  State}, Phys. Rev. D78 (2008) 117502.
\newblock \href {http://arxiv.org/abs/0808.2344} {\path{arXiv:0808.2344}},
  \href {http://dx.doi.org/10.1103/PhysRevD.78.117502}
  {\path{doi:10.1103/PhysRevD.78.117502}}.

\bibitem{Jiao:2009ra}
C.-K. Jiao, W.~Chen, H.-X. Chen, S.-L. Zhu, {The Possible $J^{PC} = 0^{--}$
  Exotic State}, Phys. Rev. D79 (2009) 114034.
\newblock \href {http://arxiv.org/abs/0905.0774} {\path{arXiv:0905.0774}},
  \href {http://dx.doi.org/10.1103/PhysRevD.79.114034}
  {\path{doi:10.1103/PhysRevD.79.114034}}.

\bibitem{Chen:2009gs}
H.-X. Chen, A.~Hosaka, H.~Toki, S.-L. Zhu, {Light Scalar Meson sigma(600) in
  QCD Sum Rule with Continuum}, Phys. Rev. D81 (2010) 114034.
\newblock \href {http://arxiv.org/abs/0912.5138} {\path{arXiv:0912.5138}},
  \href {http://dx.doi.org/10.1103/PhysRevD.81.114034}
  {\path{doi:10.1103/PhysRevD.81.114034}}.

\bibitem{Chen:2010jd}
W.~Chen, S.-L. Zhu, {The Possible $J^{PC} = 0^{--}$ Charmonium-like State},
  Phys. Rev. D81 (2010) 105018.
\newblock \href {http://arxiv.org/abs/1003.3721} {\path{arXiv:1003.3721}},
  \href {http://dx.doi.org/10.1103/PhysRevD.81.105018}
  {\path{doi:10.1103/PhysRevD.81.105018}}.

\bibitem{Chen:2011qu}
W.~Chen, Z.-X. Cai, S.-L. Zhu, {Masses of the tensor mesons with
  $J^{P}=2^{-}$}, Nucl. Phys. B887 (2014) 201--215.
\newblock \href {http://arxiv.org/abs/1107.4949} {\path{arXiv:1107.4949}},
  \href {http://dx.doi.org/10.1016/j.nuclphysb.2014.08.006}
  {\path{doi:10.1016/j.nuclphysb.2014.08.006}}.

\bibitem{Du:2012pn}
M.-L. Du, W.~Chen, X.-L. Chen, S.-L. Zhu, {The Possible $J^{PC} = 0^{+-}$
  Exotic State}, Chin.Phys. C37 (2013) 033104.
\newblock \href {http://arxiv.org/abs/1203.5199} {\path{arXiv:1203.5199}},
  \href {http://dx.doi.org/10.1088/1674-1137/37/3/033104}
  {\path{doi:10.1088/1674-1137/37/3/033104}}.

\bibitem{Chen:2012ut}
H.-X. Chen, {The ``Closed'' Chiral Symmetry and Its Application to Tetraquark},
  Eur. Phys. J. C72 (2012) 2204.
\newblock \href {http://arxiv.org/abs/1210.3399} {\path{arXiv:1210.3399}},
  \href {http://dx.doi.org/10.1140/epjc/s10052-012-2204-x}
  {\path{doi:10.1140/epjc/s10052-012-2204-x}}.

\bibitem{Chen:2013jra}
H.-X. Chen, {Chiral Structure of Vector and Axial-Vector Tetraquark Currents},
  Eur. Phys. J. C73 (2013) 2628.
\newblock \href {http://arxiv.org/abs/1311.4992} {\path{arXiv:1311.4992}},
  \href {http://dx.doi.org/10.1140/epjc/s10052-013-2628-y}
  {\path{doi:10.1140/epjc/s10052-013-2628-y}}.

\bibitem{Chen:2013gnu}
H.-X. Chen, {Chiral Structure of Scalar and Pseudoscalar Mesons}, Adv. High
  Energy Phys. 2013 (2013) 750591.
\newblock \href {http://arxiv.org/abs/1311.4434} {\path{arXiv:1311.4434}},
  \href {http://dx.doi.org/10.1155/2013/750591}
  {\path{doi:10.1155/2013/750591}}.

\bibitem{Chen:2014vha}
H.-X. Chen, E.-L. Cui, W.~Chen, T.~G. Steele, S.-L. Zhu, {QCD sum rule study of
  the $d^*(2380)$}, Phys. Rev. C91~(2) (2015) 025204.
\newblock \href {http://arxiv.org/abs/1410.0394} {\path{arXiv:1410.0394}},
  \href {http://dx.doi.org/10.1103/PhysRevC.91.025204}
  {\path{doi:10.1103/PhysRevC.91.025204}}.

\bibitem{Chen:2015fwa}
H.-X. Chen, E.-L. Cui, W.~Chen, T.~G. Steele, X.~Liu, S.-L. Zhu, {$a_1(1420)$
  resonance as a tetraquark state and its isospin partner}, Phys. Rev. D91
  (2015) 094022.
\newblock \href {http://arxiv.org/abs/1503.02597} {\path{arXiv:1503.02597}},
  \href {http://dx.doi.org/10.1103/PhysRevD.91.094022}
  {\path{doi:10.1103/PhysRevD.91.094022}}.

\bibitem{Chen:2016ymy}
H.-X. Chen, D.~Zhou, W.~Chen, X.~Liu, S.-L. Zhu, {Searching for hidden-charm
  baryonium signals in QCD sum rules}, Eur. Phys. J. C76~(11) (2016) 602.
\newblock \href {http://arxiv.org/abs/1605.07453} {\path{arXiv:1605.07453}},
  \href {http://dx.doi.org/10.1140/epjc/s10052-016-4459-0}
  {\path{doi:10.1140/epjc/s10052-016-4459-0}}.

\bibitem{Chen:2008qv}
H.-X. Chen, V.~Dmitrasinovic, A.~Hosaka, K.~Nagata, S.-L. Zhu, {Chiral
  Properties of Baryon Fields with Flavor SU(3) Symmetry}, Phys. Rev. D78
  (2008) 054021.
\newblock \href {http://arxiv.org/abs/0806.1997} {\path{arXiv:0806.1997}},
  \href {http://dx.doi.org/10.1103/PhysRevD.78.054021}
  {\path{doi:10.1103/PhysRevD.78.054021}}.

\bibitem{Chen:2009sf}
H.-X. Chen, V.~Dmitrasinovic, A.~Hosaka, {Baryon fields with $U_L(3) \times
  U_R(3)$ chiral symmetry II: Axial currents of nucleons and hyperons}, Phys.
  Rev. D81 (2010) 054002.
\newblock \href {http://arxiv.org/abs/0912.4338} {\path{arXiv:0912.4338}},
  \href {http://dx.doi.org/10.1103/PhysRevD.81.054002}
  {\path{doi:10.1103/PhysRevD.81.054002}}.

\bibitem{Chen:2010ba}
H.-X. Chen, V.~Dmitrasinovic, A.~Hosaka, {Baryon Fields with $U_L(3) times
  U_R(3)$ Chiral Symmetry III: Interactions with Chiral $(3,\bar{3})+
  (\bar{3},3)$ Spinless Mesons}, Phys. Rev. D83 (2011) 014015.
\newblock \href {http://arxiv.org/abs/1009.2422} {\path{arXiv:1009.2422}},
  \href {http://dx.doi.org/10.1103/PhysRevD.83.014015}
  {\path{doi:10.1103/PhysRevD.83.014015}}.

\bibitem{Dmitrasinovic:2011yf}
V.~Dmitrasinovic, H.-X. Chen, {Bi-local baryon interpolating fields with two
  flavours}, Eur. Phys. J. C71 (2011) 1543.
\newblock \href {http://arxiv.org/abs/1101.5906} {\path{arXiv:1101.5906}},
  \href {http://dx.doi.org/10.1140/epjc/s10052-011-1543-3}
  {\path{doi:10.1140/epjc/s10052-011-1543-3}}.

\bibitem{Chen:2011rh}
H.-X. Chen, V.~Dmitrasinovic, A.~Hosaka, {Baryons with $U_L(3) \times U_R(3)$
  Chiral Symmetry IV: Interactions with Chiral $(8,1)+(1,8)$ Vector and
  Axial-vector Mesons and Anomalous Magnetic Moments}, Phys. Rev. C85 (2012)
  055205.
\newblock \href {http://arxiv.org/abs/1109.3130} {\path{arXiv:1109.3130}},
  \href {http://dx.doi.org/10.1103/PhysRevC.85.055205}
  {\path{doi:10.1103/PhysRevC.85.055205}}.

\bibitem{Chen:2012ex}
H.-X. Chen, {Chiral Baryon Fields in the QCD Sum Rule}, Eur. Phys. J. C72
  (2012) 2180.
\newblock \href {http://arxiv.org/abs/1203.3260} {\path{arXiv:1203.3260}},
  \href {http://dx.doi.org/10.1140/epjc/s10052-012-2180-1}
  {\path{doi:10.1140/epjc/s10052-012-2180-1}}.

\bibitem{Chen:2012vs}
H.-X. Chen, {Baryon Tri-local Interpolating Fields}, Eur. Phys. J. C72 (2012)
  2129.
\newblock \href {http://arxiv.org/abs/1205.5328} {\path{arXiv:1205.5328}},
  \href {http://dx.doi.org/10.1140/epjc/s10052-012-2129-4}
  {\path{doi:10.1140/epjc/s10052-012-2129-4}}.

\bibitem{Chen:2013efa}
H.-X. Chen, V.~Dmitrasinovic, {Bilocal baryon interpolating fields with three
  flavors}, Phys. Rev. D88~(3) (2013) 036013.
\newblock \href {http://arxiv.org/abs/1309.0387} {\path{arXiv:1309.0387}},
  \href {http://dx.doi.org/10.1103/PhysRevD.88.036013}
  {\path{doi:10.1103/PhysRevD.88.036013}}.

\bibitem{Dmitrasinovic:2016hup}
V.~Dmitra{\v s}inovi{\'c}, H.-X. Chen, A.~Hosaka, {Baryon fields with $U_L(3)
  \times U_R(3)$ chiral symmetry. V. Pion-nucleon and kaon-nucleon $\sigma$
  terms}, Phys. Rev. C93~(6) (2016) 065208.
\newblock \href {http://arxiv.org/abs/1812.03414} {\path{arXiv:1812.03414}},
  \href {http://dx.doi.org/10.1103/PhysRevC.93.065208}
  {\path{doi:10.1103/PhysRevC.93.065208}}.

\bibitem{Chen:2017sbg}
H.-X. Chen, Q.~Mao, W.~Chen, X.~Liu, S.-L. Zhu, {Establishing low-lying doubly
  charmed baryons}, Phys. Rev. D96~(3) (2017) 031501, [Erratum: Phys.
  Rev.D96,no.11,119902(2017)].
\newblock \href {http://arxiv.org/abs/1707.01779} {\path{arXiv:1707.01779}},
  \href {http://dx.doi.org/10.1103/PhysRevD.96.031501,
  10.1103/PhysRevD.96.119902} {\path{doi:10.1103/PhysRevD.96.031501,
  10.1103/PhysRevD.96.119902}}.

\bibitem{Cui:2017udv}
E.-L. Cui, H.-X. Chen, W.~Chen, X.~Liu, S.-L. Zhu, {Suggested search for doubly
  charmed baryons of $J^P=3/2^+$ via their electromagnetic transitions}, Phys.
  Rev. D97~(3) (2018) 034018.
\newblock \href {http://arxiv.org/abs/1712.03615} {\path{arXiv:1712.03615}},
  \href {http://dx.doi.org/10.1103/PhysRevD.97.034018}
  {\path{doi:10.1103/PhysRevD.97.034018}}.

\bibitem{Liu:2007fg}
X.~Liu, H.-X. Chen, Y.-R. Liu, A.~Hosaka, S.-L. Zhu, {Bottom baryons}, Phys.
  Rev. D77 (2008) 014031.
\newblock \href {http://arxiv.org/abs/0710.0123} {\path{arXiv:0710.0123}},
  \href {http://dx.doi.org/10.1103/PhysRevD.77.014031}
  {\path{doi:10.1103/PhysRevD.77.014031}}.

\bibitem{Huang:2009is}
P.-Z. Huang, H.-X. Chen, S.-L. Zhu, {Light vector meson and heavy baryon strong
  interaction}, Phys. Rev. D80 (2009) 094007.
\newblock \href {http://arxiv.org/abs/0909.5551} {\path{arXiv:0909.5551}},
  \href {http://dx.doi.org/10.1103/PhysRevD.80.094007}
  {\path{doi:10.1103/PhysRevD.80.094007}}.

\bibitem{Chen:2015kpa}
H.-X. Chen, W.~Chen, Q.~Mao, A.~Hosaka, X.~Liu, S.-L. Zhu, {$P$-wave charmed
  baryons from QCD sum rules}, Phys. Rev. D91~(5) (2015) 054034.
\newblock \href {http://arxiv.org/abs/1502.01103} {\path{arXiv:1502.01103}},
  \href {http://dx.doi.org/10.1103/PhysRevD.91.054034}
  {\path{doi:10.1103/PhysRevD.91.054034}}.

\bibitem{Mao:2015gya}
Q.~Mao, H.-X. Chen, W.~Chen, A.~Hosaka, X.~Liu, S.-L. Zhu, {QCD sum rule
  calculation for P-wave bottom baryons}, Phys. Rev. D92 (2015) 114007.
\newblock \href {http://arxiv.org/abs/1510.05267} {\path{arXiv:1510.05267}},
  \href {http://dx.doi.org/10.1103/PhysRevD.92.114007}
  {\path{doi:10.1103/PhysRevD.92.114007}}.

\bibitem{Chen:2016phw}
H.-X. Chen, Q.~Mao, A.~Hosaka, X.~Liu, S.-L. Zhu, {$D$-wave charmed and
  bottomed baryons from QCD sum rules}, Phys. Rev. D94~(11) (2016) 114016.
\newblock \href {http://arxiv.org/abs/1611.02677} {\path{arXiv:1611.02677}},
  \href {http://dx.doi.org/10.1103/PhysRevD.94.114016}
  {\path{doi:10.1103/PhysRevD.94.114016}}.

\bibitem{Mao:2017wbz}
Q.~Mao, H.-X. Chen, A.~Hosaka, X.~Liu, S.-L. Zhu, {$D$-wave heavy baryons of
  the $SU(3)$ flavor $\mathbf{6}_F$}, Phys. Rev. D96~(7) (2017) 074021.
\newblock \href {http://arxiv.org/abs/1707.03712} {\path{arXiv:1707.03712}},
  \href {http://dx.doi.org/10.1103/PhysRevD.96.074021}
  {\path{doi:10.1103/PhysRevD.96.074021}}.

\bibitem{Chen:2017sci}
H.-X. Chen, Q.~Mao, W.~Chen, A.~Hosaka, X.~Liu, S.-L. Zhu, {Decay properties of
  $P$-wave charmed baryons from light-cone QCD sum rules}, Phys. Rev. D95~(9)
  (2017) 094008.
\newblock \href {http://arxiv.org/abs/1703.07703} {\path{arXiv:1703.07703}},
  \href {http://dx.doi.org/10.1103/PhysRevD.95.094008}
  {\path{doi:10.1103/PhysRevD.95.094008}}.

\bibitem{Zhou:2014ytp}
D.~Zhou, E.-L. Cui, H.-X. Chen, L.-S. Geng, X.~Liu, S.-L. Zhu, {$D$-wave
  heavy-light mesons from QCD sum rules}, Phys. Rev. D90~(11) (2014) 114035.
\newblock \href {http://arxiv.org/abs/1410.1727} {\path{arXiv:1410.1727}},
  \href {http://dx.doi.org/10.1103/PhysRevD.90.114035}
  {\path{doi:10.1103/PhysRevD.90.114035}}.

\bibitem{Zhou:2015ywa}
D.~Zhou, H.-X. Chen, L.-S. Geng, X.~Liu, S.-L. Zhu, {$F$-wave heavy-light meson
  spectroscopy in QCD sum rules and heavy quark effective theory}, Phys. Rev.
  D92~(11) (2015) 114015.
\newblock \href {http://arxiv.org/abs/1506.00766} {\path{arXiv:1506.00766}},
  \href {http://dx.doi.org/10.1103/PhysRevD.92.114015}
  {\path{doi:10.1103/PhysRevD.92.114015}}.

\bibitem{Aaij:2017nav}
R.~Aaij, et~al., {Observation of five new narrow $\Omega_c^0$ states decaying
  to $\Xi_c^+ K^-$}, Phys. Rev. Lett. 118~(18) (2017) 182001.
\newblock \href {http://arxiv.org/abs/1703.04639} {\path{arXiv:1703.04639}},
  \href {http://dx.doi.org/10.1103/PhysRevLett.118.182001}
  {\path{doi:10.1103/PhysRevLett.118.182001}}.

\bibitem{Chen:2010ze}
W.~Chen, S.-L. Zhu, {The Vector and Axial-Vector Charmonium-like States}, Phys.
  Rev. D83 (2011) 034010.
\newblock \href {http://arxiv.org/abs/1010.3397} {\path{arXiv:1010.3397}},
  \href {http://dx.doi.org/10.1103/PhysRevD.83.034010}
  {\path{doi:10.1103/PhysRevD.83.034010}}.

\bibitem{Yuan:2007sj}
C.~Z. Yuan, et~al., {Measurement of $e^+ e^- \to \pi^+ \pi^- J/\psi$
  cross-section via initial state radiation at Belle}, Phys. Rev. Lett. 99
  (2007) 182004.
\newblock \href {http://arxiv.org/abs/0707.2541} {\path{arXiv:0707.2541}},
  \href {http://dx.doi.org/10.1103/PhysRevLett.99.182004}
  {\path{doi:10.1103/PhysRevLett.99.182004}}.

\bibitem{Aubert:2005rm}
B.~Aubert, et~al., {Observation of a broad structure in the $\pi^+ \pi^-
  J/\psi$ mass spectrum around 4.26 GeV/c$^2$}, Phys. Rev. Lett. 95 (2005)
  142001.
\newblock \href {http://arxiv.org/abs/hep-ex/0506081}
  {\path{arXiv:hep-ex/0506081}}, \href
  {http://dx.doi.org/10.1103/PhysRevLett.95.142001}
  {\path{doi:10.1103/PhysRevLett.95.142001}}.

\bibitem{Aubert:2007zz}
B.~Aubert, et~al., {Evidence of a broad structure at an invariant mass of 4.32
  $GeV/c^{2}$ in the reaction $e^{+} e^{-} \to \pi^{+} \pi^{-} \psi(2S)$
  measured at BaBar}, Phys. Rev. Lett. 98 (2007) 212001.
\newblock \href {http://arxiv.org/abs/hep-ex/0610057}
  {\path{arXiv:hep-ex/0610057}}, \href
  {http://dx.doi.org/10.1103/PhysRevLett.98.212001}
  {\path{doi:10.1103/PhysRevLett.98.212001}}.

\bibitem{Pakhlova:2008vn}
G.~Pakhlova, et~al., {Observation of a near-threshold enhancement in the $e^+
  e^- \to \Lambda^+_{c} \Lambda^-_{c}$ cross section using initial-state
  radiation}, Phys. Rev. Lett. 101 (2008) 172001.
\newblock \href {http://arxiv.org/abs/0807.4458} {\path{arXiv:0807.4458}},
  \href {http://dx.doi.org/10.1103/PhysRevLett.101.172001}
  {\path{doi:10.1103/PhysRevLett.101.172001}}.

\bibitem{Wang:2007ea}
X.~L. Wang, et~al., {Observation of Two Resonant Structures in $e^+e^- \to
  \pi^+ \pi^- \psi(2S)$ via Initial State Radiation at Belle}, Phys. Rev. Lett.
  99 (2007) 142002.
\newblock \href {http://arxiv.org/abs/0707.3699} {\path{arXiv:0707.3699}},
  \href {http://dx.doi.org/10.1103/PhysRevLett.99.142002}
  {\path{doi:10.1103/PhysRevLett.99.142002}}.

\bibitem{Chen:2016otp}
H.-X. Chen, E.-L. Cui, W.~Chen, X.~Liu, T.~G. Steele, S.-L. Zhu, {QCD sum rule
  study of hidden-charm pentaquarks}, Eur. Phys. J. C76~(10) (2016) 572.
\newblock \href {http://arxiv.org/abs/1602.02433} {\path{arXiv:1602.02433}},
  \href {http://dx.doi.org/10.1140/epjc/s10052-016-4438-5}
  {\path{doi:10.1140/epjc/s10052-016-4438-5}}.

\bibitem{Xiang:2017byz}
J.-B. Xiang, H.-X. Chen, W.~Chen, X.-B. Li, X.-Q. Yao, S.-L. Zhu, {Revisiting
  hidden-charm pentaquarks from QCD sum rules}, Chin. Phys. C43 (2019) 034104.
\newblock \href {http://arxiv.org/abs/1711.01545} {\path{arXiv:1711.01545}},
  \href {http://dx.doi.org/10.1088/1674-1137/43/3/034104}
  {\path{doi:10.1088/1674-1137/43/3/034104}}.

\bibitem{Chen:2008ej}
H.-X. Chen, X.~Liu, A.~Hosaka, S.-L. Zhu, {The Y(2175) State in the QCD Sum
  Rule}, Phys. Rev. D78 (2008) 034012.
\newblock \href {http://arxiv.org/abs/0801.4603} {\path{arXiv:0801.4603}},
  \href {http://dx.doi.org/10.1103/PhysRevD.78.034012}
  {\path{doi:10.1103/PhysRevD.78.034012}}.

\bibitem{Aubert:2007ur}
B.~Aubert, et~al., {The $e^+ e^- \to K^+ K^- \pi^+ \pi^-$, $K^+ K^- \pi^0
  \pi^0$ and $K^+ K^- K^+ K^-$ cross-sections measured with initial-state
  radiation}, Phys. Rev. D76 (2007) 012008.
\newblock \href {http://arxiv.org/abs/0704.0630} {\path{arXiv:0704.0630}},
  \href {http://dx.doi.org/10.1103/PhysRevD.76.012008}
  {\path{doi:10.1103/PhysRevD.76.012008}}.

\bibitem{Shen:2009zze}
C.~P. Shen, et~al., {Observation of the $\phi(1680)$ and the $Y(2175)$ in
  $e^+e^- \to \phi \pi^+ \pi^-$}, Phys. Rev. D80 (2009) 031101.
\newblock \href {http://arxiv.org/abs/0808.0006} {\path{arXiv:0808.0006}},
  \href {http://dx.doi.org/10.1103/PhysRevD.80.031101}
  {\path{doi:10.1103/PhysRevD.80.031101}}.

\bibitem{Shen:2009mr}
C.~P. Shen, C.~Z. Yuan, {Combined fit to BaBar and Belle Data on $e^+ e^- \to
  \phi \pi^+ \pi^-$ and phi $f_0(980)$}, Chin. Phys. C34 (2010) 1045--1051.
\newblock \href {http://arxiv.org/abs/0911.1591} {\path{arXiv:0911.1591}},
  \href {http://dx.doi.org/10.1088/1674-1137/34/8/002}
  {\path{doi:10.1088/1674-1137/34/8/002}}.

\bibitem{Ablikim:2018xuz}
M.~Ablikim, et~al., {Observation and study of $J/\psi\rightarrow\phi\eta\eta'$
  at BESIII}\href {http://arxiv.org/abs/1901.00085} {\path{arXiv:1901.00085}}.

\bibitem{Chen:2019osl}
H.-X. Chen, W.~Chen, {Settling the $Z_c(4600)$ in the charged charmonium-like
  family}\href {http://arxiv.org/abs/1901.06946} {\path{arXiv:1901.06946}}.

\bibitem{Aaij:2019ipm}
R.~Aaij, et~al., {Model-independent observation of exotic contributions to
  $B^0\to J/\psi K^+\pi^-$ decays}\href {http://arxiv.org/abs/1901.05745}
  {\path{arXiv:1901.05745}}.

\bibitem{Matheus:2006xi}
R.~D. Matheus, S.~Narison, M.~Nielsen, J.~M. Richard, {Can the X(3872) be a
  $1^{++}$ four-quark state?}, Phys. Rev. D75 (2007) 014005.
\newblock \href {http://arxiv.org/abs/hep-ph/0608297}
  {\path{arXiv:hep-ph/0608297}}, \href
  {http://dx.doi.org/10.1103/PhysRevD.75.014005}
  {\path{doi:10.1103/PhysRevD.75.014005}}.

\bibitem{Navarra:2006nd}
F.~S. Navarra, M.~Nielsen, {$X(3872) \to J/\psi \pi^+ \pi^-$ and $X(3872) \to
  J/\psi \pi^+ \pi^- \pi^0$ decay widths from QCD sum rules}, Phys. Lett. B639
  (2006) 272--277.
\newblock \href {http://arxiv.org/abs/hep-ph/0605038}
  {\path{arXiv:hep-ph/0605038}}, \href
  {http://dx.doi.org/10.1016/j.physletb.2006.06.054}
  {\path{doi:10.1016/j.physletb.2006.06.054}}.

\bibitem{Lee:2007gs}
S.~H. Lee, A.~Mihara, F.~S. Navarra, M.~Nielsen, {QCD sum rules study of the
  meson $Z^+(4430)$}, Phys. Lett. B661 (2008) 28--32.
\newblock \href {http://arxiv.org/abs/0710.1029} {\path{arXiv:0710.1029}},
  \href {http://dx.doi.org/10.1016/j.physletb.2008.01.062}
  {\path{doi:10.1016/j.physletb.2008.01.062}}.

\bibitem{Bracco:2008jj}
M.~E. Bracco, S.~H. Lee, M.~Nielsen, R.~Rodrigues~da Silva, {The Meson
  $Z^+(4430)$ as a tetraquark state}, Phys. Lett. B671 (2009) 240--244.
\newblock \href {http://arxiv.org/abs/0807.3275} {\path{arXiv:0807.3275}},
  \href {http://dx.doi.org/10.1016/j.physletb.2008.12.021}
  {\path{doi:10.1016/j.physletb.2008.12.021}}.

\bibitem{Albuquerque:2008up}
R.~M. Albuquerque, M.~Nielsen, {QCD sum rules study of the $J^{PC} = 1^{--}$
  charmonium $Y$ mesons}, Nucl. Phys. A815 (2009) 53--66, [Erratum: Nucl. Phys.
  A857, 48 (2011)].
\newblock \href {http://arxiv.org/abs/0804.4817} {\path{arXiv:0804.4817}},
  \href {http://dx.doi.org/10.1016/j.nuclphysa.2011.04.001,
  10.1016/j.nuclphysa.2008.10.015} {\path{doi:10.1016/j.nuclphysa.2011.04.001,
  10.1016/j.nuclphysa.2008.10.015}}.

\bibitem{Matheus:2009vq}
R.~D. Matheus, F.~S. Navarra, M.~Nielsen, C.~M. Zanetti, {QCD Sum Rules for the
  X(3872) as a mixed molecule-charmoniun state}, Phys. Rev. D80 (2009) 056002.
\newblock \href {http://arxiv.org/abs/0907.2683} {\path{arXiv:0907.2683}},
  \href {http://dx.doi.org/10.1103/PhysRevD.80.056002}
  {\path{doi:10.1103/PhysRevD.80.056002}}.

\bibitem{Albuquerque:2009ak}
R.~M. Albuquerque, M.~E. Bracco, M.~Nielsen, {A QCD sum rule calculation for
  the Y(4140) narrow structure}, Phys. Lett. B678 (2009) 186--190.
\newblock \href {http://arxiv.org/abs/0903.5540} {\path{arXiv:0903.5540}},
  \href {http://dx.doi.org/10.1016/j.physletb.2009.06.022}
  {\path{doi:10.1016/j.physletb.2009.06.022}}.

\bibitem{Wang:2009wk}
Z.-G. Wang, {Analysis of the $X(4350)$ as a scalar $\bar c c$ and $D^*_s \bar
  D^*_s$ mixing state with QCD sum rules}, Phys. Lett. B690 (2010) 403--406.
\newblock \href {http://arxiv.org/abs/0912.4626} {\path{arXiv:0912.4626}},
  \href {http://dx.doi.org/10.1016/j.physletb.2010.05.068}
  {\path{doi:10.1016/j.physletb.2010.05.068}}.

\bibitem{Wang:2009ue}
Z.-G. Wang, {Analysis of the Y(4140) with QCD sum rules}, Eur. Phys. J. C63
  (2009) 115--122.
\newblock \href {http://arxiv.org/abs/0903.5200} {\path{arXiv:0903.5200}},
  \href {http://dx.doi.org/10.1140/epjc/s10052-009-1097-9}
  {\path{doi:10.1140/epjc/s10052-009-1097-9}}.

\bibitem{Zhang:2009st}
J.-R. Zhang, M.-Q. Huang, $(q \bar s)^{(*)}(\bar q s)^{(*)}$ molecular states
  from qcd sum rules: A view on $y(4140)$, J. Phys. G37 (2010) 025005.
\newblock \href {http://arxiv.org/abs/0905.4178} {\path{arXiv:0905.4178}},
  \href {http://dx.doi.org/10.1088/0954-3899/37/2/025005}
  {\path{doi:10.1088/0954-3899/37/2/025005}}.

\bibitem{Zhang:2010mv}
J.-R. Zhang, M.-Q. Huang, {Could $Y_{b}(10890)$ be the P-wave
  $[bq][\bar{b}\bar{q}]$ tetraquark state?}, JHEP 11 (2010) 057.
\newblock \href {http://arxiv.org/abs/1011.2815} {\path{arXiv:1011.2815}},
  \href {http://dx.doi.org/10.1007/JHEP11(2010)057}
  {\path{doi:10.1007/JHEP11(2010)057}}.

\bibitem{Zhang:2010mw}
J.-R. Zhang, M.-Q. Huang, {The $P$-wave $[cs][\bar{c}\bar{s}]$ tetraquark
  state: $Y(4260)$ or $Y(4660)$?}, Phys. Rev. D83 (2011) 036005.
\newblock \href {http://arxiv.org/abs/1011.2818} {\path{arXiv:1011.2818}},
  \href {http://dx.doi.org/10.1103/PhysRevD.83.036005}
  {\path{doi:10.1103/PhysRevD.83.036005}}.

\bibitem{Zhang:2011jja}
J.-R. Zhang, M.~Zhong, M.-Q. Huang, {Could $Z_{b}(10610)$ be a $B^{*}\bar{B}$
  molecular state?}, Phys. Lett. B704 (2011) 312--315.
\newblock \href {http://arxiv.org/abs/1105.5472} {\path{arXiv:1105.5472}},
  \href {http://dx.doi.org/10.1016/j.physletb.2011.09.039}
  {\path{doi:10.1016/j.physletb.2011.09.039}}.

\bibitem{Chen:2012pe}
W.~Chen, S.-L. Zhu, {Spin-1 charmonium-like states in QCD sum rule}, EPJ Web
  Conf. 20 (2012) 01003.
\newblock \href {http://arxiv.org/abs/1209.4748} {\path{arXiv:1209.4748}},
  \href {http://dx.doi.org/10.1051/epjconf/20122001003}
  {\path{doi:10.1051/epjconf/20122001003}}.

\bibitem{Chen:2013omd}
W.~Chen, T.~G. Steele, M.-L. Du, S.-L. Zhu, {$D^*\bar D^*$ molecule
  interpretation of $Z_c(4025)$}, Eur. Phys. J. C74~(2) (2014) 2773.
\newblock \href {http://arxiv.org/abs/1308.5060} {\path{arXiv:1308.5060}},
  \href {http://dx.doi.org/10.1140/epjc/s10052-014-2773-y}
  {\path{doi:10.1140/epjc/s10052-014-2773-y}}.

\bibitem{Chen:2013wva}
W.~Chen, W.-Z. Deng, J.~He, N.~Li, X.~Liu, Z.-G. Luo, Z.-F. Sun, S.-L. Zhu,
  {XYZ States}, PoS Hadron2013 (2013) 005.
\newblock \href {http://arxiv.org/abs/1311.3763} {\path{arXiv:1311.3763}},
  \href {http://dx.doi.org/10.22323/1.205.0005}
  {\path{doi:10.22323/1.205.0005}}.

\bibitem{Wang:2013exa}
Z.-G. Wang, {Analysis of the $Z_c(4020)$, $Z_c(4025)$, $Y(4360)$ and $Y(4660)$
  as vector tetraquark states with QCD sum rules}, Eur. Phys. J. C74~(5) (2014)
  2874.
\newblock \href {http://arxiv.org/abs/1311.1046} {\path{arXiv:1311.1046}},
  \href {http://dx.doi.org/10.1140/epjc/s10052-014-2874-7}
  {\path{doi:10.1140/epjc/s10052-014-2874-7}}.

\bibitem{Wang:2013daa}
Z.-G. Wang, T.~Huang, {Possible assignments of the $X(3872)$, $Z_c(3900)$ and
  $Z_b(10610)$ as axial-vector molecular states}, Eur. Phys. J. C74~(5) (2014)
  2891.
\newblock \href {http://arxiv.org/abs/1312.7489} {\path{arXiv:1312.7489}},
  \href {http://dx.doi.org/10.1140/epjc/s10052-014-2891-6}
  {\path{doi:10.1140/epjc/s10052-014-2891-6}}.

\bibitem{Chen:2014fza}
W.~Chen, T.~G. Steele, S.-L. Zhu, {Heavy tetraquark states and quarkonium
  hybrids}, The Universe 2 (2014) 13--40.
\newblock \href {http://arxiv.org/abs/1403.7457} {\path{arXiv:1403.7457}}.

\bibitem{Chen:2015fsa}
W.~Chen, T.~G. Steele, H.-X. Chen, S.-L. Zhu, {$Z_c(4200)^+$ decay width as a
  charmonium-like tetraquark state}, Eur. Phys. J. C75~(8) (2015) 358.
\newblock \href {http://arxiv.org/abs/1501.03863} {\path{arXiv:1501.03863}},
  \href {http://dx.doi.org/10.1140/epjc/s10052-015-3578-3}
  {\path{doi:10.1140/epjc/s10052-015-3578-3}}.

\bibitem{Chen:2015ata}
W.~Chen, T.~G. Steele, H.-X. Chen, S.-L. Zhu, {Mass spectra of $Z_c$ and $Z_b$
  exotic states as hadron molecules}, Phys. Rev. D92~(5) (2015) 054002.
\newblock \href {http://arxiv.org/abs/1505.05619} {\path{arXiv:1505.05619}},
  \href {http://dx.doi.org/10.1103/PhysRevD.92.054002}
  {\path{doi:10.1103/PhysRevD.92.054002}}.

\bibitem{Wang:2016mmg}
Z.-G. Wang, {Tetraquark state candidates: $Y(4260)$, $Y(4360)$, $Y(4660)$ and
  $Z_c(4020/4025)$}, Eur. Phys. J. C76~(7) (2016) 387.
\newblock \href {http://arxiv.org/abs/1601.05541} {\path{arXiv:1601.05541}},
  \href {http://dx.doi.org/10.1140/epjc/s10052-016-4238-y}
  {\path{doi:10.1140/epjc/s10052-016-4238-y}}.

\bibitem{Huang:2016rro}
Z.-R. Huang, W.~Chen, T.~G. Steele, Z.-F. Zhang, H.-Y. Jin, {Investigation of
  the light four-quark states with exotic $J^{PC}=0^{--}$}, Phys. Rev. D95~(7)
  (2017) 076017.
\newblock \href {http://arxiv.org/abs/1610.02081} {\path{arXiv:1610.02081}},
  \href {http://dx.doi.org/10.1103/PhysRevD.95.076017}
  {\path{doi:10.1103/PhysRevD.95.076017}}.

\bibitem{Chen:2016mqt}
W.~Chen, H.-X. Chen, X.~Liu, T.~G. Steele, S.-L. Zhu, {Decoding the $X(5568)$
  as a fully open-flavor $su\bar b\bar d$ tetraquark state}, Phys. Rev. Lett.
  117~(2) (2016) 022002.
\newblock \href {http://arxiv.org/abs/1602.08916} {\path{arXiv:1602.08916}},
  \href {http://dx.doi.org/10.1103/PhysRevLett.117.022002}
  {\path{doi:10.1103/PhysRevLett.117.022002}}.

\bibitem{Chen:2017rhl}
W.~Chen, H.-X. Chen, X.~Liu, T.~G. Steele, S.-L. Zhu, {Open-flavor charm and
  bottom $sq\bar q\bar Q$ and $qq\bar q\bar Q$ tetraquark states}, Phys. Rev.
  D95~(11) (2017) 114005.
\newblock \href {http://arxiv.org/abs/1705.10088} {\path{arXiv:1705.10088}},
  \href {http://dx.doi.org/10.1103/PhysRevD.95.114005}
  {\path{doi:10.1103/PhysRevD.95.114005}}.

\bibitem{Fu:2018ngx}
Y.-C. Fu, Z.-R. Huang, Z.-F. Zhang, W.~Chen, {Exotic tetraquark states with
  $J^{PC}=0^{+-}$}, Phys. Rev. D99~(1) (2019) 014025.
\newblock \href {http://arxiv.org/abs/1811.03333} {\path{arXiv:1811.03333}},
  \href {http://dx.doi.org/10.1103/PhysRevD.99.014025}
  {\path{doi:10.1103/PhysRevD.99.014025}}.

\bibitem{Zhang:2018mnm}
J.-R. Zhang, {Revisiting $D_{s0}^{*}(2317)$ as a $0^{+}$ tetraquark state from
  QCD sum rules}, Phys. Lett. B789 (2019) 432--437.
\newblock \href {http://arxiv.org/abs/1801.08725} {\path{arXiv:1801.08725}},
  \href {http://dx.doi.org/10.1016/j.physletb.2019.01.001}
  {\path{doi:10.1016/j.physletb.2019.01.001}}.

\bibitem{Chen:2017dpy}
W.~Chen, H.-X. Chen, X.~Liu, T.~G. Steele, S.-L. Zhu, {Mass spectra for $qc\bar
  q\bar c$, $sc\bar s\bar c$, $qb\bar q\bar b$, $sb\bar s\bar b$ tetraquark
  states with $J^{PC}=0^{++}$ and $2^{++}$}, Phys. Rev. D96~(11) (2017) 114017.
\newblock \href {http://arxiv.org/abs/1706.09731} {\path{arXiv:1706.09731}},
  \href {http://dx.doi.org/10.1103/PhysRevD.96.114017}
  {\path{doi:10.1103/PhysRevD.96.114017}}.

\bibitem{Wang:2017lbl}
Z.-G. Wang, {Analysis of the mass and width of the $X^*(3860)$ with QCD sum
  rules}, Eur. Phys. J. A53~(10) (2017) 192.
\newblock \href {http://arxiv.org/abs/1704.04111} {\path{arXiv:1704.04111}},
  \href {http://dx.doi.org/10.1140/epja/i2017-12390-6}
  {\path{doi:10.1140/epja/i2017-12390-6}}.

\bibitem{Wang:2016ujn}
Z.-G. Wang, {Reanalysis of the $X(3915)$, $X(4500)$ and $X(4700)$ with QCD sum
  rules}, Eur. Phys. J. A53~(2) (2017) 19.
\newblock \href {http://arxiv.org/abs/1607.04840} {\path{arXiv:1607.04840}},
  \href {http://dx.doi.org/10.1140/epja/i2017-12208-7}
  {\path{doi:10.1140/epja/i2017-12208-7}}.

\bibitem{Wang:2016tzr}
Z.-G. Wang, {Reanalysis of $X$(4140) as axial-vector tetraquark state with QCD
  sum rules}, Eur. Phys. J. C76~(12) (2016) 657.
\newblock \href {http://arxiv.org/abs/1607.00701} {\path{arXiv:1607.00701}},
  \href {http://dx.doi.org/10.1140/epjc/s10052-016-4515-9}
  {\path{doi:10.1140/epjc/s10052-016-4515-9}}.

\bibitem{Agaev:2017foq}
S.~S. Agaev, K.~Azizi, H.~Sundu, {Exploring the resonances $X(4140)$ and
  $X(4274)$ through their decay channels}, Phys. Rev. D95~(11) (2017) 114003.
\newblock \href {http://arxiv.org/abs/1703.10323} {\path{arXiv:1703.10323}},
  \href {http://dx.doi.org/10.1103/PhysRevD.95.114003}
  {\path{doi:10.1103/PhysRevD.95.114003}}.

\bibitem{Wang:2016dcb}
Z.-G. Wang, {Analysis of the mass and width of the $Y(4274)$ as axialvector
  molecule-like state}, Eur. Phys. J. C77~(3) (2017) 174.
\newblock \href {http://arxiv.org/abs/1612.00195} {\path{arXiv:1612.00195}},
  \href {http://dx.doi.org/10.1140/epjc/s10052-017-4751-7}
  {\path{doi:10.1140/epjc/s10052-017-4751-7}}.

\bibitem{Chen:2016jxd}
W.~Chen, H.-X. Chen, X.~Liu, T.~G. Steele, S.-L. Zhu, {Hunting for exotic
  doubly hidden-charm/bottom tetraquark states}, Phys. Lett. B773 (2017)
  247--251.
\newblock \href {http://arxiv.org/abs/1605.01647} {\path{arXiv:1605.01647}},
  \href {http://dx.doi.org/10.1016/j.physletb.2017.08.034}
  {\path{doi:10.1016/j.physletb.2017.08.034}}.

\bibitem{Chen:2018cqz}
W.~Chen, H.-X. Chen, X.~Liu, T.~G. Steele, S.-L. Zhu, {Doubly
  hidden-charm/bottom $QQ\bar Q\bar Q$ tetraquark states}, EPJ Web Conf. 182
  (2018) 02028.
\newblock \href {http://arxiv.org/abs/1803.02522} {\path{arXiv:1803.02522}},
  \href {http://dx.doi.org/10.1051/epjconf/201818202028}
  {\path{doi:10.1051/epjconf/201818202028}}.

\bibitem{Wang:2017jtz}
Z.-G. Wang, {Analysis of the $QQ\bar{Q}\bar{Q}$ tetraquark states with QCD sum
  rules}, Eur. Phys. J. C77~(7) (2017) 432.
\newblock \href {http://arxiv.org/abs/1701.04285} {\path{arXiv:1701.04285}},
  \href {http://dx.doi.org/10.1140/epjc/s10052-017-4997-0}
  {\path{doi:10.1140/epjc/s10052-017-4997-0}}.

\bibitem{Wang:2018poa}
Z.-G. Wang, Z.-Y. Di, {Analysis of the vector and axialvector
  $QQ\bar{Q}\bar{Q}$ tetraquark states with QCD sum rules}\href
  {http://arxiv.org/abs/1807.08520} {\path{arXiv:1807.08520}}.

\bibitem{Jiang:2017tdc}
J.-F. Jiang, W.~Chen, S.-L. Zhu, {Triply heavy $QQ\bar Q\bar q$ tetraquark
  states}, Phys. Rev. D96~(9) (2017) 094022.
\newblock \href {http://arxiv.org/abs/1708.00142} {\path{arXiv:1708.00142}},
  \href {http://dx.doi.org/10.1103/PhysRevD.96.094022}
  {\path{doi:10.1103/PhysRevD.96.094022}}.

\bibitem{Du:2012wp}
M.-L. Du, W.~Chen, X.-L. Chen, S.-L. Zhu, {Exotic $QQ\bar{q}\bar{q}$,
  $QQ\bar{q}\bar{s}$ and $QQ\bar{s}\bar{s}$ states}, Phys. Rev. D87~(1) (2013)
  014003.
\newblock \href {http://arxiv.org/abs/1209.5134} {\path{arXiv:1209.5134}},
  \href {http://dx.doi.org/10.1103/PhysRevD.87.014003}
  {\path{doi:10.1103/PhysRevD.87.014003}}.

\bibitem{Navarra:2007yw}
F.~S. Navarra, M.~Nielsen, S.~H. Lee, {QCD sum rules study of $Q Q \bar u \bar
  d$ mesons}, Phys. Lett. B649 (2007) 166--172.
\newblock \href {http://arxiv.org/abs/hep-ph/0703071}
  {\path{arXiv:hep-ph/0703071}}, \href
  {http://dx.doi.org/10.1016/j.physletb.2007.04.010}
  {\path{doi:10.1016/j.physletb.2007.04.010}}.

\bibitem{Chen:2013aba}
W.~Chen, T.~G. Steele, S.-L. Zhu, {Exotic open-flavor $bc\bar{q}\bar{q}$,
  $bc\bar{s}\bar{s}$ and $qc\bar{q}\bar{b}$, $sc\bar{s}\bar{b}$ tetraquark
  states}, Phys. Rev. D89~(5) (2014) 054037.
\newblock \href {http://arxiv.org/abs/1310.8337} {\path{arXiv:1310.8337}},
  \href {http://dx.doi.org/10.1103/PhysRevD.89.054037}
  {\path{doi:10.1103/PhysRevD.89.054037}}.

\bibitem{Wang:2017dtg}
Z.-G. Wang, Z.-H. Yan, {Analysis of the scalar, axialvector, vector, tensor
  doubly charmed tetraquark states with QCD sum rules}, Eur. Phys. J. C78~(1)
  (2018) 19.
\newblock \href {http://arxiv.org/abs/1710.02810} {\path{arXiv:1710.02810}},
  \href {http://dx.doi.org/10.1140/epjc/s10052-017-5507-0}
  {\path{doi:10.1140/epjc/s10052-017-5507-0}}.

\bibitem{Agaev:2018vag}
S.~S. Agaev, K.~Azizi, B.~Barsbay, H.~Sundu, {The doubly charmed pseudoscalar
  tetraquarks $T_{cc;\bar{s} \bar{s}}^{++}$ and $T_{cc;\bar{d} \bar{s}}^{++}$},
  Nucl. Phys. B939 (2019) 130--144.
\newblock \href {http://arxiv.org/abs/1806.04447} {\path{arXiv:1806.04447}},
  \href {http://dx.doi.org/10.1016/j.nuclphysb.2018.12.021}
  {\path{doi:10.1016/j.nuclphysb.2018.12.021}}.

\bibitem{Agaev:2018khe}
S.~S. Agaev, K.~Azizi, B.~Barsbay, H.~Sundu, {Weak decays of the axial-vector
  tetraquark $T_{bb;\bar{u} \bar{d}}^{-}$}, Phys. Rev. D99~(3) (2019) 033002.
\newblock \href {http://arxiv.org/abs/1809.07791} {\path{arXiv:1809.07791}},
  \href {http://dx.doi.org/10.1103/PhysRevD.99.033002}
  {\path{doi:10.1103/PhysRevD.99.033002}}.

\bibitem{Sundu:2019feu}
H.~Sundu, S.~S. Agaev, K.~Azizi, {Semileptonic decays of the scalar tetraquark
  $Z_{bc;\overline{u} \overline{d}}^{0}$}\href
  {http://arxiv.org/abs/1903.05931} {\path{arXiv:1903.05931}}.

\bibitem{Azizi:2016dhy}
K.~Azizi, Y.~Sarac, H.~Sundu, {Analysis of $P_c^+(4380)$ and $P_c^+(4450)$ as
  pentaquark states in the molecular picture with QCD sum rules}, Phys. Rev.
  D95~(9) (2017) 094016.
\newblock \href {http://arxiv.org/abs/1612.07479} {\path{arXiv:1612.07479}},
  \href {http://dx.doi.org/10.1103/PhysRevD.95.094016}
  {\path{doi:10.1103/PhysRevD.95.094016}}.

\bibitem{Azizi:2017bgs}
K.~Azizi, Y.~Sarac, H.~Sundu, {Hidden Bottom Pentaquark States with Spin 3/2
  and 5/2}, Phys. Rev. D96~(9) (2017) 094030.
\newblock \href {http://arxiv.org/abs/1707.01248} {\path{arXiv:1707.01248}},
  \href {http://dx.doi.org/10.1103/PhysRevD.96.094030}
  {\path{doi:10.1103/PhysRevD.96.094030}}.

\bibitem{Azizi:2018bdv}
K.~Azizi, Y.~Sarac, H.~Sundu, {Strong decay of $P_c(4380)$ pentaquark in a
  molecular picture}, Phys. Lett. B782 (2018) 694--701.
\newblock \href {http://arxiv.org/abs/1802.01384} {\path{arXiv:1802.01384}},
  \href {http://dx.doi.org/10.1016/j.physletb.2018.06.022}
  {\path{doi:10.1016/j.physletb.2018.06.022}}.

\bibitem{Ozdem:2018qeh}
U.~Ozdem, K.~Azizi, {Electromagnetic multipole moments of the $P_c^+(4380)$
  pentaquark in light-cone QCD}, Eur. Phys. J. C78~(5) (2018) 379.
\newblock \href {http://arxiv.org/abs/1803.06831} {\path{arXiv:1803.06831}},
  \href {http://dx.doi.org/10.1140/epjc/s10052-018-5873-2}
  {\path{doi:10.1140/epjc/s10052-018-5873-2}}.

\bibitem{Azizi:2018jky}
K.~Azizi, U.~{\"o}zdem, {The electromagnetic multipole moments of the possible
  charm-strange pentaquarks in light-cone QCD}, Eur. Phys. J. C78~(9) (2018)
  698.
\newblock \href {http://arxiv.org/abs/1807.06503} {\path{arXiv:1807.06503}},
  \href {http://dx.doi.org/10.1140/epjc/s10052-018-6187-0}
  {\path{doi:10.1140/epjc/s10052-018-6187-0}}.

\bibitem{Wang:2015epa}
Z.-G. Wang, {Analysis of $P_c(4380)$ and $P_c(4450)$ as pentaquark states in
  the diquark model with QCD sum rules}, Eur. Phys. J. C76~(2) (2016) 70.
\newblock \href {http://arxiv.org/abs/1508.01468} {\path{arXiv:1508.01468}},
  \href {http://dx.doi.org/10.1140/epjc/s10052-016-3920-4}
  {\path{doi:10.1140/epjc/s10052-016-3920-4}}.

\bibitem{Wang:2015ava}
Z.-G. Wang, T.~Huang, {Analysis of the ${\frac{1}{2}}^{\pm }$ pentaquark states
  in the diquark model with QCD sum rules}, Eur. Phys. J. C76~(1) (2016) 43.
\newblock \href {http://arxiv.org/abs/1508.04189} {\path{arXiv:1508.04189}},
  \href {http://dx.doi.org/10.1140/epjc/s10052-016-3880-8}
  {\path{doi:10.1140/epjc/s10052-016-3880-8}}.

\bibitem{Wang:2018waa}
Z.-G. Wang, {Analysis of the $\bar{D}\Sigma_c^*$, $\bar{D}^{*}\Sigma_c$ and
  $\bar{D}^{*}\Sigma_c^*$ pentaquark molecular states with QCD sum rules}\href
  {http://arxiv.org/abs/1806.10384} {\path{arXiv:1806.10384}}.

\bibitem{Wang:2018lhz}
Z.-G. Wang, {Analysis of the doubly heavy baryon states and pentaquark states
  with QCD sum rules}, Eur. Phys. J. C78~(10) (2018) 826.
\newblock \href {http://arxiv.org/abs/1808.09820} {\path{arXiv:1808.09820}},
  \href {http://dx.doi.org/10.1140/epjc/s10052-018-6300-4}
  {\path{doi:10.1140/epjc/s10052-018-6300-4}}.

\bibitem{MartinezTorres:2007sr}
A.~Martinez~Torres, K.~P. Khemchandani, E.~Oset, {Three body resonances in two
  meson-one baryon systems}, Phys. Rev. C77 (2008) 042203.
\newblock \href {http://arxiv.org/abs/0706.2330} {\path{arXiv:0706.2330}},
  \href {http://dx.doi.org/10.1103/PhysRevC.77.042203}
  {\path{doi:10.1103/PhysRevC.77.042203}}.

\bibitem{MartinezTorres:2008gy}
A.~Martinez~Torres, K.~P. Khemchandani, L.~S. Geng, M.~Napsuciale, E.~Oset,
  {The X(2175) as a resonant state of the phi K anti-K system}, Phys. Rev. D78
  (2008) 074031.
\newblock \href {http://arxiv.org/abs/0801.3635} {\path{arXiv:0801.3635}},
  \href {http://dx.doi.org/10.1103/PhysRevD.78.074031}
  {\path{doi:10.1103/PhysRevD.78.074031}}.

\bibitem{Epelbaum:2002vt}
E.~Epelbaum, A.~Nogga, W.~Gloeckle, H.~Kamada, U.~G. Meissner, H.~Witala,
  {Three nucleon forces from chiral effective field theory}, Phys. Rev. C66
  (2002) 064001.
\newblock \href {http://arxiv.org/abs/nucl-th/0208023}
  {\path{arXiv:nucl-th/0208023}}, \href
  {http://dx.doi.org/10.1103/PhysRevC.66.064001}
  {\path{doi:10.1103/PhysRevC.66.064001}}.

\bibitem{Epelbaum:2000mx}
E.~Epelbaum, H.~Kamada, A.~Nogga, H.~Witala, W.~Gloeckle, U.-G. Meissner, {The
  Three nucleon and four nucleon systems from chiral effective field theory},
  Phys. Rev. Lett. 86 (2001) 4787--4790.
\newblock \href {http://arxiv.org/abs/nucl-th/0007057}
  {\path{arXiv:nucl-th/0007057}}, \href
  {http://dx.doi.org/10.1103/PhysRevLett.86.4787}
  {\path{doi:10.1103/PhysRevLett.86.4787}}.

\bibitem{Bernard:2007sp}
V.~Bernard, E.~Epelbaum, H.~Krebs, U.-G. Meissner, {Subleading contributions to
  the chiral three-nucleon force. I. Long-range terms}, Phys. Rev. C77 (2008)
  064004.
\newblock \href {http://arxiv.org/abs/0712.1967} {\path{arXiv:0712.1967}},
  \href {http://dx.doi.org/10.1103/PhysRevC.77.064004}
  {\path{doi:10.1103/PhysRevC.77.064004}}.

\bibitem{Bernard:2011zr}
V.~Bernard, E.~Epelbaum, H.~Krebs, U.~G. Meissner, {Subleading contributions to
  the chiral three-nucleon force II: Short-range terms and relativistic
  corrections}, Phys. Rev. C84 (2011) 054001.
\newblock \href {http://arxiv.org/abs/1108.3816} {\path{arXiv:1108.3816}},
  \href {http://dx.doi.org/10.1103/PhysRevC.84.054001}
  {\path{doi:10.1103/PhysRevC.84.054001}}.

\bibitem{Petschauer:2015elq}
S.~Petschauer, N.~Kaiser, J.~Haidenbauer, U.-G. Meissner, W.~Weise, {Leading
  three-baryon forces from SU(3) chiral effective field theory}, Phys. Rev.
  C93~(1) (2016) 014001.
\newblock \href {http://arxiv.org/abs/1511.02095} {\path{arXiv:1511.02095}},
  \href {http://dx.doi.org/10.1103/PhysRevC.93.014001}
  {\path{doi:10.1103/PhysRevC.93.014001}}.

\bibitem{Meissner:2014dea}
U.-G. Meissner, G.~R¨ªos, A.~Rusetsky, {Spectrum of three-body bound states in
  a finite volume}, Phys. Rev. Lett. 114~(9) (2015) 091602, [Erratum: Phys.
  Rev. Lett.117,no.6,069902(2016)].
\newblock \href {http://arxiv.org/abs/1412.4969} {\path{arXiv:1412.4969}},
  \href {http://dx.doi.org/10.1103/PhysRevLett.117.069902,
  10.1103/PhysRevLett.114.091602} {\path{doi:10.1103/PhysRevLett.117.069902,
  10.1103/PhysRevLett.114.091602}}.

\bibitem{Mai:2017bge}
M.~Mai, M.~D{\"o}ring, {Three-body Unitarity in the Finite Volume}, Eur. Phys.
  J. A53~(12) (2017) 240.
\newblock \href {http://arxiv.org/abs/1709.08222} {\path{arXiv:1709.08222}},
  \href {http://dx.doi.org/10.1140/epja/i2017-12440-1}
  {\path{doi:10.1140/epja/i2017-12440-1}}.

\bibitem{Meng:2017jgx}
Y.~Meng, C.~Liu, U.-G. Meissner, A.~Rusetsky, {Three-particle bound states in a
  finite volume: unequal masses and higher partial waves}, Phys. Rev. D98~(1)
  (2018) 014508.
\newblock \href {http://arxiv.org/abs/1712.08464} {\path{arXiv:1712.08464}},
  \href {http://dx.doi.org/10.1103/PhysRevD.98.014508}
  {\path{doi:10.1103/PhysRevD.98.014508}}.

\bibitem{Doring:2018xxx}
M.~D{\"o}ring, H.~W. Hammer, M.~Mai, J.-Y. Pang, A.~Rusetsky, J.~Wu,
  {Three-body spectrum in a finite volume: the role of cubic symmetry}, Phys.
  Rev. D97~(11) (2018) 114508.
\newblock \href {http://arxiv.org/abs/1802.03362} {\path{arXiv:1802.03362}},
  \href {http://dx.doi.org/10.1103/PhysRevD.97.114508}
  {\path{doi:10.1103/PhysRevD.97.114508}}.

\bibitem{Mai:2018djl}
M.~Mai, M.~Doring, {Finite-Volume Spectrum of $\pi^+\pi^+$ and
  $\pi^+\pi^+\pi^+$ Systems}, Phys. Rev. Lett. 122~(6) (2019) 062503.
\newblock \href {http://arxiv.org/abs/1807.04746} {\path{arXiv:1807.04746}},
  \href {http://dx.doi.org/10.1103/PhysRevLett.122.062503}
  {\path{doi:10.1103/PhysRevLett.122.062503}}.

\bibitem{Briceno:2018mlh}
R.~A. Brice{\~n}o, M.~T. Hansen, S.~R. Sharpe, {Numerical study of the
  relativistic three-body quantization condition in the isotropic
  approximation}, Phys. Rev. D98~(1) (2018) 014506.
\newblock \href {http://arxiv.org/abs/1803.04169} {\path{arXiv:1803.04169}},
  \href {http://dx.doi.org/10.1103/PhysRevD.98.014506}
  {\path{doi:10.1103/PhysRevD.98.014506}}.

\bibitem{Briceno:2017tce}
R.~A. Brice{\~n}o, M.~T. Hansen, S.~R. Sharpe, {Relating the finite-volume
  spectrum and the two-and-three-particle $S$ matrix for relativistic systems
  of identical scalar particles}, Phys. Rev. D95~(7) (2017) 074510.
\newblock \href {http://arxiv.org/abs/1701.07465} {\path{arXiv:1701.07465}},
  \href {http://dx.doi.org/10.1103/PhysRevD.95.074510}
  {\path{doi:10.1103/PhysRevD.95.074510}}.

\bibitem{Briceno:2012rv}
R.~A. Briceno, Z.~Davoudi, {Three-particle scattering amplitudes from a finite
  volume formalism}, Phys. Rev. D87~(9) (2013) 094507.
\newblock \href {http://arxiv.org/abs/1212.3398} {\path{arXiv:1212.3398}},
  \href {http://dx.doi.org/10.1103/PhysRevD.87.094507}
  {\path{doi:10.1103/PhysRevD.87.094507}}.

\bibitem{Polejaeva:2012ut}
K.~Polejaeva, A.~Rusetsky, {Three particles in a finite volume}, Eur. Phys. J.
  A48 (2012) 67.
\newblock \href {http://arxiv.org/abs/1203.1241} {\path{arXiv:1203.1241}},
  \href {http://dx.doi.org/10.1140/epja/i2012-12067-8}
  {\path{doi:10.1140/epja/i2012-12067-8}}.

\bibitem{Guo:2017crd}
P.~Guo, V.~Gasparian, {Numerical approach for finite volume three-body
  interaction}, Phys. Rev. D97~(1) (2018) 014504.
\newblock \href {http://arxiv.org/abs/1709.08255} {\path{arXiv:1709.08255}},
  \href {http://dx.doi.org/10.1103/PhysRevD.97.014504}
  {\path{doi:10.1103/PhysRevD.97.014504}}.

\bibitem{Guo:2017ism}
P.~Guo, V.~Gasparian, {An solvable three-body model in finite volume}, Phys.
  Lett. B774 (2017) 441--445.
\newblock \href {http://arxiv.org/abs/1701.00438} {\path{arXiv:1701.00438}},
  \href {http://dx.doi.org/10.1016/j.physletb.2017.10.009}
  {\path{doi:10.1016/j.physletb.2017.10.009}}.

\bibitem{Kreuzer:2008bi}
S.~Kreuzer, H.~W. Hammer, {Efimov physics in a finite volume}, Phys. Lett. B673
  (2009) 260--263.
\newblock \href {http://arxiv.org/abs/0811.0159} {\path{arXiv:0811.0159}},
  \href {http://dx.doi.org/10.1016/j.physletb.2009.02.035}
  {\path{doi:10.1016/j.physletb.2009.02.035}}.

\bibitem{Jansen:2015lha}
M.~Jansen, H.~W. Hammer, Y.~Jia, {Finite volume corrections to the binding
  energy of the X(3872)}, Phys. Rev. D92~(11) (2015) 114031.
\newblock \href {http://arxiv.org/abs/1505.04099} {\path{arXiv:1505.04099}},
  \href {http://dx.doi.org/10.1103/PhysRevD.92.114031}
  {\path{doi:10.1103/PhysRevD.92.114031}}.

\bibitem{Bour:2012hn}
S.~Bour, H.~W. Hammer, D.~Lee, U.-G. Meissner, {Benchmark calculations for
  elastic fermion-dimer scattering}, Phys. Rev. C86 (2012) 034003.
\newblock \href {http://arxiv.org/abs/1206.1765} {\path{arXiv:1206.1765}},
  \href {http://dx.doi.org/10.1103/PhysRevC.86.034003}
  {\path{doi:10.1103/PhysRevC.86.034003}}.

\bibitem{Hammer:2017kms}
H.~W. Hammer, J.~Y. Pang, A.~Rusetsky, {Three particle quantization condition
  in a finite volume: 2. general formalism and the analysis of data}, JHEP 10
  (2017) 115.
\newblock \href {http://arxiv.org/abs/1707.02176} {\path{arXiv:1707.02176}},
  \href {http://dx.doi.org/10.1007/JHEP10(2017)115}
  {\path{doi:10.1007/JHEP10(2017)115}}.

\bibitem{Romero-Lopez:2018rcb}
F.~Romero-L{\'o}pez, A.~Rusetsky, C.~Urbach, {Two- and three-body interactions
  in $\varphi ^4$ theory from lattice simulations}, Eur. Phys. J. C78~(10)
  (2018) 846.
\newblock \href {http://arxiv.org/abs/1806.02367} {\path{arXiv:1806.02367}},
  \href {http://dx.doi.org/10.1140/epjc/s10052-018-6325-8}
  {\path{doi:10.1140/epjc/s10052-018-6325-8}}.

\bibitem{Pang:2019dfe}
J.-Y. Pang, J.-J. Wu, H.~W. Hammer, U.-G. Meissner, A.~Rusetsky, {Energy shift
  of the three-particle system in a finite volume}\href
  {http://arxiv.org/abs/1902.01111} {\path{arXiv:1902.01111}}.

\bibitem{Faddeev:1960su}
L.~D. Faddeev, {Scattering theory for a three particle system}, Sov. Phys. JETP
  12 (1961) 1014--1019, [Zh. Eksp. Teor. Fiz.39,1459(1960)].

\bibitem{Glockle}
W.~Gl{\"o}ckle, The Quantum Mechanical Few-Body Problem, Springer-Verlag, 1983.

\bibitem{Stadler:1991zz}
A.~Stadler, W.~Glockle, P.~U. Sauer, {Faddeev equations with three-nucleon
  force in momentum space}, Phys. Rev. C44 (1991) 2319--2327.
\newblock \href {http://dx.doi.org/10.1103/PhysRevC.44.2319}
  {\path{doi:10.1103/PhysRevC.44.2319}}.

\bibitem{Garcilazo:2016ylj}
H.~Garcilazo, A.~Valcarce, {Deeply bound $\Xi$ tribaryon}, Phys. Rev. C93~(3)
  (2016) 034001.
\newblock \href {http://arxiv.org/abs/1605.04108} {\path{arXiv:1605.04108}},
  \href {http://dx.doi.org/10.1103/PhysRevC.93.034001}
  {\path{doi:10.1103/PhysRevC.93.034001}}.

\bibitem{Garcilazo:2019igo}
H.~Garcilazo, A.~Valcarce, {$\Omega NN$ and $\Omega\Omega N$ states}, Phys.
  Rev. C99~(1) (2019) 014001.
\newblock \href {http://arxiv.org/abs/1901.05678} {\path{arXiv:1901.05678}},
  \href {http://dx.doi.org/10.1103/PhysRevC.99.014001}
  {\path{doi:10.1103/PhysRevC.99.014001}}.

\bibitem{Shevchenko:2007ke}
N.~V. Shevchenko, A.~Gal, J.~Mares, J.~Revai, {anti-K NN quasi-bound state and
  the anti-K N interaction: Coupled-channel Faddeev calculations of the anti-K
  NN - pi Sigma N system}, Phys. Rev. C76 (2007) 044004.
\newblock \href {http://arxiv.org/abs/0706.4393} {\path{arXiv:0706.4393}},
  \href {http://dx.doi.org/10.1103/PhysRevC.76.044004}
  {\path{doi:10.1103/PhysRevC.76.044004}}.

\bibitem{Oller:1997ti}
J.~A. Oller, E.~Oset, {Chiral symmetry amplitudes in the S wave isoscalar and
  isovector channels and the $\sigma$, f$_0$(980), a$_0$(980) scalar mesons},
  Nucl. Phys. A620 (1997) 438--456, [Erratum: Nucl. Phys.A652,407(1999)].
\newblock \href {http://arxiv.org/abs/hep-ph/9702314}
  {\path{arXiv:hep-ph/9702314}}, \href
  {http://dx.doi.org/10.1016/S0375-9474(99)00427-3,
  10.1016/S0375-9474(97)00160-7} {\path{doi:10.1016/S0375-9474(99)00427-3,
  10.1016/S0375-9474(97)00160-7}}.

\bibitem{Roca:2005nm}
L.~Roca, E.~Oset, J.~Singh, {Low lying axial-vector mesons as dynamically
  generated resonances}, Phys. Rev. D72 (2005) 014002.
\newblock \href {http://arxiv.org/abs/hep-ph/0503273}
  {\path{arXiv:hep-ph/0503273}}, \href
  {http://dx.doi.org/10.1103/PhysRevD.72.014002}
  {\path{doi:10.1103/PhysRevD.72.014002}}.

\bibitem{Khemchandani:2008rk}
K.~P. Khemchandani, A.~Martinez~Torres, E.~Oset, {The $N^*(1710)$ as a
  resonance in the $\pi \pi N$ system}, Eur. Phys. J. A37 (2008) 233--243.
\newblock \href {http://arxiv.org/abs/0804.4670} {\path{arXiv:0804.4670}},
  \href {http://dx.doi.org/10.1140/epja/i2008-10625-3}
  {\path{doi:10.1140/epja/i2008-10625-3}}.

\bibitem{MartinezTorres:2009xb}
A.~Martinez~Torres, K.~P. Khemchandani, D.~Gamermann, E.~Oset, {The $Y(4260)$
  as a $J/psi K \bar K$ system}, Phys. Rev. D80 (2009) 094012.
\newblock \href {http://arxiv.org/abs/0906.5333} {\path{arXiv:0906.5333}},
  \href {http://dx.doi.org/10.1103/PhysRevD.80.094012}
  {\path{doi:10.1103/PhysRevD.80.094012}}.

\bibitem{MartinezTorres:2018zbl}
A.~Martinez~Torres, K.~P. Khemchandani, L.-S. Geng, {Bound state formation in
  the $DDK$ system}\href {http://arxiv.org/abs/1809.01059}
  {\path{arXiv:1809.01059}}.

\bibitem{Gal:2011yp}
A.~Gal, H.~Garcilazo, {Coupled channel Faddeev calculations of a (Kbar-N-pi)
  quasibound state}, Nucl. Phys. A864 (2011) 153--166.
\newblock \href {http://arxiv.org/abs/1103.4757} {\path{arXiv:1103.4757}},
  \href {http://dx.doi.org/10.1016/j.nuclphysa.2011.06.022}
  {\path{doi:10.1016/j.nuclphysa.2011.06.022}}.

\bibitem{Valderrama:2018knt}
M.~P. Valderrama, {Three-Body $J^P = 0^+,1^+,2^+$ $B^* B^* \bar{K}$ Bound
  States}, Phys. Rev. D98~(1) (2018) 014022.
\newblock \href {http://arxiv.org/abs/1805.05100} {\path{arXiv:1805.05100}},
  \href {http://dx.doi.org/10.1103/PhysRevD.98.014022}
  {\path{doi:10.1103/PhysRevD.98.014022}}.

\bibitem{Garcilazo:2018rwu}
H.~Garcilazo, A.~Valcarce, {$T_{bbb}$: a three $B-$meson bound state}, Phys.
  Lett. B784 (2018) 169--172.
\newblock \href {http://arxiv.org/abs/1808.00226} {\path{arXiv:1808.00226}},
  \href {http://dx.doi.org/10.1016/j.physletb.2018.07.055}
  {\path{doi:10.1016/j.physletb.2018.07.055}}.

\bibitem{Chand:1962ec}
R.~Chand, R.~H. Dalitz, {Charge-independence in K- -deuterium capture
  reactions}, Annals Phys. 20 (1962) 1--19.
\newblock \href {http://dx.doi.org/10.1016/0003-4916(62)90113-6}
  {\path{doi:10.1016/0003-4916(62)90113-6}}.

\bibitem{Barrett:1999cw}
R.~C. Barrett, A.~Deloff, {Strong interaction effects in kaonic deuterium},
  Phys. Rev. C60 (1999) 025201.
\newblock \href {http://dx.doi.org/10.1103/PhysRevC.60.025201}
  {\path{doi:10.1103/PhysRevC.60.025201}}.

\bibitem{Deloff:1999gc}
A.~Deloff, {Eta d and K- d zero energy scattering: A Faddeev approach}, Phys.
  Rev. C61 (2000) 024004.
\newblock \href {http://dx.doi.org/10.1103/PhysRevC.61.024004}
  {\path{doi:10.1103/PhysRevC.61.024004}}.

\bibitem{Kamalov:2000iy}
S.~S. Kamalov, E.~Oset, A.~Ramos, {Chiral unitary approach to the K- deuteron
  scattering length}, Nucl. Phys. A690 (2001) 494--508.
\newblock \href {http://arxiv.org/abs/nucl-th/0010054}
  {\path{arXiv:nucl-th/0010054}}, \href
  {http://dx.doi.org/10.1016/S0375-9474(00)00709-0}
  {\path{doi:10.1016/S0375-9474(00)00709-0}}.

\bibitem{Roca:2010tf}
L.~Roca, E.~Oset, {A description of the f2(1270), rho3(1690), f4(2050),
  rho5(2350) and f6(2510) resonances as multi-rho(770) states}, Phys. Rev. D82
  (2010) 054013.
\newblock \href {http://arxiv.org/abs/1005.0283} {\path{arXiv:1005.0283}},
  \href {http://dx.doi.org/10.1103/PhysRevD.82.054013}
  {\path{doi:10.1103/PhysRevD.82.054013}}.

\bibitem{Sun:2011fr}
B.-X. Sun, H.-X. Chen, E.~Oset, {N-rho-rho and Delta-rho-rho molecules with
  $J^P=5/2^{+}$ and $J^P=7/2^{+}$}, Eur. Phys. J. A47 (2011) 127.
\newblock \href {http://arxiv.org/abs/1107.0209} {\path{arXiv:1107.0209}},
  \href {http://dx.doi.org/10.1140/epja/i2011-11127-y}
  {\path{doi:10.1140/epja/i2011-11127-y}}.

\bibitem{Oset:2009vf}
E.~Oset, A.~Ramos, {Dynamically generated resonances from the vector
  octet-baryon octet interaction}, Eur. Phys. J. A44 (2010) 445--454.
\newblock \href {http://arxiv.org/abs/0905.0973} {\path{arXiv:0905.0973}},
  \href {http://dx.doi.org/10.1140/epja/i2010-10957-3}
  {\path{doi:10.1140/epja/i2010-10957-3}}.

\bibitem{Cutkosky:1979fy}
R.~E. Cutkosky, C.~P. Forsyth, R.~E. Hendrick, R.~L. Kelly, {Pion - Nucleon
  Partial Wave Amplitudes}, Phys. Rev. D20 (1979) 2839.
\newblock \href {http://dx.doi.org/10.1103/PhysRevD.20.2839}
  {\path{doi:10.1103/PhysRevD.20.2839}}.

\bibitem{YamagataSekihara:2010qk}
J.~Yamagata-Sekihara, L.~Roca, E.~Oset, {On the nature of the $K^*_2(1430)$,
  $K^*_3(1780)$, $K^*_4(2045)$, $K^*_5(2380)$ and $K^*6$ as $K^*$ -
  multi-$\rho$ states}, Phys. Rev. D82 (2010) 094017, [Erratum: Phys.
  Rev.D85,119905(2012)].
\newblock \href {http://arxiv.org/abs/1010.0525} {\path{arXiv:1010.0525}},
  \href {http://dx.doi.org/10.1103/PhysRevD.82.094017,
  10.1103/PhysRevD.85.119905} {\path{doi:10.1103/PhysRevD.82.094017,
  10.1103/PhysRevD.85.119905}}.

\bibitem{Xiao:2012dw}
C.~W. Xiao, M.~Bayar, E.~Oset, {A prediction of $D^*$-multi-$\rho$ states},
  Phys. Rev. D86 (2012) 094019.
\newblock \href {http://arxiv.org/abs/1207.4030} {\path{arXiv:1207.4030}},
  \href {http://dx.doi.org/10.1103/PhysRevD.86.094019}
  {\path{doi:10.1103/PhysRevD.86.094019}}.

\bibitem{Xie:2010ig}
J.-J. Xie, A.~Martinez~Torres, E.~Oset, {Faddeev fixed center approximation to
  the $N\bar{K}K$ system and the signature of a $N^*(1920)(1/2^+)$ state},
  Phys. Rev. C83 (2011) 065207.
\newblock \href {http://arxiv.org/abs/1010.6164} {\path{arXiv:1010.6164}},
  \href {http://dx.doi.org/10.1103/PhysRevC.83.065207}
  {\path{doi:10.1103/PhysRevC.83.065207}}.

\bibitem{Liang:2013yta}
W.~Liang, C.~W. Xiao, E.~Oset, {Study of $¦ÇK\overline{K}$ and
  $¦Ç¡äK\overline{K}$ with the fixed center approximation to Faddeev
  equations}, Phys. Rev. D88~(11) (2013) 114024.
\newblock \href {http://arxiv.org/abs/1309.7310} {\path{arXiv:1309.7310}},
  \href {http://dx.doi.org/10.1103/PhysRevD.88.114024}
  {\path{doi:10.1103/PhysRevD.88.114024}}.

\bibitem{Debastiani:2017vhv}
V.~R. Debastiani, J.~M. Dias, E.~Oset, {Study of the $DKK$ and $DK\bar{K}$
  systems}, Phys. Rev. D96~(1) (2017) 016014.
\newblock \href {http://arxiv.org/abs/1705.09257} {\path{arXiv:1705.09257}},
  \href {http://dx.doi.org/10.1103/PhysRevD.96.016014}
  {\path{doi:10.1103/PhysRevD.96.016014}}.

\bibitem{Bayar:2011qj}
M.~Bayar, J.~Yamagata-Sekihara, E.~Oset, {The $\bar{K}NN$ system with chiral
  dynamics}, Phys. Rev. C84 (2011) 015209.
\newblock \href {http://arxiv.org/abs/1102.2854} {\path{arXiv:1102.2854}},
  \href {http://dx.doi.org/10.1103/PhysRevC.84.015209}
  {\path{doi:10.1103/PhysRevC.84.015209}}.

\bibitem{Dias:2017miz}
J.~M. Dias, V.~R. Debastiani, L.~Roca, S.~Sakai, E.~Oset, {On the binding of
  the $BD\bar{D}$ and $BDD$ systems}, Phys. Rev. D96~(9) (2017) 094007.
\newblock \href {http://arxiv.org/abs/1709.01372} {\path{arXiv:1709.01372}},
  \href {http://dx.doi.org/10.1103/PhysRevD.96.094007}
  {\path{doi:10.1103/PhysRevD.96.094007}}.

\bibitem{Ren:2018pcd}
X.-L. Ren, B.~B. Malabarba, L.-S. Geng, K.~P. Khemchandani,
  A.~Mart¨ªnez~Torres, {$K^*$ mesons with hidden charm arising from $KX(3872)$
  and $KZ_c(3900)$ dynamics}, Phys. Lett. B785 (2018) 112--117.
\newblock \href {http://arxiv.org/abs/1805.08330} {\path{arXiv:1805.08330}},
  \href {http://dx.doi.org/10.1016/j.physletb.2018.08.034}
  {\path{doi:10.1016/j.physletb.2018.08.034}}.

\bibitem{Skyrme:1961vq}
T.~H.~R. Skyrme, {A Nonlinear field theory}, Proc. Roy. Soc. Lond. A260 (1961)
  127--138.
\newblock \href {http://dx.doi.org/10.1098/rspa.1961.0018}
  {\path{doi:10.1098/rspa.1961.0018}}.

\bibitem{Witten:1979kh}
E.~Witten, {Baryons in the 1/n Expansion}, Nucl. Phys. B160 (1979) 57--115.
\newblock \href {http://dx.doi.org/10.1016/0550-3213(79)90232-3}
  {\path{doi:10.1016/0550-3213(79)90232-3}}.

\bibitem{Adkins:1983ya}
G.~S. Adkins, C.~R. Nappi, E.~Witten, {Static Properties of Nucleons in the
  Skyrme Model}, Nucl. Phys. B228 (1983) 552.
\newblock \href {http://dx.doi.org/10.1016/0550-3213(83)90559-X}
  {\path{doi:10.1016/0550-3213(83)90559-X}}.

\bibitem{Brodsky:1988ip}
S.~J. Brodsky, J.~R. Ellis, M.~Karliner, {Chiral Symmetry and the Spin of the
  Proton}, Phys. Lett. B206 (1988) 309--315.
\newblock \href {http://dx.doi.org/10.1016/0370-2693(88)91511-0}
  {\path{doi:10.1016/0370-2693(88)91511-0}}.

\bibitem{Weigel:1995cz}
H.~Weigel, {Baryons as three flavor solitons}, Int. J. Mod. Phys. A11 (1996)
  2419--2544.
\newblock \href {http://arxiv.org/abs/hep-ph/9509398}
  {\path{arXiv:hep-ph/9509398}}, \href
  {http://dx.doi.org/10.1142/S0217751X96001218}
  {\path{doi:10.1142/S0217751X96001218}}.

\bibitem{Diakonov:1986yh}
D.~Diakonov, V.~{\relax Yu}. Petrov, {Chiral Theory of Nucleons}, JETP Lett. 43
  (1986) 75--77, [Pisma Zh. Eksp. Teor. Fiz.43,57(1986)].

\bibitem{Reinhardt:1989fv}
H.~Reinhardt, R.~Wunsch, {Topological Solitons of the {Nambu-Jona-Lasinio}
  Model}, Phys. Lett. B230 (1989) 93--98.
\newblock \href {http://dx.doi.org/10.1016/0370-2693(89)91659-6}
  {\path{doi:10.1016/0370-2693(89)91659-6}}.

\bibitem{Meissner:1989kq}
T.~Meissner, F.~Grummer, K.~Goeke, {Solitons in the {Nambu-Jona-Lasinio}
  Model}, Phys. Lett. B227 (1989) 296--300.
\newblock \href {http://dx.doi.org/10.1016/0370-2693(89)90932-5}
  {\path{doi:10.1016/0370-2693(89)90932-5}}.

\bibitem{Wakamatsu:1990ud}
M.~Wakamatsu, H.~Yoshiki, {A Chiral quark model of the nucleon}, Nucl. Phys.
  A524 (1991) 561--600.
\newblock \href {http://dx.doi.org/10.1016/0375-9474(91)90263-6}
  {\path{doi:10.1016/0375-9474(91)90263-6}}.

\bibitem{Diakonov:1997sj}
D.~Diakonov, {Chiral quark - soliton model}, in: {Nonperturbative quantum field
  physics. Proceedings, Advanced School, Peniscola, Spain, June 2-6, 1997},
  1997, pp. 1--55.
\newblock \href {http://arxiv.org/abs/hep-ph/9802298}
  {\path{arXiv:hep-ph/9802298}}.

\bibitem{Weigel:2008zz}
H.~Weigel, {Chiral Soliton Models for Baryons}, Lect. Notes Phys. 743 (2008)
  1--274.

\bibitem{Ma:2016npf}
Y.-L. Ma, M.~Harada, {Lecture notes on the Skyrme model}\href
  {http://arxiv.org/abs/1604.04850} {\path{arXiv:1604.04850}}.

\bibitem{Witten:1983tw}
E.~Witten, {Global Aspects of Current Algebra}, Nucl. Phys. B223 (1983)
  422--432.
\newblock \href {http://dx.doi.org/10.1016/0550-3213(83)90063-9}
  {\path{doi:10.1016/0550-3213(83)90063-9}}.

\bibitem{Wess:1971yu}
J.~Wess, B.~Zumino, {Consequences of anomalous Ward identities}, Phys. Lett.
  37B (1971) 95--97.
\newblock \href {http://dx.doi.org/10.1016/0370-2693(71)90582-X}
  {\path{doi:10.1016/0370-2693(71)90582-X}}.

\bibitem{Manohar:1984ys}
A.~V. Manohar, {Equivalence of the Chiral Soliton and Quark Models in Large N},
  Nucl. Phys. B248 (1984) 19, [,19(1984)].
\newblock \href {http://dx.doi.org/10.1016/0550-3213(84)90584-4}
  {\path{doi:10.1016/0550-3213(84)90584-4}}.

\bibitem{Chemtob:1985ar}
M.~Chemtob, {Skyrme Model of Baryon Octet and Decuplet}, Nucl. Phys. B256
  (1985) 600--608.
\newblock \href {http://dx.doi.org/10.1016/0550-3213(85)90409-2}
  {\path{doi:10.1016/0550-3213(85)90409-2}}.

\bibitem{Mazur:1984yf}
P.~O. Mazur, M.~A. Nowak, M.~Praszalowicz, {SU(3) Extension of the Skyrme
  Model}, Phys. Lett. 147B (1984) 137--140.
\newblock \href {http://dx.doi.org/10.1016/0370-2693(84)90608-7}
  {\path{doi:10.1016/0370-2693(84)90608-7}}.

\bibitem{Jain:1984gp}
S.~Jain, S.~R. Wadia, {Large $N$ Baryons: Collective Coordinates of the
  Topological Soliton in SU(3) Chiral Model}, Nucl. Phys. B258 (1985) 713,
  [,713(1984)].
\newblock \href {http://dx.doi.org/10.1016/0550-3213(85)90632-7}
  {\path{doi:10.1016/0550-3213(85)90632-7}}.

\bibitem{Praszalowicz}
M.~Praszalowicz, in: {Proceedings of the Workshop on Skyrmions and Anomalies,
  World Scientific, Singapore}, 1987, p. 112.

\bibitem{Praszalowicz:2003ik}
M.~Praszalowicz, {Pentaquark in the Skyrme model}, Phys. Lett. B575 (2003)
  234--241.
\newblock \href {http://arxiv.org/abs/hep-ph/0308114}
  {\path{arXiv:hep-ph/0308114}}, \href
  {http://dx.doi.org/10.1016/j.physletb.2003.09.049}
  {\path{doi:10.1016/j.physletb.2003.09.049}}.

\bibitem{Schechter:1999hg}
J.~Schechter, H.~Weigel, {The Skyrme model for baryons} (1999) 337--369\href
  {http://arxiv.org/abs/hep-ph/9907554} {\path{arXiv:hep-ph/9907554}}.

\bibitem{Wuest:1987rc}
E.~Wuest, G.~E. Brown, A.~D. Jackson, {Topological Chiral Bags in a Baryonic
  Environment}, Nucl. Phys. A468 (1987) 450--472.
\newblock \href {http://dx.doi.org/10.1016/0375-9474(87)90178-3}
  {\path{doi:10.1016/0375-9474(87)90178-3}}.

\bibitem{Klebanov:1985qi}
I.~R. Klebanov, {Nuclear Matter in the Skyrme Model}, Nucl. Phys. B262 (1985)
  133--143.
\newblock \href {http://dx.doi.org/10.1016/0550-3213(85)90068-9}
  {\path{doi:10.1016/0550-3213(85)90068-9}}.

\bibitem{Lee:2003aq}
H.-J. Lee, B.-Y. Park, D.-P. Min, M.~Rho, V.~Vento, {A Unified approach to high
  density: Pion fluctuations in skyrmion matter}, Nucl. Phys. A723 (2003)
  427--446.
\newblock \href {http://arxiv.org/abs/hep-ph/0302019}
  {\path{arXiv:hep-ph/0302019}}, \href
  {http://dx.doi.org/10.1016/S0375-9474(03)01452-0}
  {\path{doi:10.1016/S0375-9474(03)01452-0}}.

\bibitem{Ma:2012kb}
Y.-L. Ma, Y.~Oh, G.-S. Yang, M.~Harada, H.~K. Lee, B.-Y. Park, M.~Rho, {Hidden
  Local Symmetry and Infinite Tower of Vector Mesons for Baryons}, Phys. Rev.
  D86 (2012) 074025.
\newblock \href {http://arxiv.org/abs/1206.5460} {\path{arXiv:1206.5460}},
  \href {http://dx.doi.org/10.1103/PhysRevD.86.074025}
  {\path{doi:10.1103/PhysRevD.86.074025}}.

\bibitem{Ma:2012zm}
Y.-L. Ma, G.-S. Yang, Y.~Oh, M.~Harada, {Skyrmions with vector mesons in the
  hidden local symmetry approach}, Phys. Rev. D87~(3) (2013) 034023.
\newblock \href {http://arxiv.org/abs/1209.3554} {\path{arXiv:1209.3554}},
  \href {http://dx.doi.org/10.1103/PhysRevD.87.034023}
  {\path{doi:10.1103/PhysRevD.87.034023}}.

\bibitem{Ma:2013ela}
Y.-L. Ma, M.~Harada, H.~K. Lee, Y.~Oh, B.-Y. Park, M.~Rho, {Dense baryonic
  matter in conformally-compensated hidden local symmetry: Vector manifestation
  and chiral symmetry restoration}, Phys. Rev. D90~(3) (2014) 034015.
\newblock \href {http://arxiv.org/abs/1308.6476} {\path{arXiv:1308.6476}},
  \href {http://dx.doi.org/10.1103/PhysRevD.90.034015}
  {\path{doi:10.1103/PhysRevD.90.034015}}.

\bibitem{Ma:2013ooa}
Y.-L. Ma, M.~Harada, H.~K. Lee, Y.~Oh, B.-Y. Park, M.~Rho, {Dense baryonic
  matter in the hidden local symmetry approach: Half-skyrmions and nucleon
  mass}, Phys. Rev. D88~(1) (2013) 014016, [Erratum: Phys.
  Rev.D88,no.7,079904(2013)].
\newblock \href {http://arxiv.org/abs/1304.5638} {\path{arXiv:1304.5638}},
  \href {http://dx.doi.org/10.1103/PhysRevD.88.014016,
  10.1103/PhysRevD.88.079904} {\path{doi:10.1103/PhysRevD.88.014016,
  10.1103/PhysRevD.88.079904}}.

\bibitem{Callan:1985hy}
C.~G. Callan, Jr., I.~R. Klebanov, {Bound State Approach to Strangeness in the
  Skyrme Model}, Nucl. Phys. B262 (1985) 365--382.
\newblock \href {http://dx.doi.org/10.1016/0550-3213(85)90292-5}
  {\path{doi:10.1016/0550-3213(85)90292-5}}.

\bibitem{Nowak:1992um}
M.~A. Nowak, M.~Rho, I.~Zahed, {Chiral effective action with heavy quark
  symmetry}, Phys. Rev. D48 (1993) 4370--4374.
\newblock \href {http://arxiv.org/abs/hep-ph/9209272}
  {\path{arXiv:hep-ph/9209272}}, \href
  {http://dx.doi.org/10.1103/PhysRevD.48.4370}
  {\path{doi:10.1103/PhysRevD.48.4370}}.

\bibitem{Bardeen:1993ae}
W.~A. Bardeen, C.~T. Hill, {Chiral dynamics and heavy quark symmetry in a
  solvable toy field theoretic model}, Phys. Rev. D49 (1994) 409--425.
\newblock \href {http://arxiv.org/abs/hep-ph/9304265}
  {\path{arXiv:hep-ph/9304265}}, \href
  {http://dx.doi.org/10.1103/PhysRevD.49.409}
  {\path{doi:10.1103/PhysRevD.49.409}}.

\bibitem{Nowak:2004jg}
M.~A. Nowak, M.~Praszalowicz, M.~Sadzikowski, J.~Wasiluk, {Chiral doublers of
  heavy light baryons}, Phys. Rev. D70 (2004) 031503.
\newblock \href {http://arxiv.org/abs/hep-ph/0403184}
  {\path{arXiv:hep-ph/0403184}}, \href
  {http://dx.doi.org/10.1103/PhysRevD.70.031503}
  {\path{doi:10.1103/PhysRevD.70.031503}}.

\bibitem{Harada:2012dm}
M.~Harada, Y.-L. Ma, {Chiral partner structure of heavy baryons from the bound
  state approach with hidden local symmetry}, Phys. Rev. D87~(5) (2013) 056007.
\newblock \href {http://arxiv.org/abs/1212.5079} {\path{arXiv:1212.5079}},
  \href {http://dx.doi.org/10.1103/PhysRevD.87.056007}
  {\path{doi:10.1103/PhysRevD.87.056007}}.

\bibitem{Diakonov:1987ty}
D.~Diakonov, V.~{\relax Yu}. Petrov, P.~V. Pobylitsa, {A Chiral Theory of
  Nucleons}, Nucl. Phys. B306 (1988) 809.
\newblock \href {http://dx.doi.org/10.1016/0550-3213(88)90443-9}
  {\path{doi:10.1016/0550-3213(88)90443-9}}.

\bibitem{Christov:1995vm}
C.~Christov, A.~Blotz, H.-C. Kim, P.~Pobylitsa, T.~Watabe, T.~Meissner,
  E.~Ruiz~Arriola, K.~Goeke, {Baryons as nontopological chiral solitons}, Prog.
  Part. Nucl. Phys. 37 (1996) 91--191.
\newblock \href {http://arxiv.org/abs/hep-ph/9604441}
  {\path{arXiv:hep-ph/9604441}}, \href
  {http://dx.doi.org/10.1016/0146-6410(96)00057-9}
  {\path{doi:10.1016/0146-6410(96)00057-9}}.

\bibitem{Goeke:2004ht}
K.~Goeke, H.-C. Kim, M.~Praszalowicz, G.-S. Yang, {Pentaquarks: Review on
  models and solitonic calculations of antidecuplet magnetic moments}, Prog.
  Part. Nucl. Phys. 55 (2005) 350--373.
\newblock \href {http://arxiv.org/abs/hep-ph/0411195}
  {\path{arXiv:hep-ph/0411195}}, \href
  {http://dx.doi.org/10.1016/j.ppnp.2005.01.028}
  {\path{doi:10.1016/j.ppnp.2005.01.028}}.

\bibitem{Blotz:1992br}
A.~Blotz, K.~Goke, N.~W. Park, D.~Diakonov, V.~Petrov, P.~V. Pobylitsa,
  {Strange baryons in the solitonic sector of the Nambu-Jona-Lasinio model},
  Phys. Lett. B287 (1992) 29--34.
\newblock \href {http://dx.doi.org/10.1016/0370-2693(92)91871-6}
  {\path{doi:10.1016/0370-2693(92)91871-6}}.

\bibitem{Blotz:1992pw}
A.~Blotz, D.~Diakonov, K.~Goeke, N.~W. Park, V.~Petrov, P.~V. Pobylitsa, {The
  SU(3) Nambu-Jona-Lasinio soliton in the collective quantization formulation},
  Nucl. Phys. A555 (1993) 765--792.
\newblock \href {http://dx.doi.org/10.1016/0375-9474(93)90505-R}
  {\path{doi:10.1016/0375-9474(93)90505-R}}.

\bibitem{Yang:2018gju}
G.-S. Yang, H.-C. Kim, {New narrow nucleon resonances $N^{\ast}(1685)$ and
  $N^{\ast}(1726)$ within the chiral quark-soliton model}\href
  {http://arxiv.org/abs/1809.07489} {\path{arXiv:1809.07489}}.

\bibitem{Jaffe:2004qj}
R.~L. Jaffe, {The width of the Theta+ exotic baryon in the chiral soliton
  model}, Eur. Phys. J. C35 (2004) 221--222.
\newblock \href {http://arxiv.org/abs/hep-ph/0401187}
  {\path{arXiv:hep-ph/0401187}}, \href
  {http://dx.doi.org/10.1140/epjc/s2004-01815-4}
  {\path{doi:10.1140/epjc/s2004-01815-4}}.

\bibitem{Yang:2010fm}
G.-S. Yang, H.-C. Kim, {Mass splittings of SU(3) baryons within a chiral
  soliton model}, Prog. Theor. Phys. 128 (2012) 397--413.
\newblock \href {http://arxiv.org/abs/1010.3792} {\path{arXiv:1010.3792}},
  \href {http://dx.doi.org/10.1143/PTP.128.397}
  {\path{doi:10.1143/PTP.128.397}}.

\bibitem{Yang:2011qe}
G.-S. Yang, H.-C. Kim, {Mass splittings of the baryon decuplet and antidecuplet
  with the second-order flavor symmetry breakings within a chiral soliton
  model}, J. Korean Phys. Soc. 61 (2012) 1956--1964.
\newblock \href {http://arxiv.org/abs/1102.1786} {\path{arXiv:1102.1786}},
  \href {http://dx.doi.org/10.3938/jkps.61.1956}
  {\path{doi:10.3938/jkps.61.1956}}.

\bibitem{Kuznetsov:2006kt}
V.~Kuznetsov, et~al., {Evidence for a narrow structure at W~1.68-GeV in eta
  photoproduction on the neutron}, Phys. Lett. B647 (2007) 23--29.
\newblock \href {http://arxiv.org/abs/hep-ex/0606065}
  {\path{arXiv:hep-ex/0606065}}, \href
  {http://dx.doi.org/10.1016/j.physletb.2007.01.041}
  {\path{doi:10.1016/j.physletb.2007.01.041}}.

\bibitem{Praszalowicz:2007zza}
M.~Praszalowicz, K.~Goeke, {Exotic eikosiheptaplet in the chiral quark soliton
  model}, Phys. Rev. D76 (2007) 096003.
\newblock \href {http://dx.doi.org/10.1103/PhysRevD.76.096003}
  {\path{doi:10.1103/PhysRevD.76.096003}}.

\bibitem{Luscher:1990ck}
M.~Luscher, U.~Wolff, {How to Calculate the Elastic Scattering Matrix in
  Two-dimensional Quantum Field Theories by Numerical Simulation}, Nucl. Phys.
  B339 (1990) 222--252.
\newblock \href {http://dx.doi.org/10.1016/0550-3213(90)90540-T}
  {\path{doi:10.1016/0550-3213(90)90540-T}}.

\bibitem{Blossier:2009kd}
B.~Blossier, M.~Della~Morte, G.~von Hippel, T.~Mendes, R.~Sommer, {On the
  generalized eigenvalue method for energies and matrix elements in lattice
  field theory}, JHEP 04 (2009) 094.
\newblock \href {http://arxiv.org/abs/0902.1265} {\path{arXiv:0902.1265}},
  \href {http://dx.doi.org/10.1088/1126-6708/2009/04/094}
  {\path{doi:10.1088/1126-6708/2009/04/094}}.

\bibitem{Bulava:2009jb}
J.~M. Bulava, et~al., {Excited State Nucleon Spectrum with Two Flavors of
  Dynamical Fermions}, Phys. Rev. D79 (2009) 034505.
\newblock \href {http://arxiv.org/abs/0901.0027} {\path{arXiv:0901.0027}},
  \href {http://dx.doi.org/10.1103/PhysRevD.79.034505}
  {\path{doi:10.1103/PhysRevD.79.034505}}.

\bibitem{Dudek:2007wv}
J.~J. Dudek, R.~G. Edwards, N.~Mathur, D.~G. Richards, {Charmonium excited
  state spectrum in lattice QCD}, Phys. Rev. D77 (2008) 034501.
\newblock \href {http://arxiv.org/abs/0707.4162} {\path{arXiv:0707.4162}},
  \href {http://dx.doi.org/10.1103/PhysRevD.77.034501}
  {\path{doi:10.1103/PhysRevD.77.034501}}.

\bibitem{Mahbub:2013ala}
M.~S. Mahbub, W.~Kamleh, D.~B. Leinweber, P.~J. Moran, A.~G. Williams,
  {Structure and Flow of the Nucleon Eigenstates in Lattice QCD}, Phys. Rev.
  D87~(9) (2013) 094506.
\newblock \href {http://arxiv.org/abs/1302.2987} {\path{arXiv:1302.2987}},
  \href {http://dx.doi.org/10.1103/PhysRevD.87.094506}
  {\path{doi:10.1103/PhysRevD.87.094506}}.

\bibitem{Kiratidis:2016hda}
A.~L. Kiratidis, W.~Kamleh, D.~B. Leinweber, Z.-W. Liu, F.~M. Stokes, A.~W.
  Thomas, {Search for low-lying lattice QCD eigenstates in the Roper regime},
  Phys. Rev. D95~(7) (2017) 074507.
\newblock \href {http://arxiv.org/abs/1608.03051} {\path{arXiv:1608.03051}},
  \href {http://dx.doi.org/10.1103/PhysRevD.95.074507}
  {\path{doi:10.1103/PhysRevD.95.074507}}.

\bibitem{Kiratidis:2015vpa}
A.~L. Kiratidis, W.~Kamleh, D.~B. Leinweber, B.~J. Owen, {Lattice baryon
  spectroscopy with multi-particle interpolators}, Phys. Rev. D91 (2015)
  094509.
\newblock \href {http://arxiv.org/abs/1501.07667} {\path{arXiv:1501.07667}},
  \href {http://dx.doi.org/10.1103/PhysRevD.91.094509}
  {\path{doi:10.1103/PhysRevD.91.094509}}.

\bibitem{Luscher:1986pf}
M.~Luscher, {Volume Dependence of the Energy Spectrum in Massive Quantum Field
  Theories. 2. Scattering States}, Commun. Math. Phys. 105 (1986) 153--188.
\newblock \href {http://dx.doi.org/10.1007/BF01211097}
  {\path{doi:10.1007/BF01211097}}.

\bibitem{Luscher:1985dn}
M.~Luscher, {Volume Dependence of the Energy Spectrum in Massive Quantum Field
  Theories. 1. Stable Particle States}, Commun. Math. Phys. 104 (1986) 177.
\newblock \href {http://dx.doi.org/10.1007/BF01211589}
  {\path{doi:10.1007/BF01211589}}.

\bibitem{Luscher:1991cf}
M.~Luscher, {Signatures of unstable particles in finite volume}, Nucl. Phys.
  B364 (1991) 237--251.
\newblock \href {http://dx.doi.org/10.1016/0550-3213(91)90584-K}
  {\path{doi:10.1016/0550-3213(91)90584-K}}.

\bibitem{Luscher:1990ux}
M.~Luscher, {Two particle states on a torus and their relation to the
  scattering matrix}, Nucl. Phys. B354 (1991) 531--578.
\newblock \href {http://dx.doi.org/10.1016/0550-3213(91)90366-6}
  {\path{doi:10.1016/0550-3213(91)90366-6}}.

\bibitem{Briceno:2017max}
R.~A. Briceno, J.~J. Dudek, R.~D. Young, {Scattering processes and resonances
  from lattice QCD}, Rev. Mod. Phys. 90~(2) (2018) 025001.
\newblock \href {http://arxiv.org/abs/1706.06223} {\path{arXiv:1706.06223}},
  \href {http://dx.doi.org/10.1103/RevModPhys.90.025001}
  {\path{doi:10.1103/RevModPhys.90.025001}}.

\bibitem{Doring:2011vk}
M.~Doring, U.-G. Meissner, E.~Oset, A.~Rusetsky, {Unitarized Chiral
  Perturbation Theory in a finite volume: Scalar meson sector}, Eur. Phys. J.
  A47 (2011) 139.
\newblock \href {http://arxiv.org/abs/1107.3988} {\path{arXiv:1107.3988}},
  \href {http://dx.doi.org/10.1140/epja/i2011-11139-7}
  {\path{doi:10.1140/epja/i2011-11139-7}}.

\bibitem{Doring:2011ip}
M.~Doring, J.~Haidenbauer, U.-G. Meissner, A.~Rusetsky, {Dynamical
  coupled-channel approaches on a momentum lattice}, Eur. Phys. J. A47 (2011)
  163.
\newblock \href {http://arxiv.org/abs/1108.0676} {\path{arXiv:1108.0676}},
  \href {http://dx.doi.org/10.1140/epja/i2011-11163-7}
  {\path{doi:10.1140/epja/i2011-11163-7}}.

\bibitem{Chiu:2005ey}
T.-W. Chiu, T.-H. Hsieh, {Y(4260) on the lattice}, Phys. Rev. D73 (2006)
  094510.
\newblock \href {http://arxiv.org/abs/hep-lat/0512029}
  {\path{arXiv:hep-lat/0512029}}, \href
  {http://dx.doi.org/10.1103/PhysRevD.73.094510}
  {\path{doi:10.1103/PhysRevD.73.094510}}.

\bibitem{Lacock:1996ny}
P.~Lacock, C.~Michael, P.~Boyle, P.~Rowland, {Hybrid mesons from quenched QCD},
  Phys. Lett. B401 (1997) 308--312.
\newblock \href {http://arxiv.org/abs/hep-lat/9611011}
  {\path{arXiv:hep-lat/9611011}}, \href
  {http://dx.doi.org/10.1016/S0370-2693(97)00384-5}
  {\path{doi:10.1016/S0370-2693(97)00384-5}}.

\bibitem{Manke:1998yg}
T.~Manke, I.~T. Drummond, R.~R. Horgan, H.~P. Shanahan, {Heavy hybrids from
  NRQCD}, Phys. Rev. D57 (1998) 3829--3832.
\newblock \href {http://arxiv.org/abs/hep-lat/9710083}
  {\path{arXiv:hep-lat/9710083}}, \href
  {http://dx.doi.org/10.1103/PhysRevD.57.3829}
  {\path{doi:10.1103/PhysRevD.57.3829}}.

\bibitem{Juge:1999aw}
K.~J. Juge, J.~Kuti, C.~J. Morningstar, {The Heavy hybrid spectrum from NRQCD
  and the Born-Oppenheimer approximation}, Nucl. Phys. Proc. Suppl. 83 (2000)
  304--306.
\newblock \href {http://arxiv.org/abs/hep-lat/9909165}
  {\path{arXiv:hep-lat/9909165}}, \href
  {http://dx.doi.org/10.1016/S0920-5632(00)91655-4}
  {\path{doi:10.1016/S0920-5632(00)91655-4}}.

\bibitem{Drummond:1999db}
I.~T. Drummond, N.~A. Goodman, R.~R. Horgan, H.~P. Shanahan, L.~C. Storoni,
  {Spin effects in heavy hybrid mesons on an anisotropic lattice}, Phys. Lett.
  B478 (2000) 151--160.
\newblock \href {http://arxiv.org/abs/hep-lat/9912041}
  {\path{arXiv:hep-lat/9912041}}, \href
  {http://dx.doi.org/10.1016/S0370-2693(00)00225-2}
  {\path{doi:10.1016/S0370-2693(00)00225-2}}.

\bibitem{Liu:2012ze}
L.~Liu, G.~Moir, M.~Peardon, S.~M. Ryan, C.~E. Thomas, P.~Vilaseca, J.~J.
  Dudek, R.~G. Edwards, B.~Joo, D.~G. Richards, {Excited and exotic charmonium
  spectroscopy from lattice QCD}, JHEP 07 (2012) 126.
\newblock \href {http://arxiv.org/abs/1204.5425} {\path{arXiv:1204.5425}},
  \href {http://dx.doi.org/10.1007/JHEP07(2012)126}
  {\path{doi:10.1007/JHEP07(2012)126}}.

\bibitem{Luo:2005zg}
X.-Q. Luo, Y.~Liu, {Gluonic excitation of non-exotic hybrid charmonium from
  lattice QCD}, Phys. Rev. D74 (2006) 034502, [Erratum: Phys.
  Rev.D74,039902(2006)].
\newblock \href {http://arxiv.org/abs/hep-lat/0512044}
  {\path{arXiv:hep-lat/0512044}}, \href
  {http://dx.doi.org/10.1103/PhysRevD.74.034502, 10.1103/PhysRevD.74.039902}
  {\path{doi:10.1103/PhysRevD.74.034502, 10.1103/PhysRevD.74.039902}}.

\bibitem{Chen:2013zia}
W.~Chen, R.~T. Kleiv, T.~G. Steele, B.~Bulthuis, D.~Harnett, J.~Ho,
  T.~Richards, S.-L. Zhu, {Mass Spectrum of Heavy Quarkonium Hybrids}, JHEP 09
  (2013) 019.
\newblock \href {http://arxiv.org/abs/1304.4522} {\path{arXiv:1304.4522}},
  \href {http://dx.doi.org/10.1007/JHEP09(2013)019}
  {\path{doi:10.1007/JHEP09(2013)019}}.

\bibitem{Chen:2013eha}
W.~Chen, T.~G. Steele, S.-L. Zhu, {Masses of the bottom-charm hybrid $\bar bGc$
  states}, J. Phys. G41 (2014) 025003.
\newblock \href {http://arxiv.org/abs/1306.3486} {\path{arXiv:1306.3486}},
  \href {http://dx.doi.org/10.1088/0954-3899/41/2/025003}
  {\path{doi:10.1088/0954-3899/41/2/025003}}.

\bibitem{Guo:2012tg}
F.-K. Guo, U.-G. Meissner, {Light quark mass dependence in heavy quarkonium
  physics}, Phys. Rev. Lett. 109 (2012) 062001.
\newblock \href {http://arxiv.org/abs/1203.1116} {\path{arXiv:1203.1116}},
  \href {http://dx.doi.org/10.1103/PhysRevLett.109.062001}
  {\path{doi:10.1103/PhysRevLett.109.062001}}.

\bibitem{Cheung:2016bym}
G.~K.~C. Cheung, C.~O'Hara, G.~Moir, M.~Peardon, S.~M. Ryan, C.~E. Thomas,
  D.~Tims, {Excited and exotic charmonium, $D_s$ and $D$ meson spectra for two
  light quark masses from lattice QCD}, JHEP 12 (2016) 089.
\newblock \href {http://arxiv.org/abs/1610.01073} {\path{arXiv:1610.01073}},
  \href {http://dx.doi.org/10.1007/JHEP12(2016)089}
  {\path{doi:10.1007/JHEP12(2016)089}}.

\bibitem{Dudek:2010wm}
J.~J. Dudek, R.~G. Edwards, M.~J. Peardon, D.~G. Richards, C.~E. Thomas,
  {Toward the excited meson spectrum of dynamical QCD}, Phys. Rev. D82 (2010)
  034508.
\newblock \href {http://arxiv.org/abs/1004.4930} {\path{arXiv:1004.4930}},
  \href {http://dx.doi.org/10.1103/PhysRevD.82.034508}
  {\path{doi:10.1103/PhysRevD.82.034508}}.

\bibitem{Moir:2013ub}
G.~Moir, M.~Peardon, S.~M. Ryan, C.~E. Thomas, L.~Liu, {Excited spectroscopy of
  charmed mesons from lattice QCD}, JHEP 05 (2013) 021.
\newblock \href {http://arxiv.org/abs/1301.7670} {\path{arXiv:1301.7670}},
  \href {http://dx.doi.org/10.1007/JHEP05(2013)021}
  {\path{doi:10.1007/JHEP05(2013)021}}.

\bibitem{Moir:2016srx}
G.~Moir, M.~Peardon, S.~M. Ryan, C.~E. Thomas, D.~J. Wilson, {Coupled-Channel
  $D\pi$, $D\eta$ and $D_{s}\bar{K}$ Scattering from Lattice QCD}, JHEP 10
  (2016) 011.
\newblock \href {http://arxiv.org/abs/1607.07093} {\path{arXiv:1607.07093}},
  \href {http://dx.doi.org/10.1007/JHEP10(2016)011}
  {\path{doi:10.1007/JHEP10(2016)011}}.

\bibitem{Chen:2016ejo}
Y.~Chen, W.-F. Chiu, M.~Gong, L.-C. Gui, Z.~Liu, {Exotic vector charmonium and
  its leptonic decay width}, Chin. Phys. C40~(8) (2016) 081002.
\newblock \href {http://arxiv.org/abs/1604.03401} {\path{arXiv:1604.03401}},
  \href {http://dx.doi.org/10.1088/1674-1137/40/8/081002}
  {\path{doi:10.1088/1674-1137/40/8/081002}}.

\bibitem{Morningstar:1997ff}
C.~J. Morningstar, M.~J. Peardon, {Efficient glueball simulations on
  anisotropic lattices}, Phys. Rev. D56 (1997) 4043--4061.
\newblock \href {http://arxiv.org/abs/hep-lat/9704011}
  {\path{arXiv:hep-lat/9704011}}, \href
  {http://dx.doi.org/10.1103/PhysRevD.56.4043}
  {\path{doi:10.1103/PhysRevD.56.4043}}.

\bibitem{Morningstar:1999rf}
C.~J. Morningstar, M.~J. Peardon, {The Glueball spectrum from an anisotropic
  lattice study}, Phys. Rev. D60 (1999) 034509.
\newblock \href {http://arxiv.org/abs/hep-lat/9901004}
  {\path{arXiv:hep-lat/9901004}}, \href
  {http://dx.doi.org/10.1103/PhysRevD.60.034509}
  {\path{doi:10.1103/PhysRevD.60.034509}}.

\bibitem{Chen:2005mg}
Y.~Chen, et~al., {Glueball spectrum and matrix elements on anisotropic
  lattices}, Phys. Rev. D73 (2006) 014516.
\newblock \href {http://arxiv.org/abs/hep-lat/0510074}
  {\path{arXiv:hep-lat/0510074}}, \href
  {http://dx.doi.org/10.1103/PhysRevD.73.014516}
  {\path{doi:10.1103/PhysRevD.73.014516}}.

\bibitem{Chiu:2006hd}
T.-W. Chiu, T.-H. Hsieh, {$X(3872)$ in lattice QCD with exact chiral symmetry},
  Phys. Lett. B646 (2007) 95--99.
\newblock \href {http://arxiv.org/abs/hep-ph/0603207}
  {\path{arXiv:hep-ph/0603207}}, \href
  {http://dx.doi.org/10.1016/j.physletb.2007.01.019}
  {\path{doi:10.1016/j.physletb.2007.01.019}}.

\bibitem{Chiu:2006us}
T.-W. Chiu, T.-H. Hsieh, {Pseudovector meson with strangeness and
  closed-charm}, Phys. Rev. D73 (2006) 111503, [Erratum: Phys. Rev. D75, 019902
  (2007)].
\newblock \href {http://arxiv.org/abs/hep-lat/0604008}
  {\path{arXiv:hep-lat/0604008}}, \href
  {http://dx.doi.org/10.1103/PhysRevD.73.111503, 10.1103/PhysRevD.75.019902}
  {\path{doi:10.1103/PhysRevD.73.111503, 10.1103/PhysRevD.75.019902}}.

\bibitem{Prelovsek:2010kg}
S.~Prelovsek, T.~Draper, C.~B. Lang, M.~Limmer, K.-F. Liu, N.~Mathur,
  D.~Mohler, {Lattice study of light scalar tetraquarks with $I=0,2,1/2,3/2$:
  Are $\sigma$ and $\kappa$ tetraquarks?}, Phys. Rev. D82 (2010) 094507.
\newblock \href {http://arxiv.org/abs/1005.0948} {\path{arXiv:1005.0948}},
  \href {http://dx.doi.org/10.1103/PhysRevD.82.094507}
  {\path{doi:10.1103/PhysRevD.82.094507}}.

\bibitem{Guo:2013nja}
F.-K. Guo, L.~Liu, U.-G. Meissner, P.~Wang, {Tetraquarks, hadronic molecules,
  meson-meson scattering and disconnected contributions in lattice QCD}, Phys.
  Rev. D88 (2013) 074506.
\newblock \href {http://arxiv.org/abs/1308.2545} {\path{arXiv:1308.2545}},
  \href {http://dx.doi.org/10.1103/PhysRevD.88.074506}
  {\path{doi:10.1103/PhysRevD.88.074506}}.

\bibitem{Bali:2005fu}
G.~S. Bali, H.~Neff, T.~Duessel, T.~Lippert, K.~Schilling, {Observation of
  string breaking in QCD}, Phys. Rev. D71 (2005) 114513.
\newblock \href {http://arxiv.org/abs/hep-lat/0505012}
  {\path{arXiv:hep-lat/0505012}}, \href
  {http://dx.doi.org/10.1103/PhysRevD.71.114513}
  {\path{doi:10.1103/PhysRevD.71.114513}}.

\bibitem{Bali:2011dc}
G.~Bali, et~al., {Spectra of heavy-light and heavy-heavy mesons containing
  charm quarks, including higher spin states for $N_f=2+ 1$}, PoS LATTICE2011
  (2011) 135.
\newblock \href {http://arxiv.org/abs/1108.6147} {\path{arXiv:1108.6147}},
  \href {http://dx.doi.org/10.22323/1.139.0135}
  {\path{doi:10.22323/1.139.0135}}.

\bibitem{Bali:2012ua}
G.~Bali, S.~Collins, P.~Perez-Rubio, {Charmed hadron spectroscopy on the
  lattice for $N_f=2+1$ flavours}, J. Phys. Conf. Ser. 426 (2013) 012017.
\newblock \href {http://arxiv.org/abs/1212.0565} {\path{arXiv:1212.0565}},
  \href {http://dx.doi.org/10.1088/1742-6596/426/1/012017}
  {\path{doi:10.1088/1742-6596/426/1/012017}}.

\bibitem{Mohler:2012na}
D.~Mohler, S.~Prelovsek, R.~M. Woloshyn, {$D \pi$ scattering and $D$ meson
  resonances from lattice QCD}, Phys. Rev. D87~(3) (2013) 034501.
\newblock \href {http://arxiv.org/abs/1208.4059} {\path{arXiv:1208.4059}},
  \href {http://dx.doi.org/10.1103/PhysRevD.87.034501}
  {\path{doi:10.1103/PhysRevD.87.034501}}.

\bibitem{Bali:2011rd}
G.~S. Bali, S.~Collins, C.~Ehmann, {Charmonium spectroscopy and mixing with
  light quark and open charm states from $n_F$=2 lattice QCD}, Phys. Rev. D84
  (2011) 094506.
\newblock \href {http://arxiv.org/abs/1110.2381} {\path{arXiv:1110.2381}},
  \href {http://dx.doi.org/10.1103/PhysRevD.84.094506}
  {\path{doi:10.1103/PhysRevD.84.094506}}.

\bibitem{Suzuki:2005ha}
M.~Suzuki, {The $X(3872)$ boson: Molecule or charmonium}, Phys. Rev. D72 (2005)
  114013.
\newblock \href {http://arxiv.org/abs/hep-ph/0508258}
  {\path{arXiv:hep-ph/0508258}}, \href
  {http://dx.doi.org/10.1103/PhysRevD.72.114013}
  {\path{doi:10.1103/PhysRevD.72.114013}}.

\bibitem{Meng:2005er}
C.~Meng, Y.-J. Gao, K.-T. Chao, {$B\to \chi_{c1}(1P,2P)K$ decays in QCD
  factorization and $X(3872)$}, Phys. Rev. D87~(7) (2013) 074035.
\newblock \href {http://arxiv.org/abs/hep-ph/0506222}
  {\path{arXiv:hep-ph/0506222}}, \href
  {http://dx.doi.org/10.1103/PhysRevD.87.074035}
  {\path{doi:10.1103/PhysRevD.87.074035}}.

\bibitem{Prelovsek:2013cra}
S.~Prelovsek, L.~Leskovec, {Evidence for $X(3872)$ from $DD^*$ scattering on
  the lattice}, Phys. Rev. Lett. 111 (2013) 192001.
\newblock \href {http://arxiv.org/abs/1307.5172} {\path{arXiv:1307.5172}},
  \href {http://dx.doi.org/10.1103/PhysRevLett.111.192001}
  {\path{doi:10.1103/PhysRevLett.111.192001}}.

\bibitem{Lee:2014uta}
S.-h. Lee, C.~DeTar, H.~Na, D.~Mohler, {Searching for the $X(3872)$ and
  $Z_c^+(3900)$ on HISQ Lattices}\href {http://arxiv.org/abs/1411.1389}
  {\path{arXiv:1411.1389}}.

\bibitem{Padmanath:2015era}
M.~Padmanath, C.~B. Lang, S.~Prelovsek, {$X(3872)$ and $Y(4140)$ using
  diquark-antidiquark operators with lattice QCD}, Phys. Rev. D92~(3) (2015)
  034501.
\newblock \href {http://arxiv.org/abs/1503.03257} {\path{arXiv:1503.03257}},
  \href {http://dx.doi.org/10.1103/PhysRevD.92.034501}
  {\path{doi:10.1103/PhysRevD.92.034501}}.

\bibitem{Baru:2013rta}
V.~Baru, E.~Epelbaum, A.~A. Filin, C.~Hanhart, U.~G. Meissner, A.~V. Nefediev,
  {Quark mass dependence of the $X(3872)$ binding energy}, Phys. Lett. B726
  (2013) 537--543.
\newblock \href {http://arxiv.org/abs/1306.4108} {\path{arXiv:1306.4108}},
  \href {http://dx.doi.org/10.1016/j.physletb.2013.08.073}
  {\path{doi:10.1016/j.physletb.2013.08.073}}.

\bibitem{Jansen:2013cba}
M.~Jansen, H.~W. Hammer, Y.~Jia, {Light quark mass dependence of the $X(3872)$
  in an effective field theory}, Phys. Rev. D89~(1) (2014) 014033.
\newblock \href {http://arxiv.org/abs/1310.6937} {\path{arXiv:1310.6937}},
  \href {http://dx.doi.org/10.1103/PhysRevD.89.014033}
  {\path{doi:10.1103/PhysRevD.89.014033}}.

\bibitem{Garzon:2013uwa}
E.~J. Garzon, R.~Molina, A.~Hosaka, E.~Oset, {Strategies for an accurate
  determination of the $X(3872)$ energy from QCD lattice simulations}, Phys.
  Rev. D89 (2014) 014504.
\newblock \href {http://arxiv.org/abs/1310.0972} {\path{arXiv:1310.0972}},
  \href {http://dx.doi.org/10.1103/PhysRevD.89.014504}
  {\path{doi:10.1103/PhysRevD.89.014504}}.

\bibitem{Albaladejo:2013aka}
M.~Albaladejo, C.~Hidalgo-Duque, J.~Nieves, E.~Oset, {Hidden charm molecules in
  finite volume}, Phys. Rev. D88 (2013) 014510.
\newblock \href {http://arxiv.org/abs/1304.1439} {\path{arXiv:1304.1439}},
  \href {http://dx.doi.org/10.1103/PhysRevD.88.014510}
  {\path{doi:10.1103/PhysRevD.88.014510}}.

\bibitem{Li:2012cs}
N.~Li, S.-L. Zhu, {Isospin breaking, Coupled-channel effects and Diagnosis of
  $X(3872)$}, Phys. Rev. D86 (2012) 074022.
\newblock \href {http://arxiv.org/abs/1207.3954} {\path{arXiv:1207.3954}},
  \href {http://dx.doi.org/10.1103/PhysRevD.86.074022}
  {\path{doi:10.1103/PhysRevD.86.074022}}.

\bibitem{Zhao:2014gqa}
L.~Zhao, L.~Ma, S.-L. Zhu, {Spin-orbit force, recoil corrections, and possible
  $B \bar{B}^{*}$ and $D \bar{D}^{*}$ molecular states}, Phys. Rev. D89~(9)
  (2014) 094026.
\newblock \href {http://arxiv.org/abs/1403.4043} {\path{arXiv:1403.4043}},
  \href {http://dx.doi.org/10.1103/PhysRevD.89.094026}
  {\path{doi:10.1103/PhysRevD.89.094026}}.

\bibitem{Liu:2008tn}
X.~Liu, Z.-G. Luo, Y.-R. Liu, S.-L. Zhu, {$X(3872)$ and Other Possible Heavy
  Molecular States}, Eur. Phys. J. C61 (2009) 411--428.
\newblock \href {http://arxiv.org/abs/0808.0073} {\path{arXiv:0808.0073}},
  \href {http://dx.doi.org/10.1140/epjc/s10052-009-1020-4}
  {\path{doi:10.1140/epjc/s10052-009-1020-4}}.

\bibitem{Sun:2012zzd}
Z.-F. Sun, Z.-G. Luo, J.~He, X.~Liu, S.-L. Zhu, {A note on the $B^*\bar{B}$,
  $B^*\bar{B}^*$, $D^*\bar{D}$, $D^*\bar{D}^*$ molecular states}, Chin. Phys.
  C36 (2012) 194--204.
\newblock \href {http://dx.doi.org/10.1088/1674-1137/36/3/002}
  {\path{doi:10.1088/1674-1137/36/3/002}}.

\bibitem{He:2014nya}
J.~He, {Study of the $B\bar{B}^*/D\bar{D}^*$ bound states in a Bethe-Salpeter
  approach}, Phys. Rev. D90~(7) (2014) 076008.
\newblock \href {http://arxiv.org/abs/1409.8506} {\path{arXiv:1409.8506}},
  \href {http://dx.doi.org/10.1103/PhysRevD.90.076008}
  {\path{doi:10.1103/PhysRevD.90.076008}}.

\bibitem{Prelovsek:2013xba}
S.~Prelovsek, L.~Leskovec, {Search for $Z^{+}_{c}$(3900) in the $1^{+-}$
  Channel on the Lattice}, Phys. Lett. B727 (2013) 172--176.
\newblock \href {http://arxiv.org/abs/1308.2097} {\path{arXiv:1308.2097}},
  \href {http://dx.doi.org/10.1016/j.physletb.2013.10.009}
  {\path{doi:10.1016/j.physletb.2013.10.009}}.

\bibitem{Levkova:2010ft}
L.~Levkova, C.~DeTar, {Charm annihilation effects on the hyperfine splitting in
  charmonium}, Phys. Rev. D83 (2011) 074504.
\newblock \href {http://arxiv.org/abs/1012.1837} {\path{arXiv:1012.1837}},
  \href {http://dx.doi.org/10.1103/PhysRevD.83.074504}
  {\path{doi:10.1103/PhysRevD.83.074504}}.

\bibitem{Prelovsek:2014swa}
S.~Prelovsek, C.~B. Lang, L.~Leskovec, D.~Mohler, {Study of the $Z_c^+$ channel
  using lattice QCD}, Phys. Rev. D91~(1) (2015) 014504.
\newblock \href {http://arxiv.org/abs/1405.7623} {\path{arXiv:1405.7623}},
  \href {http://dx.doi.org/10.1103/PhysRevD.91.014504}
  {\path{doi:10.1103/PhysRevD.91.014504}}.

\bibitem{Chen:2013coa}
D.-Y. Chen, X.~Liu, T.~Matsuki, {Reproducing the $Z_c(3900)$ structure through
  the initial-single-pion-emission mechanism}, Phys. Rev. D88~(3) (2013)
  036008.
\newblock \href {http://arxiv.org/abs/1304.5845} {\path{arXiv:1304.5845}},
  \href {http://dx.doi.org/10.1103/PhysRevD.88.036008}
  {\path{doi:10.1103/PhysRevD.88.036008}}.

\bibitem{Swanson:2014tra}
E.~Swanson, {$Z_b$ and $Z_c$ Exotic States as Coupled Channel Cusps}, Phys.
  Rev. D91~(3) (2015) 034009.
\newblock \href {http://arxiv.org/abs/1409.3291} {\path{arXiv:1409.3291}},
  \href {http://dx.doi.org/10.1103/PhysRevD.91.034009}
  {\path{doi:10.1103/PhysRevD.91.034009}}.

\bibitem{Guerrieri:2014nxa}
A.~L. Guerrieri, M.~Papinutto, A.~Pilloni, A.~D. Polosa, N.~Tantalo, {Flavored
  tetraquark spectroscopy}, PoS LATTICE2014 (2015) 106.
\newblock \href {http://arxiv.org/abs/1411.2247} {\path{arXiv:1411.2247}},
  \href {http://dx.doi.org/10.22323/1.214.0106}
  {\path{doi:10.22323/1.214.0106}}.

\bibitem{Chen:2014afa}
Y.~Chen, et~al., {Low-energy scattering of the $(D\bar{D}^*)^\pm$ system and
  the resonance-like structure $Z_c(3900)$}, Phys. Rev. D89~(9) (2014) 094506.
\newblock \href {http://arxiv.org/abs/1403.1318} {\path{arXiv:1403.1318}},
  \href {http://dx.doi.org/10.1103/PhysRevD.89.094506}
  {\path{doi:10.1103/PhysRevD.89.094506}}.

\bibitem{Chen:2015jwa}
Y.~Chen, et~al., {Low-energy Scattering of $(D^{*}\bar{D}^{*})^\pm$ System and
  the Resonance-like Structure $Z_c(4025)$}, Phys. Rev. D92~(5) (2015) 054507.
\newblock \href {http://arxiv.org/abs/1503.02371} {\path{arXiv:1503.02371}},
  \href {http://dx.doi.org/10.1103/PhysRevD.92.054507}
  {\path{doi:10.1103/PhysRevD.92.054507}}.

\bibitem{Ikeda:2016zwx}
Y.~Ikeda, S.~Aoki, T.~Doi, S.~Gongyo, T.~Hatsuda, T.~Inoue, T.~Iritani,
  N.~Ishii, K.~Murano, K.~Sasaki, {Fate of the Tetraquark Candidate $Z_c$(3900)
  from Lattice QCD}, Phys. Rev. Lett. 117~(24) (2016) 242001.
\newblock \href {http://arxiv.org/abs/1602.03465} {\path{arXiv:1602.03465}},
  \href {http://dx.doi.org/10.1103/PhysRevLett.117.242001}
  {\path{doi:10.1103/PhysRevLett.117.242001}}.

\bibitem{Ikeda:2017mee}
Y.~Ikeda, {The tetraquark candidate $Z_c$(3900) from dynamical lattice QCD
  simulations}, J. Phys. G45~(2) (2018) 024002.
\newblock \href {http://arxiv.org/abs/1706.07300} {\path{arXiv:1706.07300}},
  \href {http://dx.doi.org/10.1088/1361-6471/aa9afd}
  {\path{doi:10.1088/1361-6471/aa9afd}}.

\bibitem{Aoki:2011gt}
S.~Aoki, N.~Ishii, T.~Doi, T.~Hatsuda, Y.~Ikeda, T.~Inoue, K.~Murano,
  H.~Nemura, K.~Sasaki, {Extraction of Hadron Interactions above Inelastic
  Threshold in Lattice QCD}, Proc. Japan Acad. B87 (2011) 509--517.
\newblock \href {http://arxiv.org/abs/1106.2281} {\path{arXiv:1106.2281}},
  \href {http://dx.doi.org/10.2183/pjab.87.509}
  {\path{doi:10.2183/pjab.87.509}}.

\bibitem{Aoki:2012bb}
S.~Aoki, B.~Charron, T.~Doi, T.~Hatsuda, T.~Inoue, N.~Ishii, {Construction of
  energy-independent potentials above inelastic thresholds in quantum field
  theories}, Phys. Rev. D87~(3) (2013) 034512.
\newblock \href {http://arxiv.org/abs/1212.4896} {\path{arXiv:1212.4896}},
  \href {http://dx.doi.org/10.1103/PhysRevD.87.034512}
  {\path{doi:10.1103/PhysRevD.87.034512}}.

\bibitem{Sasaki:2015ifa}
K.~Sasaki, S.~Aoki, T.~Doi, T.~Hatsuda, Y.~Ikeda, T.~Inoue, N.~Ishii,
  K.~Murano, {Coupled-channel approach to strangeness $S=-2$ baryon¨Cbayron
  interactions in lattice QCD}, PTEP 2015~(11) (2015) 113B01.
\newblock \href {http://arxiv.org/abs/1504.01717} {\path{arXiv:1504.01717}},
  \href {http://dx.doi.org/10.1093/ptep/ptv144}
  {\path{doi:10.1093/ptep/ptv144}}.

\bibitem{Cheung:2017tnt}
G.~K.~C. Cheung, C.~E. Thomas, J.~J. Dudek, R.~G. Edwards, {Tetraquark
  operators in lattice QCD and exotic flavour states in the charm sector}, JHEP
  11 (2017) 033.
\newblock \href {http://arxiv.org/abs/1709.01417} {\path{arXiv:1709.01417}},
  \href {http://dx.doi.org/10.1007/JHEP11(2017)033}
  {\path{doi:10.1007/JHEP11(2017)033}}.

\bibitem{Peardon:2009gh}
M.~Peardon, J.~Bulava, J.~Foley, C.~Morningstar, J.~Dudek, R.~G. Edwards,
  B.~Joo, H.-W. Lin, D.~G. Richards, K.~J. Juge, {A Novel quark-field creation
  operator construction for hadronic physics in lattice QCD}, Phys. Rev. D80
  (2009) 054506.
\newblock \href {http://arxiv.org/abs/0905.2160} {\path{arXiv:0905.2160}},
  \href {http://dx.doi.org/10.1103/PhysRevD.80.054506}
  {\path{doi:10.1103/PhysRevD.80.054506}}.

\bibitem{Edwards:2008ja}
R.~G. Edwards, B.~Joo, H.-W. Lin, {Tuning for Three-flavors of Anisotropic
  Clover Fermions with Stout-link Smearing}, Phys. Rev. D78 (2008) 054501.
\newblock \href {http://arxiv.org/abs/0803.3960} {\path{arXiv:0803.3960}},
  \href {http://dx.doi.org/10.1103/PhysRevD.78.054501}
  {\path{doi:10.1103/PhysRevD.78.054501}}.

\bibitem{Meng:2009qt}
G.-Z. Meng, et~al., {Low-energy $D^{*+} D^0_1$ Scattering and the
  Resonance-like Structure $Z^+(4430)$}, Phys. Rev. D80 (2009) 034503.
\newblock \href {http://arxiv.org/abs/0905.0752} {\path{arXiv:0905.0752}},
  \href {http://dx.doi.org/10.1103/PhysRevD.80.034503}
  {\path{doi:10.1103/PhysRevD.80.034503}}.

\bibitem{Chen:2016lkl}
T.~Chen, et~al., {A Lattice Study of $(\bar{D}_1 D^{*})^\pm$ Near-threshold
  Scattering}, Phys. Rev. D93~(11) (2016) 114501.
\newblock \href {http://arxiv.org/abs/1602.00200} {\path{arXiv:1602.00200}},
  \href {http://dx.doi.org/10.1103/PhysRevD.93.114501}
  {\path{doi:10.1103/PhysRevD.93.114501}}.

\bibitem{Richards:1990xf}
D.~G. Richards, D.~K. Sinclair, D.~W. Sivers, {Lattice QCD simulation of meson
  exchange forces}, Phys. Rev. D42 (1990) 3191--3196.
\newblock \href {http://dx.doi.org/10.1103/PhysRevD.42.3191}
  {\path{doi:10.1103/PhysRevD.42.3191}}.

\bibitem{Mihaly:1996ue}
A.~Mihaly, H.~R. Fiebig, H.~Markum, K.~Rabitsch, {Interactions between heavy -
  light mesons in lattice QCD}, Phys. Rev. D55 (1997) 3077--3081.
\newblock \href {http://dx.doi.org/10.1103/PhysRevD.55.3077}
  {\path{doi:10.1103/PhysRevD.55.3077}}.

\bibitem{Pennanen:1999xi}
P.~Pennanen, C.~Michael, A.~M. Green, {Interactions of heavy light mesons},
  Nucl. Phys. Proc. Suppl. 83 (2000) 200--202.
\newblock \href {http://arxiv.org/abs/hep-lat/9908032}
  {\path{arXiv:hep-lat/9908032}}, \href
  {http://dx.doi.org/10.1016/S0920-5632(00)91622-0}
  {\path{doi:10.1016/S0920-5632(00)91622-0}}.

\bibitem{Green:1999mf}
A.~M. Green, J.~Koponen, P.~Pennanen, {A Variational fit to the lattice energy
  of two heavy light mesons}, Phys. Rev. D61 (2000) 014014.
\newblock \href {http://arxiv.org/abs/hep-ph/9902249}
  {\path{arXiv:hep-ph/9902249}}, \href
  {http://dx.doi.org/10.1103/PhysRevD.61.014014}
  {\path{doi:10.1103/PhysRevD.61.014014}}.

\bibitem{Richard:1990wf}
J.~M. Richard, {Tetraquarks, pentaquarks and hexaquarks}, Nucl. Phys. Proc.
  Suppl. 21 (1991) 254--257.
\newblock \href {http://dx.doi.org/10.1016/0920-5632(91)90263-E}
  {\path{doi:10.1016/0920-5632(91)90263-E}}.

\bibitem{Bander:1994sp}
M.~Bander, A.~Subbaraman, {Baryons with two heavy quarks as solitons}, Phys.
  Rev. D50 (1994) R5478--R5480.
\newblock \href {http://arxiv.org/abs/hep-ph/9407309}
  {\path{arXiv:hep-ph/9407309}}, \href
  {http://dx.doi.org/10.1103/PhysRevD.50.R5478}
  {\path{doi:10.1103/PhysRevD.50.R5478}}.

\bibitem{Detmold:2007wk}
W.~Detmold, K.~Orginos, M.~J. Savage, {BB Potentials in Quenched Lattice QCD},
  Phys. Rev. D76 (2007) 114503.
\newblock \href {http://arxiv.org/abs/hep-lat/0703009}
  {\path{arXiv:hep-lat/0703009}}, \href
  {http://dx.doi.org/10.1103/PhysRevD.76.114503}
  {\path{doi:10.1103/PhysRevD.76.114503}}.

\bibitem{Bali:2010xa}
G.~Bali, M.~Hetzenegger, {Static-light meson-meson potentials}, PoS LATTICE2010
  (2010) 142.
\newblock \href {http://arxiv.org/abs/1011.0571} {\path{arXiv:1011.0571}},
  \href {http://dx.doi.org/10.22323/1.105.0142}
  {\path{doi:10.22323/1.105.0142}}.

\bibitem{Brown:2012tm}
Z.~S. Brown, K.~Orginos, {Tetraquark bound states in the heavy-light
  heavy-light system}, Phys. Rev. D86 (2012) 114506.
\newblock \href {http://arxiv.org/abs/1210.1953} {\path{arXiv:1210.1953}},
  \href {http://dx.doi.org/10.1103/PhysRevD.86.114506}
  {\path{doi:10.1103/PhysRevD.86.114506}}.

\bibitem{Wagner:2010ad}
M.~Wagner, {Forces between static-light mesons}, PoS LATTICE2010 (2010) 162.
\newblock \href {http://arxiv.org/abs/1008.1538} {\path{arXiv:1008.1538}},
  \href {http://dx.doi.org/10.22323/1.105.0162}
  {\path{doi:10.22323/1.105.0162}}.

\bibitem{Wagner:2011ev}
M.~Wagner, {Static-static-light-light tetraquarks in lattice QCD}, Acta Phys.
  Polon. Supp. 4 (2011) 747--752.
\newblock \href {http://arxiv.org/abs/1103.5147} {\path{arXiv:1103.5147}},
  \href {http://dx.doi.org/10.5506/APhysPolBSupp.4.747}
  {\path{doi:10.5506/APhysPolBSupp.4.747}}.

\bibitem{Bicudo:2016ooe}
P.~Bicudo, J.~Scheunert, M.~Wagner, {Including heavy spin effects in the
  prediction of a $\bar{b} \bar{b} u d$ tetraquark with lattice QCD
  potentials}, Phys. Rev. D95~(3) (2017) 034502.
\newblock \href {http://arxiv.org/abs/1612.02758} {\path{arXiv:1612.02758}},
  \href {http://dx.doi.org/10.1103/PhysRevD.95.034502}
  {\path{doi:10.1103/PhysRevD.95.034502}}.

\bibitem{Francis:2016hui}
A.~Francis, R.~J. Hudspith, R.~Lewis, K.~Maltman, {Lattice Prediction for
  Deeply Bound Doubly Heavy Tetraquarks}, Phys. Rev. Lett. 118~(14) (2017)
  142001.
\newblock \href {http://arxiv.org/abs/1607.05214} {\path{arXiv:1607.05214}},
  \href {http://dx.doi.org/10.1103/PhysRevLett.118.142001}
  {\path{doi:10.1103/PhysRevLett.118.142001}}.

\bibitem{Manohar:1997qy}
A.~V. Manohar, {The HQET / NRQCD Lagrangian to order alpha / m-3}, Phys. Rev.
  D56 (1997) 230--237.
\newblock \href {http://arxiv.org/abs/hep-ph/9701294}
  {\path{arXiv:hep-ph/9701294}}, \href
  {http://dx.doi.org/10.1103/PhysRevD.56.230}
  {\path{doi:10.1103/PhysRevD.56.230}}.

\bibitem{Lepage:1992tx}
G.~P. Lepage, L.~Magnea, C.~Nakhleh, U.~Magnea, K.~Hornbostel, {Improved
  nonrelativistic QCD for heavy quark physics}, Phys. Rev. D46 (1992)
  4052--4067.
\newblock \href {http://arxiv.org/abs/hep-lat/9205007}
  {\path{arXiv:hep-lat/9205007}}, \href
  {http://dx.doi.org/10.1103/PhysRevD.46.4052}
  {\path{doi:10.1103/PhysRevD.46.4052}}.

\bibitem{Thacker:1990bm}
B.~A. Thacker, G.~P. Lepage, {Heavy quark bound states in lattice QCD}, Phys.
  Rev. D43 (1991) 196--208.
\newblock \href {http://dx.doi.org/10.1103/PhysRevD.43.196}
  {\path{doi:10.1103/PhysRevD.43.196}}.

\bibitem{Francis:2018jyb}
A.~Francis, R.~J. Hudspith, R.~Lewis, K.~Maltman, {Evidence for charm-bottom
  tetraquarks and the mass dependence of heavy-light tetraquark states from
  lattice QCD}, Phys. Rev. D99~(5) (2019) 054505.
\newblock \href {http://arxiv.org/abs/1810.10550} {\path{arXiv:1810.10550}},
  \href {http://dx.doi.org/10.1103/PhysRevD.99.054505}
  {\path{doi:10.1103/PhysRevD.99.054505}}.

\bibitem{Ikeda:2013vwa}
Y.~Ikeda, B.~Charron, S.~Aoki, T.~Doi, T.~Hatsuda, T.~Inoue, N.~Ishii,
  K.~Murano, H.~Nemura, K.~Sasaki, {Charmed tetraquarks $T_{cc}$ and $T_{cs}$
  from dynamical lattice QCD simulations}, Phys. Lett. B729 (2014) 85--90.
\newblock \href {http://arxiv.org/abs/1311.6214} {\path{arXiv:1311.6214}},
  \href {http://dx.doi.org/10.1016/j.physletb.2014.01.002}
  {\path{doi:10.1016/j.physletb.2014.01.002}}.

\bibitem{Junnarkar:2018twb}
P.~Junnarkar, N.~Mathur, M.~Padmanath, {Study of doubly heavy tetraquarks in
  Lattice QCD}, Phys. Rev. D99 (2019) 034507.
\newblock \href {http://arxiv.org/abs/1810.12285} {\path{arXiv:1810.12285}},
  \href {http://dx.doi.org/10.1103/PhysRevD.99.034507}
  {\path{doi:10.1103/PhysRevD.99.034507}}.

\bibitem{Neuberger:1997fp}
H.~Neuberger, {Exactly massless quarks on the lattice}, Phys. Lett. B417 (1998)
  141--144.
\newblock \href {http://arxiv.org/abs/hep-lat/9707022}
  {\path{arXiv:hep-lat/9707022}}, \href
  {http://dx.doi.org/10.1016/S0370-2693(97)01368-3}
  {\path{doi:10.1016/S0370-2693(97)01368-3}}.

\bibitem{Neuberger:1998wv}
H.~Neuberger, {More about exactly massless quarks on the lattice}, Phys. Lett.
  B427 (1998) 353--355.
\newblock \href {http://arxiv.org/abs/hep-lat/9801031}
  {\path{arXiv:hep-lat/9801031}}, \href
  {http://dx.doi.org/10.1016/S0370-2693(98)00355-4}
  {\path{doi:10.1016/S0370-2693(98)00355-4}}.

\bibitem{Lang:2016jpk}
C.~B. Lang, D.~Mohler, S.~Prelovsek, {$B_s\pi^+$ scattering and search for
  X(5568) with lattice QCD}, Phys. Rev. D94 (2016) 074509.
\newblock \href {http://arxiv.org/abs/1607.03185} {\path{arXiv:1607.03185}},
  \href {http://dx.doi.org/10.1103/PhysRevD.94.074509}
  {\path{doi:10.1103/PhysRevD.94.074509}}.

\bibitem{Lang:2014yfa}
C.~B. Lang, L.~Leskovec, D.~Mohler, S.~Prelovsek, R.~M. Woloshyn, {Ds mesons
  with DK and D*K scattering near threshold}, Phys. Rev. D90~(3) (2014) 034510.
\newblock \href {http://arxiv.org/abs/1403.8103} {\path{arXiv:1403.8103}},
  \href {http://dx.doi.org/10.1103/PhysRevD.90.034510}
  {\path{doi:10.1103/PhysRevD.90.034510}}.

\bibitem{ElKhadra:1996mp}
A.~X. El-Khadra, A.~S. Kronfeld, P.~B. Mackenzie, {Massive fermions in lattice
  gauge theory}, Phys. Rev. D55 (1997) 3933--3957.
\newblock \href {http://arxiv.org/abs/hep-lat/9604004}
  {\path{arXiv:hep-lat/9604004}}, \href
  {http://dx.doi.org/10.1103/PhysRevD.55.3933}
  {\path{doi:10.1103/PhysRevD.55.3933}}.

\bibitem{Oktay:2008ex}
M.~B. Oktay, A.~S. Kronfeld, {New lattice action for heavy quarks}, Phys. Rev.
  D78 (2008) 014504.
\newblock \href {http://arxiv.org/abs/0803.0523} {\path{arXiv:0803.0523}},
  \href {http://dx.doi.org/10.1103/PhysRevD.78.014504}
  {\path{doi:10.1103/PhysRevD.78.014504}}.

\bibitem{Hughes:2017xie}
C.~Hughes, E.~Eichten, C.~T.~H. Davies, {Searching for beauty-fully bound
  tetraquarks using lattice nonrelativistic QCD}, Phys. Rev. D97~(5) (2018)
  054505.
\newblock \href {http://arxiv.org/abs/1710.03236} {\path{arXiv:1710.03236}},
  \href {http://dx.doi.org/10.1103/PhysRevD.97.054505}
  {\path{doi:10.1103/PhysRevD.97.054505}}.

\bibitem{Kawanai:2010ev}
T.~Kawanai, S.~Sasaki, {Charmonium-nucleon potential from lattice QCD}, Phys.
  Rev. D82 (2010) 091501.
\newblock \href {http://arxiv.org/abs/1009.3332} {\path{arXiv:1009.3332}},
  \href {http://dx.doi.org/10.1103/PhysRevD.82.091501}
  {\path{doi:10.1103/PhysRevD.82.091501}}.

\bibitem{Luscher:1996ug}
M.~Luscher, S.~Sint, R.~Sommer, P.~Weisz, U.~Wolff, {Nonperturbative O(a)
  improvement of lattice QCD}, Nucl. Phys. B491 (1997) 323--343.
\newblock \href {http://arxiv.org/abs/hep-lat/9609035}
  {\path{arXiv:hep-lat/9609035}}, \href
  {http://dx.doi.org/10.1016/S0550-3213(97)00080-1}
  {\path{doi:10.1016/S0550-3213(97)00080-1}}.

\bibitem{Aoki:2001ra}
S.~Aoki, Y.~Kuramashi, S.-i. Tominaga, {Relativistic heavy quarks on the
  lattice}, Prog. Theor. Phys. 109 (2003) 383--413.
\newblock \href {http://arxiv.org/abs/hep-lat/0107009}
  {\path{arXiv:hep-lat/0107009}}, \href {http://dx.doi.org/10.1143/PTP.109.383}
  {\path{doi:10.1143/PTP.109.383}}.

\bibitem{Yokokawa:2006td}
K.~Yokokawa, S.~Sasaki, T.~Hatsuda, A.~Hayashigaki, {First lattice study of
  low-energy charmonium-hadron interaction}, Phys. Rev. D74 (2006) 034504.
\newblock \href {http://arxiv.org/abs/hep-lat/0605009}
  {\path{arXiv:hep-lat/0605009}}, \href
  {http://dx.doi.org/10.1103/PhysRevD.74.034504}
  {\path{doi:10.1103/PhysRevD.74.034504}}.

\bibitem{Liu:2008rza}
L.~Liu, H.-W. Lin, K.~Orginos, {Charmed Hadron Interactions}, PoS LATTICE2008
  (2008) 112.
\newblock \href {http://arxiv.org/abs/0810.5412} {\path{arXiv:0810.5412}},
  \href {http://dx.doi.org/10.22323/1.066.0112}
  {\path{doi:10.22323/1.066.0112}}.

\bibitem{Beane:2014sda}
S.~R. Beane, E.~Chang, S.~D. Cohen, W.~Detmold, H.~W. Lin, K.~Orginos,
  A.~Parre{\~n}o, M.~J. Savage, {Quarkonium-Nucleus Bound States from Lattice
  QCD}, Phys. Rev. D91~(11) (2015) 114503.
\newblock \href {http://arxiv.org/abs/1410.7069} {\path{arXiv:1410.7069}},
  \href {http://dx.doi.org/10.1103/PhysRevD.91.114503}
  {\path{doi:10.1103/PhysRevD.91.114503}}.

\bibitem{Sugiura:2017vks}
T.~Sugiura, Y.~Ikeda, N.~Ishii, {Charmonium-nucleon interactions from the
  time-dependent HAL QCD method}, EPJ Web Conf. 175 (2018) 05011.
\newblock \href {http://arxiv.org/abs/1711.11219} {\path{arXiv:1711.11219}},
  \href {http://dx.doi.org/10.1051/epjconf/201817505011}
  {\path{doi:10.1051/epjconf/201817505011}}.

\bibitem{Skerbis:2018lew}
U.~Skerbis, S.~Prelovsek, {Nucleon-$J/\psi$ and nucleon-$\eta_{c}$ scattering
  in $P_{c}$ pentaquark channels from LQCD}\href
  {http://arxiv.org/abs/1811.02285} {\path{arXiv:1811.02285}}.

\bibitem{Dubynskiy:2008mq}
S.~Dubynskiy, M.~B. Voloshin, {Hadro-Charmonium}, Phys. Lett. B666 (2008)
  344--346.
\newblock \href {http://arxiv.org/abs/0803.2224} {\path{arXiv:0803.2224}},
  \href {http://dx.doi.org/10.1016/j.physletb.2008.07.086}
  {\path{doi:10.1016/j.physletb.2008.07.086}}.

\bibitem{Alberti:2016dru}
M.~Alberti, G.~S. Bali, S.~Collins, F.~Knechtli, G.~Moir, W.~S{\"o}ldner,
  {Hadroquarkonium from lattice QCD}, Phys. Rev. D95~(7) (2017) 074501.
\newblock \href {http://arxiv.org/abs/1608.06537} {\path{arXiv:1608.06537}},
  \href {http://dx.doi.org/10.1103/PhysRevD.95.074501}
  {\path{doi:10.1103/PhysRevD.95.074501}}.

\bibitem{Ma:2003zk}
J.~P. Ma, Z.~G. Si, {Factorization approach for inclusive production of doubly
  heavy baryon}, Phys. Lett. B568 (2003) 135--145.
\newblock \href {http://arxiv.org/abs/hep-ph/0305079}
  {\path{arXiv:hep-ph/0305079}}, \href
  {http://dx.doi.org/10.1016/j.physletb.2003.06.064}
  {\path{doi:10.1016/j.physletb.2003.06.064}}.

\bibitem{Jin:2014nva}
Y.~Jin, S.-Y. Li, Y.-R. Liu, Z.-G. Si, T.~Yao, {Search for a doubly charmed
  hadron at B factories}, Phys. Rev. D89~(9) (2014) 094006.
\newblock \href {http://arxiv.org/abs/1401.6652} {\path{arXiv:1401.6652}},
  \href {http://dx.doi.org/10.1103/PhysRevD.89.094006}
  {\path{doi:10.1103/PhysRevD.89.094006}}.

\bibitem{Li:2017ghe}
S.-Y. Li, Y.-R. Liu, Y.-N. Liu, Z.-G. Si, X.-F. Zhang, {Hidden-charm Pentaquark
  Production at $e^+e^-$ Colliders}, Commun. Theor. Phys. 69~(3) (2018) 291.
\newblock \href {http://arxiv.org/abs/1706.04765} {\path{arXiv:1706.04765}},
  \href {http://dx.doi.org/10.1088/0253-6102/69/3/291}
  {\path{doi:10.1088/0253-6102/69/3/291}}.

\bibitem{Han:2009jw}
W.~Han, S.-Y. Li, Y.-H. Shang, F.-L. Shao, T.~Yao, {Exotic hadron production in
  quark combination model}, Phys. Rev. C80 (2009) 035202.
\newblock \href {http://arxiv.org/abs/0906.2473} {\path{arXiv:0906.2473}},
  \href {http://dx.doi.org/10.1103/PhysRevC.80.035202}
  {\path{doi:10.1103/PhysRevC.80.035202}}.

\bibitem{Cho:2010db}
S.~Cho, et~al., {Multi-quark hadrons from Heavy Ion Collisions}, Phys. Rev.
  Lett. 106 (2011) 212001.
\newblock \href {http://arxiv.org/abs/1011.0852} {\path{arXiv:1011.0852}},
  \href {http://dx.doi.org/10.1103/PhysRevLett.106.212001}
  {\path{doi:10.1103/PhysRevLett.106.212001}}.

\bibitem{Cho:2011ew}
S.~Cho, et~al., {Studying Exotic Hadrons in Heavy Ion Collisions}, Phys. Rev.
  C84 (2011) 064910.
\newblock \href {http://arxiv.org/abs/1107.1302} {\path{arXiv:1107.1302}},
  \href {http://dx.doi.org/10.1103/PhysRevC.84.064910}
  {\path{doi:10.1103/PhysRevC.84.064910}}.

\bibitem{Cho:2017dcy}
S.~Cho, et~al., {Exotic Hadrons from Heavy Ion Collisions}, Prog. Part. Nucl.
  Phys. 95 (2017) 279--322.
\newblock \href {http://arxiv.org/abs/1702.00486} {\path{arXiv:1702.00486}},
  \href {http://dx.doi.org/10.1016/j.ppnp.2017.02.002}
  {\path{doi:10.1016/j.ppnp.2017.02.002}}.

\bibitem{Liu:2016sip}
X.~Liu, H.-W. Ke, X.~Liu, X.-Q. Li, {Exploring open-charm decay mode
  $\Lambda_c\bar{\Lambda}_c$ of charmonium-like state $Y(4630)$,~}\href
  {http://arxiv.org/abs/1601.00762} {\path{arXiv:1601.00762}}.

\bibitem{Xiao:2018iez}
L.-Y. Xiao, X.-Z. Weng, Q.-F. L¨¹, X.-H. Zhong, S.-L. Zhu, {A new decay mode of
  higher charmonium}, Eur. Phys. J. C78~(7) (2018) 605.
\newblock \href {http://arxiv.org/abs/1805.07096} {\path{arXiv:1805.07096}},
  \href {http://dx.doi.org/10.1140/epjc/s10052-018-6087-3}
  {\path{doi:10.1140/epjc/s10052-018-6087-3}}.

\bibitem{Ackleh:1991dy}
E.~S. Ackleh, T.~Barnes, {Two photon widths of singlet positronium and
  quarkonium with arbitrary total angular momentum}, Phys. Rev. D45 (1992)
  232--240.
\newblock \href {http://dx.doi.org/10.1103/PhysRevD.45.232}
  {\path{doi:10.1103/PhysRevD.45.232}}.

\bibitem{Wong:2001td}
C.-Y. Wong, E.~S. Swanson, T.~Barnes, {Heavy quarkonium dissociation
  cross-sections in relativistic heavy ion collisions}, Phys. Rev. C65 (2002)
  014903, [Erratum: Phys. Rev.C66,029901(2002)].
\newblock \href {http://arxiv.org/abs/nucl-th/0106067}
  {\path{arXiv:nucl-th/0106067}}, \href
  {http://dx.doi.org/10.1103/PhysRevC.66.029901, 10.1103/PhysRevC.65.014903}
  {\path{doi:10.1103/PhysRevC.66.029901, 10.1103/PhysRevC.65.014903}}.

\bibitem{Barnes:2003dg}
T.~Barnes, E.~S. Swanson, C.~Y. Wong, X.~M. Xu, {Dissociation cross-sections of
  ground state and excited charmonia with light mesons in the quark model},
  Phys. Rev. C68 (2003) 014903.
\newblock \href {http://arxiv.org/abs/nucl-th/0302052}
  {\path{arXiv:nucl-th/0302052}}, \href
  {http://dx.doi.org/10.1103/PhysRevC.68.014903}
  {\path{doi:10.1103/PhysRevC.68.014903}}.

\bibitem{Wang:2018pwi}
G.-J. Wang, X.-H. Liu, L.~Ma, X.~Liu, X.-L. Chen, W.-Z. Deng, S.-L. Zhu, {The
  strong decay patterns of $Z_c$ and $Z_b$ states in the relativized quark
  model}\href {http://arxiv.org/abs/1811.10339} {\path{arXiv:1811.10339}}.

\bibitem{Guo:2016iej}
X.-D. Guo, D.-Y. Chen, H.-W. Ke, X.~Liu, X.-Q. Li, {Study on the rare decays of
  $Y(4630)$ induced by final state interactions}, Phys. Rev. D93~(5) (2016)
  054009.
\newblock \href {http://arxiv.org/abs/1602.02222} {\path{arXiv:1602.02222}},
  \href {http://dx.doi.org/10.1103/PhysRevD.93.054009}
  {\path{doi:10.1103/PhysRevD.93.054009}}.

\bibitem{Zhou:2015uva}
Z.-Y. Zhou, Z.~Xiao, H.-Q. Zhou, {Could the $X(3915)$ and the $X(3930)$ Be the
  Same Tensor State?}, Phys. Rev. Lett. 115~(2) (2015) 022001.
\newblock \href {http://arxiv.org/abs/1501.00879} {\path{arXiv:1501.00879}},
  \href {http://dx.doi.org/10.1103/PhysRevLett.115.022001}
  {\path{doi:10.1103/PhysRevLett.115.022001}}.

\bibitem{Baru:2017fgv}
V.~Baru, C.~Hanhart, A.~V. Nefediev, {Can X(3915) be the tensor partner of the
  X(3872)?}, JHEP 06 (2017) 010.
\newblock \href {http://arxiv.org/abs/1703.01230} {\path{arXiv:1703.01230}},
  \href {http://dx.doi.org/10.1007/JHEP06(2017)010}
  {\path{doi:10.1007/JHEP06(2017)010}}.

\bibitem{Albaladejo:2017blx}
M.~Albaladejo, F.-K. Guo, C.~Hanhart, U.-G. Meissner, J.~Nieves, A.~Nogga,
  Z.~Yang, {Note on X(3872) production at hadron colliders and its molecular
  structure}, Chin. Phys. C41~(12) (2017) 121001.
\newblock \href {http://arxiv.org/abs/1709.09101} {\path{arXiv:1709.09101}},
  \href {http://dx.doi.org/10.1088/1674-1137/41/12/121001}
  {\path{doi:10.1088/1674-1137/41/12/121001}}.

\bibitem{Cho:2013rpa}
S.~Cho, S.~H. Lee, {Hadronic effects on the X(3872) meson abundance in heavy
  ion collisions}, Phys. Rev. C88 (2013) 054901.
\newblock \href {http://arxiv.org/abs/1302.6381} {\path{arXiv:1302.6381}},
  \href {http://dx.doi.org/10.1103/PhysRevC.88.054901}
  {\path{doi:10.1103/PhysRevC.88.054901}}.

\bibitem{Carvalho:2015nqf}
F.~Carvalho, E.~R. Cazaroto, V.~P. Gon{\c c}alves, F.~S. Navarra, {Tetraquark
  Production in Double Parton Scattering}, Phys. Rev. D93~(3) (2016) 034004,
  [Phys. Rev.D93,034004(2016)].
\newblock \href {http://arxiv.org/abs/1511.05209} {\path{arXiv:1511.05209}},
  \href {http://dx.doi.org/10.1103/PhysRevD.93.034004}
  {\path{doi:10.1103/PhysRevD.93.034004}}.

\bibitem{Wang:2015rcz}
W.~Wang, Q.~Zhao, {Decipher the short-distance component of $X(3872)$ in $B_c$
  decays}, Phys. Lett. B755 (2016) 261--264.
\newblock \href {http://arxiv.org/abs/1512.03123} {\path{arXiv:1512.03123}},
  \href {http://dx.doi.org/10.1016/j.physletb.2016.02.012}
  {\path{doi:10.1016/j.physletb.2016.02.012}}.

\bibitem{Hsiao:2016vck}
Y.~K. Hsiao, C.~Q. Geng, {Search for $XYZ$ states in $\Lambda_b$ decays at the
  LHCb}, Phys. Lett. B757 (2016) 47--49.
\newblock \href {http://arxiv.org/abs/1602.01236} {\path{arXiv:1602.01236}},
  \href {http://dx.doi.org/10.1016/j.physletb.2016.03.055}
  {\path{doi:10.1016/j.physletb.2016.03.055}}.

\bibitem{Hsiao:2016pml}
Y.~K. Hsiao, C.~Q. Geng, {Branching fractions of $B_{(c)}$ decays involving
  $J/\psi$ and $X(3872)$}, Chin. Phys. C41~(1) (2017) 013101.
\newblock \href {http://arxiv.org/abs/1607.02718} {\path{arXiv:1607.02718}},
  \href {http://dx.doi.org/10.1088/1674-1137/41/1/013101}
  {\path{doi:10.1088/1674-1137/41/1/013101}}.

\bibitem{Wang:2016ydp}
Z.-H. Wang, Y.~Zhang, T.-h. Wang, Y.~Jiang, G.-L. Wang, {The Production of
  $X(3940)$ and $X(4160)$ in $B_c$ decays}, J. Phys. G43~(10) (2016) 105002.
\newblock \href {http://arxiv.org/abs/1605.09091} {\path{arXiv:1605.09091}},
  \href {http://dx.doi.org/10.1088/0954-3899/43/10/105002}
  {\path{doi:10.1088/0954-3899/43/10/105002}}.

\bibitem{Wang:2016mqb}
Z.-H. Wang, Y.~Zhang, L.~Jiang, T.-H. Wang, Y.~Jiang, G.-L. Wang, {The Strong
  Decays of X(3940) and X(4160)}, Eur. Phys. J. C77~(1) (2017) 43.
\newblock \href {http://arxiv.org/abs/1608.05201} {\path{arXiv:1608.05201}},
  \href {http://dx.doi.org/10.1140/epjc/s10052-017-4596-0}
  {\path{doi:10.1140/epjc/s10052-017-4596-0}}.

\bibitem{delAmoSanchez:2010jr}
P.~del Amo~Sanchez, et~al., {Evidence for the decay $X(3872) \to J/\psi
  \omega$}, Phys. Rev. D82 (2010) 011101.
\newblock \href {http://arxiv.org/abs/1005.5190} {\path{arXiv:1005.5190}},
  \href {http://dx.doi.org/10.1103/PhysRevD.82.011101}
  {\path{doi:10.1103/PhysRevD.82.011101}}.

\bibitem{Liu:2015cah}
X.-H. Liu, M.~Oka, {Searching for charmoniumlike states with hidden
  $s\bar{s}$}, Phys. Rev. D93~(5) (2016) 054032.
\newblock \href {http://arxiv.org/abs/1512.05474} {\path{arXiv:1512.05474}},
  \href {http://dx.doi.org/10.1103/PhysRevD.93.054032}
  {\path{doi:10.1103/PhysRevD.93.054032}}.

\bibitem{Abreu:2016dfe}
L.~M. Abreu, {Analysis of $X(3872)$ production via Heavy-Meson Effective
  Theory}, PTEP 2016~(10) (2016) 103B01.
\newblock \href {http://arxiv.org/abs/1608.08165} {\path{arXiv:1608.08165}},
  \href {http://dx.doi.org/10.1093/ptep/ptw138}
  {\path{doi:10.1093/ptep/ptw138}}.

\bibitem{Abreu:2016qci}
L.~M. Abreu, K.~P. Khemchandani, A.~Martinez~Torres, F.~S. Navarra, M.~Nielsen,
  {$X(3872)$ production and absorption in a hot hadron gas}, Phys. Lett. B761
  (2016) 303--309.
\newblock \href {http://arxiv.org/abs/1604.07716} {\path{arXiv:1604.07716}},
  \href {http://dx.doi.org/10.1016/j.physletb.2016.08.050}
  {\path{doi:10.1016/j.physletb.2016.08.050}}.

\bibitem{Abreu:2016xlr}
L.~M. Abreu, A.~Lafayette~Vasconcellos, {Production of $Z_b^{(')}$ states in
  heavy-meson effective theory}, Phys. Rev. D94~(9) (2016) 096009.
\newblock \href {http://dx.doi.org/10.1103/PhysRevD.94.096009}
  {\path{doi:10.1103/PhysRevD.94.096009}}.

\bibitem{Abreu:2017nuc}
L.~M. Abreu, K.~P. Khemchandani, A.~Mart¨ªnez~Torres, F.~S. Navarra,
  M.~Nielsen, A.~L. Vasconcellos, {Production and absorption of exotic
  bottomoniumlike states in high energy heavy ion collisions}, Phys. Rev.
  D95~(9) (2017) 096002.
\newblock \href {http://arxiv.org/abs/1704.08781} {\path{arXiv:1704.08781}},
  \href {http://dx.doi.org/10.1103/PhysRevD.95.096002}
  {\path{doi:10.1103/PhysRevD.95.096002}}.

\bibitem{Abreu:2017pos}
L.~M. Abreu, F.~S. Navarra, M.~Nielsen, A.~L. Vasconcellos, {$Z_b(10610)$ in a
  hadronic medium}, Eur. Phys. J. C78~(9) (2018) 752.
\newblock \href {http://arxiv.org/abs/1711.05205} {\path{arXiv:1711.05205}},
  \href {http://dx.doi.org/10.1140/epjc/s10052-018-6182-5}
  {\path{doi:10.1140/epjc/s10052-018-6182-5}}.

\bibitem{Jin:2016vjn}
Y.~Jin, S.-Y. Li, Y.-R. Liu, L.~Meng, Z.-G. Si, X.-F. Zhang, {Exotic hadron
  bound state production at hadron colliders}, Chin. Phys. C41~(8) (2017)
  083106.
\newblock \href {http://arxiv.org/abs/1610.04411} {\path{arXiv:1610.04411}},
  \href {http://dx.doi.org/10.1088/1674-1137/41/8/083106}
  {\path{doi:10.1088/1674-1137/41/8/083106}}.

\bibitem{Chen:2016iua}
D.-Y. Chen, {Where are $\chi _{cJ}(3P)$ ?}, Eur. Phys. J. C76~(12) (2016) 671.
\newblock \href {http://arxiv.org/abs/1611.00109} {\path{arXiv:1611.00109}},
  \href {http://dx.doi.org/10.1140/epjc/s10052-016-4531-9}
  {\path{doi:10.1140/epjc/s10052-016-4531-9}}.

\bibitem{Kang:2016jxw}
X.-W. Kang, J.~A. Oller, {Different pole structures in line shapes of the
  $X(3872)$}, Eur. Phys. J. C77~(6) (2017) 399.
\newblock \href {http://arxiv.org/abs/1612.08420} {\path{arXiv:1612.08420}},
  \href {http://dx.doi.org/10.1140/epjc/s10052-017-4961-z}
  {\path{doi:10.1140/epjc/s10052-017-4961-z}}.

\bibitem{Wang:2017mrt}
E.~Wang, J.-J. Xie, L.-S. Geng, E.~Oset, {Analysis of the $B^+\to J/\psi \phi
  K^+$ data at low $J/\psi \phi$ invariant masses and the $X(4140)$ and
  $X(4160)$ resonances}, Phys. Rev. D97~(1) (2018) 014017.
\newblock \href {http://arxiv.org/abs/1710.02061} {\path{arXiv:1710.02061}},
  \href {http://dx.doi.org/10.1103/PhysRevD.97.014017}
  {\path{doi:10.1103/PhysRevD.97.014017}}.

\bibitem{Dai:2018tgo}
L.~R. Dai, J.~M. Dias, E.~Oset, {Disclosing $D^*\bar{D}^*$ molecular states in
  the $B_c^- \rightarrow \pi ^- J/\psi \omega $ decay}, Eur. Phys. J. C78~(3)
  (2018) 210.
\newblock \href {http://arxiv.org/abs/1801.07091} {\path{arXiv:1801.07091}},
  \href {http://dx.doi.org/10.1140/epjc/s10052-018-5702-7}
  {\path{doi:10.1140/epjc/s10052-018-5702-7}}.

\bibitem{Ikeno:2018ugx}
N.~Ikeno, M.~Bayar, E.~Oset, {Semileptonic decay of $B^{-}_c$ into $X(3930)$,
  $X(3940)$, $X(4160)$}, Eur. Phys. J. C78~(5) (2018) 429.
\newblock \href {http://arxiv.org/abs/1803.11226} {\path{arXiv:1803.11226}},
  \href {http://dx.doi.org/10.1140/epjc/s10052-018-5898-6}
  {\path{doi:10.1140/epjc/s10052-018-5898-6}}.

\bibitem{Goncalves:2018hiw}
V.~P. Goncalves, B.~D. Moreira, {Probing the $X(4350)$ in $\gamma \gamma$
  interactions at the LHC}, Eur. Phys. J. C79~(1) (2019) 7.
\newblock \href {http://arxiv.org/abs/1809.08125} {\path{arXiv:1809.08125}},
  \href {http://dx.doi.org/10.1140/epjc/s10052-018-6517-2}
  {\path{doi:10.1140/epjc/s10052-018-6517-2}}.

\bibitem{Braaten:2018eov}
E.~Braaten, L.-P. He, K.~Ingles, {Predictive Solution to the $X(3872)$ Collider
  Production Puzzle}\href {http://arxiv.org/abs/1811.08876}
  {\path{arXiv:1811.08876}}.

\bibitem{Braaten:2019sxh}
E.~Braaten, L.-P. He, K.~Ingles, {Production of $X(3872)$ Accompanied by a Pion
  at Hadron Colliders}\href {http://arxiv.org/abs/1903.04355}
  {\path{arXiv:1903.04355}}.

\bibitem{Braaten:2019yua}
E.~Braaten, L.-P. He, K.~Ingles, {Production of $X(3872)$ Accompanied by a Pion
  in $B$ Meson Decay}\href {http://arxiv.org/abs/1902.03259}
  {\path{arXiv:1902.03259}}.

\bibitem{Andronic:2019wva}
A.~Andronic, P.~Braun-Munzinger, M.~K. K{\"o}hler, K.~Redlich, J.~Stachel,
  {Transverse momentum distributions of charmonium states with the statistical
  hadronization model}\href {http://arxiv.org/abs/1901.09200}
  {\path{arXiv:1901.09200}}.

\bibitem{Guo:2019qcn}
F.-K. Guo, {Novel method for precisely measuring the $X(3872)$ mass}\href
  {http://arxiv.org/abs/1902.11221} {\path{arXiv:1902.11221}}.

\bibitem{Wu:2016dws}
Q.~Wu, G.~Li, F.~Shao, Q.~Wang, R.~Wang, Y.~Zhang, Y.~Zheng, {Production of
  $X_b$ in $\Upsilon(5S, 6S)\to \gamma X_b$ Radiative Decays}, Adv. High Energy
  Phys. 2016 (2016) 3729050.
\newblock \href {http://arxiv.org/abs/1606.05118} {\path{arXiv:1606.05118}},
  \href {http://dx.doi.org/10.1155/2016/3729050}
  {\path{doi:10.1155/2016/3729050}}.

\bibitem{Gonzalez:2016fsr}
P.~Gonz{\'a}lez, {A quark model study of strong decays of $X\left( 3915\right)
  $}, J. Phys. G44~(7) (2017) 075004.
\newblock \href {http://arxiv.org/abs/1611.03710} {\path{arXiv:1611.03710}},
  \href {http://dx.doi.org/10.1088/1361-6471/aa6d8a}
  {\path{doi:10.1088/1361-6471/aa6d8a}}.

\bibitem{Chen:2018xok}
X.~Chen, X.~L{\"u}, R.~Shi, X.~Guo, Q.~Wang, {Radiative decay of hadronic
  molecule state for quarks}\href {http://arxiv.org/abs/1810.07347}
  {\path{arXiv:1810.07347}}.

\bibitem{Chen:2018fbs}
X.~Chen, X.~L{\"u}, {Decay width of hadronic molecule structure for quarks},
  Phys. Rev. D97~(11) (2018) 114005.
\newblock \href {http://dx.doi.org/10.1103/PhysRevD.97.114005}
  {\path{doi:10.1103/PhysRevD.97.114005}}.

\bibitem{Chen:2015bft}
D.-Y. Chen, X.~Liu, X.-Q. Li, H.-W. Ke, {Unified Fano-like interference picture
  for charmoniumlike states Y(4008), Y(4260) and Y(4360)}, Phys. Rev. D93
  (2016) 014011.
\newblock \href {http://arxiv.org/abs/1512.04157} {\path{arXiv:1512.04157}},
  \href {http://dx.doi.org/10.1103/PhysRevD.93.014011}
  {\path{doi:10.1103/PhysRevD.93.014011}}.

\bibitem{Chen:2017uof}
D.-Y. Chen, X.~Liu, T.~Matsuki, {Interference effect as resonance killer of
  newly observed charmoniumlike states $Y(4320)$ and $Y(4390)$}, Eur. Phys. J.
  C78~(2) (2018) 136.
\newblock \href {http://arxiv.org/abs/1708.01954} {\path{arXiv:1708.01954}},
  \href {http://dx.doi.org/10.1140/epjc/s10052-018-5635-1}
  {\path{doi:10.1140/epjc/s10052-018-5635-1}}.

\bibitem{Wang:2019mhs}
J.-Z. Wang, D.-Y. Chen, X.~Liu, T.~Matsuki, {Constructing $J/\psi$ family with
  updated data of charmoniumlike $Y$ states}\href
  {http://arxiv.org/abs/1903.07115} {\path{arXiv:1903.07115}}.

\bibitem{Liu:2016nbm}
X.~Liu, H.-W. Ke, X.~Liu, X.-Q. Li, {Study of structures and dynamical decay
  mechanisms for multiquark systems}, Phys. Rev. D93~(7) (2016) 074013.
\newblock \href {http://arxiv.org/abs/1602.00226} {\path{arXiv:1602.00226}},
  \href {http://dx.doi.org/10.1103/PhysRevD.93.074013}
  {\path{doi:10.1103/PhysRevD.93.074013}}.

\bibitem{Wang:2016fhj}
Y.-Y. Wang, Q.-F. L{\"u}, E.~Wang, D.-M. li, {Role of $Y(4630)$ in the
  $p\bar{p}\rightarrow\Lambda_c\bar{\Lambda}_c$ reaction near threshold}, Phys.
  Rev. D94 (2016) 014025.
\newblock \href {http://arxiv.org/abs/1604.01553} {\path{arXiv:1604.01553}},
  \href {http://dx.doi.org/10.1103/PhysRevD.94.014025}
  {\path{doi:10.1103/PhysRevD.94.014025}}.

\bibitem{Wang:2017sxq}
J.-Z. Wang, H.~Xu, J.-J. Xie, X.~Liu, {Production of the charmoniumlike state
  Y(4220) through the $p\bar{p} \to Y(4220) \pi^0$ reaction}, Phys. Rev.
  D96~(9) (2017) 094004.
\newblock \href {http://arxiv.org/abs/1710.08738} {\path{arXiv:1710.08738}},
  \href {http://dx.doi.org/10.1103/PhysRevD.96.094004}
  {\path{doi:10.1103/PhysRevD.96.094004}}.

\bibitem{Cleven:2016qbn}
M.~Cleven, Q.~Zhao, {Cross section line shape of $e^+e^-\to\chi_{c0}\omega$
  around the $Y(4260)$ mass region}, Phys. Lett. B768 (2017) 52--56.
\newblock \href {http://arxiv.org/abs/1611.04408} {\path{arXiv:1611.04408}},
  \href {http://dx.doi.org/10.1016/j.physletb.2017.02.041}
  {\path{doi:10.1016/j.physletb.2017.02.041}}.

\bibitem{Xue:2017xpu}
S.-R. Xue, H.-J. Jing, F.-K. Guo, Q.~Zhao, {Disentangling the role of the
  $Y(4260)$ in $e^+e^-\to D^*\bar{D}^*$ and $D_s^*\bar{D}_s^*$ via line shape
  studies}, Phys. Lett. B779 (2018) 402--408.
\newblock \href {http://arxiv.org/abs/1708.06961} {\path{arXiv:1708.06961}},
  \href {http://dx.doi.org/10.1016/j.physletb.2018.02.027}
  {\path{doi:10.1016/j.physletb.2018.02.027}}.

\bibitem{Dai:2017fwx}
L.-Y. Dai, J.~Haidenbauer, U.~G. Meissner, {Re-examining the $X(4630)$
  resonance in the reaction $e^+e^-\rightarrow \Lambda^+_c\bar\Lambda^-_c$},
  Phys. Rev. D96~(11) (2017) 116001.
\newblock \href {http://arxiv.org/abs/1710.03142} {\path{arXiv:1710.03142}},
  \href {http://dx.doi.org/10.1103/PhysRevD.96.116001}
  {\path{doi:10.1103/PhysRevD.96.116001}}.

\bibitem{Zhang:2018zog}
J.~Zhang, L.~Yuan, R.~Wang, {Study on the resonant parameters of $Y(4220)$ and
  $Y(4390)$}, Adv. High Energy Phys. 2018 (2018) 5428734.
\newblock \href {http://arxiv.org/abs/1805.03565} {\path{arXiv:1805.03565}},
  \href {http://dx.doi.org/10.1155/2018/5428734}
  {\path{doi:10.1155/2018/5428734}}.

\bibitem{Piotrowska:2018rzl}
M.~Piotrowska, F.~Giacosa, P.~Kovacs, {Can the $\psi(4040)$ explain the peak
  associated with $Y(4008)$?}, Eur. Phys. J. C79~(2) (2019) 98.
\newblock \href {http://arxiv.org/abs/1810.03495} {\path{arXiv:1810.03495}},
  \href {http://dx.doi.org/10.1140/epjc/s10052-019-6615-9}
  {\path{doi:10.1140/epjc/s10052-019-6615-9}}.

\bibitem{Coito:2019cts}
S.~Coito, F.~Giacosa, {On the origin of the $Y(4260)$}\href
  {http://arxiv.org/abs/1902.09268} {\path{arXiv:1902.09268}}.

\bibitem{Chen:2019mgp}
Y.-H. Chen, L.-Y. Dai, F.-K. Guo, B.~Kubis, {On the nature of the $Y(4260)$: a
  light-quark perspective}\href {http://arxiv.org/abs/1902.10957}
  {\path{arXiv:1902.10957}}.

\bibitem{Wang:2013cya}
Q.~Wang, C.~Hanhart, Q.~Zhao, {Decoding the riddle of $Y(4260)$ and
  $Z_c(3900)$}, Phys. Rev. Lett. 111~(13) (2013) 132003.
\newblock \href {http://arxiv.org/abs/1303.6355} {\path{arXiv:1303.6355}},
  \href {http://dx.doi.org/10.1103/PhysRevLett.111.132003}
  {\path{doi:10.1103/PhysRevLett.111.132003}}.

\bibitem{Chen:2016byt}
D.-Y. Chen, Y.-B. Dong, M.-T. Li, W.-L. Wang, {Pionic transition from Y(4260)
  to Z$_{c}$(3900) in a hadronic molecular scenario}, Eur. Phys. J. A52~(10)
  (2016) 310.
\newblock \href {http://dx.doi.org/10.1140/epja/i2016-16310-0}
  {\path{doi:10.1140/epja/i2016-16310-0}}.

\bibitem{Chen:2018fsi}
D.-Y. Chen, J.~He, C.-Q. Pang, Z.-Y. Zhou, {$Y(4320)$ and $Y(4390)$ as the
  candidate for $\psi(3^3D_1)$ charmonium}\href
  {http://arxiv.org/abs/1804.00614} {\path{arXiv:1804.00614}}.

\bibitem{Gong:2016hlt}
Q.-R. Gong, Z.-H. Guo, C.~Meng, G.-Y. Tang, Y.-F. Wang, H.-Q. Zheng,
  {$Z_c(3900)$ as a $D\bar{D}^*$ molecule from the pole counting rule}, Phys.
  Rev. D94~(11) (2016) 114019.
\newblock \href {http://arxiv.org/abs/1604.08836} {\path{arXiv:1604.08836}},
  \href {http://dx.doi.org/10.1103/PhysRevD.94.114019}
  {\path{doi:10.1103/PhysRevD.94.114019}}.

\bibitem{Morgan:1992ge}
D.~Morgan, {Pole counting and resonance classification}, Nucl. Phys. A543
  (1992) 632--644.
\newblock \href {http://dx.doi.org/10.1016/0375-9474(92)90550-4}
  {\path{doi:10.1016/0375-9474(92)90550-4}}.

\bibitem{Au:1986vs}
K.~L. Au, D.~Morgan, M.~R. Pennington, {Meson Dynamics Beyond the Quark Model:
  A Study of Final State Interactions}, Phys. Rev. D35 (1987) 1633.
\newblock \href {http://dx.doi.org/10.1103/PhysRevD.35.1633}
  {\path{doi:10.1103/PhysRevD.35.1633}}.

\bibitem{Gong:2016jzb}
Q.-R. Gong, J.-L. Pang, Y.-F. Wang, H.-Q. Zheng, {The $Z_c(3900)$ peak does not
  come from the ¡°triangle singularity¡±}, Eur. Phys. J. C78~(4) (2018) 276.
\newblock \href {http://arxiv.org/abs/1612.08159} {\path{arXiv:1612.08159}},
  \href {http://dx.doi.org/10.1140/epjc/s10052-018-5690-7}
  {\path{doi:10.1140/epjc/s10052-018-5690-7}}.

\bibitem{Albaladejo:2016jsg}
M.~Albaladejo, P.~Fernandez-Soler, J.~Nieves, {$Z_c(3900)$: Confronting theory
  and lattice simulations}, Eur. Phys. J. C76~(10) (2016) 573.
\newblock \href {http://arxiv.org/abs/1606.03008} {\path{arXiv:1606.03008}},
  \href {http://dx.doi.org/10.1140/epjc/s10052-016-4427-8}
  {\path{doi:10.1140/epjc/s10052-016-4427-8}}.

\bibitem{Pilloni:2016obd}
A.~Pilloni, C.~Fernandez-Ramirez, A.~Jackura, V.~Mathieu, M.~Mikhasenko,
  J.~Nys, A.~P. Szczepaniak, {Amplitude analysis and the nature of the
  Z$_c$(3900)}, Phys. Lett. B772 (2017) 200--209.
\newblock \href {http://arxiv.org/abs/1612.06490} {\path{arXiv:1612.06490}},
  \href {http://dx.doi.org/10.1016/j.physletb.2017.06.030}
  {\path{doi:10.1016/j.physletb.2017.06.030}}.

\bibitem{Yang:2017nde}
Z.~Yang, Q.~Wang, U.-G. Meissner, {Isospin analysis of $B\to D^*\bar{D}K$ and
  the absence of the $Z_c(3900)$ in $B$ decays}, Phys. Lett. B775 (2017)
  50--53.
\newblock \href {http://arxiv.org/abs/1706.00960} {\path{arXiv:1706.00960}},
  \href {http://dx.doi.org/10.1016/j.physletb.2017.10.049}
  {\path{doi:10.1016/j.physletb.2017.10.049}}.

\bibitem{He:2017lhy}
J.~He, D.-Y. Chen, {$Z_c(3900)/Z_c(3885)$ as a virtual state from $\pi
  J/\psi-\bar{D}^*D$ interaction}, Eur. Phys. J. C78~(2) (2018) 94.
\newblock \href {http://arxiv.org/abs/1712.05653} {\path{arXiv:1712.05653}},
  \href {http://dx.doi.org/10.1140/epjc/s10052-018-5580-z}
  {\path{doi:10.1140/epjc/s10052-018-5580-z}}.

\bibitem{Zhao:2018xrd}
Q.~Zhao, {Some insights into the newly observed $Z_c(4100)$ in $B^0\to \eta_c
  K^+ \pi^-$ by LHCb}\href {http://arxiv.org/abs/1811.05357}
  {\path{arXiv:1811.05357}}.

\bibitem{Cao:2018vmv}
X.~Cao, J.-P. Dai, {The spin parity of $Z_c^-$(4100), $Z_1^+$(4050) and
  $Z_2^+$(4250)}\href {http://arxiv.org/abs/1811.06434}
  {\path{arXiv:1811.06434}}.

\bibitem{Mizuk:2008me}
R.~Mizuk, et~al., {Observation of two resonance-like structures in the $\pi^+
  \chi_{c1}$ mass distribution in exclusive $\bar B^0 \to K^- \pi^+ \chi_{c1}$
  decays}, Phys. Rev. D78 (2008) 072004.
\newblock \href {http://arxiv.org/abs/0806.4098} {\path{arXiv:0806.4098}},
  \href {http://dx.doi.org/10.1103/PhysRevD.78.072004}
  {\path{doi:10.1103/PhysRevD.78.072004}}.

\bibitem{Nakamura:2019btl}
S.~X. Nakamura, K.~Tsushima, {$Z_c(4430)$ and $Z_c(4200)$ as triangle
  singularities}\href {http://arxiv.org/abs/1901.07385}
  {\path{arXiv:1901.07385}}.

\bibitem{Nakamura:2019emd}
S.~X. Nakamura, {Triangle singularities in $\bar{B}^0\to \chi_{c1}K^-\pi^+$
  relevant to $Z_1(4050)$ and $Z_2(4250)$}\href
  {http://arxiv.org/abs/1903.08098} {\path{arXiv:1903.08098}}.

\bibitem{Wu:2019vbk}
Q.~Wu, D.-Y. Chen, X.-J. Fan, G.~Li, {Production of $Z_c(3900$) and $Z_c(4020)$
  in $B_c$ decay}\href {http://arxiv.org/abs/1902.05737}
  {\path{arXiv:1902.05737}}.

\bibitem{Ke:2016owt}
H.-W. Ke, X.-Q. Li, {Study on decays of $Z_c(4020)$ and $Z_c(3900)$ into
  $h_c+\pi $}, Eur. Phys. J. C76~(6) (2016) 334.
\newblock \href {http://arxiv.org/abs/1601.03575} {\path{arXiv:1601.03575}},
  \href {http://dx.doi.org/10.1140/epjc/s10052-016-4183-9}
  {\path{doi:10.1140/epjc/s10052-016-4183-9}}.

\bibitem{Molnar:2019uos}
D.~A.~S. Molnar, I.~Danilkin, M.~Vanderhaeghen, {The role of charged exotic
  states in $e^+e^- \to \psi(2S) \; \pi^+ \pi ^-$}\href
  {http://arxiv.org/abs/1903.08458} {\path{arXiv:1903.08458}}.

\bibitem{Agaev:2016dev}
S.~S. Agaev, K.~Azizi, H.~Sundu, {Strong $Z_c^{+}(3900)\rightarrow J/\psi
  \pi^{+}; \eta_{c} \rho^{+}$ decays in QCD}, Phys. Rev. D93~(7) (2016) 074002.
\newblock \href {http://arxiv.org/abs/1601.03847} {\path{arXiv:1601.03847}},
  \href {http://dx.doi.org/10.1103/PhysRevD.93.074002}
  {\path{doi:10.1103/PhysRevD.93.074002}}.

\bibitem{Wu:2016ypc}
Q.~Wu, G.~Li, F.~Shao, R.~Wang, {Investigations on the charmless decay modes of
  Zc(3900) and Zc(4025)}, Phys. Rev. D94~(1) (2016) 014015.
\newblock \href {http://dx.doi.org/10.1103/PhysRevD.94.014015}
  {\path{doi:10.1103/PhysRevD.94.014015}}.

\bibitem{Goerke:2016hxf}
F.~Goerke, T.~Gutsche, M.~A. Ivanov, J.~G. Korner, V.~E. Lyubovitskij,
  P.~Santorelli, {Four-quark structure of Zc(3900), Z(4430) and Xb(5568)
  states}, Phys. Rev. D94~(9) (2016) 094017.
\newblock \href {http://arxiv.org/abs/1608.04656} {\path{arXiv:1608.04656}},
  \href {http://dx.doi.org/10.1103/PhysRevD.94.094017}
  {\path{doi:10.1103/PhysRevD.94.094017}}.

\bibitem{Liu:2014eka}
X.-H. Liu, L.~Ma, L.-P. Sun, X.~Liu, S.-L. Zhu, {Resolving the puzzling decay
  patterns of charged $Z_c$ and $Z_b$ states}, Phys. Rev. D90~(7) (2014)
  074020.
\newblock \href {http://arxiv.org/abs/1407.3684} {\path{arXiv:1407.3684}},
  \href {http://dx.doi.org/10.1103/PhysRevD.90.074020}
  {\path{doi:10.1103/PhysRevD.90.074020}}.

\bibitem{Voloshin:2019ivc}
M.~B. Voloshin, {Radiative and pionic transitions $Z_c(4020)^0 \to X(3872)
  \gamma$ and $Z_c(4020)^\pm \to X(3872) \pi^\pm$}\href
  {http://arxiv.org/abs/1902.01281} {\path{arXiv:1902.01281}}.

\bibitem{Chen:2015jgl}
Y.-H. Chen, J.~T. Daub, F.-K. Guo, B.~Kubis, U.-G. Meissner, B.-S. Zou, {Effect
  of $Z_b$ states on $\Upsilon(3S)\to\Upsilon(1S)\pi\pi$ decays}, Phys. Rev.
  D93~(3) (2016) 034030.
\newblock \href {http://arxiv.org/abs/1512.03583} {\path{arXiv:1512.03583}},
  \href {http://dx.doi.org/10.1103/PhysRevD.93.034030}
  {\path{doi:10.1103/PhysRevD.93.034030}}.

\bibitem{Chen:2016mjn}
Y.-H. Chen, M.~Cleven, J.~T. Daub, F.-K. Guo, C.~Hanhart, B.~Kubis, U.-G.
  Meissner, B.-S. Zou, {Effects of $Z_b$ states and bottom meson loops on
  $\Upsilon(4S) \to \Upsilon(1S,2S) \pi^+\pi^-$ transitions}, Phys. Rev.
  D95~(3) (2017) 034022.
\newblock \href {http://arxiv.org/abs/1611.00913} {\path{arXiv:1611.00913}},
  \href {http://dx.doi.org/10.1103/PhysRevD.95.034022}
  {\path{doi:10.1103/PhysRevD.95.034022}}.

\bibitem{Guo:2016bjq}
F.~K. Guo, C.~Hanhart, {\relax Yu}.~S. Kalashnikova, P.~Matuschek, R.~V. Mizuk,
  A.~V. Nefediev, Q.~Wang, J.~L. Wynen, {Interplay of quark and meson degrees
  of freedom in near-threshold states: A practical parametrization for line
  shapes}, Phys. Rev. D93~(7) (2016) 074031.
\newblock \href {http://arxiv.org/abs/1602.00940} {\path{arXiv:1602.00940}},
  \href {http://dx.doi.org/10.1103/PhysRevD.93.074031}
  {\path{doi:10.1103/PhysRevD.93.074031}}.

\bibitem{Bondar:2016pox}
A.~E. Bondar, M.~B. Voloshin, {$\Upsilon(6S)$ and triangle singularity in
  $e^+e^- \to B_1(5721) \bar B \to Z_b(10610) \, \pi$}, Phys. Rev. D93~(9)
  (2016) 094008.
\newblock \href {http://arxiv.org/abs/1603.08436} {\path{arXiv:1603.08436}},
  \href {http://dx.doi.org/10.1103/PhysRevD.93.094008}
  {\path{doi:10.1103/PhysRevD.93.094008}}.

\bibitem{Xiao:2017uve}
C.-J. Xiao, D.-Y. Chen, {Analysis of the hidden bottom decays of Zb(10610) and
  Zb(10650) via final state interaction}, Phys. Rev. D96~(1) (2017) 014035.
\newblock \href {http://dx.doi.org/10.1103/PhysRevD.96.014035}
  {\path{doi:10.1103/PhysRevD.96.014035}}.

\bibitem{Xiao:2018kfx}
C.-J. Xiao, D.-Y. Chen, Y.-B. Dong, W.~Zuo, T.~Matsuki, {Understanding the
  $\eta_c\rho$ decay mode of $Z_c^{(\prime)}$ via final state
  interactions}\href {http://arxiv.org/abs/1811.04688}
  {\path{arXiv:1811.04688}}.

\bibitem{Yuan:2018inv}
C.-Z. Yuan, {The XYZ states revisited}, Int. J. Mod. Phys. A33~(21) (2018)
  1830018.
\newblock \href {http://arxiv.org/abs/1808.01570} {\path{arXiv:1808.01570}},
  \href {http://dx.doi.org/10.1142/S0217751X18300181}
  {\path{doi:10.1142/S0217751X18300181}}.

\bibitem{Voloshin:2017gnc}
M.~B. Voloshin, {Mixing model for bottomoniumlike $Z_b$ resonances}, Phys. Rev.
  D96~(9) (2017) 094024.
\newblock \href {http://arxiv.org/abs/1707.00565} {\path{arXiv:1707.00565}},
  \href {http://dx.doi.org/10.1103/PhysRevD.96.094024}
  {\path{doi:10.1103/PhysRevD.96.094024}}.

\bibitem{Wang:2018jlv}
Q.~Wang, V.~Baru, A.~A. Filin, C.~Hanhart, A.~V. Nefediev, J.~L. Wynen, {Line
  shapes of the $Z_b(10610)$ and $Z_b(10650)$ in the elastic and inelastic
  channels revisited}, Phys. Rev. D98~(7) (2018) 074023.
\newblock \href {http://arxiv.org/abs/1805.07453} {\path{arXiv:1805.07453}},
  \href {http://dx.doi.org/10.1103/PhysRevD.98.074023}
  {\path{doi:10.1103/PhysRevD.98.074023}}.

\bibitem{Baru:2019xnh}
V.~Baru, E.~Epelbaum, A.~A. Filin, C.~Hanhart, A.~V. Nefediev, Q.~Wang, {Spin
  partners $W_{bJ}$ from the line shapes of the $Z_b(10610)$ and
  $Z_b(10650)$}\href {http://arxiv.org/abs/1901.10319}
  {\path{arXiv:1901.10319}}.

\bibitem{Wu:2018xaa}
Q.~Wu, D.-Y. Chen, F.-K. Guo, {Production of the $Z_b^{(\prime)}$ states from
  the $\Upsilon(5S,6S)$ decays}, Phys. Rev. D99~(3) (2019) 034022.
\newblock \href {http://arxiv.org/abs/1810.09696} {\path{arXiv:1810.09696}},
  \href {http://dx.doi.org/10.1103/PhysRevD.99.034022}
  {\path{doi:10.1103/PhysRevD.99.034022}}.

\bibitem{Goerke:2017svb}
F.~Goerke, T.~Gutsche, M.~A. Ivanov, J.~G. K{\"o}rner, V.~E. Lyubovitskij,
  {$Z_b(10610)$ and $Z_b'(10650)$ decays in a covariant quark model}, Phys.
  Rev. D96~(5) (2017) 054028.
\newblock \href {http://arxiv.org/abs/1707.00539} {\path{arXiv:1707.00539}},
  \href {http://dx.doi.org/10.1103/PhysRevD.96.054028}
  {\path{doi:10.1103/PhysRevD.96.054028}}.

\bibitem{Voloshin:2018pqn}
M.~B. Voloshin, {Radiative and $\rho$ transitions between a heavy quarkonium
  and isovector four-quark states}, Phys. Rev. D98~(3) (2018) 034025.
\newblock \href {http://arxiv.org/abs/1806.05651} {\path{arXiv:1806.05651}},
  \href {http://dx.doi.org/10.1103/PhysRevD.98.034025}
  {\path{doi:10.1103/PhysRevD.98.034025}}.

\bibitem{DelFabbro:2004ta}
A.~Del~Fabbro, D.~Janc, M.~Rosina, D.~Treleani, {Production and detection of
  doubly charmed tetraquarks}, Phys. Rev. D71 (2005) 014008.
\newblock \href {http://arxiv.org/abs/hep-ph/0408258}
  {\path{arXiv:hep-ph/0408258}}, \href
  {http://dx.doi.org/10.1103/PhysRevD.71.014008}
  {\path{doi:10.1103/PhysRevD.71.014008}}.

\bibitem{Esposito:2013fma}
A.~Esposito, M.~Papinutto, A.~Pilloni, A.~D. Polosa, N.~Tantalo, {Doubly
  charmed tetraquarks in $B_c$ and $\Xi_{bc}$ decays}, Phys. Rev. D88~(5)
  (2013) 054029.
\newblock \href {http://arxiv.org/abs/1307.2873} {\path{arXiv:1307.2873}},
  \href {http://dx.doi.org/10.1103/PhysRevD.88.054029}
  {\path{doi:10.1103/PhysRevD.88.054029}}.

\bibitem{Hong:2018mpk}
J.~Hong, S.~Cho, T.~Song, S.~H. Lee, {Hadronic effects on the
  $cc\bar{q}\bar{q}$ tetraquark state in relativistic heavy ion collisions},
  Phys. Rev. C98~(1) (2018) 014913.
\newblock \href {http://arxiv.org/abs/1804.05336} {\path{arXiv:1804.05336}},
  \href {http://dx.doi.org/10.1103/PhysRevC.98.014913}
  {\path{doi:10.1103/PhysRevC.98.014913}}.

\bibitem{Ali:2018ifm}
A.~Ali, A.~Y. Parkhomenko, Q.~Qin, W.~Wang, {Prospects of discovering stable
  double-heavy tetraquarks at a Tera-$Z$ factory}, Phys. Lett. B782 (2018)
  412--420.
\newblock \href {http://arxiv.org/abs/1805.02535} {\path{arXiv:1805.02535}},
  \href {http://dx.doi.org/10.1016/j.physletb.2018.05.055}
  {\path{doi:10.1016/j.physletb.2018.05.055}}.

\bibitem{Ali:2018xfq}
A.~Ali, Q.~Qin, W.~Wang, {Discovery potential of stable and near-threshold
  doubly heavy tetraquarks at the LHC}, Phys. Lett. B785 (2018) 605--609.
\newblock \href {http://arxiv.org/abs/1806.09288} {\path{arXiv:1806.09288}},
  \href {http://dx.doi.org/10.1016/j.physletb.2018.09.018}
  {\path{doi:10.1016/j.physletb.2018.09.018}}.

\bibitem{Gershon:2018gda}
T.~Gershon, A.~Poluektov, {Displaced B$_{c}^{-}$ mesons as an inclusive
  signature of weakly decaying double beauty hadrons}, JHEP 01 (2019) 019.
\newblock \href {http://arxiv.org/abs/1810.06657} {\path{arXiv:1810.06657}},
  \href {http://dx.doi.org/10.1007/JHEP01(2019)019}
  {\path{doi:10.1007/JHEP01(2019)019}}.

\bibitem{Ridgway:2019zks}
A.~K. Ridgway, M.~B. Wise, {An Estimate of the Inclusive Branching Ratio to
  ${\bar B}_c$ in $\Xi_{bbq}$ Decay}\href {http://arxiv.org/abs/1902.04582}
  {\path{arXiv:1902.04582}}.

\bibitem{Jin:2013bra}
Y.~Jin, S.-Y. Li, Z.-G. Si, Z.-J. Yang, T.~Yao, {Colour connections of four
  quark $Q\bar{Q}Q'\bar{Q}'$ system and doubly heavy baryon production in
  $e^{+}e^{-}$ annihilation}, Phys. Lett. B727 (2013) 468--473.
\newblock \href {http://arxiv.org/abs/1309.5849} {\path{arXiv:1309.5849}},
  \href {http://dx.doi.org/10.1016/j.physletb.2013.10.070}
  {\path{doi:10.1016/j.physletb.2013.10.070}}.

\bibitem{Eichten:2017ual}
E.~Eichten, Z.~Liu, {Would a Deeply Bound $b\bar b b\bar b$ Tetraquark Meson be
  Observed at the LHC?}\href {http://arxiv.org/abs/1709.09605}
  {\path{arXiv:1709.09605}}.

\bibitem{Vega-Morales:2017pmm}
R.~Vega-Morales, R.~Vega-Morales, {Golden Probe of the di$-\Upsilon$
  Threshold}\href {http://arxiv.org/abs/1710.02738} {\path{arXiv:1710.02738}}.

\bibitem{Li:2019uch}
G.~Li, X.-F. Wang, Y.~Xing, {Fully Heavy Tetraquark ${bb \bar c \bar c}$:
  Lifetimes and Weak Decays}\href {http://arxiv.org/abs/1902.05805}
  {\path{arXiv:1902.05805}}.

\bibitem{Roca:2015dva}
L.~Roca, J.~Nieves, E.~Oset, {LHCb pentaquark as a
  $\bar{D}^*\Sigma_c-\bar{D}^*\Sigma_c^*$ molecular state}, Phys. Rev. D92~(9)
  (2015) 094003.
\newblock \href {http://arxiv.org/abs/1507.04249} {\path{arXiv:1507.04249}},
  \href {http://dx.doi.org/10.1103/PhysRevD.92.094003}
  {\path{doi:10.1103/PhysRevD.92.094003}}.

\bibitem{Roca:2016tdh}
L.~Roca, E.~Oset, {On the hidden charm pentaquarks in $\Lambda _b \rightarrow
  J/\psi K^- p$ decay}, Eur. Phys. J. C76~(11) (2016) 591.
\newblock \href {http://arxiv.org/abs/1602.06791} {\path{arXiv:1602.06791}},
  \href {http://dx.doi.org/10.1140/epjc/s10052-016-4407-z}
  {\path{doi:10.1140/epjc/s10052-016-4407-z}}.

\bibitem{Xiao:2016ogq}
C.~W. Xiao, {$J/\psi N$ interactions revisited and $\Lambda_b^0 \to J/\psi K^-
  (\pi^-) p$ decays}, Phys. Rev. D95~(1) (2017) 014006.
\newblock \href {http://arxiv.org/abs/1609.02712} {\path{arXiv:1609.02712}},
  \href {http://dx.doi.org/10.1103/PhysRevD.95.014006}
  {\path{doi:10.1103/PhysRevD.95.014006}}.

\bibitem{Lu:2016roh}
J.-X. Lu, E.~Wang, J.-J. Xie, L.-S. Geng, E.~Oset, {The $\Lambda_{b}\rightarrow
  J/\psi K^{0}\Lambda$ reaction and a hidden-charm pentaquark state with
  strangeness}, Phys. Rev. D93 (2016) 094009.
\newblock \href {http://arxiv.org/abs/1601.00075} {\path{arXiv:1601.00075}},
  \href {http://dx.doi.org/10.1103/PhysRevD.93.094009}
  {\path{doi:10.1103/PhysRevD.93.094009}}.

\bibitem{Huang:2016tcr}
Y.~Huang, J.-J. Xie, J.~He, X.~Chen, H.-F. Zhang, {Photoproduction of
  hidden-charm states in the $\gamma p \to \bar D^{*0} \Lambda^+_c$ reaction
  near threshold}, Chin. Phys. C40~(12) (2016) 124104.
\newblock \href {http://arxiv.org/abs/1604.05969} {\path{arXiv:1604.05969}},
  \href {http://dx.doi.org/10.1088/1674-1137/40/12/124104}
  {\path{doi:10.1088/1674-1137/40/12/124104}}.

\bibitem{Cheng:2016ddp}
C.~Cheng, X.-Y. Wang, {The production of neutral $N^*(11052)$ resonance with
  hidden beauty from $\pi^-p$ scattering}, Adv. High Energy Phys. 2017 (2017)
  9398732.
\newblock \href {http://arxiv.org/abs/1612.08822} {\path{arXiv:1612.08822}},
  \href {http://dx.doi.org/10.1155/2017/9398732}
  {\path{doi:10.1155/2017/9398732}}.

\bibitem{Xie:2017gwc}
J.-J. Xie, W.-H. Liang, E.~Oset, {Hidden charm pentaquark and $\Lambda(1405)$
  in the $\Lambda^0_b \to \eta_c K^- p (\pi \Sigma)$ reaction}, Phys. Lett.
  B777 (2018) 447--452.
\newblock \href {http://arxiv.org/abs/1711.01710} {\path{arXiv:1711.01710}},
  \href {http://dx.doi.org/10.1016/j.physletb.2017.12.064}
  {\path{doi:10.1016/j.physletb.2017.12.064}}.

\bibitem{Wang:2016vxa}
R.-Q. Wang, J.~Song, K.-J. Sun, L.-W. Chen, G.~Li, F.-L. Shao, {Hidden-charm
  pentaquark states in heavy ion collisions at energies available at the CERN
  Large Hadron Collider}, Phys. Rev. C94~(4) (2016) 044913.
\newblock \href {http://arxiv.org/abs/1601.02835} {\path{arXiv:1601.02835}},
  \href {http://dx.doi.org/10.1103/PhysRevC.94.044913}
  {\path{doi:10.1103/PhysRevC.94.044913}}.

\bibitem{Schmidt:2016cmd}
I.~Schmidt, M.~Siddikov, {Production of pentaquarks in $pA$-collisions}, Phys.
  Rev. D93~(9) (2016) 094005.
\newblock \href {http://arxiv.org/abs/1601.05621} {\path{arXiv:1601.05621}},
  \href {http://dx.doi.org/10.1103/PhysRevD.93.094005}
  {\path{doi:10.1103/PhysRevD.93.094005}}.

\bibitem{Liu:2016dli}
X.-H. Liu, M.~Oka, {Understanding the nature of heavy pentaquarks and searching
  for them in pion-induced reactions}, Nucl. Phys. A954 (2016) 352--364.
\newblock \href {http://arxiv.org/abs/1602.07069} {\path{arXiv:1602.07069}},
  \href {http://dx.doi.org/10.1016/j.nuclphysa.2016.04.040}
  {\path{doi:10.1016/j.nuclphysa.2016.04.040}}.

\bibitem{Guo:2016bkl}
F.-K. Guo, U.~G. Meissner, J.~Nieves, Z.~Yang, {Remarks on the $P_c$ structures
  and triangle singularities}, Eur. Phys. J. A52~(10) (2016) 318.
\newblock \href {http://arxiv.org/abs/1605.05113} {\path{arXiv:1605.05113}},
  \href {http://dx.doi.org/10.1140/epja/i2016-16318-4}
  {\path{doi:10.1140/epja/i2016-16318-4}}.

\bibitem{Bayar:2016ftu}
M.~Bayar, F.~Aceti, F.-K. Guo, E.~Oset, {A Discussion on Triangle Singularities
  in the $\Lambda_b \to J/\psi K^{-} p$ Reaction}, Phys. Rev. D94~(7) (2016)
  074039.
\newblock \href {http://arxiv.org/abs/1609.04133} {\path{arXiv:1609.04133}},
  \href {http://dx.doi.org/10.1103/PhysRevD.94.074039}
  {\path{doi:10.1103/PhysRevD.94.074039}}.

\bibitem{Kim:2016cxr}
S.-H. Kim, H.-C. Kim, A.~Hosaka, {Heavy pentaquark states $P_c(4380)$ and
  $P_c(4450)$ in the $J/\psi$ production induced by pion beams off the
  nucleon}, Phys. Lett. B763 (2016) 358--364.
\newblock \href {http://arxiv.org/abs/1605.02919} {\path{arXiv:1605.02919}},
  \href {http://dx.doi.org/10.1016/j.physletb.2016.10.061}
  {\path{doi:10.1016/j.physletb.2016.10.061}}.

\bibitem{Karliner:2017qje}
M.~Karliner, J.~L. Rosner, {$J/\psi N$ photoproduction on deuterium as a test
  for exotic baryons}\href {http://arxiv.org/abs/1705.07691}
  {\path{arXiv:1705.07691}}.

\bibitem{Zhou:2017bhq}
P.~Zhou, Y.~K. Hsiao, C.~Q. Geng, {Pentaquarks in semileptonic $\Lambda_b$
  decays}, Annals Phys. 383 (2017) 278--284.
\newblock \href {http://arxiv.org/abs/1706.02460} {\path{arXiv:1706.02460}},
  \href {http://dx.doi.org/10.1016/j.aop.2017.06.001}
  {\path{doi:10.1016/j.aop.2017.06.001}}.

\bibitem{Pilloni:2018kwm}
A.~Pilloni, J.~Nys, M.~Mikhasenko, M.~Albaladejo, C.~Fern¨¢ndez-Ram{\'i}rez,
  A.~Jackura, V.~Mathieu, N.~Sherrill, T.~Skwarnicki, A.~P. Szczepaniak, {What
  is the right formalism to search for resonances? II. The pentaquark chain},
  Eur. Phys. J. C78~(9) (2018) 727.
\newblock \href {http://arxiv.org/abs/1805.02113} {\path{arXiv:1805.02113}},
  \href {http://dx.doi.org/10.1140/epjc/s10052-018-6177-2}
  {\path{doi:10.1140/epjc/s10052-018-6177-2}}.

\bibitem{Paryev:2018fyv}
E.~{\relax Ya}. Paryev, {\relax Yu}.~T. Kiselev, {The role of hidden-charm
  pentaquark resonance $P^+_c$(4450) , in $J/\psi$ photoproduction on nuclei
  near threshold}, Nucl. Phys. A978 (2018) 201--213.
\newblock \href {http://arxiv.org/abs/1810.01715} {\path{arXiv:1810.01715}},
  \href {http://dx.doi.org/10.1016/j.nuclphysa.2018.08.009}
  {\path{doi:10.1016/j.nuclphysa.2018.08.009}}.

\bibitem{Voloshin:2019wxx}
M.~B. Voloshin, {Hidden-charm pentaquark formation in antiproton - deuterium
  collisions}\href {http://arxiv.org/abs/1903.04422} {\path{arXiv:1903.04422}}.

\bibitem{Lu:2016nnt}
Q.-F. Lu, Y.-B. Dong, {Strong decay mode $J/\psi p$ of hidden charm pentaquark
  states $P_c^+(4380)$ and $P_c^+(4450)$ in $\Sigma_c \bar{D}^*$ molecular
  scenario}, Phys. Rev. D93~(7) (2016) 074020.
\newblock \href {http://arxiv.org/abs/1603.00559} {\path{arXiv:1603.00559}},
  \href {http://dx.doi.org/10.1103/PhysRevD.93.074020}
  {\path{doi:10.1103/PhysRevD.93.074020}}.

\bibitem{Shen:2016tzq}
C.-W. Shen, F.-K. Guo, J.-J. Xie, B.-S. Zou, {Disentangling the hadronic
  molecule nature of the $P_c(4380)$ pentaquark-like structure}, Nucl. Phys.
  A954 (2016) 393--405.
\newblock \href {http://arxiv.org/abs/1603.04672} {\path{arXiv:1603.04672}},
  \href {http://dx.doi.org/10.1016/j.nuclphysa.2016.04.034}
  {\path{doi:10.1016/j.nuclphysa.2016.04.034}}.

\bibitem{Lin:2017mtz}
Y.-H. Lin, C.-W. Shen, F.-K. Guo, B.-S. Zou, {Decay behaviors of the $P_c$
  hadronic molecules}, Phys. Rev. D95~(11) (2017) 114017.
\newblock \href {http://arxiv.org/abs/1703.01045} {\path{arXiv:1703.01045}},
  \href {http://dx.doi.org/10.1103/PhysRevD.95.114017}
  {\path{doi:10.1103/PhysRevD.95.114017}}.

\bibitem{Lin:2018kcc}
Y.-H. Lin, C.-W. Shen, B.-S. Zou, {Decay behavior of the strange and beauty
  partners of $P_c$ hadronic molecules}, Nucl. Phys. A980 (2018) 21--31.
\newblock \href {http://arxiv.org/abs/1805.06843} {\path{arXiv:1805.06843}},
  \href {http://dx.doi.org/10.1016/j.nuclphysa.2018.10.001}
  {\path{doi:10.1016/j.nuclphysa.2018.10.001}}.

\bibitem{Eides:2018lqg}
M.~I. Eides, V.~{\relax Yu}. Petrov, {Decays of pentaquarks in hadrocharmonium
  and molecular scenarios}, Phys. Rev. D98~(11) (2018) 114037.
\newblock \href {http://arxiv.org/abs/1811.01691} {\path{arXiv:1811.01691}},
  \href {http://dx.doi.org/10.1103/PhysRevD.98.114037}
  {\path{doi:10.1103/PhysRevD.98.114037}}.

\bibitem{Jin:2016cpv}
Y.~Jin, S.-Y. Li, S.-Q. Li, {New $B_{s}^0\pi^\pm$ and $D_s^\pm \pi^\pm$ states
  in high energy multiproduction process}, Phys. Rev. D94~(1) (2016) 014023.
\newblock \href {http://arxiv.org/abs/1603.03250} {\path{arXiv:1603.03250}},
  \href {http://dx.doi.org/10.1103/PhysRevD.94.014023}
  {\path{doi:10.1103/PhysRevD.94.014023}}.

\bibitem{Cerri:2018ypt}
A.~Cerri, et~al., {Opportunities in Flavour Physics at the HL-LHC and
  HE-LHC}\href {http://arxiv.org/abs/1812.07638} {\path{arXiv:1812.07638}}.

\bibitem{Ablikim:2019faj}
M.~Ablikim, et~al., {Observation of $e^{+}e^{-}\rightarrow
  \pi^{+}\pi^{-}\psi(3770)$ and $D_{1}(2420)^{0}\bar{D}^{0} + c.c.$}\href
  {http://arxiv.org/abs/1903.08126} {\path{arXiv:1903.08126}}.

\bibitem{Ablikim:2019apl}
M.~Ablikim, et~al., {Cross section measurements of $e^+ e^-\to\omega\chi_{c0}$
  from $\sqrt{s}=$4.178 to 4.278 GeV}\href {http://arxiv.org/abs/1903.02359}
  {\path{arXiv:1903.02359}}.

\bibitem{Aaij:2019kcg}
R.~Aaij, et~al., {Observation of $CP$ violation in charm decays}\href
  {http://arxiv.org/abs/1903.08726} {\path{arXiv:1903.08726}}.

\bibitem{Kou:2018nap}
W.~Altmannshofer, et~al., {The Belle II Physics Book}\href
  {http://arxiv.org/abs/1808.10567} {\path{arXiv:1808.10567}}.

\end{thebibliography}
